\documentclass[structabstract]{aa}

\usepackage{graphicx}
\usepackage{amsmath}
\usepackage{txfonts, longtable, dcolumn, ctable, verbatim, lscape, multicol, url}
\usepackage[T1]{fontenc}
\usepackage{natbib}
\bibpunct{(}{)}{;}{a}{}{,}

\usepackage{graphicx,subcaption,caption}
\bibpunct{(}{)}{;}{a}{}{,}

\begin{document}

   \title{Volumes and bulk densities of forty asteroids from ADAM shape modeling}

   \author{J. Hanu{\v s}
	  \inst{1,2,3*}
           \and
          M.~Viikinkoski\inst{4}
           \and
	  F.~Marchis\inst{5}
           \and           
          J. {\v D}urech\inst{3}
           \and
          M.~Kaasalainen\inst{4}
           \and
          M.~Delbo'\inst{2}
           \and
          D.~Herald\inst{6}
           \and
          E.~Frappa\inst{7}
           \and
          T.~Hayamizu\inst{8}
           \and
          S.~Kerr\inst{9}
           \and
          S.~Preston\inst{9}
           \and
          B.~Timerson\inst{9}
	   \and
	  D.~Dunham\inst{9}
	   \and
          J.~Talbot\inst{10}
          }

   \institute{
	     Centre National d'\'Etudes Spatiales, 2 place Maurice Quentin, 75039 Paris cedex 01, France\\
	     $^*$\email{hanus.home@gmail.com}
	 \and
	     Universit\' e C\^ ote d'Azur, OCA, CNRS, Lagrange, France
	 \and
	     Astronomical Institute, Faculty of Mathematics and Physics, Charles University, V~Hole{\v s}ovi{\v c}k{\'a}ch 2, 18000 Prague, Czech Republic
         \and
	     Department of Mathematics, Tampere University of Technology, PO Box 553, 33101, Tampere, Finland
	 \and
	     SETI Institute, Carl Sagan Center, 189 Bernado Avenue, Mountain View CA 94043, USA
	 \and
	     RASNZ Occultation Section, 3 Lupin Pl., Murrumbateman, NSW 2582, Australia 
	 \and
	     Euraster, 8 route de Soulomes, 46240 Labastide-Murat, France 
	 \and
	     JOIN / Japan Occultation Infomation Network 
	 \and
	     International Occultation Timing Association (IOTA) 
	 \and
	     RASNZ Occultation Section, 3 Hughes Street, Waikanae Beach, Kapiti Coast, 5036, New Zealand 
}

   \date{Received x-x-2016 / Accepted x-x-2016}
 
  \abstract
   {Disk-integrated photometric data of asteroids do not contain accurate information on shape details or size scale. Additional data such as disk-resolved images or stellar occultation measurements further constrain asteroid shapes and allow size estimates.}
   {We aim to use all available disk-resolved images of about forty asteroids obtained by the Near-InfraRed Camera (Nirc2) mounted on the W.M. Keck II telescope together with the disk-integrated photometry and stellar occultation measurements to determine their volumes. We can then use the volume, in combination with the known mass, to derive the bulk density.}
   {We download and process all asteroid disk-resolved images obtained by the Nirc2 that are available in the Keck Observatory Archive (KOA). We combine optical disk-integrated data and stellar occultation profiles with the disk-resolved images and use the All-Data Asteroid Modeling (ADAM) algorithm for the shape and size modeling. Our approach provides constraints on the expected uncertainty in the volume and size as well. }
   {We present shape models and volume for 41 asteroids. For 35 asteroids, the knowledge of their mass estimates from the literature allowed us to derive their bulk densities. We clearly see a trend of lower bulk densities for primitive objects (C-complex) than for S-complex asteroids. The range of densities in the X-complex is large, suggesting various compositions. Moreover, we identified a few objects with rather peculiar bulk densities, which is likely a hint of their poor mass estimates. Asteroid masses determined from the Gaia astrometric observations should further refine most of the density estimates. }
   {}
 

  \titlerunning{Bulk densities of asteroids based on ADAM}
  \maketitle

\section{Introduction}\label{sec:introduction}


Density and internal structure belong to the most important characteristics of asteroids, which are also some of the least constrained. Moreover, when compared with the densities of meteorites one can deduce the nature of asteroid interiors. These physical properties of asteroids reflect the accretional and collisional environment of the early solar system. On top of that, because some asteroids are analogs to the building blocks that formed the terrestrial planets 4.56 Gyr ago, the density and internal structures of minor bodies inform us about the formation conditions and evolution processes of planets and the solar system as a whole. To determine the density directly, we need both the mass and the volume of the object. The current density estimates are mostly governed by the knowledge of these two properties. On the other hand, indirect density measurements based on photometric observations of mutual eclipses of small binary near-Earth asteroids (NEAs) \citep[e.g.,][]{Scheirich2009} do not require the mass nor the size. However, the achieved accuracy of such density estimates is usually much lower when compared with the direct measurements. Additionally, the typical objects size ranges of these methods are different as well. 

The majority of reported mass estimates are based on orbit deflections during close encounters \citep[e.g.,][]{Michalak2000, Michalak2001, Pitjeva2001, Konopliv2006, Mouret2009, Zielenbach2011} and planetary ephemeris \citep[e.g.,][]{Baer2008, Baer2011, Fienga2008, Fienga2009, Fienga2011, Fienga2014, Folkner2009}. These methods give accurate masses for the largest asteroids (within a few percent), but the accuracy gets worse very quickly with decreasing size/mass of the objects. The astrometric observations of the ESA's Gaia satellite promise a significant improvement of the poor knowledge of the mass. More specifically, Gaia will constrain masses for $\sim$150 asteroids \citep[$\sim$50 with an accuracy below 10\%,][]{Mouret2007, Mouret2008} by the orbit deflection method. The advantage of Gaia masses is in the uniqueness of the mission, which should result in a comprehensive sample with well described biases (e.g., the current mass estimates are strongly biased towards the inner main belt). The list of asteroids, for which the masses will be most likely determined, is already known. \citet{Carry2012b} analyzed available mass estimates for $\sim$250 asteroids and concluded that only about a half of them have a precision better than 20\%, although some values might be still affected by systematic errors. The second most accurate mass determinations so far (after those determined by the spacecraft tracking method) are based on the study of multiple systems \citep[e.g.,][]{Marchis2008, Marchis2008a, Marchis2013a, Fang2012} and reach a typical uncertainty of 10--15\%. Masses based on planetary ephemeris can be often inconsistent with those derived from the satellite orbits, which is the indication that masses from planetary ephemeris should be treated with caution.

Determining the volume to a similar uncertainty level as the mass ($<$20\%) is very challenging. The density is proportional to the mass and inversely proportional to the cube of the asteroid size, so one needs a relative size uncertainty three times smaller than of the mass estimate to contribute with the same relative uncertainty to the density uncertainty as the mass. The most frequent method for the size determination is the fitting of the thermal infrared observations (usually from IRAS, WISE, Spitzer or AKARI satellites) by simple thermal models such as the Standard Thermal Model \citep[STM,][]{Lebofsky1986} or the Near-Earth Asteroid Thermal Model \citep[NEATM,][]{Harris1998} assuming a spherical shape model. This size is often called the radiometric diameter and it corresponds to the surface-equivalent diameter\footnote{Surface-equivalent diameter is a diameter of a sphere that has the same surface as the surface of the shape model.}. Because thermal models usually assume a spherical shape model, the surface equivalent diameter equals the volume-equivalent diameter\footnote{Volume-equivalent diameter is a diameter of a sphere that has the same volume as the volume of the shape model.}. Reported size uncertainties for individual asteroids are usually very small \citep[within a few percent,][]{Masiero2011}, however, they are not realistic \citep{Usui2014}. Indeed, the uncertainties are dominated by the model systematics -- the spherical shape assumption is too crude and also the role of the geometry is neglected (e.g., the spin axis orientation). In the statistical sense, the sizes determined by thermal models are reliable, but could be easily off for individual objects by 10--30\% \citep{Usui2014}. This implies a density uncertainty of 30--90\%, respectively. Other size determination methods that assume a sphere or a triaxial ellipsoid for the shape model suffer by the same model systematics. Obviously, more complex shape models have to be used for the more accurate size determinations.

Several methods for reliable size determination that require lightcurve- or radar-based shape models have been already employed (the only few exceptions are the largest asteroids that can be approximated by simple  rotational ellipsoids):  
(i)~scaling the asteroid shape projections by disk-resolved images observed by the 8-10m class telescopes equipped with adaptive optics systems \citep[e.g.,][]{Marchis2006,Drummond2009a,Hanus2013b}; 
(ii)~scaling the asteroid shape projections to fit the stellar occultation measurements \citep[e.g.,][]{Durech2011}; or
(iii)~analyizing thermal infrared measurements by the means of a thermophysical modeling which allows to scale the shape from radar or lightcurve inversion to match the size information carried by the infra red radiation \citep[e.g.,][]{Muller2013a,AliLagoa2014,Rozitis2014,Emery2014,Hanus2015a,Hanus2016b}. 
The lightcurve-based shape models are usually best described as convex \citep[e.g.,][]{Kaasalainen2002c,Torppa2003,Durech2009,Hanus2011,Hanus2016a}, the radar models are reconstructed from delay-Doppler echoes, sometimes in combination with light curve data \citep[e.g.,][]{Hudson1999,Busch2011}. Size uncertainties achieved by these methods are usually below 10\%.

Recently, models combining both disk-integrated and disk-resolved data were developed \citep[e.g., KOALA and ADAM models,][]{Carry2012b,Viikinkoski2015}. With those, both shape and size are optimized \citep{Merline2013,Berthier2014,Viikinkoski2015b,Hanus2016c}. For instance, a KOALA-based shape model of asteroid (21)~Lutetia was derived from optical light curves and disk-resolved images by \citet{Carry2010b}. Moreover, this result was later confirmed by the ground-truth shape model reconstructed from images obtained by the camera on board the Rosetta space mission during its close fly-by \citep{Sierks2011}, which effectively validated the KOALA shape modeling approach \citep{Carry2012b}.

In our work, we use the ADAM algorithm for asteroid shape modeling from the disk-integrated and disk-resolved data, and stellar occultation measurements. We describe the optical data in Sec.~\ref{sec:photometry}, the disk-resolved data from the Keck II telescope in Sec.~\ref{sec:keckAO}, and the occultation measurements in Sec.~\ref{sec:occ}. The ADAM shape modeling algorithm is presented in Sec.~\ref{sec:ADAM}. We show and discuss derived shape models and corresponding volume-equivalent sizes and bulk densities in Sec.~\ref{sec:results}. Finally, we conclude our work in Sec.~\ref{sec:conclusions}.

\section{Data}\label{sec:data}

\subsection{Shape models from disk-integrated photometry}\label{sec:photometry}

In this work, we mostly focused on asteroids for which their rotation states and shape models were already derived or revised recently. We used rotation state parameters of these asteroids as initial inputs for the shape and size optimization by ADAM. 
The majority of previously published shape models, spin states and optical data are available in the Database of Asteroid Models from Inversion Techniques \citep[DAMIT\footnote{\url{http://astro.troja.mff.cuni.cz/projects/asteroids3D}},][]{Durech2010}, from where we adopted the disk-integrated light curve datasets as well. Moreover, we list adopted rotation state parameters and references to the publications in Tab.~\ref{tab:results}.

\subsection{Keck disk-resolved data}\label{sec:keckAO}

The W.M. Keck II telescope located at Maunakea in Hawaii is equipped since 2000 with an adaptive optics (AO) system and the Near-InfraRed Camera (Nirc2). This AO system provides an angular resolution close to the diffraction limit of the telescope at $\sim$2.2 $\mu$m, so $\sim$45 mas for bright targets (V$<$13.5) \citep{Wizinowich2000}. The AO system was improved several times since it was mounted. For example, the correction quality of the system was improved in 2007 \citep{vanDam2004}, resulting into reaching an angular resolution of 33 mas at shorter wavelengths ($\sim$1.6 $\mu$m).

All data obtained by the Nirc2 extending back to 2001 are available at the Keck Observatory Archive (KOA). It is possible to download the raw images with all necessary calibration and reduction files, and often also images on which basic reduction was performed. We downloaded and processed all disk-resolved images of all observed asteroids. Usually, several frames were obtained by shift-adding 3--30 frames with an exposure time of fractions of seconds to several seconds depending on the asteroid's brightness at the particular epoch. We performed the flat-field correction and we used a bad-pixel suppressing algorithm to improve the quality of the images before shift-adding them. Then, we visually checked all images and selected only those where the asteroids were resolved. Typically, we considered an asteroid as resolved if its maximum size on the image was at least $\sim$10 pixels. Also, we rejected fuzzy and saturated images, and images with various artifacts. We obtained about 500 individual images of about 80 asteroids. Finally, we deconvolved each resolved image by the AIDA algorithm \citep{Hom2007} to improve its sharpness.

Many images were already used independently in previous shape studies \citep{Marchis2006,Drummond2009a,Descamps2009,Merline2013,Hanus2013b,Berthier2014}. In Tab.~\ref{tab:ao}, we list all used disk-resolved images for each studied asteroid and by courtesy the name of the principal investigator of the scientific project within which the data were obtained. 

\subsection{Occultation data}\label{sec:occ}

Stellar occultations are publicly available in the OCCULT software\footnote{\url{http://www.lunar-occultations.com/iota/occult4.htm}} maintained by David Herald. In Tab.~\ref{tab:occ}, we list all observers that participated in each stellar occultation measurement we used for the shape modeling. To achieve a better convergence of the shape modeling, we visually examined each occultation measurement and removed chords with large uncertainties in their timings (mostly visual observations) and chords that were clearly inconsistent with the remaining ones (mostly due to the incorrect timing). The chord removal was a rather safe procedure, because the offset of the incorrect chord with respect to several close-by chords was always obvious. Moreover, such cases were quite rare. We also rejected occultation events with less then three reliable chords. 

\subsection{Asteroid masses}\label{sec:mass}

The most accurate mass estimates are based on space probe fly-by measurements or the satellite's orbits in the multiple systems. We adopted these estimates from the corresponding studies. Densities based on these masses should be the most reliable ones.

Masses derived from astrometric observations (close encounters or planetary ephemeris methods) are available for most asteroids in our sample. Moreover, multiple determinations for individual asteroids are common. However, these determinations are often inconsistent or result in an unrealistic density determination. To select the most reliable mass estimates, we decided to use values from the work of \citet{Carry2012b}, who investigated available mass estimates for $\sim$250 asteroids and presents a single value for each of them. The author also provides bulk density estimates and ranks their quality. A low rank is usually a hint that the mass estimate is not reliable. Recently, \citet{Fienga2014} computed masses for tens of asteroids from INPOP planetary ephemerides. However, several masses of multiple asteroids are inconsistent with masses from \citet{Carry2012b}. It is not obvious, which values should be the better ones. For example, masses for the (45)~Eugenia and (107)~Camilla multiple systems are clearly wrong in \citet{Fienga2014}, because their reliable mass estimates based on the satellite's orbits are too different. On the other hand, the mass of the (41)~Daphne system is consistent. Moreover, masses for several asteroids from \citet{Fienga2014} lead to more realistic bulk densities than those from \citet{Carry2012b}. So, we decided to use masses from \citet{Fienga2014} only in cases where the density would be unrealistic otherwise. All these cases are individually commented in Sec.~\ref{sec:single}. Additionally, we also comment the cases where the masses are inconsistent within each other. 

Masses based on astrometric observations of the ESA's Gaia satellite should be available in 2019. After that, our volume estimates of several asteroids studied here could be used for future bulk density refinements.

\section{All-Data Asteroid Modeling (ADAM) algorithm}\label{sec:ADAM}

Reconstructing a 3-D shape model of an asteroid from various observations is a typical ill-posed problem, since noise-corrupted observations contain only low-frequency information. To mitigate effects of ill-posedness, we use parametric shape representations combined with several regularization methods.

While the reconstruction can be made well-behaved in the sense that the optimization process converges to a shape model, there is also the problem of uniqueness: Often it is not obvious whether features present in the shape model are supported by data or if they are artifacts caused by the parameterization and regularization methods. The chance for these spurious features can be alleviated by the use of several different parameterizations and regularization methods: It is conceivable that all the representations should produce similar shapes if the solution is well constrained by the data. Therefore, in this article, we derive shape models for asteroids using two different parametric shape representations -- subdivision surfaces and octanoids \citep[see,][]{Viikinkoski2015}. If the resulting shape models for the asteroid are significantly different, we conclude that the available data are not sufficient for reliable reconstruction and discard the model.

The procedure used in this article for shape reconstruction is called ADAM \citep{Viikinkoski2015}. It is an universal inversion technique for various disk-resolved data types. ADAM facilitates the usage of adaptive optics images directly, without requiring deconvolution or boundary extraction. The software used in this article is freely available on the web\footnote{\url{https://github.com/matvii/adam}}.

Utilizing the Levenberg-Marquardt optimization algorithm, ADAM minimizes an objective function

\[\chi^2:=\chi^2_{LC}+\lambda_{AO}\chi^2_{AO}+\lambda_{OC}\chi^2_{OC}+\sum_{i}\lambda_i\gamma_i^2,\]
where terms $\chi^2_{LC}$, $\chi^2_{AO}$ and $\chi^2_{OC}$ are, respectively, model fit to light curves, adaptive optics images, and stellar occultation chords. The last sum corresponds to regularization functions measuring the smoothness and complexity of the mesh.  

The formulation of terms $\chi^2_{LC}$ and $\chi^2_{AO}$ is covered in \citep{Viikinkoski2015}, and the theoretical foundations of stellar occultations relating to the shape reconstruction of asteroids are well established in \cite{Durech2011}, so we describe here how the goodness-of-fit measure $\chi^2_{OC}$ for occultation chords is implemented in ADAM.

As an asteroid occults a star, its shadow travels on the surface of the Earth. The positions of the observers, together with the disappearance and reappearance times of the star, determine a chord on the fundamental plane, which is the plane perpendicular to the line determined by the asteroid and the star.
Given the fundamental plane determined by the occultation, we project the shape model represented by a triangular mesh $\mathcal{M}$ into the plane by using an orthogonal projection $P$. To form the goodness-of-fit measure $\chi^2_{OC}$, we must first define a reasonable distance function $d(\mathcal{C},P\mathcal{M})$.

Let $\mathcal{C}$ be the occultation chord with the endpoints $p_1$ and $p_2$ on the plane, and $\mathcal{L}$ the line determined by the chord. We consider the case where the line $\mathcal{L}$ intersects the boundary of the projected shape model $P\mathcal{M}$ at two points $q_1$ and $q_2$. Assuming the points are ordered so that the vectors $p_1-p_2$ and $q_1-q_2$ are parallel, we set

\[d(\mathcal{C},P\mathcal{M})=\|q_1-p_1\|^2+\|q_2-p_2\|^2.\]

If the line does not intersect the projected shape, let $\delta$ be the perpendicular distance from the line $\mathcal{L}$ to the closest vertex in $P\mathcal{M}$. We define

\[d(\mathcal{C},P\mathcal{M})=2\|p_2-p_1\|^2L(\delta),\]
where \[L(x)=\frac{1}{1+e^{-kx}}\] is the logistic function with the parameter $k$.

For the negative chords (i.e. chords along of which no occultation is observed), we use a slightly different approach. We set \[d(\mathcal{C},P\mathcal{M})=\gamma\cdot (1-L(\delta)),\] 
where $\gamma$ is a constant weight, and $\delta$ is defined as follows: If the chord $\mathcal{C}$ intersects $P\mathcal{M}$, let $\delta_1$ be the distance to the farthest vertex on the positive side of the line $\mathcal{L}$, and similarly let $\delta_2$ be the distance to the farthest vertex on the negative side of the line. We set \[\delta=-\min\{\delta_1,\delta_2\}.\]

If the chord does not intersect $P\mathcal{M}$, let $\delta$ be the perpendicular distance from the line to the closest vertex on $P\mathcal{M}$. 

The idea here is that if the negative chord intersects the projected shape, the distance function attains its maximum value $\gamma$. The weight $\gamma$ is chosen large enough to ensure that an optimization step causing an intersection is rejected. The logistic function is used instead of the step function to make the distance function differentiable.

Given an occultation event consisting of $n$ chords $\mathcal{C}_i$, we define

 \[\chi^2_{OC}:=\sum_i d(\mathcal{C}_i,P\mathcal{M}+(O_x,O_y)),\]
where $(O_x,O_y)$ is the offset from the projection origin, to be determined during the optimization.

\section{Results and discussions}\label{sec:results}

Here we present shape models of asteroids based on the ADAM shape modeling algorithm. All derived shape models as well as all their optical disk-integrated and disk-resolved data, and occultation measurements are available on-line in the DAMIT database. Our observation datasets always contain all three types of these data. The uncertainties in the spin vector determinations were estimated from the differences between the solutions based on the usage of the two different shape supports in ADAM (i.e, subdivission surfaces and octanoids, see Section~\ref{sec:ADAM}), and the usage of raw and deconvolved versions of the disk-resolved data. These uncertainties are usually 2--5 degrees.  

To estimate the size, first, we computed the volume from the scaled shape model and estimated its uncertainty from the differences between the solutions based on the usage of different shape supports in ADAM the same way as for the pole solution. Then, we computed the corresponding volume-equivalent diameter and its possible range from the volume. We only report the volume-equivalent size with its 1$\sigma$ uncertainty in Tab.~\ref{tab:densities}, however, the volume can be easily accessed based on this size.


The bulk density in Tab.~\ref{tab:densities} is then the ratio between the mass and the volume, and its uncertainty was computed from the propagation of the volume and mass uncertainties.

\subsection{Shape models of primaries in multiple systems}\label{sec:primaries}

Several main-belt binaries that consist of a large primary ($\gtrsim$100 km) and a few-kilometer sized secondary (or even 2 satellites) were discovered in the images obtained by the 8-10m class telescopes equipped with AO systems during the last decade. Usually, tens of large asteroids were surveyed during a few-year campaign and their close proximity was searched for a potential presence of a satellite. Once a satellite was detected, the system was then imaged in other epochs, so the satellite's orbit could be constrained. Fortunately, some of the primaries were large enough to be resolved, often during multiple distant epochs (apparitions). On the other hand, single objects were usually observed only once or twice, so mostly single- or double-apparition observations are available for them.

\subsubsection{Comparison with previously modeled primaries}

Asteroids with multiple disk-resolved images were natural candidates for shape modeling, so several shape models have already been published for those. All these shape models are based on methods that use the 2D contours extracted from the disk-resolved images. Such approach is sensitive to the boundary condition applied when extracting the contour. Shape models of asteroids (22)~Kalliope, (87)~Sylvia, (93)~Minerva, and (216)~Kleopatra have been previously derived \citep{Descamps2008,Berthier2014,Marchis2013a,Kaasalainen2012}. As the first step, we decided to validate our modeling approach on these asteroids -- we have similar or even larger optical, disk-resolved and stellar occultation datasets for them. We present shape models for all four asteroids and reproduce well the previous results. 

Before we start listing the asteroids one by one, we would like to note that there are plenty of previous shape and spin pole studies for each asteroid in our work, including single-epoch methods assuming triaxial ellipsoids as well as a more general modeling by the lightcurve inversion method. Below, we mostly comment on the shape modeling results based on the lightcurve inversion of optical photometry and neglect most other studies for the sake of simplicity. Moreover, mostly spin states based on lightcurve inversion were used as necessary initial inputs for the shape modeling by ADAM, because the one-apparition ellipsoidal shape and spin solutions lack the necessary precision in the sidereal rotation period. 

\paragraph{22 Kalliope} 
Reliable size and bulk density of Kalliope have already been derived from observations of mutual events by \citet{Descamps2008}. Our shape model and size based on 102 light curves, 23 disk-resolved images and one occultation is in an agreement with the previous results (e.g., the difference in the pole orientation is only one degree). Our size (161$\pm$6~km) is slightly smaller (compared to 166$\pm$3~km), but still within the small uncertainties. So, we derived a rather similar bulk density of (3.7$\pm$0.4)~g\,cm$^{-3}$ that is consistent with the M-type taxonomic classification of Kalliope.

\paragraph{87 Sylvia} 
As Sylvia is a multiple system, a large number of 22 disk-resolved images could be used for the shape modeling. Together with 55 optical light curves and 2 occultation measurements, the ADAM modeling resulted in a reliable shape model and size that is in a perfect agreement with an independent shape model derived by the KOALA algorithm from a similar dataset by \citet{Berthier2014}. The size of (273$\pm$5)~km combined with the mass estimate gave us a bulk density of (1.39$\pm$0.08)~g\,cm$^{-3}$. Sylvia is the only P-type asteroid in our sample and its bulk density is one of the most precise so far estimated for an asteroid. Most of the C-complex asteroids have similar bulk densities as Sylvia.

\paragraph{93 Minerva} 
Size and bulk density of the C-type asteroid Minerva based on optical light curves, disk-resolved images and occultation data have already been determined by \citet{Marchis2013a}. We used a similar optical dataset and a subset of disk-resolved data and derived a shape model and size (159$\pm$3~km) that are consistent with those of \citet{Marchis2013a} (154$\pm$6~km). The difference in the pole orientation is only four degrees. Our size estimate is slightly higher, which resulted in a smaller, however, more consistent with respect to its taxonomic type (i.e., C-complex), bulk density of (1.59$\pm$0.27)~g\,cm$^{-3}$. 

\paragraph{216 Kleopatra}
So far, the disk-resolved images of Kleopatra have not been sufficient for a proper shape model \citet{Kaasalainen2012}. Using all available data, we could derive a model with ADAM. Only one pole solution of ($\lambda$,$\beta$)$\sim$(73, 21)$^{\circ}$ is consistent with the AO data. Our ADAM model is based on 55 optical light curves, 14 AO images and three occultations. All occultations consist of multiple chords that sample most of the shape projection. The issue with the shape model is that there are no AO nor occultation data that were obtained at a view close to the pole. Closest is the 2016 occultation, $\sim$70 degrees above equator. Similarly there is one AO image 70 degrees below the equator, but it is too fuzzy to be useful. We obtained a pole solution of $\sim$(74, 20)$^{\circ}$, that is in a perfect agreement with the one of \citet{Kaasalainen2012}. Our size (121$\pm$5)~km and adopted mass of \citet{Descamps2011} lead to a bulk density of (5.0$\pm$0.7)~g\,cm$^{-3}$, an unusually large value within the M-type asteroids, however, with a larger uncertainty. Additionally, \citet{Ostro2000} derived a shape model of Kleopatra from the delay-Doppler observations obtained by Arecibo. The spin state is similar to ours, however, the shape models differ. The one of \citet{Ostro2000} has a dumbbell appearance with a handle that is substantially narrower than the two lobes. In our shape model, the handle is of about the same thickness as the lobes.

\subsubsection{New shape models of primaries}\label{sec:primaries2}

\paragraph{41 Daphne}
All recent shape model studies of the C/Ch-type asteroid Daphne reported a consistent pole solution of $\sim$(200, $-30$)$^{\circ}$ \citep{Kaasalainen2002c, Durech2011, Matter2011, Hanus2016a}, which we used as an initial input for the shape modeling with ADAM. Because Daphne is a binary asteroid \citep{Conrad2008}, the number of disk-resolved images is rather high due to the attraction to the satellite's position \citep{Conrad2008b}. We also have occultation observations from two distant epochs. Our shape model fits all the AO, light curve and occultation data well, so the size is reliably constrained to be (188$\pm$5)~km. Moreover, this size is compatible with the size estimate of \citet{Matter2011} based on interpretation of interferometric data by a thermophysical model and the use of a shape model with local topography. Our size and the precise mass estimate from \citet{Carry2012b} lead to a bulk density of (1.81$\pm$0.15)~g\,cm$^{-3}$, which belongs to higher values within the C-complex asteroids.

\paragraph{45 Eugenia} 
The small moonlet 'Petit-Prince' of Eugenia was discovered by \citet{Merline1999}, which made Eugenia a target for several AO campaigns studying orbit of the moon \citep[e.g.,][]{Marchis2010}. As a consequence, 23 disk-resolved images were obtained by the Nirc2 during six different apparitions. \citet{Hanus2013b} rejected one of the mirror solutions and our shape modeling with ADAM confirmed that conclusion. The shape model fits nicely all the disk-resolved images as well as two occultation measurements, which lead to a precise size estimate of (186$\pm$4)~km. The corresponding density of (1.69$\pm$0.11)~g\,cm$^{-3}$ is consistent with typical densities of C-type asteroids. The reliable mass estimate of (5.69$\pm$0.12).10$^{18}$~kg is based on the moon's orbit \citep{Marchis2008}.

\paragraph{107 Camilla} 
The single pole solution of $\sim$(72, 51)$^{\circ}$ \citep{Torppa2003, Durech2011, Hanus2013b, Hanus2016a} is well established and we used it as an initial input for the shape modeling by ADAM. Camilla is another binary (actually triple) asteroid that was often observed by the NICR2 at Keck. So, we gathered 21 disk-resolved images obtained at 7 different apparitions. As a result, our shape and size solution that explains all the observations is well constrained. The size of (254$\pm$6)~km combined with the mass from \citet{Marchis2008} resulted in a typical C-type asteroid bulk density of (1.31$\pm$0.10)~g\,cm$^{-3}$.

\subsection{Shape models of single asteroids}\label{sec:single}

Only few single asteroids were observed during multiple epochs by the Keck AO system, namely mostly the largest ones (e.g., 2 Pallas, 52 Europa) and the space mission targets (e.g., 1 Ceres, 4 Vesta, 21 Lutetia), because most AO surveys at the Keck telescope were dedicated to the discovery of satellites and then the system follow-up. Shape models based on disk-resolved data were previously independently derived for asteroids (2)~Pallas, (16)~Psyche and (52)~Europa \citep{Carry2010a,Shepard2017,Merline2013}, so we provide our ADAM solutions for comparison and as a reliability test. For the remaining asteroids, we present their first shape model solutions from disk-resolved data and sometimes their first non-radiometric size estimates. For the majority of the asteroids studied, we used adopted mass estimates and derived their bulk densities.

\paragraph{2 Pallas} 
Our ADAM modeling started with the single pole solution from \citet{Carry2010a} as an initial input and converged to a solution that fitted nicely all optical lightcurves, 18 disk-resolved images and two occultations. Note that the occultation from the year 1983 \citep{Dunham1990} is of an exceptional quality, the projection consists of 131 chords (there are also 117 observations outside the path of Pallas shadow) and sample almost the whole projected disk. Our shape and size of (523$\pm$10)~km is consistent with the solution of \citet{Carry2010a} of (512$\pm$6)~km. Pallas is one of the three B-type asteroids in our sample and has a bulk density of (2.72$\pm$0.17)~g\,cm$^{-3}$. Clearly, such density is exceptionally high among the primitive C-complex asteroids, it is even higher than the bulk density of Ceres. This suggests a different composition of B-type asteroids than of the other C-complex subgroups.

\paragraph{5 Astraea}
We used a spin state derived by \citet{Durech2009} as an initial input for the ADAM shape modeling of Astraea. The pole ambiguity was already removed by \citet{Durech2011} based on stellar occultation measurements, the AO data further support the uniqueness of the pole solution \citep{Hanus2013b}. Our ADAM shape model is consistent with all observed data and has a size of (114$\pm$4)~km. The corresponding bulk density of (3.4$\pm$0.7)~g\,cm$^{-3}$ seems to be realistic for an S-type asteroid.

\paragraph{8 Flora}
Flora is an S-type asteroid and the largest member of the Flora collisional family. The spin state based on a lightcurve inversion was already known from \citet{Torppa2003}, and the pole ambiguity was later removed by \citet{Durech2011}. Our shape model is consistent with this pole solution and fits well all the data we posses. The size of (140$\pm$4)~km does not differ from the previous estimates. The mass estimate of \citet{Carry2012b} is based on multiple determinations that are rather consistent, which suggests that the mass might be reliable. However, the resulting bulk density is suspiciously high for an S-type asteroid -- (6.4$\pm$1.5)~g\,cm$^{-3}$. On the other hand, the recent mass estimate of \citet{Fienga2014} is by about 30\% smaller, which effectively reduced the bulk density by 30\% as well (listed in Tab.~\ref{tab:densities}). Moreover, the most recent INPOP solution (Fienga, private communication) suggests an even smaller mass that results in a bulk density of (3.3$\pm$0.7)~g\,cm$^{-3}$. Such a density is typical within S-type asteroids. Mass determination based on Gaia astrometric measurements should resolve this discrepancy. Although it is unlikely that the high bulk density is realistic, it would suggest that the interior of Flora should be metal rich possibly indicating a differentiated body, maybe a core of the parent body of the Flora family.

\paragraph{10 Hygiea}
Both the mirror pole solutions \citep{Kaasalainen2002c} are consistent with the AO and occultation data. Although Hygiea is one of the largest and most massive asteroid, only two disk-resolved images are available. Despite that, we computed the average volume-equivalent diameter of Hygiea of (411$\pm$20)~km. The density of (2.4$\pm$0.4)~g\,cm$^{-3}$ is rather high considering Hygiea is a C-type asteroid. Additional disk-integrated data should further constrain the size estimate, remove the pole ambiguity, and consequently confirm or refine the bulk density. 

\paragraph{11 Parthenope}
The pole solution of $\sim$(127, 15)$^{\circ}$ fits the disk-resolved data slightly better, so it is preferred. However, the mirror solution still gives a reasonable fit and cannot be fully rejected. A huge number of optical data (138 light curves) on one side, and only one occultation and one AO image on the other are available for Parthenope. We computed a volume-equivalent diameter of (156$\pm$5)~km and a typical S-type bulk density of (3.0$\pm$0.4)~g\,cm$^{-3}$. We provide the first non-radiometric size estimate of Parthenope.

\paragraph{13 Egeria}
Both pole for Egeria from \citet{Hanus2011} agree with the AO data. By fitting optical, AO and occultation data by ADAM, we provide the first non-radiometric size of Egeria -- (205$\pm$6)~km. Unfortunately, the mass estimate of \citet{Carry2012b} is affected by a 50\% uncertainty, so we rather used the mass estimate of \citet{Fienga2014}. The average bulk density is then (2.1$\pm$0.6)~g\,cm$^{-3}$. We expect a slightly lower value for this G- or Ch-type asteroid, however, the uncertainty is rather large.

\paragraph{16 Psyche}
The pole ambiguity was already removed by \citet{Drummond2008} and our ADAM modeling is consistent with this pole solution. A large dataset of 118 light curves, 7 AO images and 2 occultations reliably constrained the shape and size (225$\pm$4~km). The bulk density of (4.6$\pm$1.3)~g\,cm$^{-3}$ based on the mass from \citet{Carry2012b} is a slightly larger value than the typical values for M-type asteroids. However, the mass from \citet{Fienga2014} is lower by almost 10\% and has a significantly smaller uncertainty. Moreover, it provides a more consistent bulk density of (3.7$\pm$0.6)~g\,cm$^{-3}$. A metal-rich composition is often proposed as a reason for such high values, and the high radar albedo further supports this idea. Our solution is in a perfect agreement with the recent shape model of Psyche based on delay-Doppler, optical, AO and occultation data \citep{Shepard2017}.

\paragraph{18 Melpomene}
As an initial input for rotation state parameters for the ADAM modeling of Melpomene's shape we used values from \citet{Hanus2016a}. Our shape and size solution fits all six disk-resolved images and the occultation measurements. We present the first non-radiometric size estimate for Melpomene of (146$\pm$3)~km. The main limitation for the bulk density of (2.0$\pm$0.8)~g\,cm$^{-3}$ comes from the poor mass accuracy. The quoted density seems to be a little small for an S-type asteroid, but we have a large uncertainty.


\paragraph{29 Amphitrite}
All recent shape and spin state studies of the S-type asteroid Amphitrite \citep{Drummond1988b,Drummond1991,DeAngelis1995b,Kaasalainen2002b,Hanus2016a} report only one pole solution with ecliptic coordinates $\sim$(140, $-20$)$^{\circ}$. However, this shape solution is not consistent with the AO images and stellar occultation measurements. On the other hand, the mirror solution with (322, $-29$)$^{\circ}$ fits the AO images and occultation chords nicely, so this is the correct one. The size between different shape supports does not vary much and it is estimated to be (204$\pm$3)~km. Combination with the mass leads to a bulk density of (2.9$\pm$0.5)~g\,cm$^{-3}$, which is consistent with the typical range within S-type asteroids.

\paragraph{39 Laetitia}
The unique pole solution of \citet{Kaasalainen2002b} and \citet{Hanus2016a} has been confirmed by agreeing with the AO and stellar occultation data. We also present a size estimate that is consistent with previous values \citep{Durech2011,Hanus2013b}. The size (164$\pm$3~km) and the adopted mass estimates lead to a rather low density of (2.0$\pm$0.5)~g\,cm$^{-3}$ for this S-type asteroid.

\paragraph{43 Ariadne}
The only lightcurve-based shape model of Ariadne was computed by \citet{Kaasalainen2002b}. Moreover, only one pole solution was consistent with the optical data. Our ADAM shape model fits all the optical light curves, AO and occultation data well. Unfortunately, our size (59$\pm$4~km) and an adopted mass from \citet{Carry2012b} provide an unrealistic bulk density of (11.3$\pm$2.6)~g\,cm$^{-3}$. Most likely, the mass estimate of this S-type asteroid is wrong. 

\paragraph{51 Nemausa}
First lightcurve-based shape model of Nemausa was derived in \citet{Hanus2016a}, from where we adopted the rotation state as an initial input for the shape modeling by ADAM. We successfully removed the pole ambiguity and derived a reliable shape solution that fits nicely all the available data. We present the first non-radiometric size of this Ch-type asteroid -- (144$\pm$3)~km that leads to a bulk density of (1.6$\pm$0.6)~g\,cm$^{-3}$. This value is consistent with those for other C-complex asteroids.

\paragraph{52 Europa}
Our shape and spin state based on 49 optical light curves, 25 AO images and 4 occultation are in a perfect agreement with the solution of \citet{Merline2013} derived by the KOALA algorithm from a similar dataset. So, the resulting bulk density of the C-type asteroid Europa is similar as well -- (1.5$\pm$0.4)~g\,cm$^{-3}$.

\paragraph{54 Alexandra}
The pole ambiguity of the C-type asteroid Alexandra was already removed by \citet{Durech2011} and later confirmed by \citet{Hanus2013b}. Our shape model fits well all the available data and has a size of (143$\pm$5)~km. Unfortunately, the available mass estimate from \citet{Carry2012b} has a large uncertainty, so the corresponding density of (4.0$\pm$2.3)~g\,cm$^{-3}$ is rather meaningless.

\paragraph{80 Sappho}
The pole ambiguity of Sappho was already removed by \citet{Durech2011} and later confirmed by \citet{Hanus2013b}, so we used a single pole solution as an initial input for the ADAM modeling. Unfortunately, we obtained only one low-quality disk-resolved image and one occultation, which allowed us to constrain the size only poorly to (66$\pm$8)~km. More specifically, the sizes based on raw and deconvolved AO data systematically differed by about 10\%. There is no reliable mass estimate for Sappho at the moment.

\paragraph{85 Io}
A single shape and pole solution was reported by \citet{Durech2011} and confirmed by \citet{Hanus2013b}. Our ADAM model with the size of (165$\pm$3)~km fits nicely both our two AO images and stellar occultation profiles from three epochs. Due to a poor mass estimate, we cannot draw reliable conclusions from the bulk density of (1.1$\pm$0.6)~g\,cm$^{-3}$ other than that a low density is expected for a primitive body such as Io (B-type). On the other hand, this low bulk density differs from the one of another B-type asteroid Pallas, although this could be due to the size differences. 

\paragraph{88 Thisbe}
We used a single pole solution from \citet{Hanus2016a} as an initial input for the shape modeling by ADAM, because the pole ambiguity has been already removed in \citet{Hanus2013b}. Disk-resolved data and one occultation measurement were sufficient to reliably constrain the size to (212$\pm$10)~km, which resulted in a bulk density of (3.1$\pm$0.8)~g\,cm$^{-3}$. Such a value is rather high for a primitive C-complex body, however, Thisbe is a large B-type asteroid and its bulk density is comparable to the one of another B-type asteroid Pallas. On top of that, the uncertainty in the bulk density is rather large.

\paragraph{89 Julia}
We used a single pole solution from \citet{Durech2011} as an initial input for the shape modeling by ADAM. Although the occultation and AO data are rather poor, our shape model seems to be reliable and fits all the available data. The size of (142$\pm$4)~km with the mass from \citet{Carry2012b} provide a bulk density of (4.5$\pm$1.3)~g\,cm$^{-3}$. Julia is classified as a K-type by \citet{DeMeo2009} and belongs to the S-complex. We expect a lower bulk density value, although the uncertainty is rather high. Moreover, \citet{Fienga2014} reports a significantly smaller mass, however, with a more than 50\% uncertainty. Nevertheless, this mass would place the density to ($\sim$1.5--2~g\,cm$^{-3}$, which is a rather low value. Unfortunately, it seems that the mass estimates for Julia are poor, so the bulk density cannot be reliably estimated.

\paragraph{94 Aurora}
As an initial spin state for the ADAM modeling we used pole solutions of \citet{Marciniak2011} and \citet{Hanus2016a}. One pole orientation fitted the AO and occultation data slightly better, namely $\sim$(56, 7)$^{\circ}$, however, the second pole solution cannot be rejected. We determined the first non-radiometric size estimate for Aurora to be (196$\pm$4)~km. Due to a poor mass estimate, our bulk density of (1.6$\pm$0.9)~g\,cm$^{-3}$ has a large uncertainty. On the other hand, the density falls into a typical range for C-type objects.

\paragraph{129 Antigone}
The single pole solution of $\sim$(210, 55)$^{\circ}$ derived by the lightcurve inversion \citep{Torppa2003, Hanus2016a} was confirmed by occultation measurements \citep{Durech2011} or comparison with disk-resolved data \citep{Drummond2009a,Hanus2013b}. We adopted this pole solution as an initial input for the shape modeling and derived a solution that is consistent with both occultation observations, optical light curves and 8 disk-resolved images. The size is well constrained and gives, in combination with the mass, a bulk density of (2.5$\pm$0.9)~g\,cm$^{-3}$. The large uncertainty dominated by the poor mass estimate prevents reliable interpretation of the density, although one can argue that an M-type asteroid with a significant metal component should have a higher density. On the other hand, the extreme bulk density value of 3.4~g\,cm$^{-3}$ allowed by the large uncertainty is similar to those of other M-type asteroids in our sample.

\paragraph{135 Hertha}
The available single-epoch occultation of the M-type asteroid Hertha consists of 18 well spaced chords and belongs to one of the best stellar occultation observations obtained so far. Our ADAM shape model fits the occultation as well as both disk-resolved images and optical light curves well, the pole solution is consistent with the one of \citet{Torppa2003}, and our size of (80$\pm$2)~km agrees with the previous estimate of \citet{Timerson2009}. The density of (4.5$\pm$0.7)~g\,cm$^{-3}$ suggests some contribution of iron to Hertha's composition (similar to the case of Kleopatra). The higher bulk density is consistent with some of the proposed scenarios that could explain the observed properties of the Hertha cluster \citep{Dykhuis2015}. For instance, Hertha could be a remnant of an iron interior of a partially differenciated parent body that was destroyed by a past collision. However, one should also be careful with the mass estimate based on the ephemeris method.

\paragraph{144 Vibilia}
Disk-resolved data and stellar occultation chords are consistent only with the pole solution with ecliptic coordinates (248, 56)$^{\circ}$ effectively removing the pole ambiguity of \citet{Hanus2016a}. We present the first non-radiometric size estimate of (141$\pm$3)~km. The corresponding bulk density is (3.6$\pm$0.9)~g\,cm$^{-3}$, which is surprisingly high concerning the C-complex taxonomic classification (C, Ch). Most likely, the mass estimate is not accurate. Alternatively, the interior could have larger density or a very low macroporosity.

\paragraph{165 Loreley}
The pole ambiguity of the Cb-type asteroid Loreley was already removed by \citet{Durech2011} via stellar occultation measurements and later confirmed by \citet{Hanus2013b} via disk-resolved images from Keck. Our ADAM shape model reproduces all optical, disk-resolved and stellar occultation data nicely. However, the size combined with the mass leads to an unrealistic density of (7.1$\pm$0.9)~g\,cm$^{-3}$, which likely indicates that the mass estimate is wrong and should be revised. 

\paragraph{233 Asterope}
Only one out of the two pole solutions from \citet{Hanus2016a} is consistent with the disk-resolved data and occultation measurements -- (316, 58)$^{\circ}$. We also present the first non-radiometric size for Asterope of (106$\pm$3)~km. Unfortunately, there is no reliable mass estimate for Asterope in the literature. 

\paragraph{360 Carlova}
We used the rotation state of \citet{Hanus2016a} as an input for the shape modeling by ADAM. The occultation measurements and one disk-resolved image are consistent with only one pole solution. Recently, \citet{Wang2015} derived a lightcurve-based shape model for Carlova from a larger optical dataset. Their spin solution is close to our preferred one, only about 20$^{\circ}$ distant in the ecliptic longitude. By using their pole solution as an input for ADAM, we converged to our original solution, which suggests that our photometric dataset and disk-resolved data are not consistent with the solution of \citet{Wang2015}. However, we are aware that our optical dataset is rather small (a subset off Wang's dataset). Unfortunately, there is no reliable mass estimate available for Carlova.

\paragraph{386 Siegena}
Only a single pole solution is reported by \citet{Hanus2016a} and this solution is also consistent with the AO and occultation data. Our size (first non-radiometric estimate) based on ADAM modeling is (167$\pm$5)~km and the corresponding bulk density, assuming mass from \citet{Carry2012b}, is (3.3$\pm$0.8)~g\,cm$^{-3}$, which is a rather high value considering Siegena is a C-type. The mass estimate is likely non-reliable. On the other hand, \citet{Fienga2014} computed a much lower mass, however, with a 60\% uncertainty. The corresponding bulk density would be (1.0$\pm$0.6)~g\,cm$^{-3}$, being a more realistic estimate. 

\paragraph{387 Aquitania}
Our shape and size solution is consistent with the convex shape model of \citet{Devogele2017}. Unfortunately, there is no reliable mass estimate for Aquitania.

\paragraph{409 Aspasia}
\citet{Drummond2009a} presented a tri-axial ellipsoidal shape model of this C-complex asteroid (Xc-type) as well as the size of (184$\pm$6)~km based on the disk-resolved images from Keck. However, the spin solution of \citet{Warner2008a} based on lightcurve inversion of optical data is different from the one of \citet{Drummond2009a}. The solution of \citet{Warner2008a} was later confirmed by \citet{Durech2011}, where the authors even removed the pole ambiguity. A shape model based on optical photometry and three Keck AO images was derived by \citet{Hanus2013b}. Our ADAM shape model is based on an up-to-date optical photometric dataset, nine disk-resolved images and three stellar occultations observed at three different epochs and results in a reliable size estimate of (164$\pm$3)~km. The bulk density of (5.1$\pm$1.0)~g\,cm$^{-3}$ is quite high for a C-complex object, which likely suggests that the mass estimate is not fully reliable.

\paragraph{419 Aurelia}
Aurelia is an F-type asteroid in the Tholen taxonomy and its previous spin and shape solutions \citep{Hanus2016a} are ambiguous as it is common for results based on lightcurve data. We ran the ADAM algorithm with both pole solutions as initial inputs, however, both derived shape solutions reproduce reliably well both the AO and occultation data. Perhaps, the second pole solution might be preferred, but we cannot convincingly reject the first pole solution. The sizes of these two shape models are slightly different (120 and 125~km), but still consistent within their uncertainties. This results in two slightly different, but consistent density estimates. The average value of $\sim$1.8~g\,cm$^{-3}$ is reasonable for a primitive F-type asteroid.  

\paragraph{471 Papagena}
Lightcurve-based shape models of the S-type asteroid Papagena were derived by \citet{Torppa2008} and \citet{Hanus2011}. The pole ambiguity was then removed by \citet{Durech2011} based on occultation measurements. This single pole solution was confirmed by our ADAM modeling. Our size with a much smaller uncertainty of (132$\pm$4)~km is consistent with the occultation size of (137$\pm$25)~km derived by \citet{Durech2011}. Unfortunately, the available mass estimate has a large uncertainty and leads to a rather meaningless bulk density of (2.5$\pm$1.5)~g\,cm$^{-3}$.

\paragraph{532 Herculina}
\citet{Kaasalainen2002b} provided the pole solution of (288, 11)$^{\circ}$ for Herculina (S-type), while \citet{Hanus2016a} reported a value of (100, 9)$^{\circ}$. The latter solution is consistent with the disk-resolved images and occultation measurements, so it is reported as the final solution. The combination of the size (164$\pm$3~km) and mass gives a density of (3.2$\pm$0.8)~g\,cm$^{-3}$ that is typical within S-type asteroids.

\paragraph{849 Ara}
A pole ambiguous lightcurve-based shape model of Ara was independently derived by \citet{Durech2009} and \citet{Marciniak2009a}. \citet{Durech2011} used stellar occultation measurements to remove the pole ambiguity and to estimate the size (76$\pm$14~km). Our ADAM modeling provides a consistent, however, more precise diameter of (73$\pm$2)~km. There is no reliable mass estimate for Ara.

\subsection{Bulk densities}\label{sec:densities}

\begin{figure*}
    \begin{center}
        \resizebox{1.0\hsize}{!}{\includegraphics{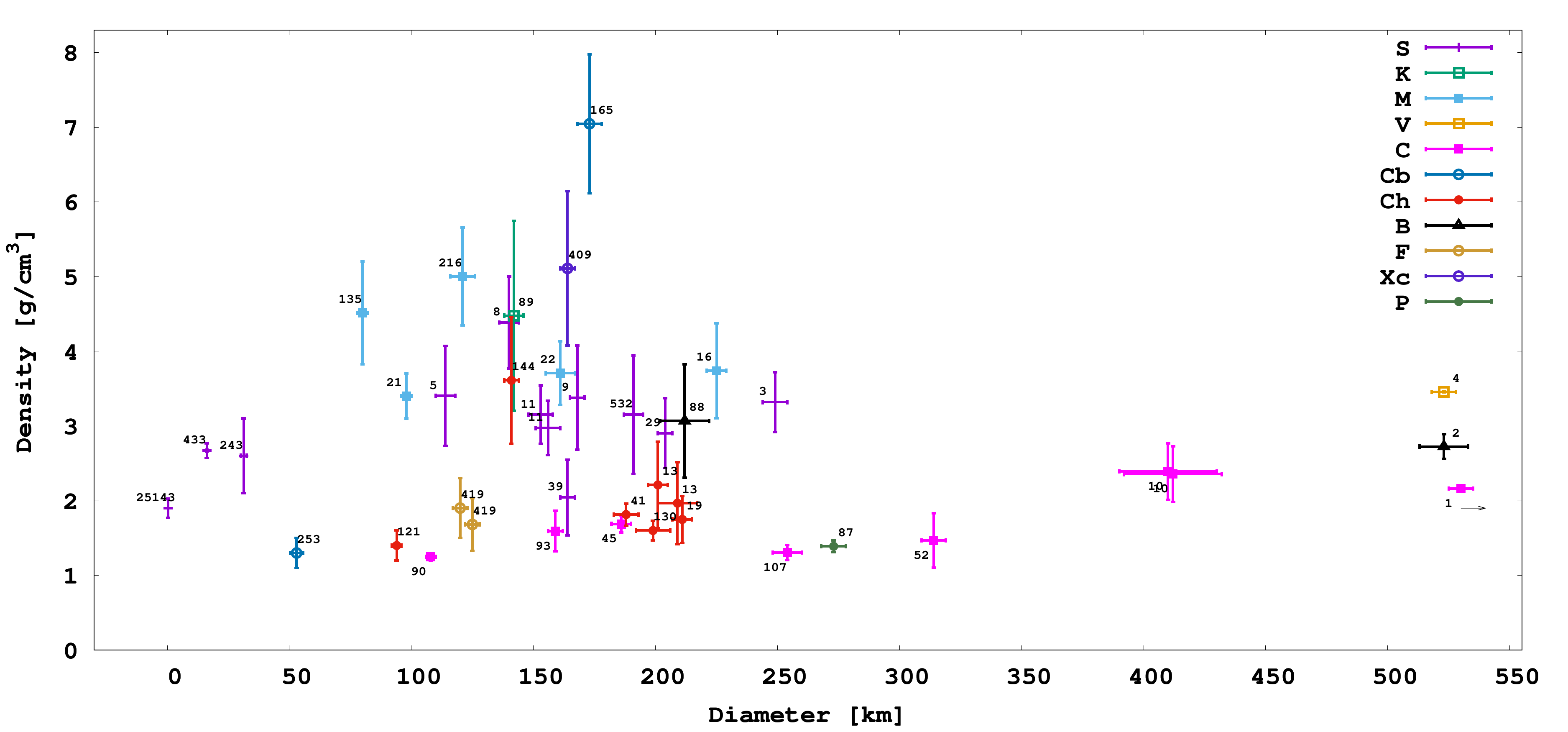}}\\
    \end{center}
    \caption{\label{fig:D_vs_rho}Size vs. bulk density dependence for asteroids of different taxonomic classes from our sample. The taxonomy is primarily based on SMASS II \citep{Bus2002}, however, we also used the Tholen classes \citep{Tholen1989} to distinguish the X-complex group. Only bulk densities with an accuracy of 30\% and better are included. We applied an offset to the position of Ceres to have a more compact plot, its true size is (941$\pm$6)~km.}
\end{figure*}

\begin{figure*}
    \begin{center}
        \resizebox{1.0\hsize}{!}{\includegraphics{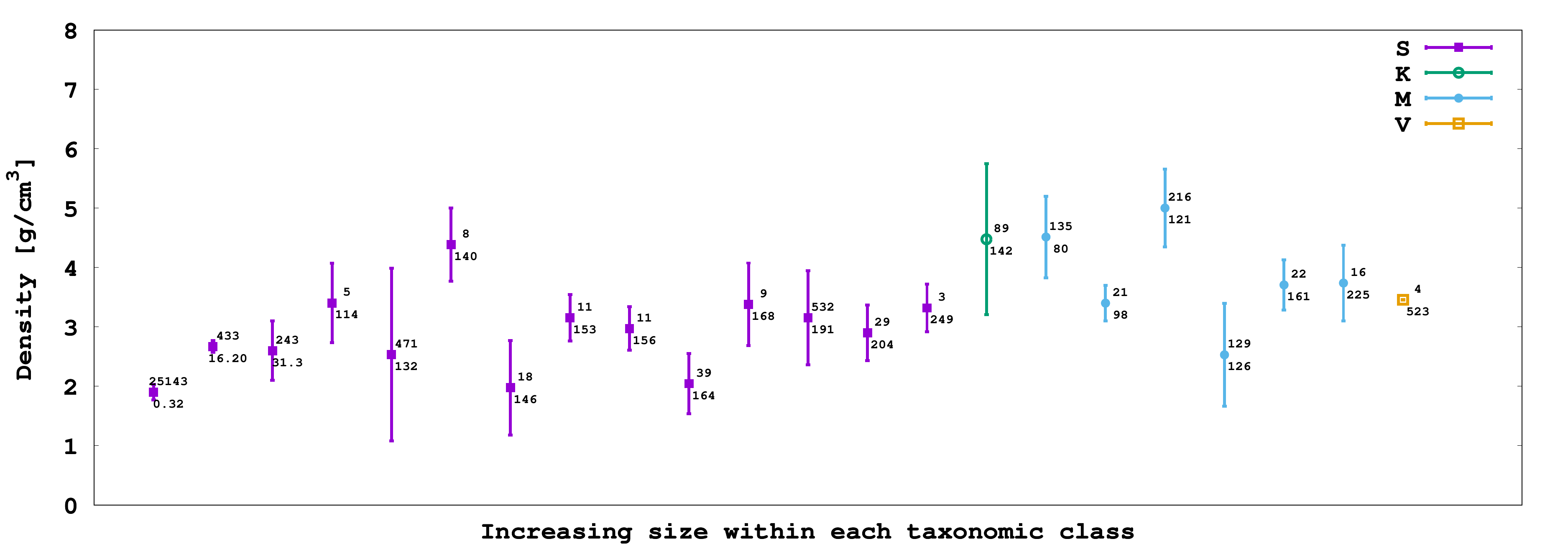}}\\
    \end{center}
    \caption{\label{fig:rhoScomplex}Bulk densities within S, K, M, and V taxonomic classes. We included asteroids with poor/non-reliable bulk density estimates as well. Asteroids in each class are ordered according to their sizes that are indicated as labels below the asteroid numbers.}
\end{figure*}

\begin{figure*}
    \begin{center}
        \resizebox{1.0\hsize}{!}{\includegraphics{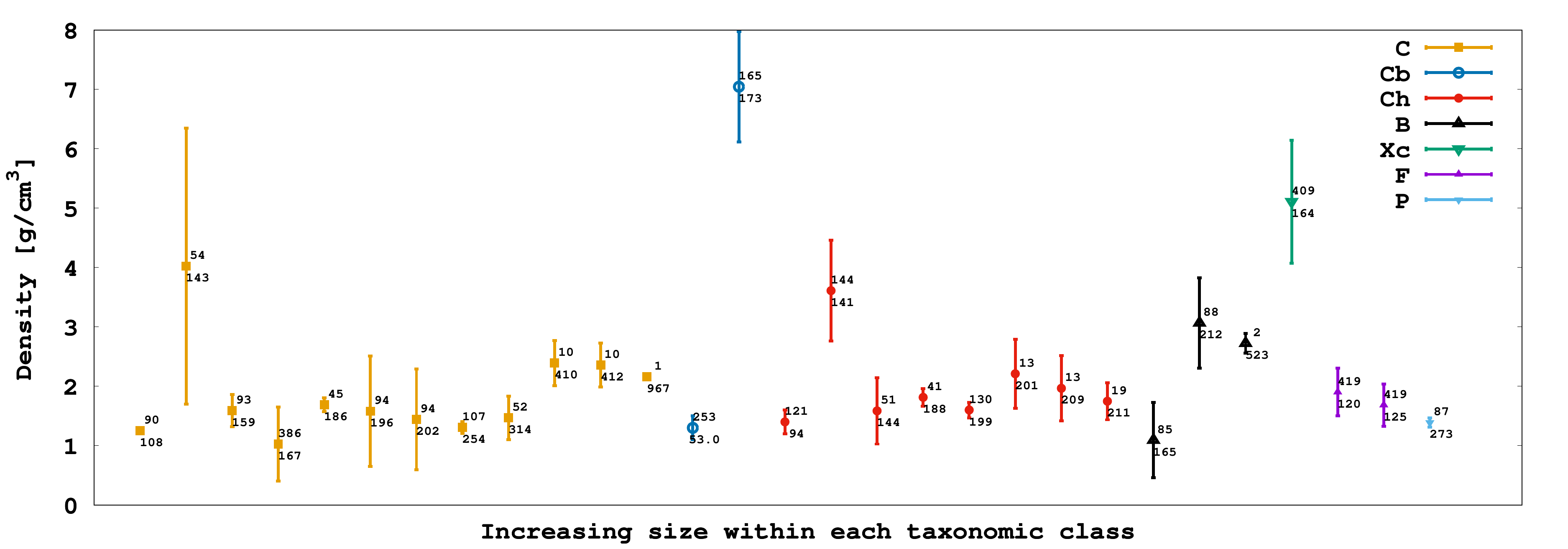}}\\
    \end{center}
    \caption{\label{fig:rhoCcomplex}Bulk densities within 'primitive' C, Cb, Ch, B, F, Xc, and P taxonomic classes. We included asteroids with poor/non-reliable bulk density estimates as well. Asteroids in each class are ordered according to their sizes that are indicated as labels below the asteroid numbers.}
\end{figure*}

To increase our sample of asteroids with density determinations, we compiled reliable bulk density estimates from the literature for several asteroids (see Tab.~\ref{tab:literature}). Most of the densities are based on data obtained during space probe flybys, or an orbit around the body.

For the majority of asteroids, we improved the precision of their size estimates leading to the volume determinations with an unprecedented precision. Consequently, the uncertainties of many our density determinations are governed by the uncertainty of the mass determination. The masses based on satellites' orbits are the most accurate ones, so densities based on those should be reliable as well. On the other hand, poor/non-reliable mass estimates based on planetary ephemeris prevent us to draw reliable conclusions on the bulk densities or could bias our results. It is clear that a few of our density determinations are not realistic, most likely due to an incorrect mass estimate -- (43)~Ariadne, (144)~Vibilia, (165)~Loreley, and (409)~Aspasia. Additionally, several other density estimates are suspicious -- (18)~Melpomene, (39)~Laetitia, (88)~Thisbe, (129)~Antigone, and (135)~Hertha, or with too large uncertainties -- (54)~Alexandra, (89)~Julia, (386)~Siegena, and (471)~Papagena. On the other hand, the peculiar density estimates, if true, might suggest that these objects are of a different nature than the others of similar taxonomic type, and so imply open questions on their origin. However, the more likely explanation are the incorrect mass estimates. Currently, there is no way how to validate the mass estimates that yield peculiar bulk densities, but this should change soon due to Gaia astrometric measurements that will provide reliable masses for tens of asteroids.

In Fig.~\ref{fig:D_vs_rho}, we plot the size vs. bulk density relationship for different taxonomic types. We included only densities with an accuracy of 30\% and better. Immediately, it is obvious that C-complex asteroids have densities between 1 and 2 g\,cm$^{-3}$ with a very weak trend of increasing with size and with the exception of the three largest asteroids, which densities are larger than 2 g\,cm$^{-3}$. On the other hand, S-complex asteroids have bulk densities between 2.5 and 3.5 g\,cm$^{-3}$ and M-type asteroids between 3 and 5 g\,cm$^{-3}$. Several outliers are discussed earlier, and their masses are probably inaccurate.

Figs.~\ref{fig:rhoScomplex} and \ref{fig:rhoCcomplex} show the bulk densities within several taxonomic classes. The first figure represents the groups with higher bulk densities -- S, K, M, and V, the second figure includes the 'primitive' asteroids with lower densities -- C, Cb, Ch, B, F, Xc, and P. We plot all asteroids, even those with poor/non-reliable bulk densities. Asteroids in each class are ordered according to their size.

The three smallest S-type asteroids ($D<35$~km) all have smaller bulk densities than most of the larger S-type asteroids. This could be due to larger macroporosity. If we exclude the asteroids (8)~Flora, (18)~Melpomene and (39)~Laetitia, all bulk densities for asteroids larger than 100~km have consistent values between 3.0 and 3.5 g\,cm$^{-3}$. Unfortunately, our sample does not contain asteroids with sizes between 35 and 110 km. There are six M-type asteroids in our sample, which bulk densities span a large range of values between 2.5 and 5.0 g\,cm$^{-3}$. The bulk densities of the asteroids (16)~Psyche, (21)~Lutetia, and (22)~Kalliope are similar ($\sim$3.5 g\,cm$^{-3}$), slightly higher than the densities of S-type asteroids, comparable to the bulk density of Mars (3.93 g\,cm$^{-3}$) and even larger than of the Moon (3.35 g\,cm$^{-3}$). The densities of the asteroids (129)~Antigone and (135)~Hertha are lower and larger, respectively, than the densities of the other three M-type asteroids, however, still comparable considering their larger uncertainties. The bulk density of the asteroid (216)~Kleopatra is the only incompatible value within M-type asteroids. Such a high density (5.0$\pm$0.7 g\,cm$^{-3}$) suggests a significant metallic contribution to the composition. Moreover, bulk densities of M-type asteroids are similar to the density of asteroid (4)~Vesta that is believed to be differentiated. So far, M-type asteroids are objects with the largest bulk density within the asteroid population, which is consistent with the general consensus that they could represent the remnants of planetesimal's metal-rich cores.

If we ignore the poor density estimates in Fig.~\ref{fig:rhoCcomplex}, where we included the 'primitive' taxonomic classes, we see that almost all asteroids have bulk densities $\sim$1.5 g\,cm$^{-3}$. There is no obvious difference within the various classes. The main exceptions are asteroids (1)~Ceres, (2)~Pallas, and (10)~Hygiea. Ceres and Hygiea are C-types, and it is believed that Ceres is a differentiated body, which naturally explains the larger bulk density. There is not much known about Hygiea, its bulk density seems to be even larger than that of Ceres, however, there might still be some the small systematics in the size and mass determinations. On the other hand, even possible refinements in these properties would not likely place the bulk density below $\sim$2.0 g\,cm$^{-3}$. For now, the bulk density of Hygiea seems to be similar to that of Ceres, which suggests that Hygiea could be a differentiated body. The bulk density of Pallas is larger than that of Ceres, but note that Pallas is a B-type asteroid, so at least its surface composition is different. We have only two other B-type asteroids in our sample, however, both of them have large uncertainties. Moreover, the density of (88)~Thisbe could be unrealistic due to the mass estimate. The more reliable density seems to be that of asteroid (85)~Io that is consistent with the other similarly-sized primitive asteroids. Currently, we cannot distinguish if all the B-type asteroids have bulk densities similar to that of Pallas, or if their densities are rather low with the large value of Pallas as an (possibly differentiated) exception. 

\section{Conclusions}\label{sec:conclusions}

We derived shape models and volumes for 41 asteroids by the ADAM algorithm from the inversion of their optical light curves, disk-integrated images from Nirc2 at the Keck II telescope and stellar occultation measurements. For 36 asteroids, the knowledge of their mass estimates from the literature allowed us to derive their bulk densities.

We present an analysis of derived bulk densities with respect to the different taxonomic classes. We observe a consistency within the S- and M-type objects, only the smallest S-type objects ($<35$~km) have systematically lower bulk densities, probably due to larger macroporosity. On top of that, only few largest primitive (C-complex) asteroids have significantly larger bulk densities compared to the remaining asteroids in the sample. This majority exhibits a rather narrow range of density values around $\sim$1.5 g\,cm$^{-3}$. The three largest members of the C-complex are or could be differentiated.  

Our high precision in the volume, thus consequently in the volume-equivalent diameter as well, was achieved mostly due to the usage of stellar occultations in the shape modeling. The advantage of the occultations is that they essentially provide direct measurements of the size along the star path behind the asteroid's projection. On the other hand, disk-resolved images often have a low resolution (i.e., the projection is represented by only few pixels) and the disk boundary is dependent on the regularization weights of the shape modeling or deconvolution algorithms. As a consequence, the size uncertainty is usually larger if we do not use stellar occultations in the shape modeling. We would like to stress out that our results were only feasible due to the contribution of hundreds of observers that participated in the occultation campaigns (see Tab.~\ref{tab:occ}). 

The main limitation of the bulk density determination is the poor knowledge of the mass. Astrometric observations from the ESA's Gaia satellite should partly solve this issue by providing accurate masses for about hundred asteroids. Moreover, good occultation measurements are important for a reliable size estimate, so this domain should benefit from the work of the occultation community. In our future work, we will also model other asteroids for which only light curve and AO data are available. There are tens of such objects. 

\begin{acknowledgements}
JH greatly appreciates the CNES post-doctoral fellowship program. JD was supported by the grant 15-04816S of the Czech Science Foundation.

This research has made use of the Keck Observatory Archive (KOA), which is operated by the W. M. Keck Observatory and the NASA Exoplanet Science Institute (NExScI), under contract with the National Aeronautics and Space Administration.

\end{acknowledgements}

\bibliography{mybib}
\bibliographystyle{aa}

\onecolumn
\scriptsize{
\longtab{1}{
\begin{longtable}{r@{\,\,\,}l rrrr rrr rrr}
\caption{\label{tab:results}Rotation state parameters $\lambda_{\mathrm{a}}$, $\beta_{\mathrm{a}}$, $P_{\mathrm{a}}$ with a reference to the corresponding publication that we used as initial inputs for the modeling with ADAM, rotation state parameters $\lambda$, $\beta$, $P$ derived by the ADAM algorithm, and the number of available light curves $N_{\mathrm{lc}}$, disk-resolved images $N_{\mathrm{ao}}$ and stellar occultations $N_{\mathrm{occ}}$.}\\
\hline 
\multicolumn{2}{c} {Asteroid} & \multicolumn{1}{c} {$\lambda_{\mathrm{a}}$} & \multicolumn{1}{c} {$\beta_{\mathrm{a}}$} & \multicolumn{1}{c} {$P_{\mathrm{a}}$} & \multicolumn{1}{c} {Reference} & \multicolumn{1}{c} {$\lambda$} & \multicolumn{1}{c} {$\beta$} & \multicolumn{1}{c} {$P$} & \multicolumn{1}{c} {$N_{\mathrm{lc}}$} & \multicolumn{1}{c} {$N_{\mathrm{ao}}$} & \multicolumn{1}{c} {$N_{\mathrm{occ}}$} \\
\multicolumn{2}{l} {} & [deg] & [deg] & \multicolumn{1}{c} {[hours]} &  & [deg] & [deg] & \multicolumn{1}{c} {[hours]} &  &  & \\
\hline\hline

\endfirsthead
\caption{continued.}\\

\hline
\multicolumn{2}{c} {Asteroid} & \multicolumn{1}{c} {$\lambda_{\mathrm{a}}$} & \multicolumn{1}{c} {$\beta_{\mathrm{a}}$} & \multicolumn{1}{c} {$P_{\mathrm{a}}$} & \multicolumn{1}{c} {Reference} & \multicolumn{1}{c} {$\lambda$} & \multicolumn{1}{c} {$\beta$} & \multicolumn{1}{c} {$P$} & \multicolumn{1}{c} {$N_{\mathrm{lc}}$} & \multicolumn{1}{c} {$N_{\mathrm{ao}}$} & \multicolumn{1}{c} {$N_{\mathrm{occ}}$} \\
\multicolumn{2}{l} {} & [deg] & [deg] & \multicolumn{1}{c} {[hours]} &  & [deg] & [deg] & \multicolumn{1}{c} {[hours]} &  &  & \\
\hline\hline
\endhead
\hline
\endfoot
\hline
  2 &       Pallas &   31 & $-$16 & 7.81322 &        \citet{Carry2010a} &   30$\pm$3 & $-$13$\pm$2 & 7.81322 &  61 & 18 &  2 \\
  5 &      Astraea &  126 &  40 & 16.80061 &        \citet{Durech2009} &  125$\pm$3 &  39$\pm$3 & 16.80060 &  24 &  2 &  1 \\
  8 &        Flora &  335 &  $-$5 & 12.86667 &        \citet{Hanus2013b} &  342$\pm$5 &  $-$6$\pm$6 & 12.86667 &  54 &  6 &  1 \\
  9 &        Metis &  185 &  24 & 5.079176 &        \citet{Hanus2013b} &  182$\pm$2 &  20$\pm$2 & 5.079176 &  34 &  8 &  2 \\
 10 &       Hygiea &  312 & $-$42 & 27.65907 &         \citet{Hanus2011} &  303$\pm$3 & $-$35$\pm$2 & 27.65906 &  26 &  2 &  2 \\
 10 &       Hygiea &  122 & $-$44 & 27.65905 &         \citet{Hanus2011} &  115$\pm$2 & $-$36$\pm$4 & 27.65906 &  26 &  2 &  2 \\
 11 &   Parthenope &  312 &  15 & 13.72205 &        \citet{Hanus2013a} &  312$\pm$3 &  16$\pm$4 & 13.72205 & 138 &  1 &  1 \\
 11 &   Parthenope &  129 &  14 & 13.72205 &        \citet{Hanus2013a} &  127$\pm$4 &  15$\pm$3 & 13.72205 & 138 &  1 &  1 \\
 13 &       Egeria &   44 &  21 & 7.046671 &         \citet{Hanus2011} &  50$\pm$10 &   20$\pm$10 & 7.046673 &  13 &  1 &  1 \\
 13 &       Egeria &  238 &  11 & 7.046673 &         \citet{Hanus2011} &  232$\pm$2 &   7$\pm$2 & 7.046673 &  13 &  1 &  1 \\
 16 &       Psyche &   32 &  $-$7 & 4.195948 &        \citet{Hanus2016a} &   28$\pm$4 &   $-$6$\pm$3 & 4.195948 & 118 &  7 &  2 \\
 18 &    Melpomene &   11 &  14 & 11.57031 &        \citet{Hanus2016a} &   11$\pm$3 &  16$\pm$6 & 11.570306 &  64 &  6 &  1 \\
 19 &      Fortuna &   96 &  56 & 7.44322 &        \citet{Hanus2016a} &   97$\pm$2 &  69$\pm$2 & 7.443222 &  48 &  4 &  2 \\
 22 &     Kalliope &  196 &   3 & 4.14820 &      \citet{Descamps2008} &  195$\pm$3 &   2$\pm$3 & 4.14820 & 102 & 23 &  1 \\
 29 &   Amphitrite &  322 & $-$29 & 5.390119 &        \citet{Hanus2016a} &  323$\pm$2 & $-$26$\pm$2 & 5.390119 &  66 &  7 &  1 \\
 39 &     Laetitia &  322 &  30 & 5.138238 &        \citet{Hanus2016a} &  323$\pm$2 &  32$\pm$2 & 5.138238 &  68 &  3 &  1 \\
 41 &       Daphne &  198 & $-$32 & 5.98798 &        \citet{Hanus2016a} &  195$\pm$3 & $-$32$\pm$3 & 5.987981 &  33 &  7 &  2 \\
 43 &      Ariadne &  253 & $-$15 & 5.76199 &  \citet{Kaasalainen2002b} &  252$\pm$3 &  $-$9$\pm$3 & 5.76199 &  45 &  1 &  1 \\
 45 &      Eugenia &  125 & $-$34 & 5.699151 &        \citet{Hanus2016a} &  127$\pm$2 & $-$36$\pm$2 & 5.699152 & 101 & 23 &  1 \\
 51 &      Nemausa &  169 & $-$62 & 7.784840 &        \citet{Hanus2016a} &  169$\pm$4 & $-$64$\pm$5 & 7.784840 &  58 &  3 &  2 \\
 51 &      Nemausa &  347 & $-$68 & 7.784841 &        \citet{Hanus2016a} &    \multicolumn{3}{c} {Rejected} &  58 &  3 &  2 \\
 52 &       Europa &  254 &  37 & 5.629962 &       \citet{Merline2013} &  254$\pm$7 &  36$\pm$4 & 5.629957 &  49 & 25 &  4 \\
 54 &    Alexandra &  152 &  19 & 7.022641 &        \citet{Hanus2016a} &  155$\pm$4 &  17$\pm$3 & 7.022642 &  38 &  2 &  1 \\
 80 &       Sappho &  194 & $-$26 & 14.03087 &        \citet{Durech2009} &  195$\pm$2 & $-$22$\pm$3 & 14.03086 &  16 &  2 &  1 \\
 85 &           Io &   95 & $-$65 & 6.874783 &        \citet{Durech2011} &   90$\pm$3 & $-$68$\pm$2 & 6.874783 &  29 &  2 &  3 \\
 87 &       Sylvia &   70 &  69 & 5.18364 &      \citet{Berthier2014} &   72$\pm$3 &  67$\pm$3 & 5.183641 &  55 & 22 &  2 \\
 88 &       Thisbe &   82 &  69 & 6.04131 &        \citet{Hanus2016a} &   79$\pm$4 &  68$\pm$3 & 6.04132 &  28 &  2 &  1 \\
 89 &        Julia &    8 & $-$13 & 11.38834 &        \citet{Durech2011} &   15$\pm$4 & $-$13$\pm$5 & 11.38834 &  31 &  1 &  2 \\
 93 &      Minerva &   21 &  21 & 5.981767 &      \citet{Marchis2013a} &   20$\pm$3 &  21$\pm$4 & 5.981768 &  34 &  4 &  2 \\
 94 &       Aurora &  242 &  $-$7 & 7.22619 &        \citet{Hanus2016a} &  244$\pm$2 &   3$\pm$5 & 7.22619 &  22 &  2 &  2 \\
 94 &       Aurora &   65 &   9 & 7.226191 &        \citet{Hanus2016a} &   56$\pm$3 &   7$\pm$5 & 7.226188 &  22 &  2 &  2 \\
107 &      Camilla &   72 &  51 & 4.843928 &        \citet{Hanus2016a} &   75$\pm$2 &  55$\pm$2 & 4.843928 &  34 & 21 &  1 \\
129 &     Antigone &  211 &  55 & 4.957154 &        \citet{Hanus2016a} &  198$\pm$6 &  58$\pm$6 & 4.957156 &  52 &  9 &  2 \\
135 &       Hertha &  272 &  52 & 8.40060 &        \citet{Torppa2003} &  277$\pm$3 &  53$\pm$3 & 8.40060 &  30 &  2 &  1 \\
144 &      Vibilia &  248 &  56 & 13.82516 &        \citet{Hanus2016a} &  250$\pm$3 &  58$\pm$5 & 13.82517 &  43 &  2 &  3 \\
144 &      Vibilia &   54 &  48 & 13.82517 &        \citet{Hanus2016a} &    \multicolumn{3}{c} {Rejected} &  43 &  2 &  3 \\
165 &      Loreley &  174 &  29 & 7.224390 &        \citet{Durech2011} &  178$\pm$3 &  31$\pm$3 & 7.224390 &  30 &  4 &  1 \\
216 &    Kleopatra &   73 &  21 & 5.385280 &   \citet{Kaasalainen2012} &   74$\pm$2 &  20$\pm$2 & 5.385280 &  55 & 14 &  3 \\
233 &     Asterope &  322 &  59 & 19.69803 &        \citet{Hanus2016a} &  316$\pm$8 &  58$\pm$3 & 19.69803 &  13 &  1 &  1 \\
233 &     Asterope &  132 &  36 & 19.69806 &        \citet{Hanus2016a} &    \multicolumn{3}{c} {Rejected} &  13 &  1 &  1 \\
360 &      Carlova &  142 &  67 & 6.189596 &        \citet{Hanus2016a} &    \multicolumn{3}{c} {Rejected} &   9 &  2 &  1 \\
360 &      Carlova &    3 &  56 & 6.189595 &        \citet{Hanus2016a} &  355$\pm$7 &  56$\pm$5 & 6.189594 &   9 &  2 &  1 \\
386 &      Siegena &  289 &  25 & 9.765030 &        \citet{Hanus2016a} &  287$\pm$2 &  26$\pm$3 & 9.76503 &  83 &  1 &  2 \\
387 &    Aquitania &  142 &  51 & 24.14012 &      \citet{Devogele2017} &  123$\pm$5 &  46$\pm$5 & 24.14012 &  27 &  7 &  1 \\
409 &      Aspasia &    2 &  28 & 9.02145 &        \citet{Hanus2016a} &    4$\pm$2 &  30$\pm$2 & 9.02145 &  22 &  9 &  3 \\
419 &      Aurelia &    0 &  48 & 16.78093 &        \citet{Hanus2016a} &  354$\pm$5 &  43$\pm$4 & 16.78091 &  47 &  1 &  1 \\
419 &      Aurelia &  174 &  42 & 16.78090 &        \citet{Hanus2016a} &  173$\pm$3 &  35$\pm$3 & 16.78090 &  47 &  1 &  1 \\
471 &     Papagena &  223 &  67 & 7.115394 &        \citet{Durech2011} &  221$\pm$5 &  62$\pm$8 & 7.115390 &  13 &  1 &  2 \\
532 &    Herculina &  100 &   9 & 9.40494 &        \citet{Hanus2016a} &  103$\pm$4 &  11$\pm$4 & 9.40494 &  74 &  4 &  1 \\
849 &          Ara &  223 & $-$40 & 4.116391 &     \citet{Marciniak2009a} &  223$\pm$3 & $-$41$\pm$3 & 4.116391 &  23 &  0 &  2 \\
\hline
\end{longtable}
}
}

\onecolumn
\scriptsize{
\longtab{2}{
\begin{longtable}{r@{\,\,\,}l rr r rr rr r}
\caption{\label{tab:densities}Bulk density estimates based on our volume estimated by the ADAM shape modeling from combined optical light curves, disk-resolved images and stellar occultations.}\\
\hline
 \multicolumn{2}{c} {Asteroid} & $D_{\mathrm{a}}$ & Reference & $D$ & $M$ & Reference & T1 & T2 & $\rho$ \\
 \multicolumn{2}{c} {} & [km] &  & [km] & [10$^{18}$ kg] &  &  &  &  [gcm$^{-3}$] \\ \hline\hline

\endfirsthead
\caption{continued.}\\

\hline
 \multicolumn{2}{c} {Asteroid} & $D_{\mathrm{a}}$ & Reference & $D$ & $M$ & Reference & T1 & T2 & $\rho$ \\
 \multicolumn{2}{c} {} & [km] &  & [km] & [10$^{18}$ kg] &  &  &  &  [gcm$^{-3}$] \\ \hline\hline
\endhead
\hline
\endfoot
  2 &       Pallas &  512$\pm$6 &        \citet{Carry2010a} &   523$\pm$10 &  204.0$\pm$4.0 &        \citet{Carry2012b} &  B &  B & 2.72$\pm$0.17 \\
  5 &      Astraea &  115$\pm$6 &        \citet{Durech2011} &   114$\pm$4 &  2.6$\pm$0.4 &        \citet{Carry2012b} &  S &  S & 3.4$\pm$0.7 \\
  8 &        Flora &  140$\pm$7 &        \citet{Durech2011} &   140$\pm$4 &  6.3$\pm$0.7 &        \citet{Fienga2014} &  S &  $-$ & 4.4$\pm$0.6 \\
  9 &        Metis &  153$\pm$11 &        \citet{Hanus2013b} &   168$\pm$3 &  8.4$\pm$1.7 &        \citet{Carry2012b} &  S &  $-$ & 3.4$\pm$0.7 \\
 10 &       Hygiea &  443$\pm$45 &        \citet{Durech2011} &   412$\pm$20 &  86.3$\pm$5.2 &        \citet{Carry2012b} &  C &  C & 2.3$\pm$0.4 \\
 10 &       Hygiea &  351$\pm$27 &        \citet{Durech2011} &   410$\pm$20 &  86.3$\pm$5.2 &        \citet{Carry2012b} &  C &  C & 2.4$\pm$0.4 \\
 11 &   Parthenope &  151$\pm$5 &        \citet{Carry2012b} &   153$\pm$5 &  5.9$\pm$0.5 &        \citet{Carry2012b} &  S & Sk & 3.2$\pm$0.4 \\
 11 &   Parthenope &  151$\pm$5 &        \citet{Carry2012b} &   156$\pm$5 &  5.9$\pm$0.5 &        \citet{Carry2012b} &  S & Sk & 3.0$\pm$0.4 \\
\end{longtable}
\tablefoot{
The table gives previous size (surface- or volume-equivalent diameter) estimate $D_\mathrm{a}$ and its reference, volume-equivalent diameter $D$ of the shape solution derived here by ADAM, adopted mass $M$ and its reference, the Tholen (T1) and SMASS II (T2) taxonomic classes, and our density determination $\rho$.}
}
}
\onecolumn
\scriptsize{
\longtab{3}{
\begin{longtable}{r@{\,\,\,}l cc cc ll}
\caption{\label{tab:literature}Bulk density estimates compiled from the literature. Most of these bulk densities are based on data obtained during space probe flybys, or an orbit around the body.}\\
\hline
 \multicolumn{2}{c} {Asteroid} & $D_\mathrm{a}$ & Reference & $\rho$ & Reference & \multicolumn{1}{c} {T1} & \multicolumn{1}{c} {T2} \\
 \multicolumn{2}{c} {} & [km] &  & [gcm$^{-3}$] &  & \multicolumn{2}{c} {} \\ \hline\hline

\endfirsthead
\caption{continued.}\\

\hline
 \multicolumn{2}{c} {Asteroid} & $D_\mathrm{a}$ & Reference & $\rho$ & Reference & \multicolumn{1}{c} {T1} & \multicolumn{1}{c} {T2} \\
 \multicolumn{2}{c} {} & [km] &  & [gcm$^{-3}$] &  & \multicolumn{2}{c} {} \\ \hline\hline
\endhead
\hline
\endfoot
1   & Ceres     & 941$\pm$6  & \citet{Drummond2014}  &   2.163$\pm$0.008   & \citet{Park2016}  &    G & C   \\
3   & Juno      & 249$\pm$5  & \citet{Viikinkoski2015b} & 3.32$\pm$0.40    & \citet{Viikinkoski2015b}  & S & Sk   \\
4   & Vesta     & 523$\pm$5  & \citet{Russell2012}  &     3.456$\pm$0.005   & \citet{Russell2012}  & V & V   \\
21  & Lutetia   & 98$\pm$2   & \citet{Sierks2011}  &      3.4$\pm$0.3     & \citet{Sierks2011}  &  M & Xk   \\
90  & Antiope   & 108$\pm$2  & \citet{Descamps2007}  &    1.25$\pm$0.05    & \citet{Descamps2007} & C & C   \\
121 & Hermione  & 94$\pm$2   & \citet{Marchis2009}  &     1.4$\pm$0.2     & \citet{Marchis2009}  & C & Ch   \\
130 & Elektra   & 199$\pm$7  & \citet{Hanus2016c}  &      1.60$\pm$0.13    & \citet{Hanus2016c}  &  G & Ch   \\
243 & Ida       & 31.3$\pm$1.2& \citet{Archinal2011}  &    2.6$\pm$0.5     & \citet{Belton1994}  &  S & S   \\
\hline
\end{longtable}
\tablefoot{
The table gives previous size (volume-equivalent diameter) estimate $D_\mathrm{a}$ and its reference, adopted density $\rho$ and its reference, and the Tholen (T1) and SMASS II (T2) taxonomic classes.}
}
}

\begin{appendix}
\section{Online tables and figures}
\onecolumn
\scriptsize{
\begin{longtable}{rr rr rr rrr}
\caption{\label{tab:ao}List of disk-resolve images obtained by the NIRC2 at Keck II telescope used for the shape modeling with ADAM. For each observation, the table gives the epoch, filter, exposure time, airmass, R.A. and Dec of the asteroid, distance to the Earth $r$ and the reference or the PI of the project within which were the data obtained.}\\
\hline
\multicolumn{1}{c} {Date} & \multicolumn{1}{c} {UT} & \multicolumn{1}{c} {Filter} & \multicolumn{1}{c} {Exp} & \multicolumn{1}{c} {Airmass} & \multicolumn{1}{c} {R.A.} & \multicolumn{1}{c} {Dec} & \multicolumn{1}{c} {$r$} & Reference or PI \\ \hline\hline
\endfirsthead
\caption{continued.}\\

\hline
\multicolumn{1}{c} {Date} & \multicolumn{1}{c} {UT} & \multicolumn{1}{c} {Filter} & \multicolumn{1}{c} {Exp} & \multicolumn{1}{c} {Airmass} & \multicolumn{1}{c} {R.A.} & \multicolumn{1}{c} {Dec} & \multicolumn{1}{c} {$r$} & Reference or PI \\
\hline\hline
\endhead
\hline
\endfoot
\multicolumn{9}{c} {   2          Pallas} \\
  2002-05-08 &     14:58:36 &     H &   1.0 & 1.09 & 21 15 47 & 12 35 35 & 3.38 &               Margot \\
  2003-10-10 &     11:58:46 &    Kp &  20.0 & 1.28 & 01 56 36 & $-$16 17 33 & 1.80 &             Engineering \\
  2003-10-12 &     09:12:42 &    Kp &   2.0 & 1.40 & 01 55 17 & $-$16 47 35 & 1.80 &             Engineering \\
  2003-10-12 &     11:11:03 &    Kp &   2.0 & 1.25 & 01 55 12 & $-$16 49 18 & 1.80 &             Engineering \\
  2003-10-12 &     11:27:02 &    Kp &   2.0 & 1.26 & 01 55 12 & $-$16 49 18 & 1.80 &             Engineering \\
  2006-08-16 &     06:48:36 &    Ks &   1.0 & 1.00 & 18 02 59 & 17 10 55 & 2.76 &             Nelson \\
  2006-08-16 &     07:29:28 &    Ks &   1.0 & 1.42 & 17 14 00 & $-$19 42 18 & 2.76 &             Nelson \\
\hline
\end{longtable}
}

\twocolumn
\newpage
\onecolumn
\begin{longtable}{l}
\caption{\label{tab:occ}Occultations.}\\
\hline
 Observer \\ \hline\hline

\endfirsthead
\caption{continued.}\\

\hline
 Observer \\ \hline\hline
\endhead
\hline
\endfoot
\multicolumn{1}{c} {\textbf{(5) Astraea 2008-06-06}} \\
Milan Kapka, SK                       \\
Petr Zeleny, CZ                       \\
Jan Mocek, CZ                         \\
Milan Antos, CZ                       \\
Josef Durech, CZ                      \\
Michael Krocil, CZ                    \\
Peter Kusnirak, CZ                    \\
\hline
 \end{longtable}
\clearpage
\twocolumn
\begin{figure}[tbp]
    \centering
        \resizebox{0.24\hsize}{!}{\includegraphics{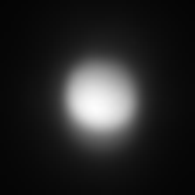}}\resizebox{0.24\hsize}{!}{\includegraphics{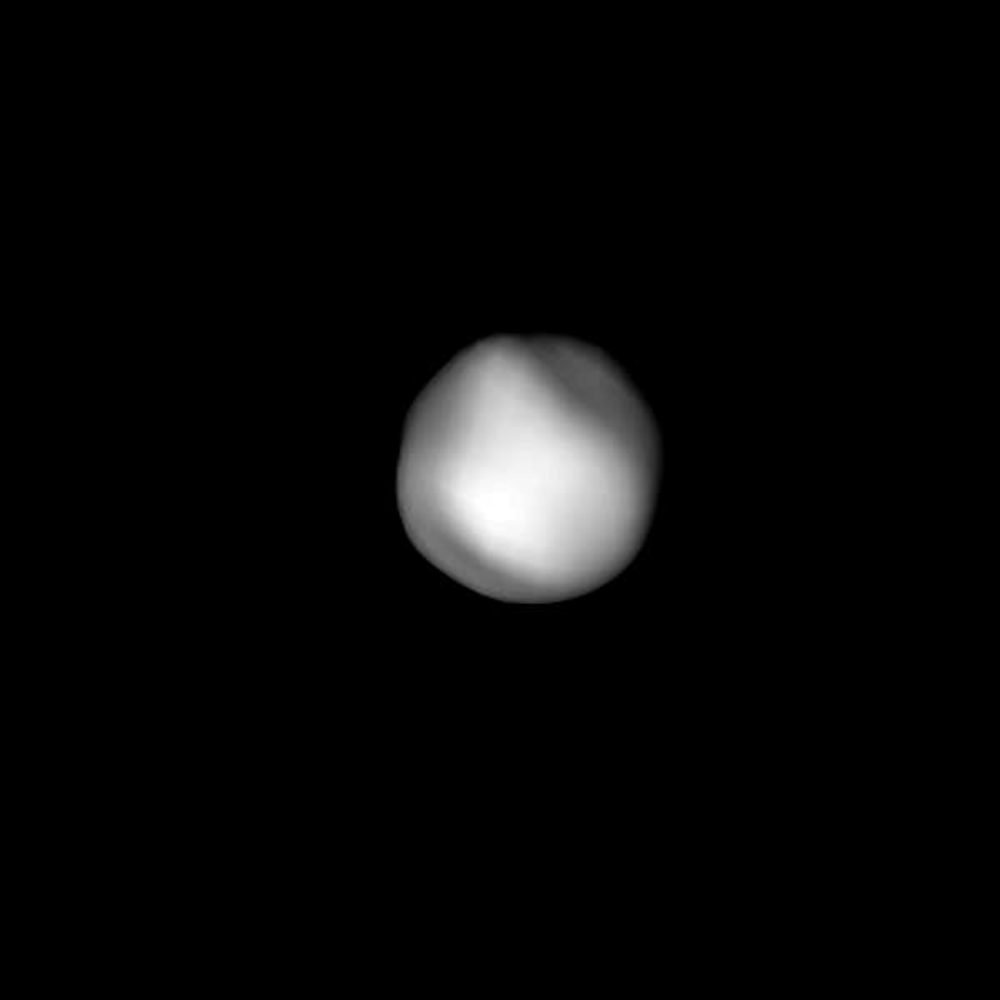}}\resizebox{0.24\hsize}{!}{\includegraphics{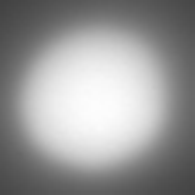}}\resizebox{0.24\hsize}{!}{\includegraphics{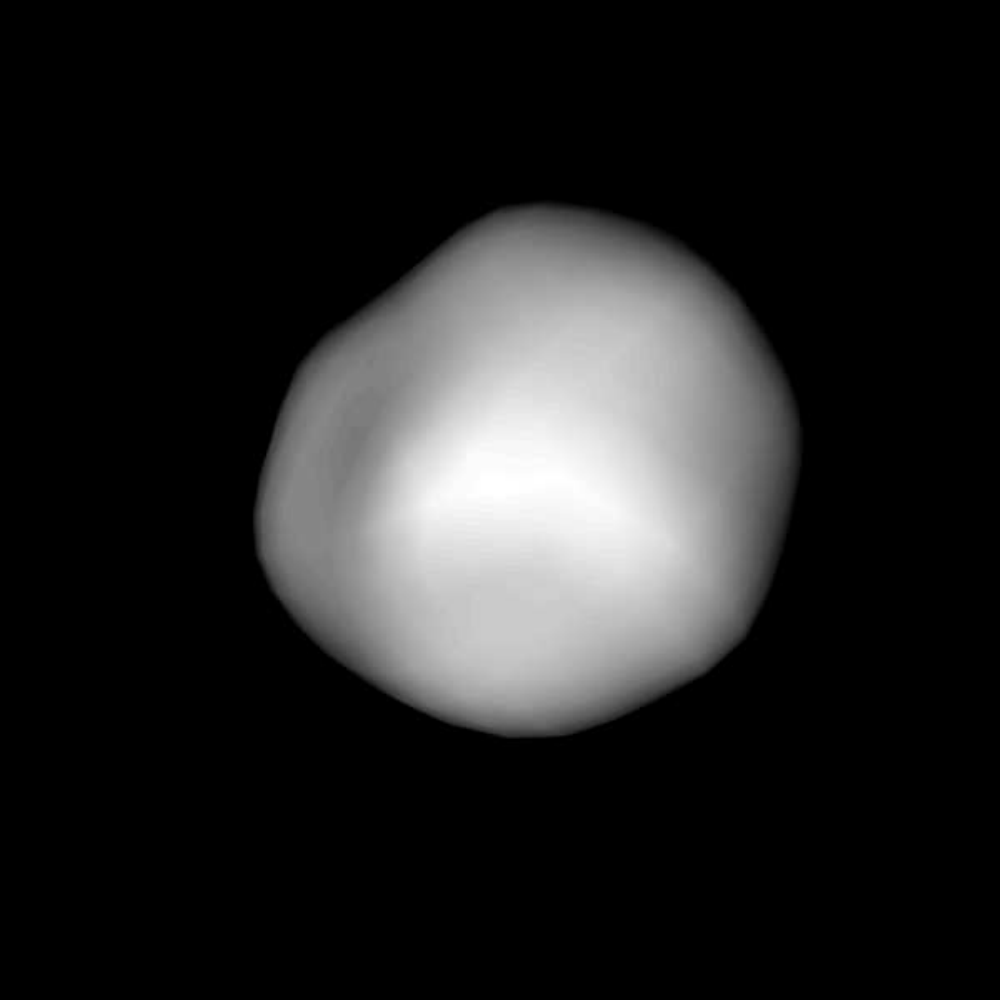}}\\
        \resizebox{0.24\hsize}{!}{\includegraphics{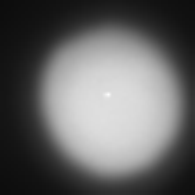}}\resizebox{0.24\hsize}{!}{\includegraphics{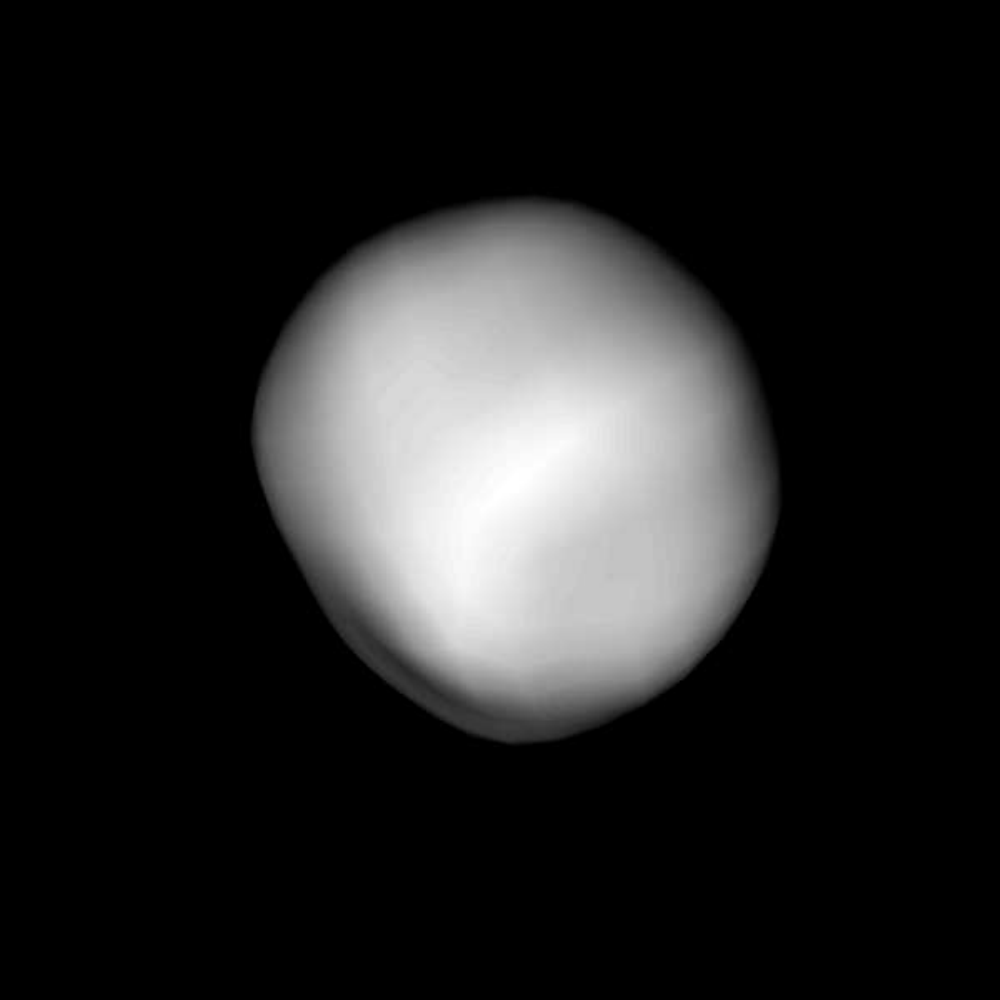}}\resizebox{0.24\hsize}{!}{\includegraphics{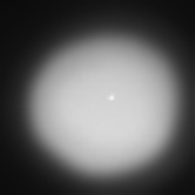}}\resizebox{0.24\hsize}{!}{\includegraphics{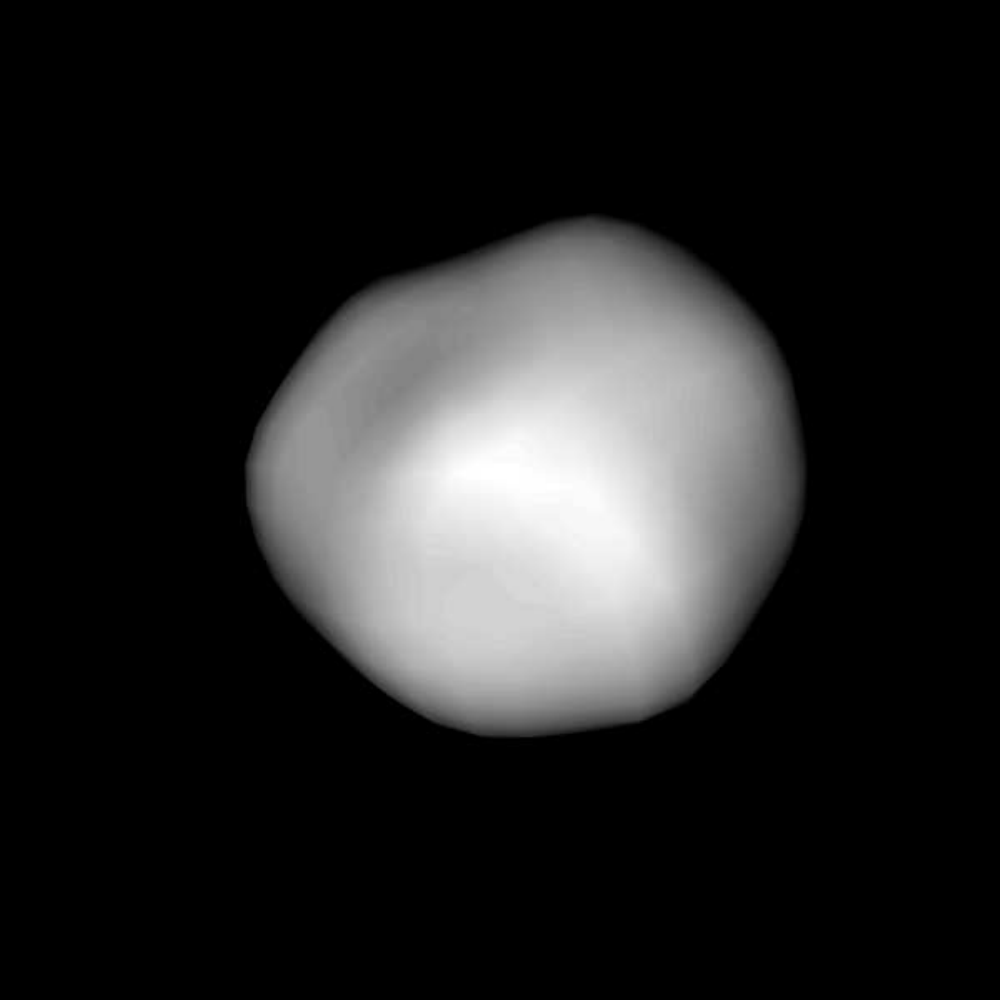}}\\
        \resizebox{0.24\hsize}{!}{\includegraphics{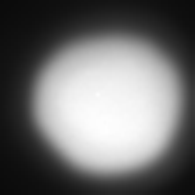}}\resizebox{0.24\hsize}{!}{\includegraphics{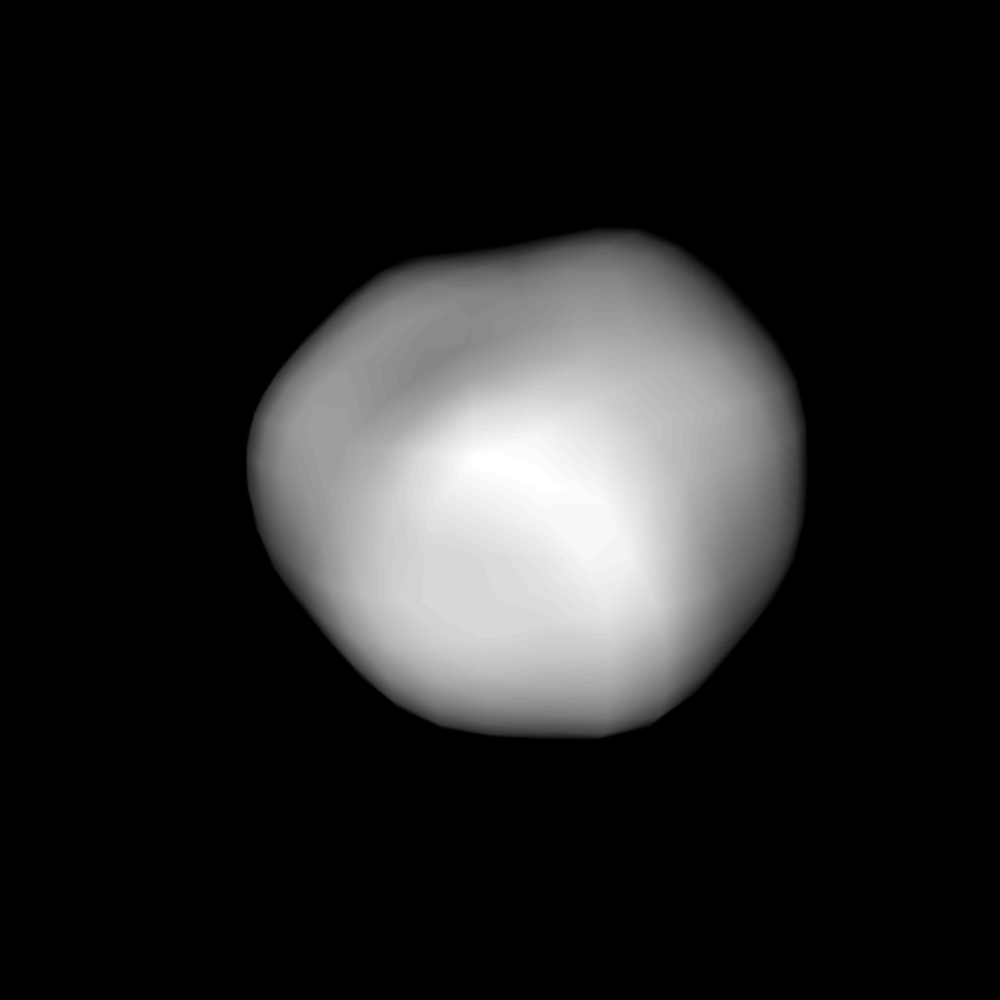}}\resizebox{0.24\hsize}{!}{\includegraphics{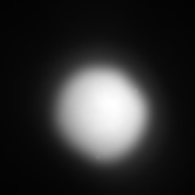}}\resizebox{0.24\hsize}{!}{\includegraphics{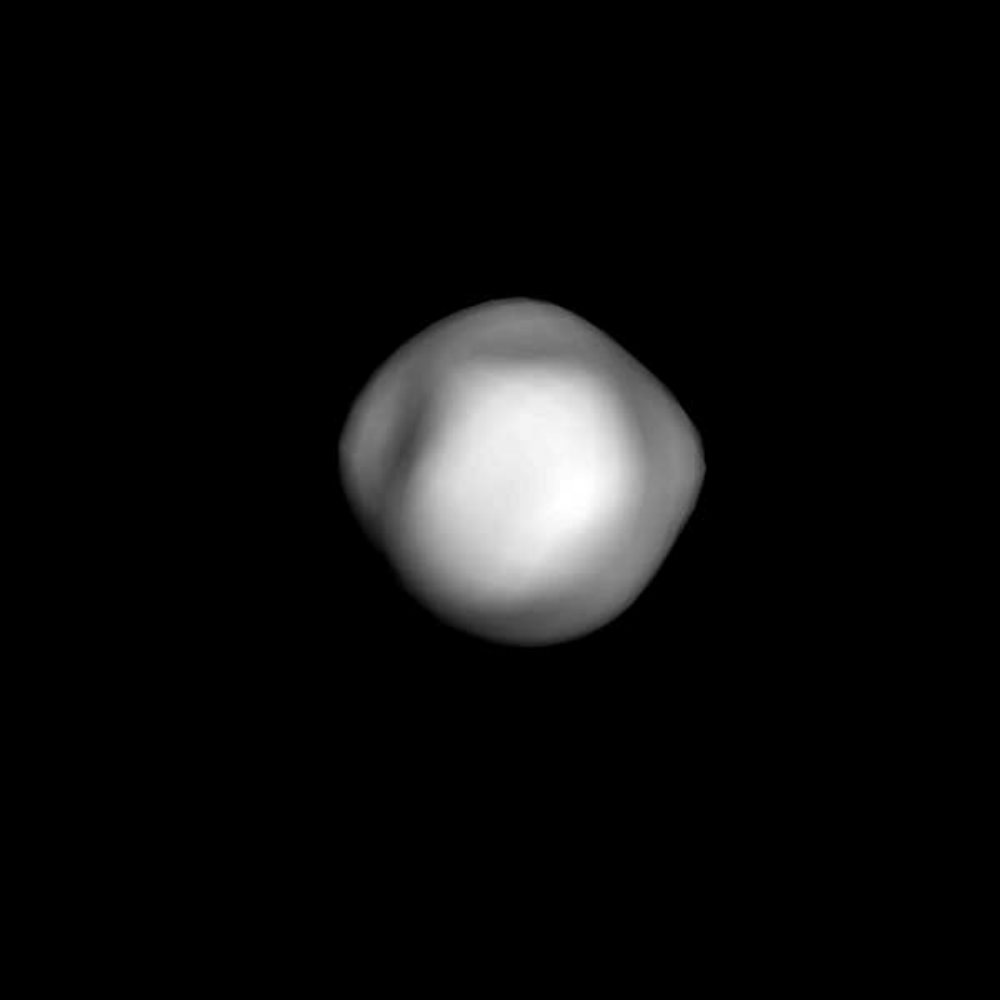}}\\
        \resizebox{0.24\hsize}{!}{\includegraphics{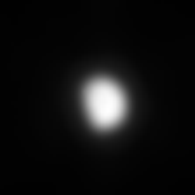}}\resizebox{0.24\hsize}{!}{\includegraphics{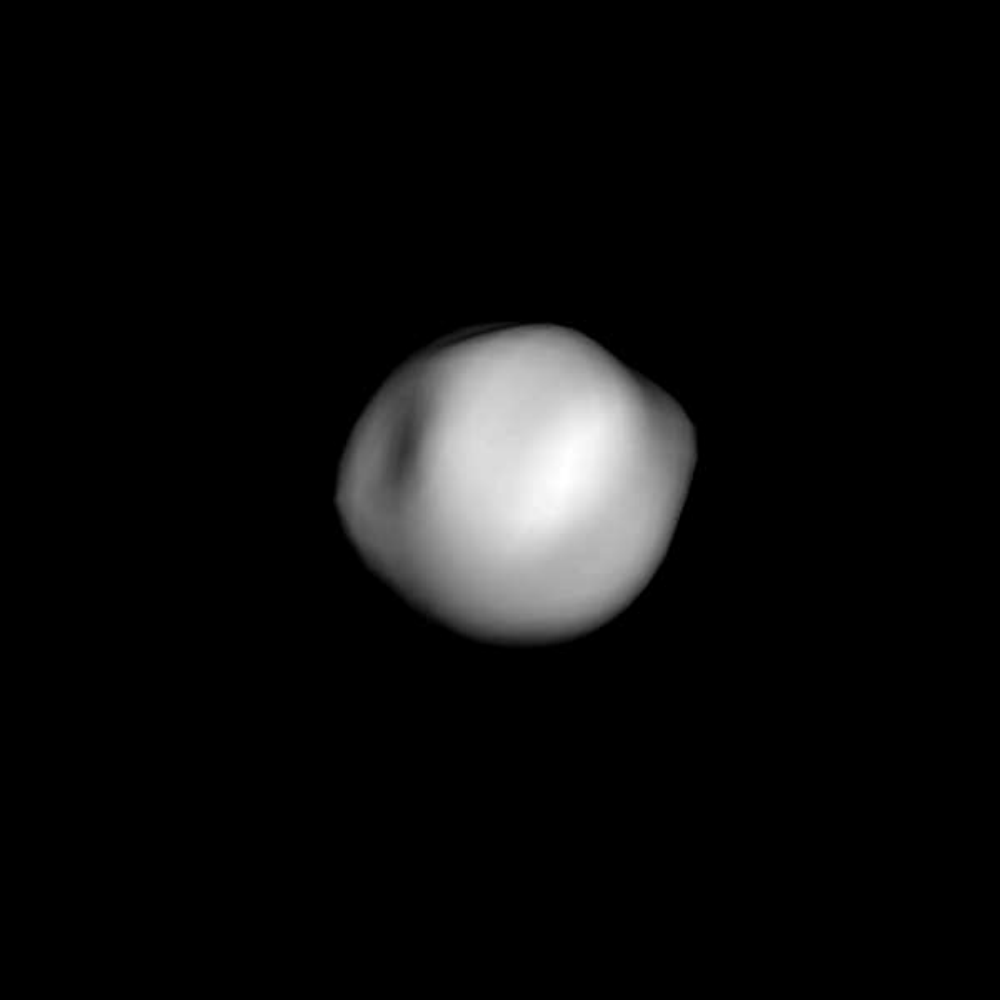}}\resizebox{0.24\hsize}{!}{\includegraphics{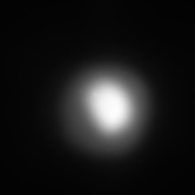}}\resizebox{0.24\hsize}{!}{\includegraphics{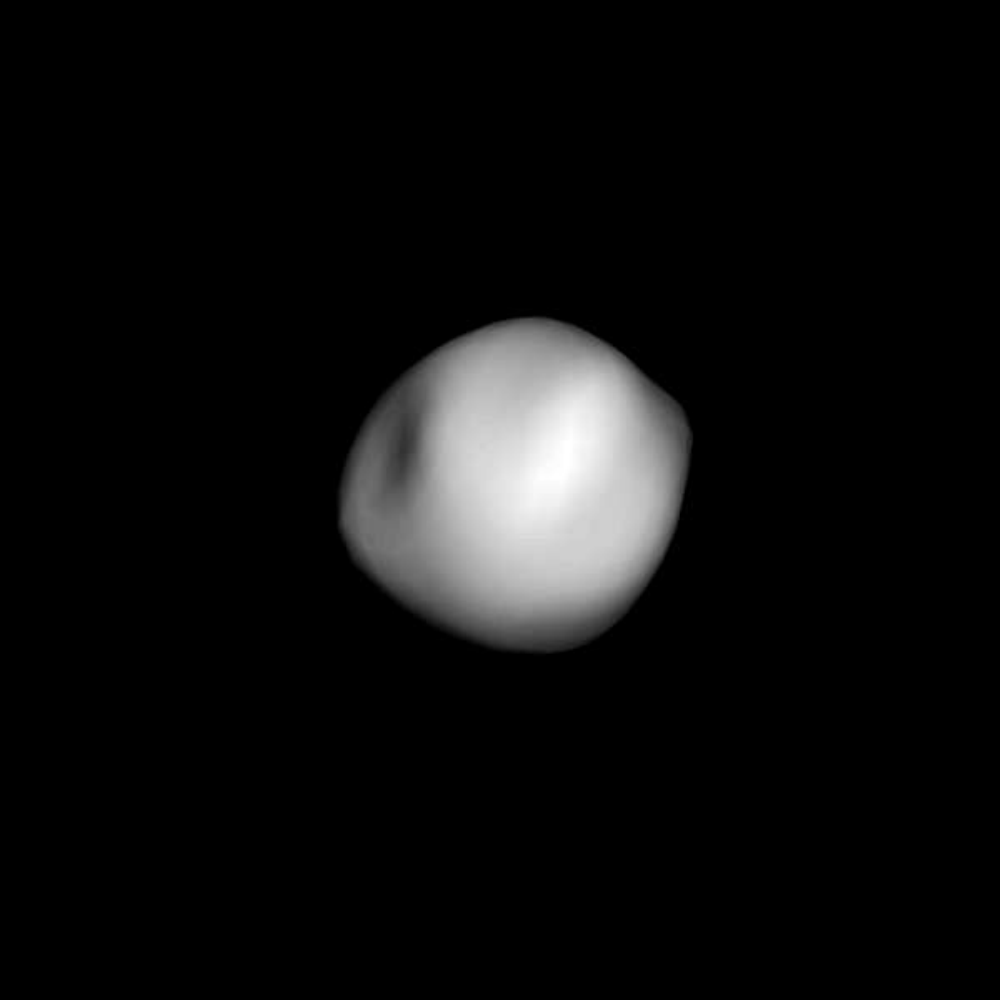}}\\
        \resizebox{0.24\hsize}{!}{\includegraphics{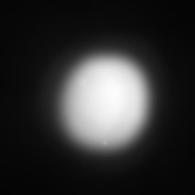}}\resizebox{0.24\hsize}{!}{\includegraphics{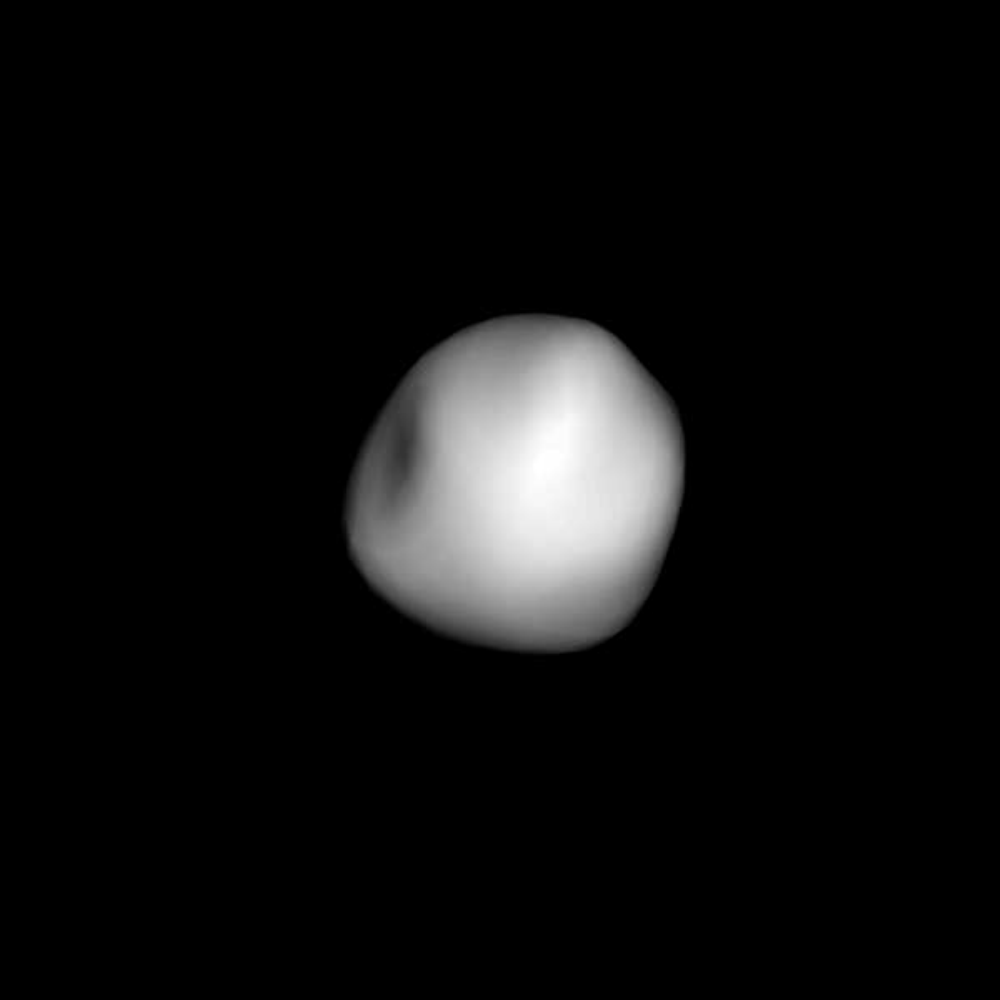}}\resizebox{0.24\hsize}{!}{\includegraphics{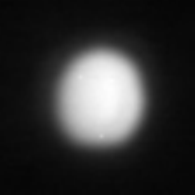}}\resizebox{0.24\hsize}{!}{\includegraphics{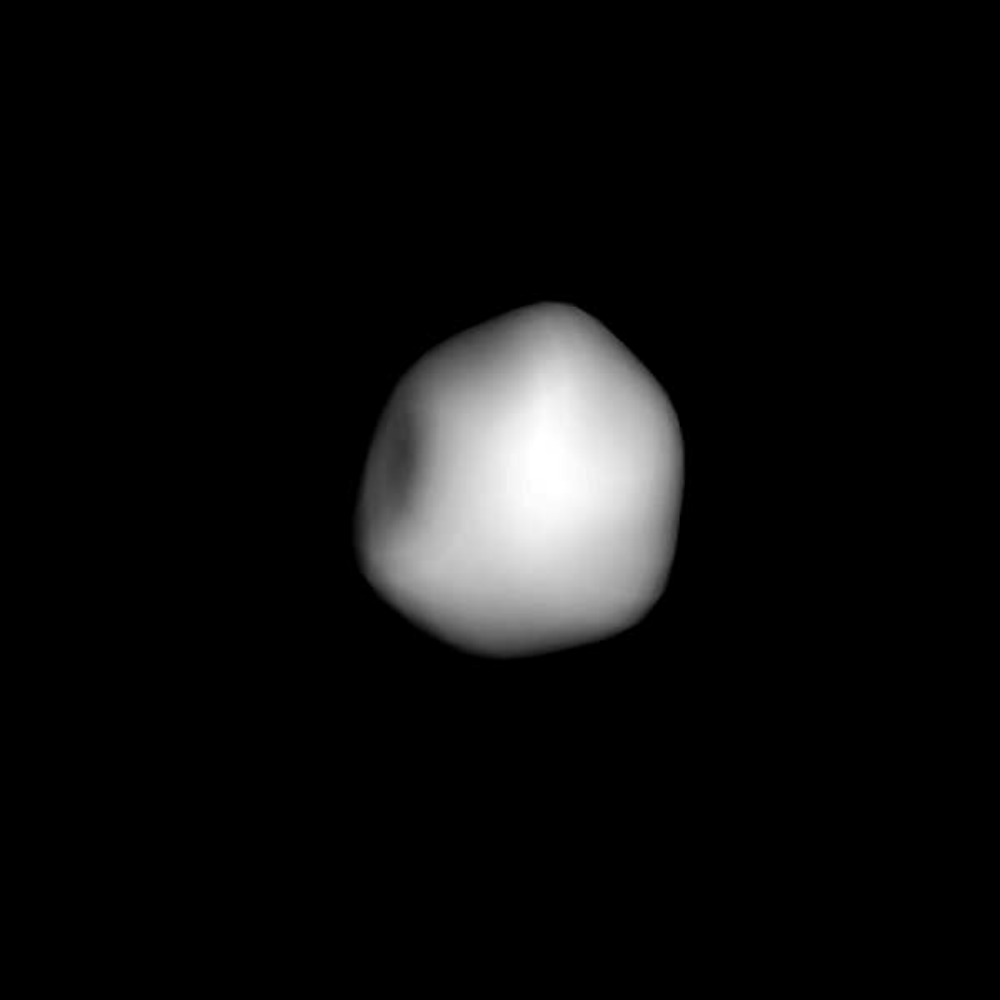}}\\
        \resizebox{0.24\hsize}{!}{\includegraphics{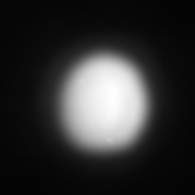}}\resizebox{0.24\hsize}{!}{\includegraphics{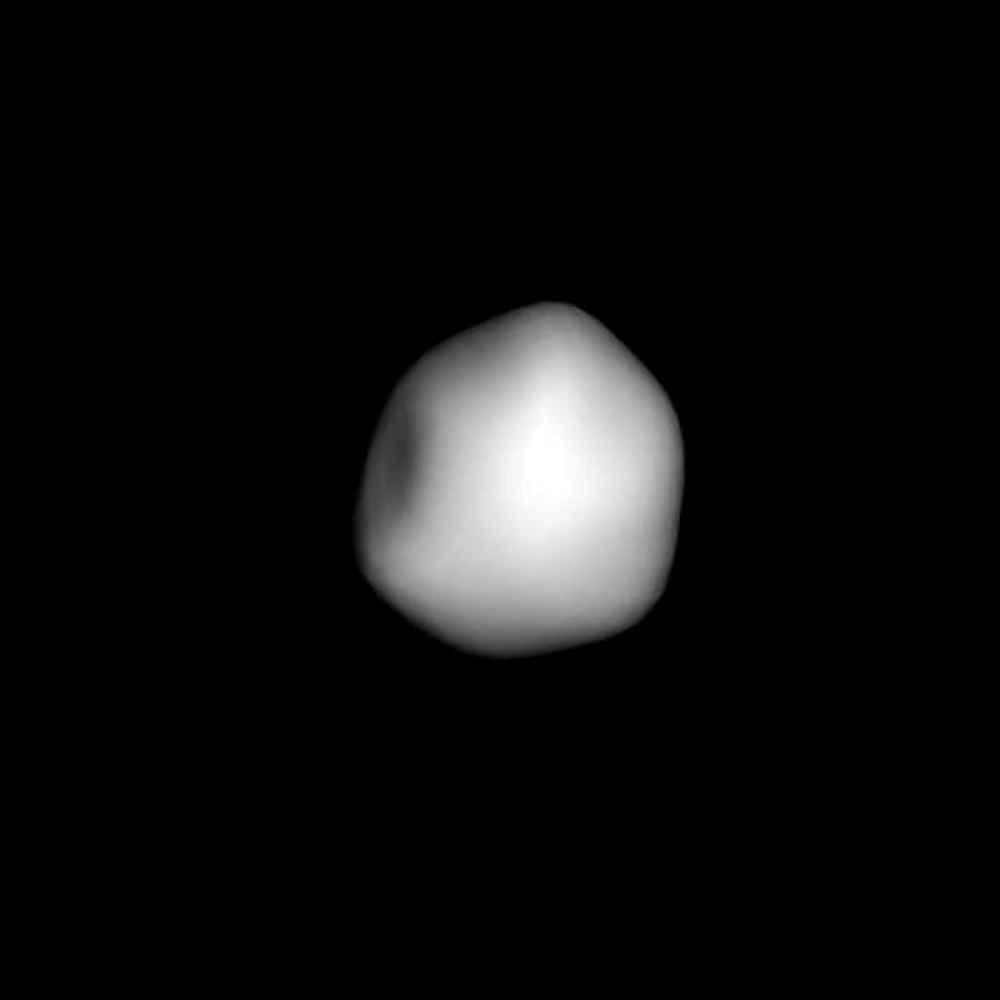}}\resizebox{0.24\hsize}{!}{\includegraphics{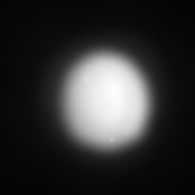}}\resizebox{0.24\hsize}{!}{\includegraphics{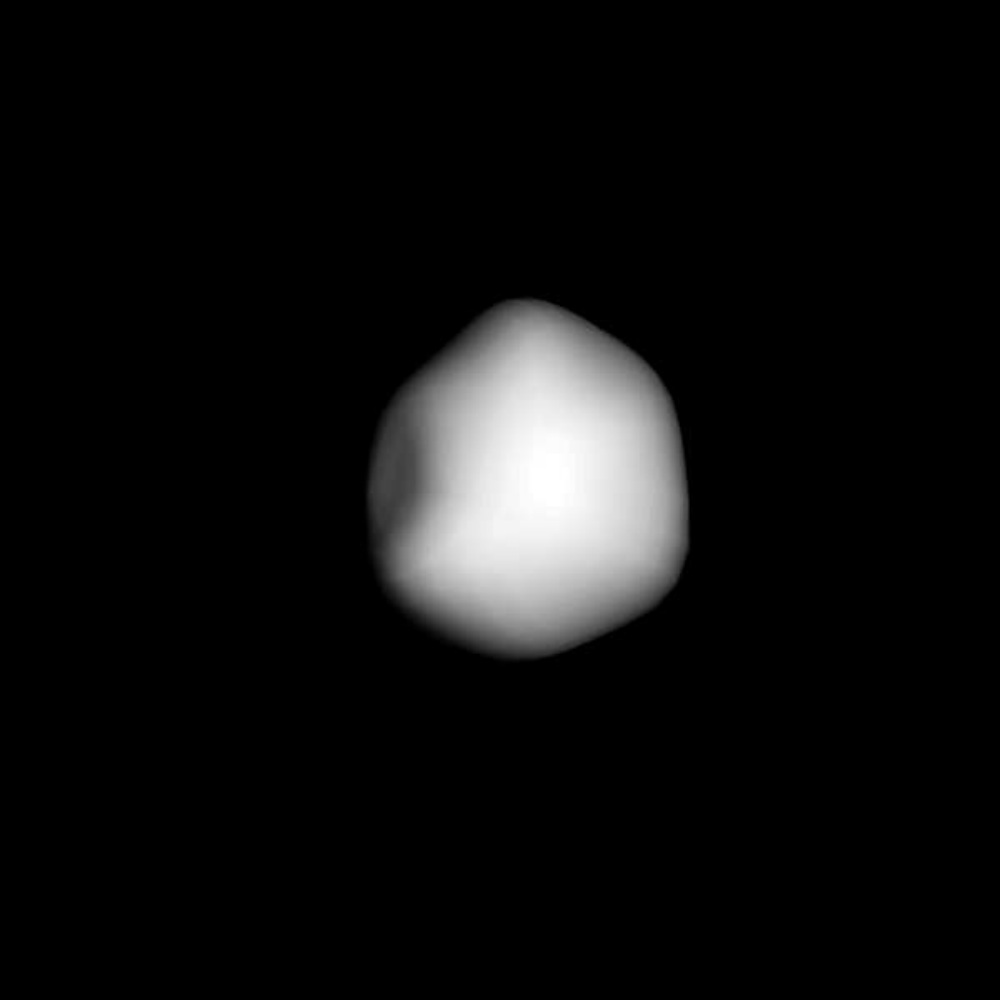}}\\
        \resizebox{0.24\hsize}{!}{\includegraphics{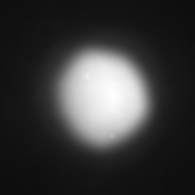}}\resizebox{0.24\hsize}{!}{\includegraphics{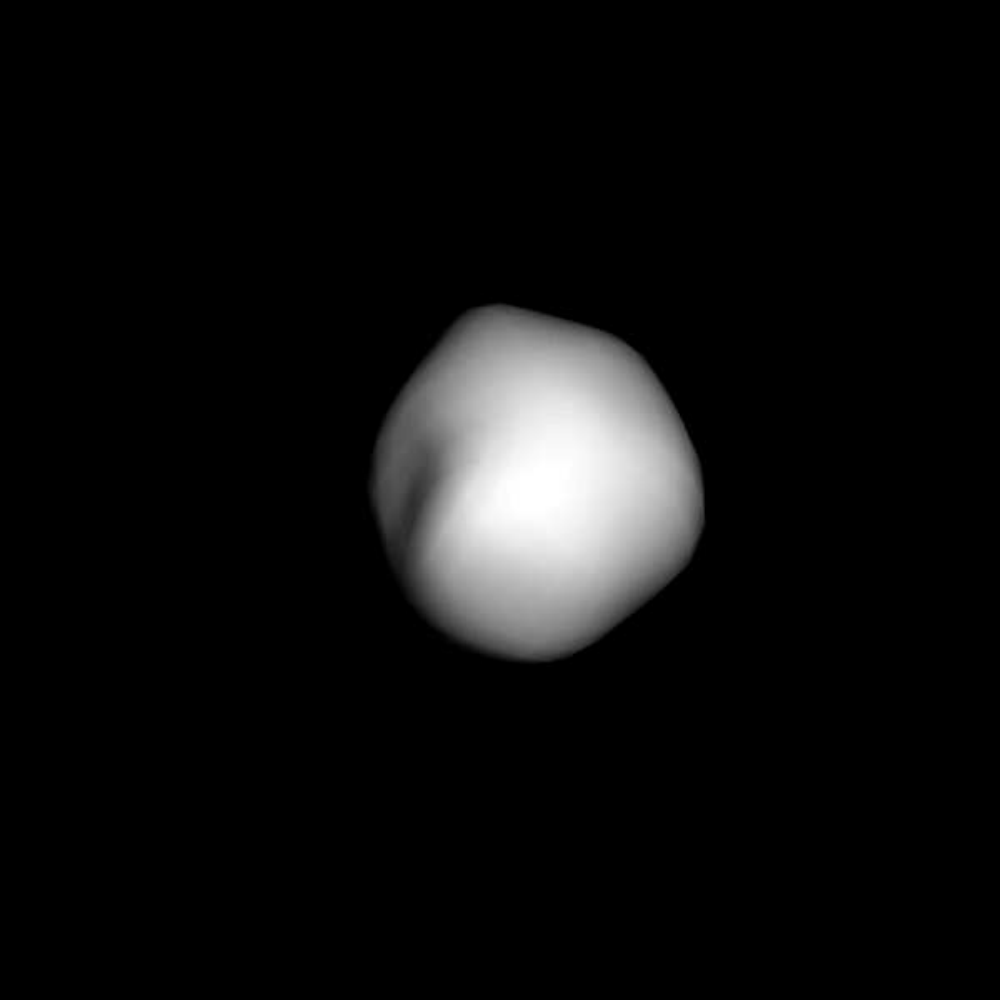}}\resizebox{0.24\hsize}{!}{\includegraphics{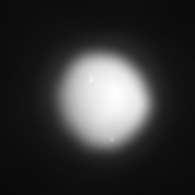}}\resizebox{0.24\hsize}{!}{\includegraphics{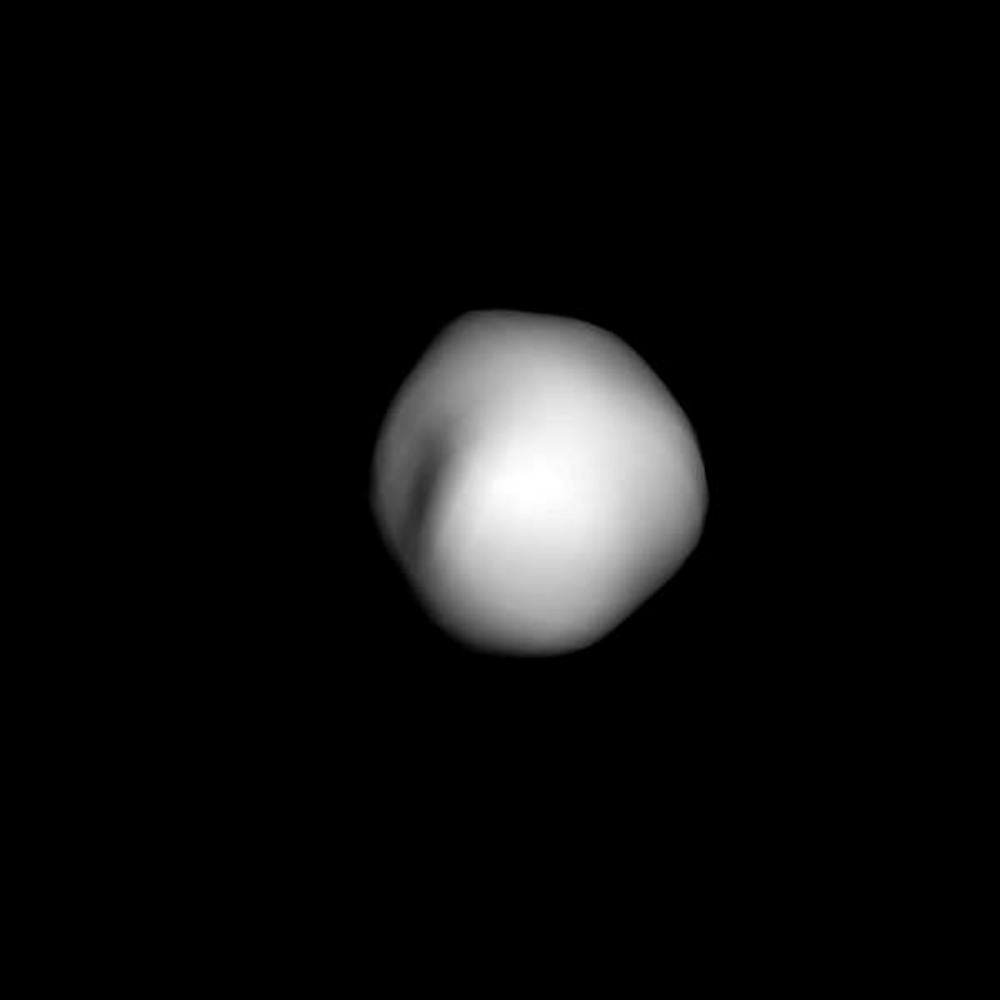}}\\
        \resizebox{0.24\hsize}{!}{\includegraphics{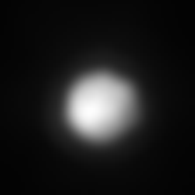}}\resizebox{0.24\hsize}{!}{\includegraphics{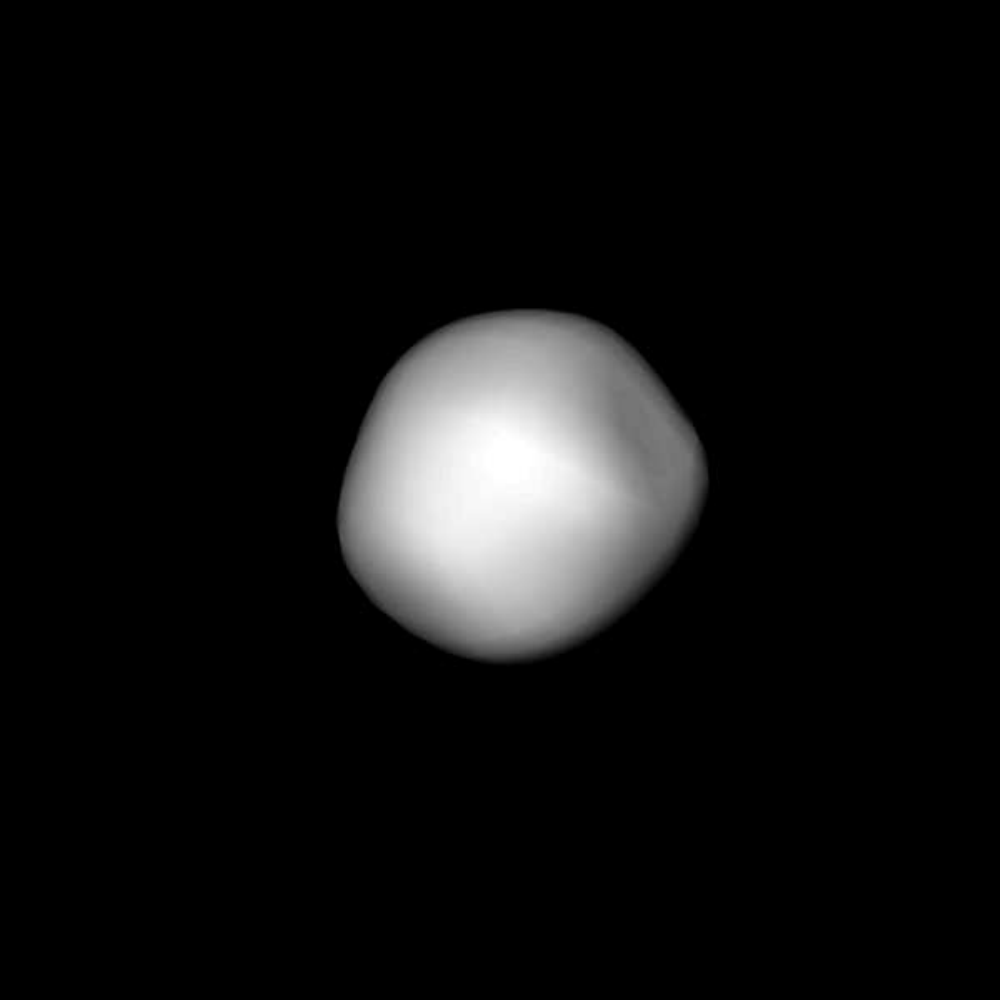}}\resizebox{0.24\hsize}{!}{\includegraphics{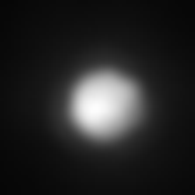}}\resizebox{0.24\hsize}{!}{\includegraphics{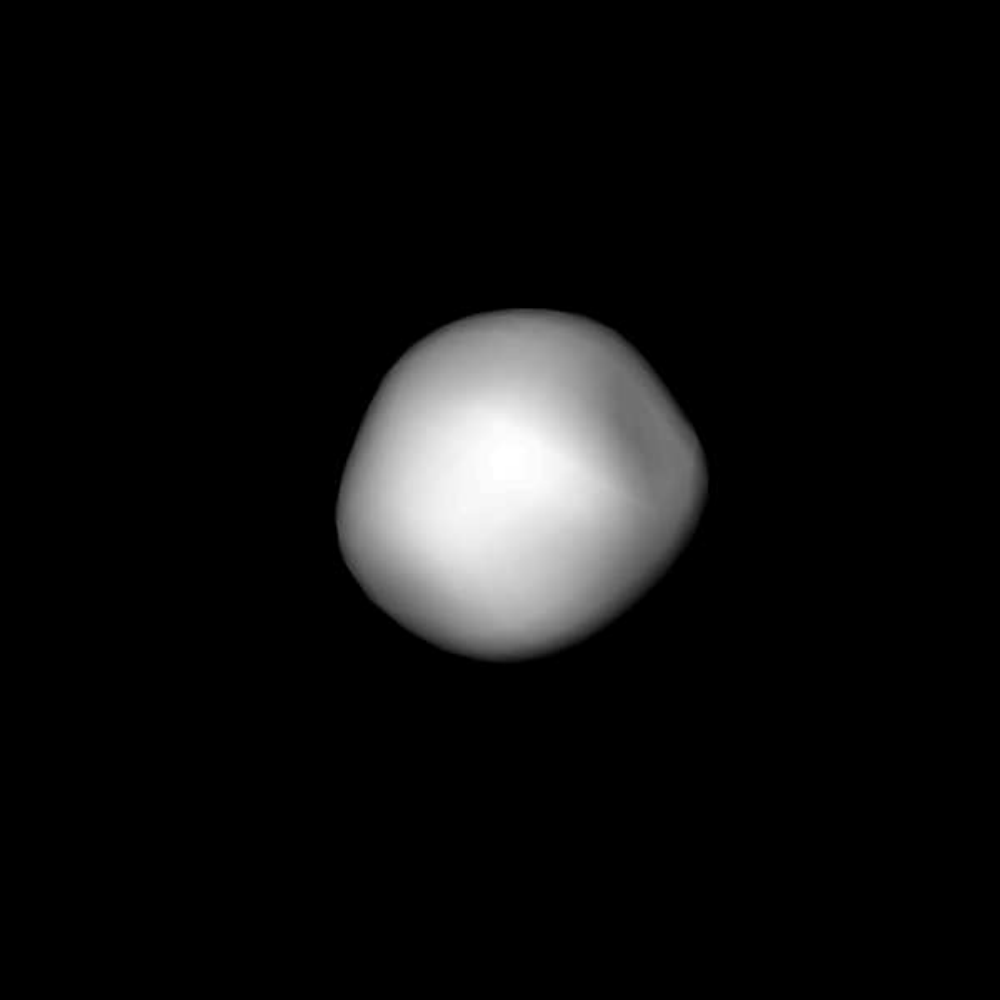}}\\
        \resizebox{0.24\hsize}{!}{\includegraphics{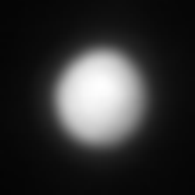}}\resizebox{0.24\hsize}{!}{\includegraphics{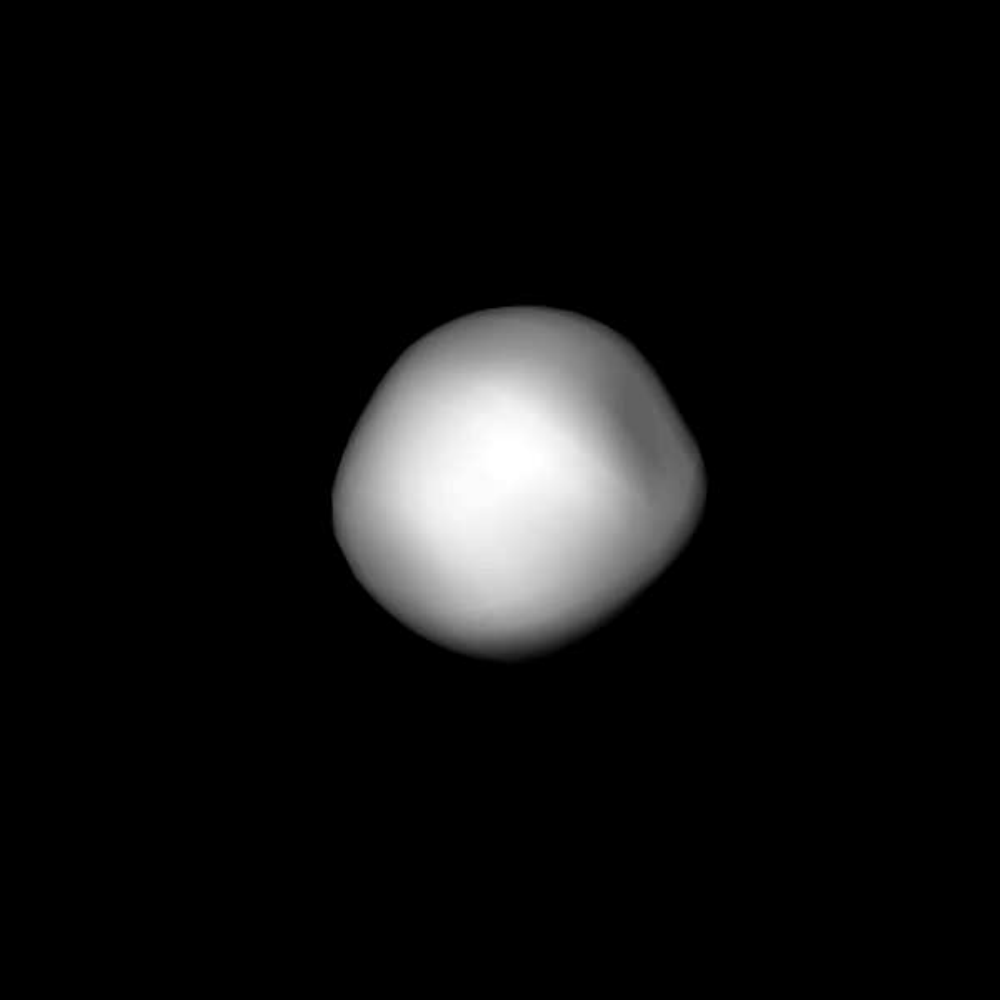}}\resizebox{0.24\hsize}{!}{\includegraphics{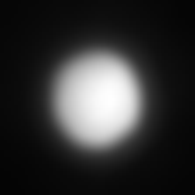}}\resizebox{0.24\hsize}{!}{\includegraphics{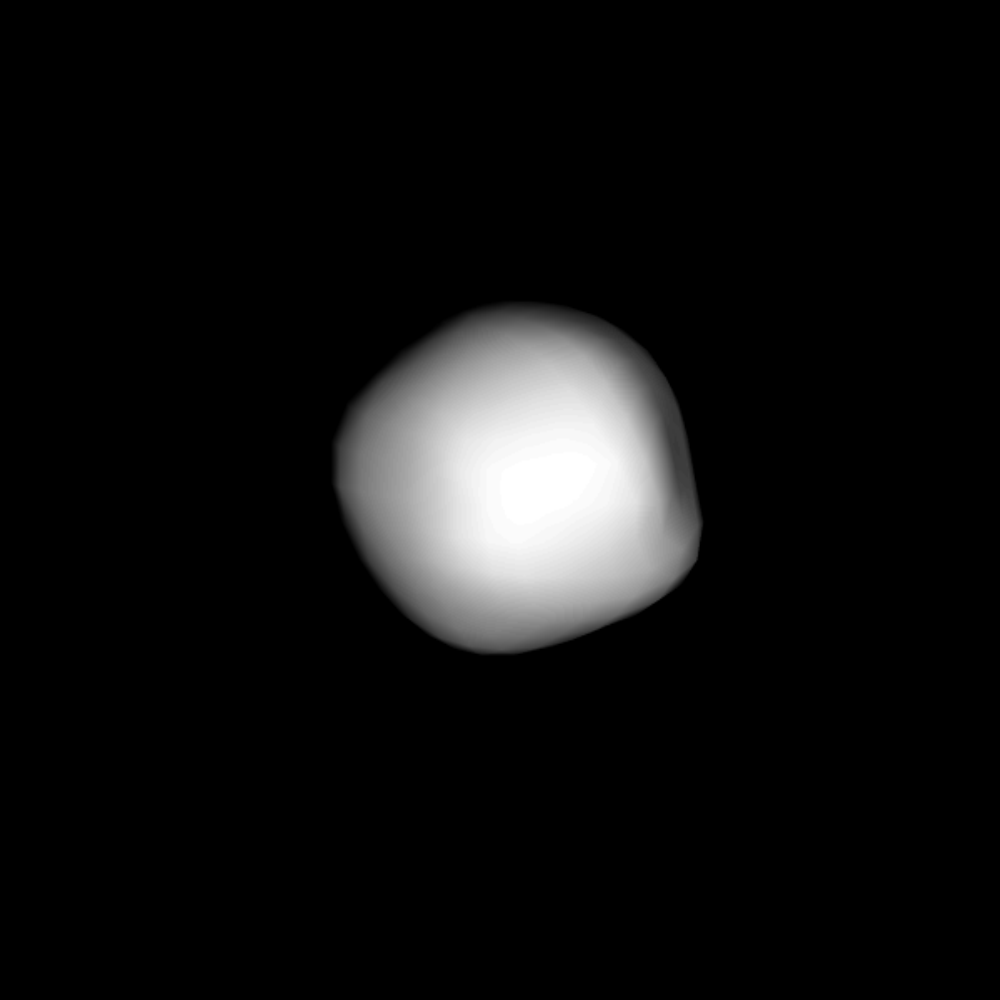}}\\
    \caption{\label{fig:2}Comparison between model projections and corresponding AO images for asteroid (2) Pallas.}
\end{figure}

\begin{figure}[tbp]
    \centering
        \resizebox{0.24\hsize}{!}{\includegraphics{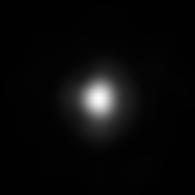}}\resizebox{0.24\hsize}{!}{\includegraphics{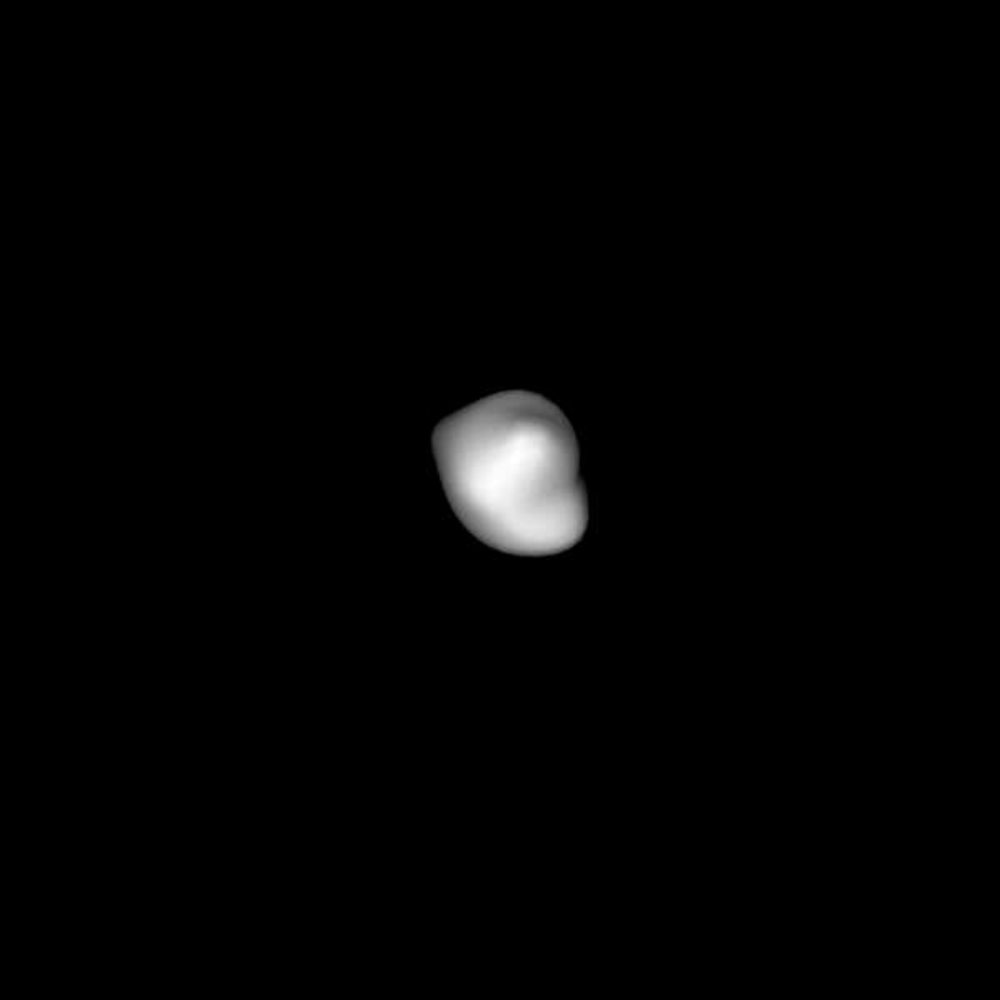}}\resizebox{0.24\hsize}{!}{\includegraphics{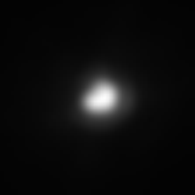}}\resizebox{0.24\hsize}{!}{\includegraphics{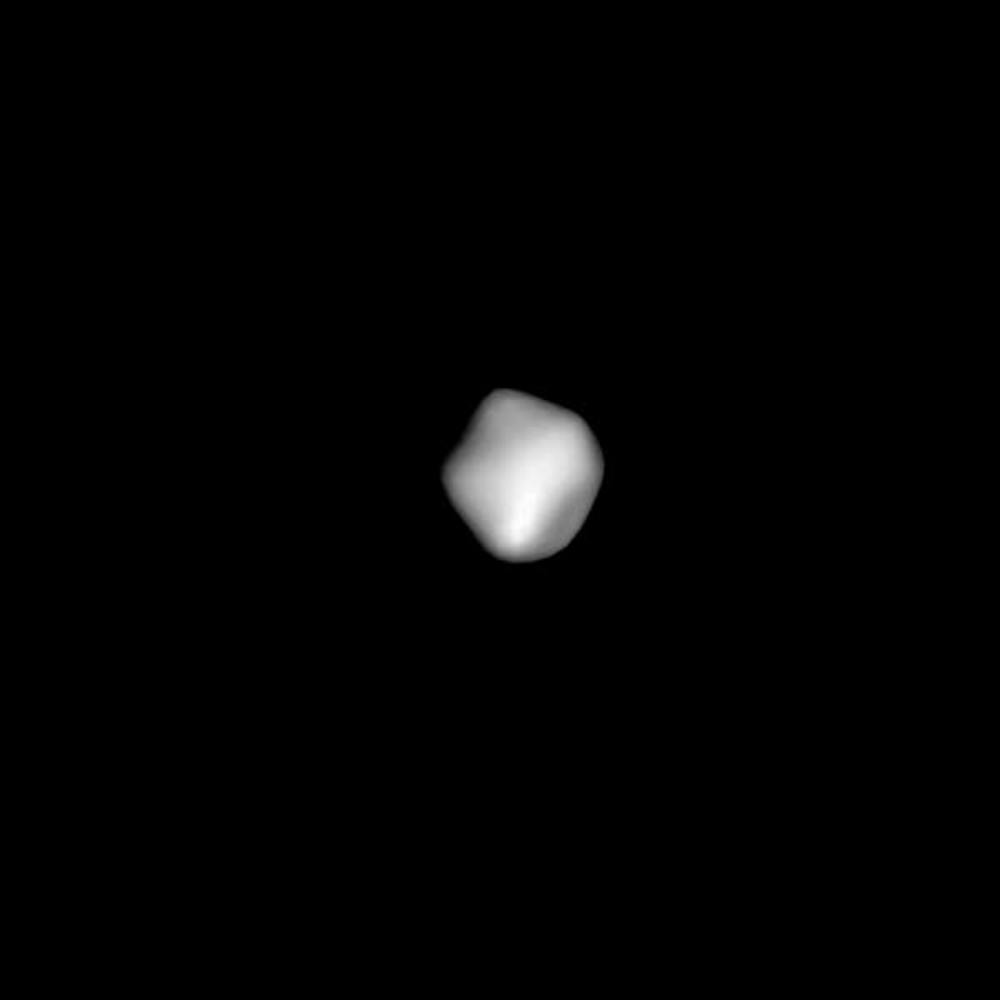}}\\
    \caption{\label{fig:5}Comparison between model projections and corresponding AO images for asteroid (5) Astraea.}
\end{figure}

\begin{figure}[tbp]
    \centering
        \resizebox{0.24\hsize}{!}{\includegraphics{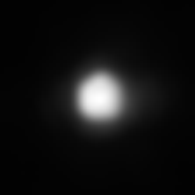}}\resizebox{0.24\hsize}{!}{\includegraphics{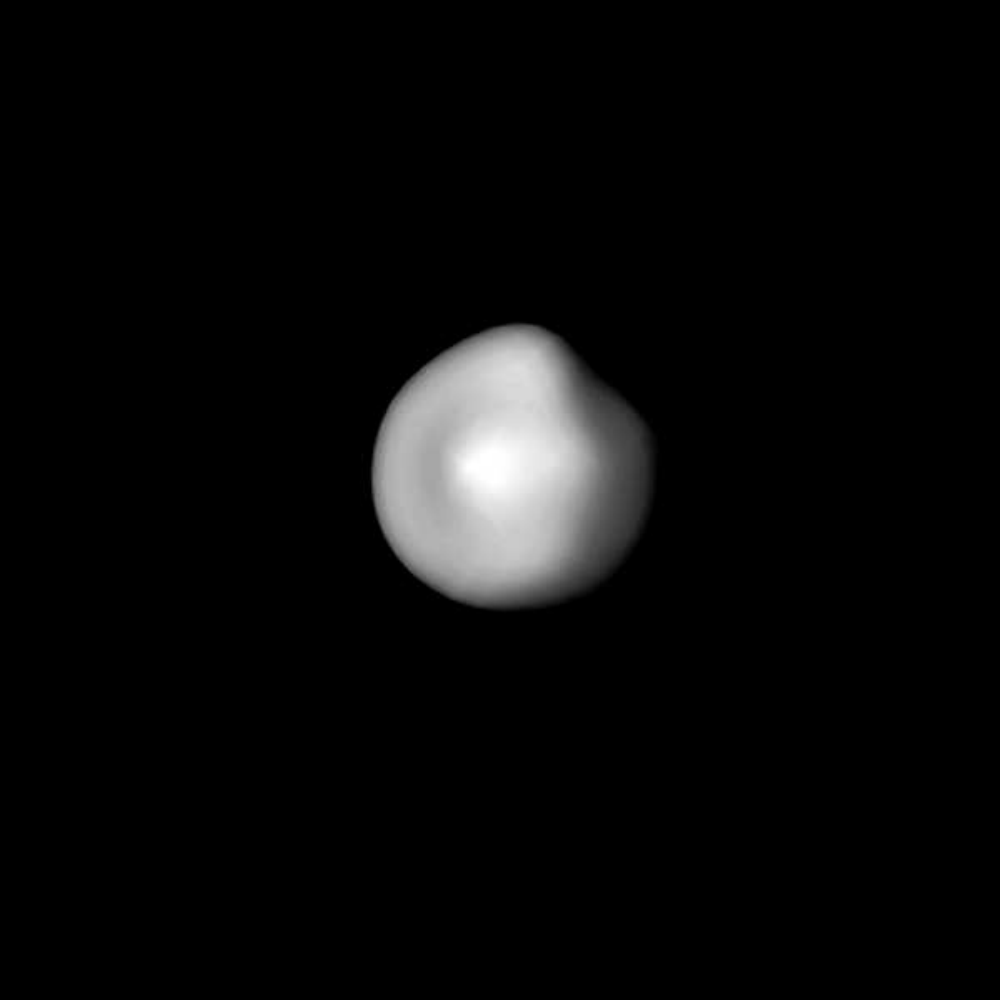}}\resizebox{0.24\hsize}{!}{\includegraphics{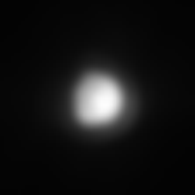}}\resizebox{0.24\hsize}{!}{\includegraphics{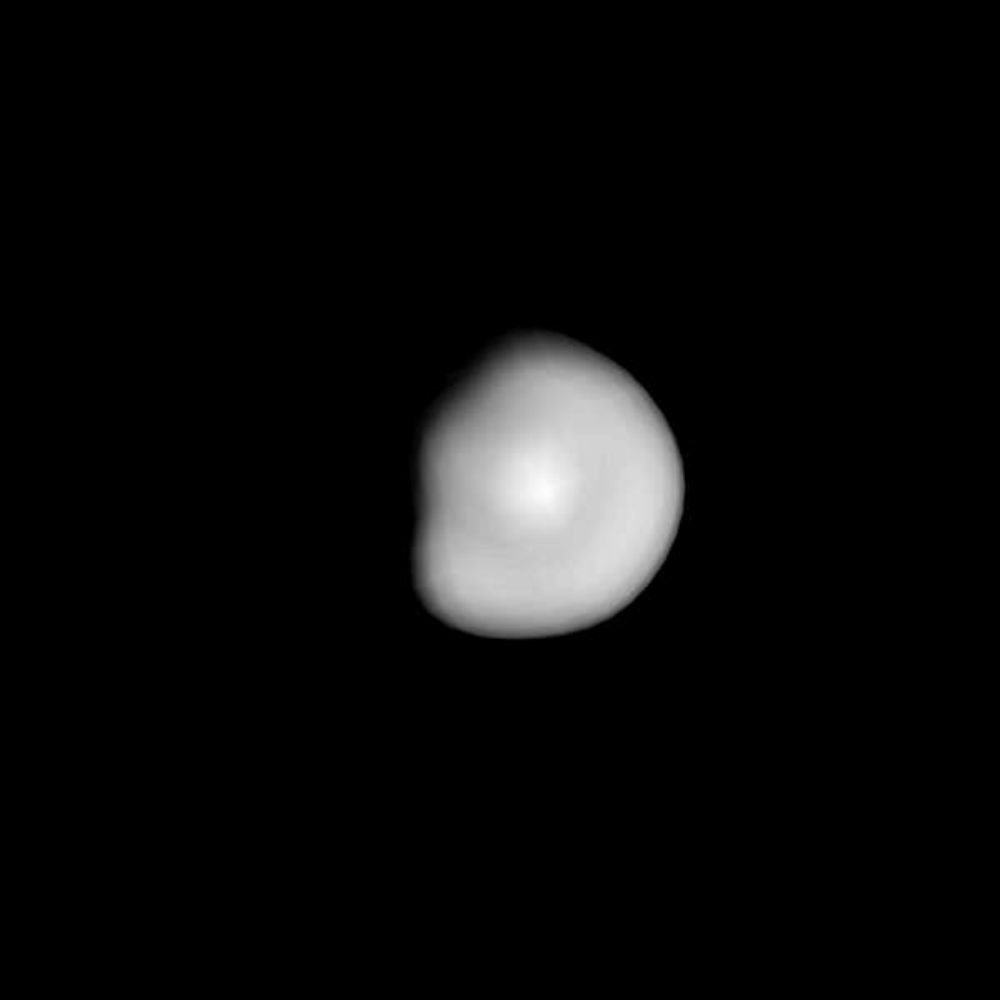}}\\
        \resizebox{0.24\hsize}{!}{\includegraphics{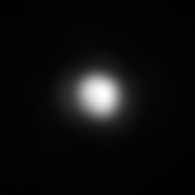}}\resizebox{0.24\hsize}{!}{\includegraphics{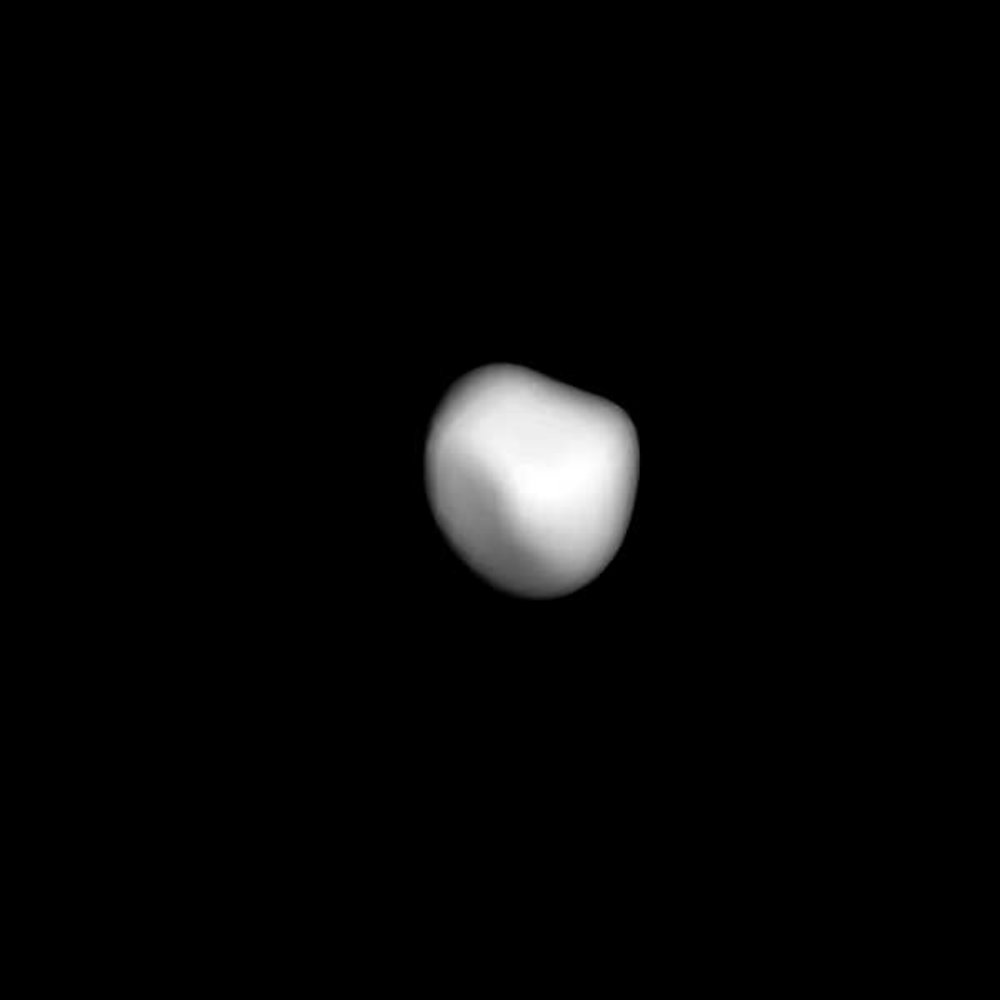}}\resizebox{0.24\hsize}{!}{\includegraphics{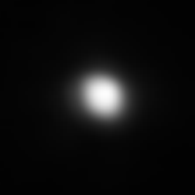}}\resizebox{0.24\hsize}{!}{\includegraphics{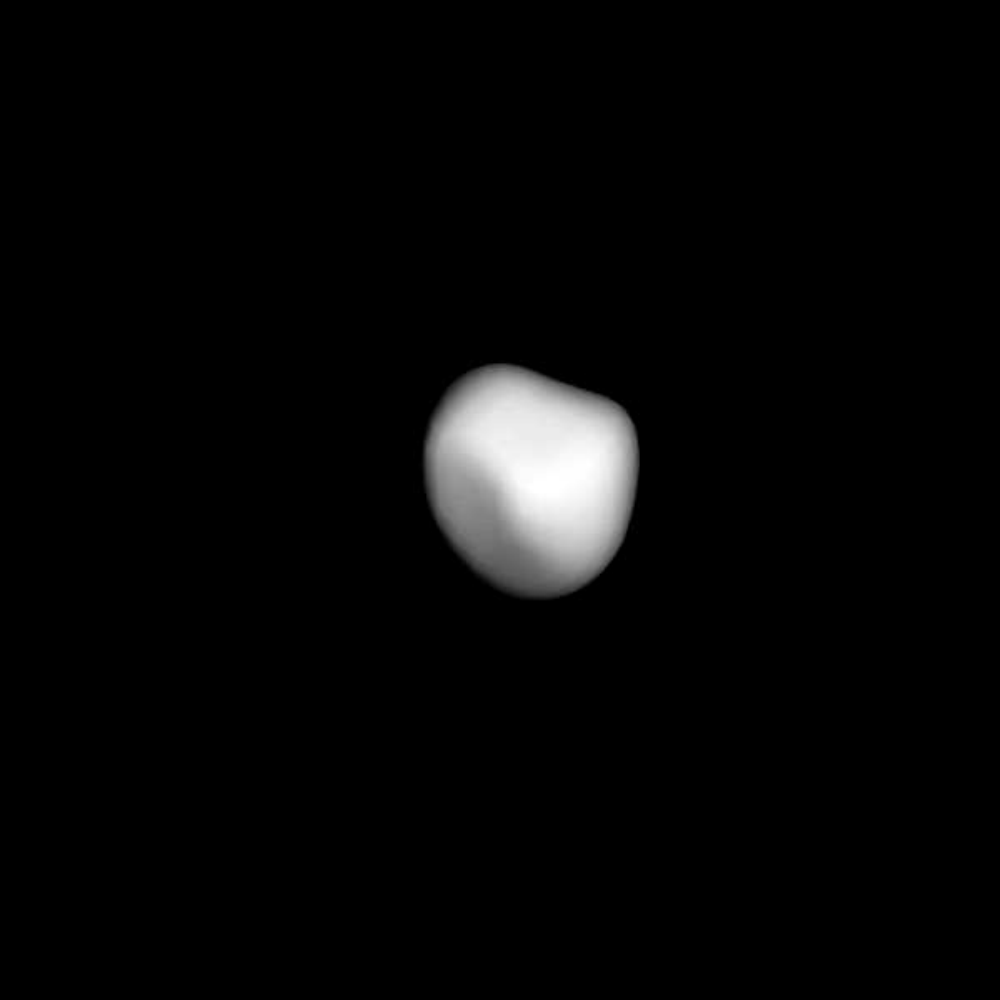}}\\
        \resizebox{0.24\hsize}{!}{\includegraphics{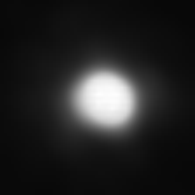}}\resizebox{0.24\hsize}{!}{\includegraphics{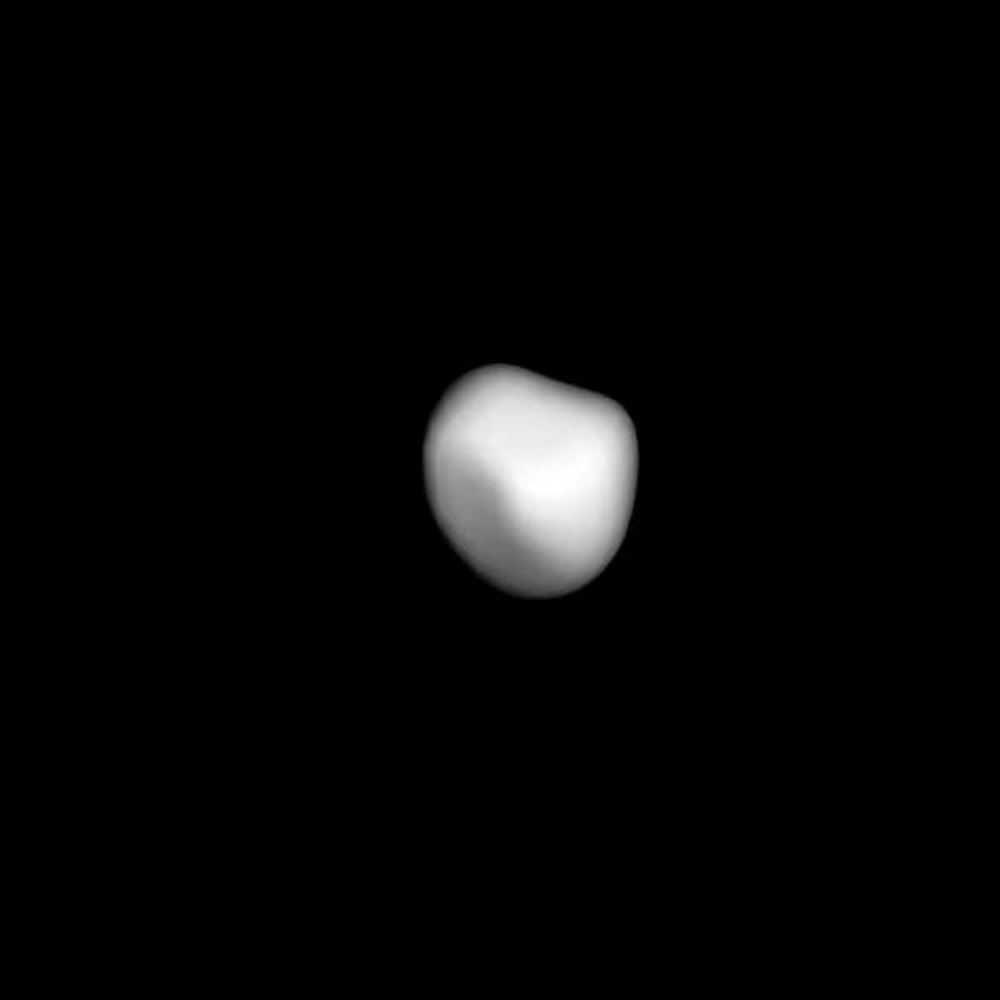}}\resizebox{0.24\hsize}{!}{\includegraphics{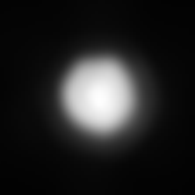}}\resizebox{0.24\hsize}{!}{\includegraphics{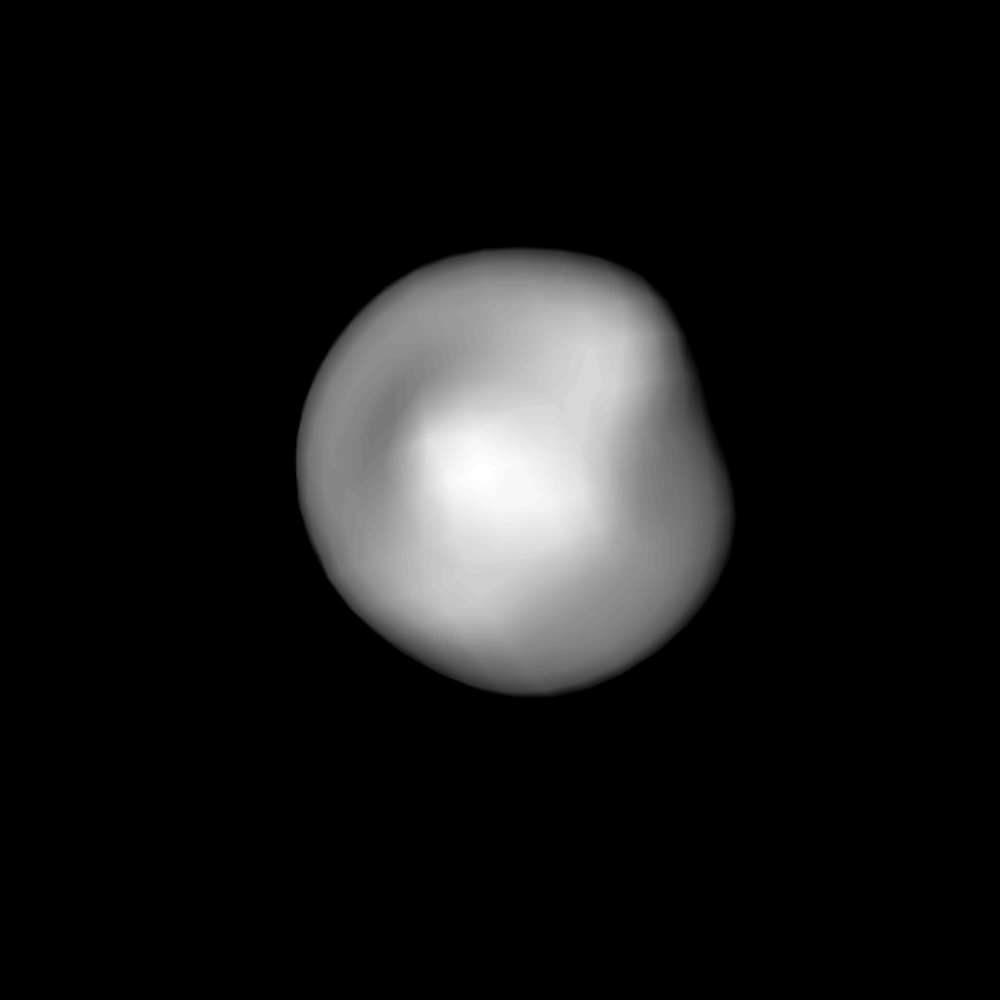}}\\
    \caption{\label{fig:8}Comparison between model projections and corresponding AO images for asteroid (8) Flora.}
\end{figure}

\begin{figure}[tbp]
    \centering
        \resizebox{0.24\hsize}{!}{\includegraphics{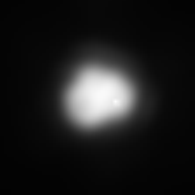}}\resizebox{0.24\hsize}{!}{\includegraphics{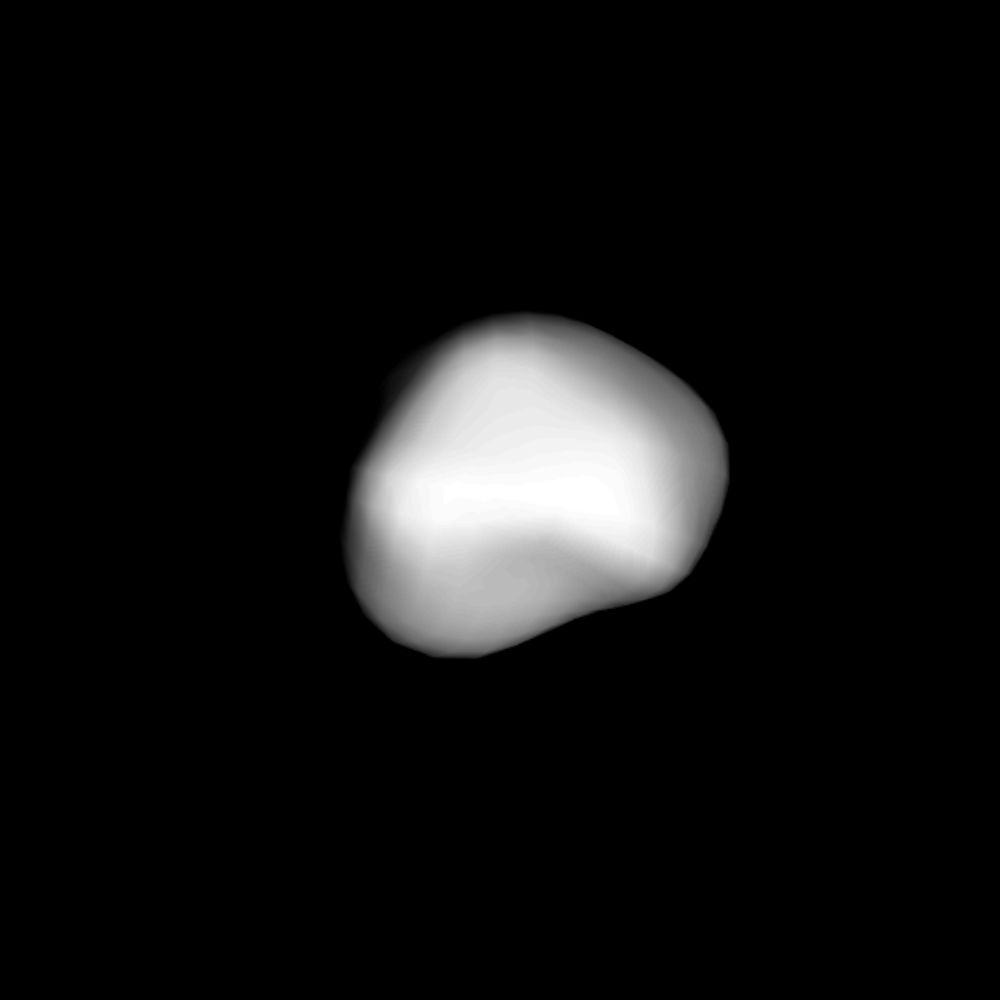}}\resizebox{0.24\hsize}{!}{\includegraphics{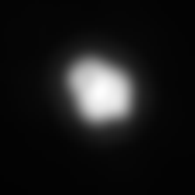}}\resizebox{0.24\hsize}{!}{\includegraphics{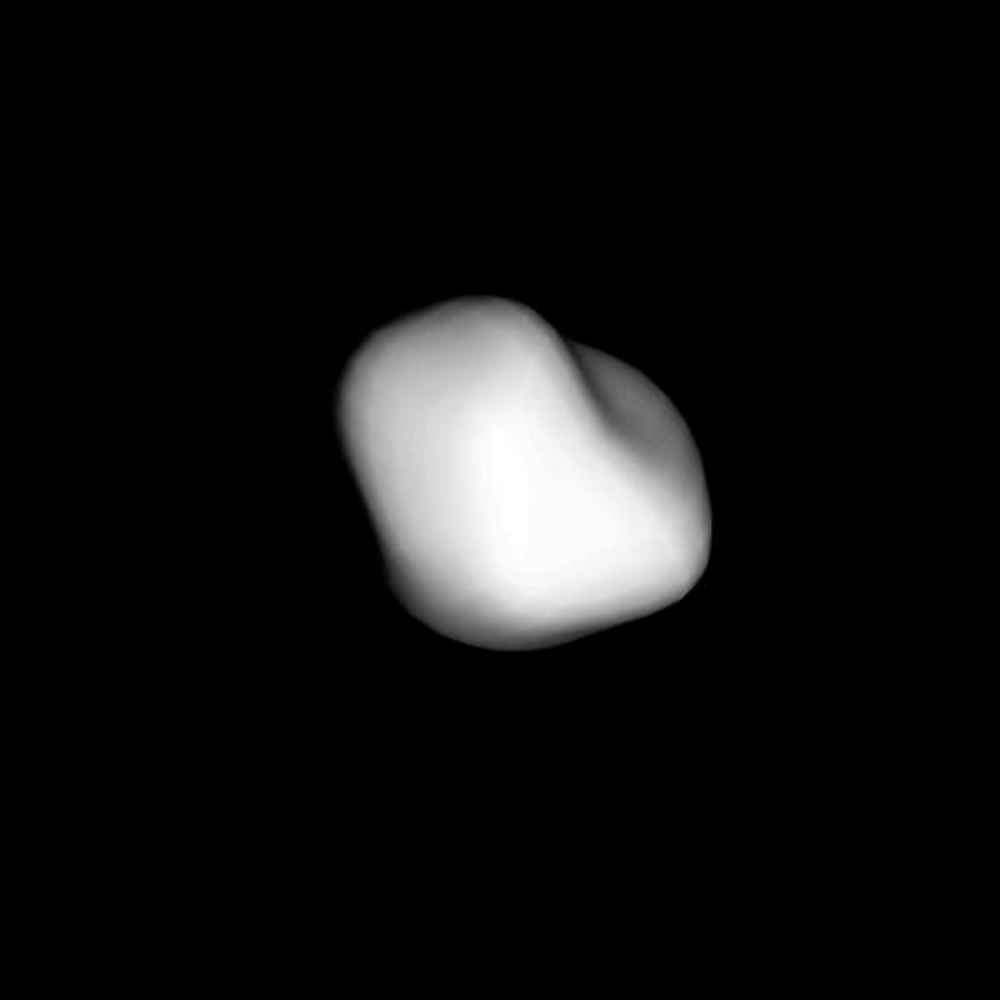}}\\
        \resizebox{0.24\hsize}{!}{\includegraphics{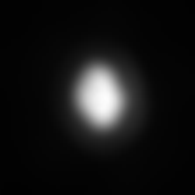}}\resizebox{0.24\hsize}{!}{\includegraphics{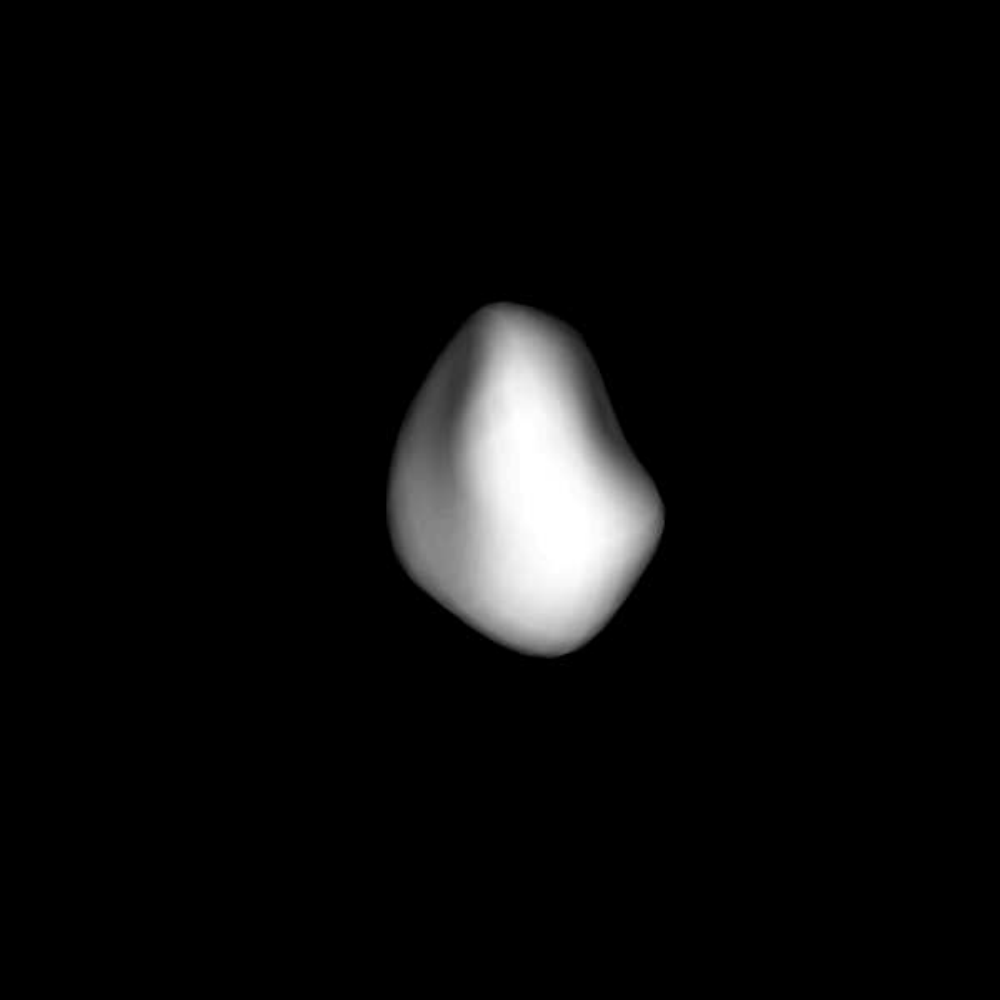}}\resizebox{0.24\hsize}{!}{\includegraphics{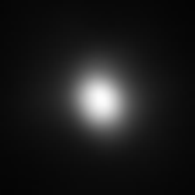}}\resizebox{0.24\hsize}{!}{\includegraphics{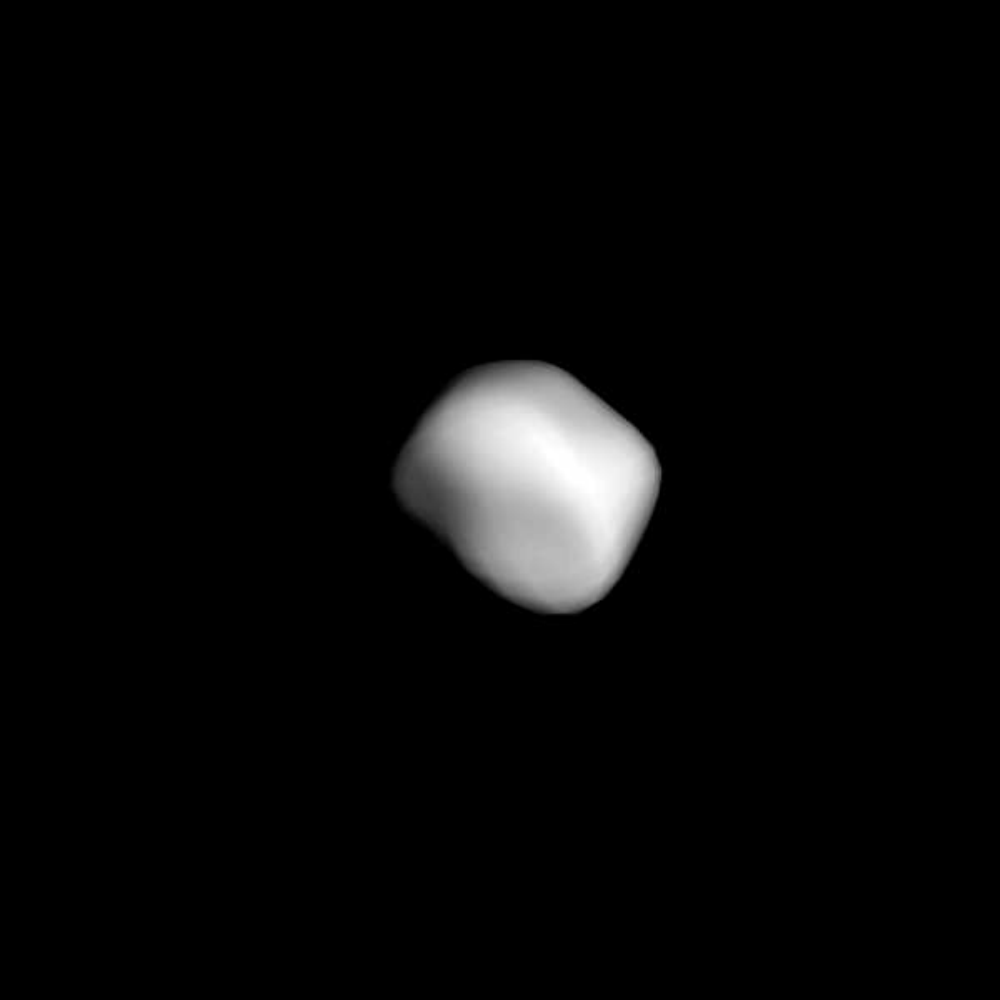}}\\
        \resizebox{0.24\hsize}{!}{\includegraphics{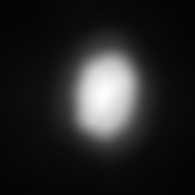}}\resizebox{0.24\hsize}{!}{\includegraphics{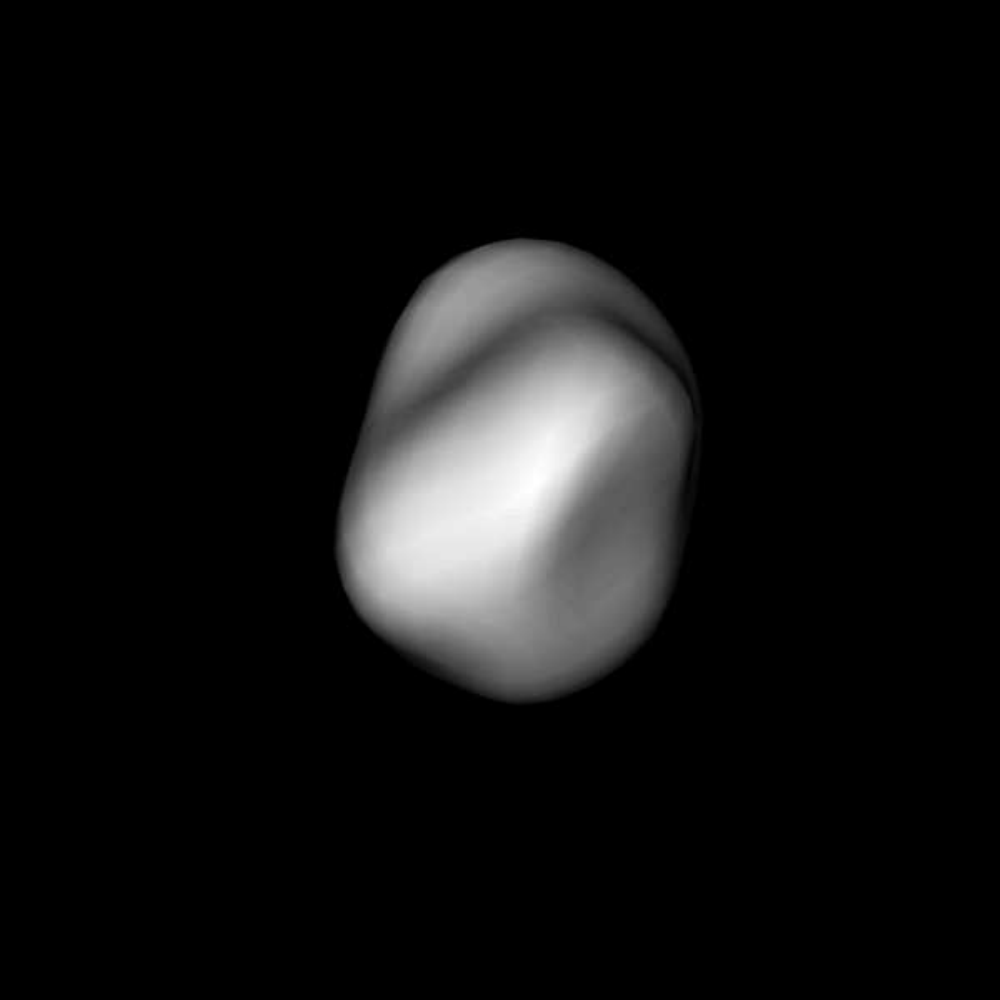}}\resizebox{0.24\hsize}{!}{\includegraphics{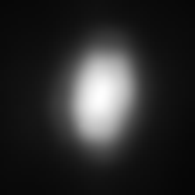}}\resizebox{0.24\hsize}{!}{\includegraphics{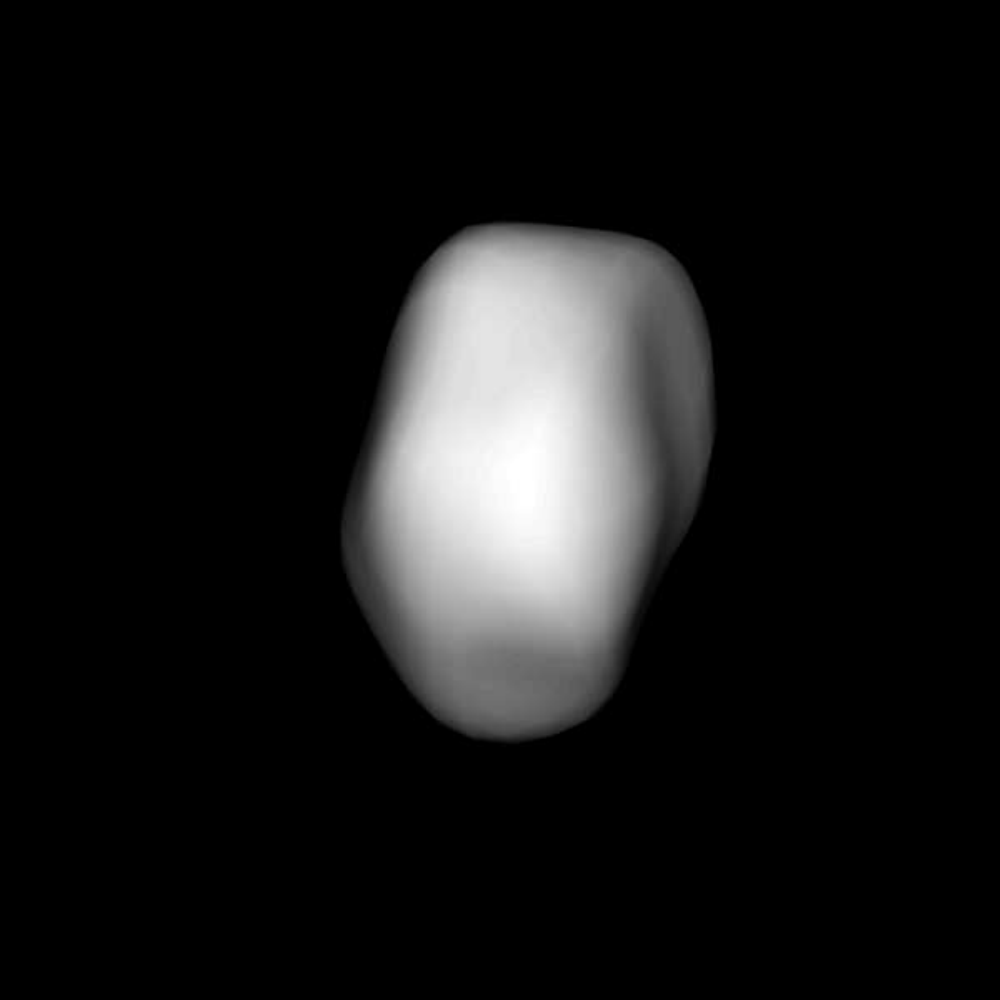}}\\
        \resizebox{0.24\hsize}{!}{\includegraphics{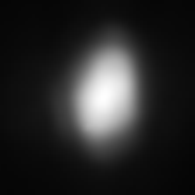}}\resizebox{0.24\hsize}{!}{\includegraphics{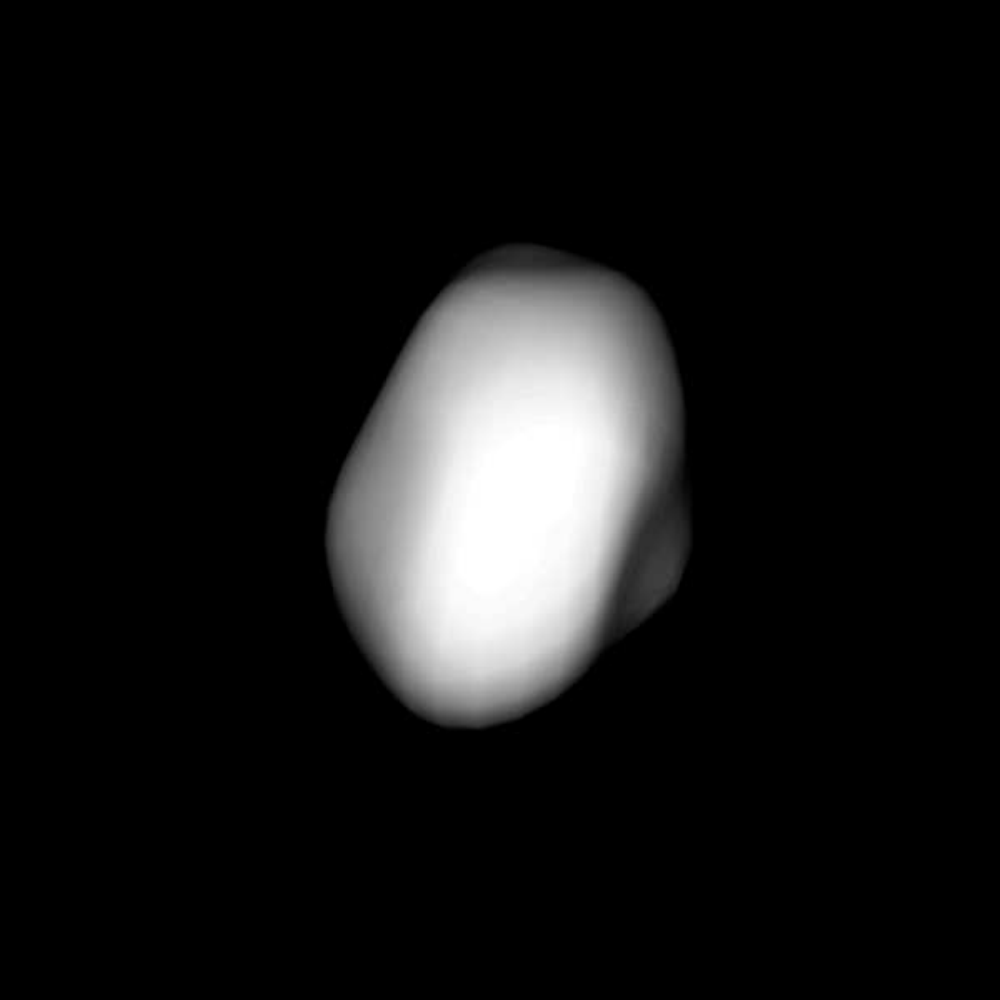}}\resizebox{0.24\hsize}{!}{\includegraphics{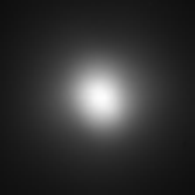}}\resizebox{0.24\hsize}{!}{\includegraphics{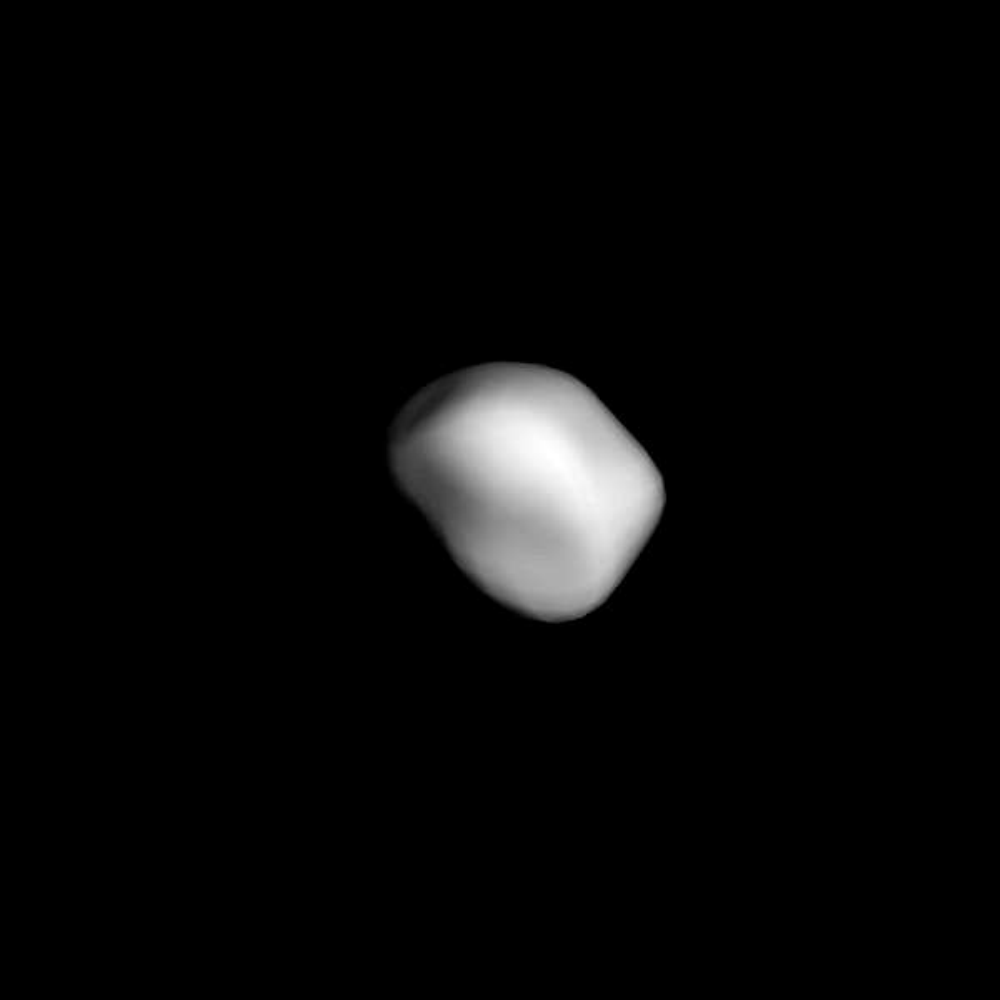}}\\
    \caption{\label{fig:9}Comparison between model projections and corresponding AO images for asteroid (9) Metis.}
\end{figure}

\begin{figure}[tbp]
    \centering
        \resizebox{0.24\hsize}{!}{\includegraphics{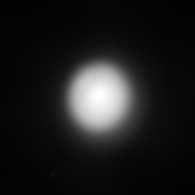}}\resizebox{0.24\hsize}{!}{\includegraphics{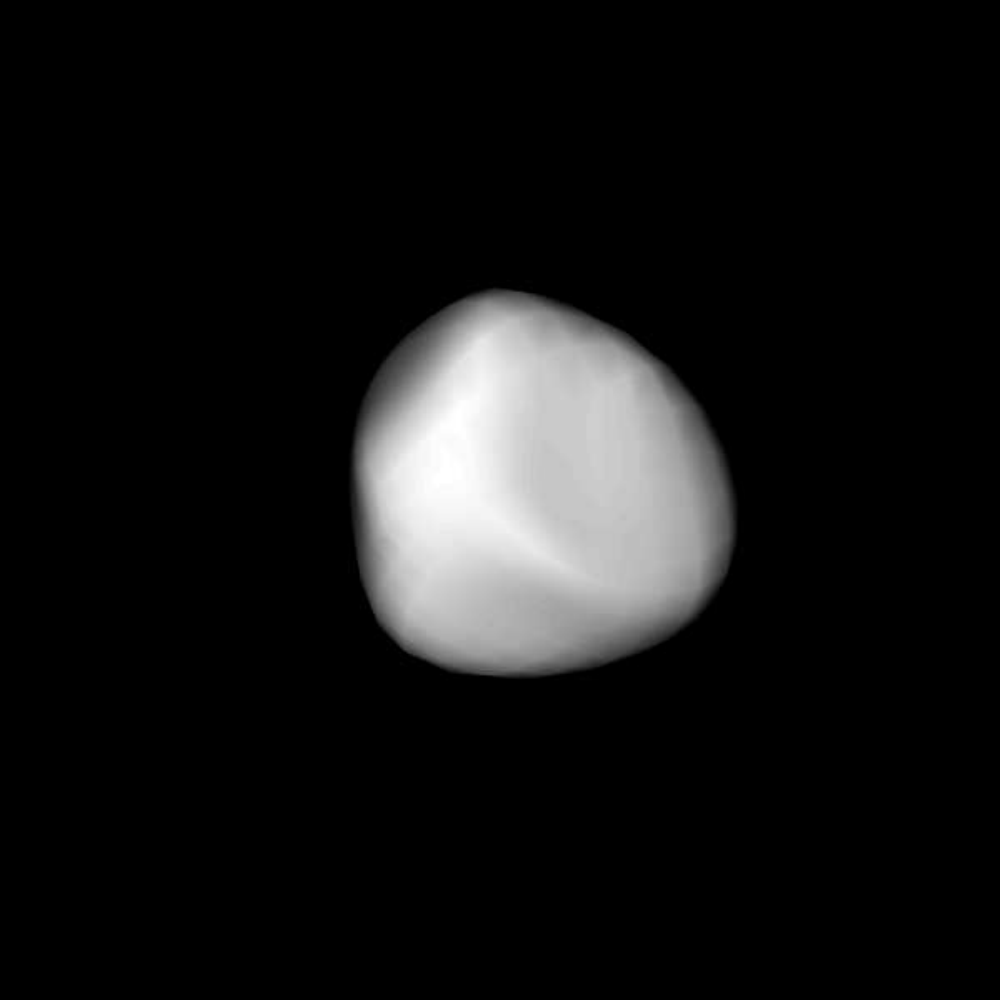}}\resizebox{0.24\hsize}{!}{\includegraphics{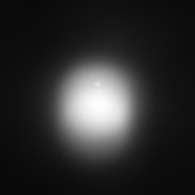}}\resizebox{0.24\hsize}{!}{\includegraphics{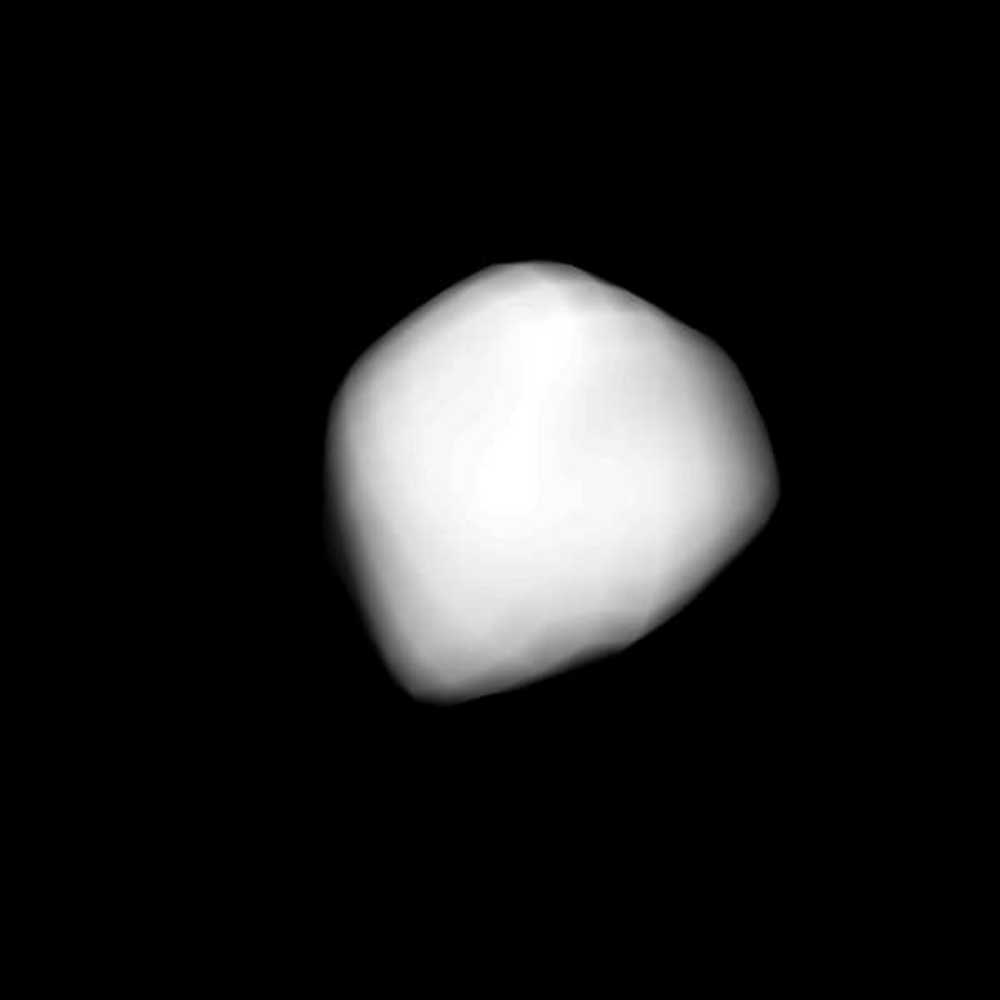}}\\
    \caption{\label{fig:10}Comparison between model projections and corresponding AO images for asteroid (10) Hygiea.}
\end{figure}

\begin{figure}[tbp]
    \centering
        \resizebox{0.24\hsize}{!}{\includegraphics{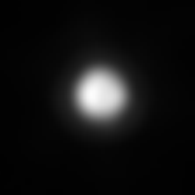}}\resizebox{0.24\hsize}{!}{\includegraphics{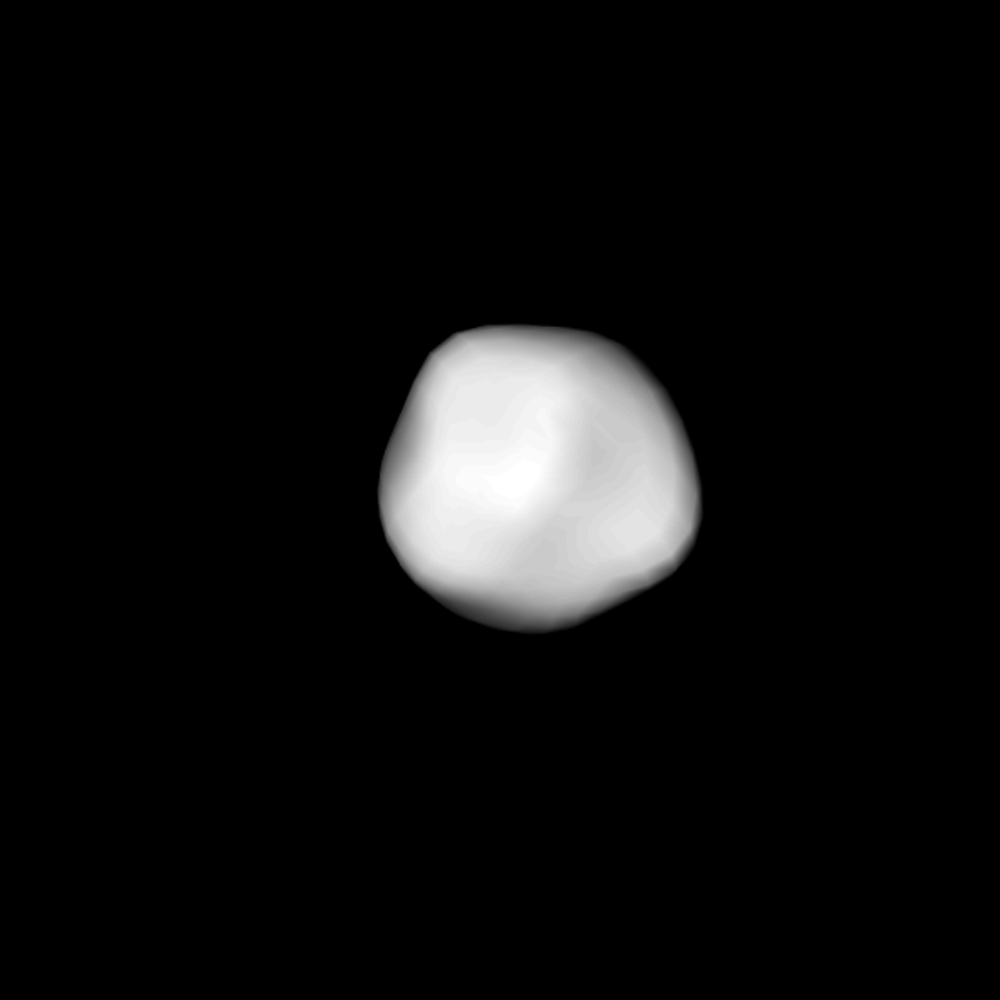}}\\
    \caption{\label{fig:11}Comparison between model projections and corresponding AO images for asteroid (11) Parthenope.}
\end{figure}

\begin{figure}[tbp]
    \centering
        \resizebox{0.24\hsize}{!}{\includegraphics{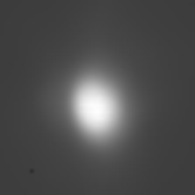}}\resizebox{0.24\hsize}{!}{\includegraphics{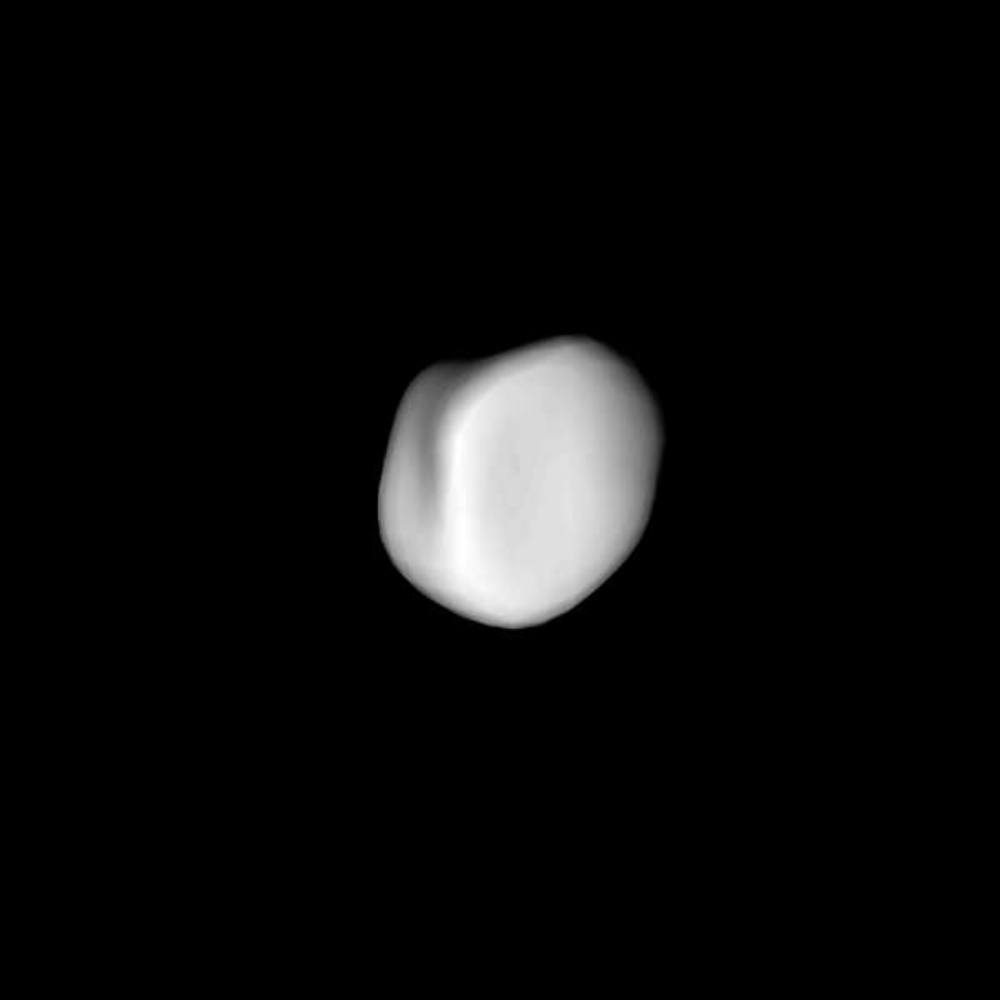}}\\
    \caption{\label{fig:13}Comparison between model projections and corresponding AO images for asteroid (13) Egeria.}
\end{figure}

\begin{figure}[tbp]
    \centering
        \resizebox{0.24\hsize}{!}{\includegraphics{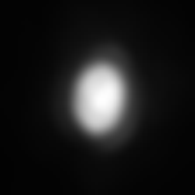}}\resizebox{0.24\hsize}{!}{\includegraphics{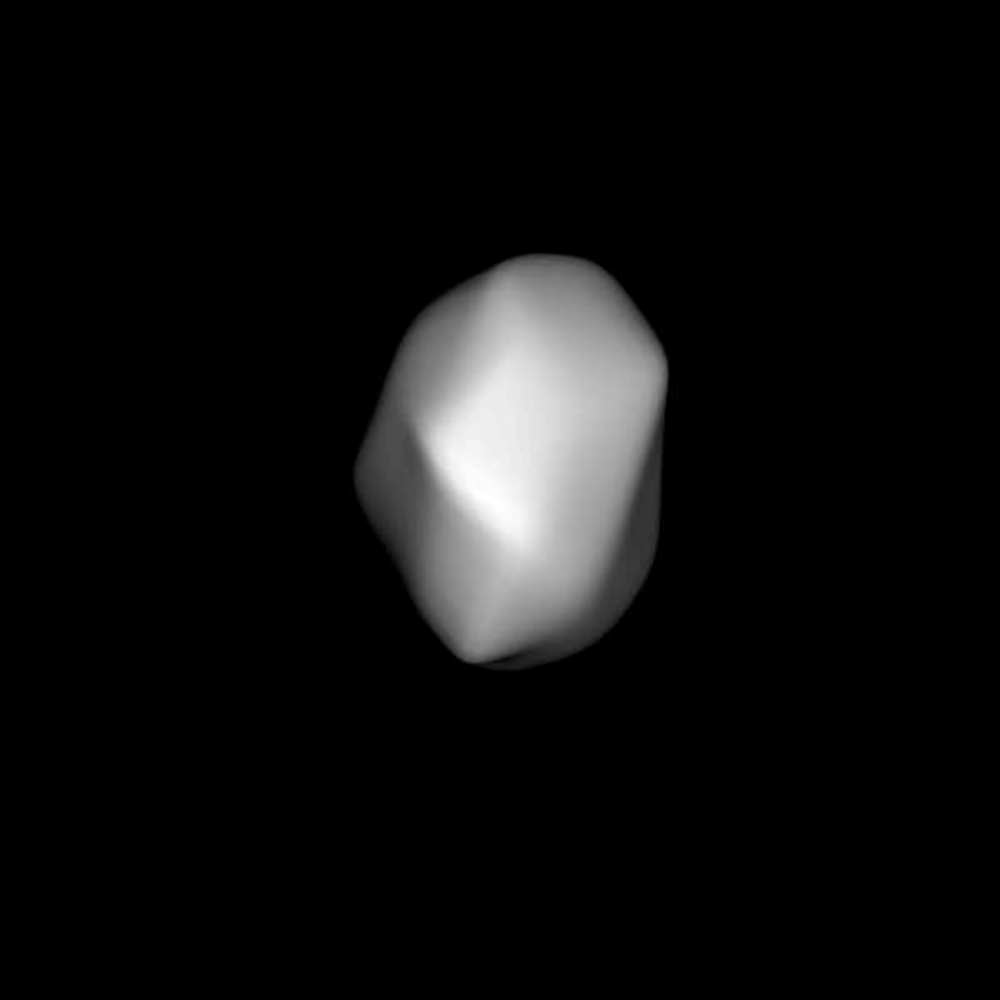}}\resizebox{0.24\hsize}{!}{\includegraphics{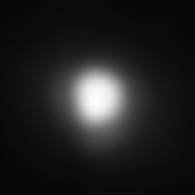}}\resizebox{0.24\hsize}{!}{\includegraphics{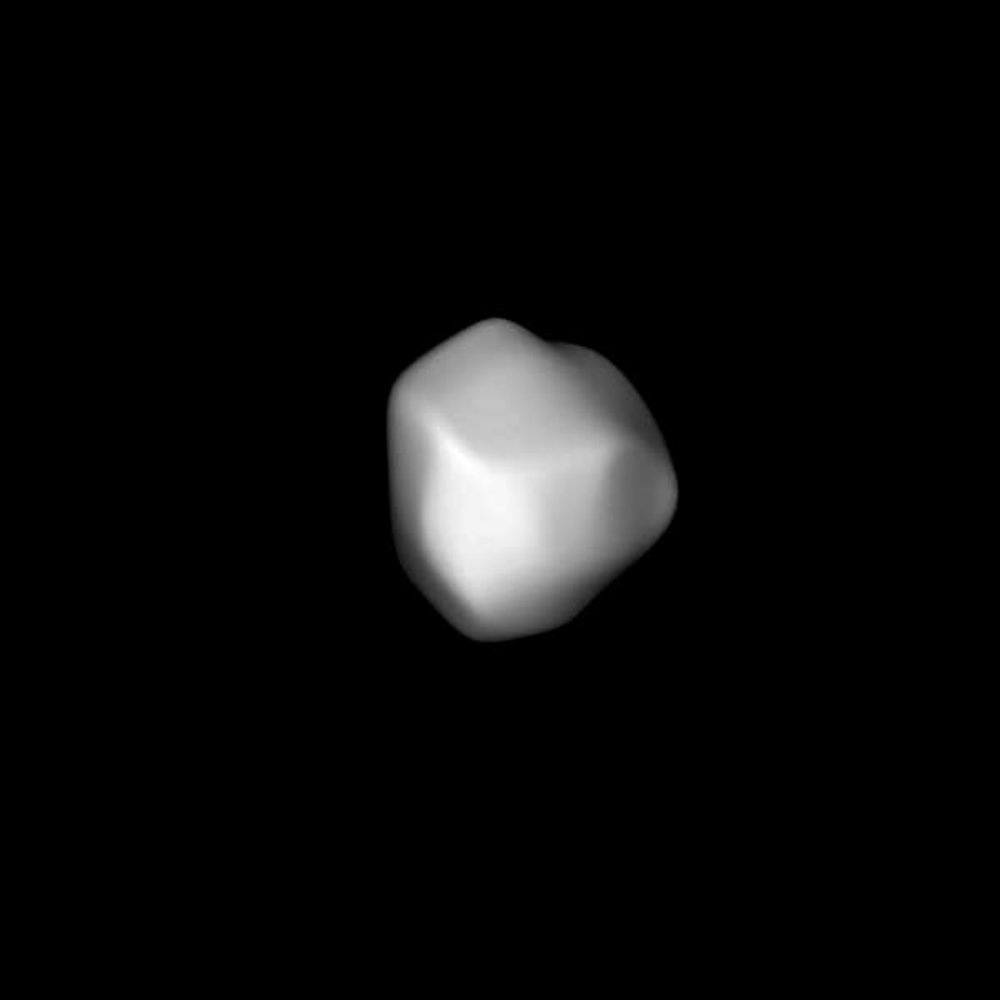}}\\
        \resizebox{0.24\hsize}{!}{\includegraphics{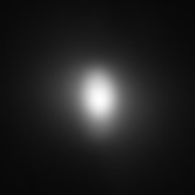}}\resizebox{0.24\hsize}{!}{\includegraphics{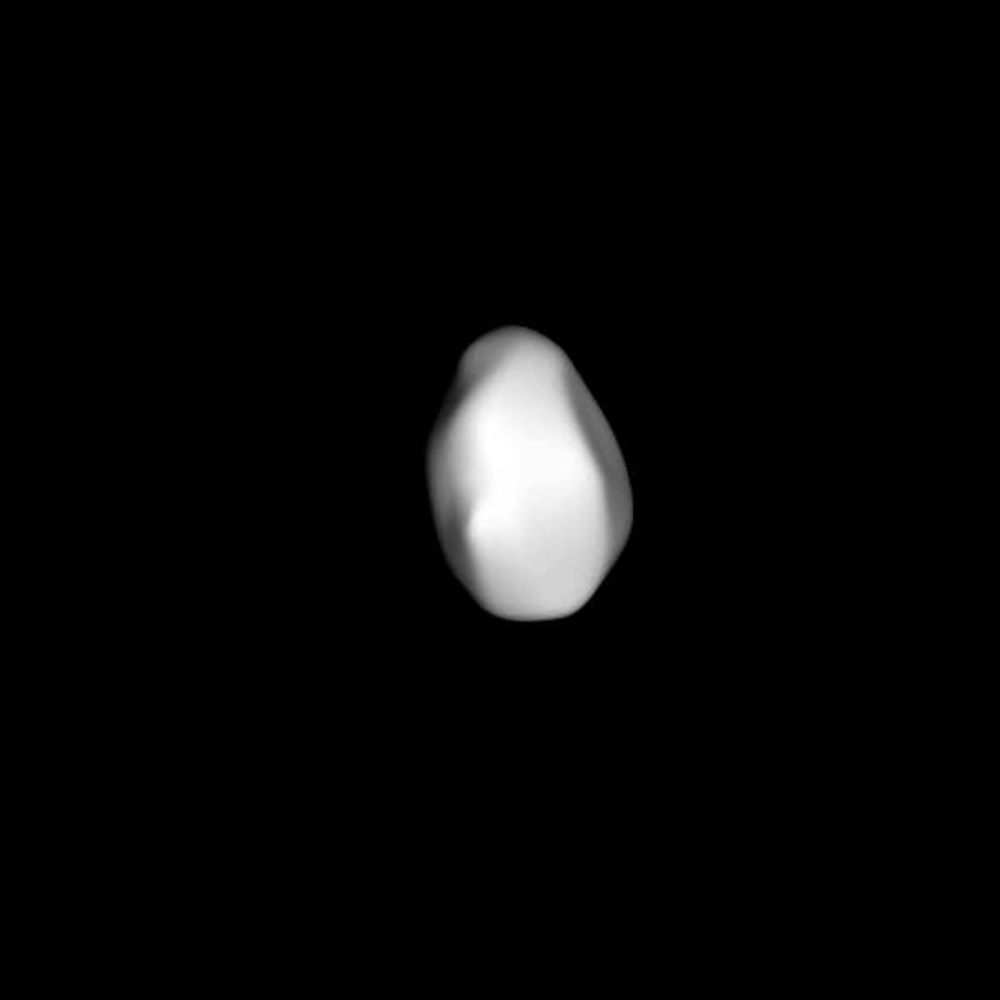}}\resizebox{0.24\hsize}{!}{\includegraphics{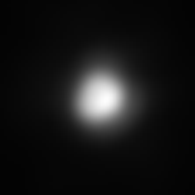}}\resizebox{0.24\hsize}{!}{\includegraphics{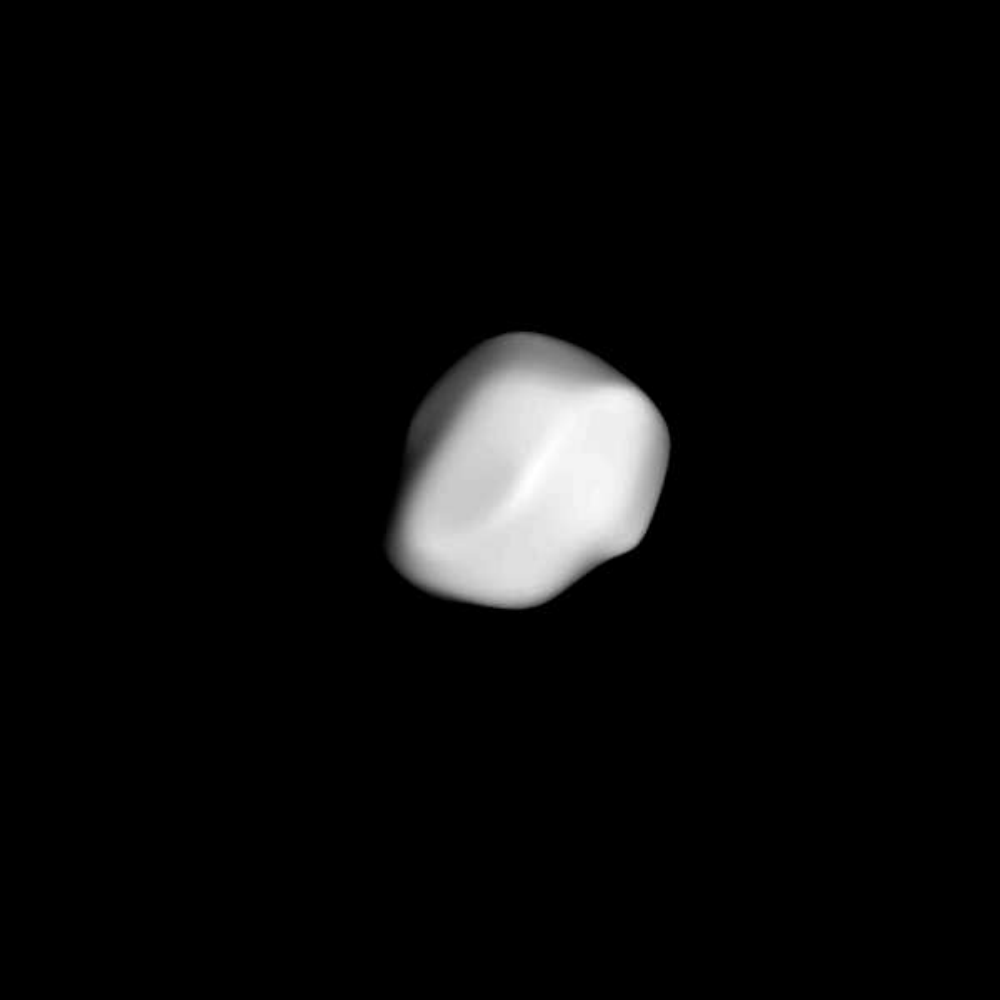}}\\
        \resizebox{0.24\hsize}{!}{\includegraphics{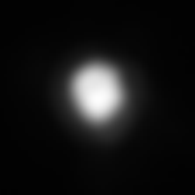}}\resizebox{0.24\hsize}{!}{\includegraphics{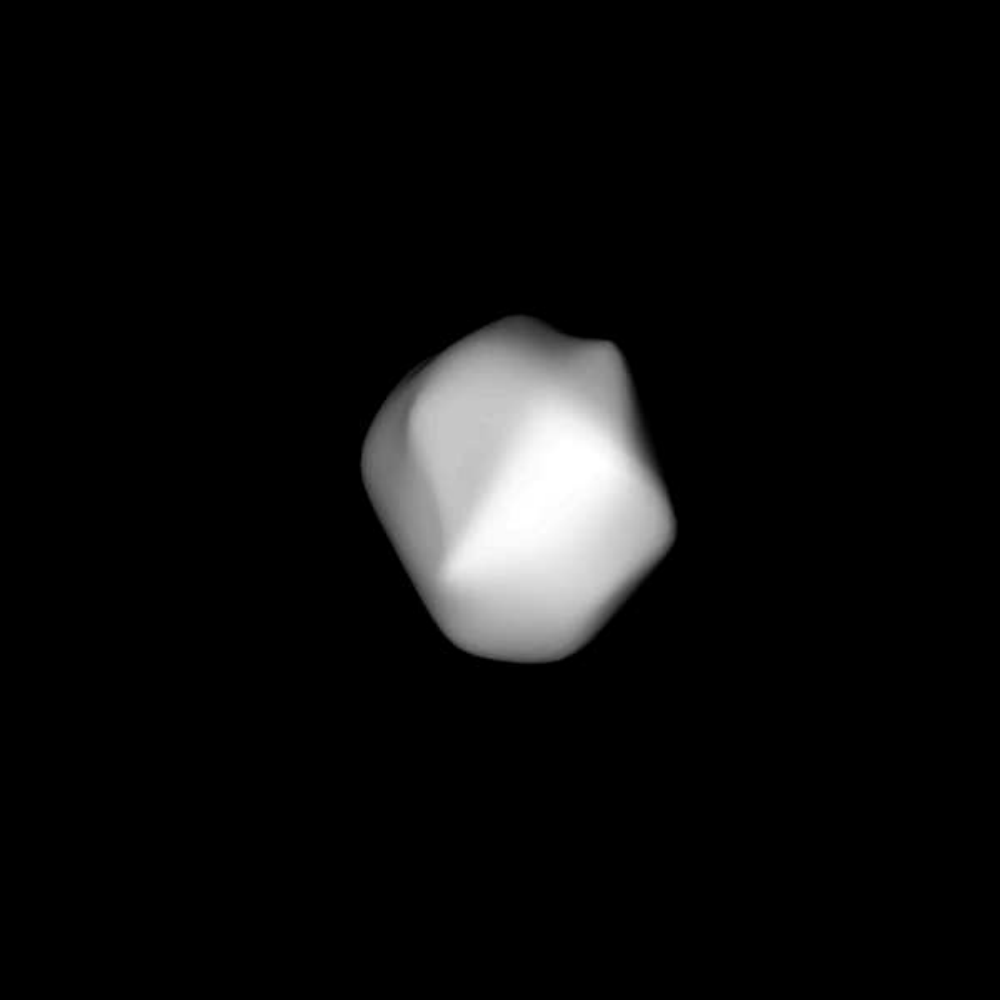}}\resizebox{0.24\hsize}{!}{\includegraphics{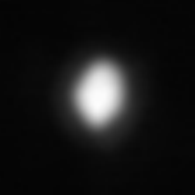}}\resizebox{0.24\hsize}{!}{\includegraphics{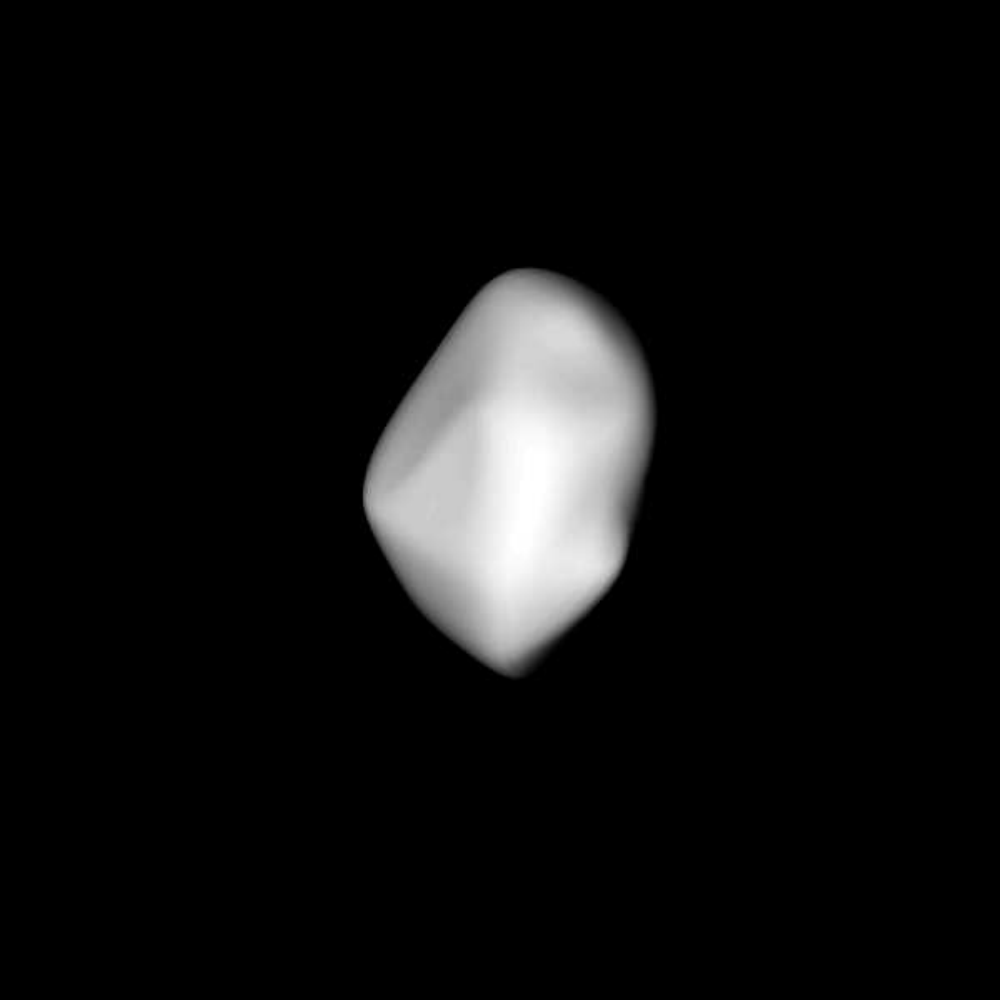}}\\
        \resizebox{0.24\hsize}{!}{\includegraphics{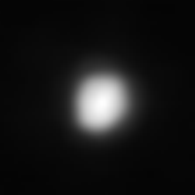}}\resizebox{0.24\hsize}{!}{\includegraphics{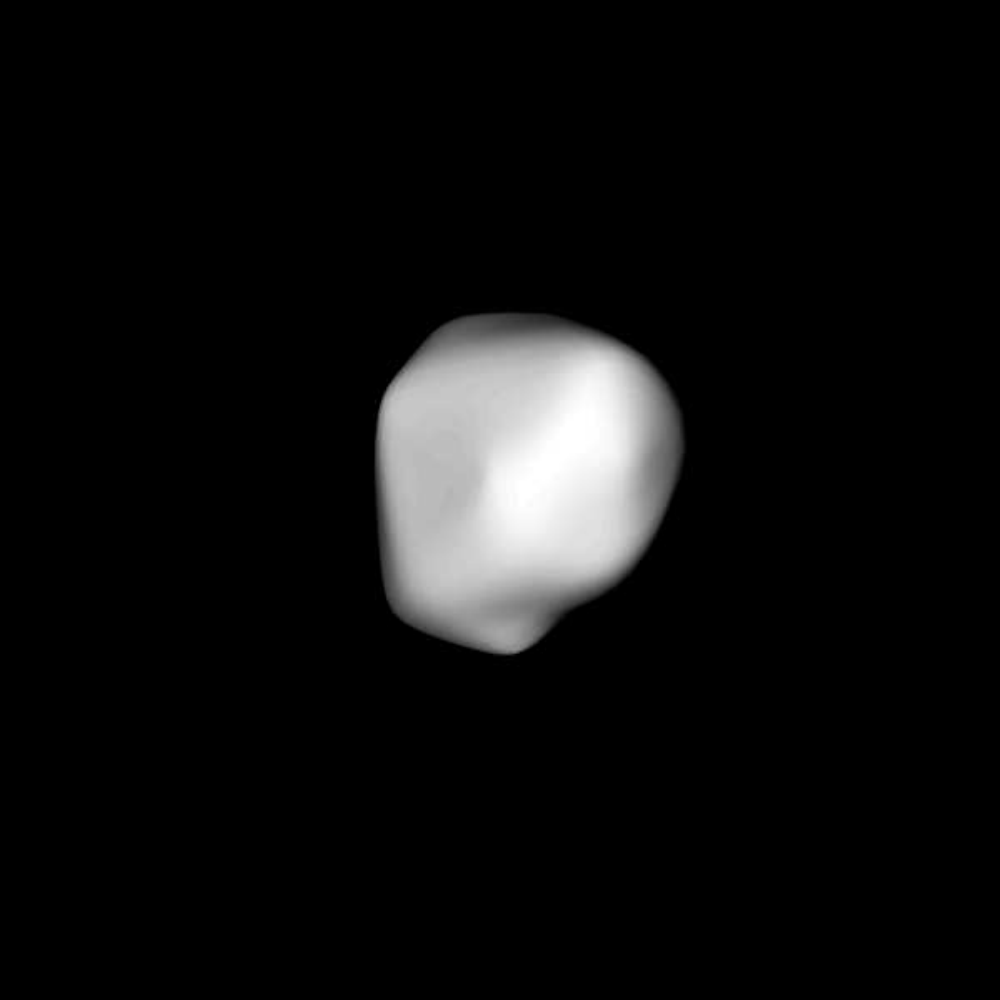}}\\
    \caption{\label{fig:16}Comparison between model projections and corresponding AO images for asteroid (16) Psyche.}
\end{figure}

\begin{figure}[tbp]
    \centering
        \resizebox{0.24\hsize}{!}{\includegraphics{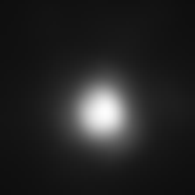}}\resizebox{0.24\hsize}{!}{\includegraphics{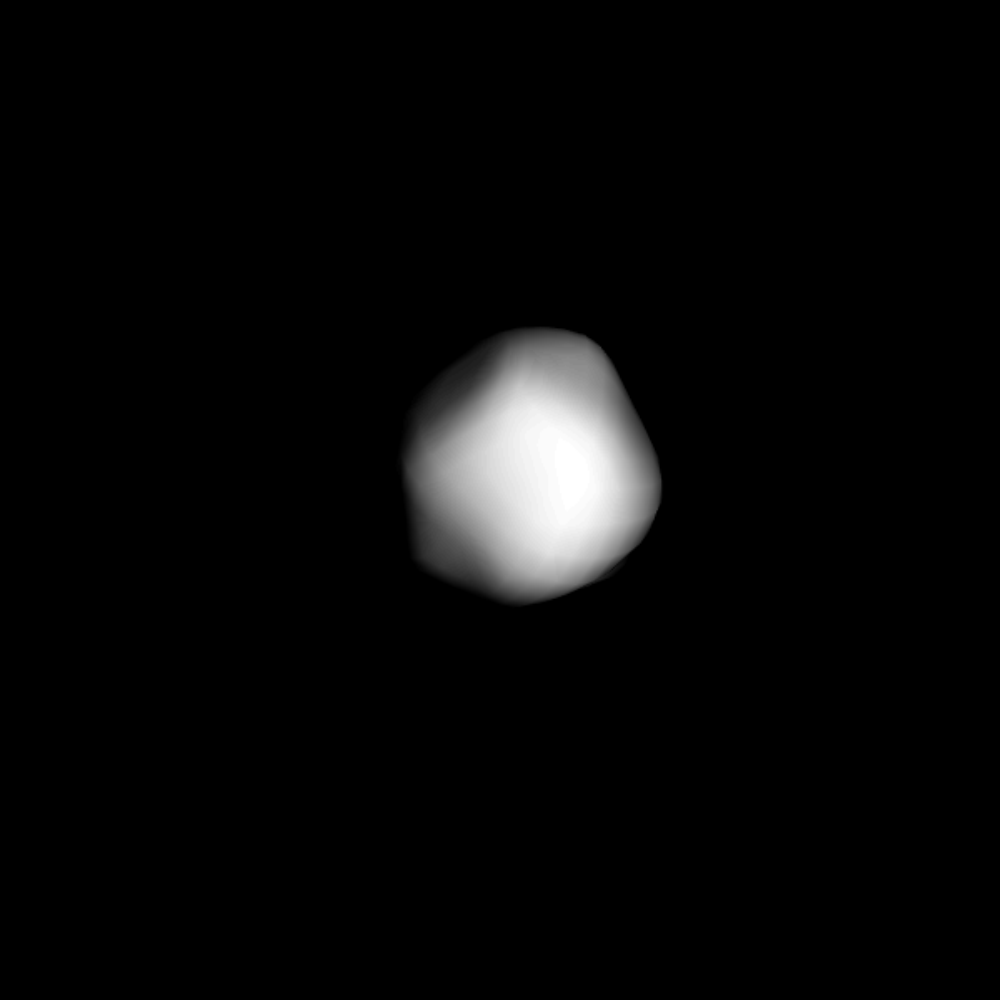}}\resizebox{0.24\hsize}{!}{\includegraphics{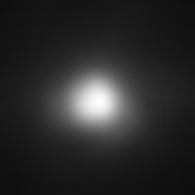}}\resizebox{0.24\hsize}{!}{\includegraphics{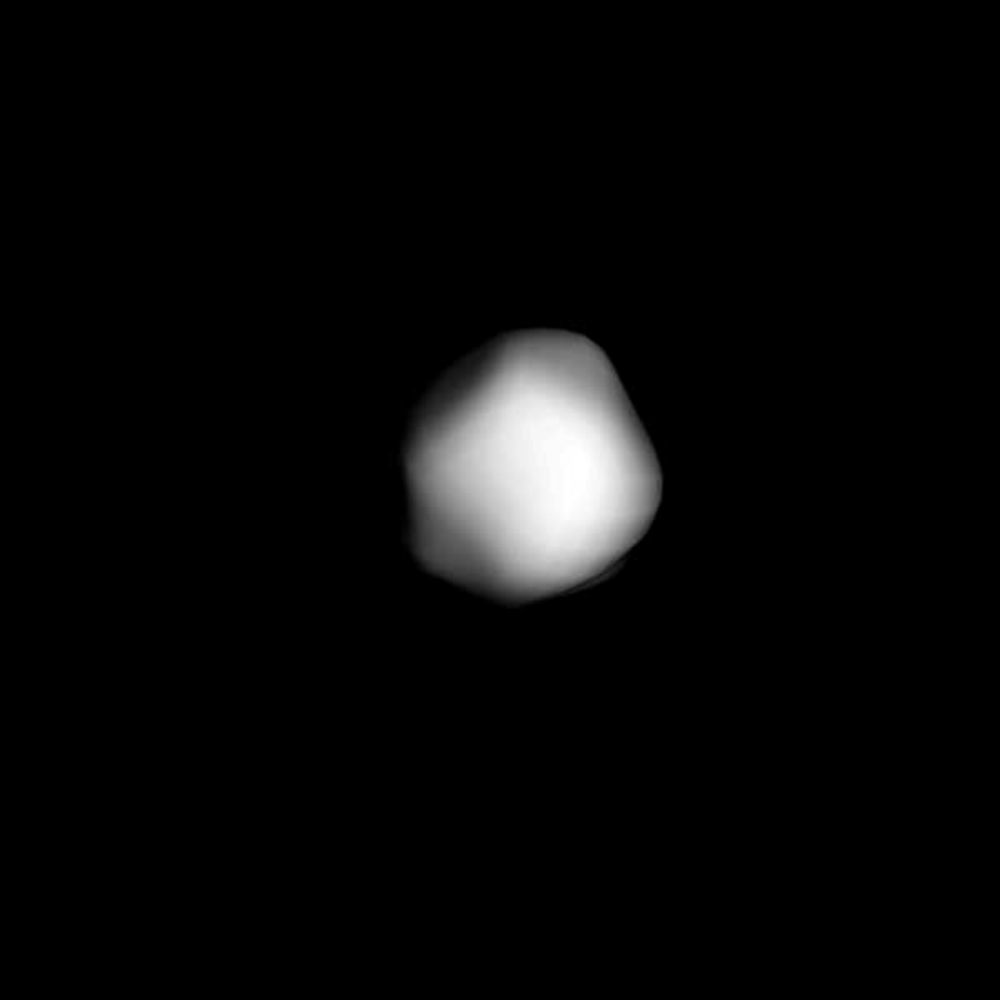}}\\
        \resizebox{0.24\hsize}{!}{\includegraphics{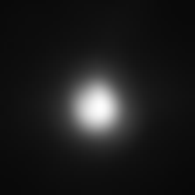}}\resizebox{0.24\hsize}{!}{\includegraphics{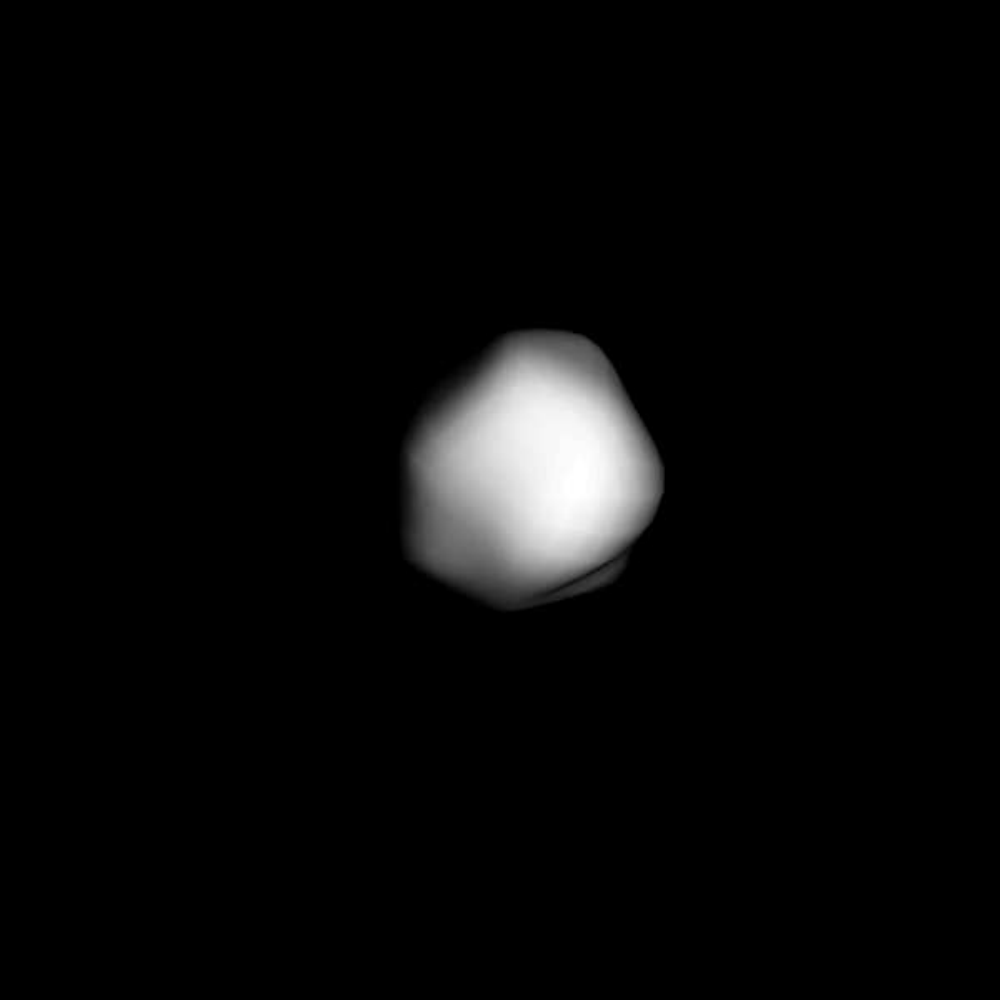}}\resizebox{0.24\hsize}{!}{\includegraphics{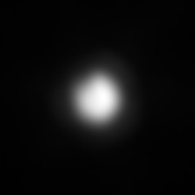}}\resizebox{0.24\hsize}{!}{\includegraphics{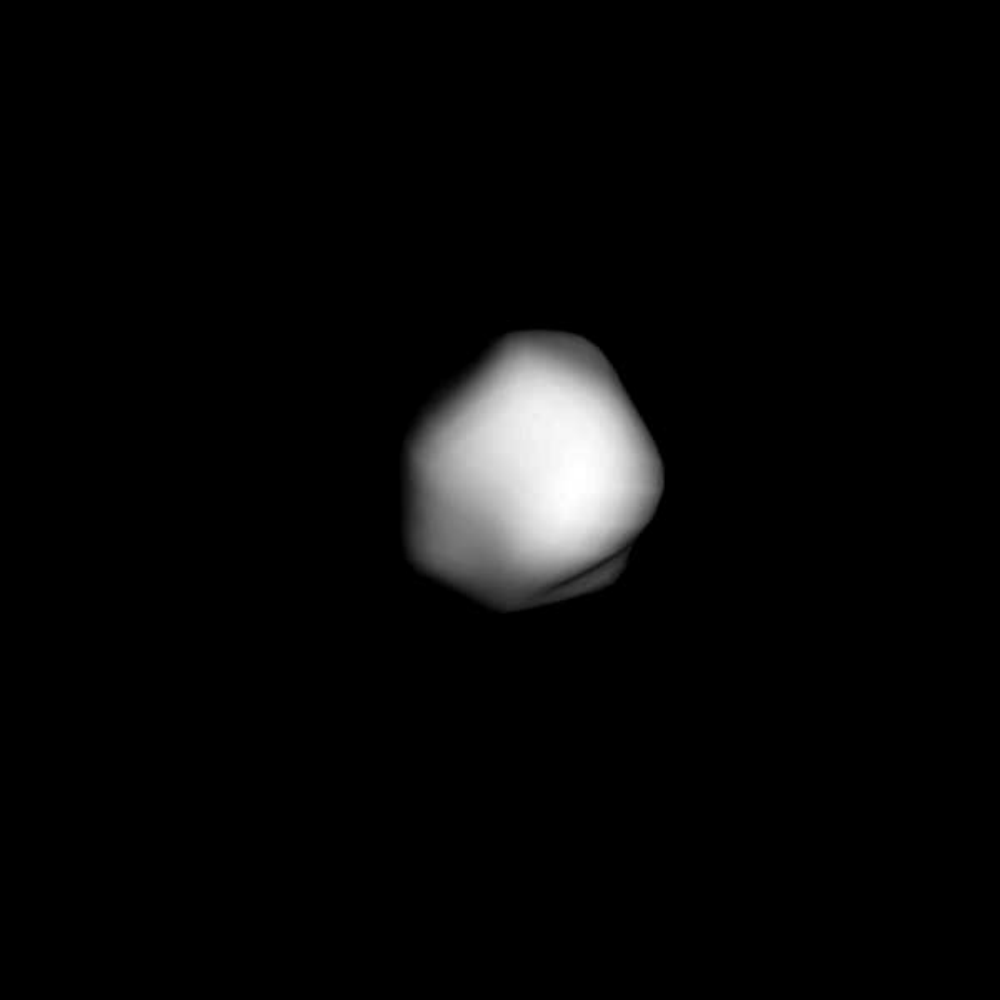}}\\
        \resizebox{0.24\hsize}{!}{\includegraphics{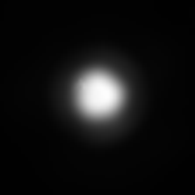}}\resizebox{0.24\hsize}{!}{\includegraphics{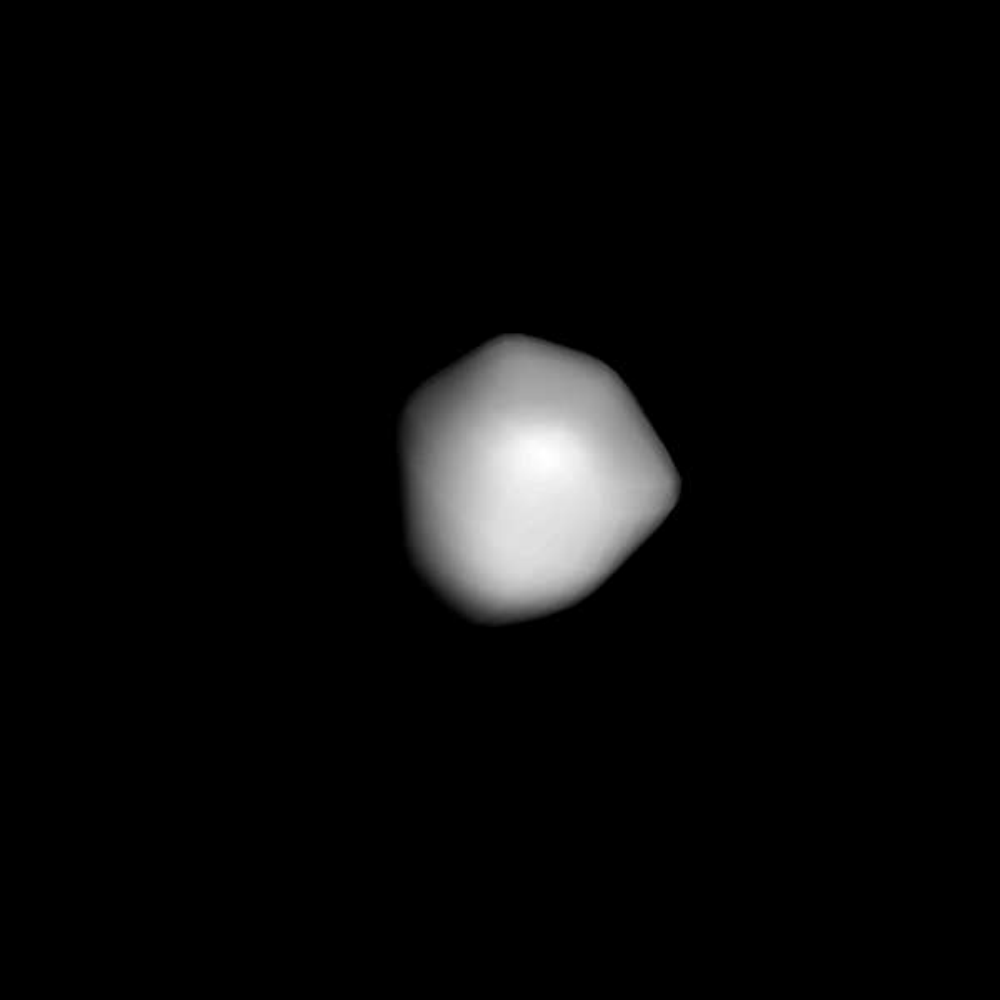}}\resizebox{0.24\hsize}{!}{\includegraphics{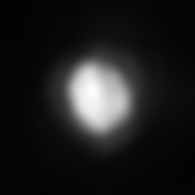}}\resizebox{0.24\hsize}{!}{\includegraphics{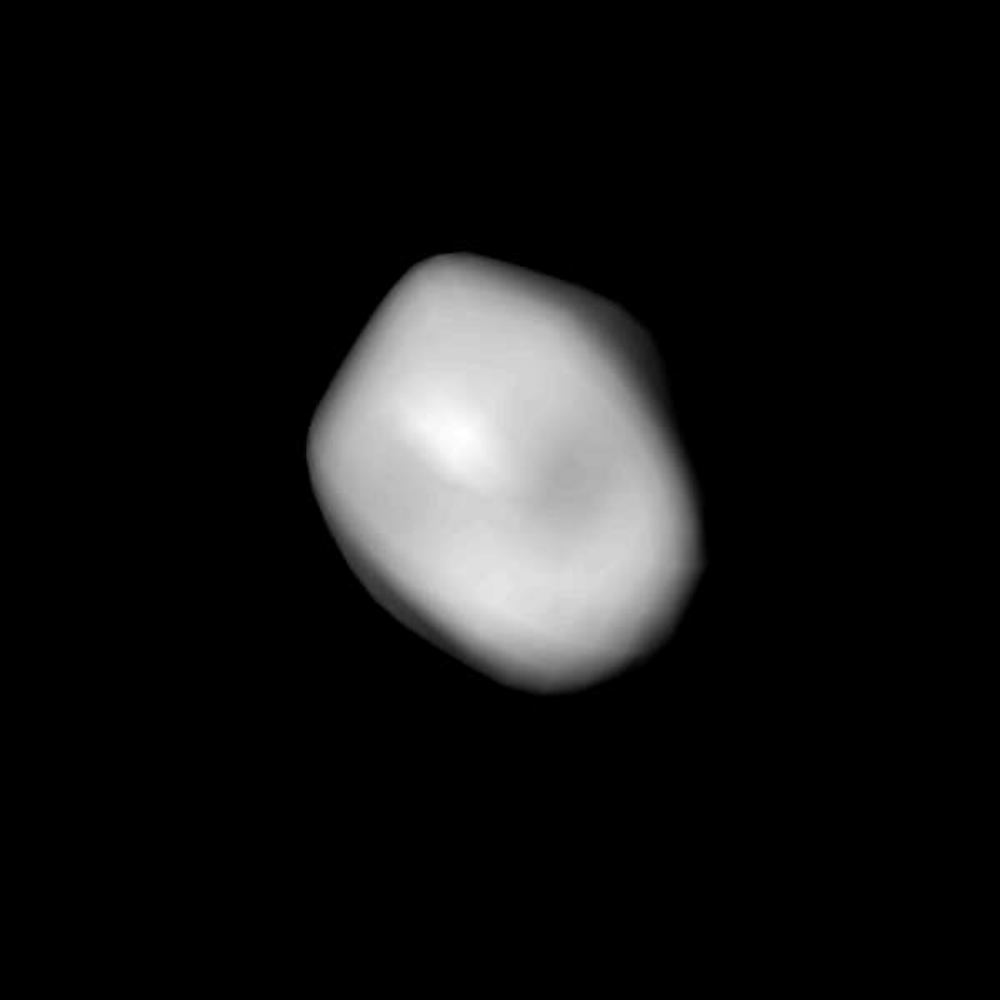}}\\
    \caption{\label{fig:18}Comparison between model projections and corresponding AO images for asteroid (18) Melpomene.}
\end{figure}

\begin{figure}[tbp]
    \centering
        \resizebox{0.24\hsize}{!}{\includegraphics{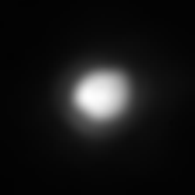}}\resizebox{0.24\hsize}{!}{\includegraphics{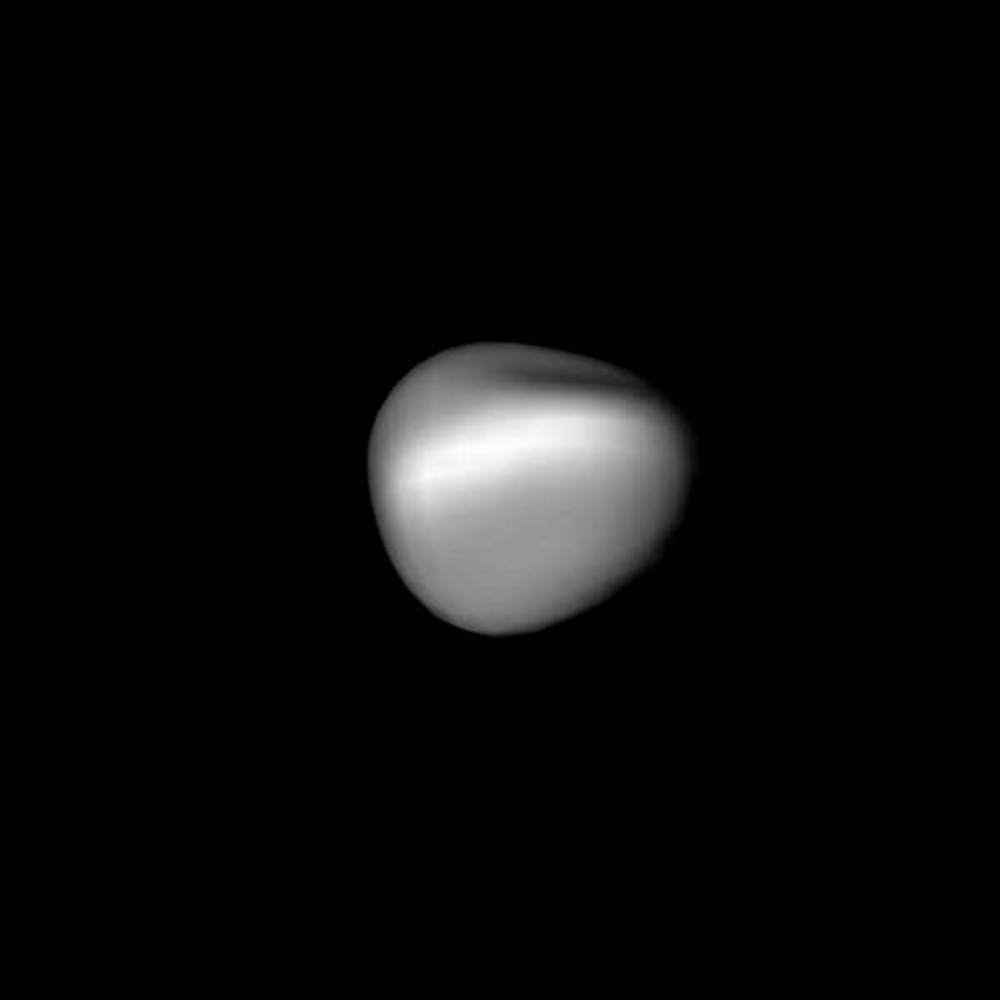}}\resizebox{0.24\hsize}{!}{\includegraphics{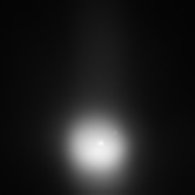}}\resizebox{0.24\hsize}{!}{\includegraphics{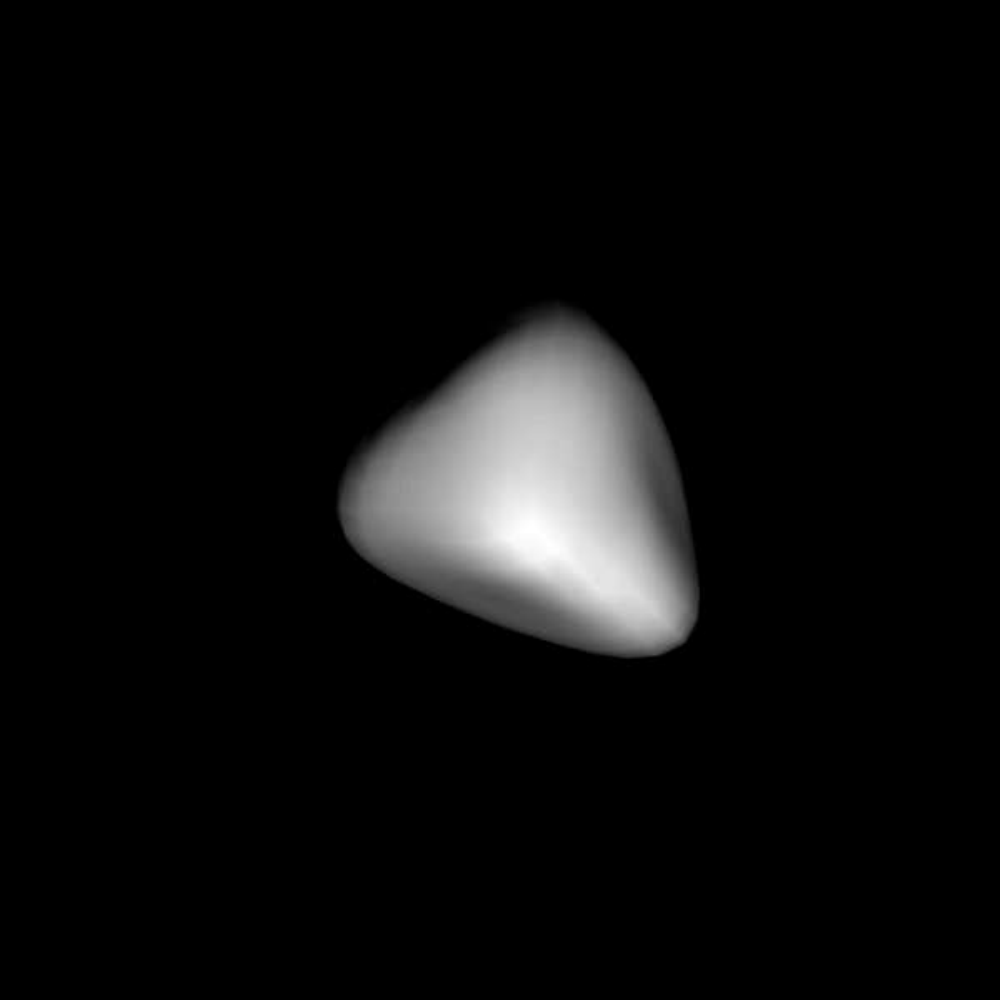}}\\
        \resizebox{0.24\hsize}{!}{\includegraphics{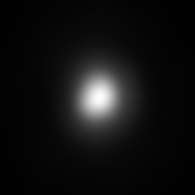}}\resizebox{0.24\hsize}{!}{\includegraphics{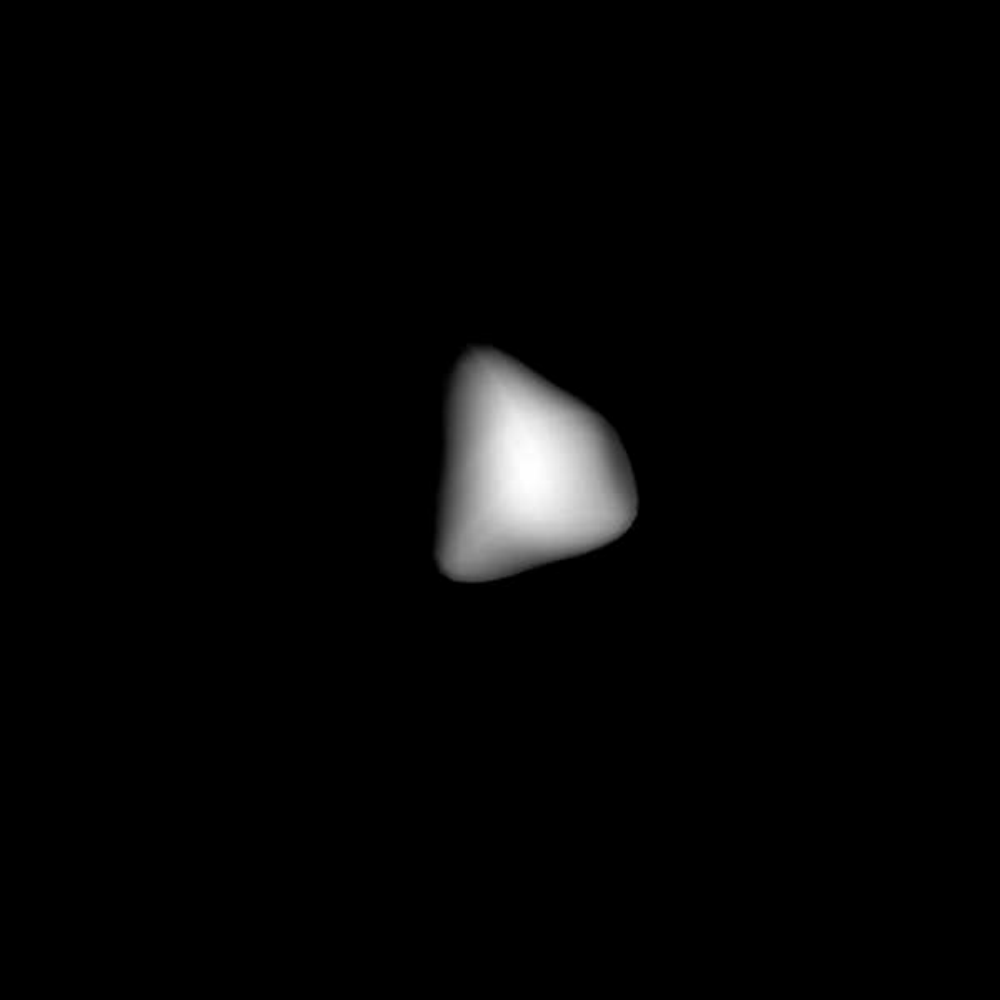}}\resizebox{0.24\hsize}{!}{\includegraphics{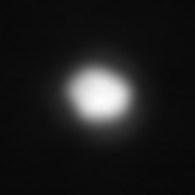}}\resizebox{0.24\hsize}{!}{\includegraphics{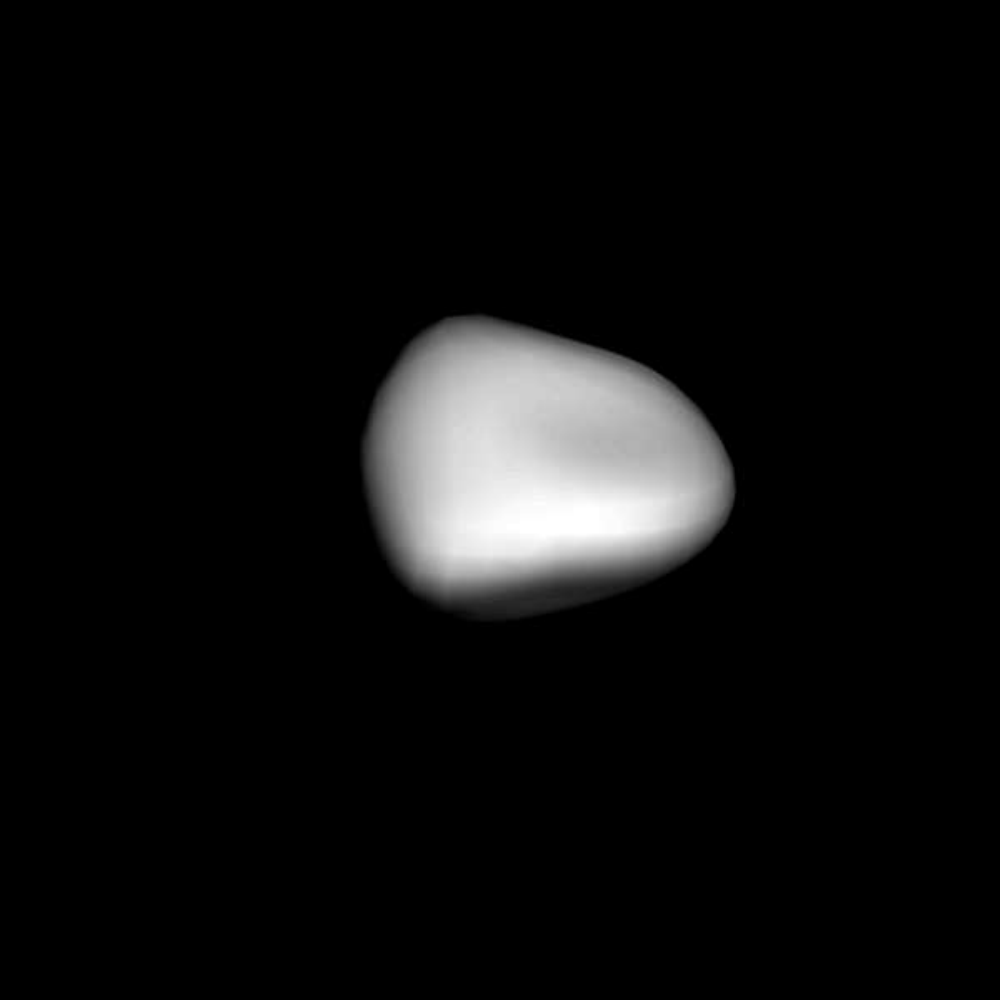}}\\
    \caption{\label{fig:19}Comparison between model projections and corresponding AO images for asteroid (19) Fortuna.}
\end{figure}


\clearpage

\begin{figure}[tbp]
    \centering
        \resizebox{0.24\hsize}{!}{\includegraphics{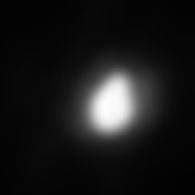}}\resizebox{0.24\hsize}{!}{\includegraphics{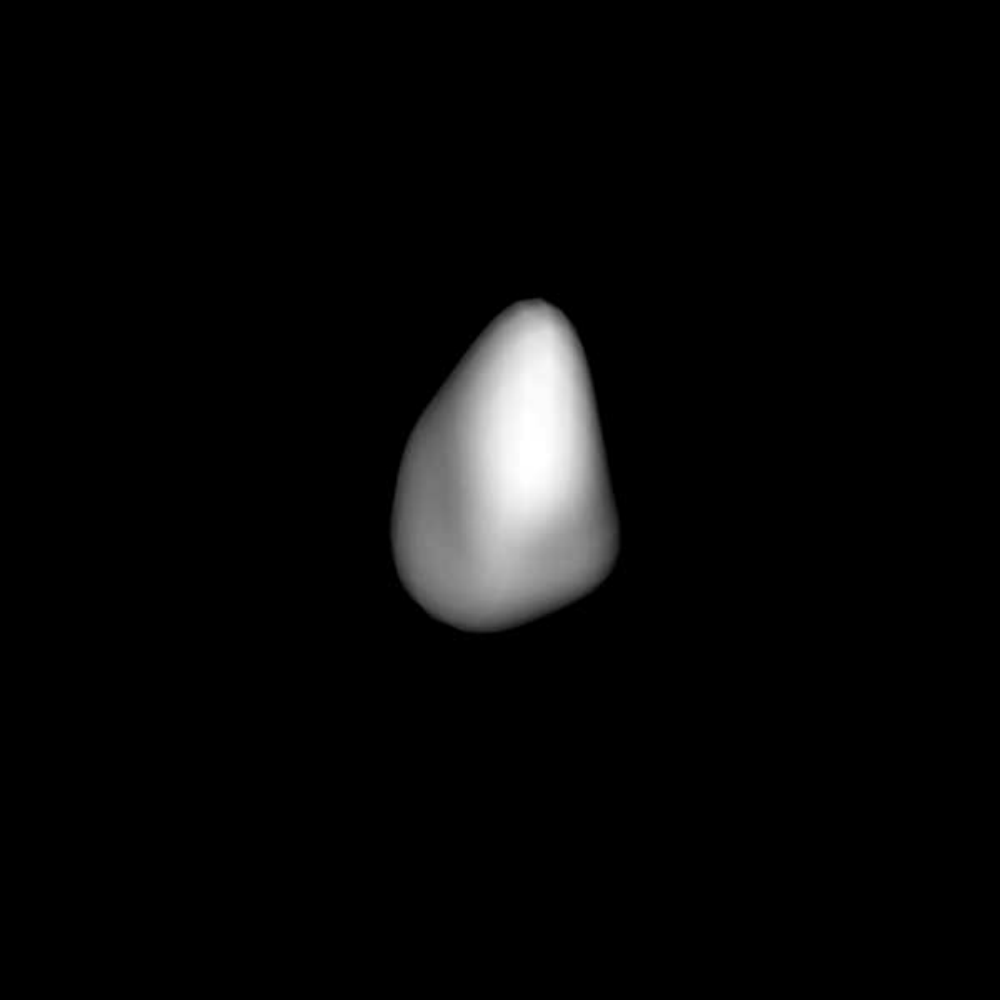}}\resizebox{0.24\hsize}{!}{\includegraphics{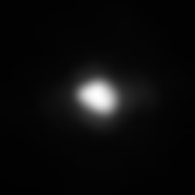}}\resizebox{0.24\hsize}{!}{\includegraphics{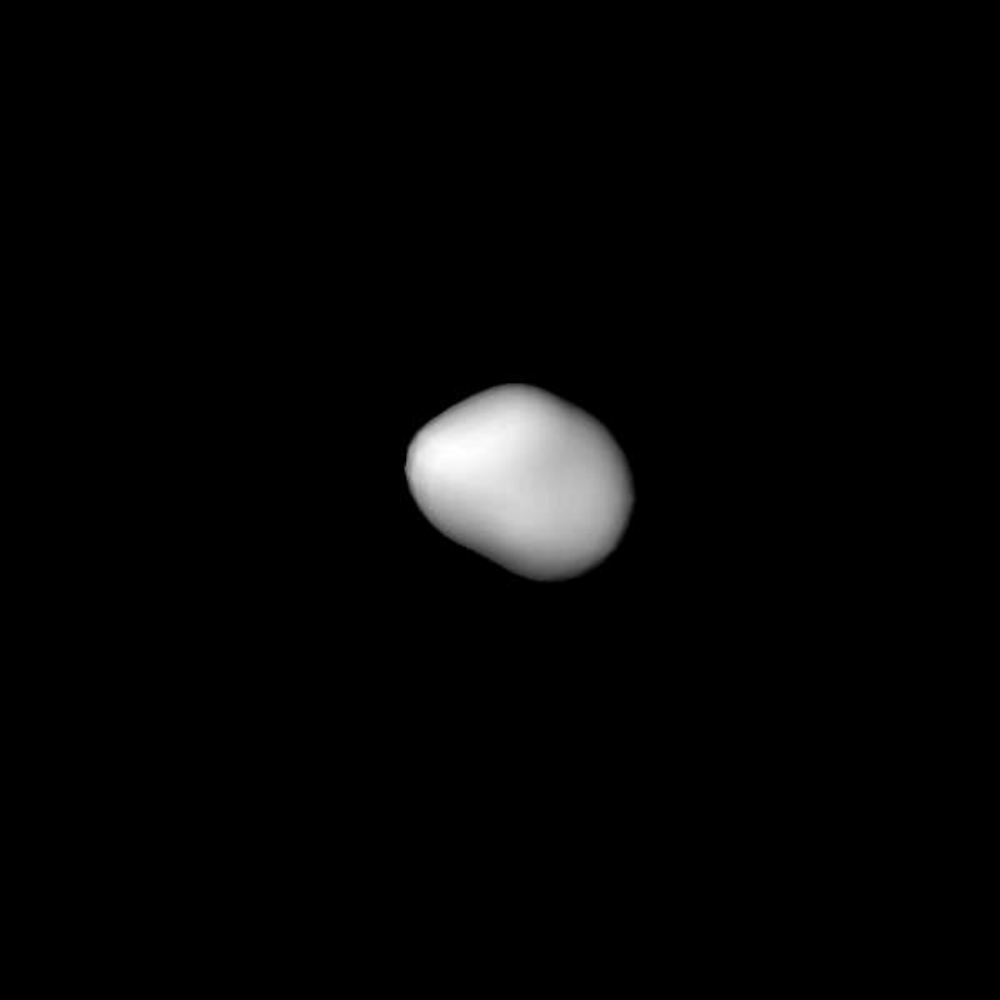}}\\
        \resizebox{0.24\hsize}{!}{\includegraphics{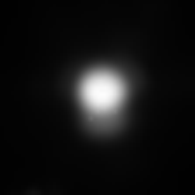}}\resizebox{0.24\hsize}{!}{\includegraphics{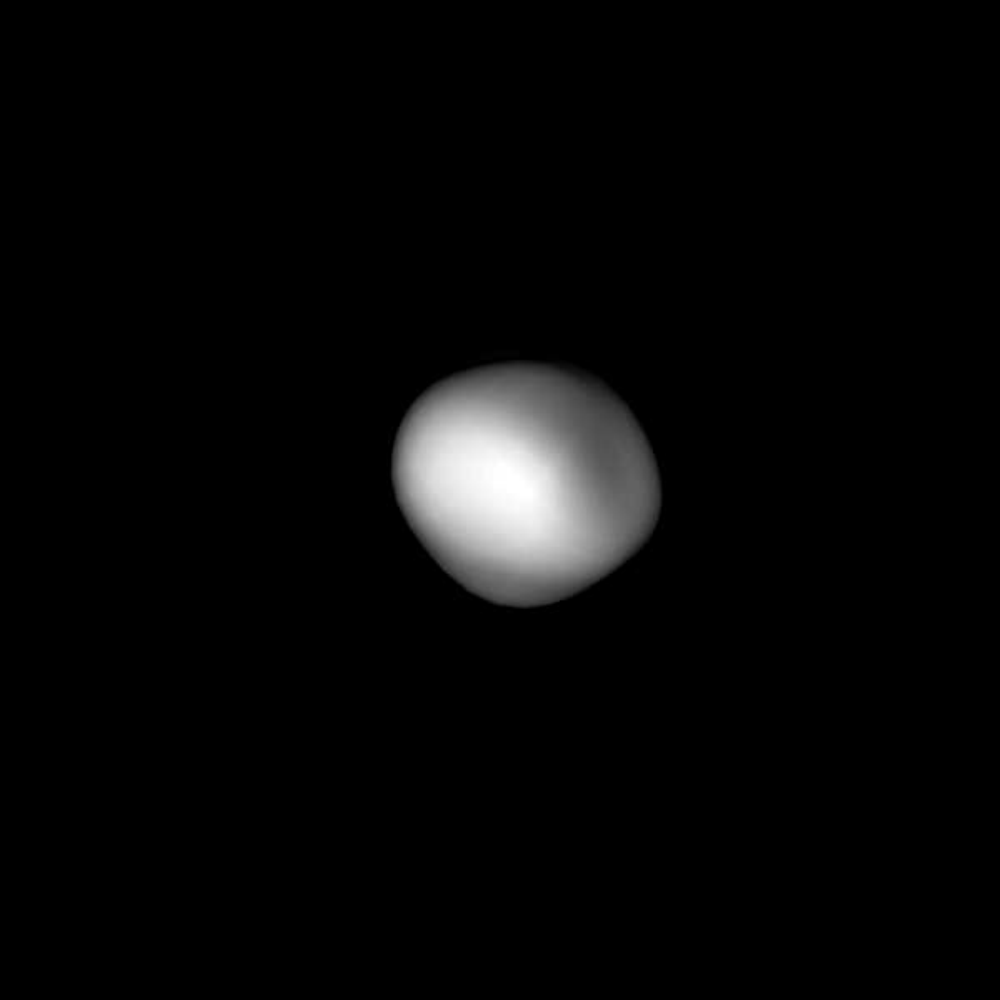}}\resizebox{0.24\hsize}{!}{\includegraphics{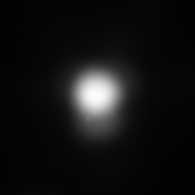}}\resizebox{0.24\hsize}{!}{\includegraphics{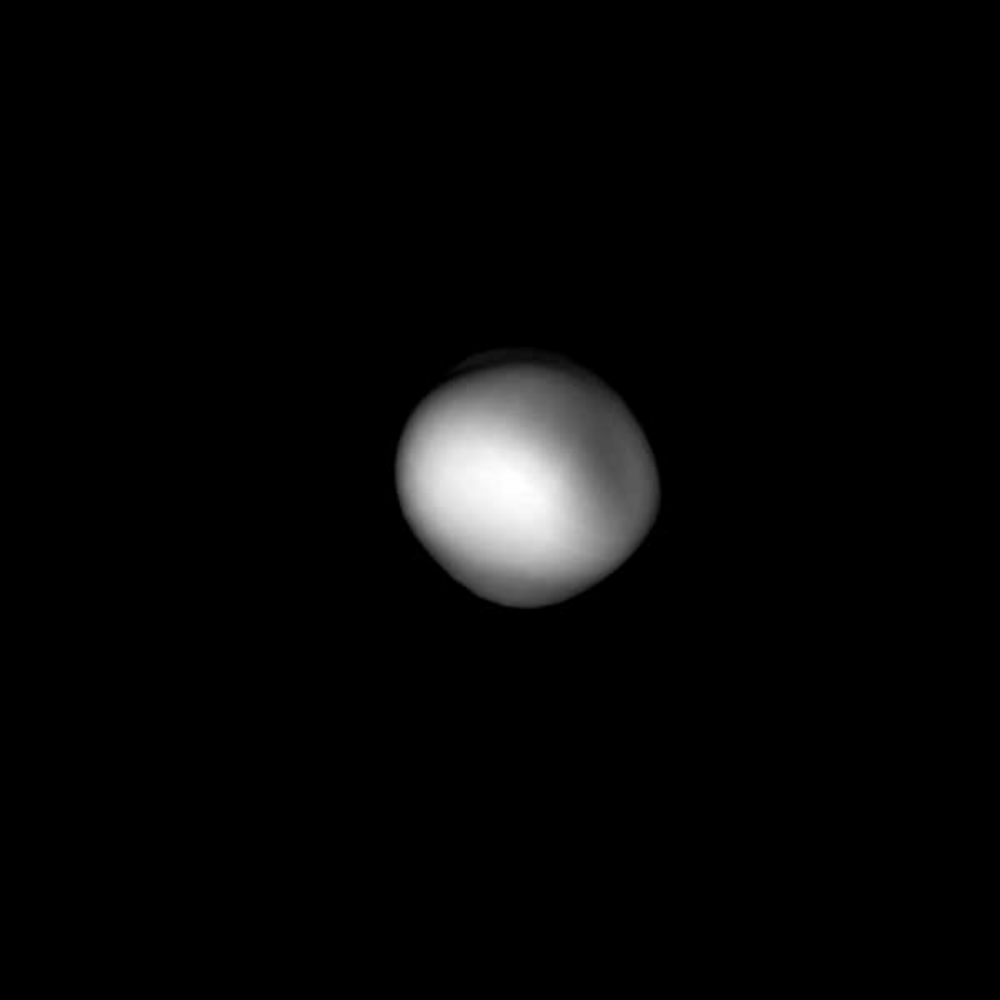}}\\
        \resizebox{0.24\hsize}{!}{\includegraphics{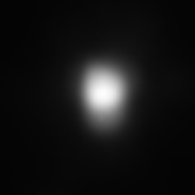}}\resizebox{0.24\hsize}{!}{\includegraphics{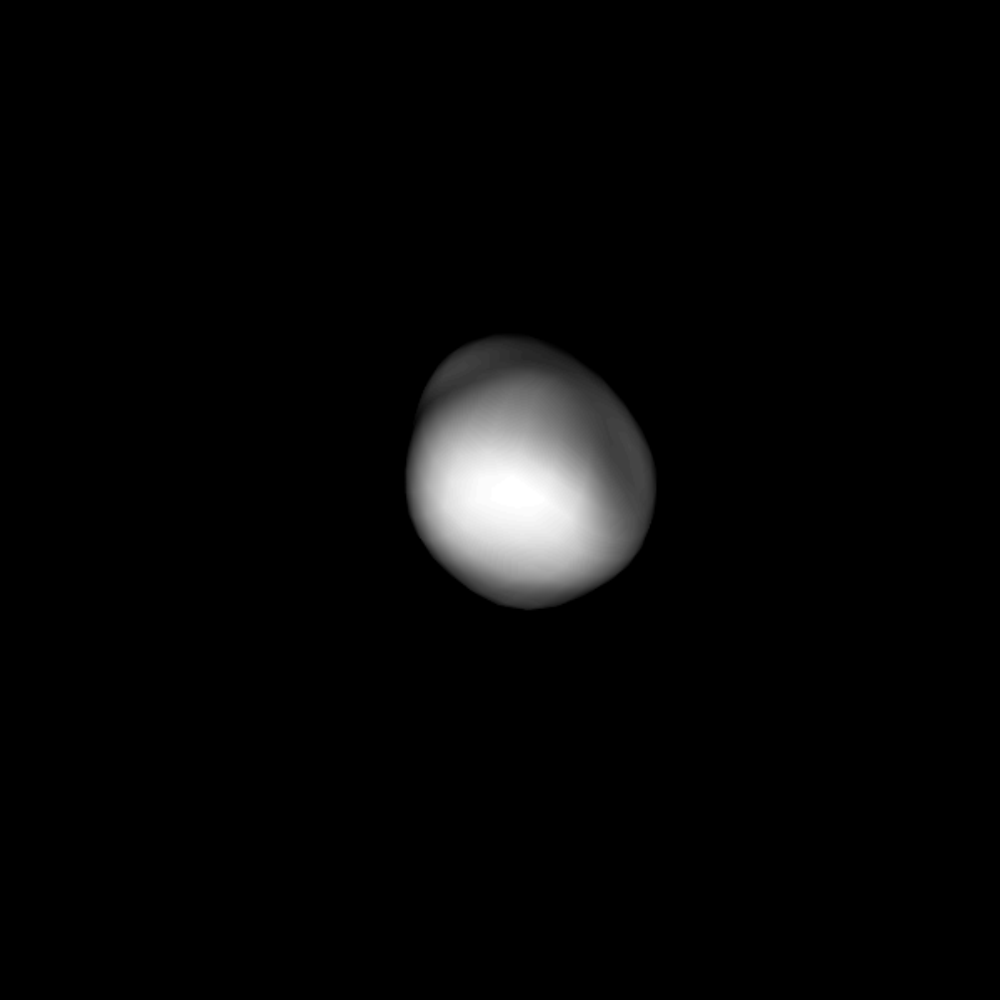}}\resizebox{0.24\hsize}{!}{\includegraphics{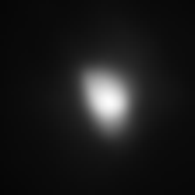}}\resizebox{0.24\hsize}{!}{\includegraphics{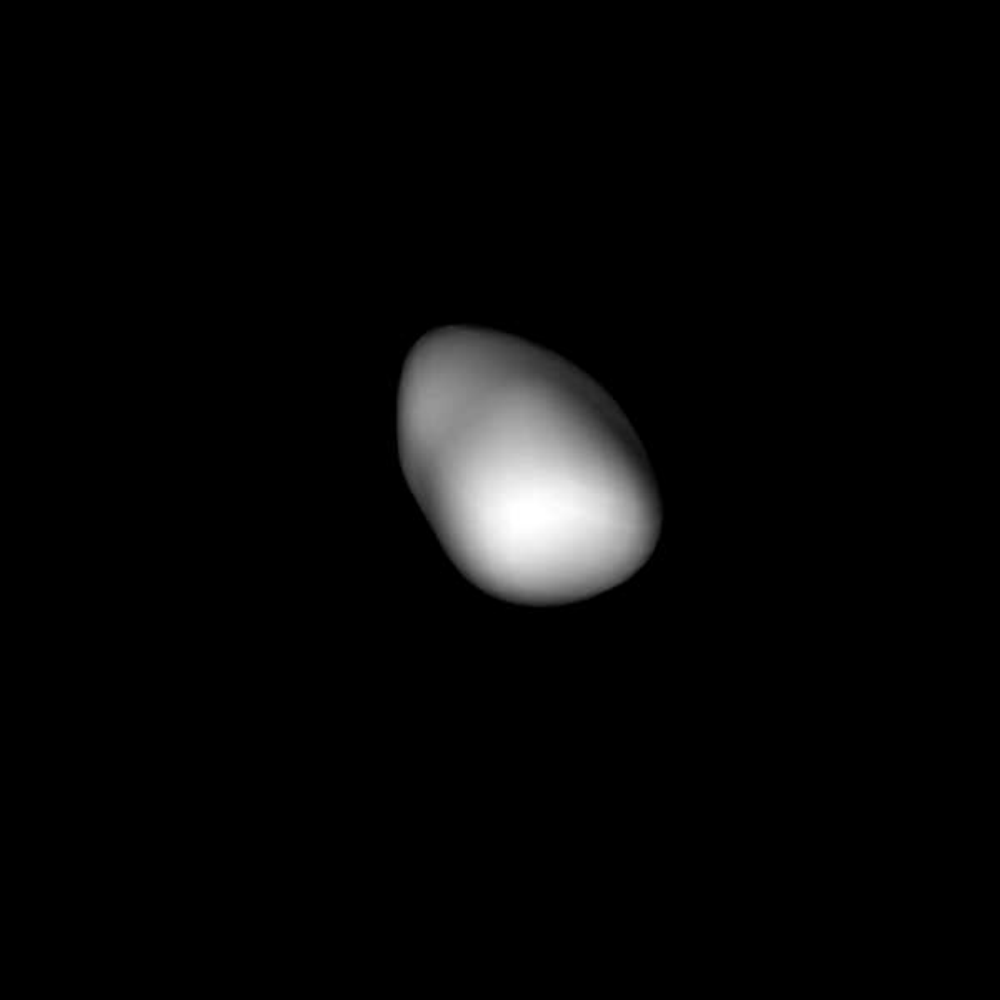}}\\
        \resizebox{0.24\hsize}{!}{\includegraphics{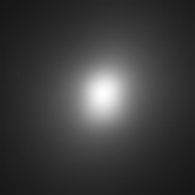}}\resizebox{0.24\hsize}{!}{\includegraphics{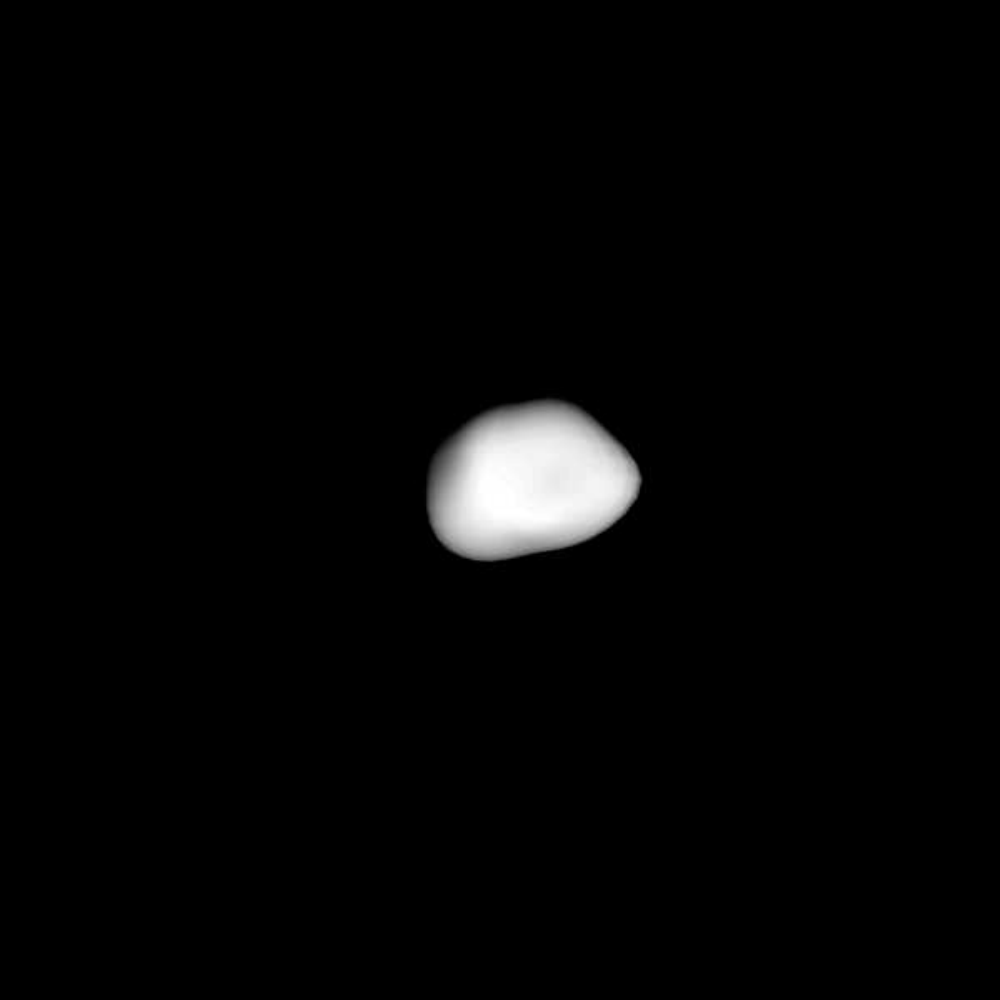}}\resizebox{0.24\hsize}{!}{\includegraphics{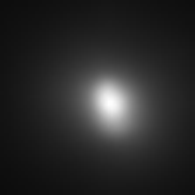}}\resizebox{0.24\hsize}{!}{\includegraphics{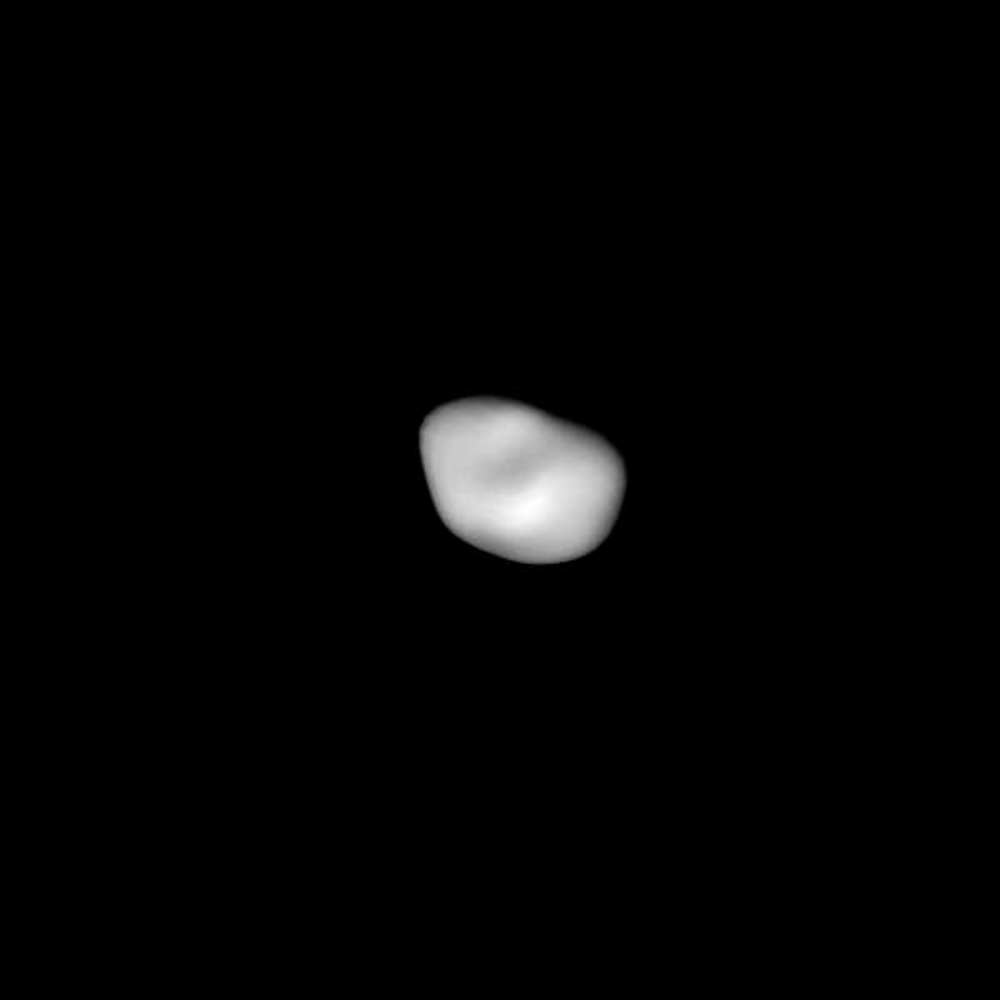}}\\
        \resizebox{0.24\hsize}{!}{\includegraphics{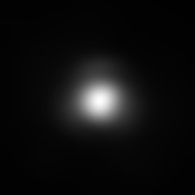}}\resizebox{0.24\hsize}{!}{\includegraphics{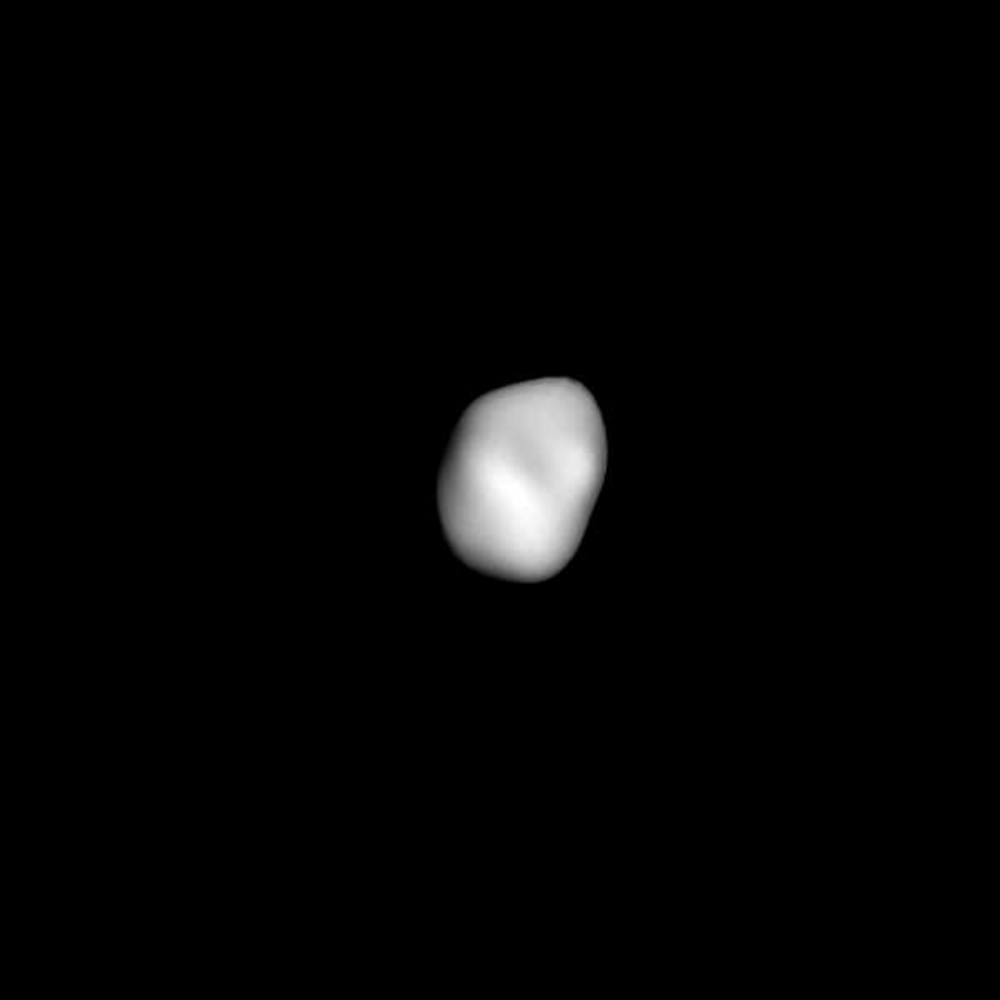}}\resizebox{0.24\hsize}{!}{\includegraphics{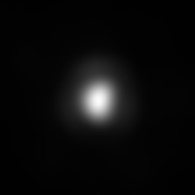}}\resizebox{0.24\hsize}{!}{\includegraphics{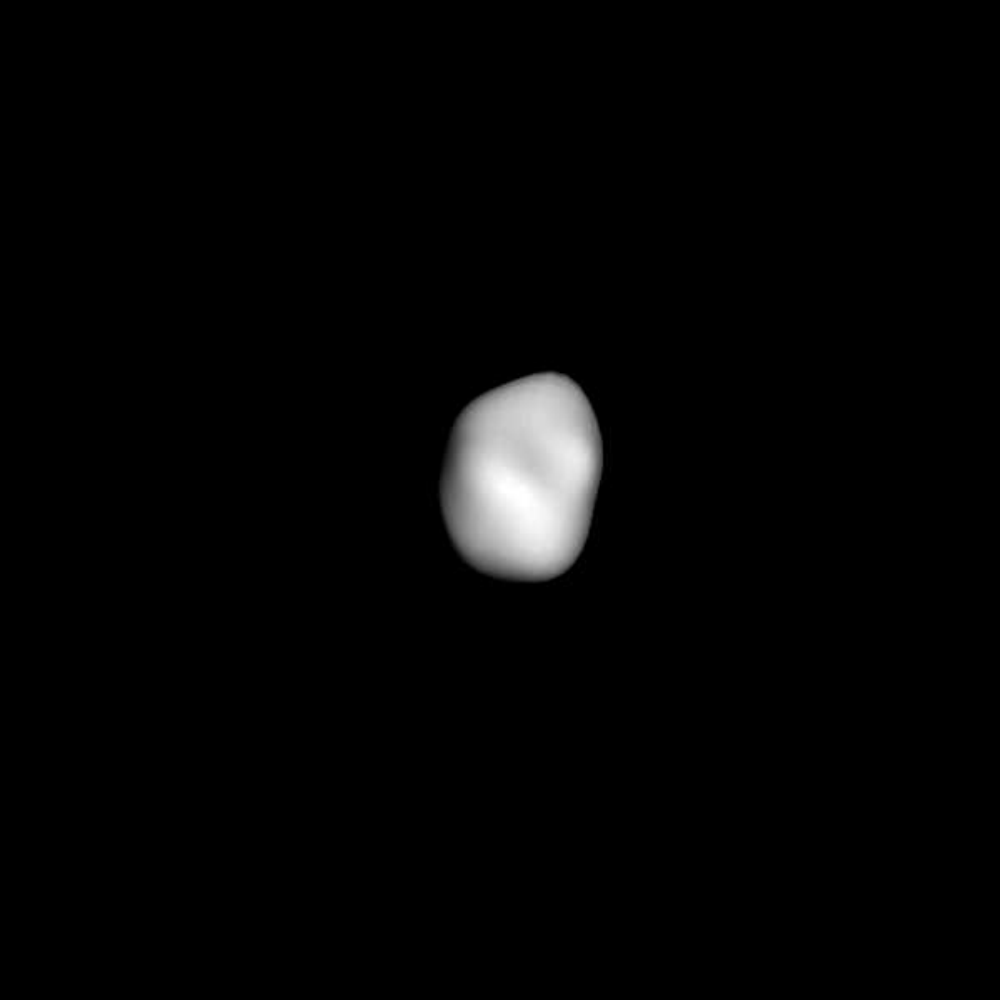}}\\
        \resizebox{0.24\hsize}{!}{\includegraphics{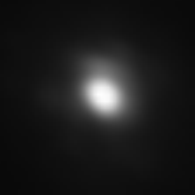}}\resizebox{0.24\hsize}{!}{\includegraphics{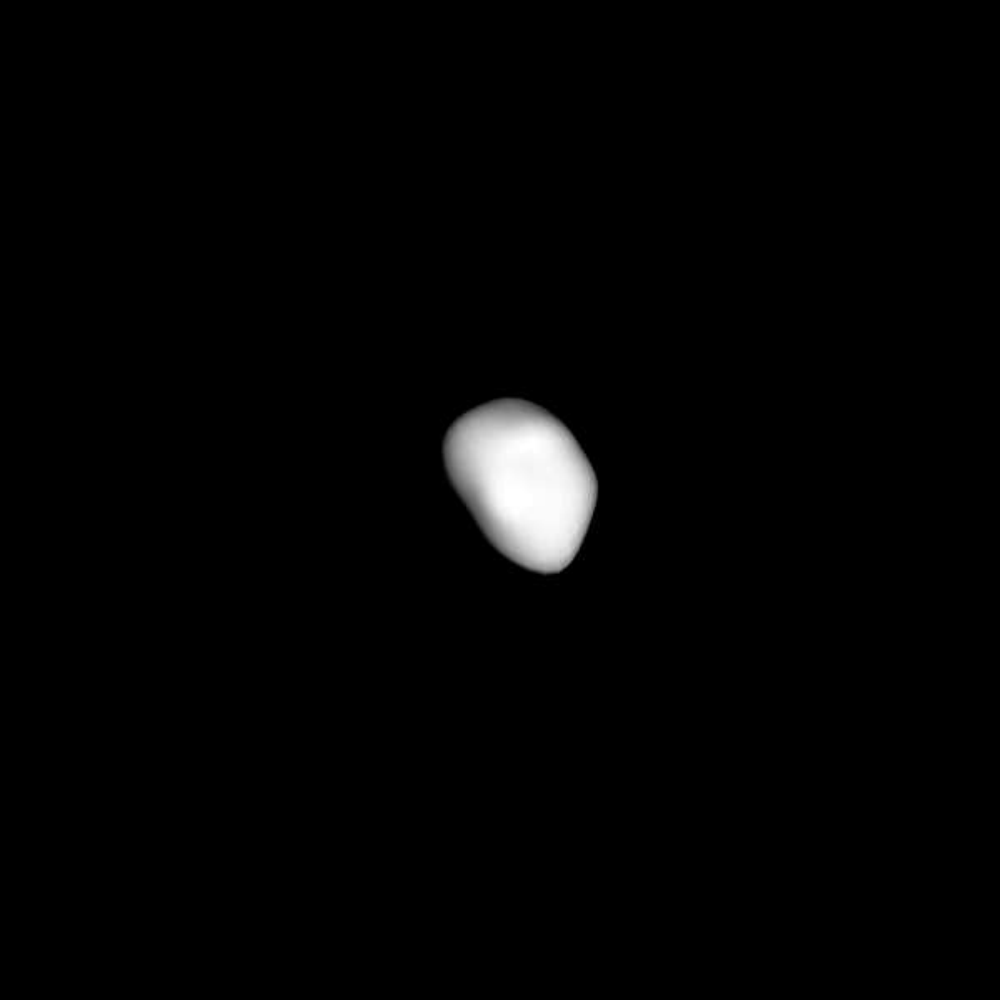}}\resizebox{0.24\hsize}{!}{\includegraphics{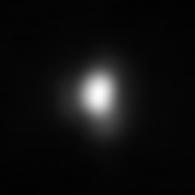}}\resizebox{0.24\hsize}{!}{\includegraphics{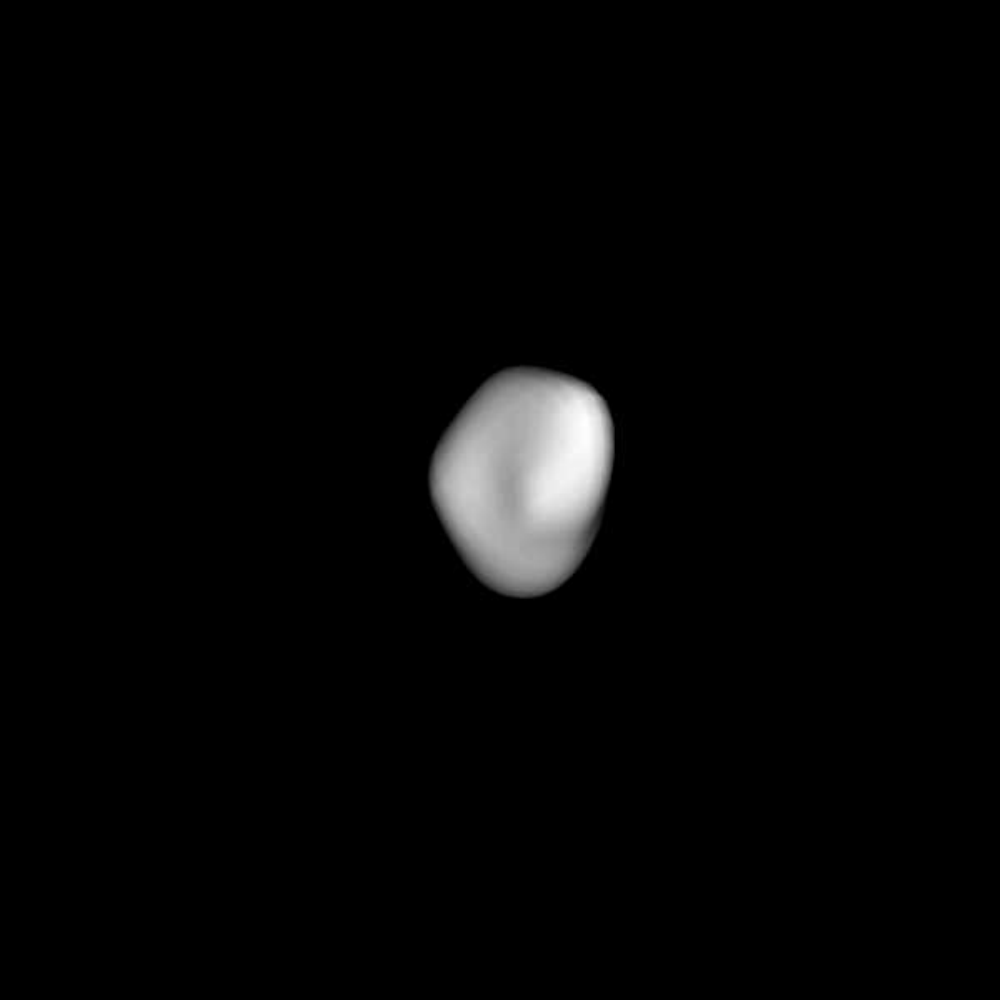}}\\
        \resizebox{0.24\hsize}{!}{\includegraphics{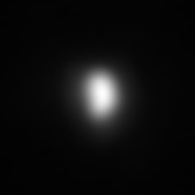}}\resizebox{0.24\hsize}{!}{\includegraphics{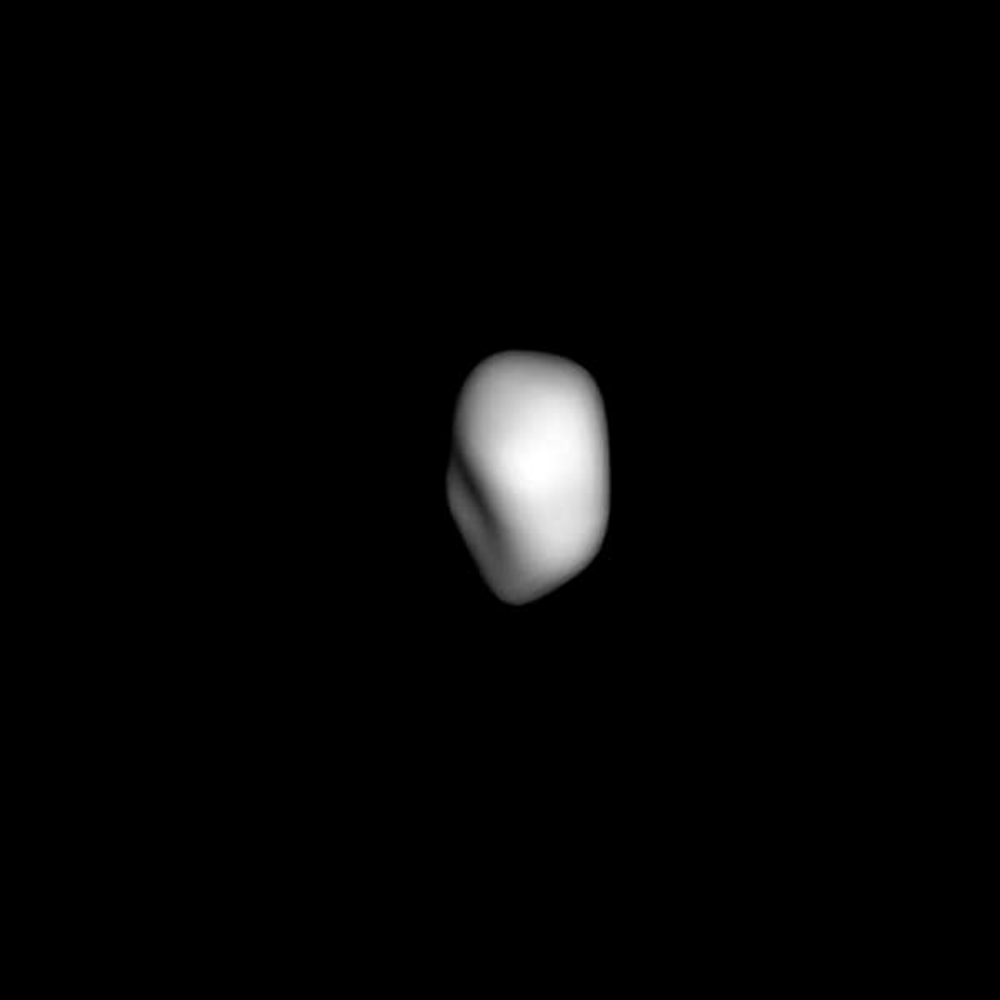}}\resizebox{0.24\hsize}{!}{\includegraphics{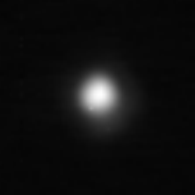}}\resizebox{0.24\hsize}{!}{\includegraphics{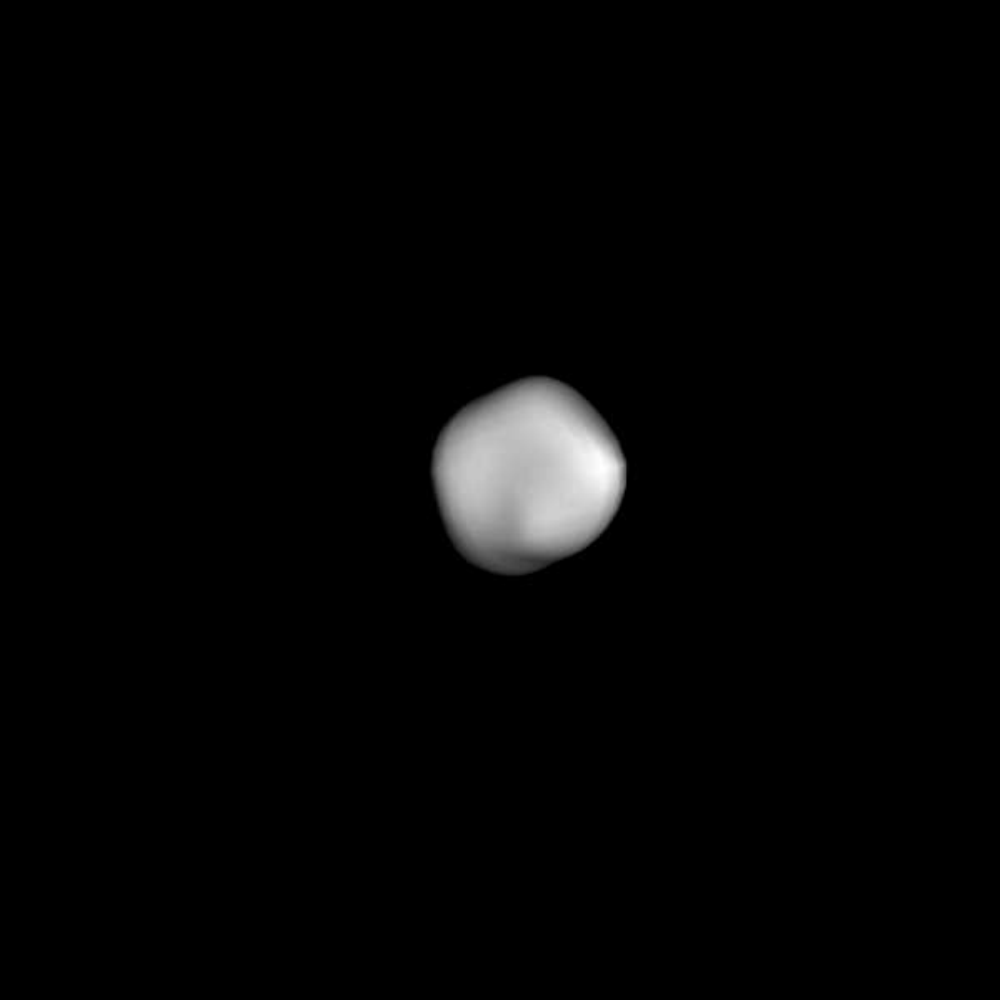}}\\
        \resizebox{0.24\hsize}{!}{\includegraphics{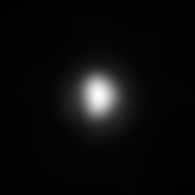}}\resizebox{0.24\hsize}{!}{\includegraphics{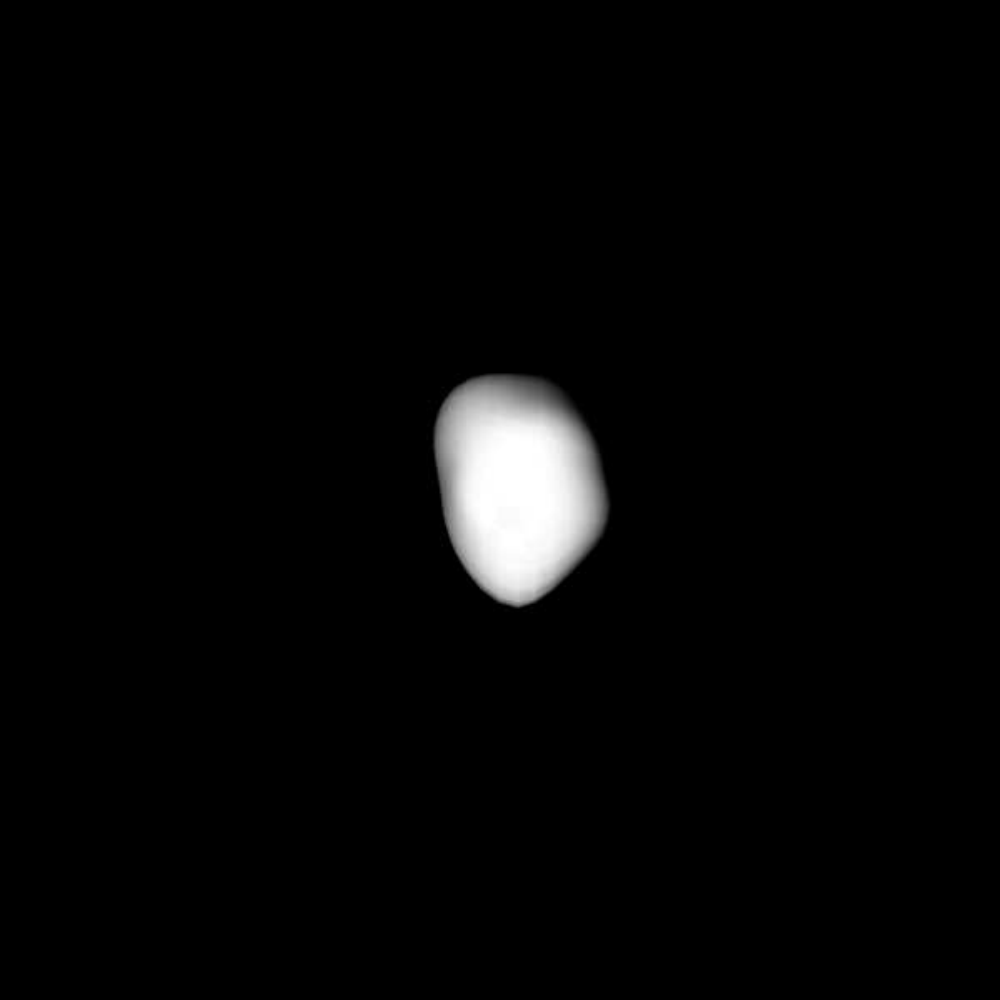}}\resizebox{0.24\hsize}{!}{\includegraphics{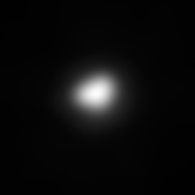}}\resizebox{0.24\hsize}{!}{\includegraphics{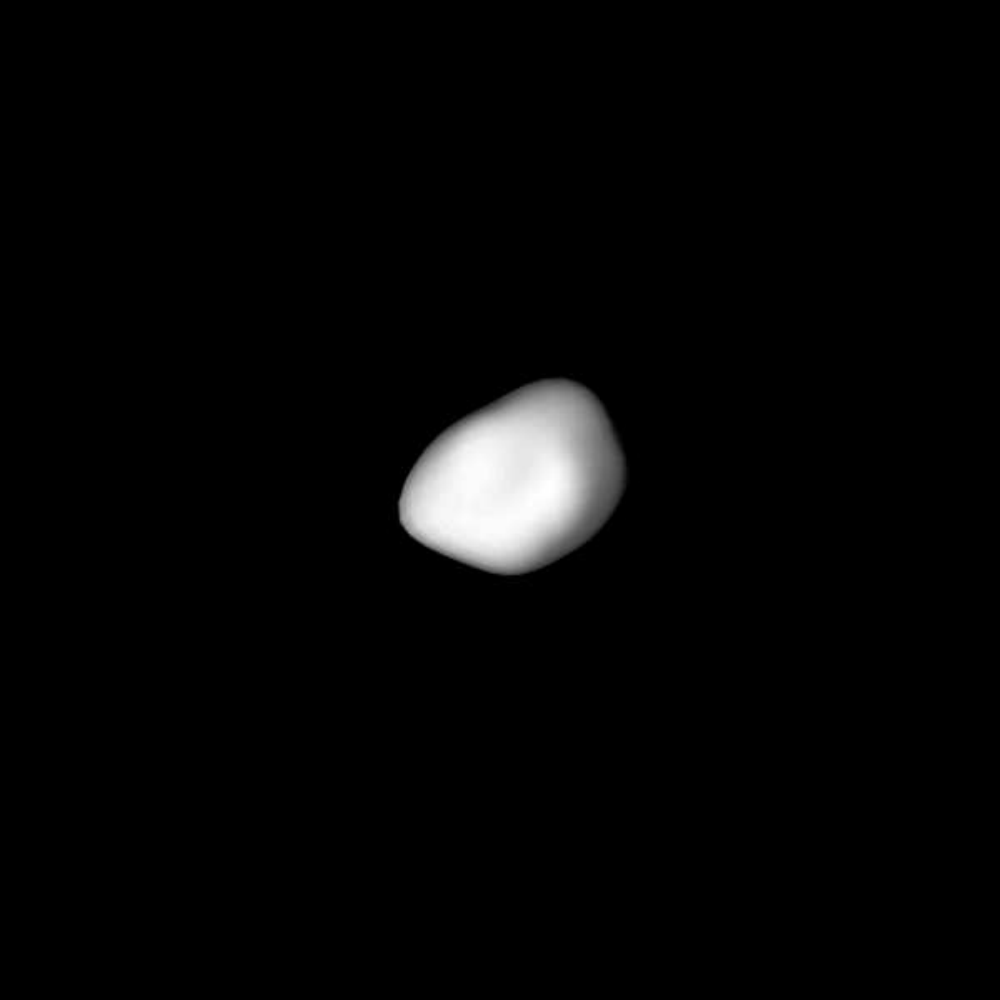}}\\
    \caption{\label{fig:22a}Comparison between model projections and corresponding AO images for asteroid (22) Kalliope (first part).}
\end{figure}

\begin{figure}[tbp]
    \centering
        \resizebox{0.24\hsize}{!}{\includegraphics{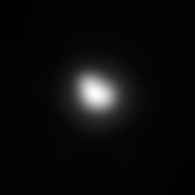}}\resizebox{0.24\hsize}{!}{\includegraphics{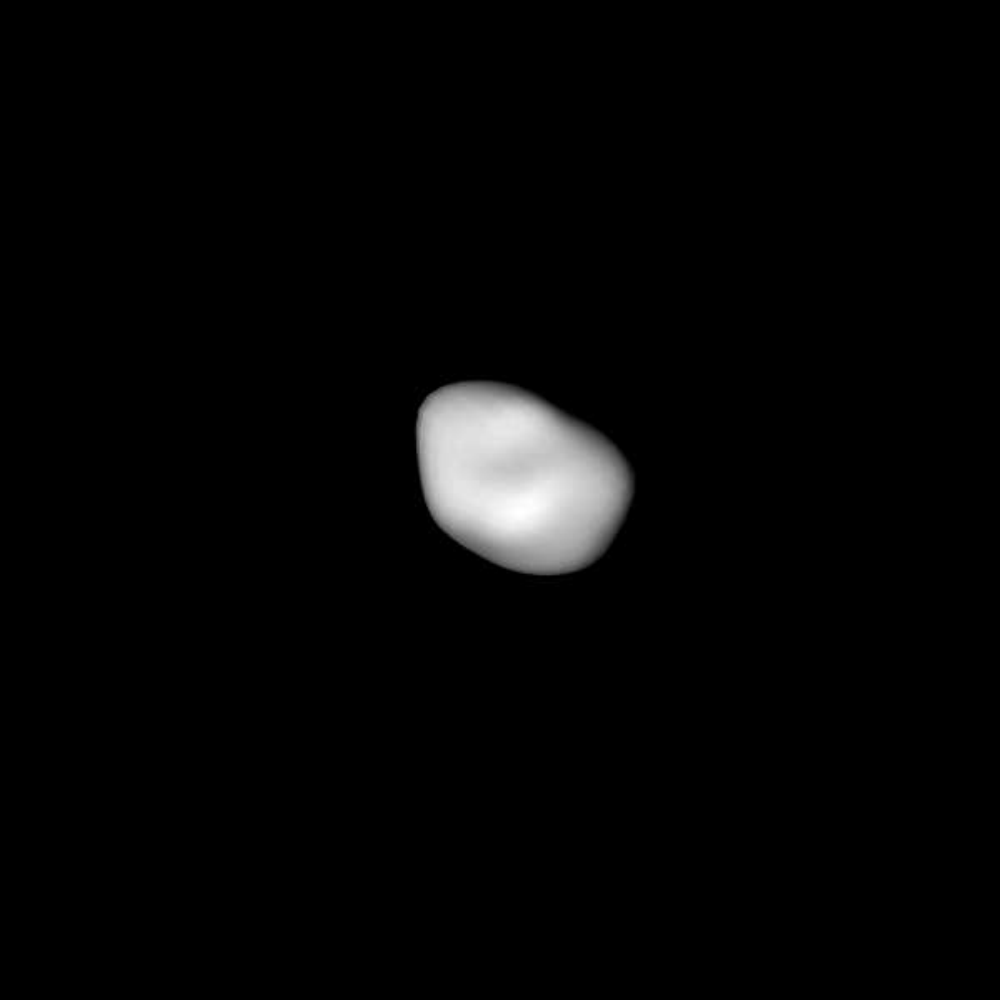}}\resizebox{0.24\hsize}{!}{\includegraphics{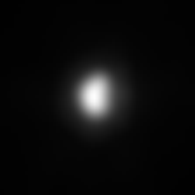}}\resizebox{0.24\hsize}{!}{\includegraphics{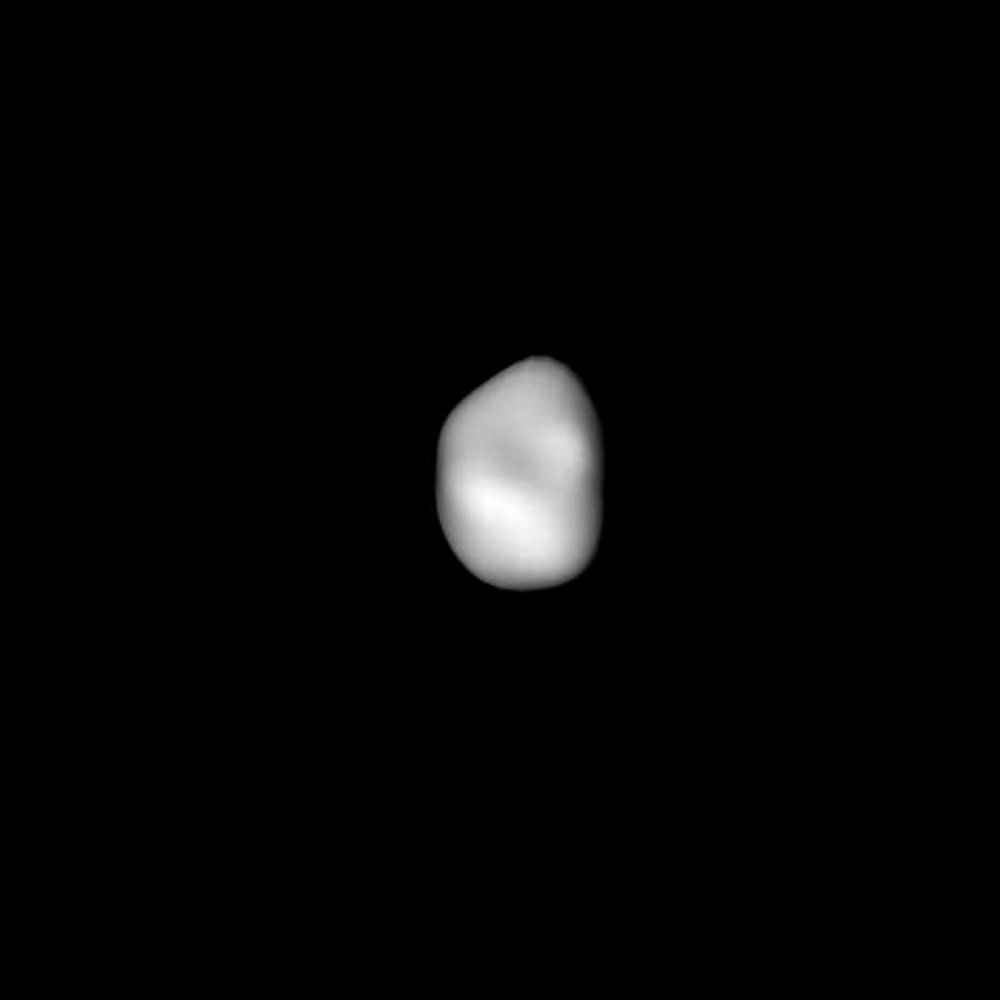}}\\
        \resizebox{0.24\hsize}{!}{\includegraphics{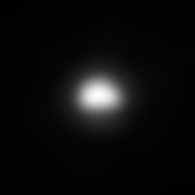}}\resizebox{0.24\hsize}{!}{\includegraphics{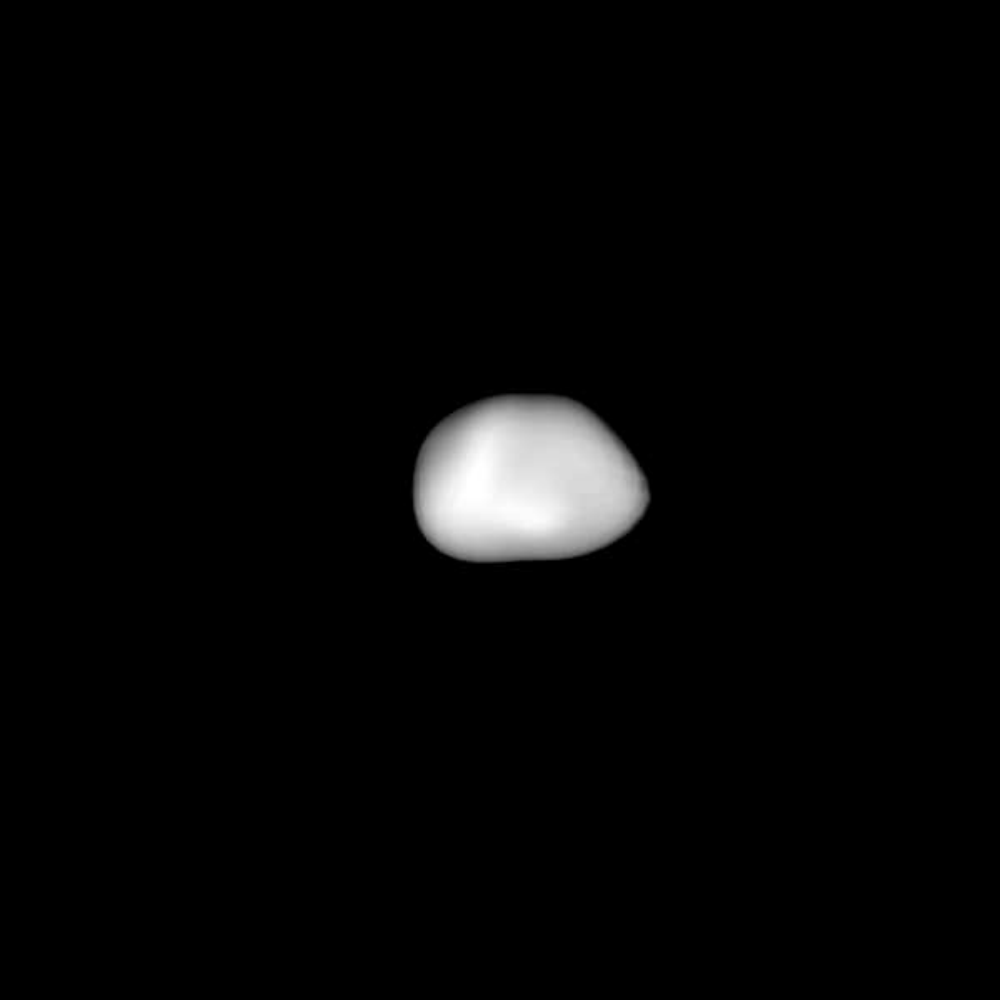}}\resizebox{0.24\hsize}{!}{\includegraphics{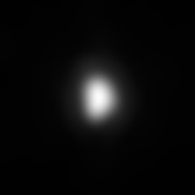}}\resizebox{0.24\hsize}{!}{\includegraphics{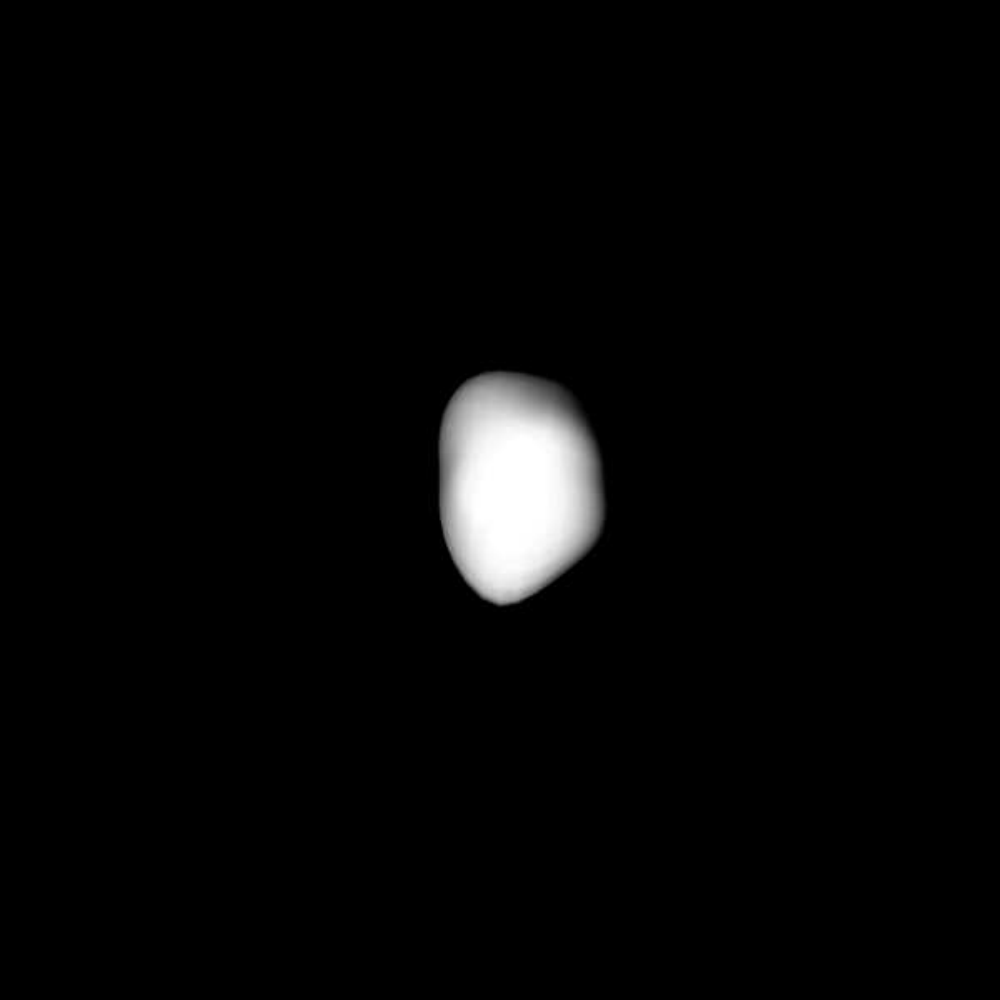}}\\
        \resizebox{0.24\hsize}{!}{\includegraphics{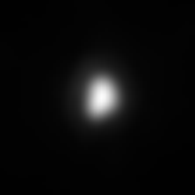}}\resizebox{0.24\hsize}{!}{\includegraphics{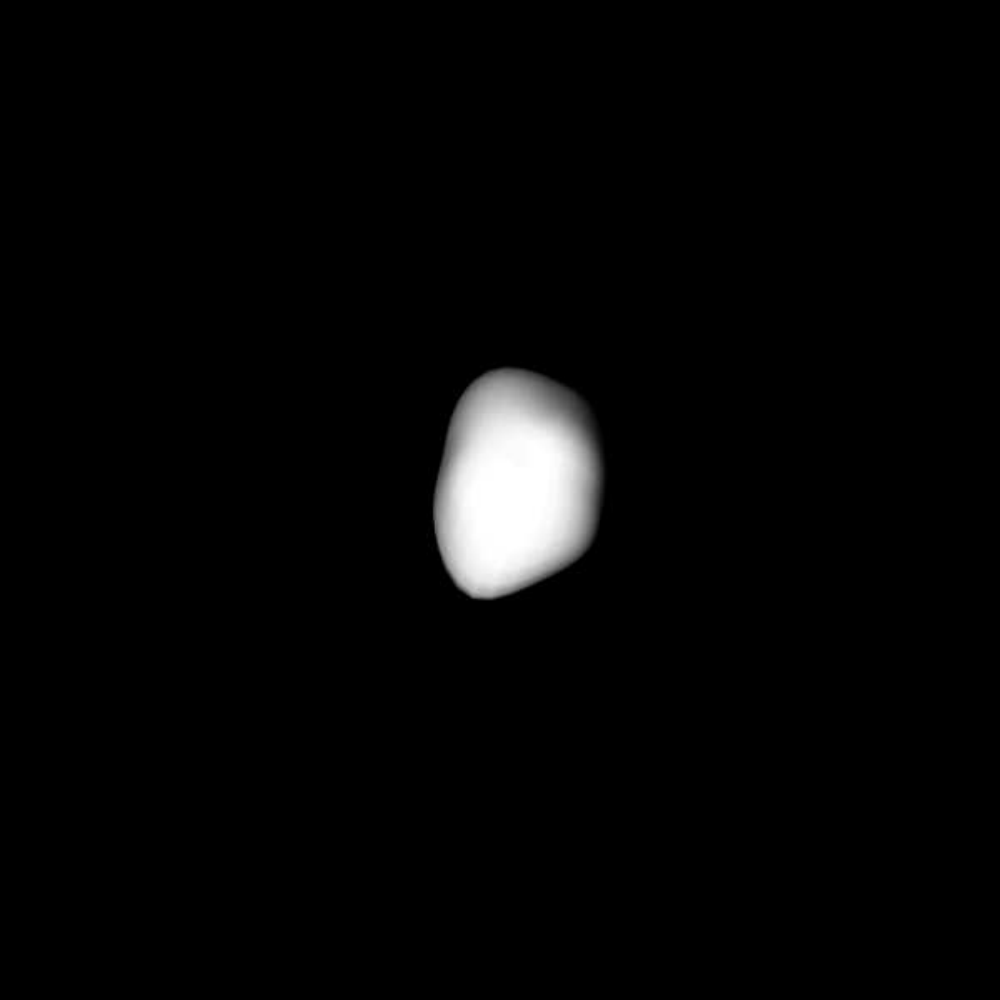}}\resizebox{0.24\hsize}{!}{\includegraphics{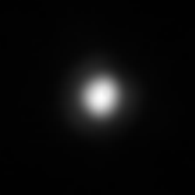}}\resizebox{0.24\hsize}{!}{\includegraphics{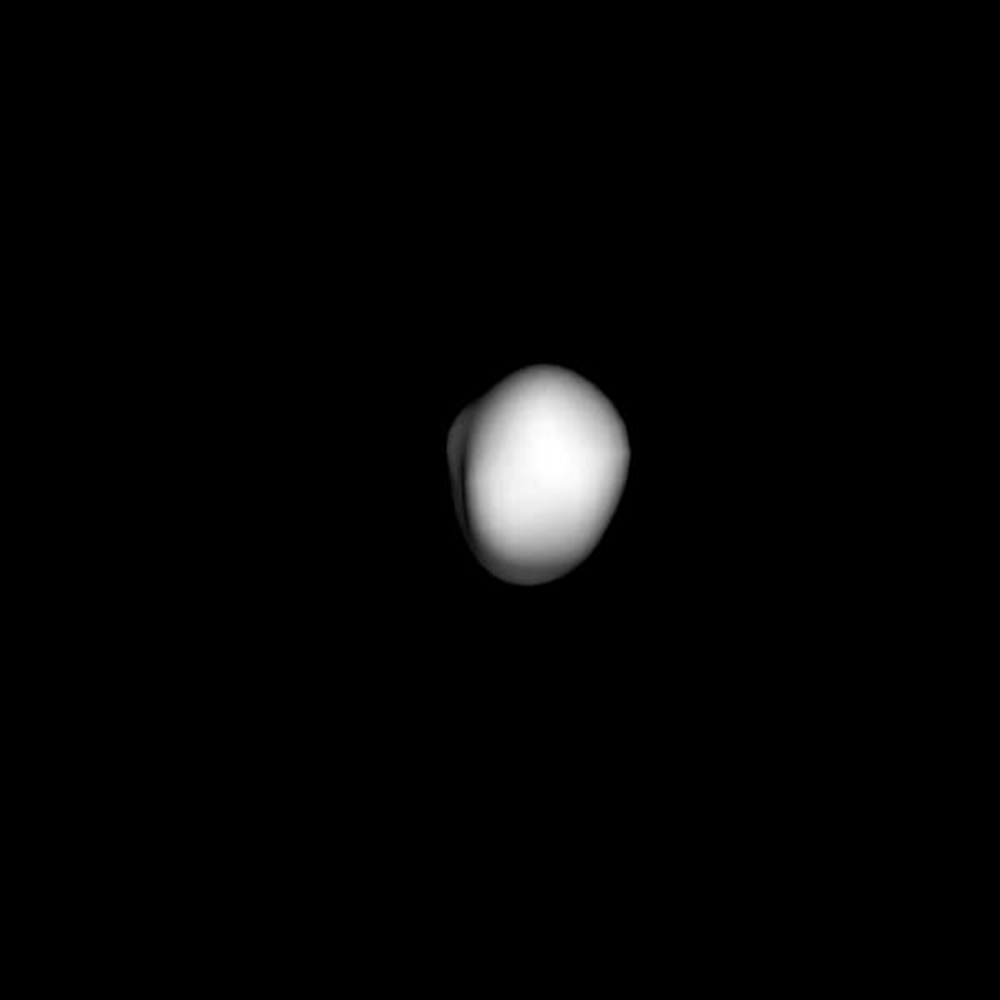}}\\
        \resizebox{0.24\hsize}{!}{\includegraphics{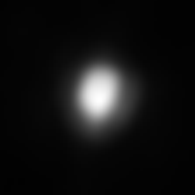}}\resizebox{0.24\hsize}{!}{\includegraphics{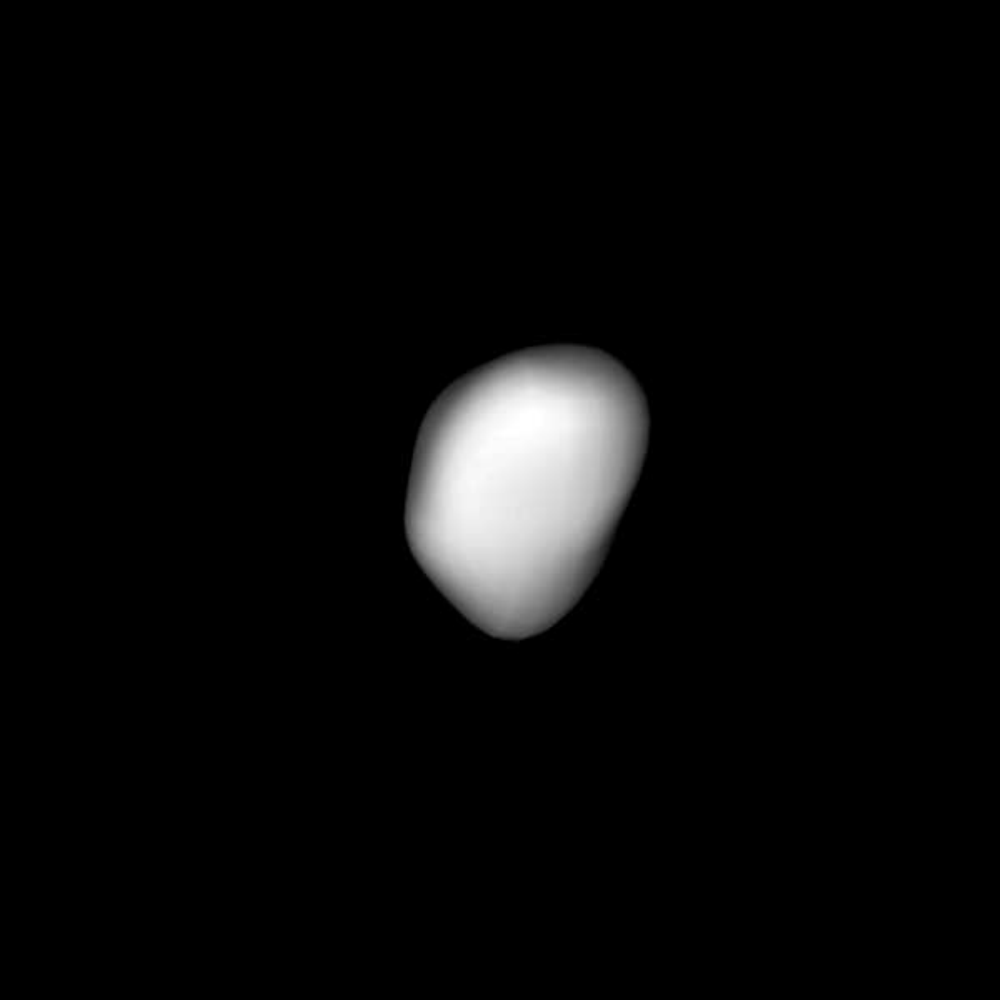}}\\
    \caption{\label{fig:22b}Comparison between model projections and corresponding AO images for asteroid (22) Kalliope (second part).}
\end{figure}

\begin{figure}[tbp]
    \centering
        \resizebox{0.24\hsize}{!}{\includegraphics{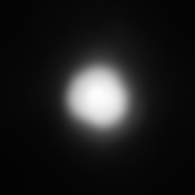}}\resizebox{0.24\hsize}{!}{\includegraphics{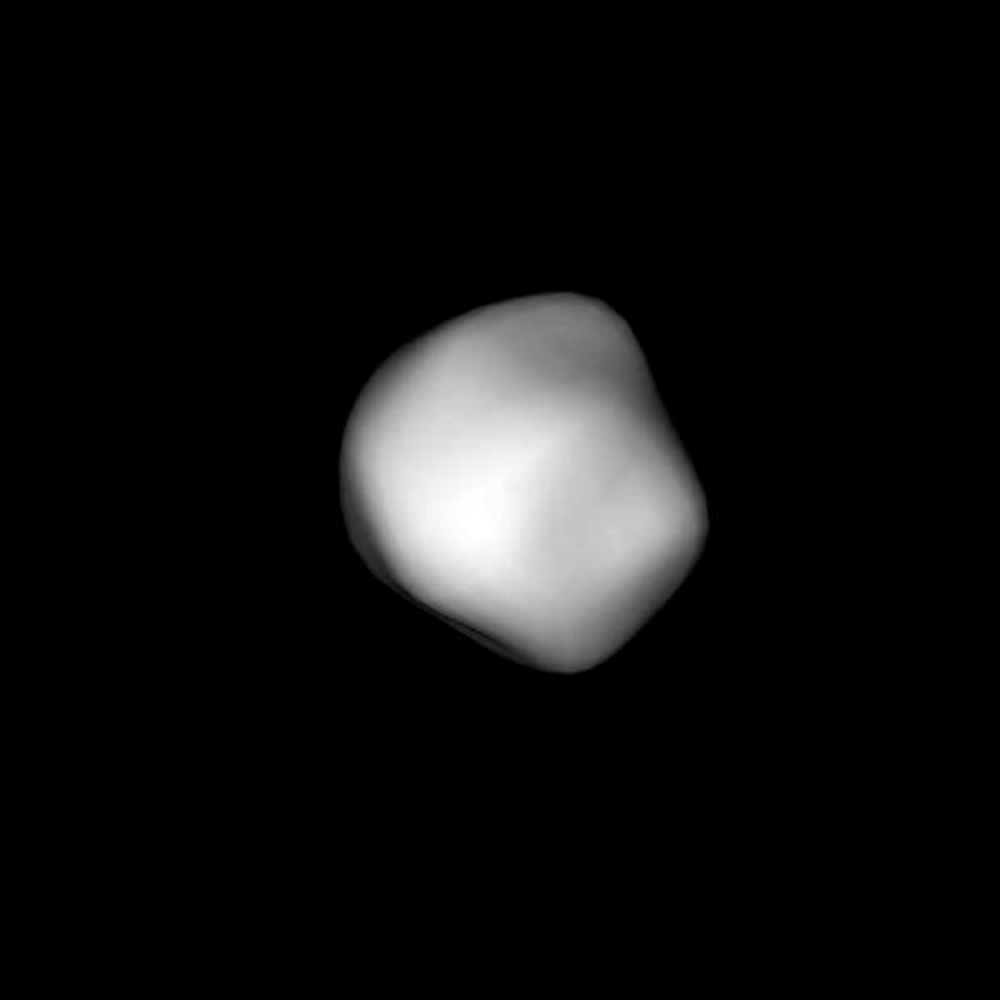}}\resizebox{0.24\hsize}{!}{\includegraphics{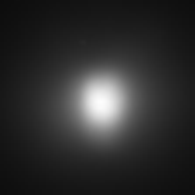}}\resizebox{0.24\hsize}{!}{\includegraphics{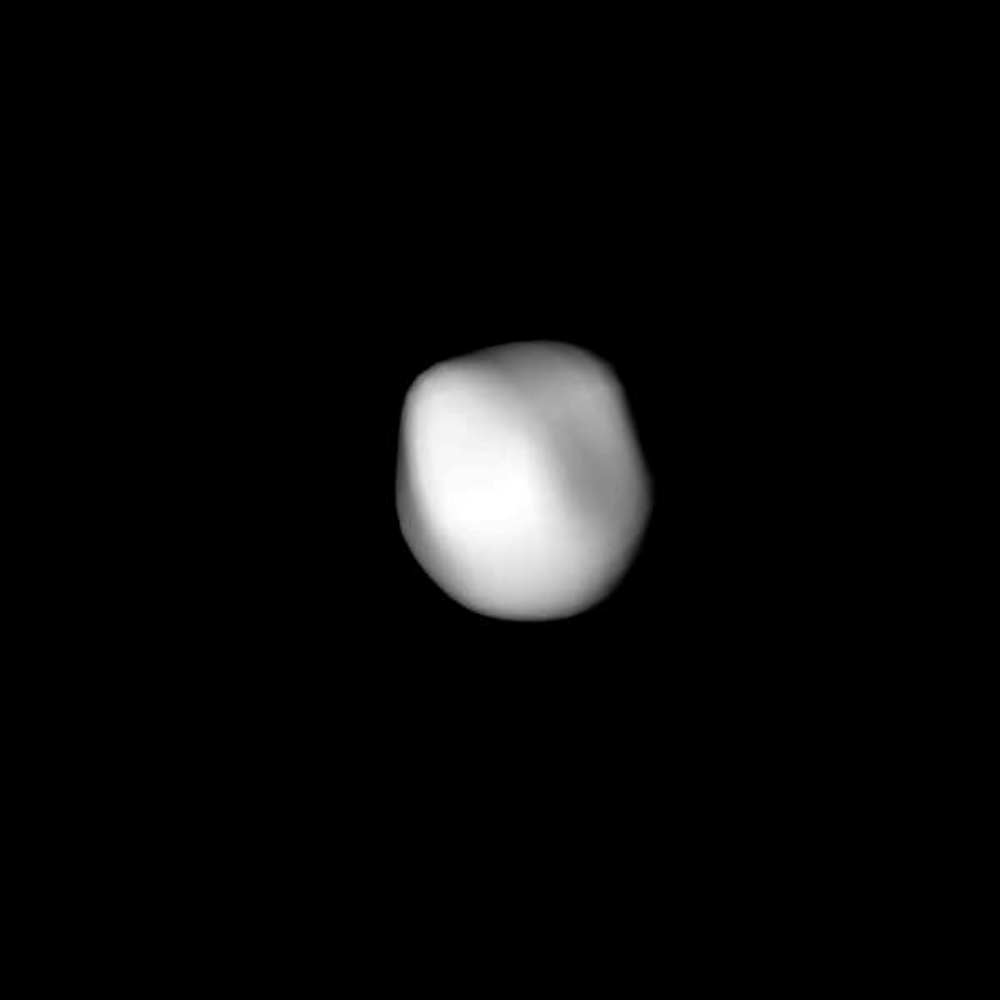}}\\
        \resizebox{0.24\hsize}{!}{\includegraphics{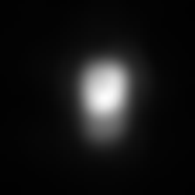}}\resizebox{0.24\hsize}{!}{\includegraphics{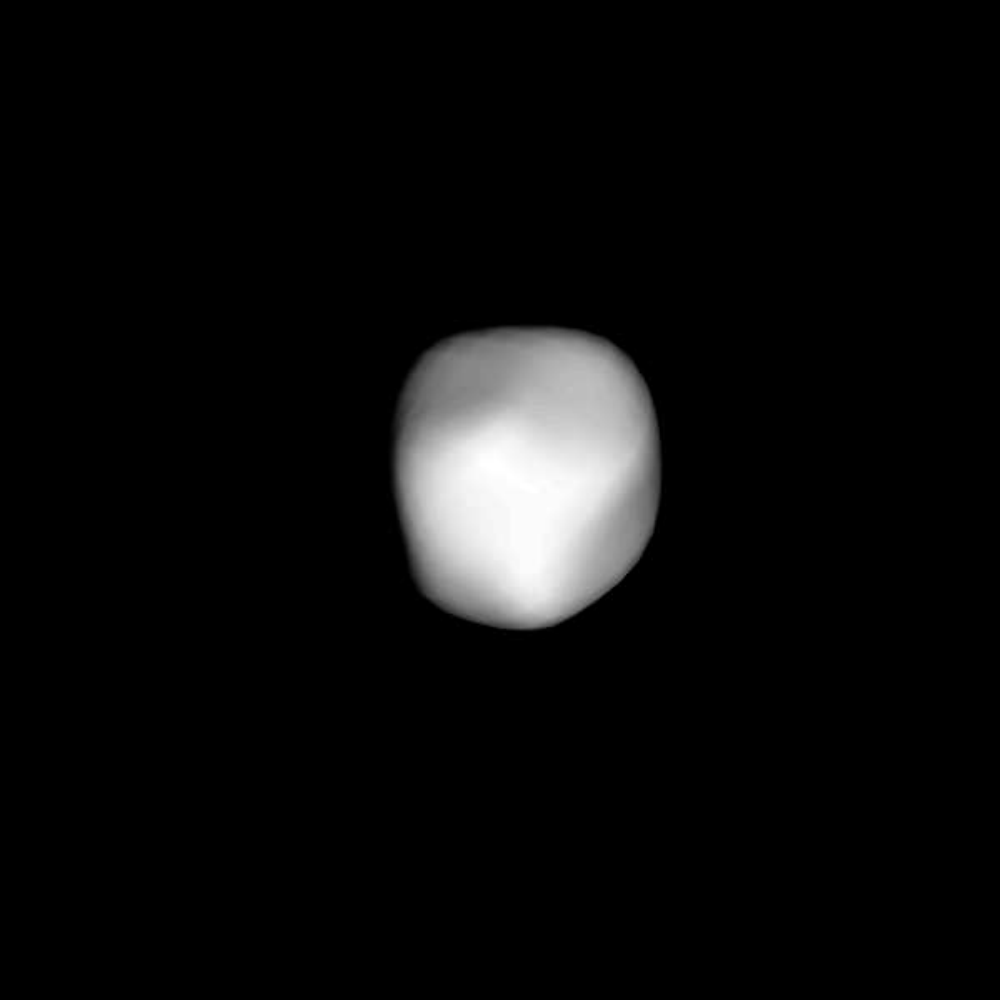}}\resizebox{0.24\hsize}{!}{\includegraphics{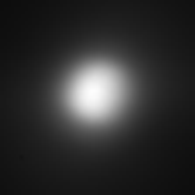}}\resizebox{0.24\hsize}{!}{\includegraphics{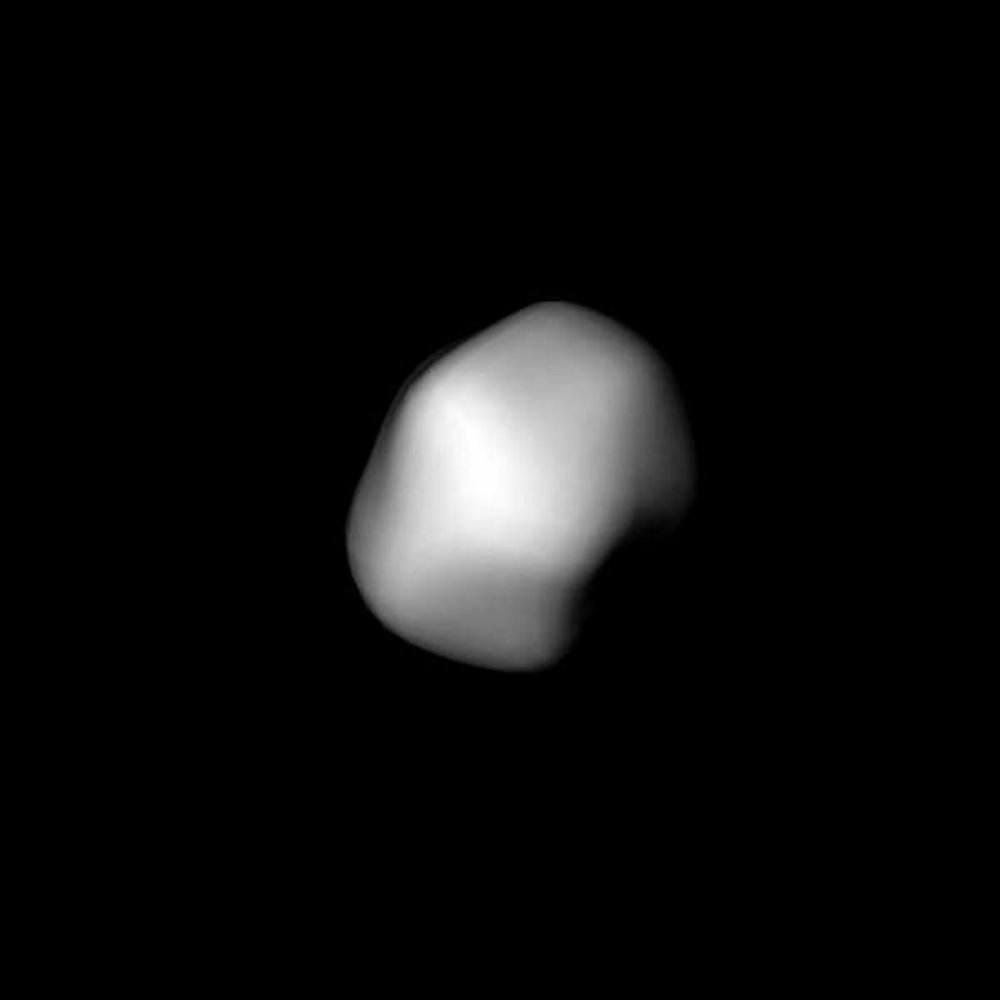}}\\
        \resizebox{0.24\hsize}{!}{\includegraphics{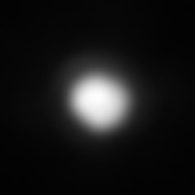}}\resizebox{0.24\hsize}{!}{\includegraphics{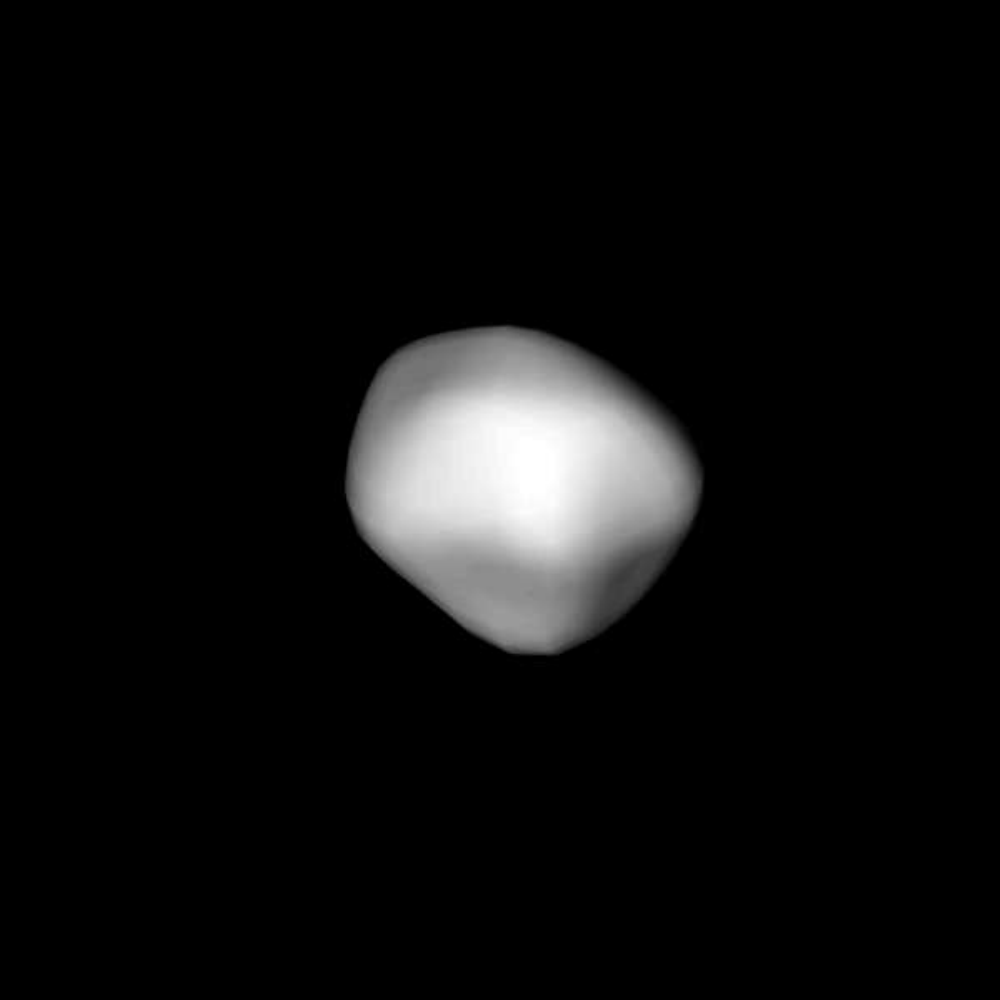}}\resizebox{0.24\hsize}{!}{\includegraphics{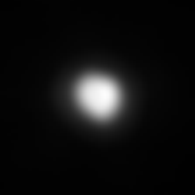}}\resizebox{0.24\hsize}{!}{\includegraphics{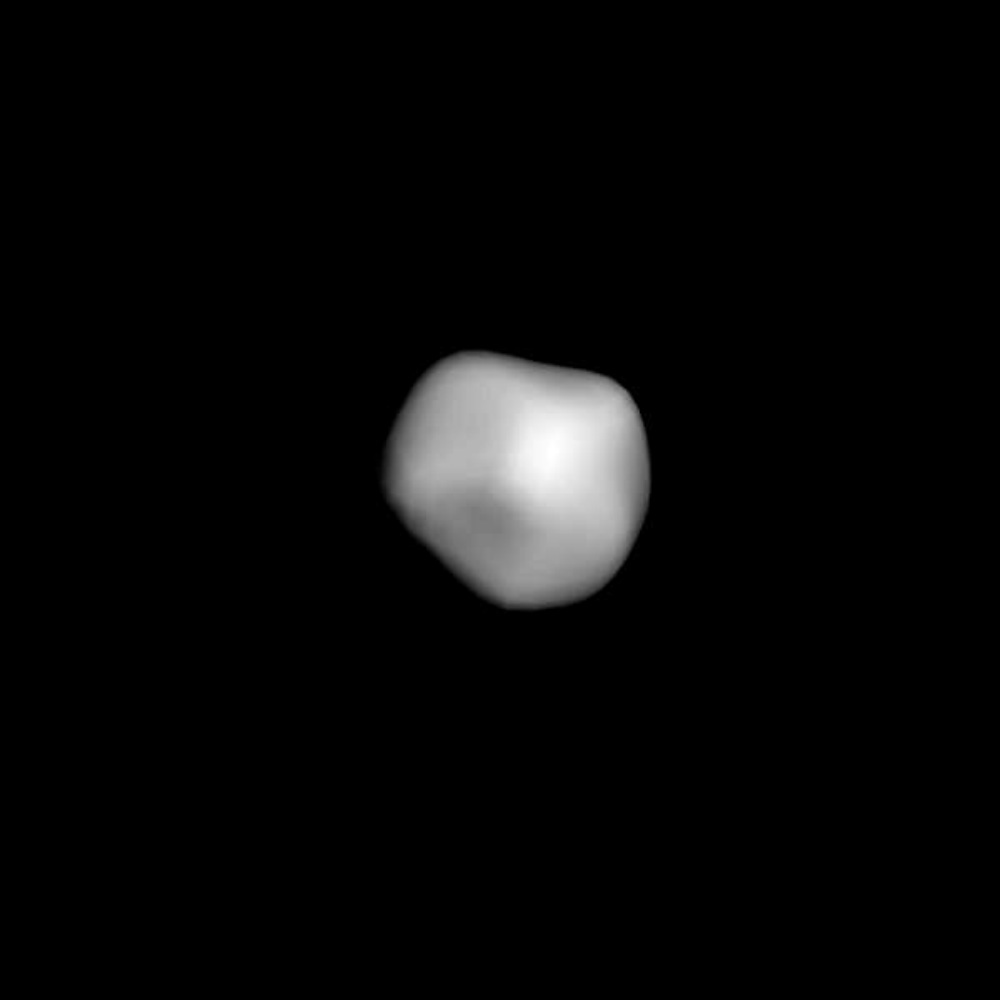}}\\
        \resizebox{0.24\hsize}{!}{\includegraphics{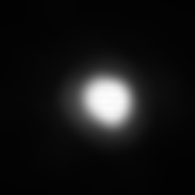}}\resizebox{0.24\hsize}{!}{\includegraphics{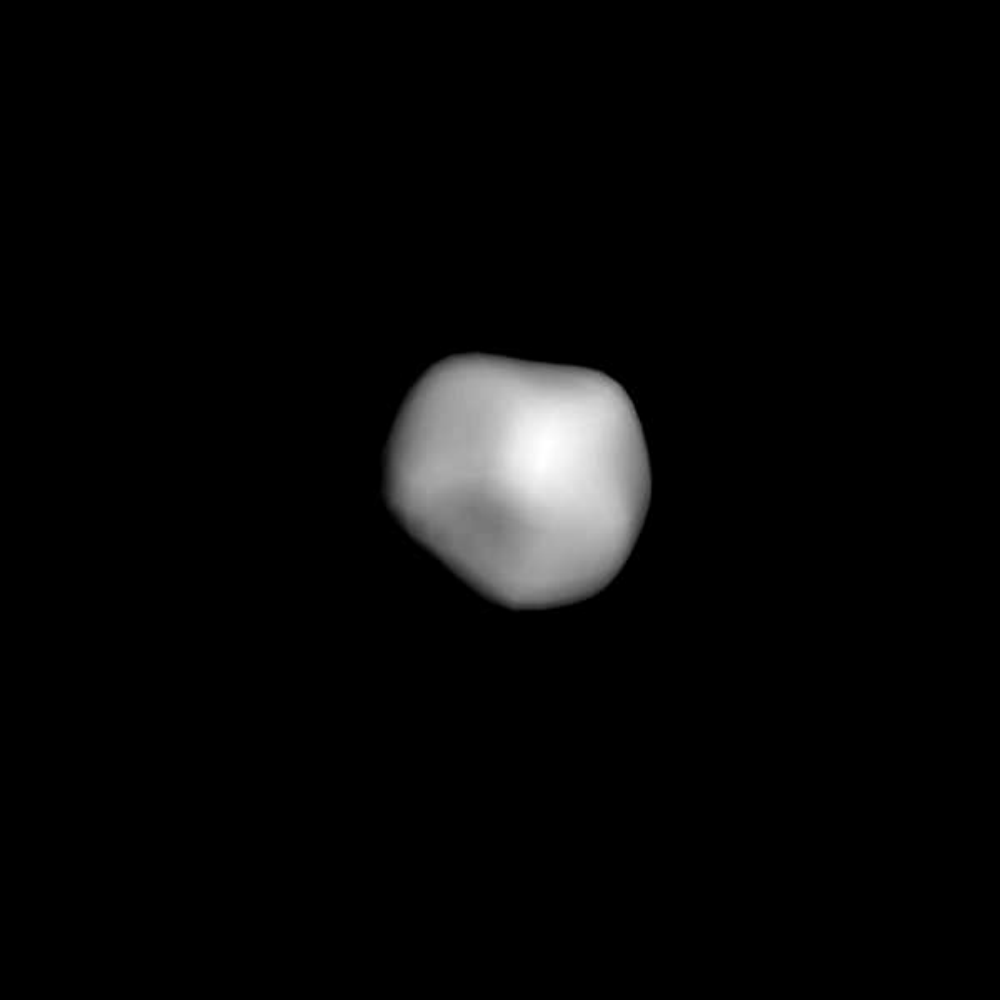}}\\
    \caption{\label{fig:29}Comparison between model projections and corresponding AO images for asteroid (29) Amphitrite.}
\end{figure}

\begin{figure}[tbp]
    \centering
        \resizebox{0.24\hsize}{!}{\includegraphics{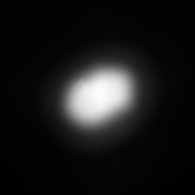}}\resizebox{0.24\hsize}{!}{\includegraphics{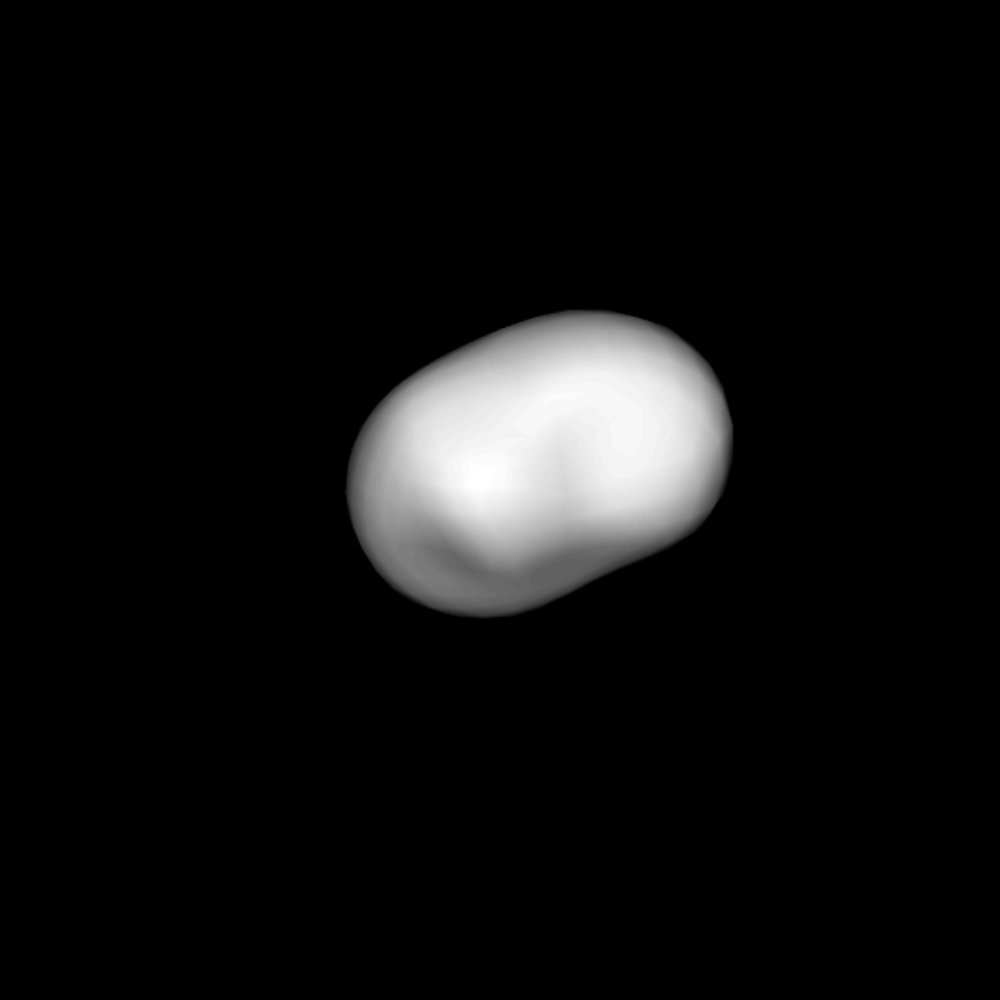}}\resizebox{0.24\hsize}{!}{\includegraphics{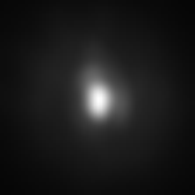}}\resizebox{0.24\hsize}{!}{\includegraphics{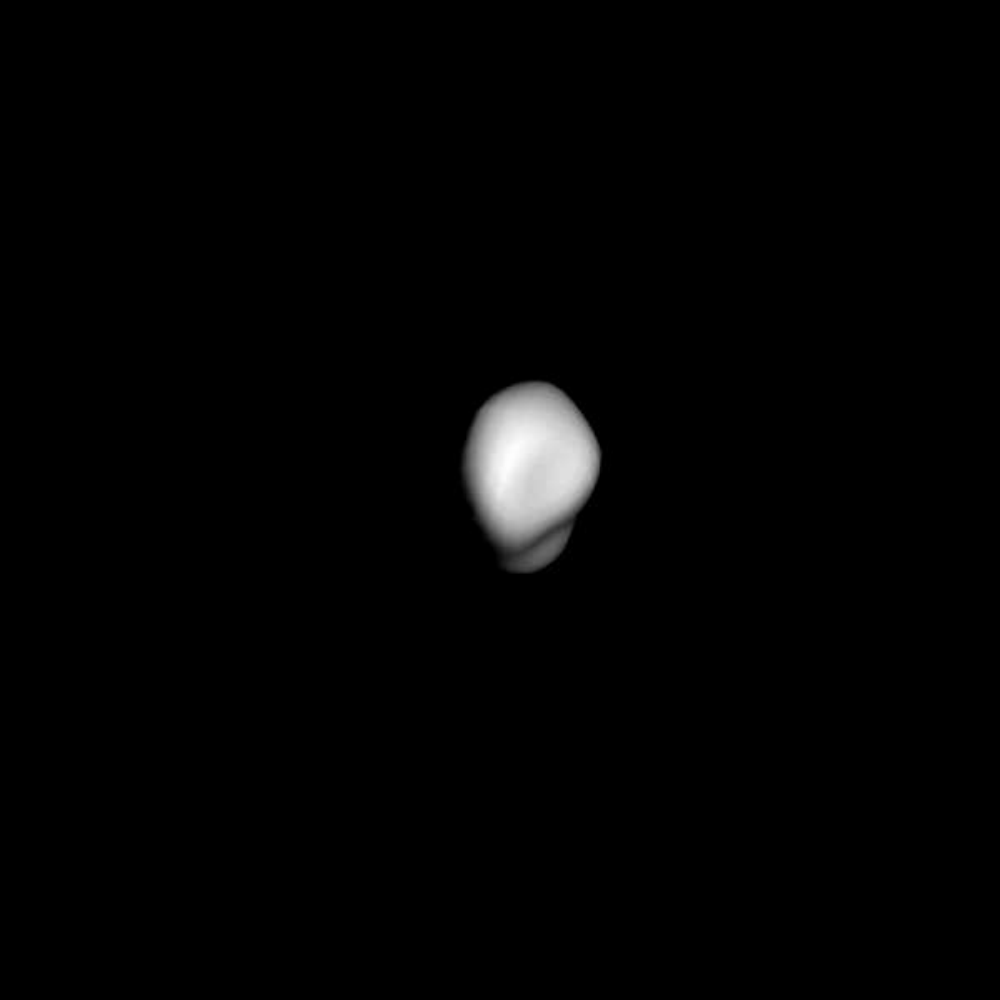}}\\
        \resizebox{0.24\hsize}{!}{\includegraphics{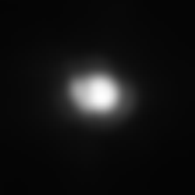}}\resizebox{0.24\hsize}{!}{\includegraphics{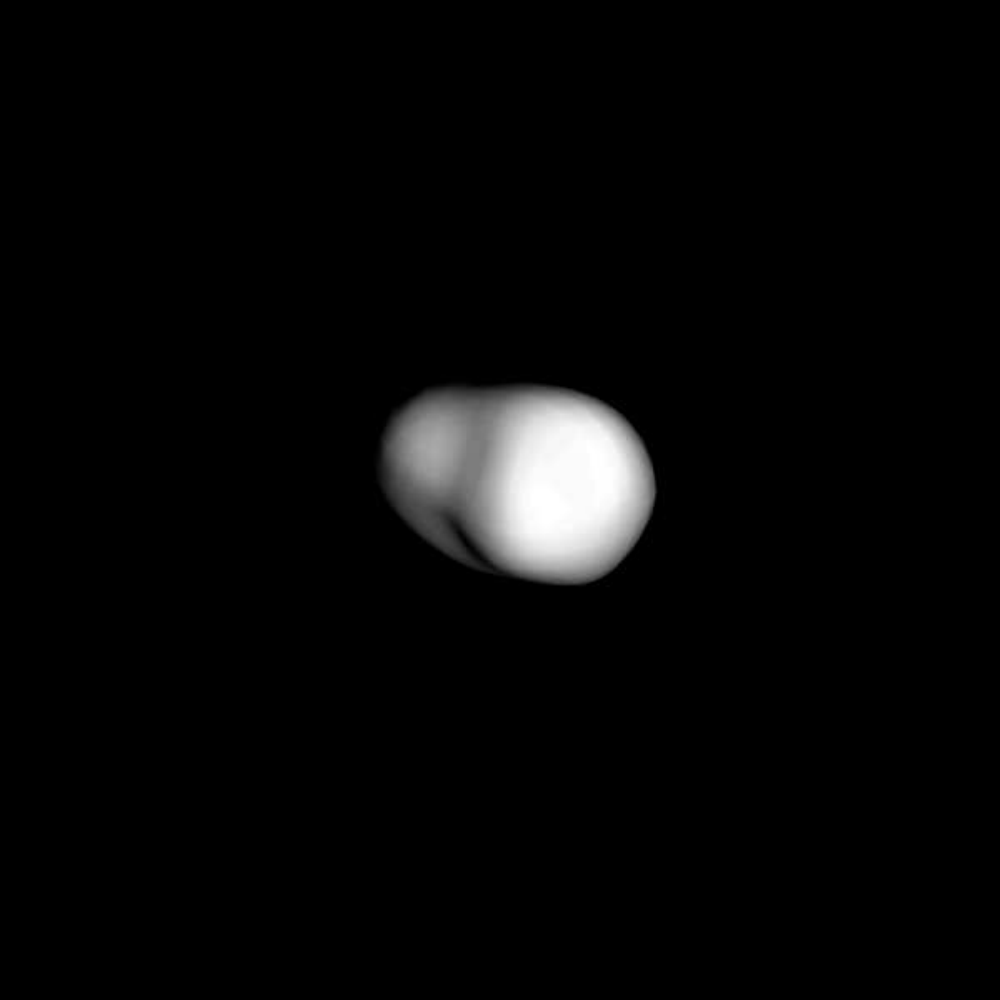}}\\
    \caption{\label{fig:39}Comparison between model projections and corresponding AO images for asteroid (39) Laetitia.}
\end{figure}

\begin{figure}[tbp]
    \centering
        \resizebox{0.24\hsize}{!}{\includegraphics{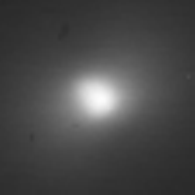}}\resizebox{0.24\hsize}{!}{\includegraphics{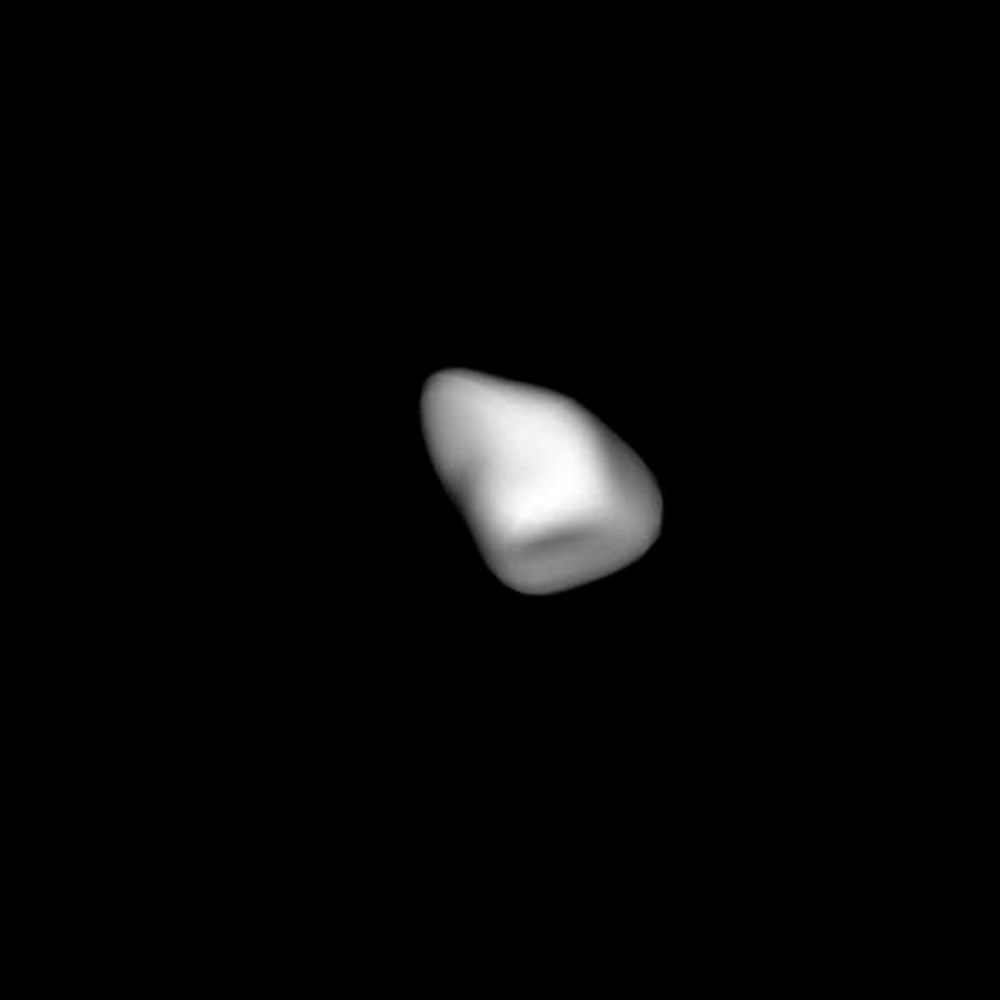}}\resizebox{0.24\hsize}{!}{\includegraphics{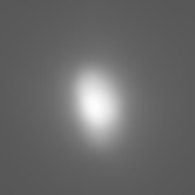}}\resizebox{0.24\hsize}{!}{\includegraphics{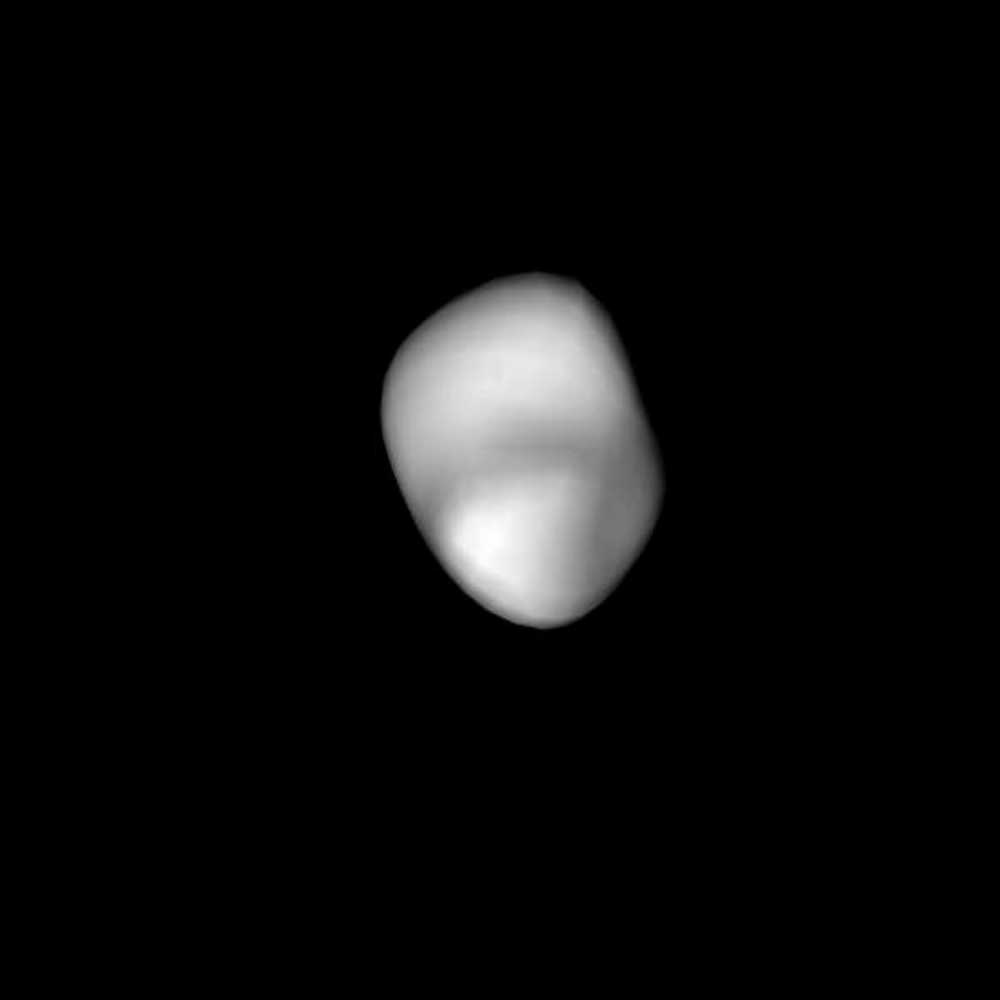}}\\
        \resizebox{0.24\hsize}{!}{\includegraphics{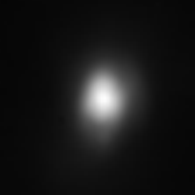}}\resizebox{0.24\hsize}{!}{\includegraphics{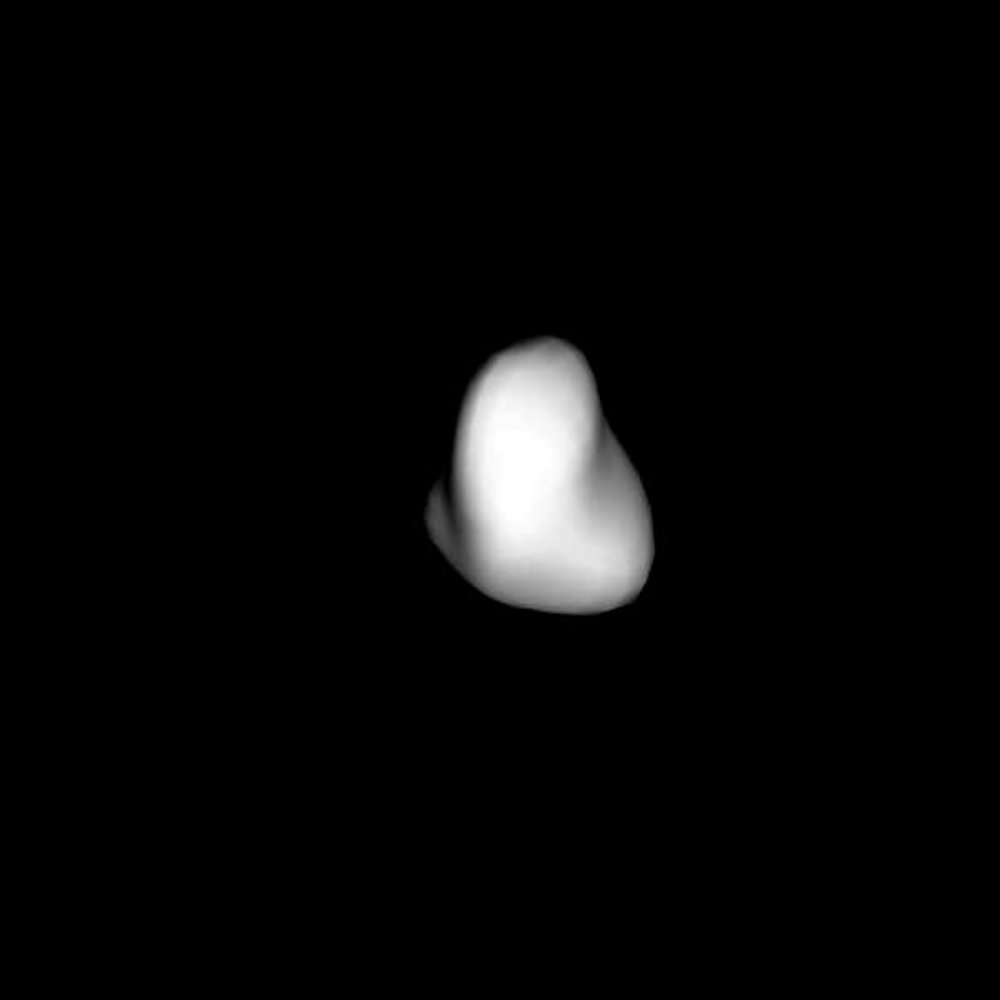}}\resizebox{0.24\hsize}{!}{\includegraphics{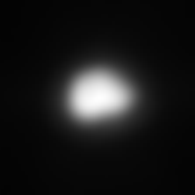}}\resizebox{0.24\hsize}{!}{\includegraphics{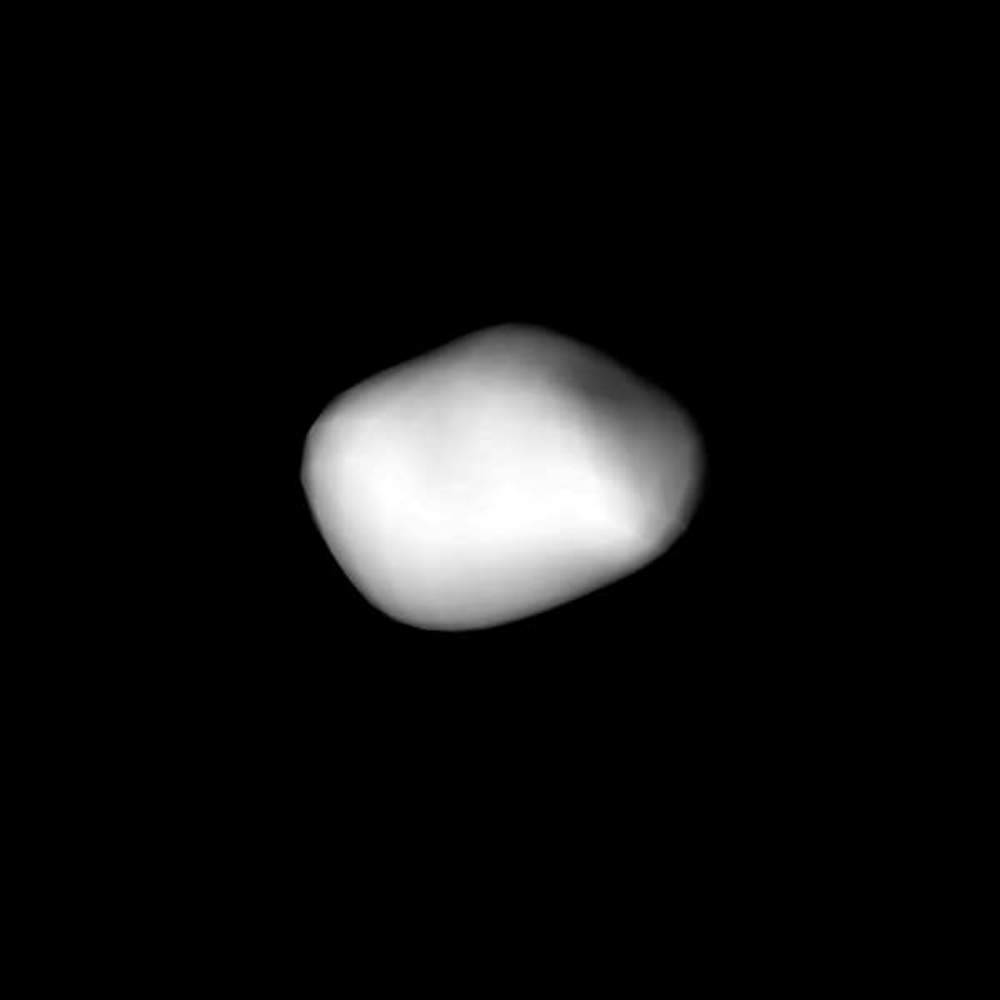}}\\
        \resizebox{0.24\hsize}{!}{\includegraphics{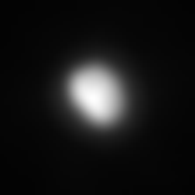}}\resizebox{0.24\hsize}{!}{\includegraphics{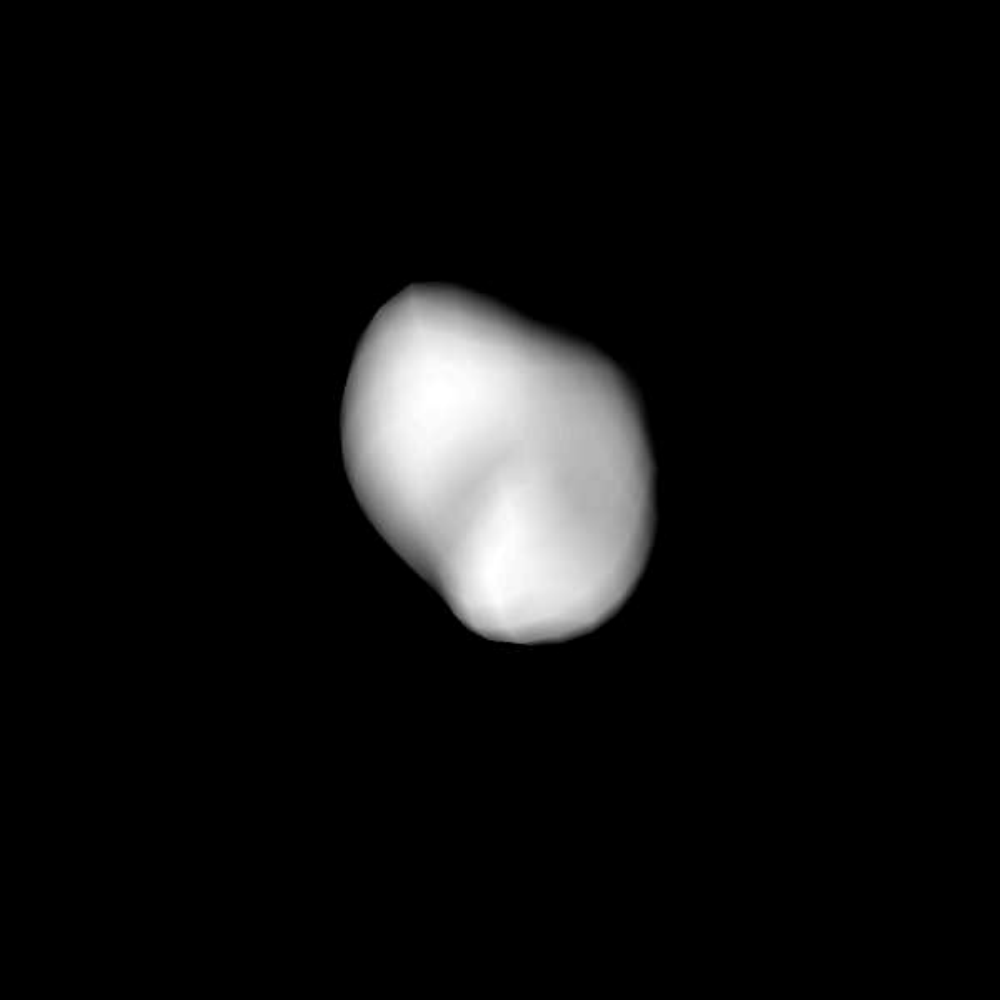}}\resizebox{0.24\hsize}{!}{\includegraphics{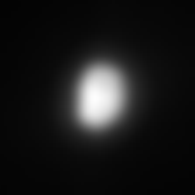}}\resizebox{0.24\hsize}{!}{\includegraphics{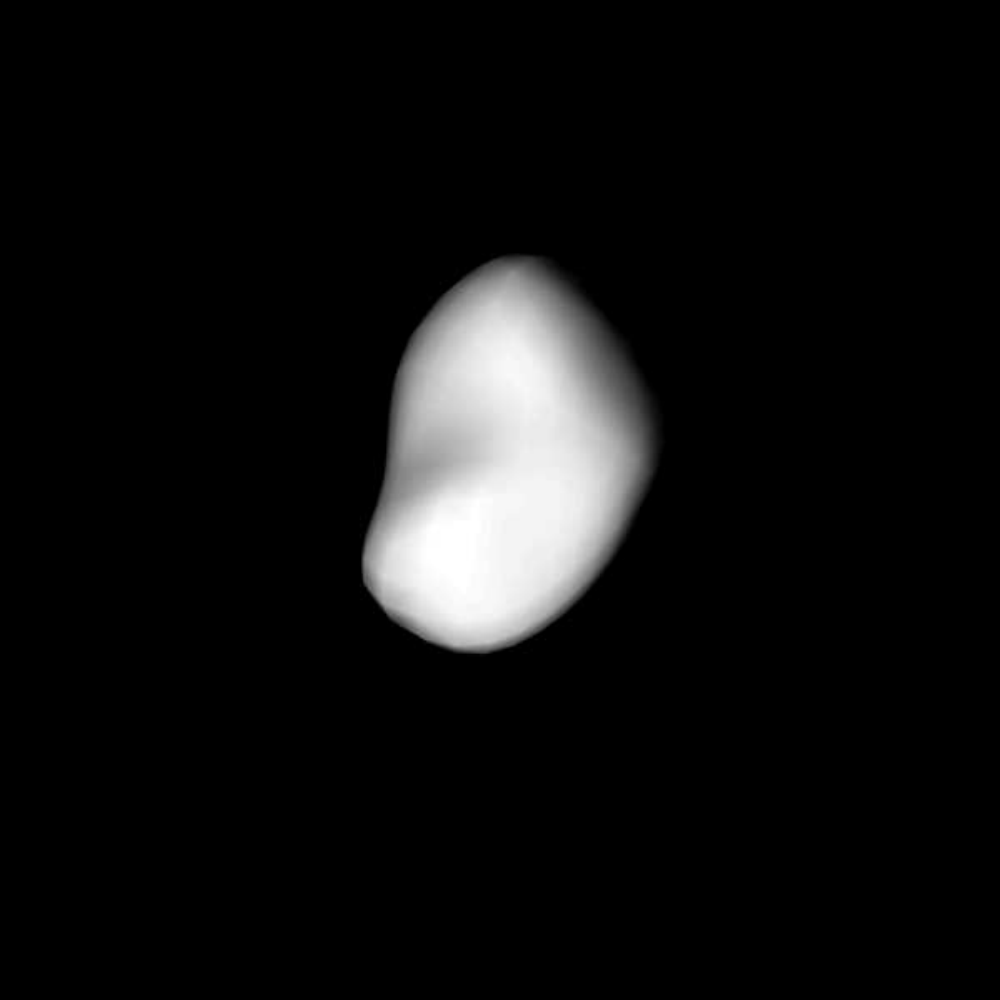}}\\
        \resizebox{0.24\hsize}{!}{\includegraphics{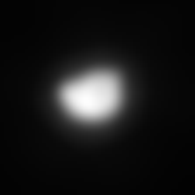}}\resizebox{0.24\hsize}{!}{\includegraphics{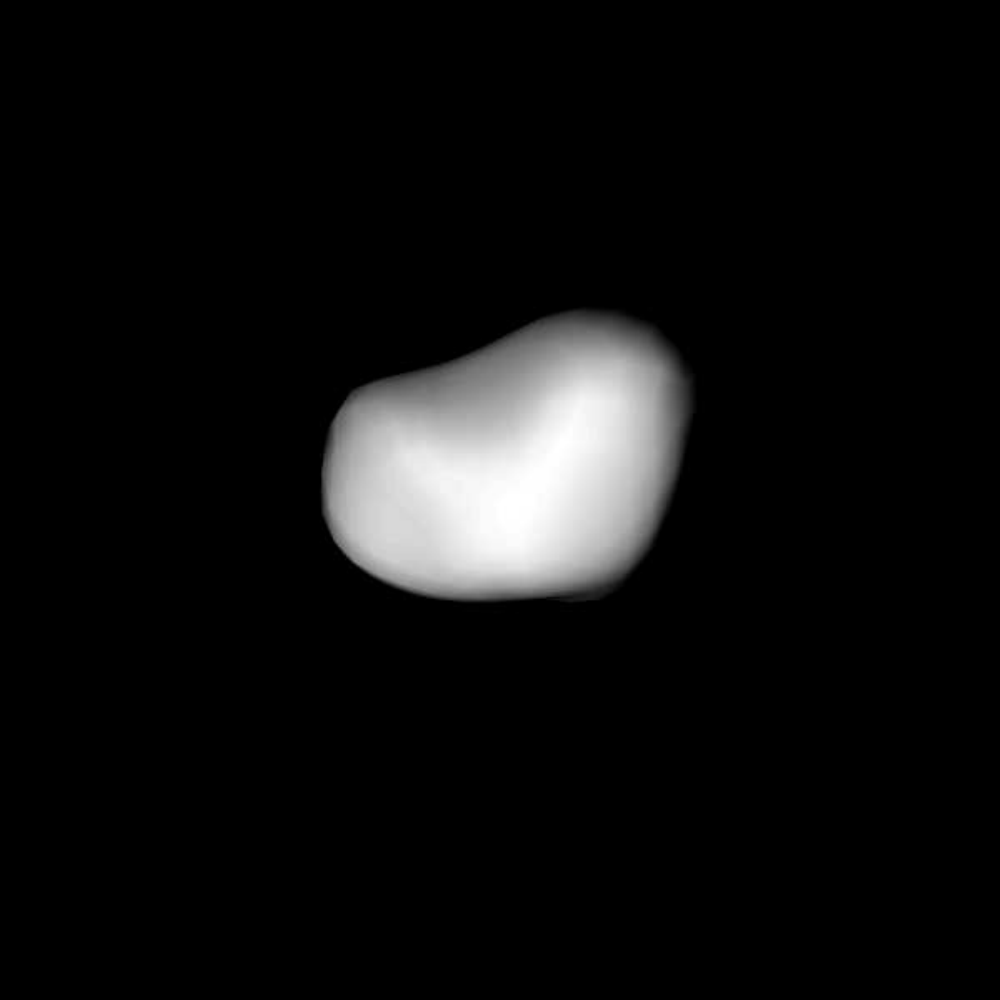}}\\
    \caption{\label{fig:41}Comparison between model projections and corresponding AO images for asteroid (41) Daphne.}
\end{figure}

\begin{figure}[tbp]
    \centering
        \resizebox{0.24\hsize}{!}{\includegraphics{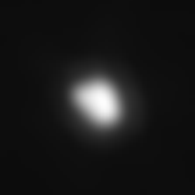}}\resizebox{0.24\hsize}{!}{\includegraphics{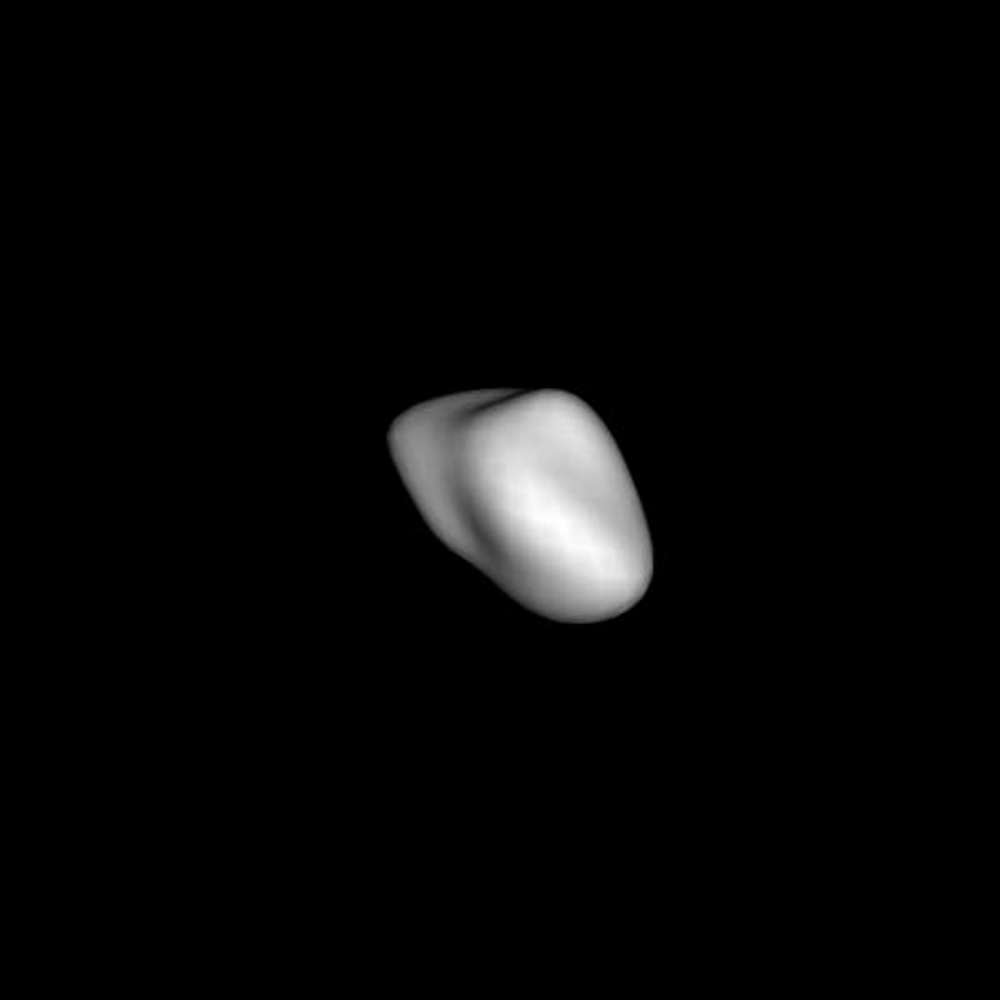}}\\
    \caption{\label{fig:43}Comparison between model projections and corresponding AO images for asteroid (43) Ariadne.}
\end{figure}

\clearpage

\begin{figure}[tbp]
    \centering
        \resizebox{0.24\hsize}{!}{\includegraphics{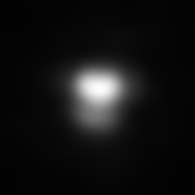}}\resizebox{0.24\hsize}{!}{\includegraphics{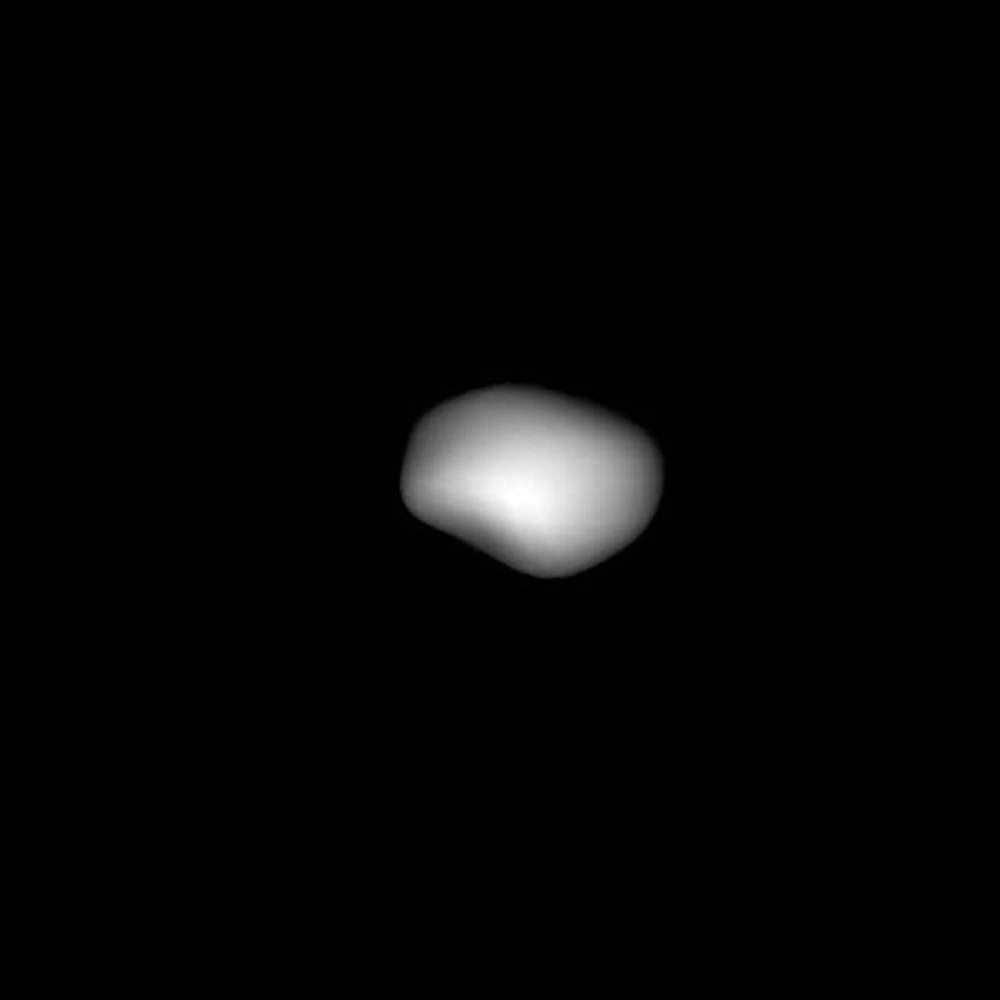}}\resizebox{0.24\hsize}{!}{\includegraphics{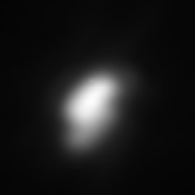}}\resizebox{0.24\hsize}{!}{\includegraphics{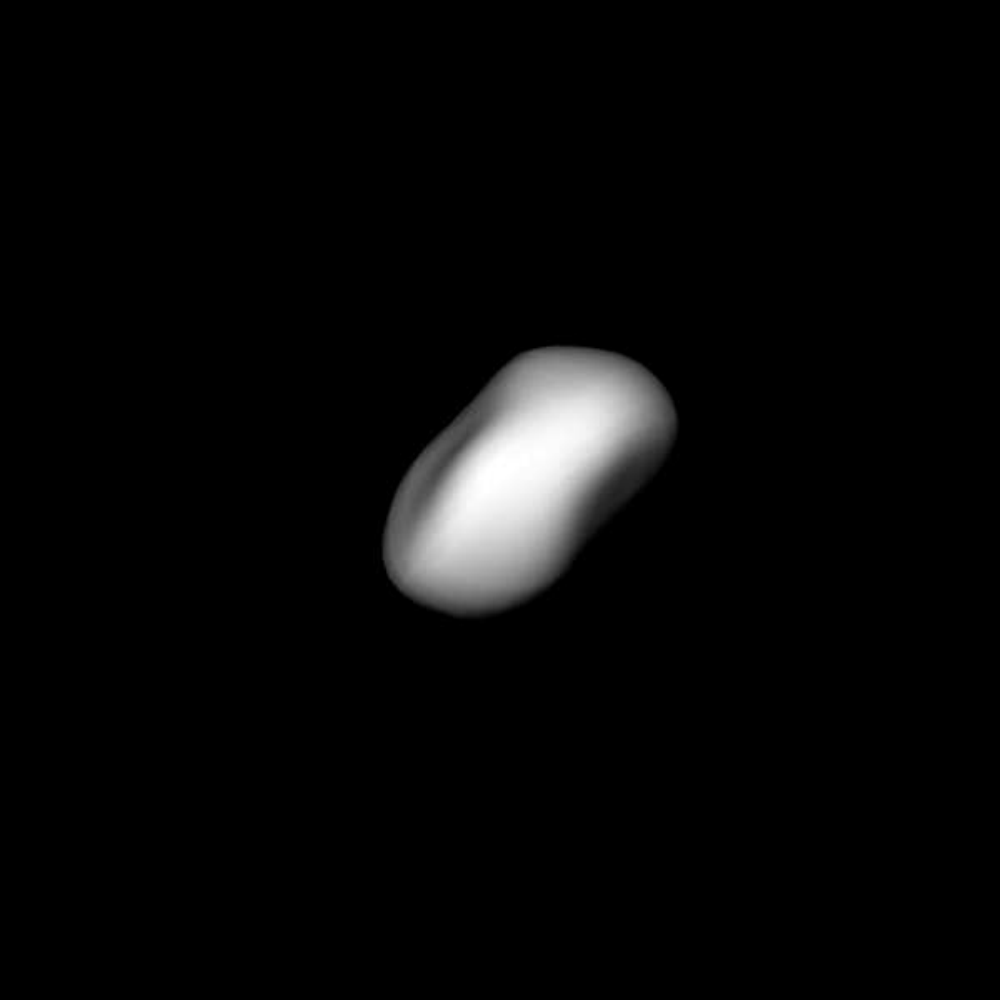}}\\
        \resizebox{0.24\hsize}{!}{\includegraphics{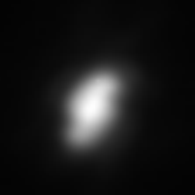}}\resizebox{0.24\hsize}{!}{\includegraphics{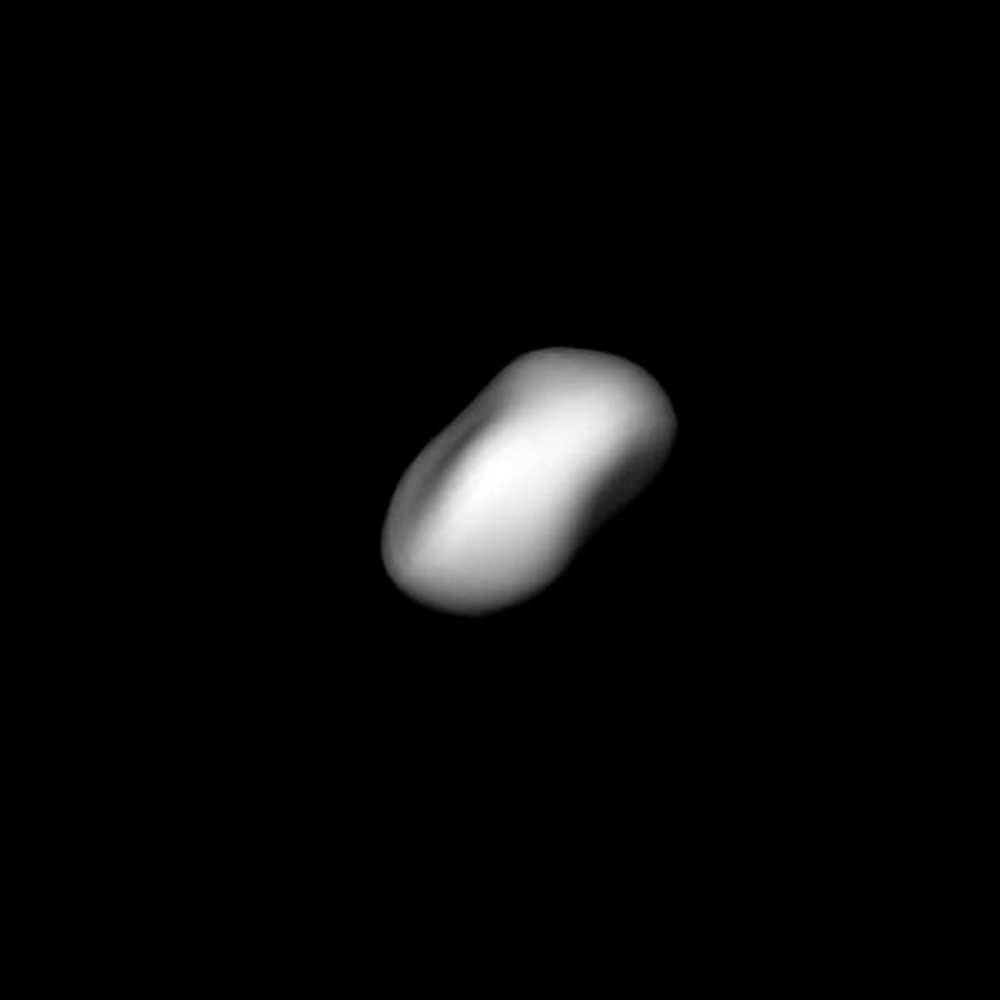}}\resizebox{0.24\hsize}{!}{\includegraphics{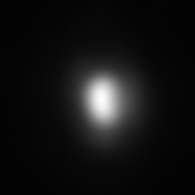}}\resizebox{0.24\hsize}{!}{\includegraphics{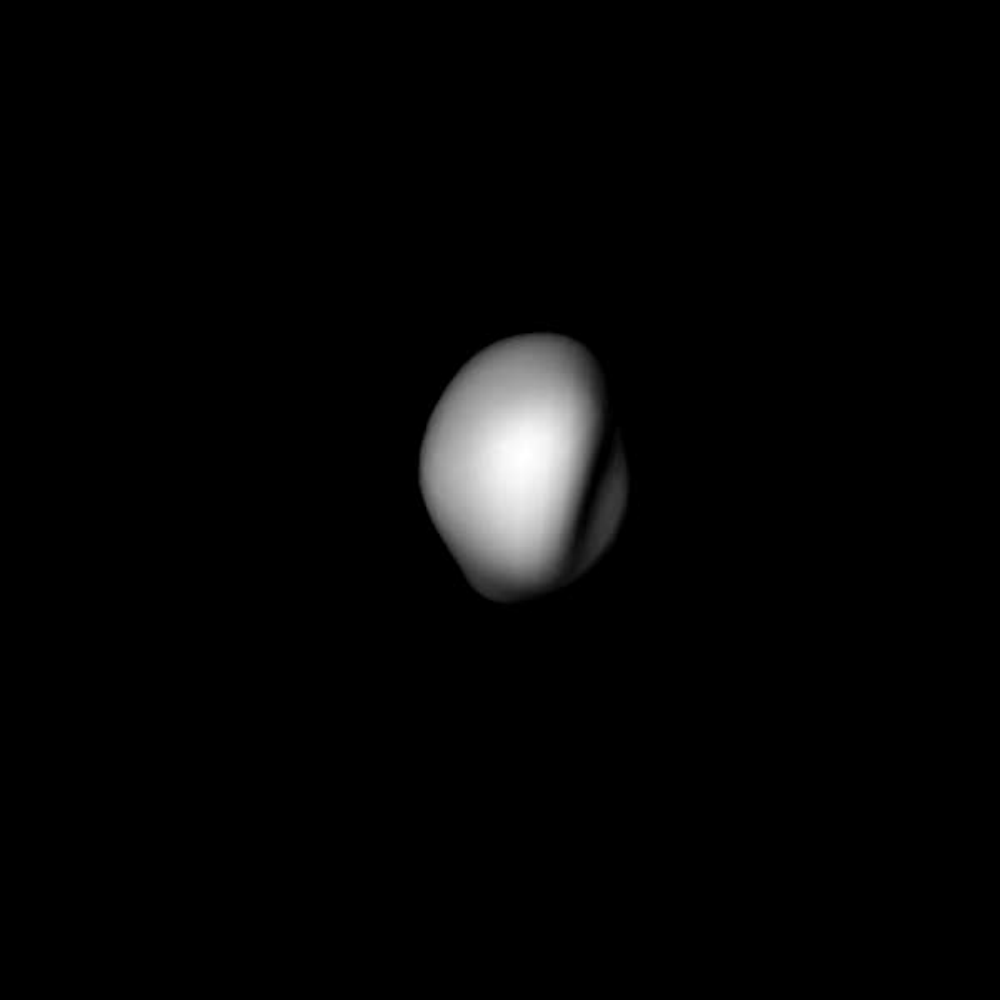}}\\
        \resizebox{0.24\hsize}{!}{\includegraphics{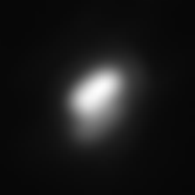}}\resizebox{0.24\hsize}{!}{\includegraphics{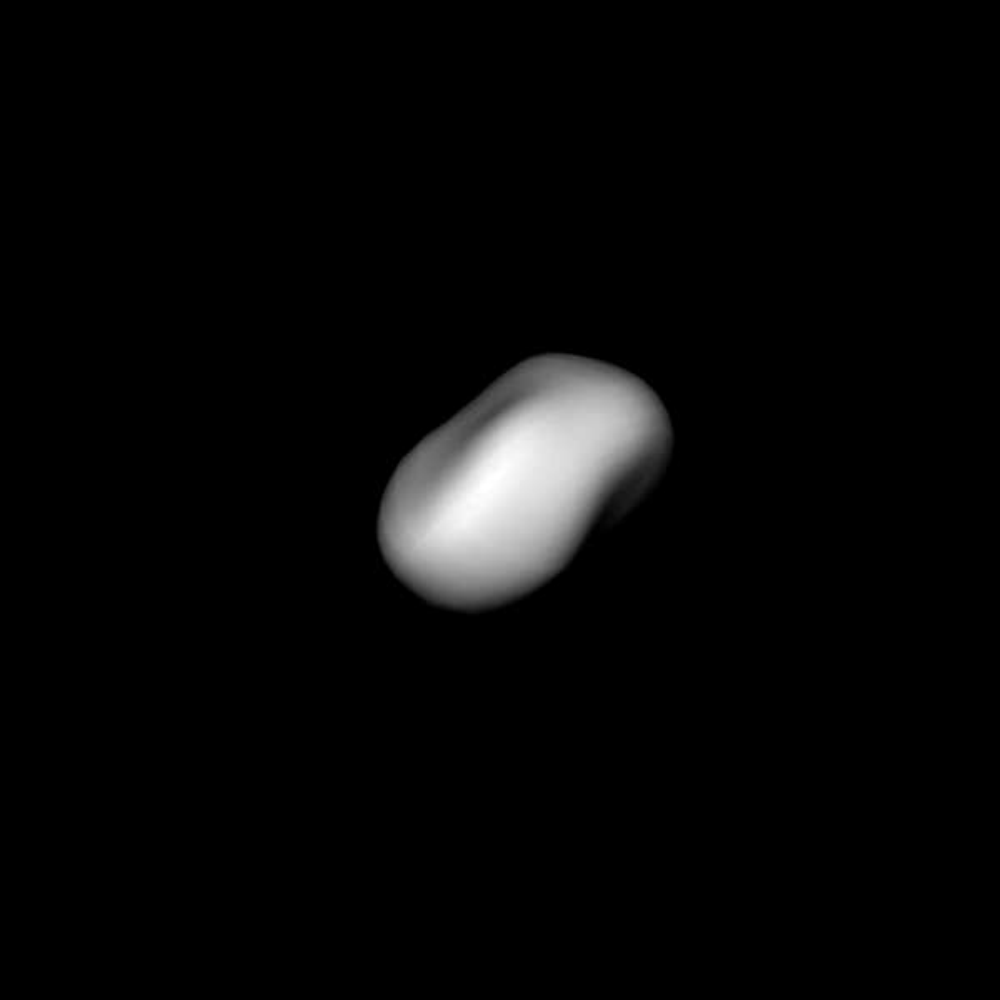}}\resizebox{0.24\hsize}{!}{\includegraphics{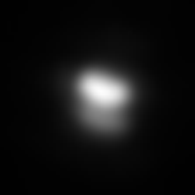}}\resizebox{0.24\hsize}{!}{\includegraphics{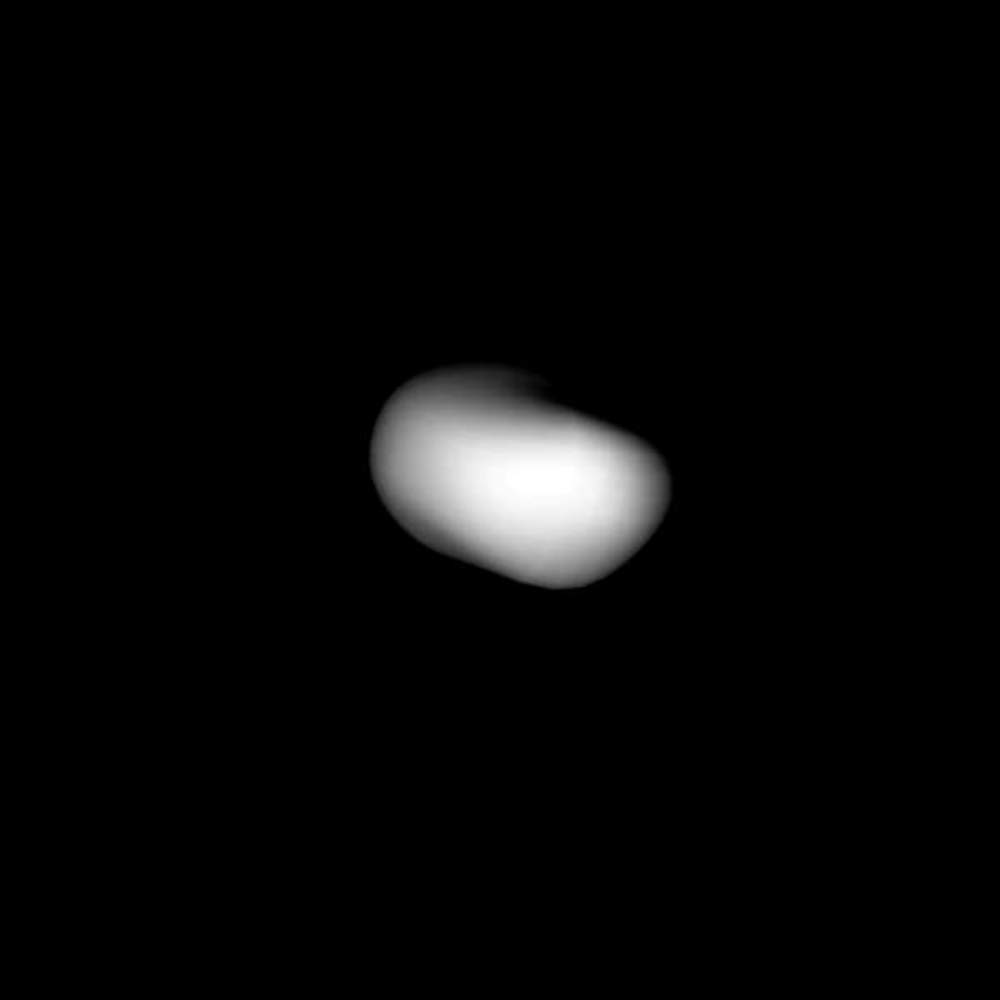}}\\
        \resizebox{0.24\hsize}{!}{\includegraphics{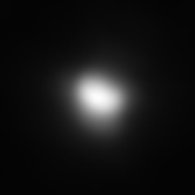}}\resizebox{0.24\hsize}{!}{\includegraphics{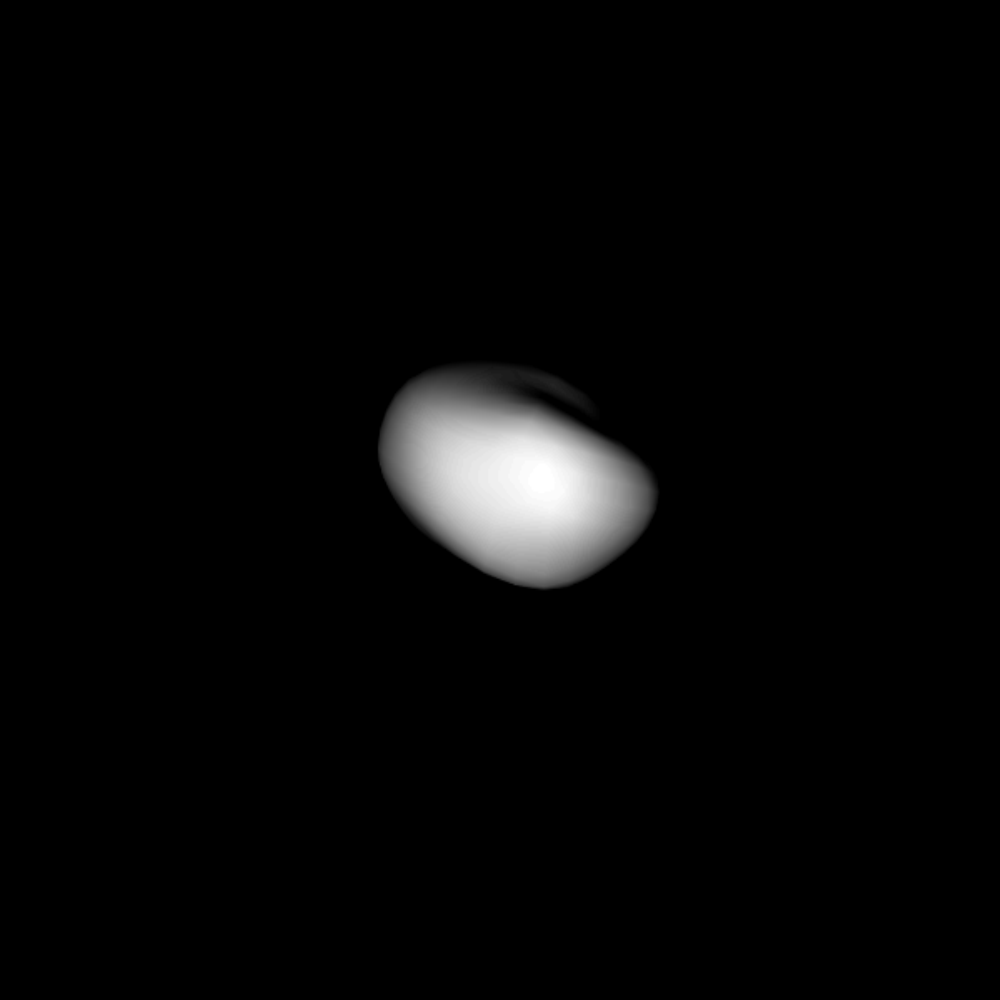}}\resizebox{0.24\hsize}{!}{\includegraphics{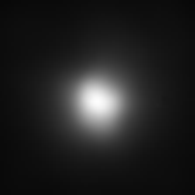}}\resizebox{0.24\hsize}{!}{\includegraphics{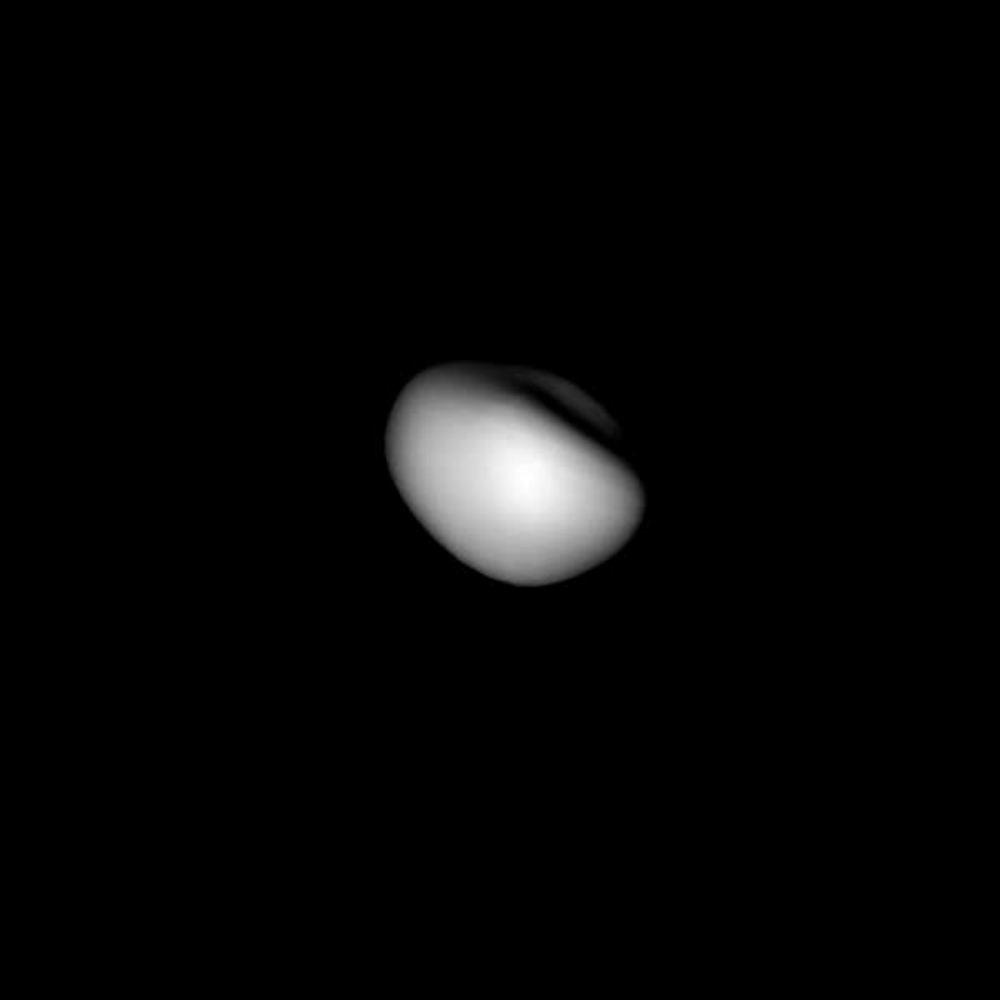}}\\
        \resizebox{0.24\hsize}{!}{\includegraphics{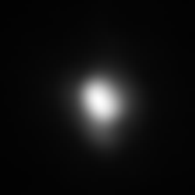}}\resizebox{0.24\hsize}{!}{\includegraphics{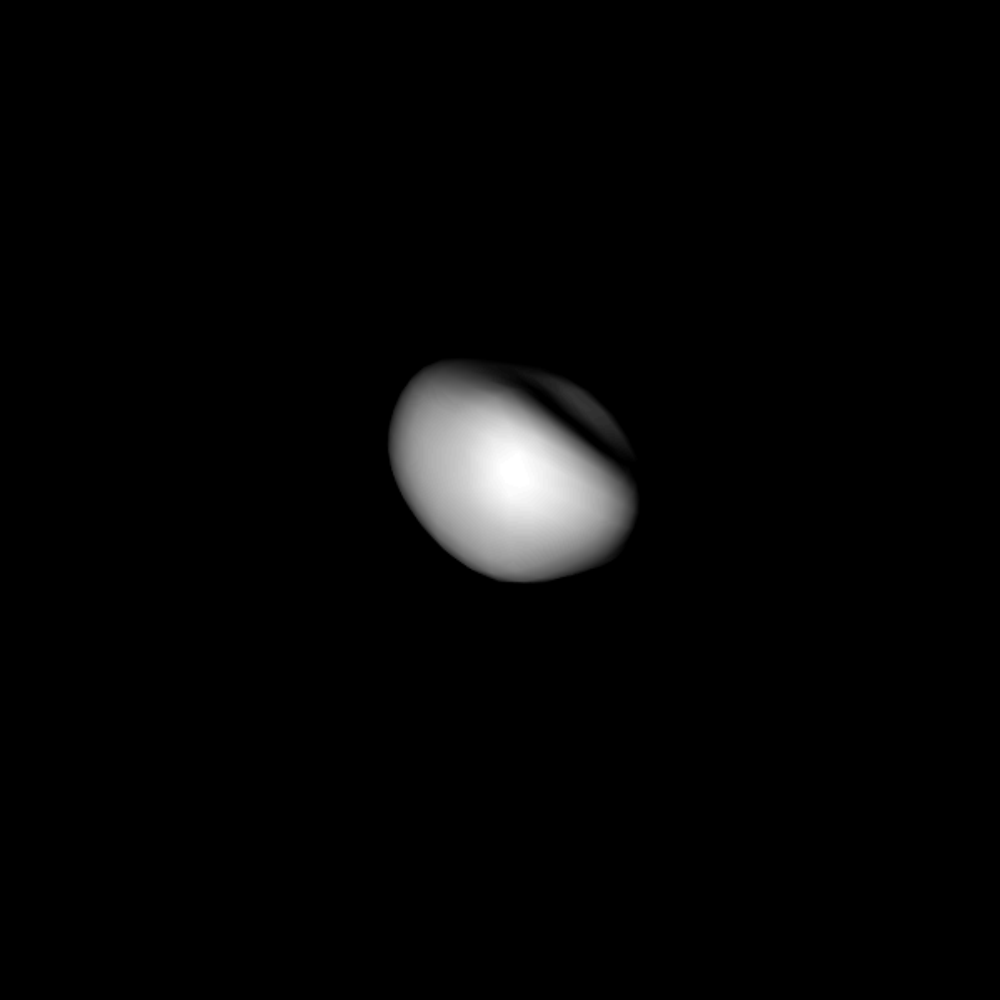}}\resizebox{0.24\hsize}{!}{\includegraphics{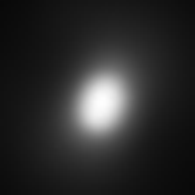}}\resizebox{0.24\hsize}{!}{\includegraphics{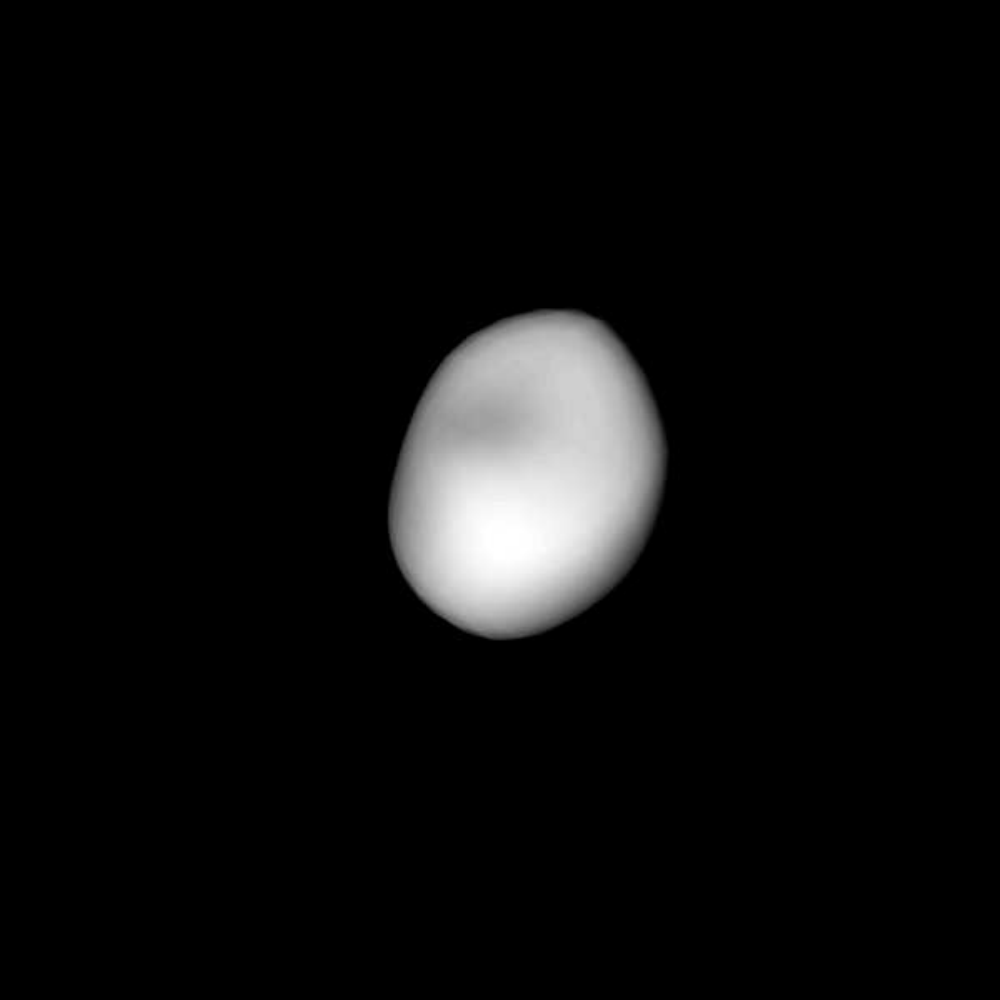}}\\
        \resizebox{0.24\hsize}{!}{\includegraphics{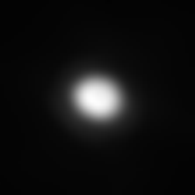}}\resizebox{0.24\hsize}{!}{\includegraphics{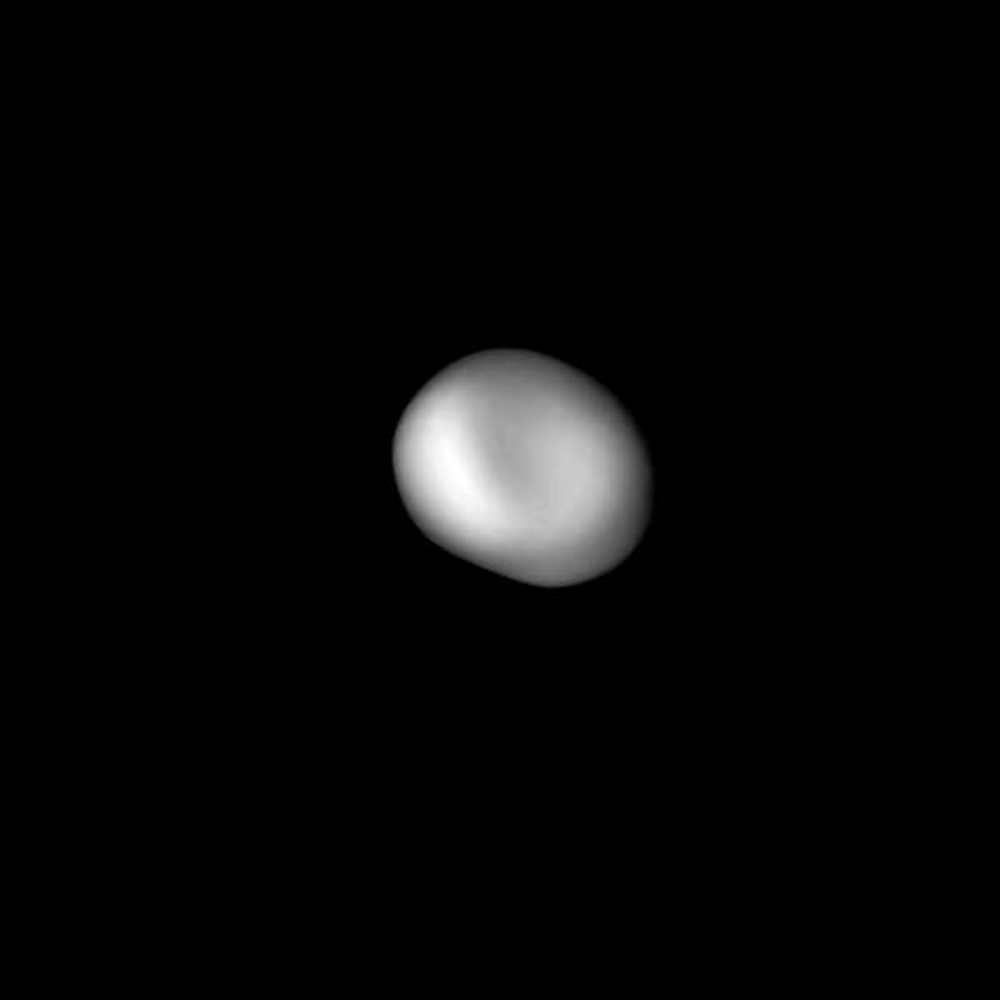}}\resizebox{0.24\hsize}{!}{\includegraphics{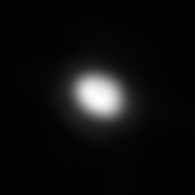}}\resizebox{0.24\hsize}{!}{\includegraphics{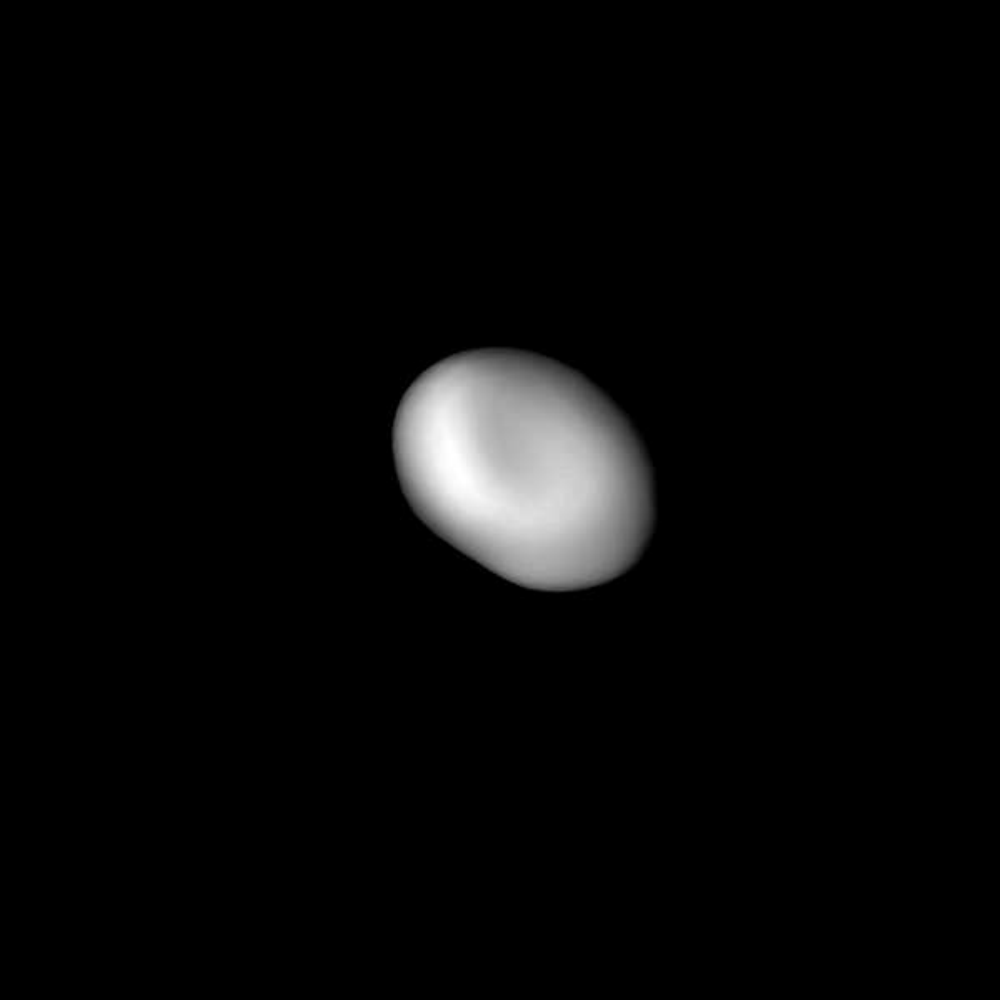}}\\
        \resizebox{0.24\hsize}{!}{\includegraphics{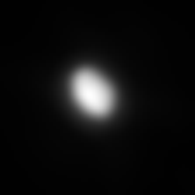}}\resizebox{0.24\hsize}{!}{\includegraphics{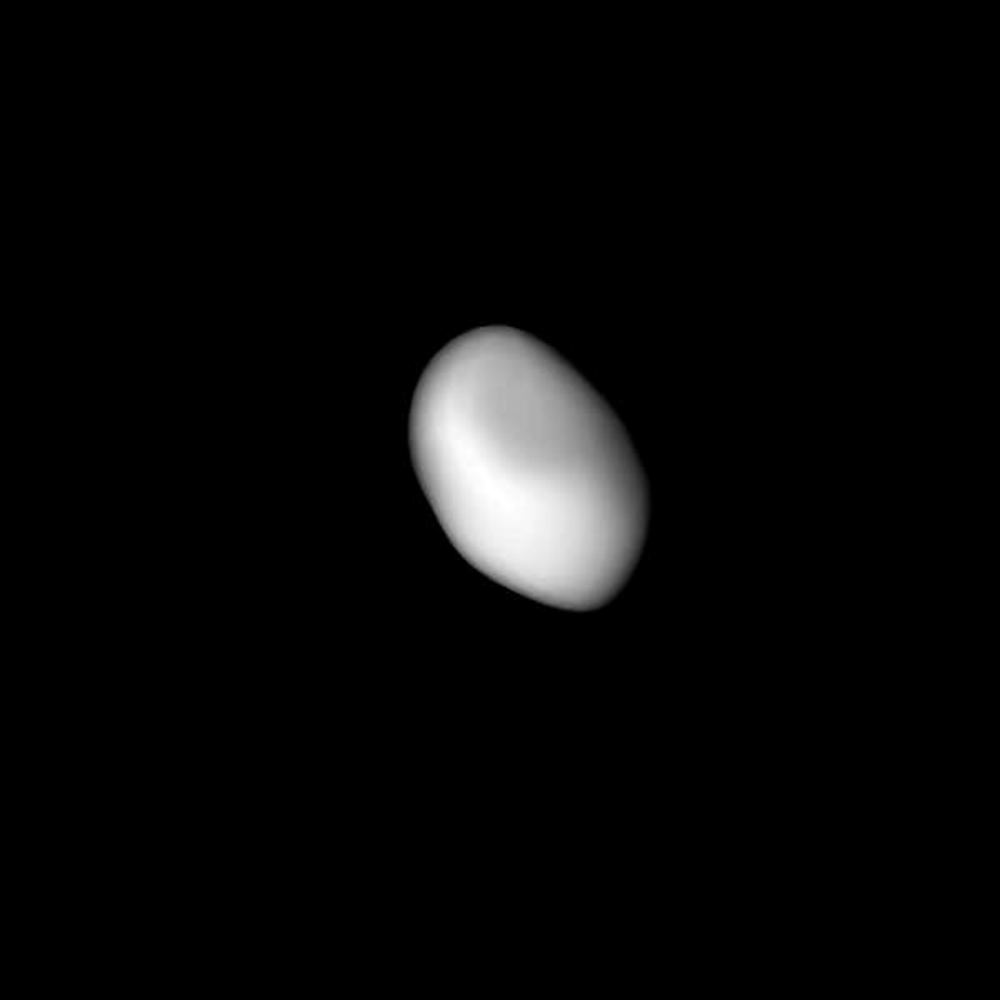}}\resizebox{0.24\hsize}{!}{\includegraphics{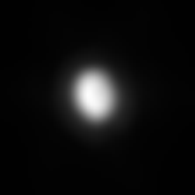}}\resizebox{0.24\hsize}{!}{\includegraphics{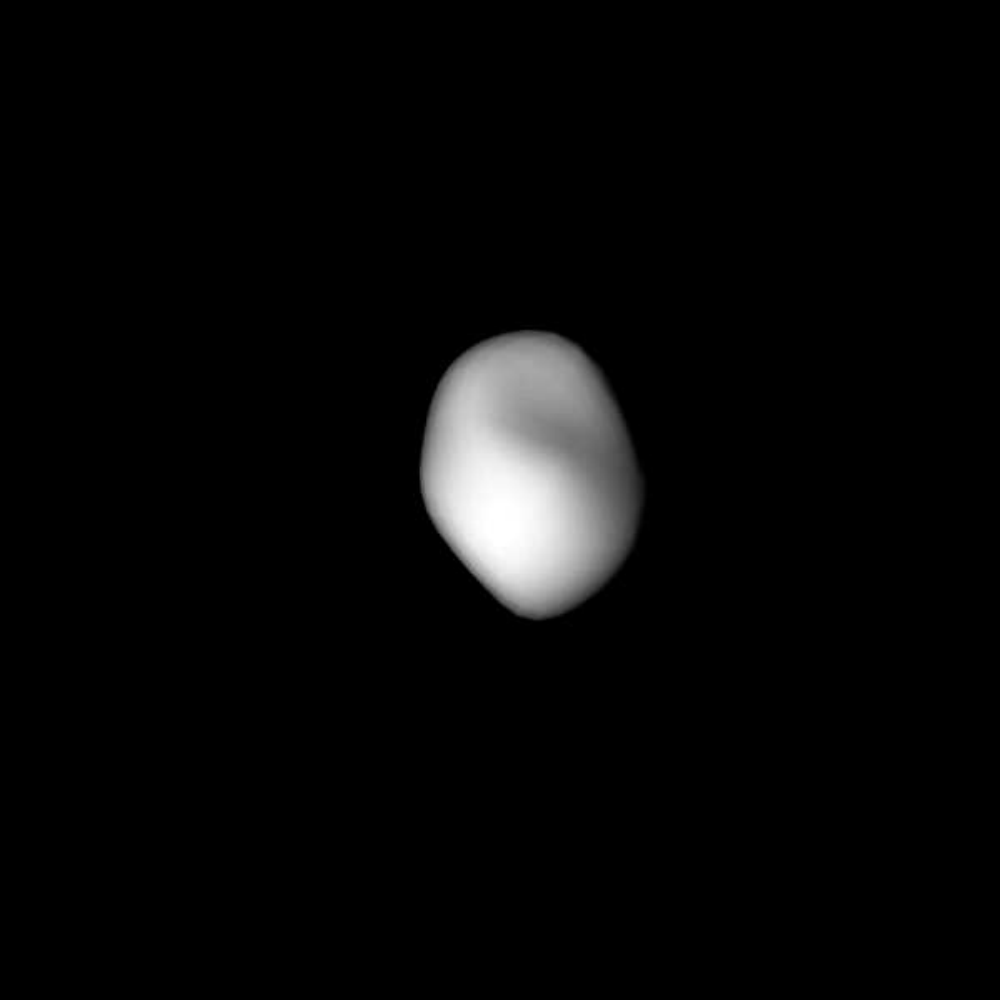}}\\
        \resizebox{0.24\hsize}{!}{\includegraphics{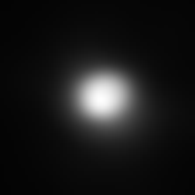}}\resizebox{0.24\hsize}{!}{\includegraphics{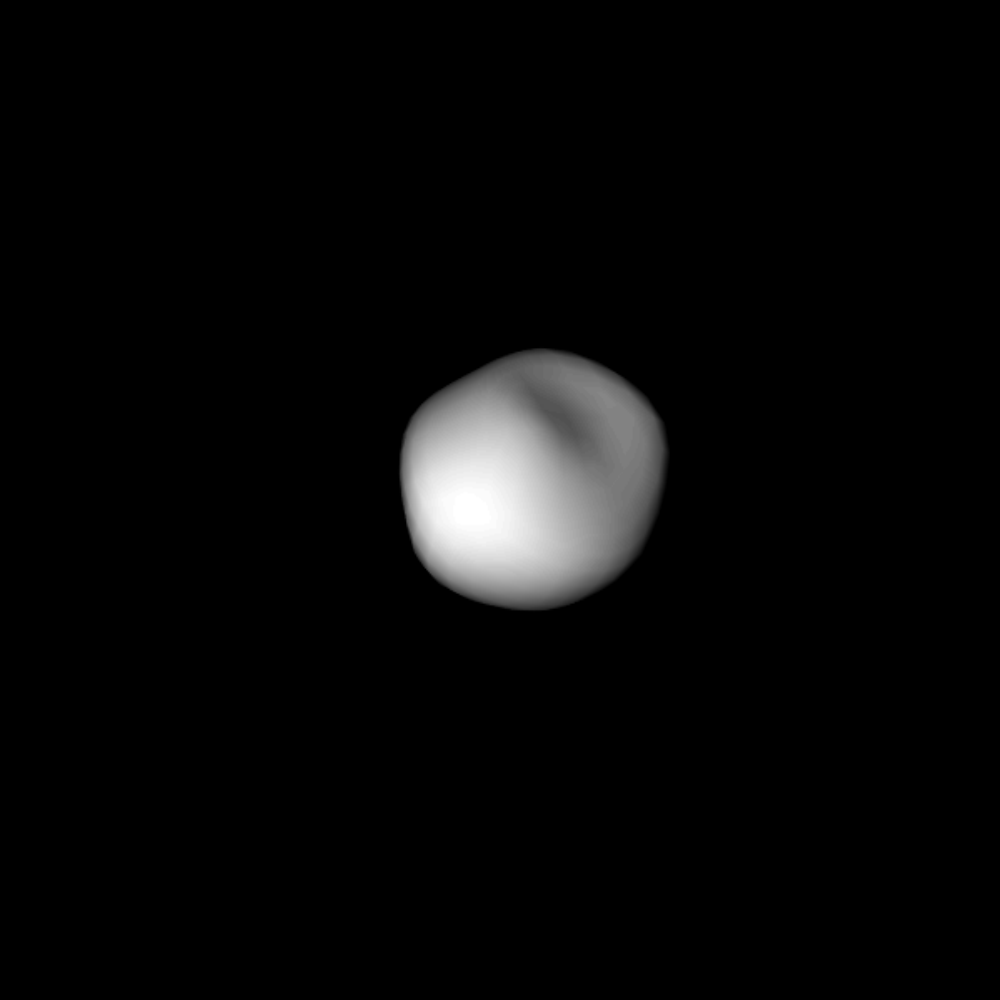}}\resizebox{0.24\hsize}{!}{\includegraphics{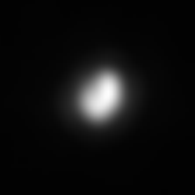}}\resizebox{0.24\hsize}{!}{\includegraphics{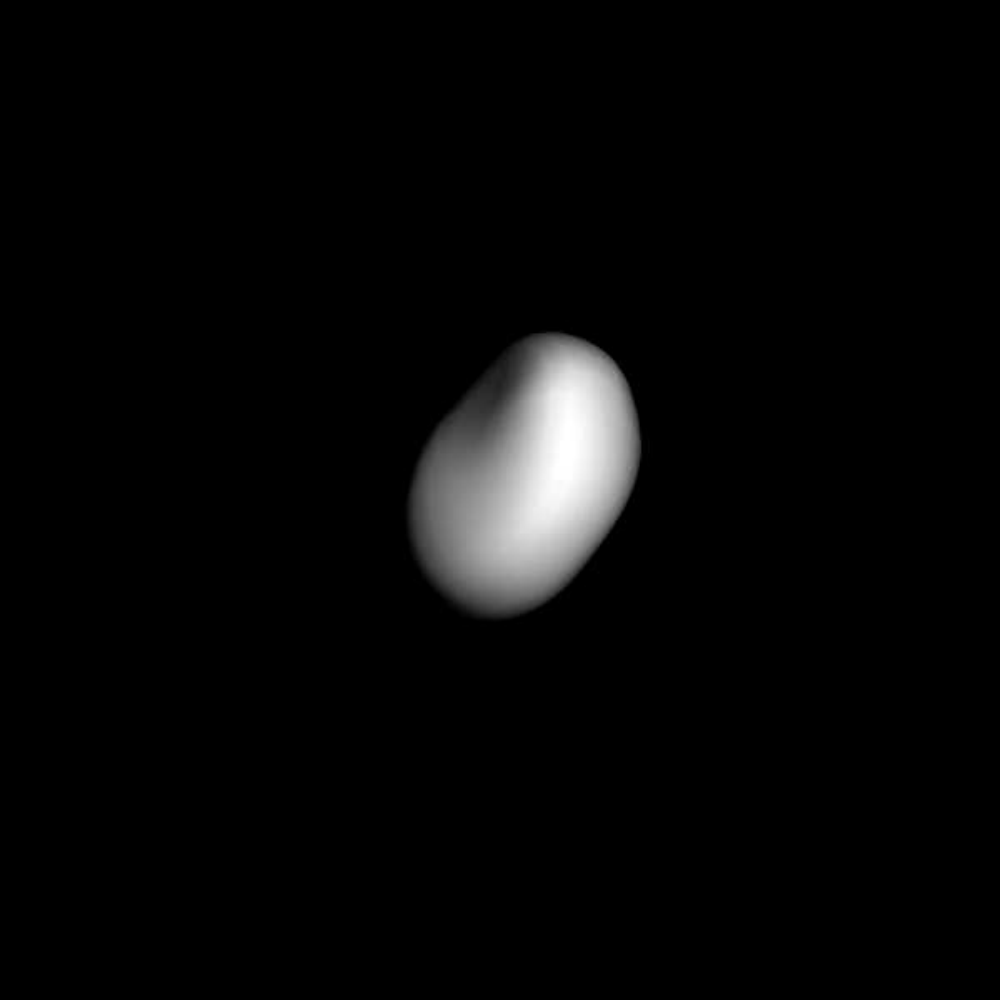}}\\
    \caption{\label{fig:45a}Comparison between model projections and corresponding AO images for asteroid (45) Eugenia (first part).}
\end{figure}

\begin{figure}[tbp]
    \centering
        \resizebox{0.24\hsize}{!}{\includegraphics{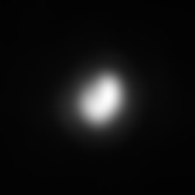}}\resizebox{0.24\hsize}{!}{\includegraphics{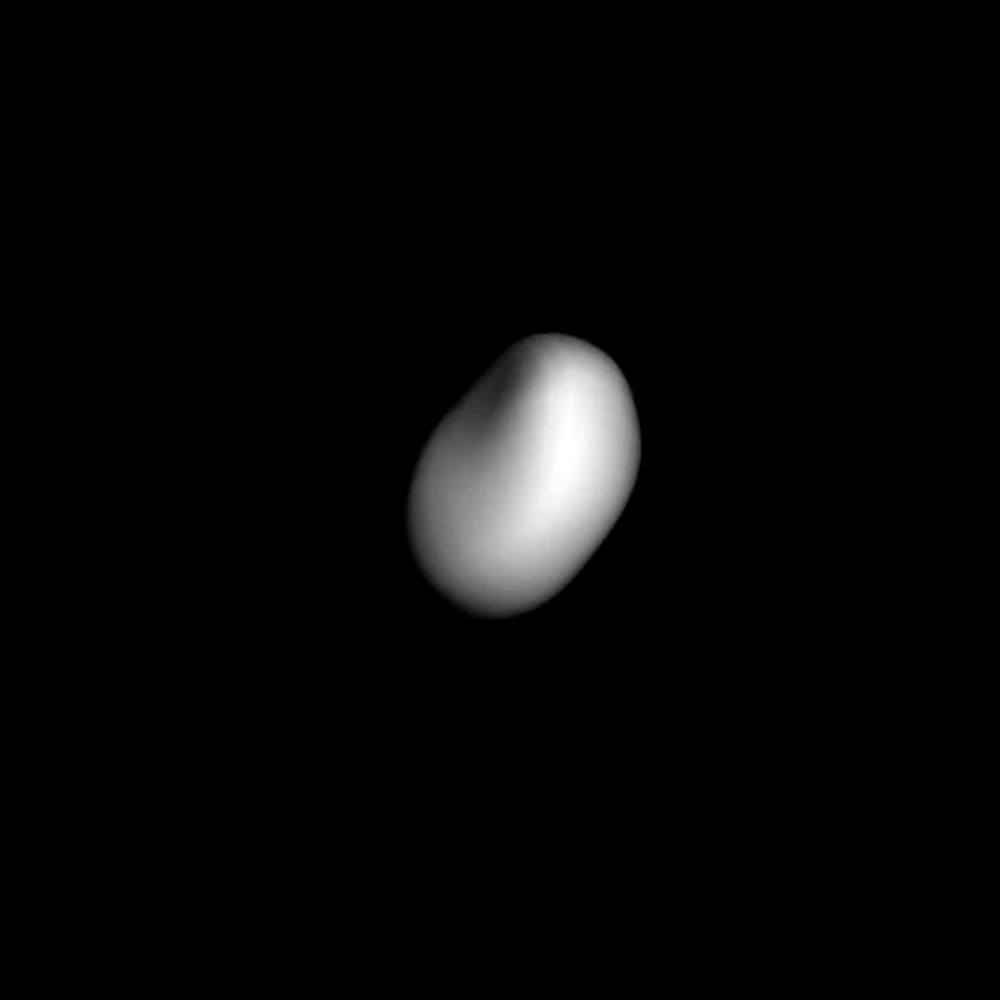}}\resizebox{0.24\hsize}{!}{\includegraphics{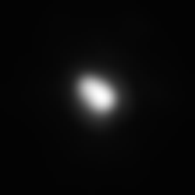}}\resizebox{0.24\hsize}{!}{\includegraphics{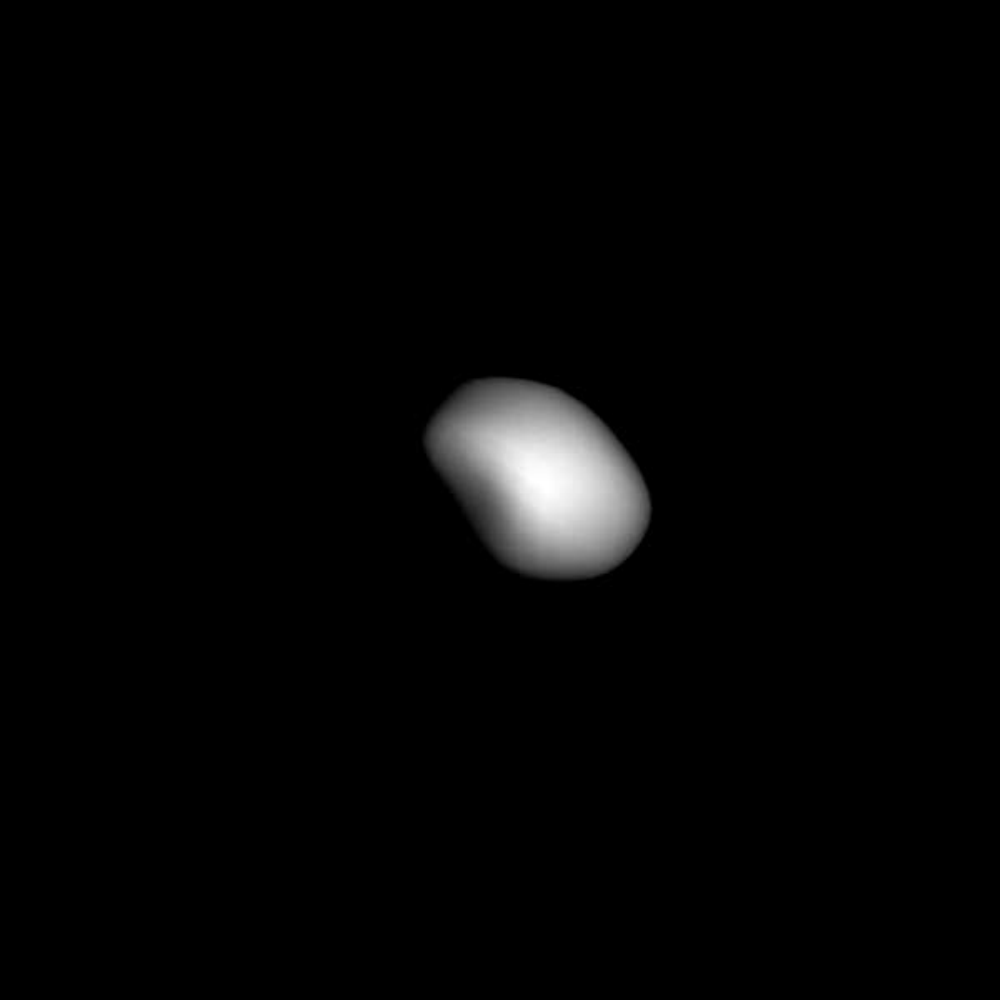}}\\
        \resizebox{0.24\hsize}{!}{\includegraphics{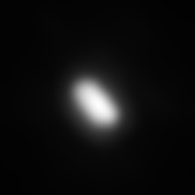}}\resizebox{0.24\hsize}{!}{\includegraphics{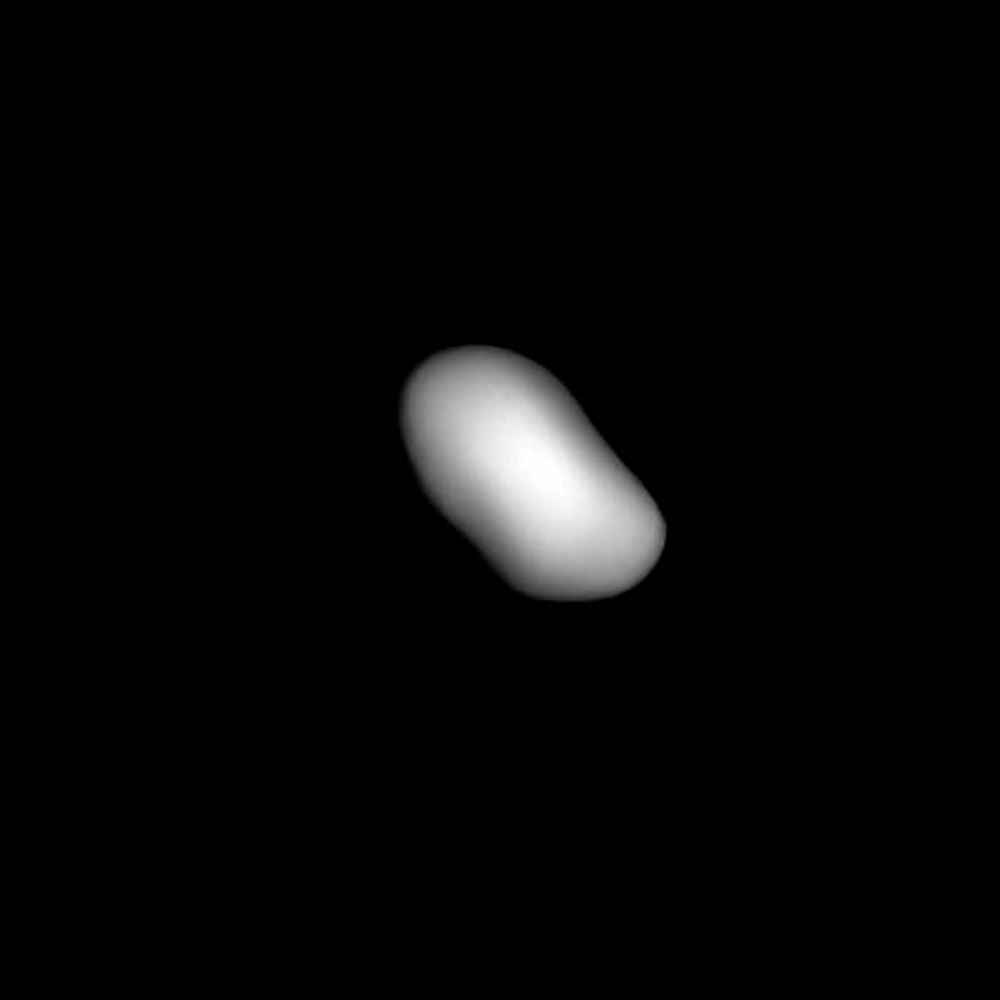}}\resizebox{0.24\hsize}{!}{\includegraphics{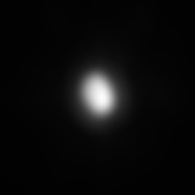}}\resizebox{0.24\hsize}{!}{\includegraphics{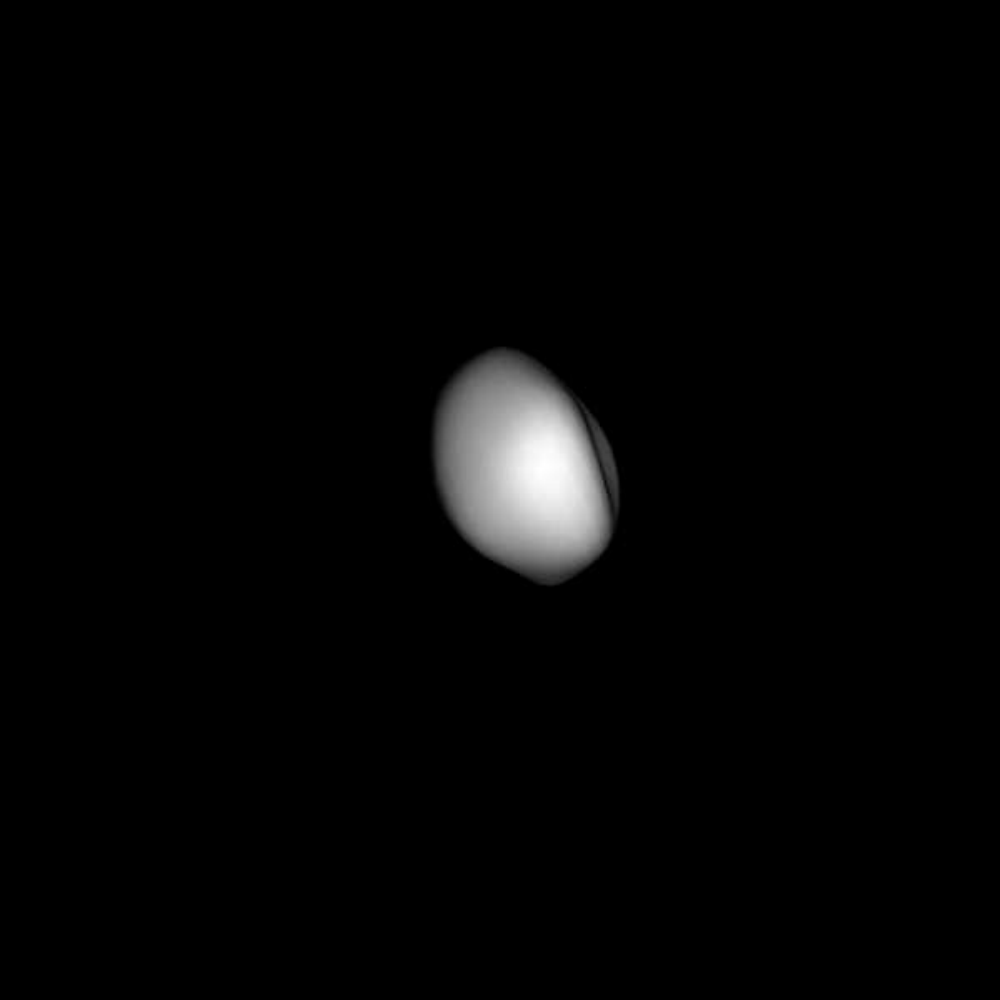}}\\
        \resizebox{0.24\hsize}{!}{\includegraphics{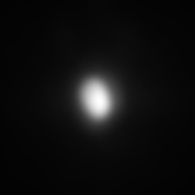}}\resizebox{0.24\hsize}{!}{\includegraphics{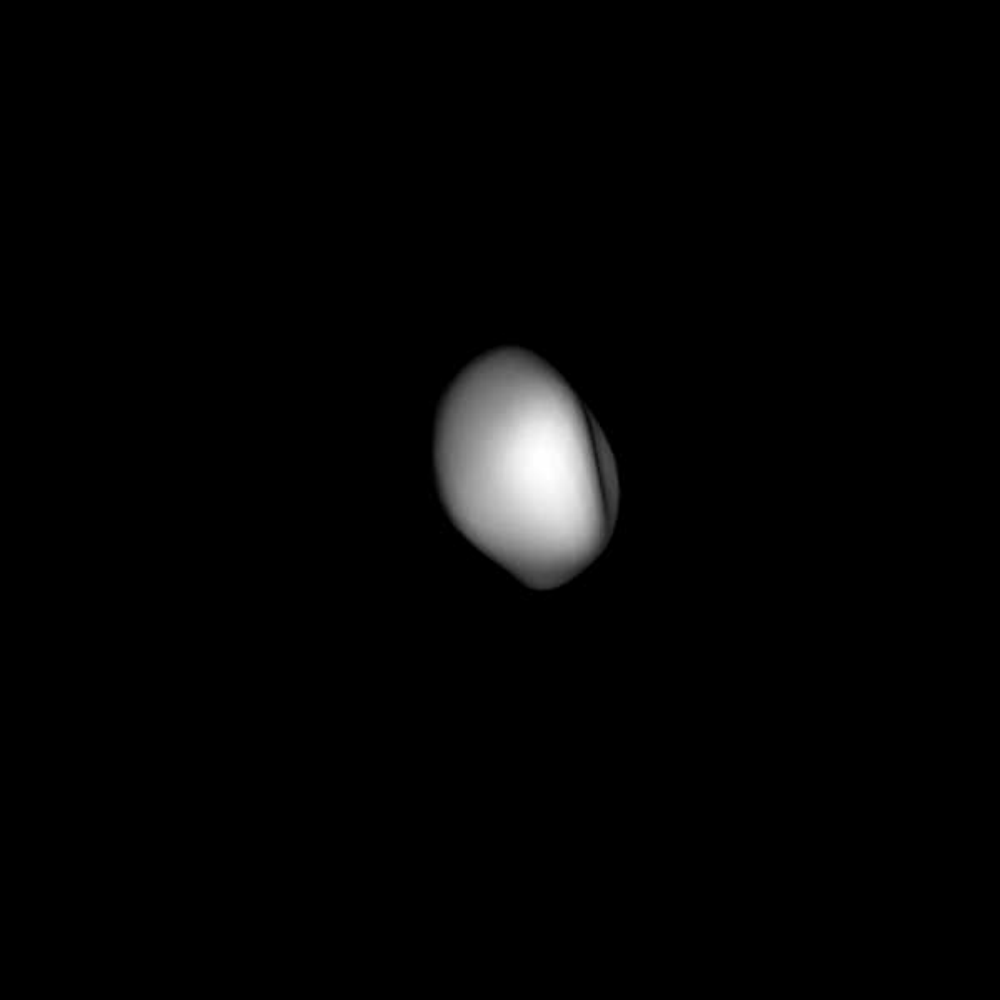}}\resizebox{0.24\hsize}{!}{\includegraphics{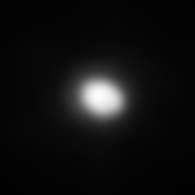}}\resizebox{0.24\hsize}{!}{\includegraphics{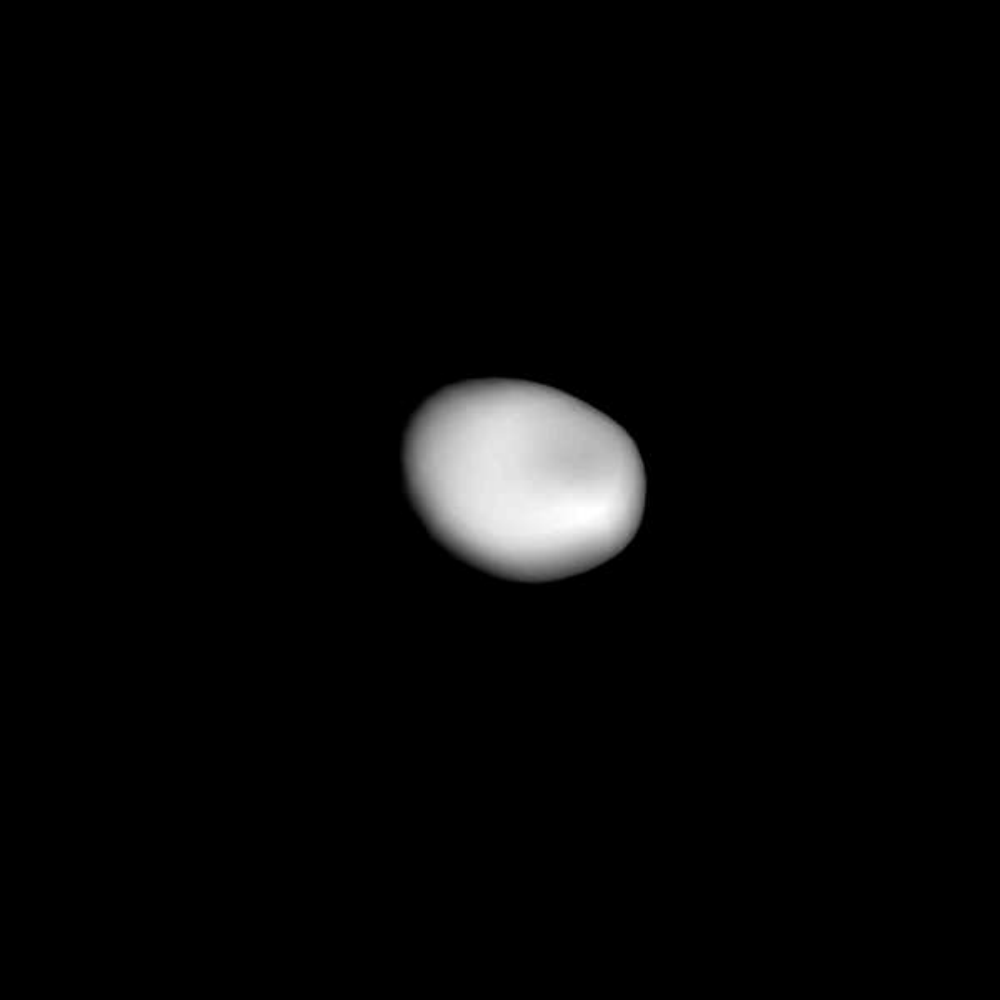}}\\
        \resizebox{0.24\hsize}{!}{\includegraphics{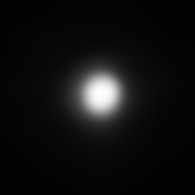}}\resizebox{0.24\hsize}{!}{\includegraphics{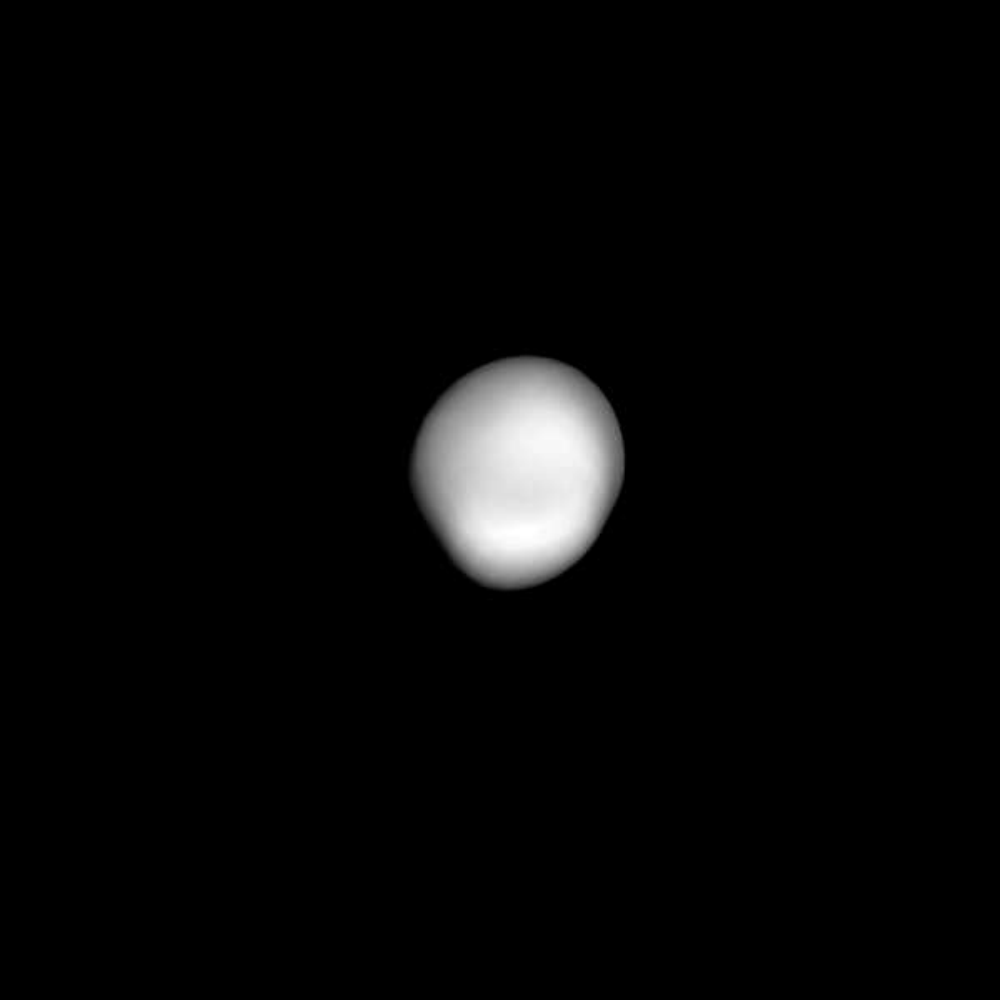}}\\
    \caption{\label{fig:45b}Comparison between model projections and corresponding AO images for asteroid (45) Eugenia (second part).}
\end{figure}

\begin{figure}[tbp]
    \centering
        \resizebox{0.24\hsize}{!}{\includegraphics{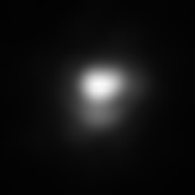}}\resizebox{0.24\hsize}{!}{\includegraphics{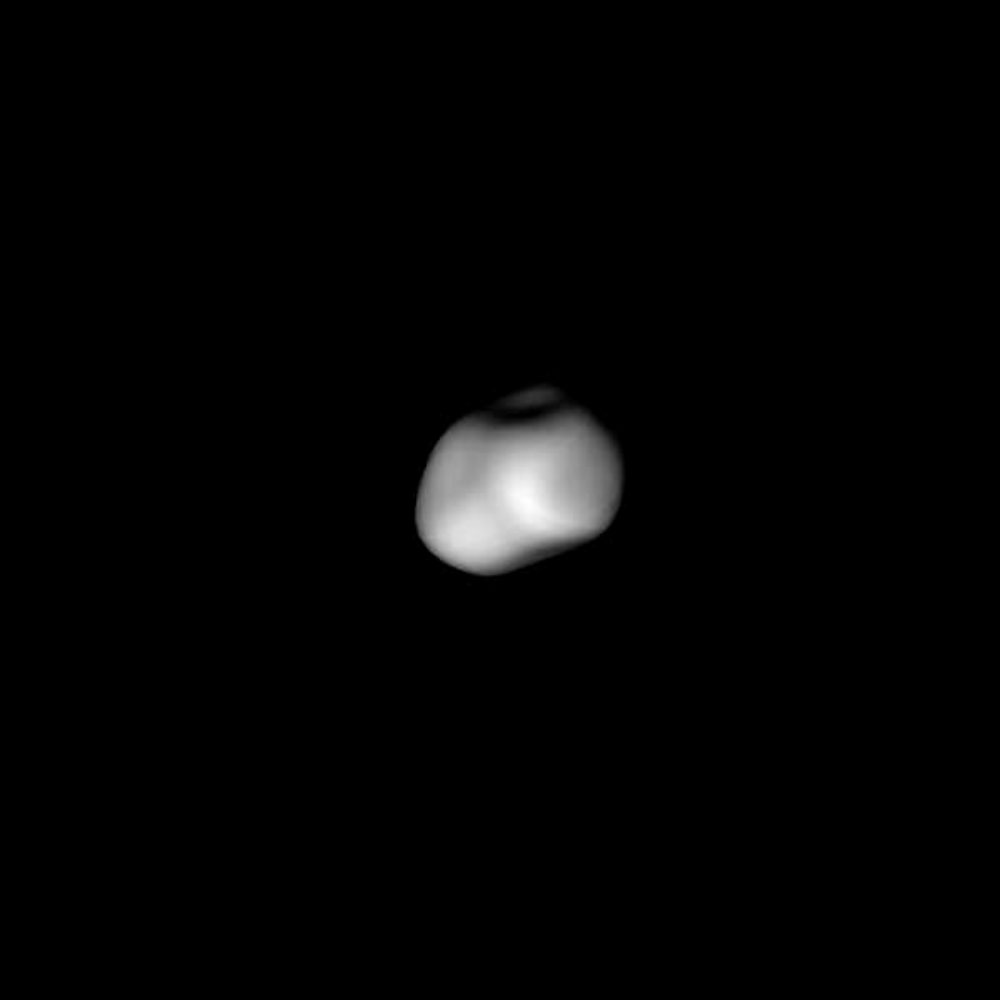}}\resizebox{0.24\hsize}{!}{\includegraphics{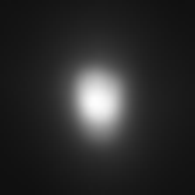}}\resizebox{0.24\hsize}{!}{\includegraphics{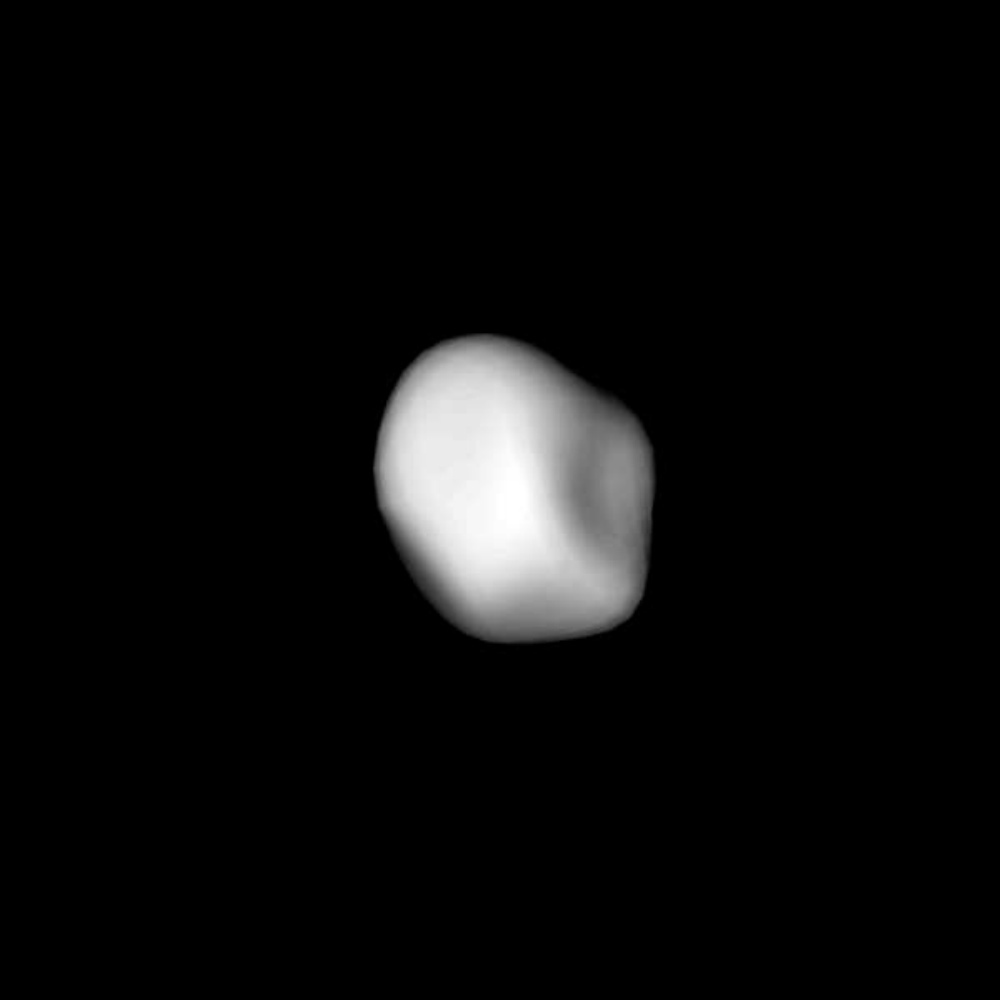}}\\
        \resizebox{0.24\hsize}{!}{\includegraphics{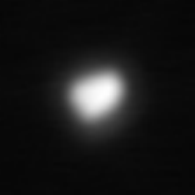}}\resizebox{0.24\hsize}{!}{\includegraphics{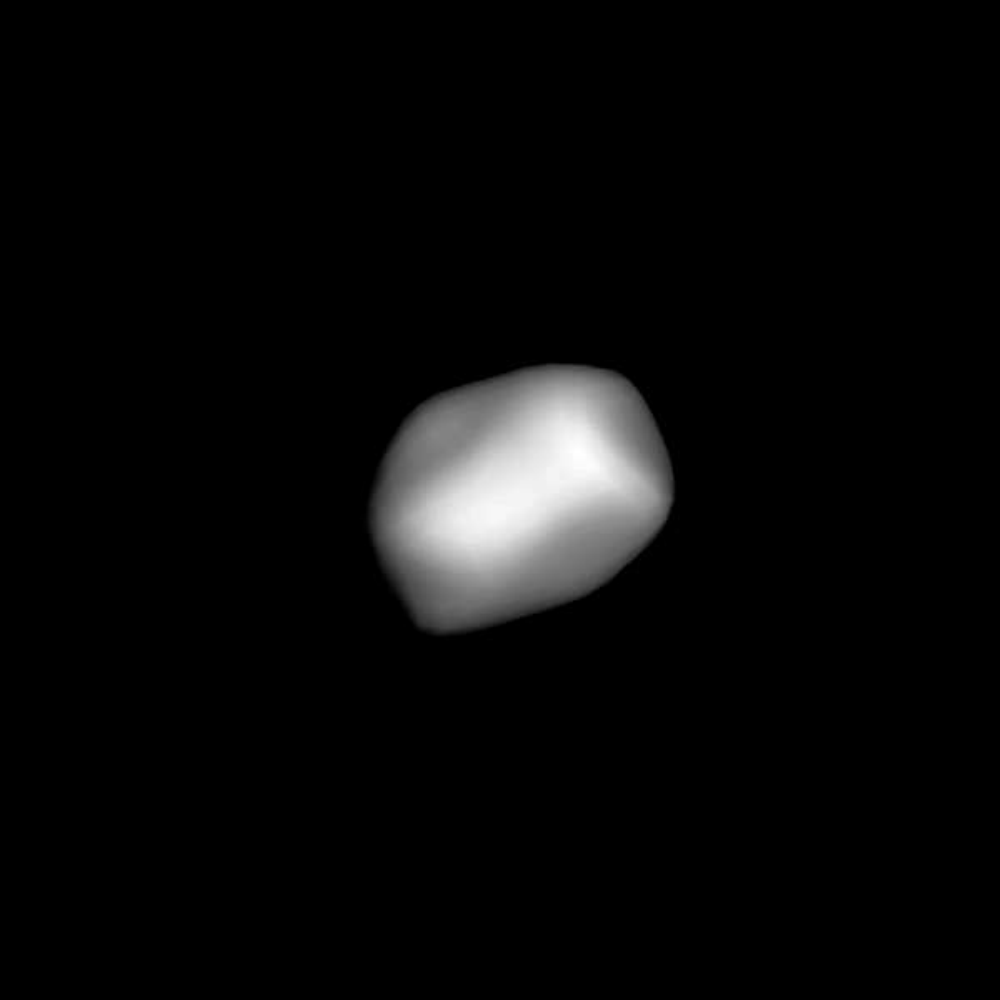}}\\
    \caption{\label{fig:51}Comparison between model projections and corresponding AO images for asteroid (51) Nemausa.}
\end{figure}

\clearpage

\begin{figure}[tbp]
    \centering
        \resizebox{0.24\hsize}{!}{\includegraphics{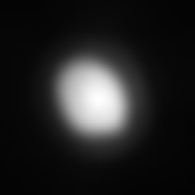}}\resizebox{0.24\hsize}{!}{\includegraphics{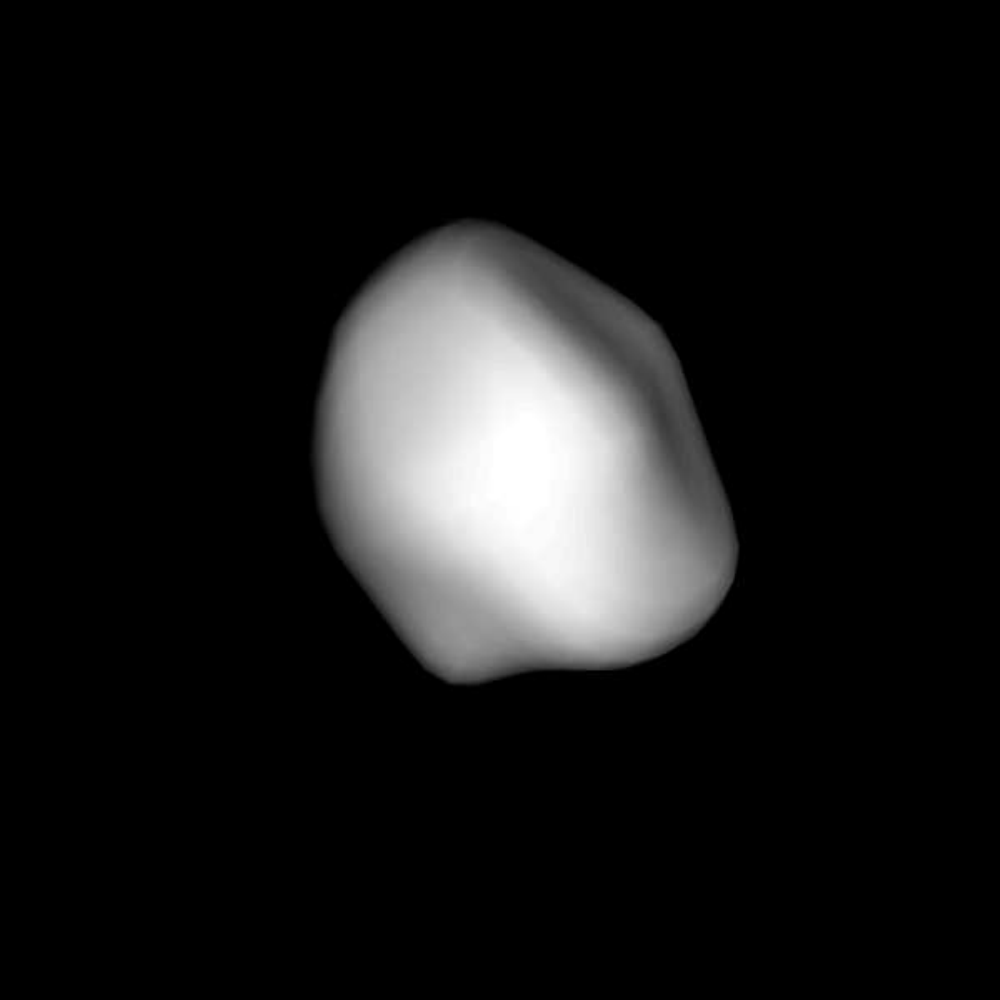}}\resizebox{0.24\hsize}{!}{\includegraphics{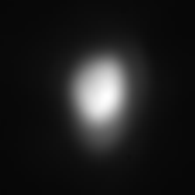}}\resizebox{0.24\hsize}{!}{\includegraphics{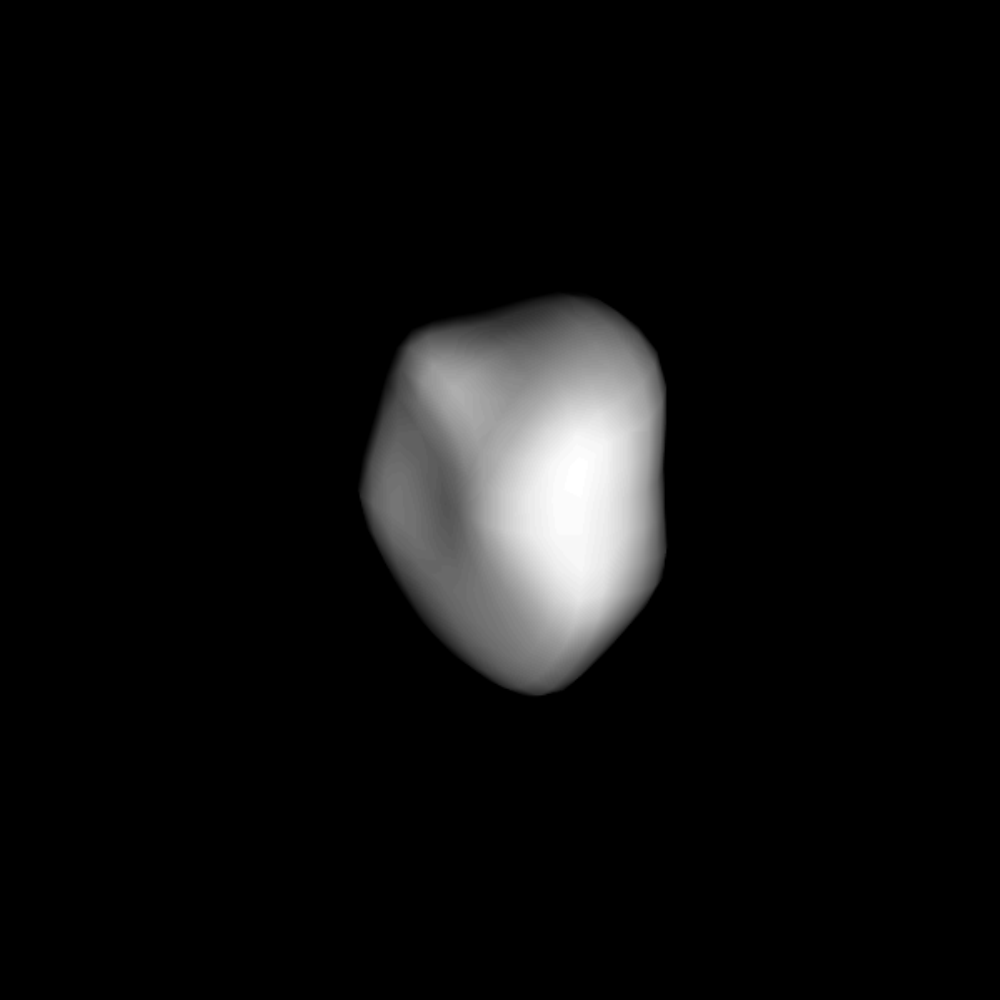}}\\
        \resizebox{0.24\hsize}{!}{\includegraphics{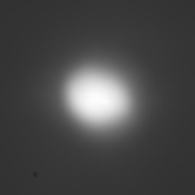}}\resizebox{0.24\hsize}{!}{\includegraphics{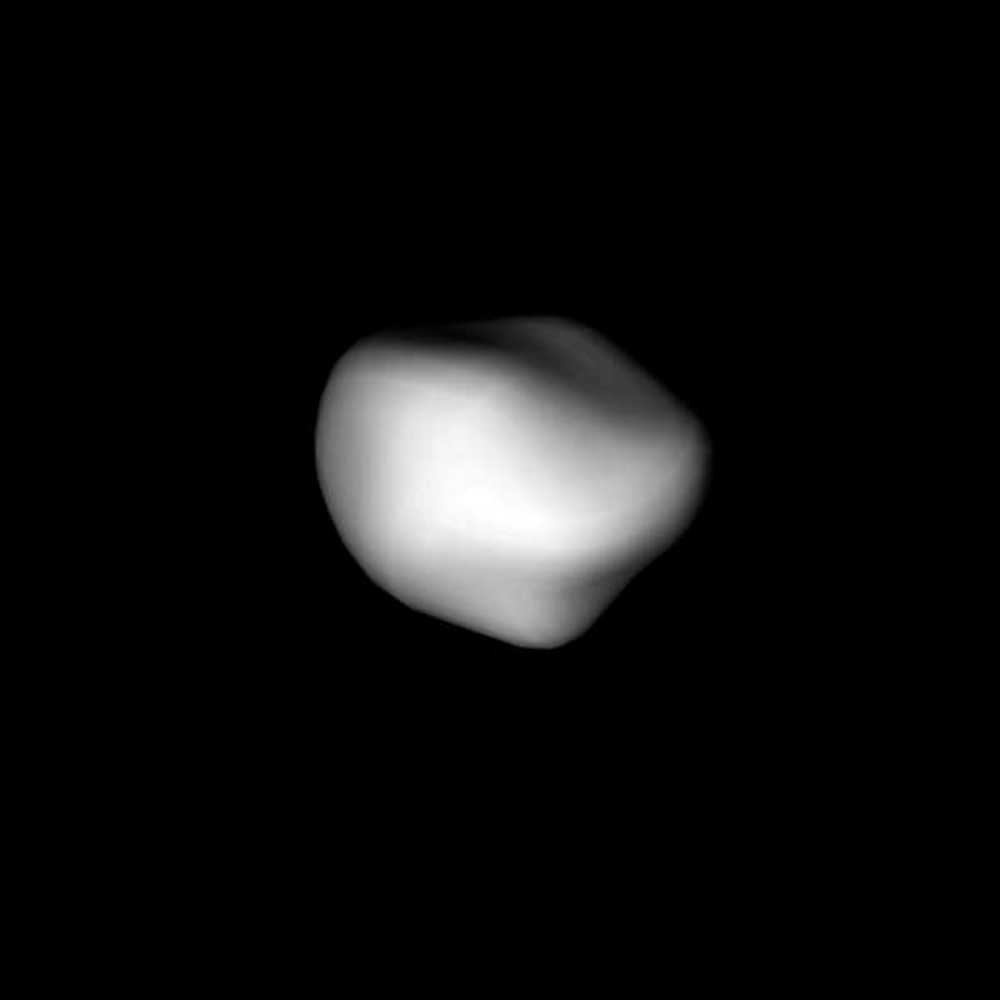}}\resizebox{0.24\hsize}{!}{\includegraphics{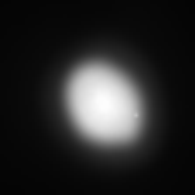}}\resizebox{0.24\hsize}{!}{\includegraphics{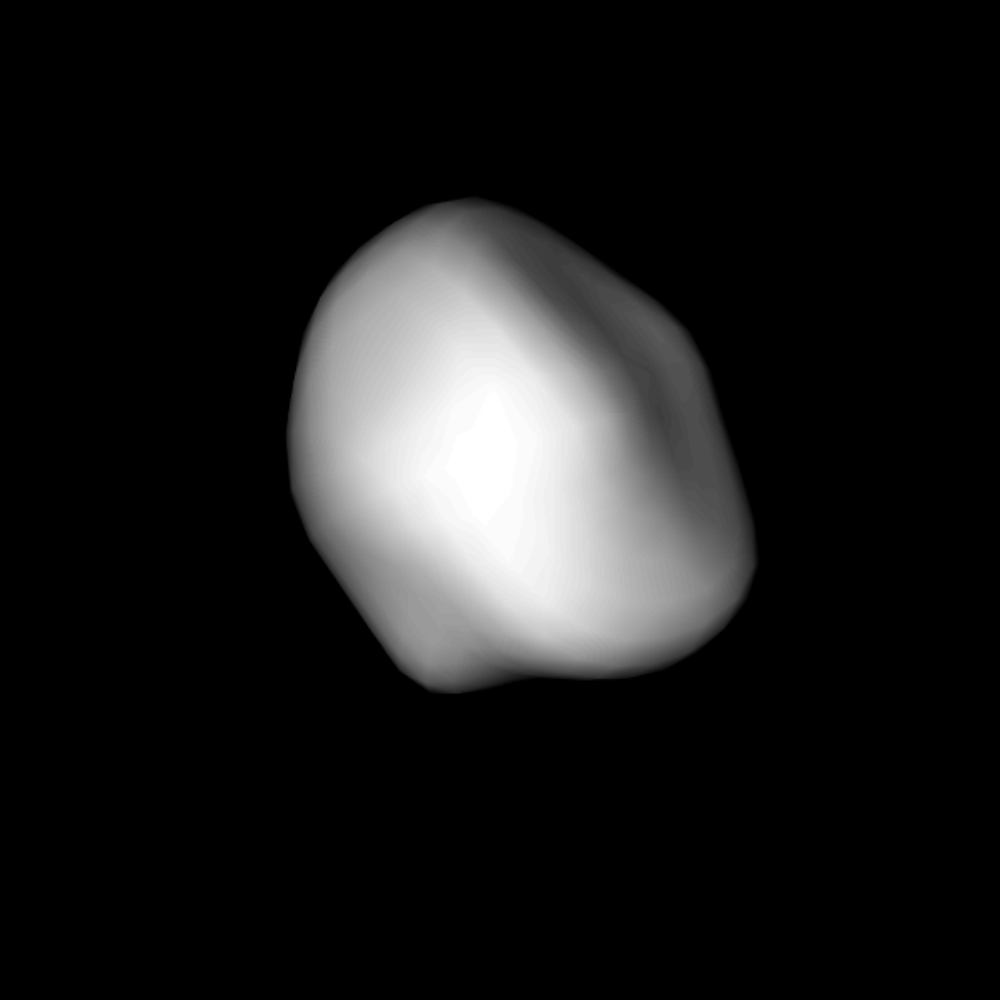}}\\
        \resizebox{0.24\hsize}{!}{\includegraphics{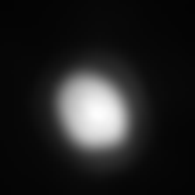}}\resizebox{0.24\hsize}{!}{\includegraphics{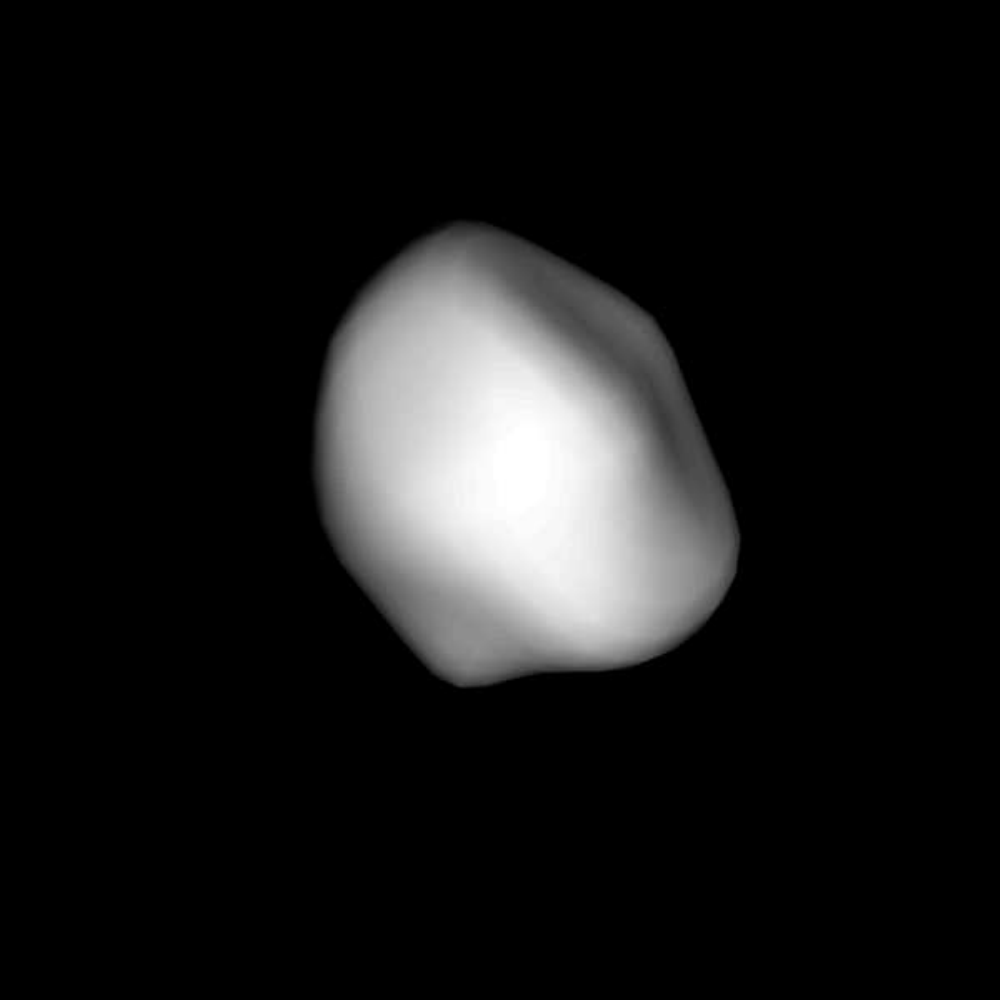}}\resizebox{0.24\hsize}{!}{\includegraphics{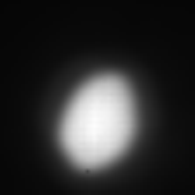}}\resizebox{0.24\hsize}{!}{\includegraphics{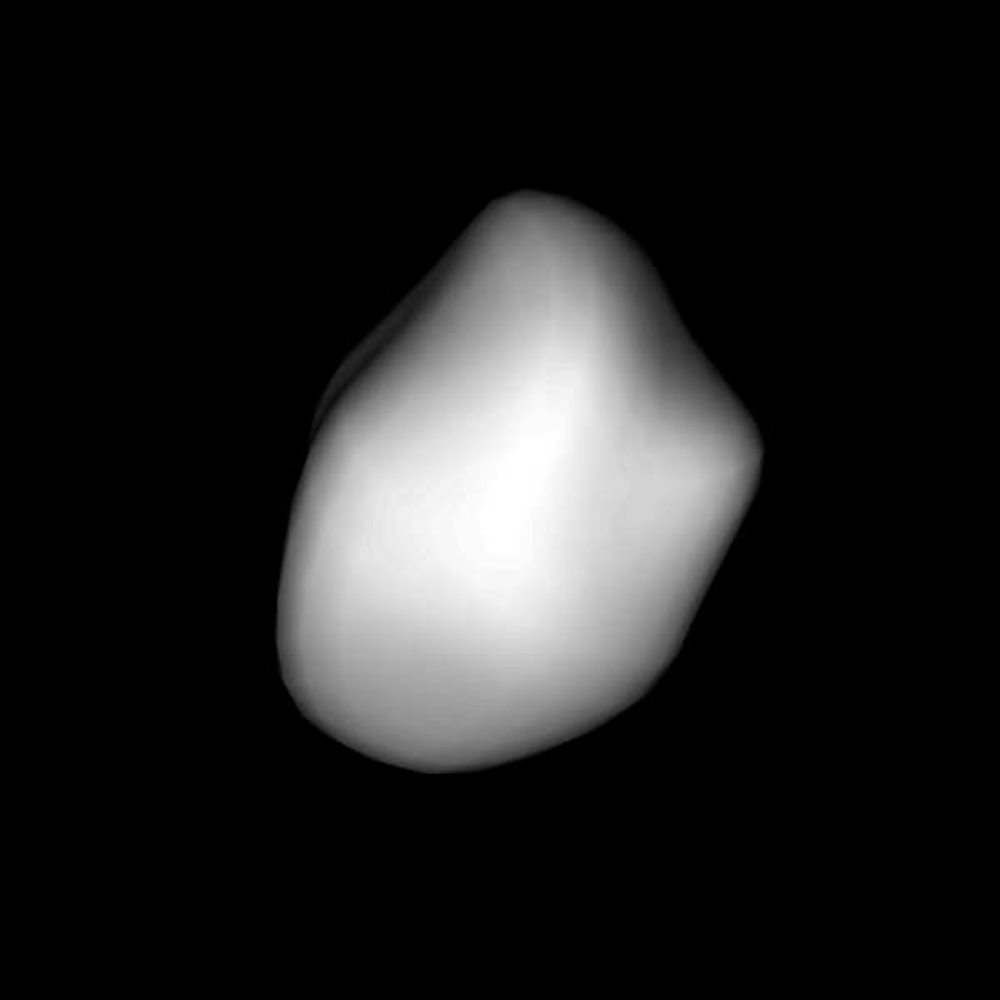}}\\
        \resizebox{0.24\hsize}{!}{\includegraphics{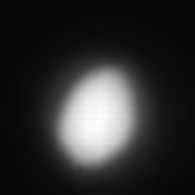}}\resizebox{0.24\hsize}{!}{\includegraphics{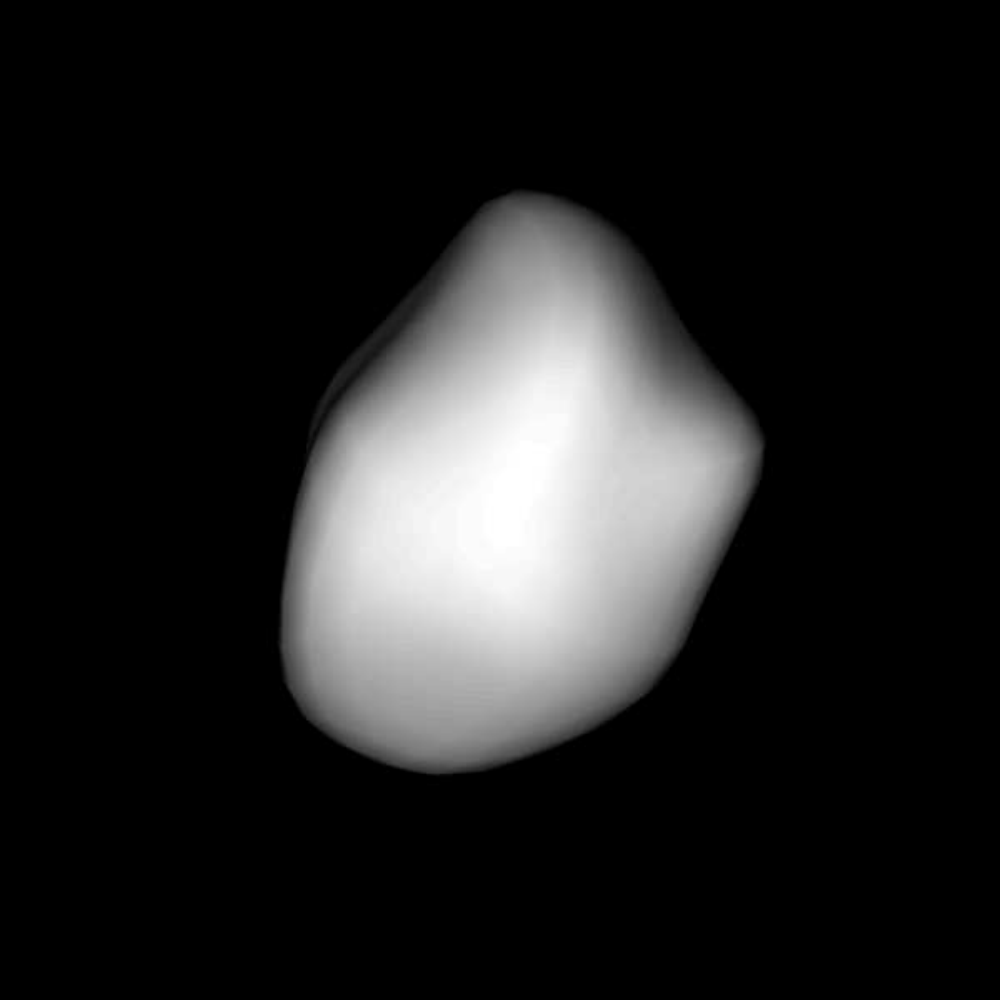}}\resizebox{0.24\hsize}{!}{\includegraphics{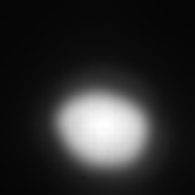}}\resizebox{0.24\hsize}{!}{\includegraphics{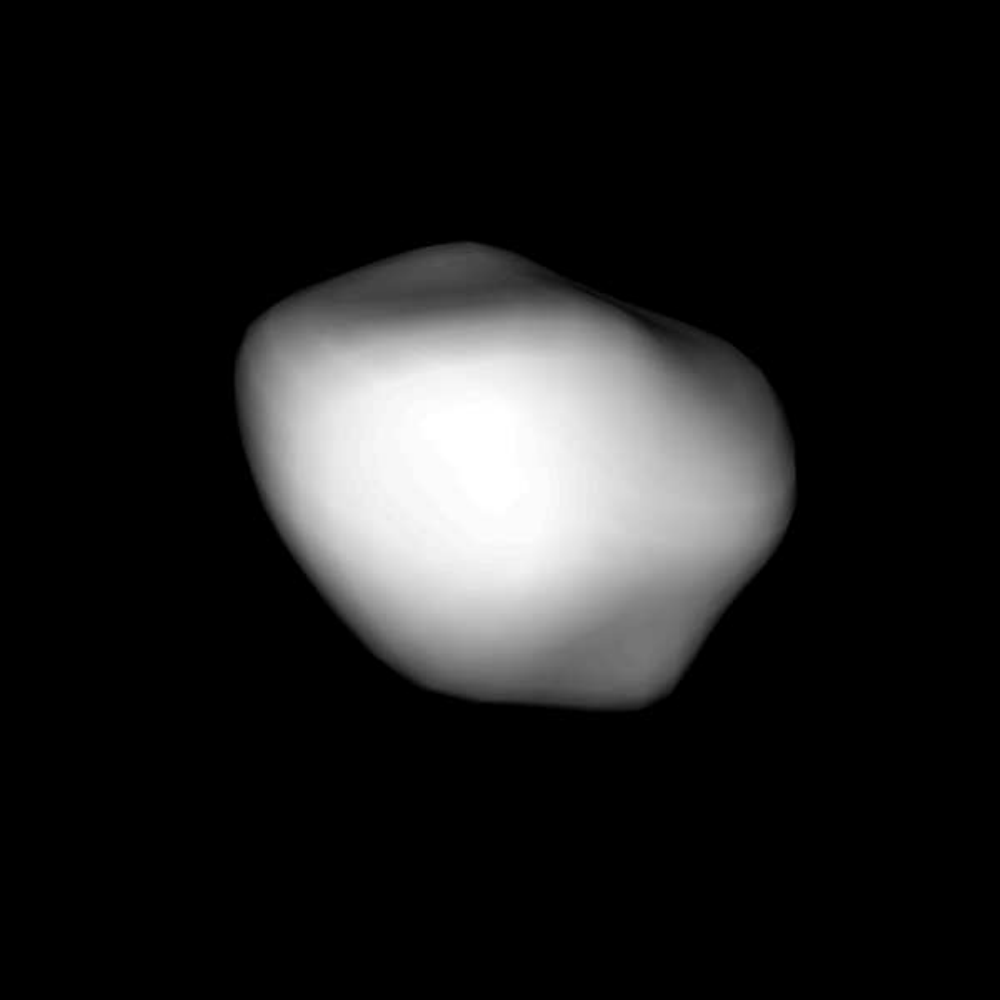}}\\
        \resizebox{0.24\hsize}{!}{\includegraphics{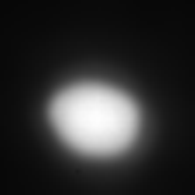}}\resizebox{0.24\hsize}{!}{\includegraphics{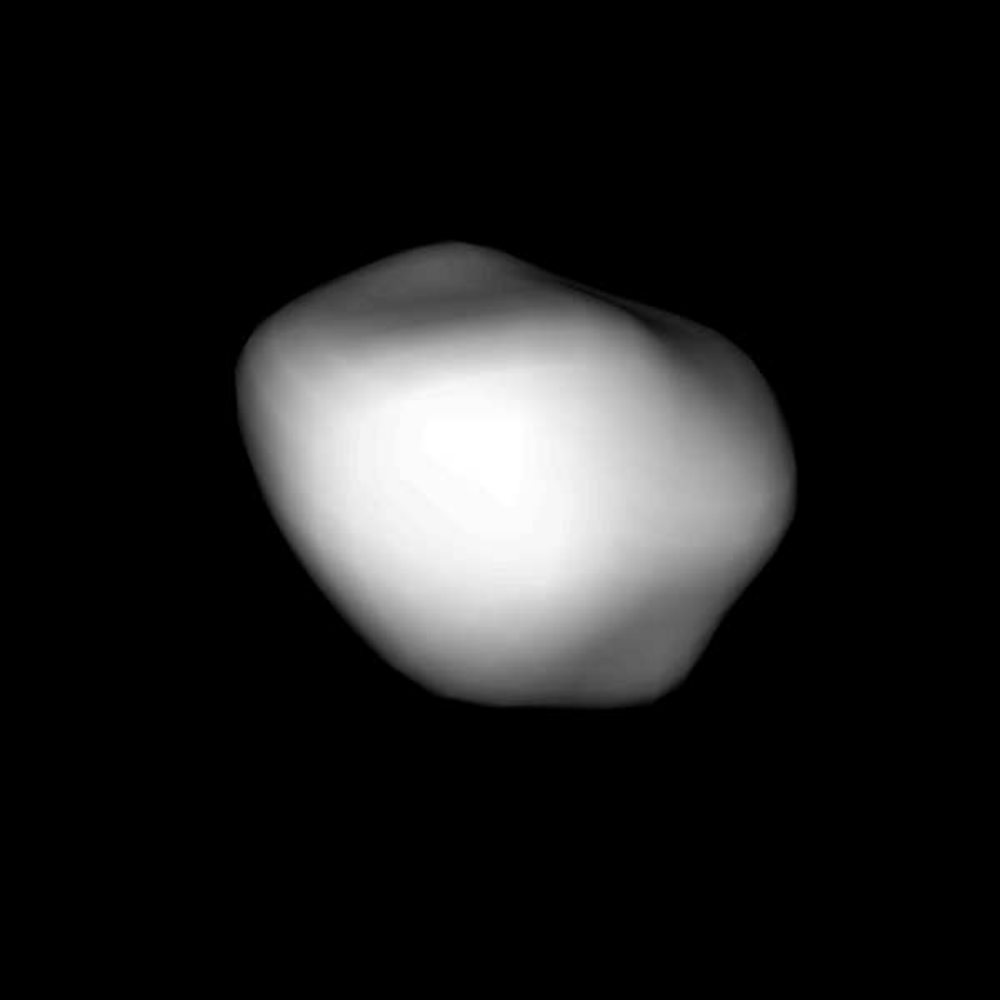}}\resizebox{0.24\hsize}{!}{\includegraphics{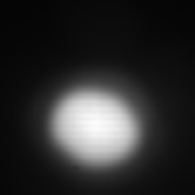}}\resizebox{0.24\hsize}{!}{\includegraphics{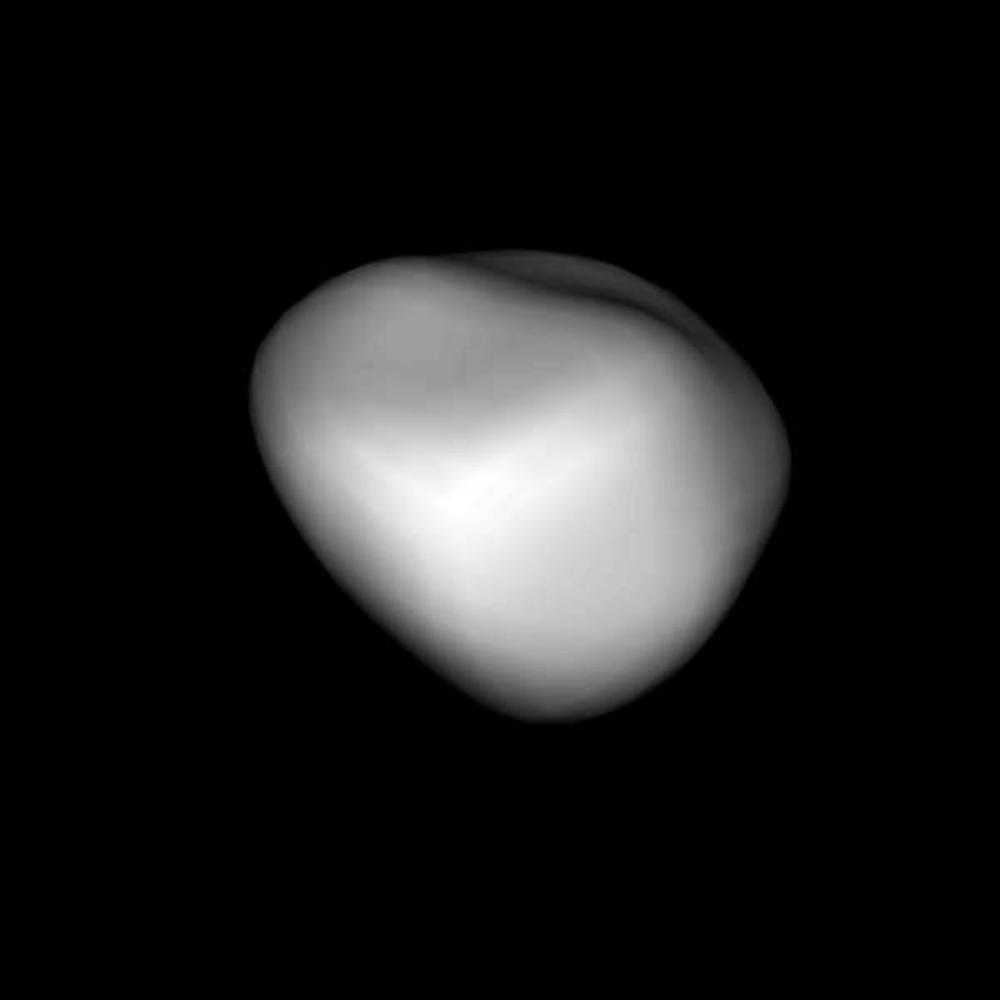}}\\
        \resizebox{0.24\hsize}{!}{\includegraphics{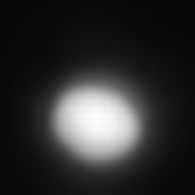}}\resizebox{0.24\hsize}{!}{\includegraphics{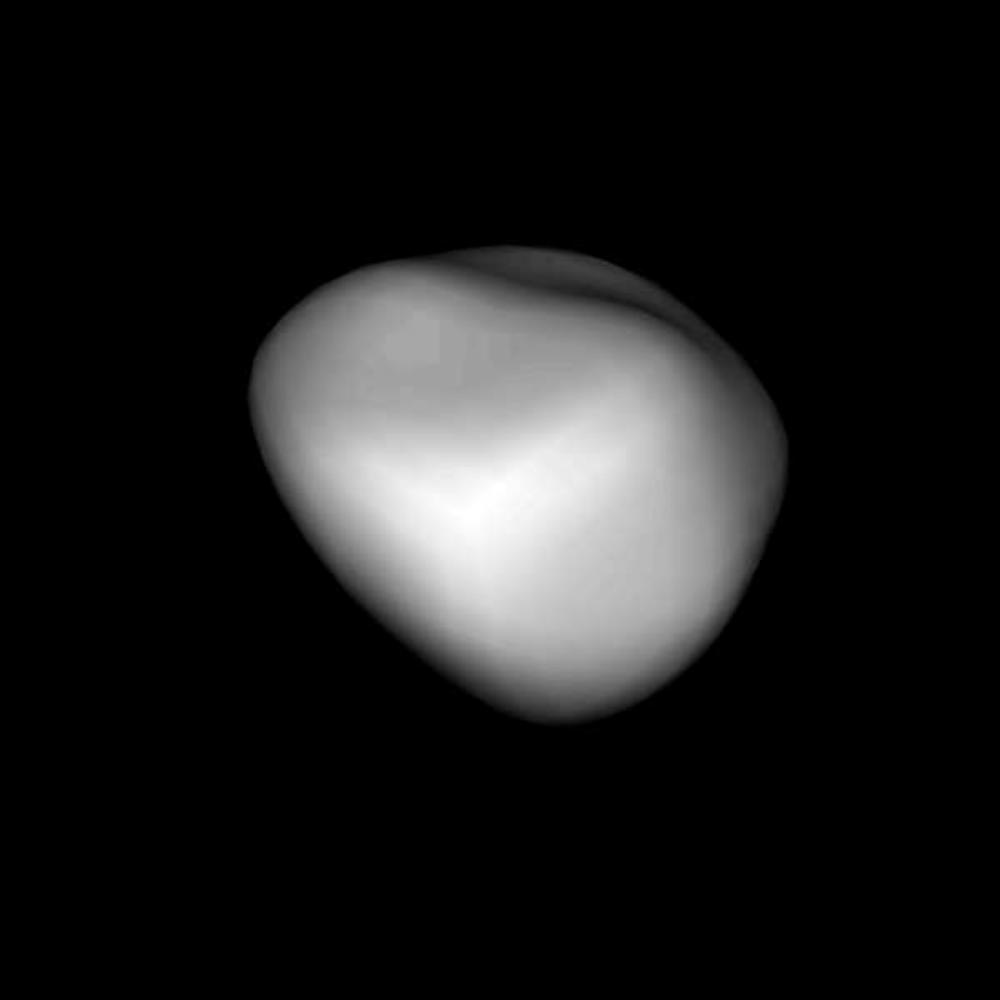}}\resizebox{0.24\hsize}{!}{\includegraphics{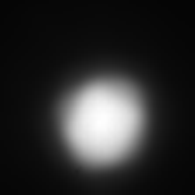}}\resizebox{0.24\hsize}{!}{\includegraphics{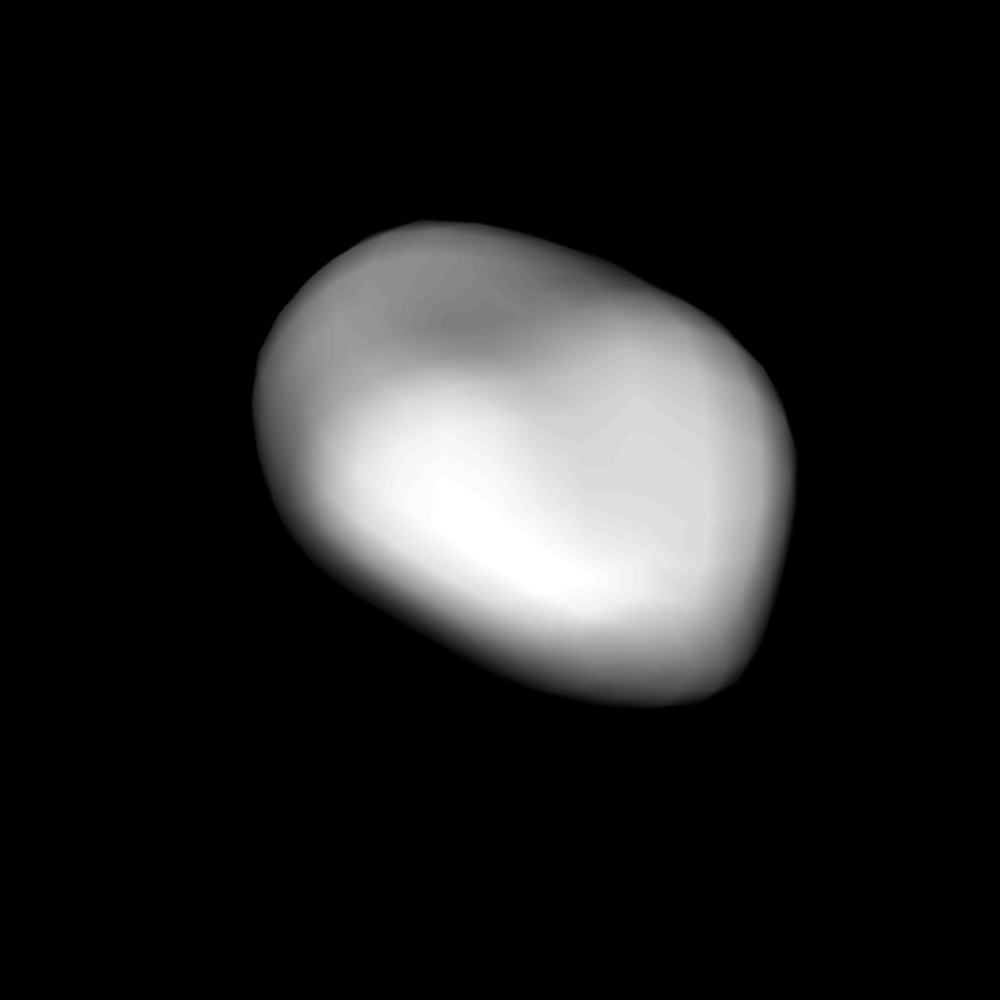}}\\
        \resizebox{0.24\hsize}{!}{\includegraphics{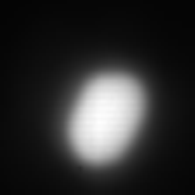}}\resizebox{0.24\hsize}{!}{\includegraphics{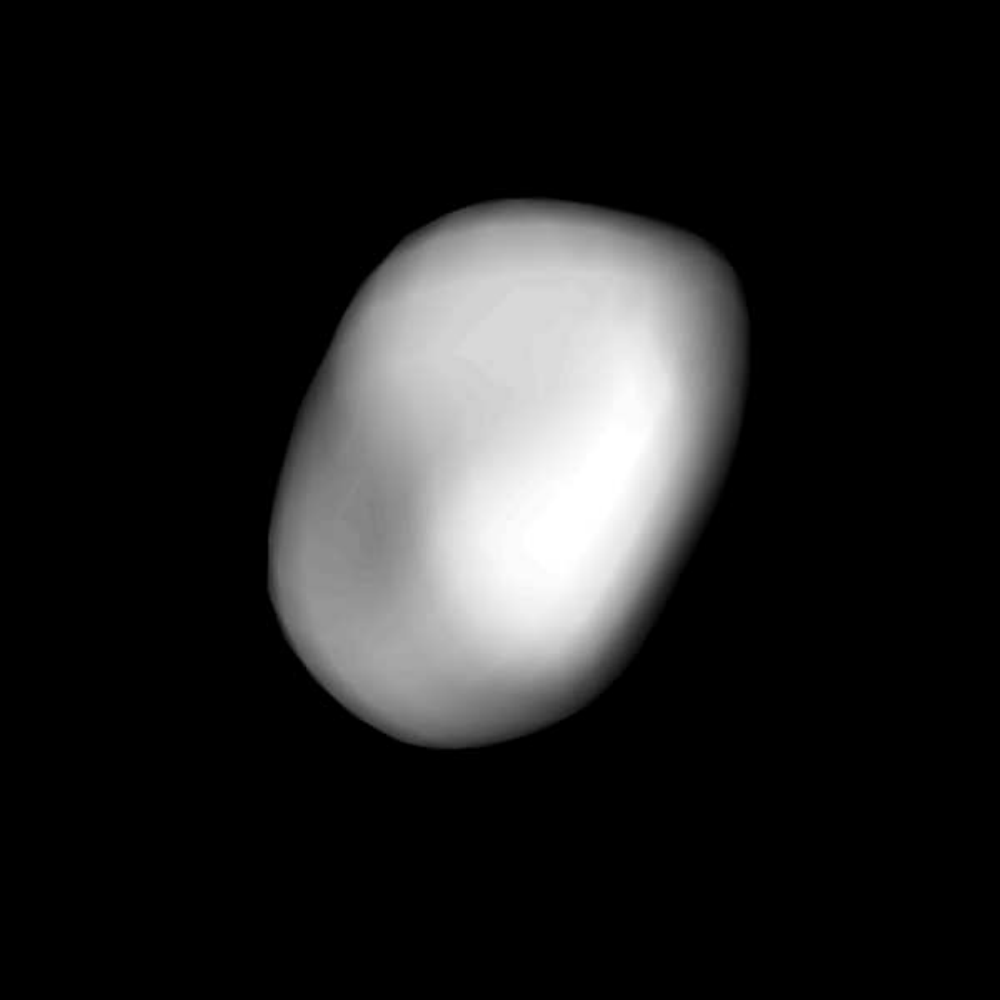}}\resizebox{0.24\hsize}{!}{\includegraphics{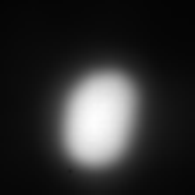}}\resizebox{0.24\hsize}{!}{\includegraphics{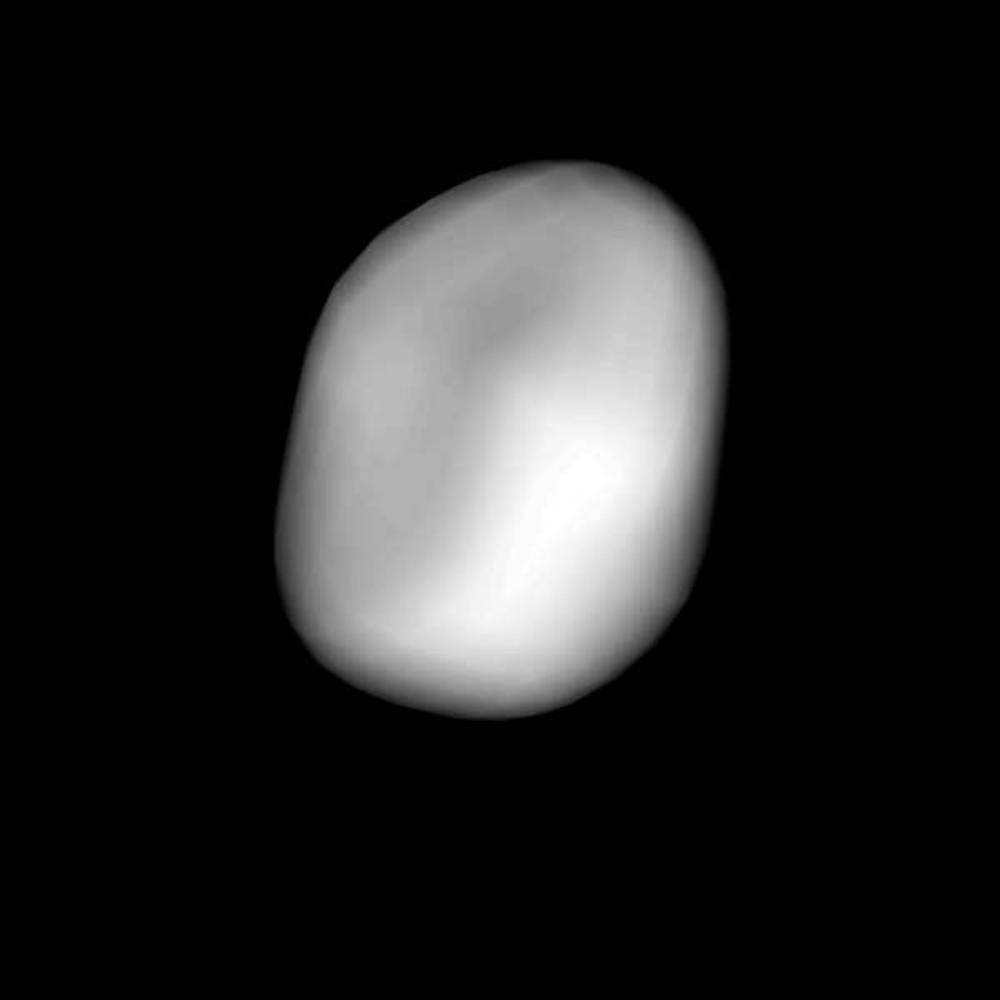}}\\
        \resizebox{0.24\hsize}{!}{\includegraphics{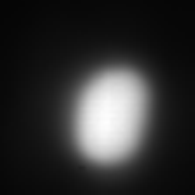}}\resizebox{0.24\hsize}{!}{\includegraphics{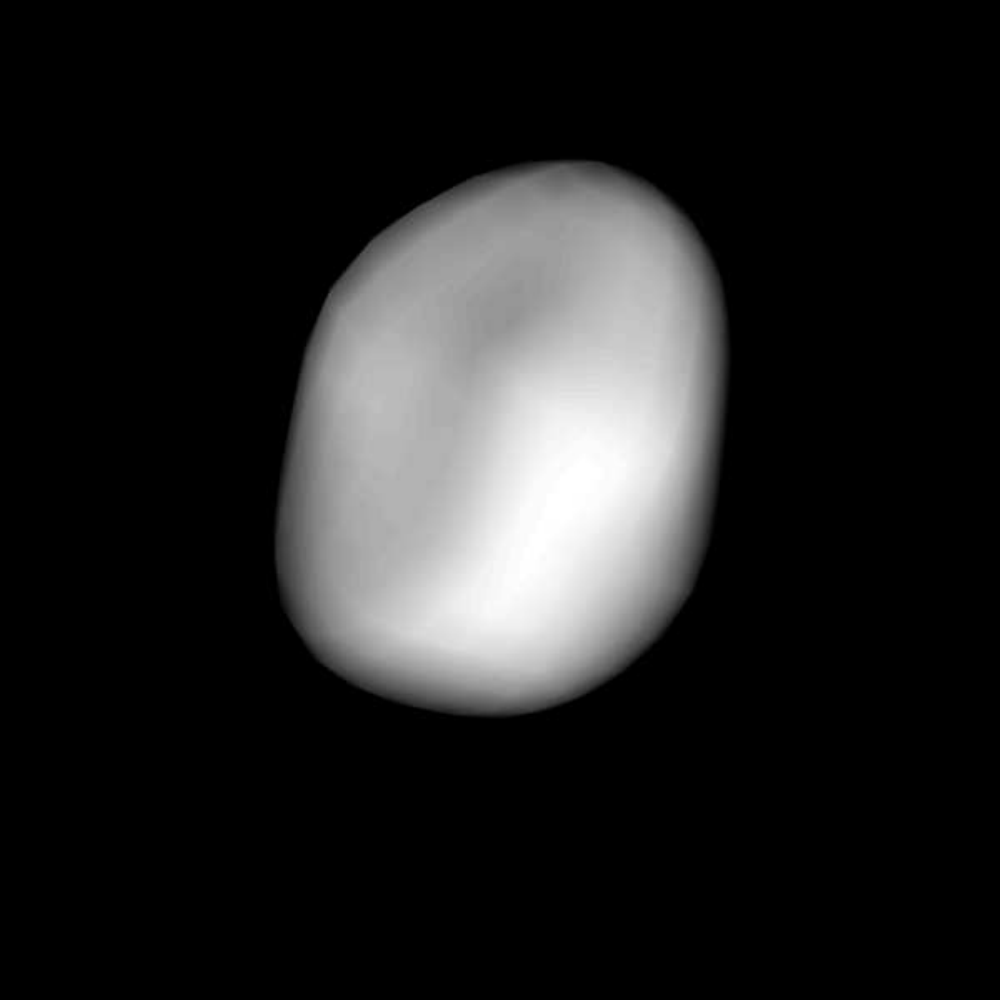}}\resizebox{0.24\hsize}{!}{\includegraphics{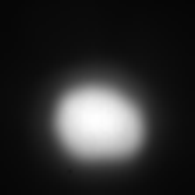}}\resizebox{0.24\hsize}{!}{\includegraphics{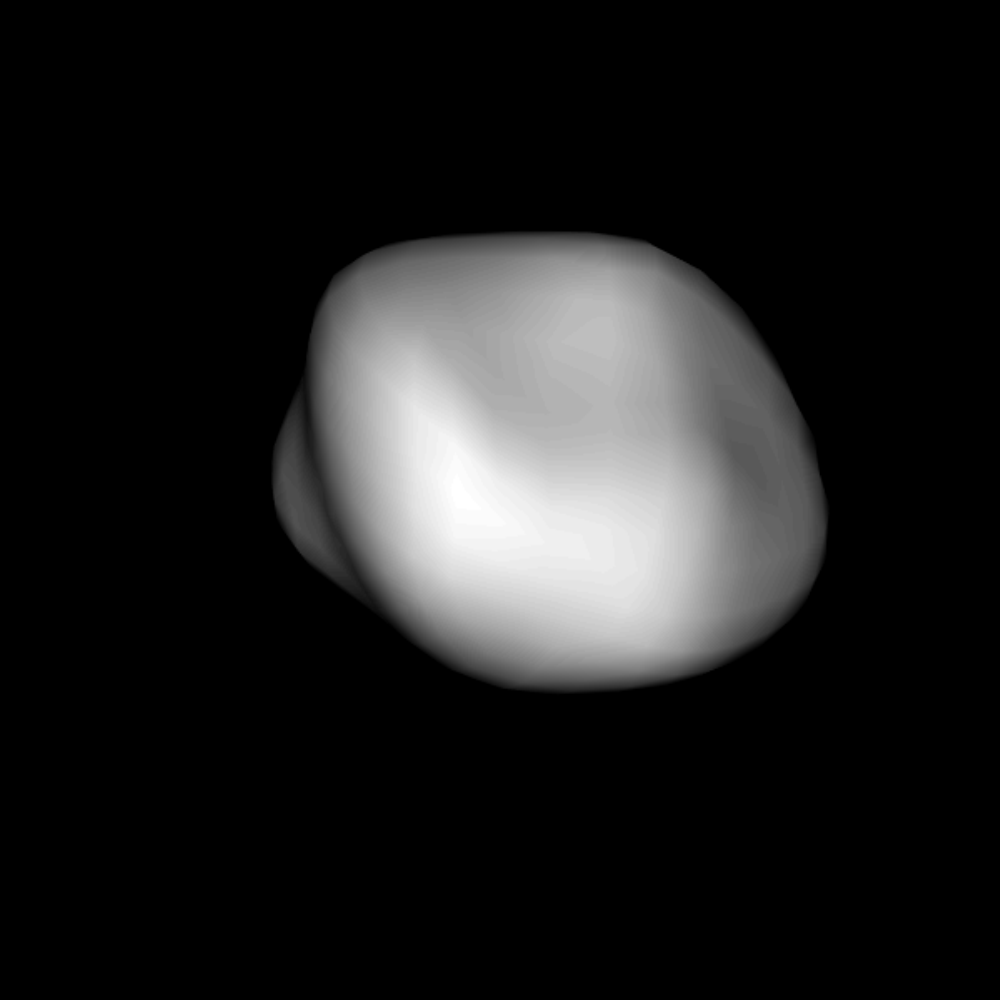}}\\
        \resizebox{0.24\hsize}{!}{\includegraphics{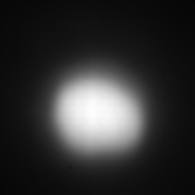}}\resizebox{0.24\hsize}{!}{\includegraphics{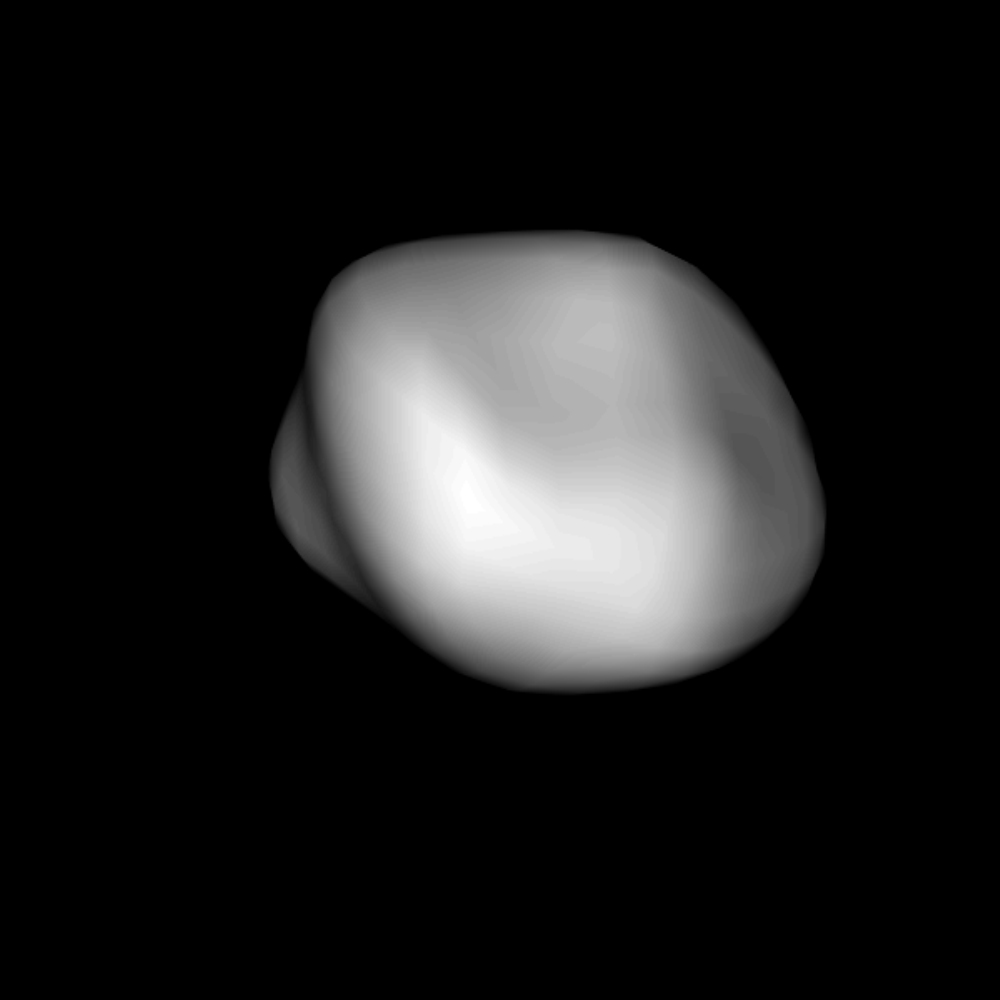}}\resizebox{0.24\hsize}{!}{\includegraphics{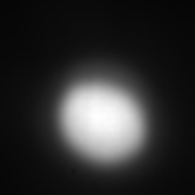}}\resizebox{0.24\hsize}{!}{\includegraphics{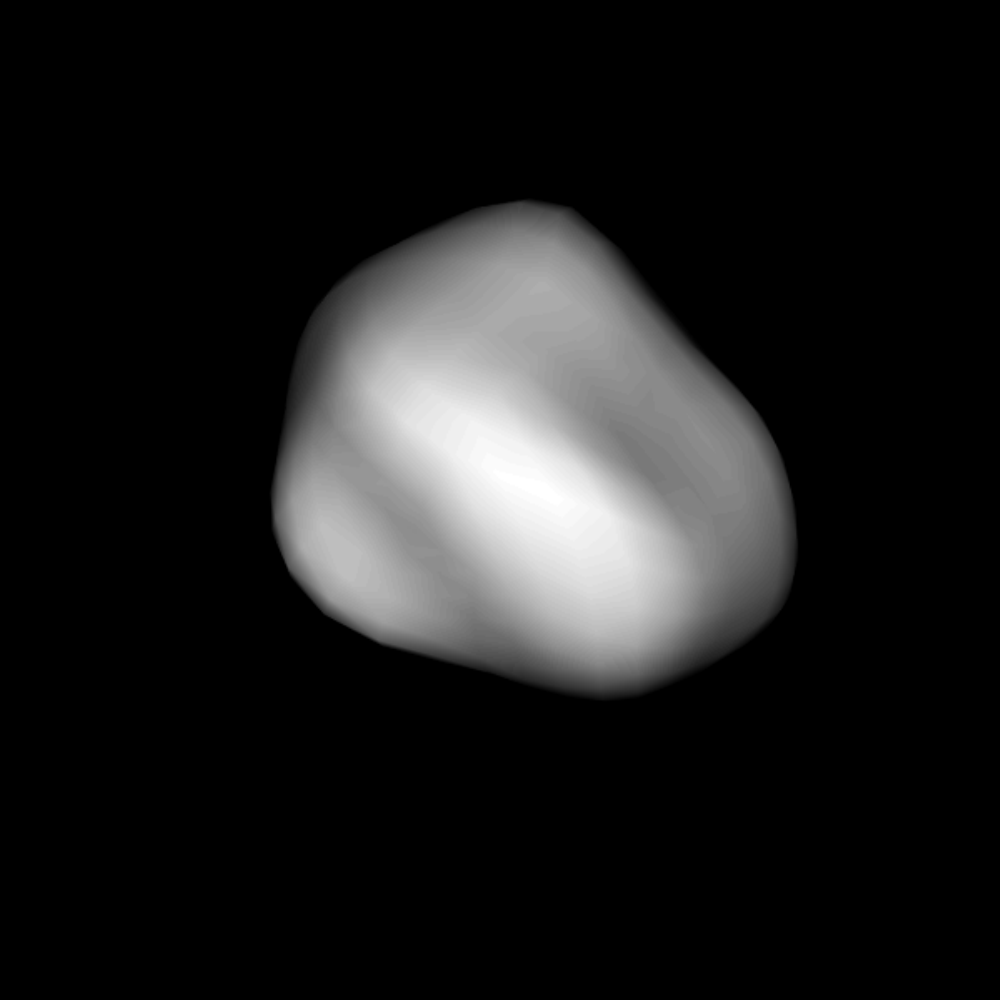}}\\
    \caption{\label{fig:52a}Comparison between model projections and corresponding AO images for asteroid (52) Europa (first part).}
\end{figure}

\begin{figure}[tbp]
    \centering
        \resizebox{0.24\hsize}{!}{\includegraphics{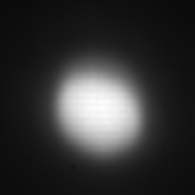}}\resizebox{0.24\hsize}{!}{\includegraphics{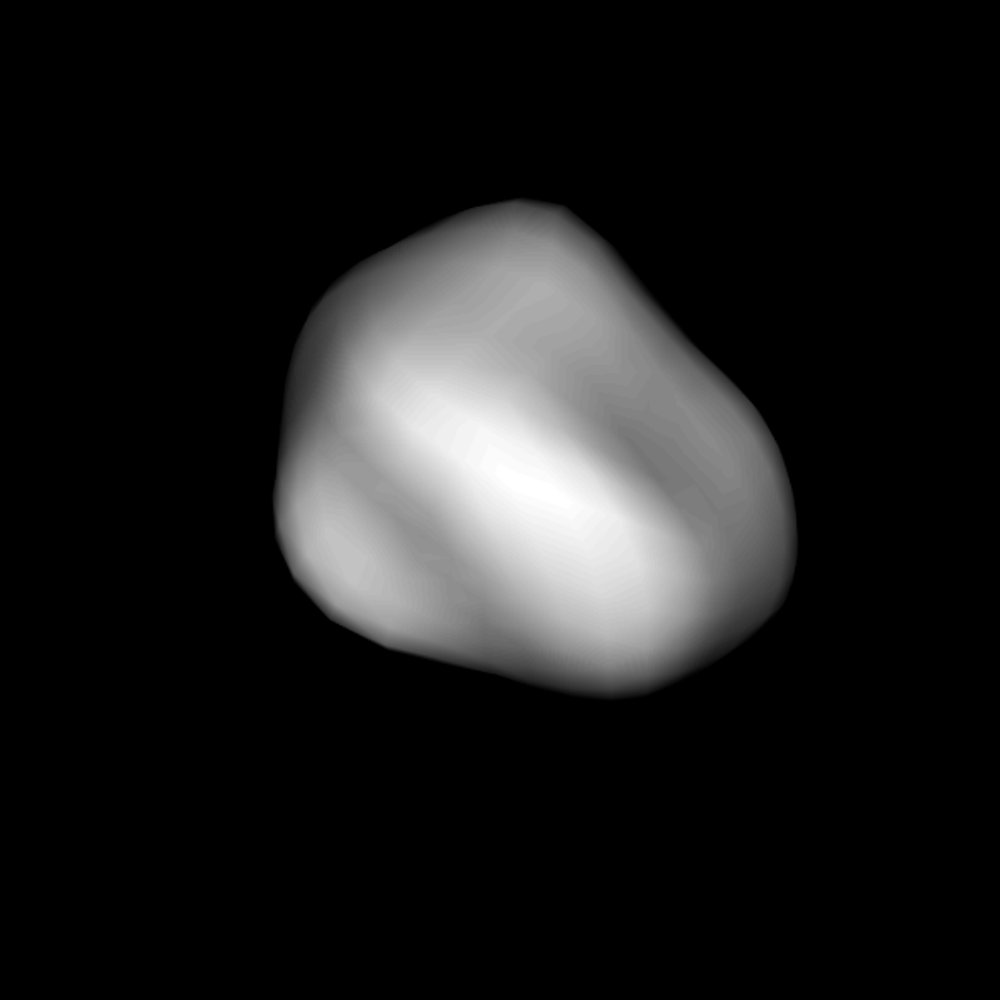}}\resizebox{0.24\hsize}{!}{\includegraphics{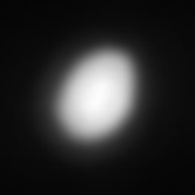}}\resizebox{0.24\hsize}{!}{\includegraphics{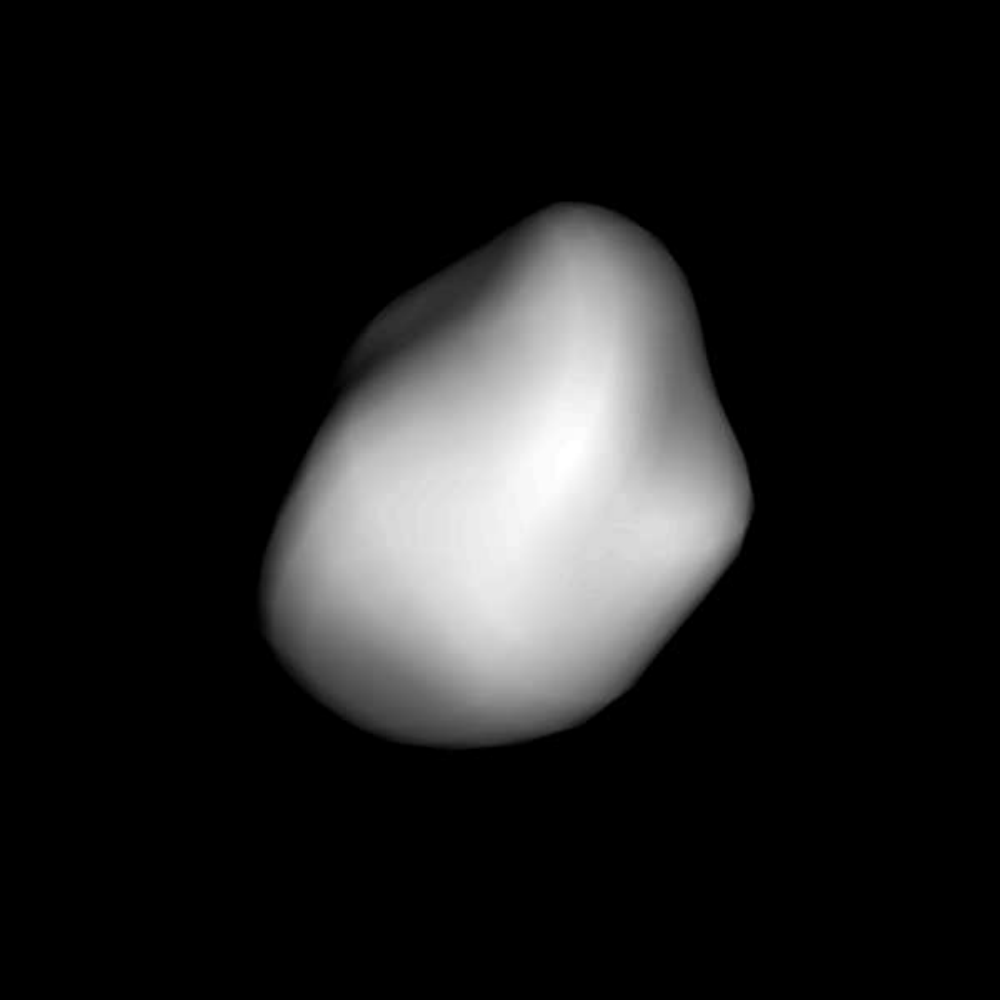}}\\
        \resizebox{0.24\hsize}{!}{\includegraphics{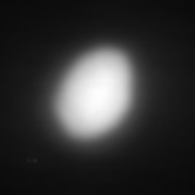}}\resizebox{0.24\hsize}{!}{\includegraphics{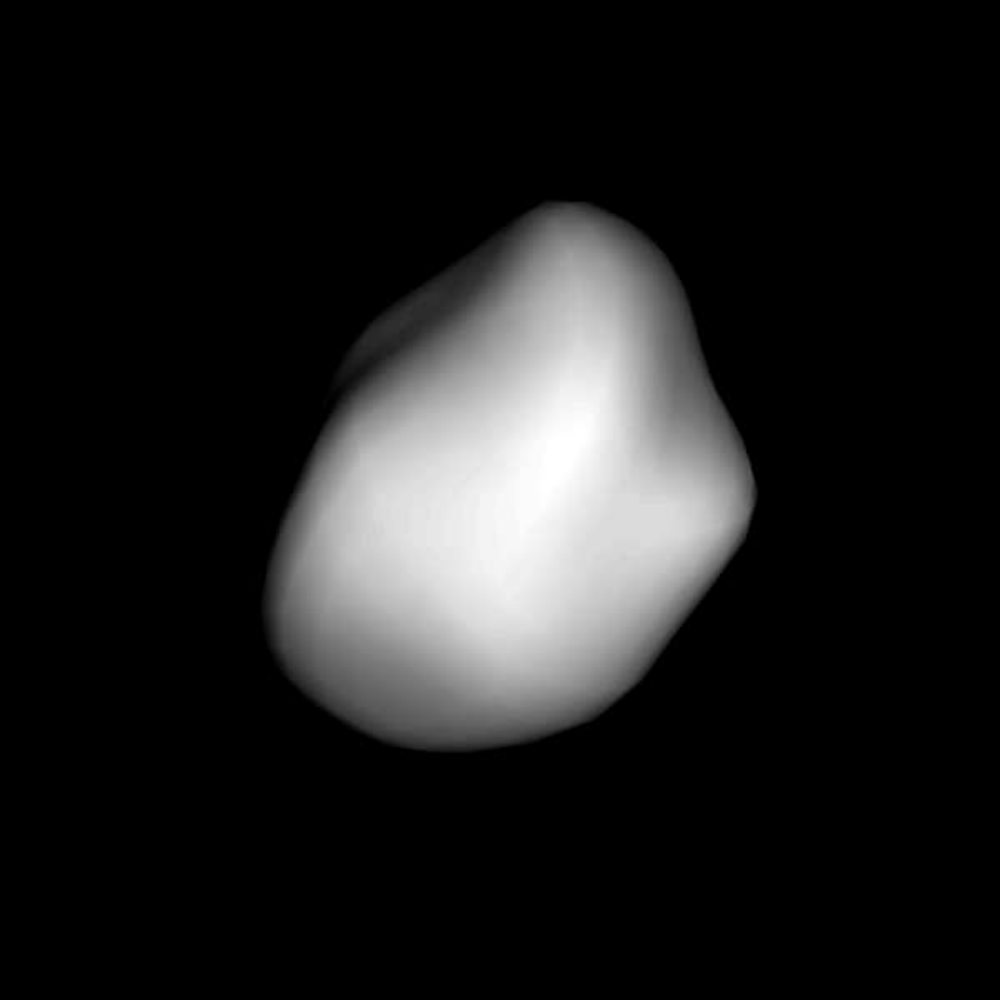}}\resizebox{0.24\hsize}{!}{\includegraphics{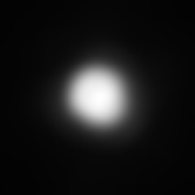}}\resizebox{0.24\hsize}{!}{\includegraphics{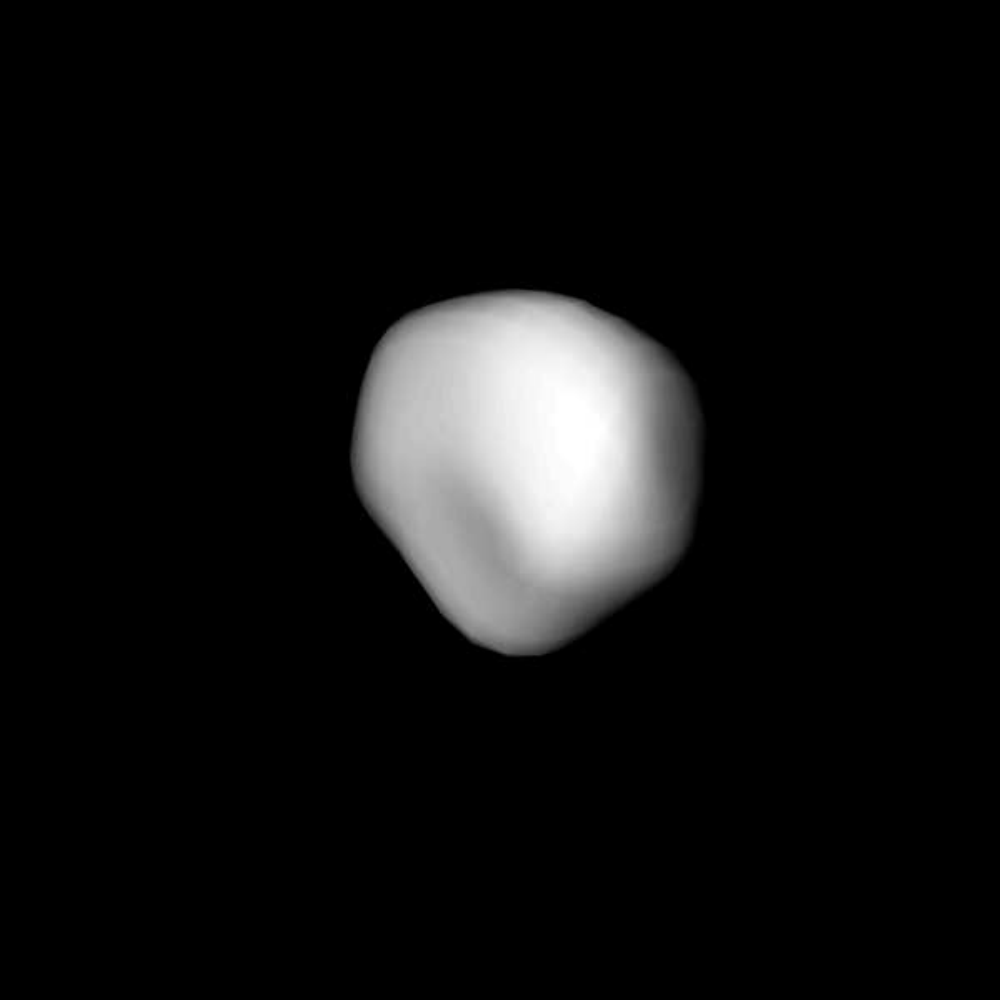}}\\
        \resizebox{0.24\hsize}{!}{\includegraphics{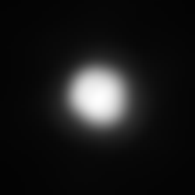}}\resizebox{0.24\hsize}{!}{\includegraphics{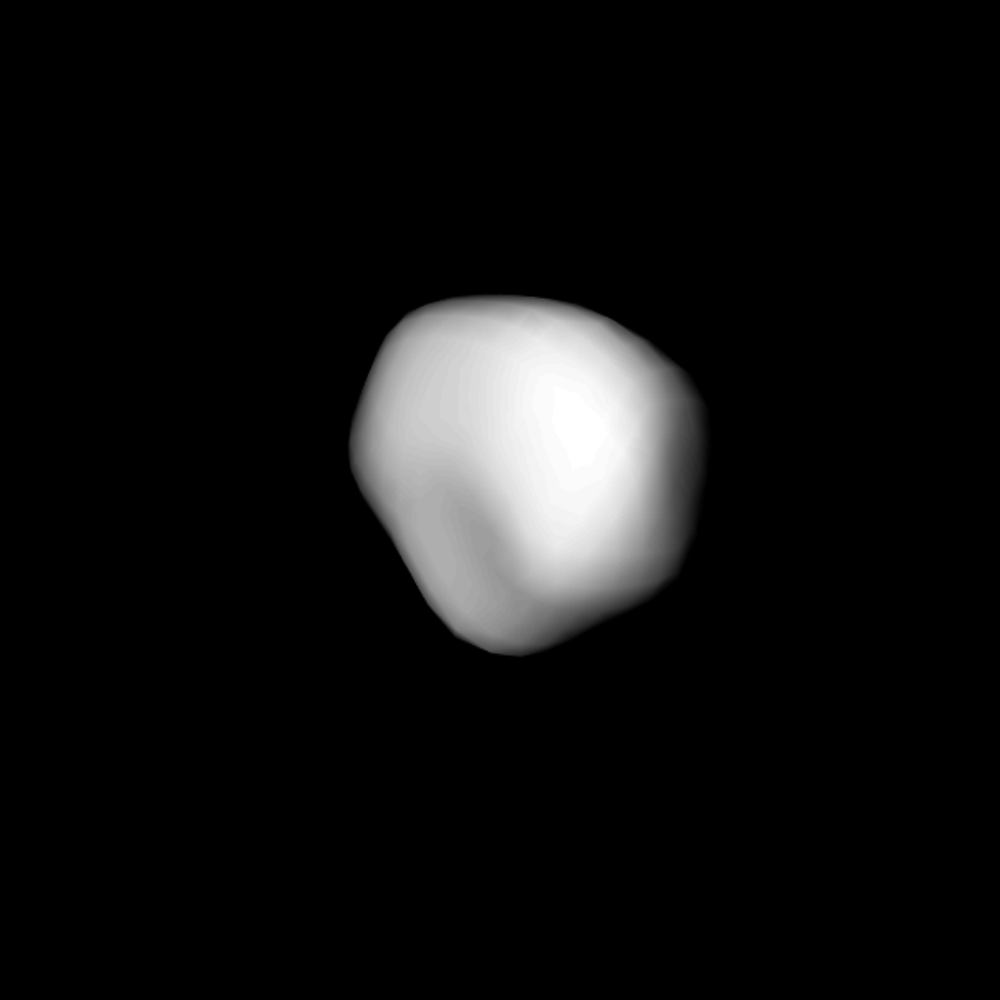}}\resizebox{0.24\hsize}{!}{\includegraphics{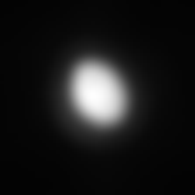}}\resizebox{0.24\hsize}{!}{\includegraphics{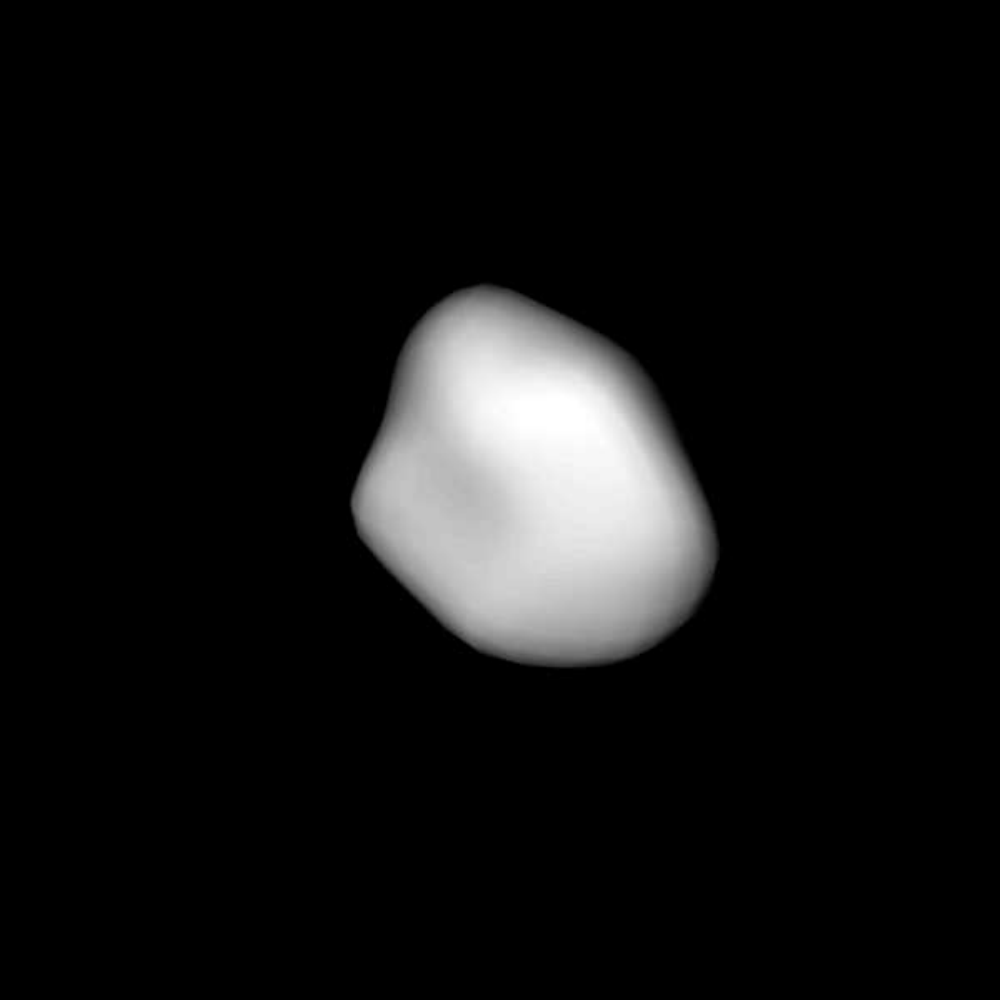}}\\
        \resizebox{0.24\hsize}{!}{\includegraphics{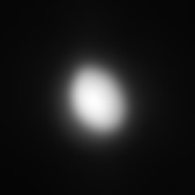}}\resizebox{0.24\hsize}{!}{\includegraphics{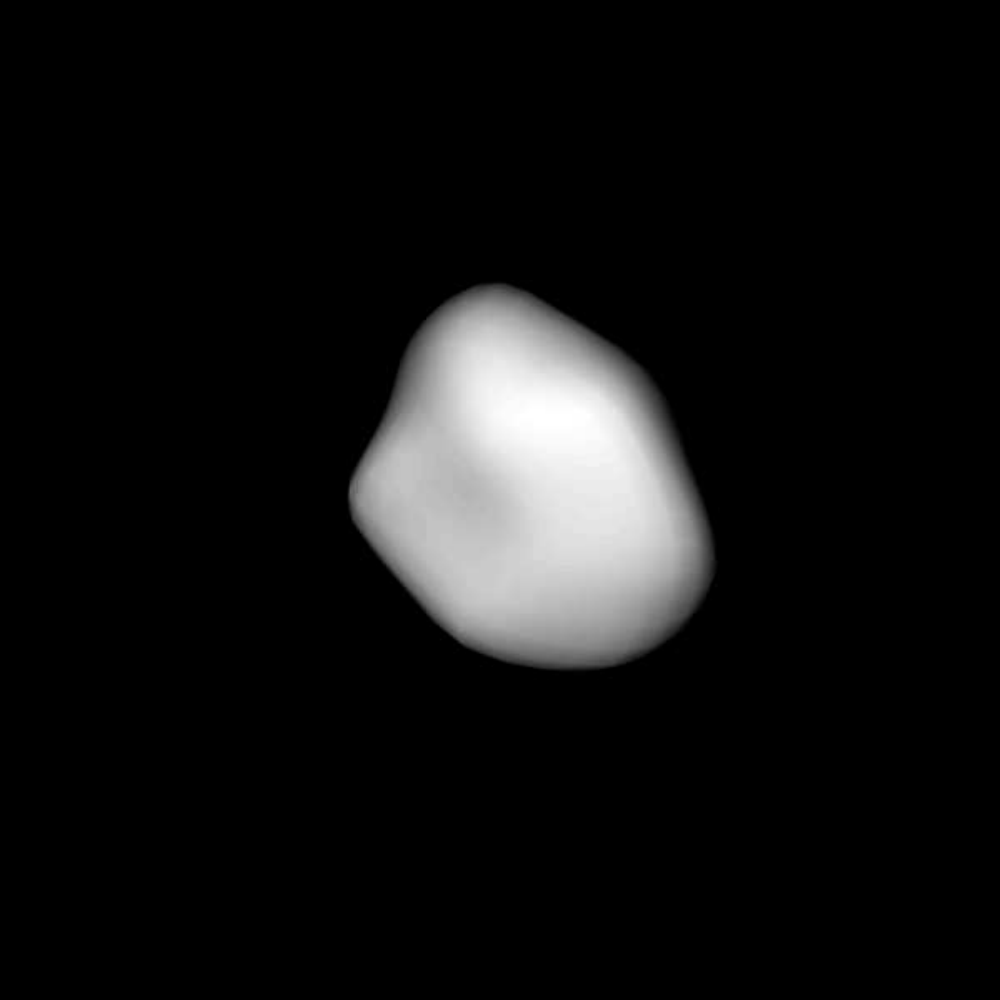}}\\
    \caption{\label{fig:52b}Comparison between model projections and corresponding AO images for asteroid (52) Europa (second part).}
\end{figure}

\begin{figure}[tbp]
    \centering
        \resizebox{0.24\hsize}{!}{\includegraphics{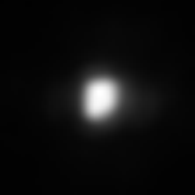}}\resizebox{0.24\hsize}{!}{\includegraphics{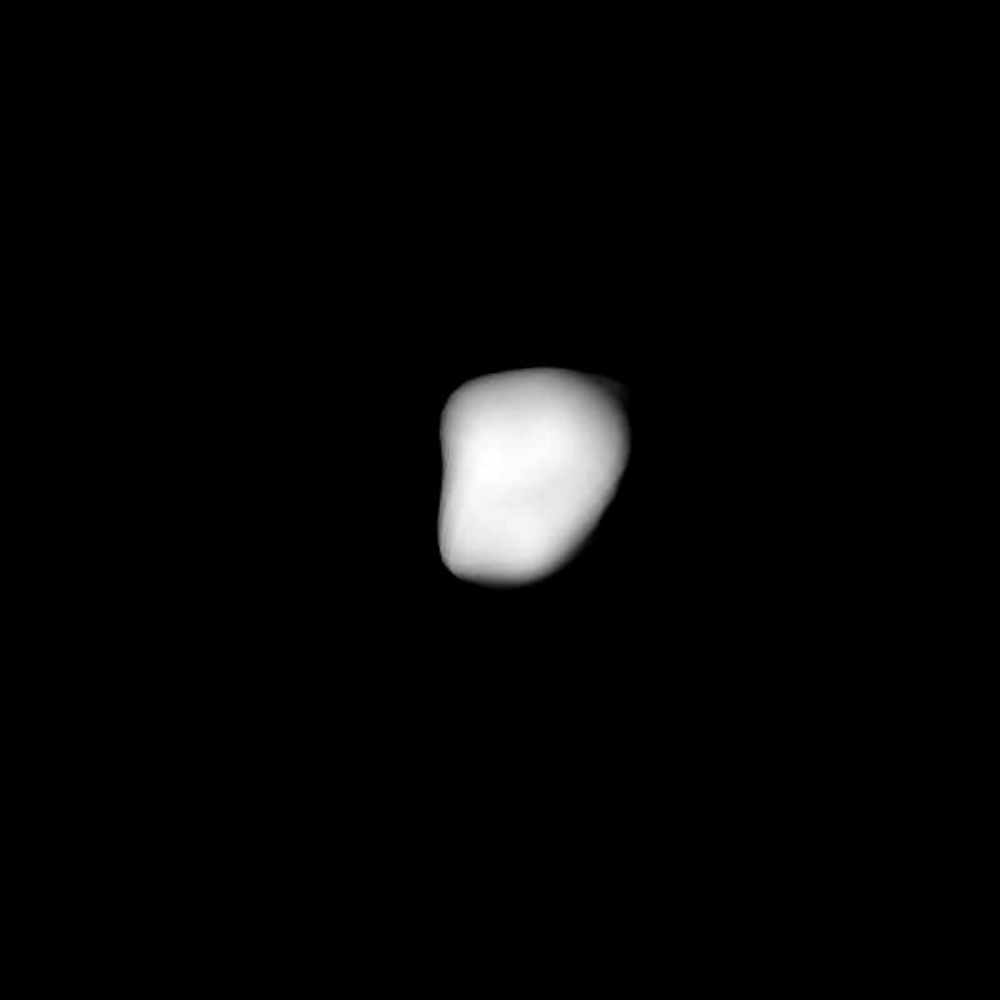}}\resizebox{0.24\hsize}{!}{\includegraphics{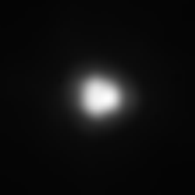}}\resizebox{0.24\hsize}{!}{\includegraphics{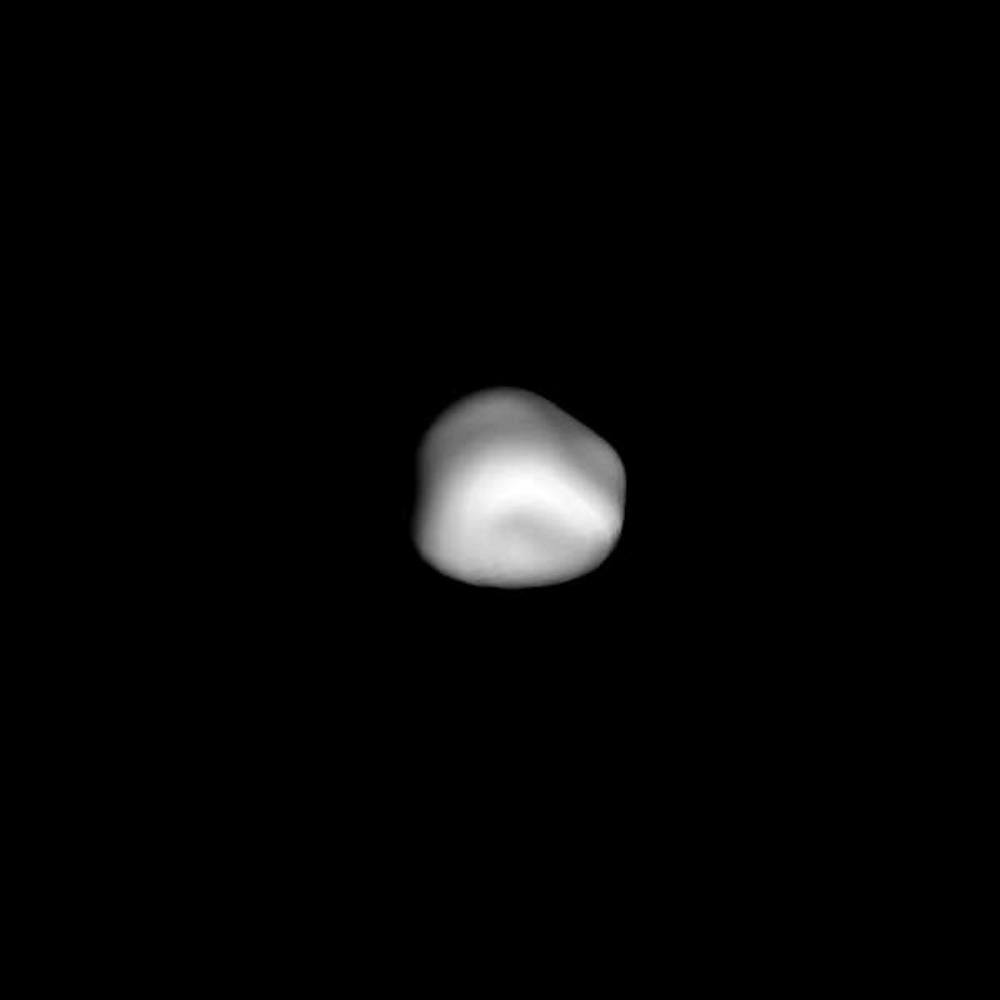}}\\
    \caption{\label{fig:54}Comparison between model projections and corresponding AO images for asteroid (54) Alexandra.}
\end{figure}

\begin{figure}[tbp]
    \centering
        \resizebox{0.24\hsize}{!}{\includegraphics{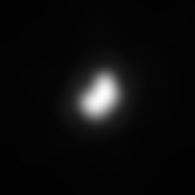}}\resizebox{0.24\hsize}{!}{\includegraphics{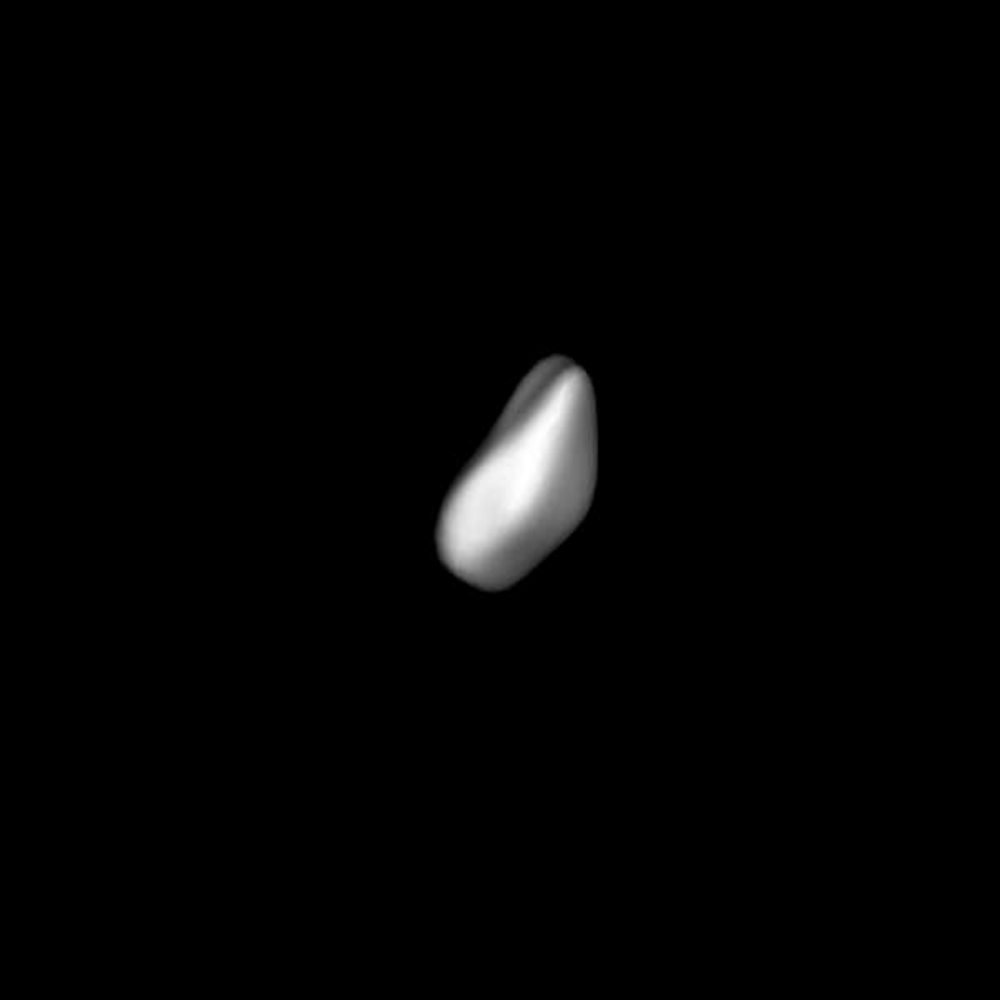}}\resizebox{0.24\hsize}{!}{\includegraphics{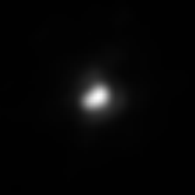}}\resizebox{0.24\hsize}{!}{\includegraphics{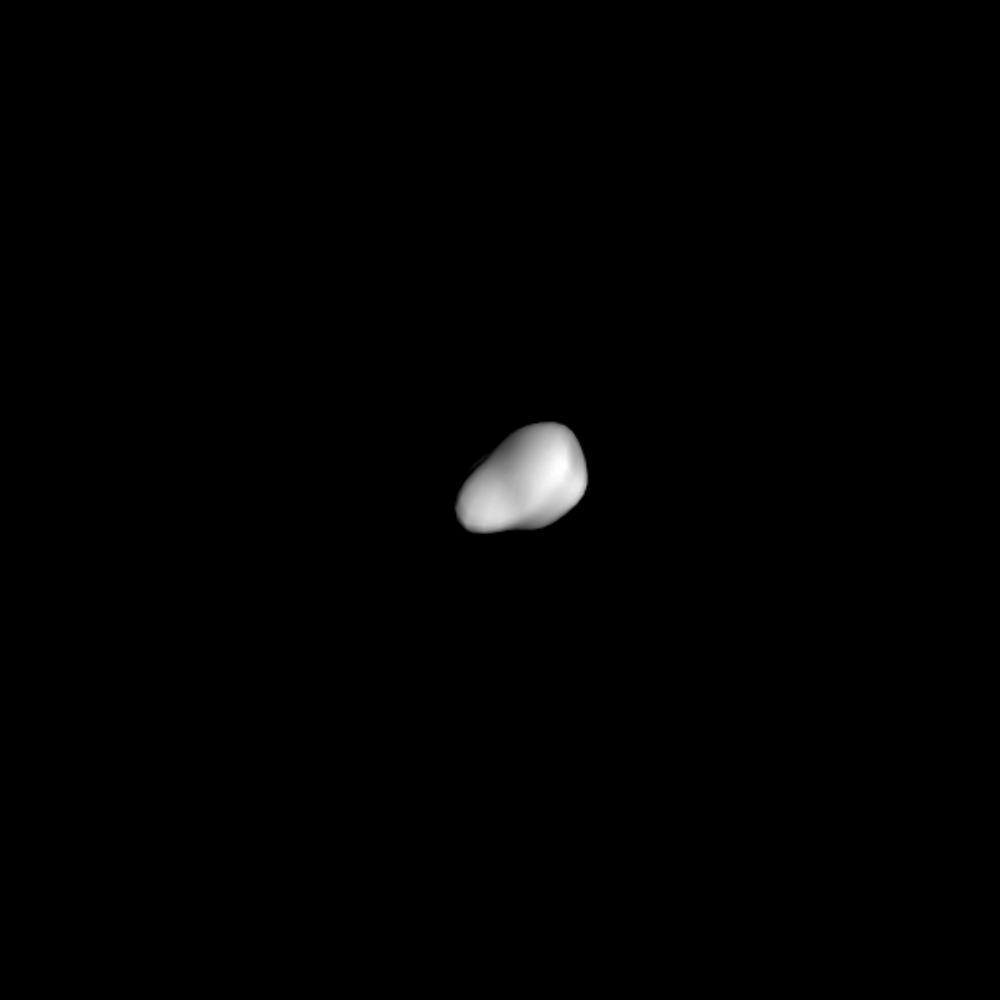}}\\
    \caption{\label{fig:80}Comparison between model projections and corresponding AO images for asteroid (80) Sappho.}
\end{figure}

\begin{figure}[tbp]
    \centering
        \resizebox{0.24\hsize}{!}{\includegraphics{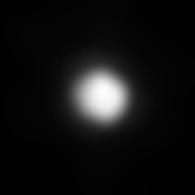}}\resizebox{0.24\hsize}{!}{\includegraphics{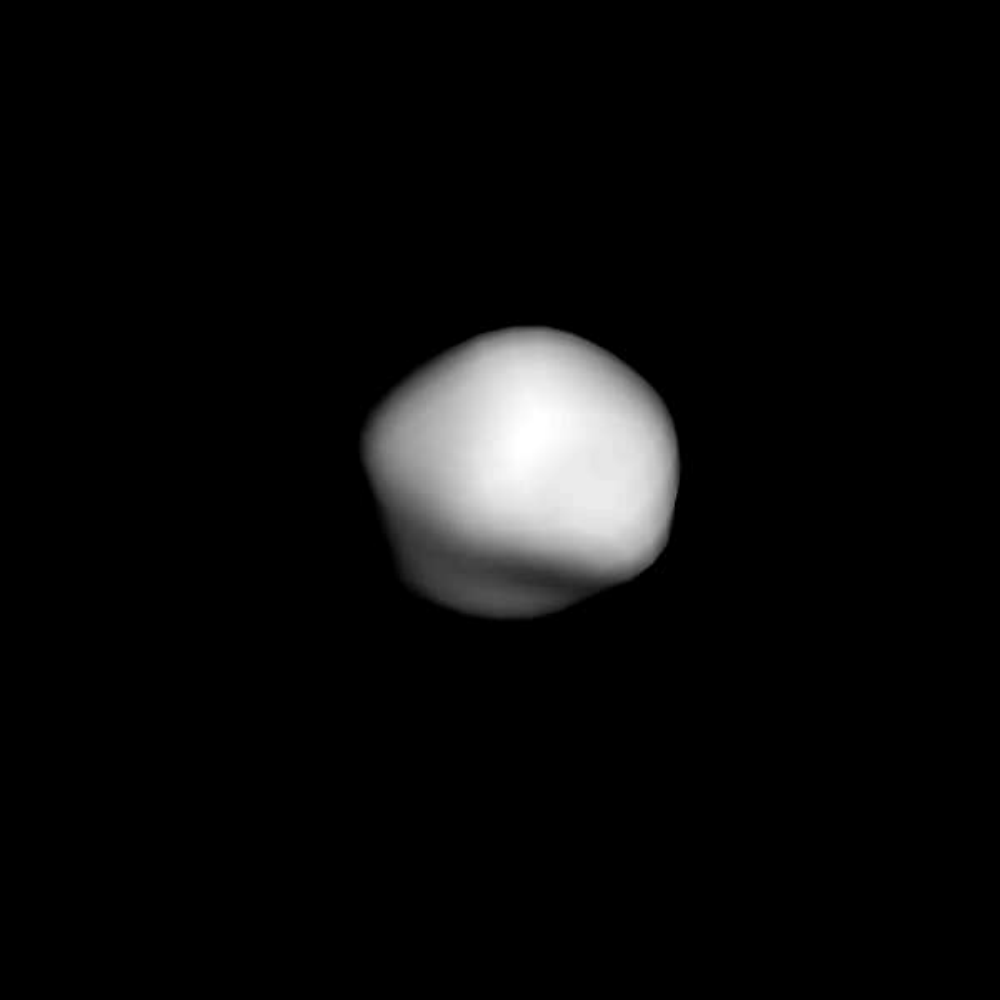}}\resizebox{0.24\hsize}{!}{\includegraphics{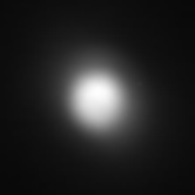}}\resizebox{0.24\hsize}{!}{\includegraphics{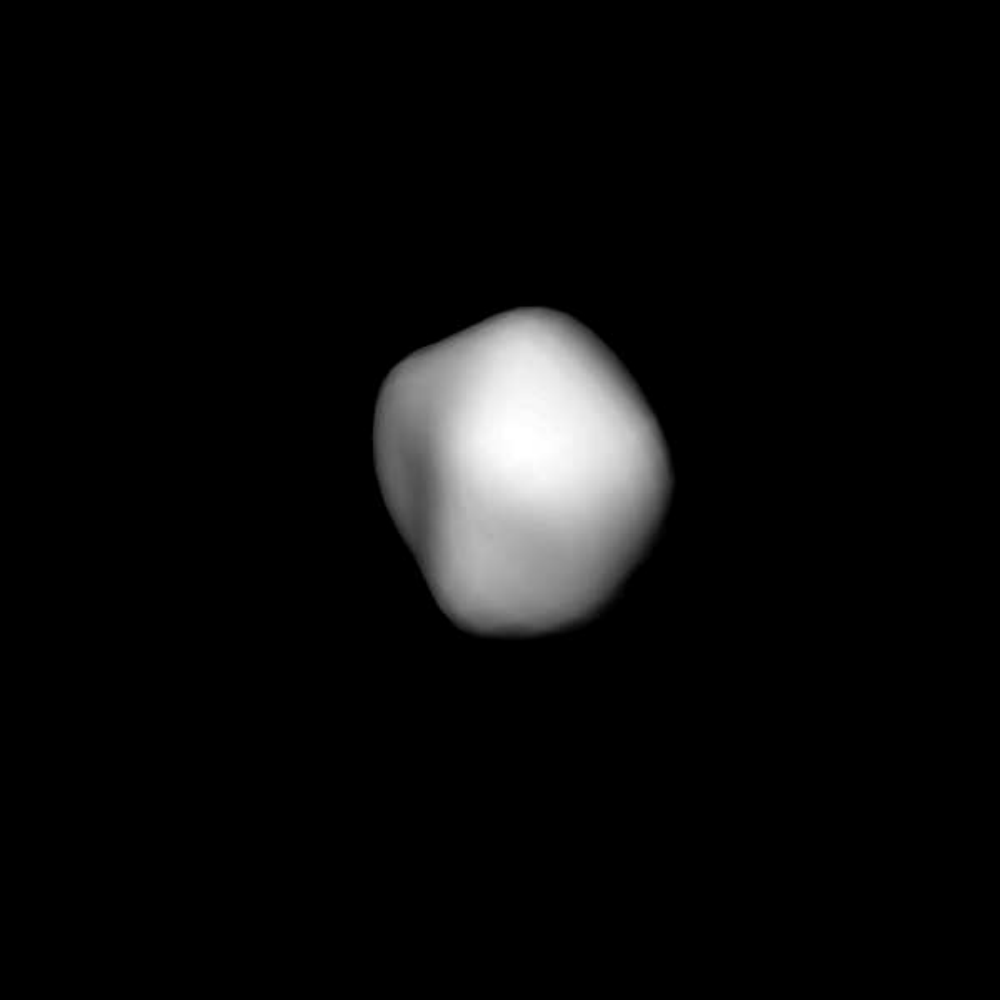}}\\
    \caption{\label{fig:85}Comparison between model projections and corresponding AO images for asteroid (85) Io.}
\end{figure}

\clearpage

\begin{figure}[tbp]
    \centering
        \resizebox{0.24\hsize}{!}{\includegraphics{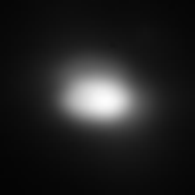}}\resizebox{0.24\hsize}{!}{\includegraphics{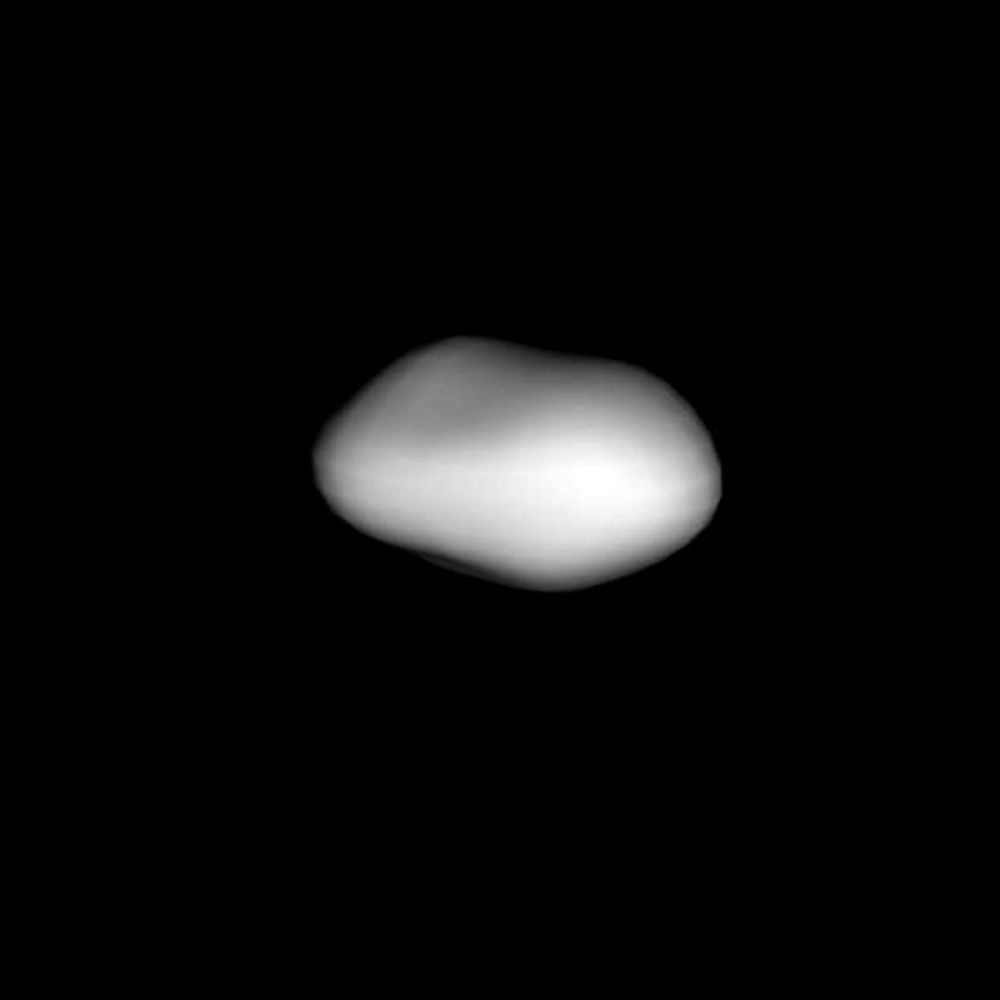}}\resizebox{0.24\hsize}{!}{\includegraphics{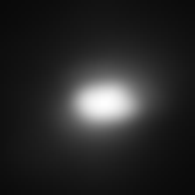}}\resizebox{0.24\hsize}{!}{\includegraphics{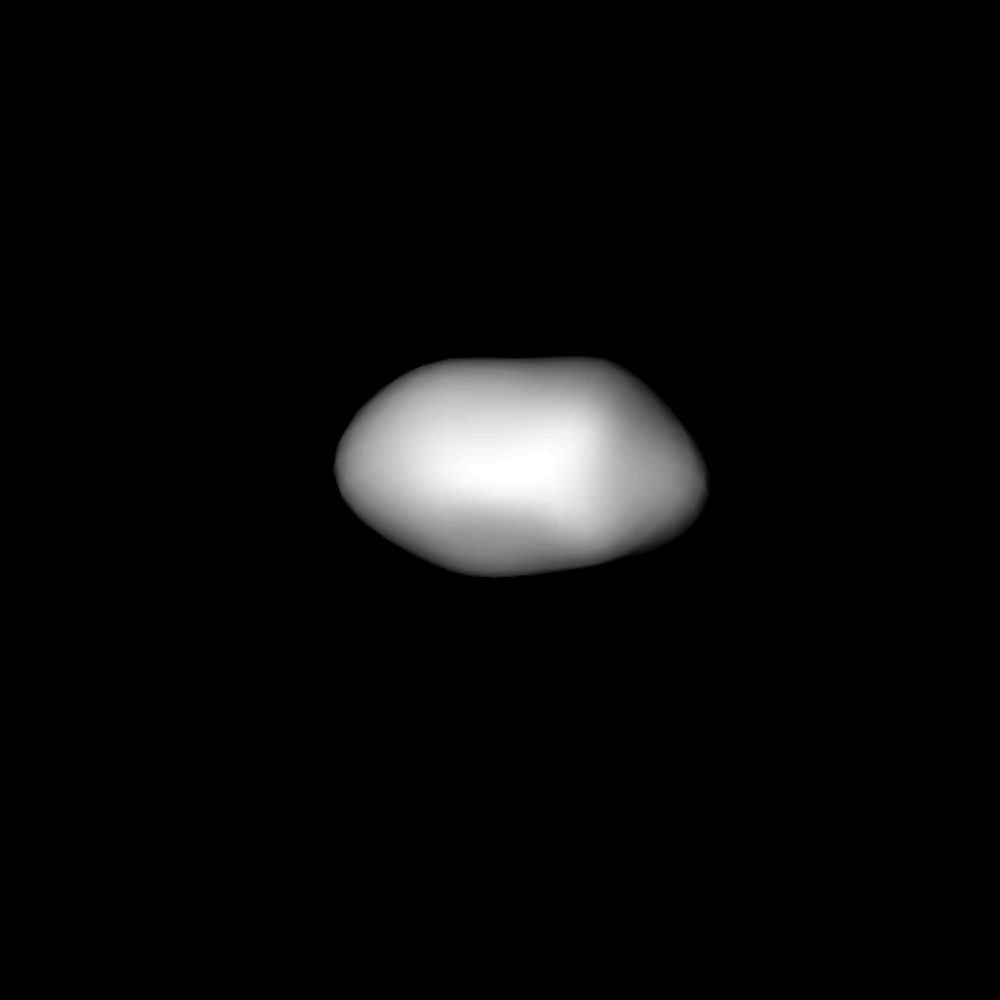}}\\
        \resizebox{0.24\hsize}{!}{\includegraphics{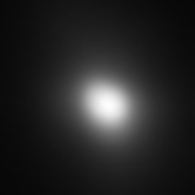}}\resizebox{0.24\hsize}{!}{\includegraphics{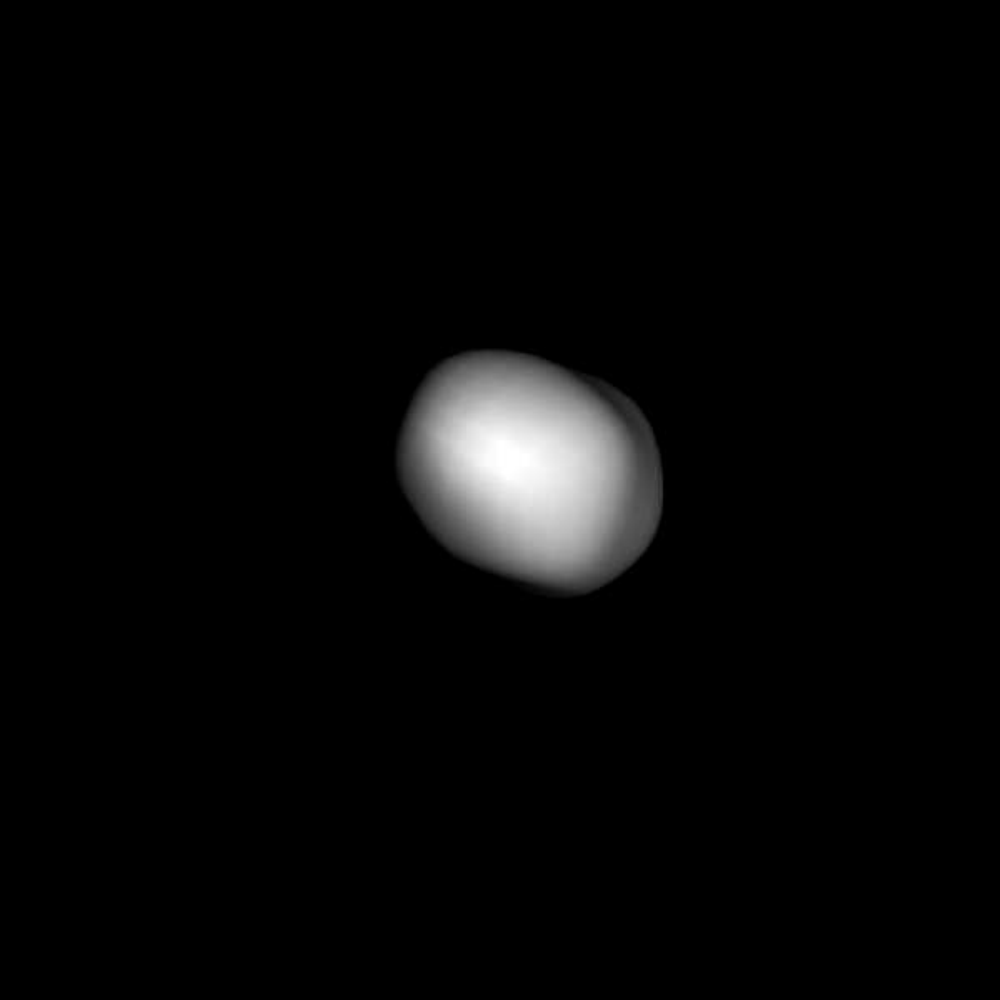}}\resizebox{0.24\hsize}{!}{\includegraphics{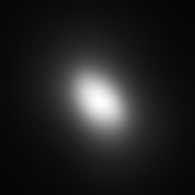}}\resizebox{0.24\hsize}{!}{\includegraphics{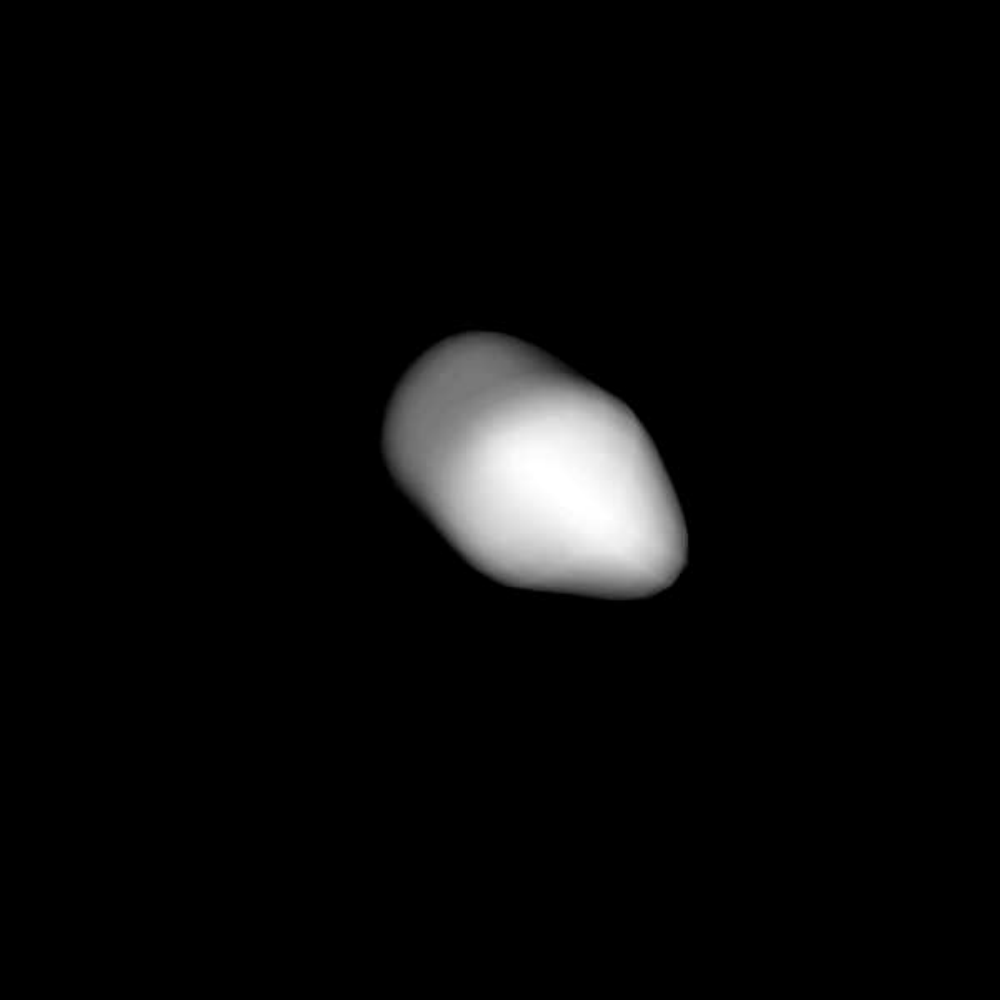}}\\
        \resizebox{0.24\hsize}{!}{\includegraphics{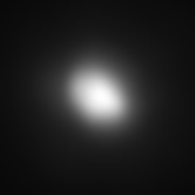}}\resizebox{0.24\hsize}{!}{\includegraphics{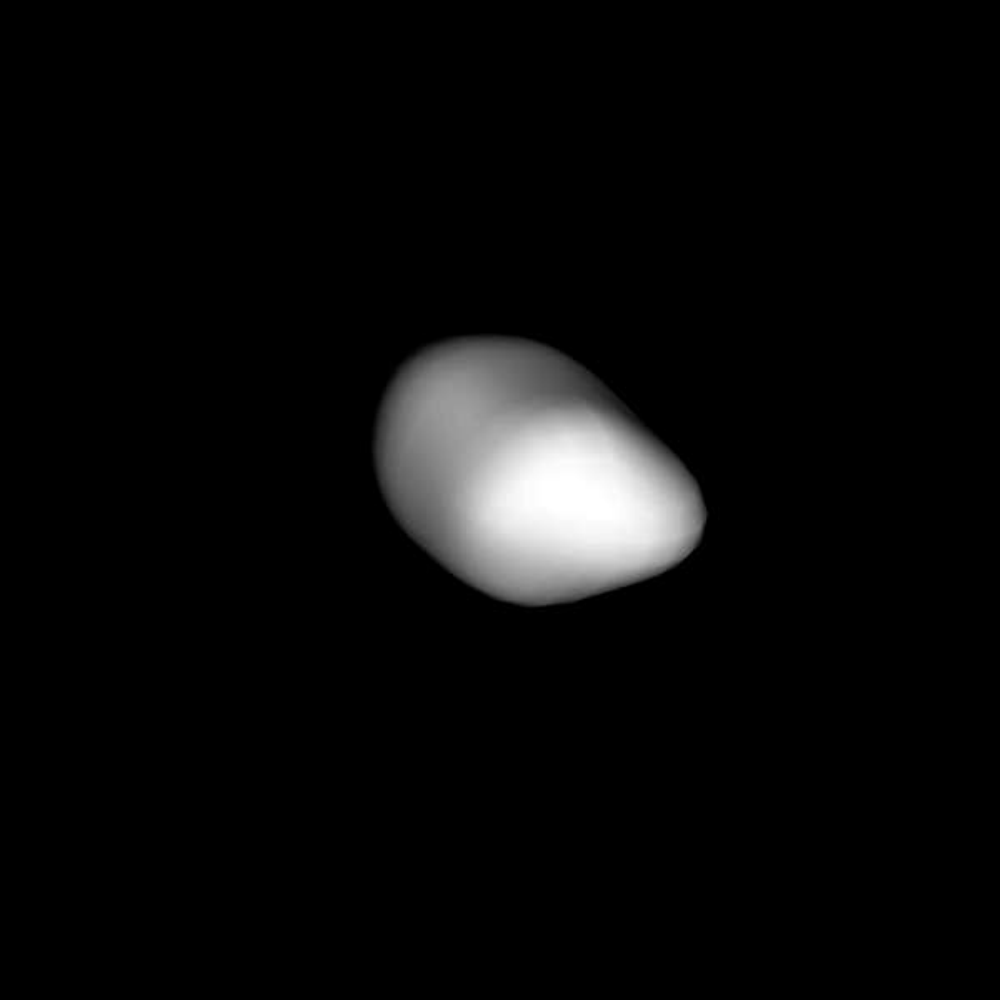}}\resizebox{0.24\hsize}{!}{\includegraphics{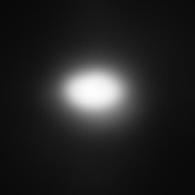}}\resizebox{0.24\hsize}{!}{\includegraphics{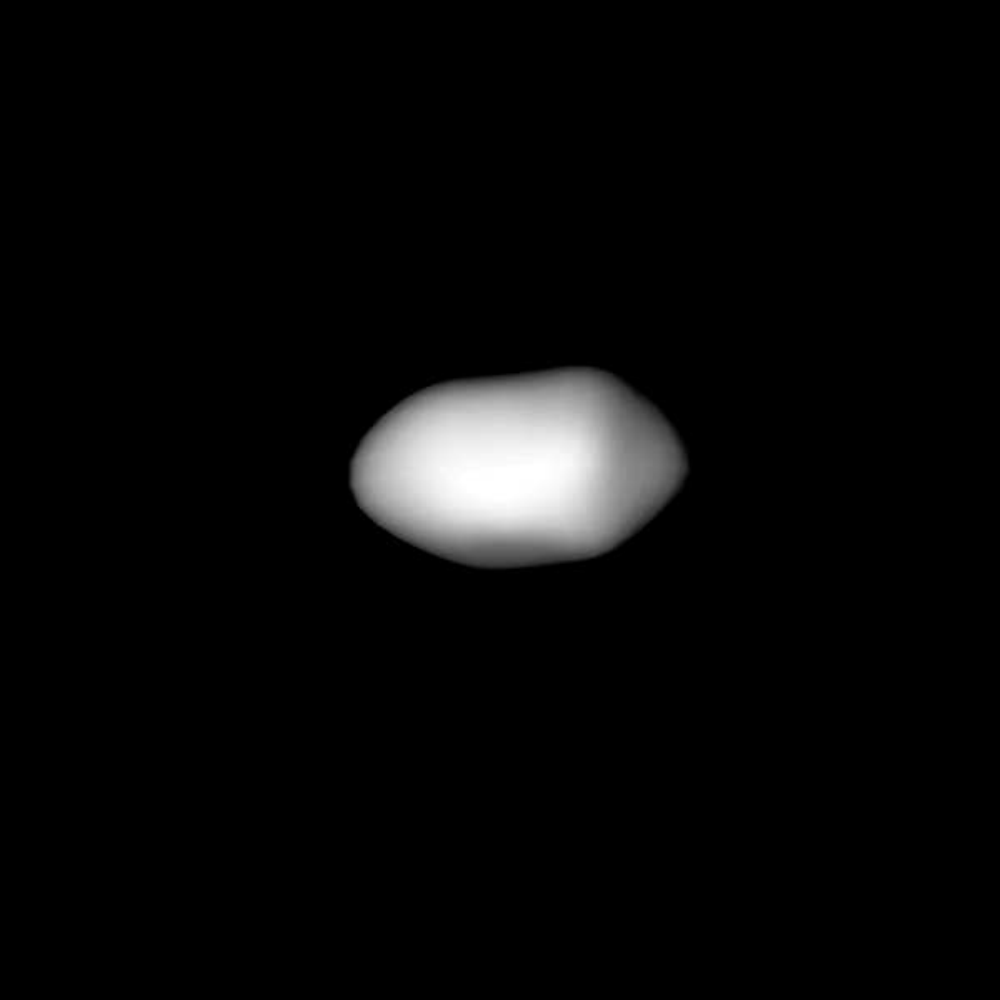}}\\
        \resizebox{0.24\hsize}{!}{\includegraphics{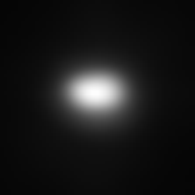}}\resizebox{0.24\hsize}{!}{\includegraphics{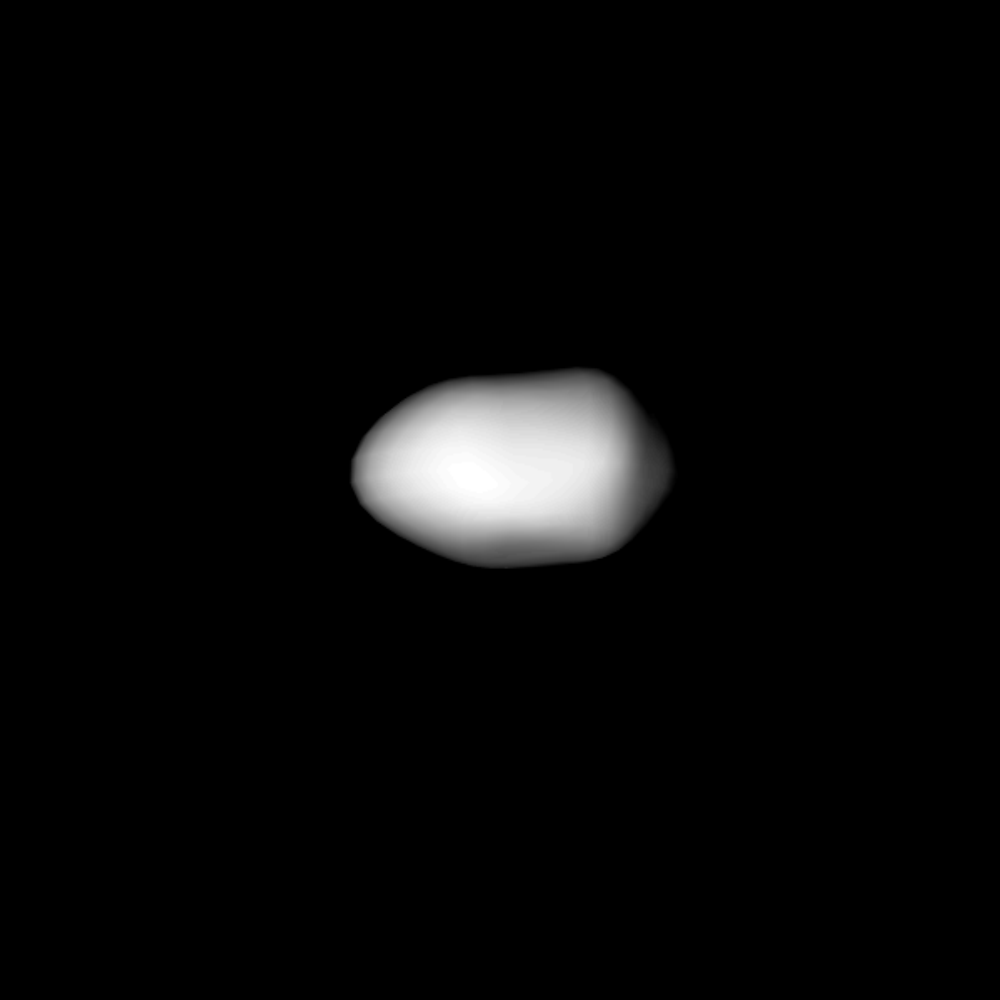}}\resizebox{0.24\hsize}{!}{\includegraphics{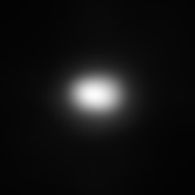}}\resizebox{0.24\hsize}{!}{\includegraphics{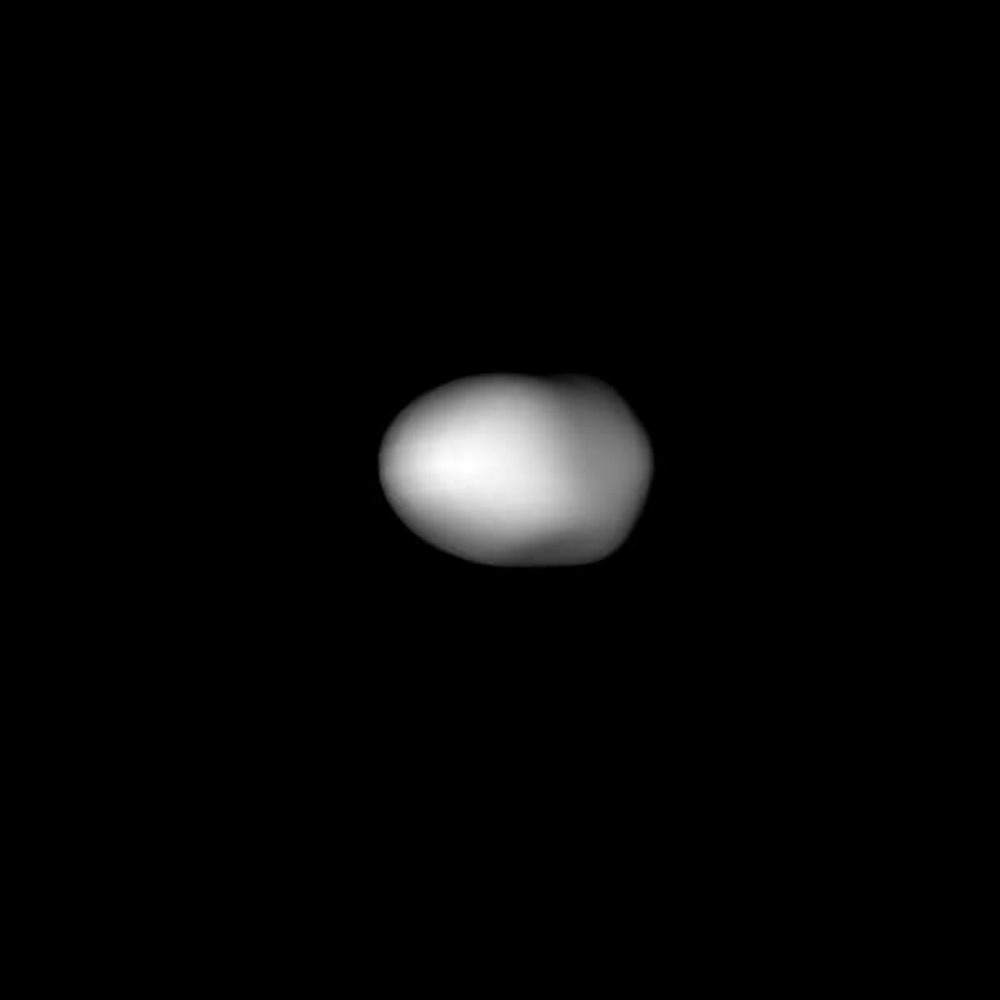}}\\
        \resizebox{0.24\hsize}{!}{\includegraphics{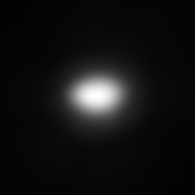}}\resizebox{0.24\hsize}{!}{\includegraphics{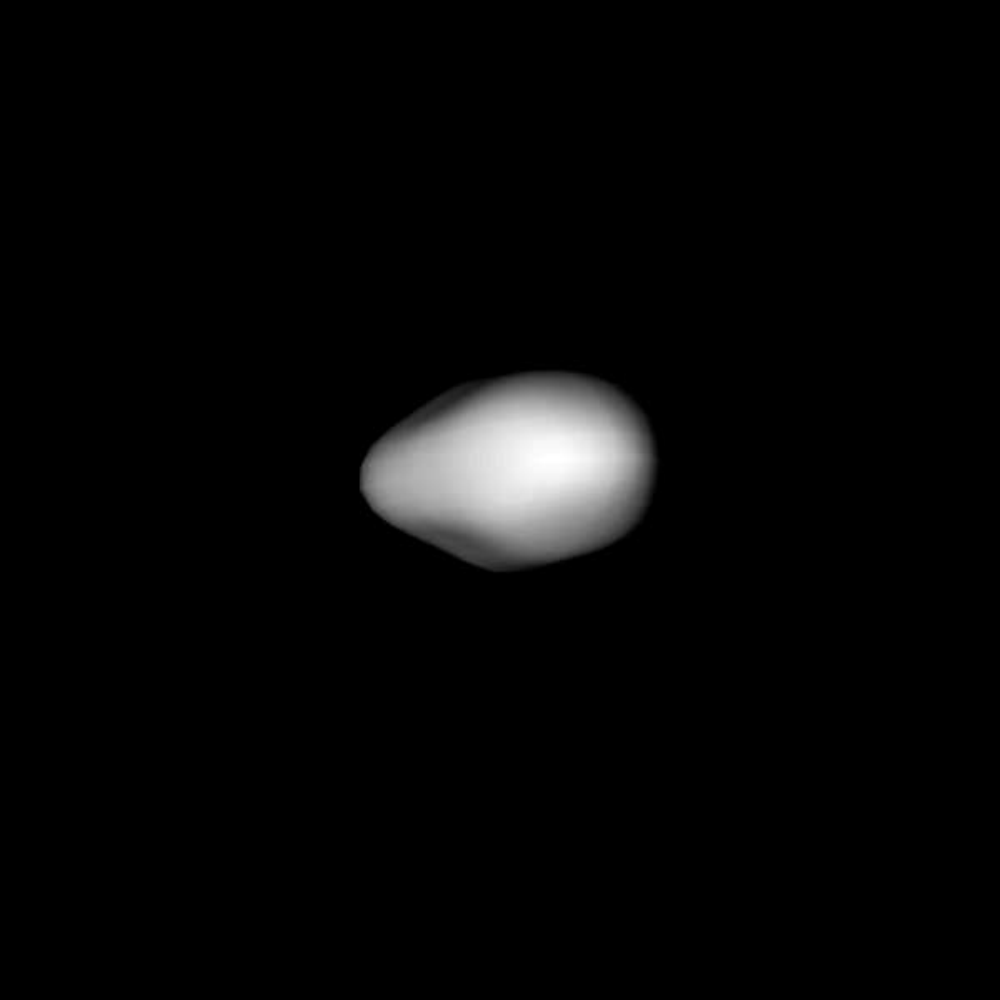}}\resizebox{0.24\hsize}{!}{\includegraphics{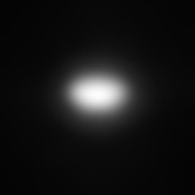}}\resizebox{0.24\hsize}{!}{\includegraphics{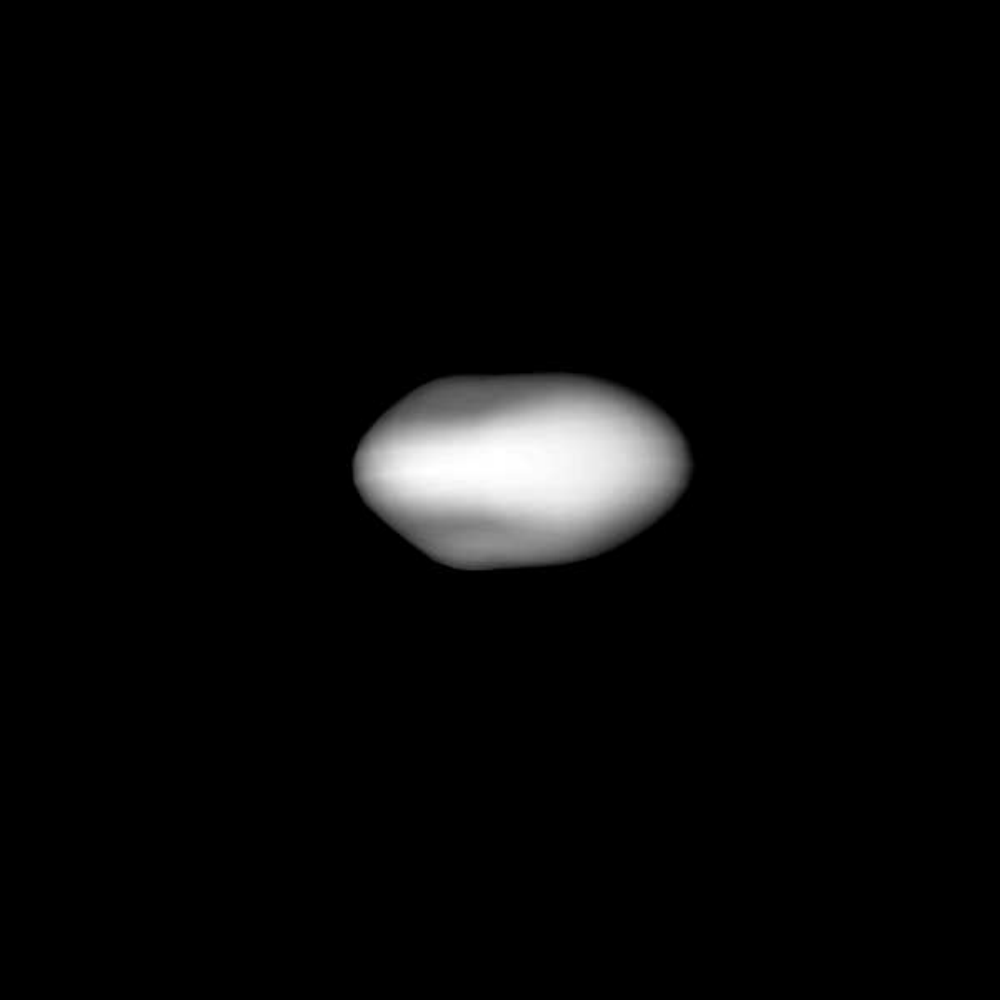}}\\
        \resizebox{0.24\hsize}{!}{\includegraphics{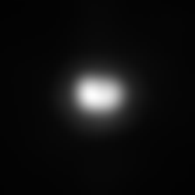}}\resizebox{0.24\hsize}{!}{\includegraphics{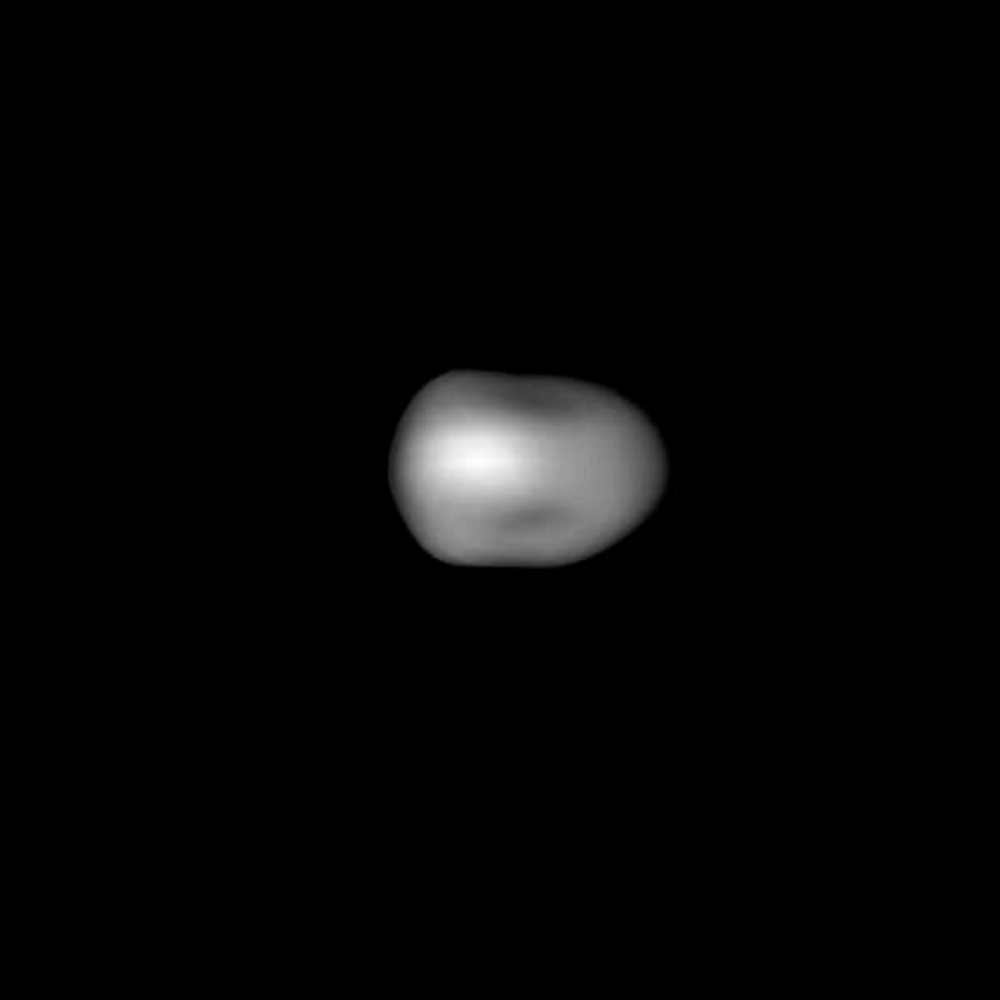}}\resizebox{0.24\hsize}{!}{\includegraphics{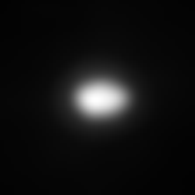}}\resizebox{0.24\hsize}{!}{\includegraphics{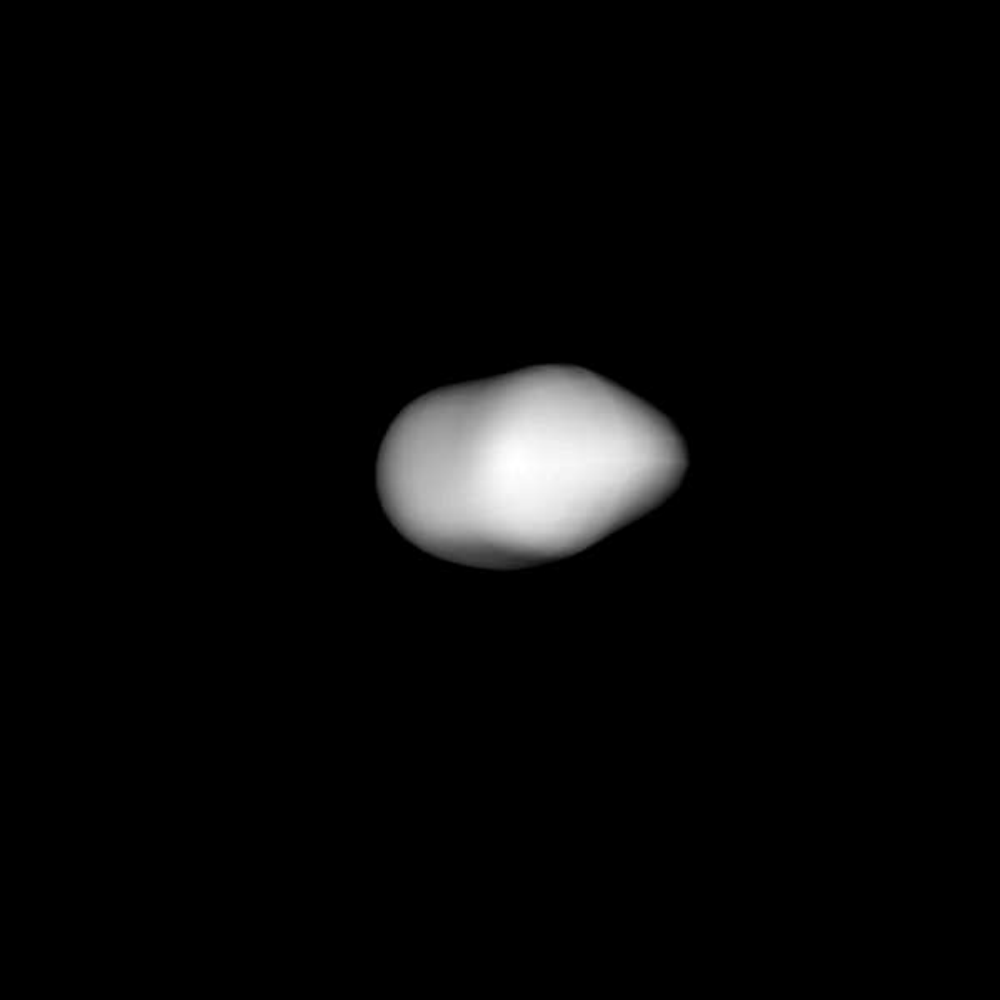}}\\
        \resizebox{0.24\hsize}{!}{\includegraphics{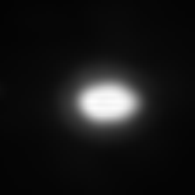}}\resizebox{0.24\hsize}{!}{\includegraphics{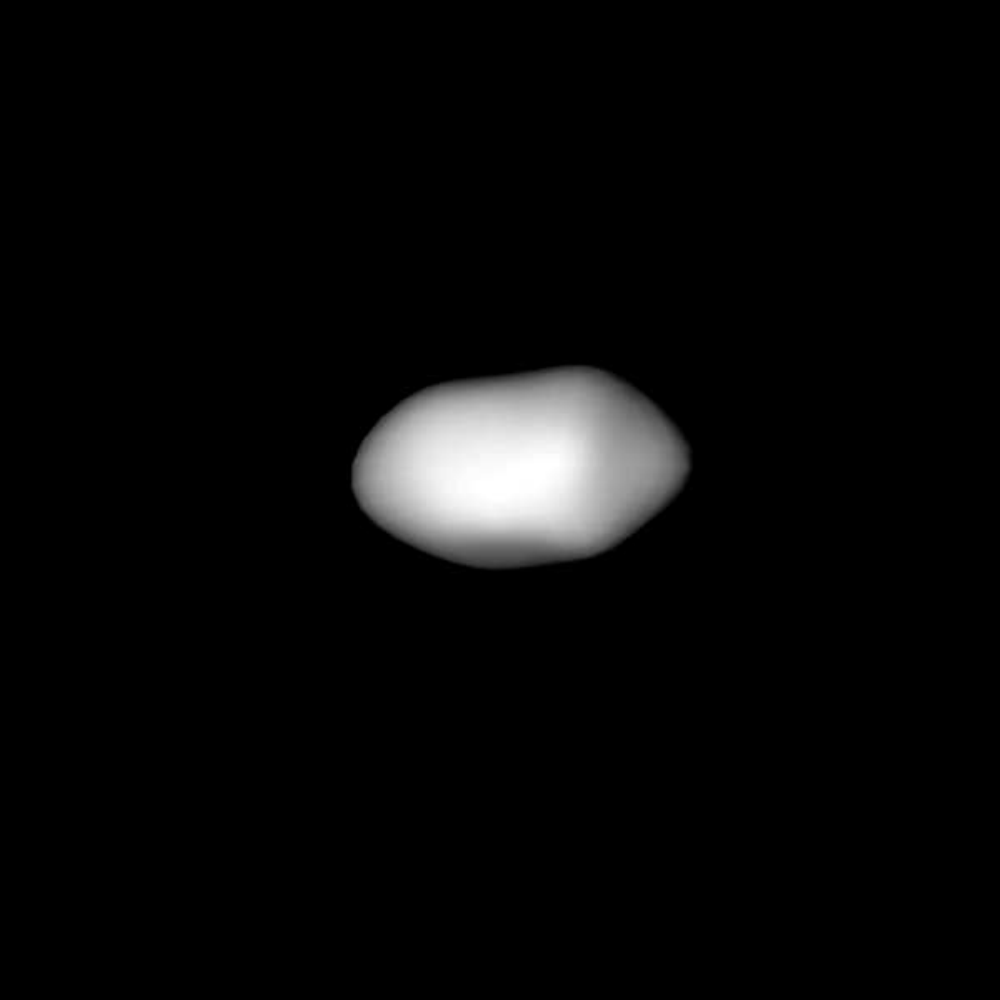}}\resizebox{0.24\hsize}{!}{\includegraphics{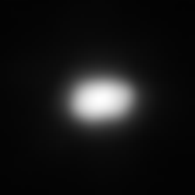}}\resizebox{0.24\hsize}{!}{\includegraphics{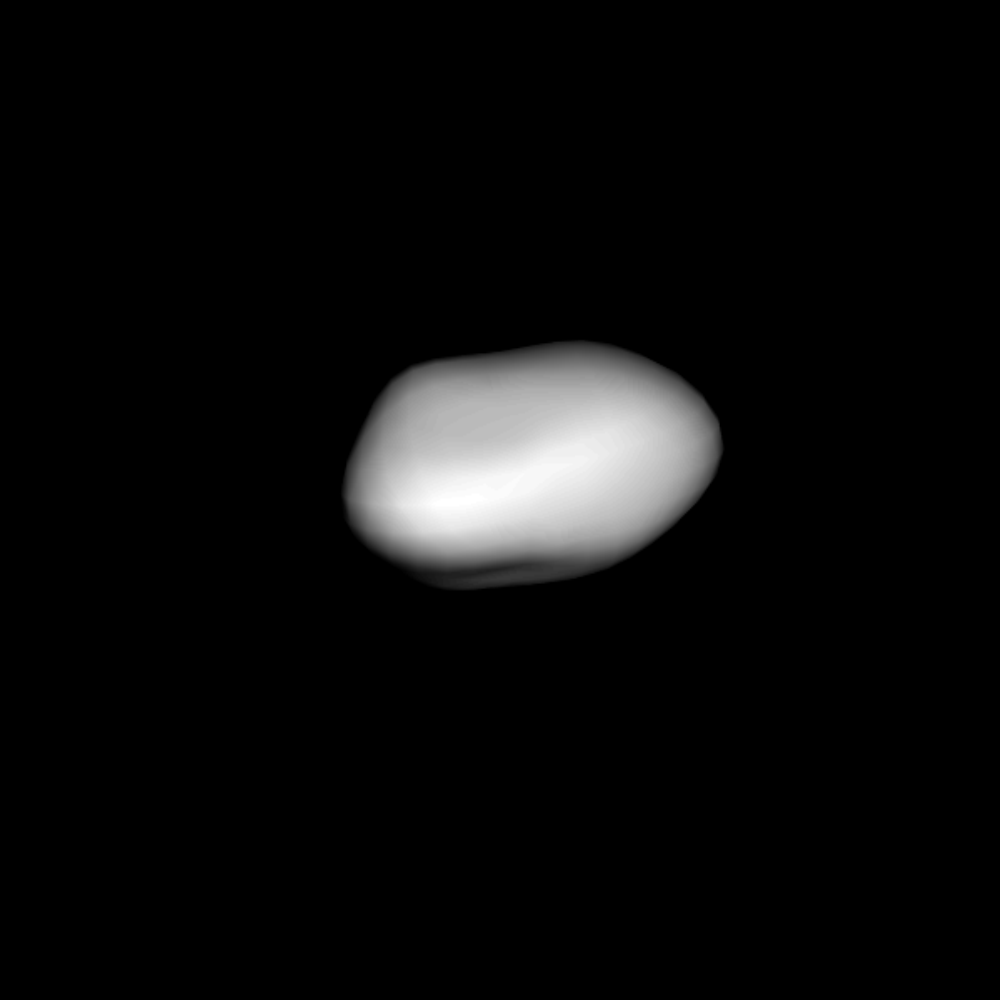}}\\
    \caption{\label{fig:87a}Comparison between model projections and corresponding AO images for asteroid (87) Sylvia (first part).}
\end{figure}

\begin{figure}[tbp]
    \centering
        \resizebox{0.24\hsize}{!}{\includegraphics{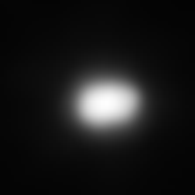}}\resizebox{0.24\hsize}{!}{\includegraphics{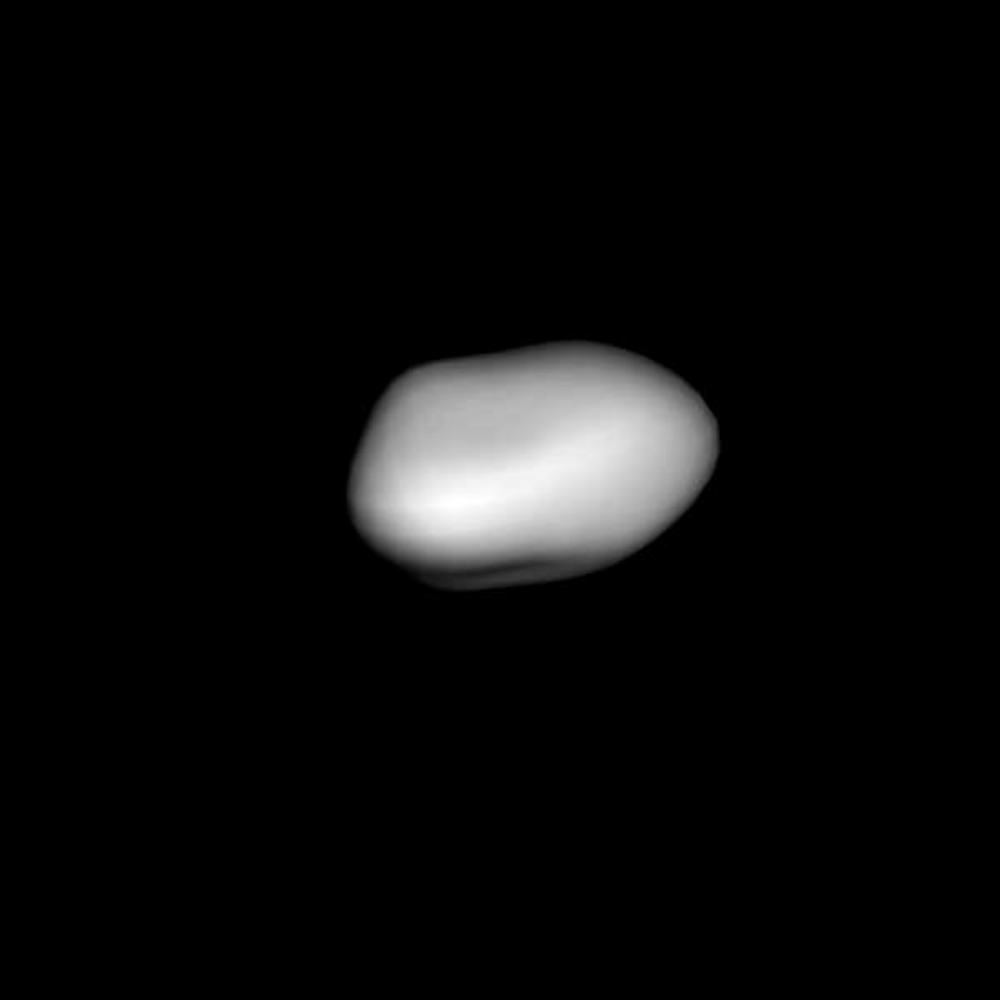}}\resizebox{0.24\hsize}{!}{\includegraphics{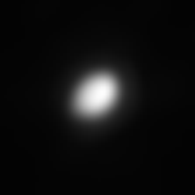}}\resizebox{0.24\hsize}{!}{\includegraphics{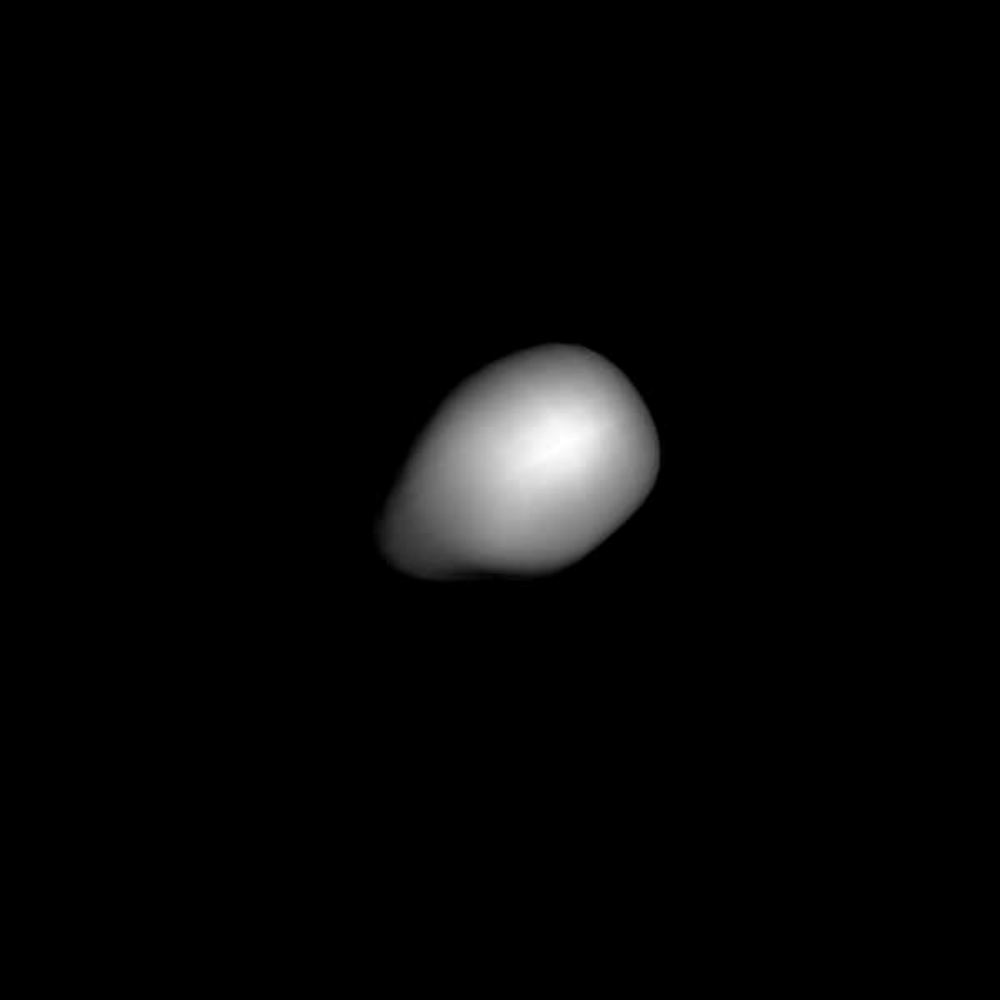}}\\
        \resizebox{0.24\hsize}{!}{\includegraphics{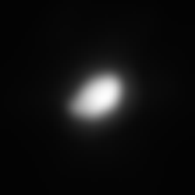}}\resizebox{0.24\hsize}{!}{\includegraphics{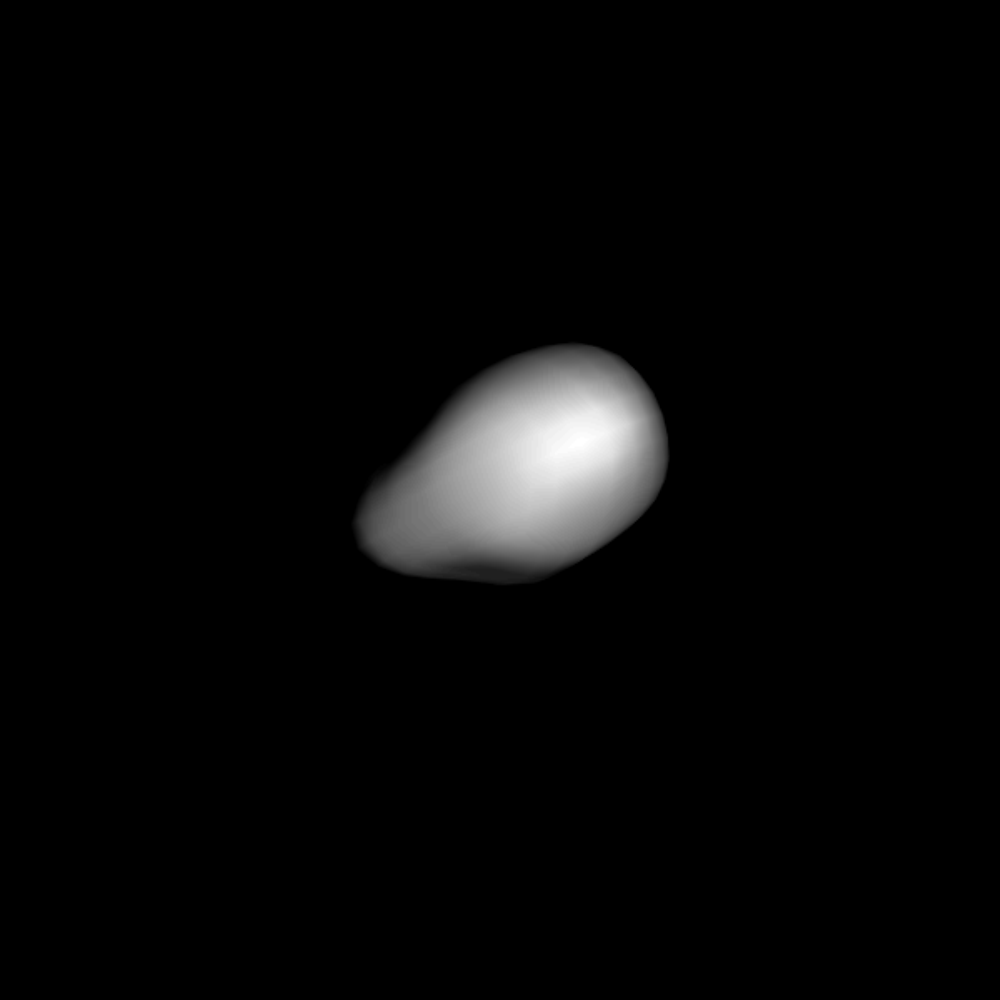}}\resizebox{0.24\hsize}{!}{\includegraphics{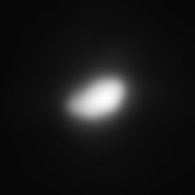}}\resizebox{0.24\hsize}{!}{\includegraphics{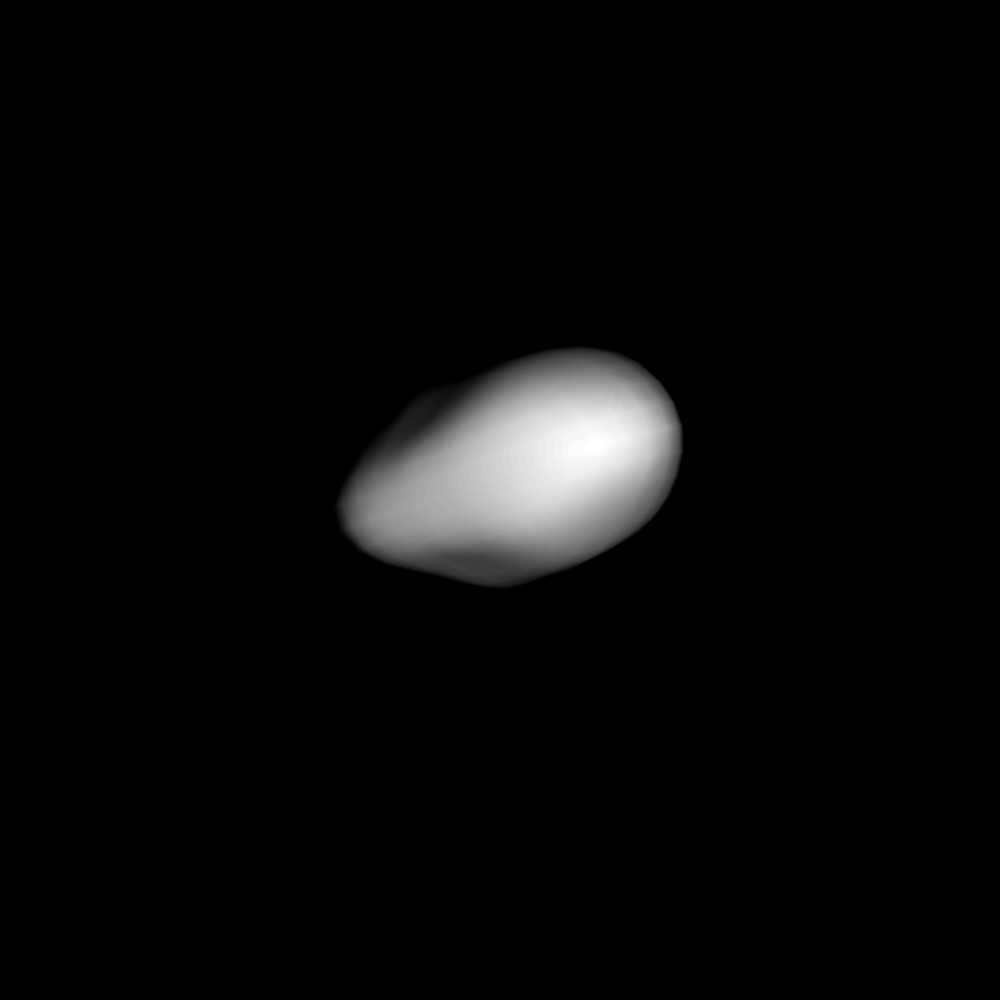}}\\
        \resizebox{0.24\hsize}{!}{\includegraphics{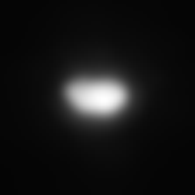}}\resizebox{0.24\hsize}{!}{\includegraphics{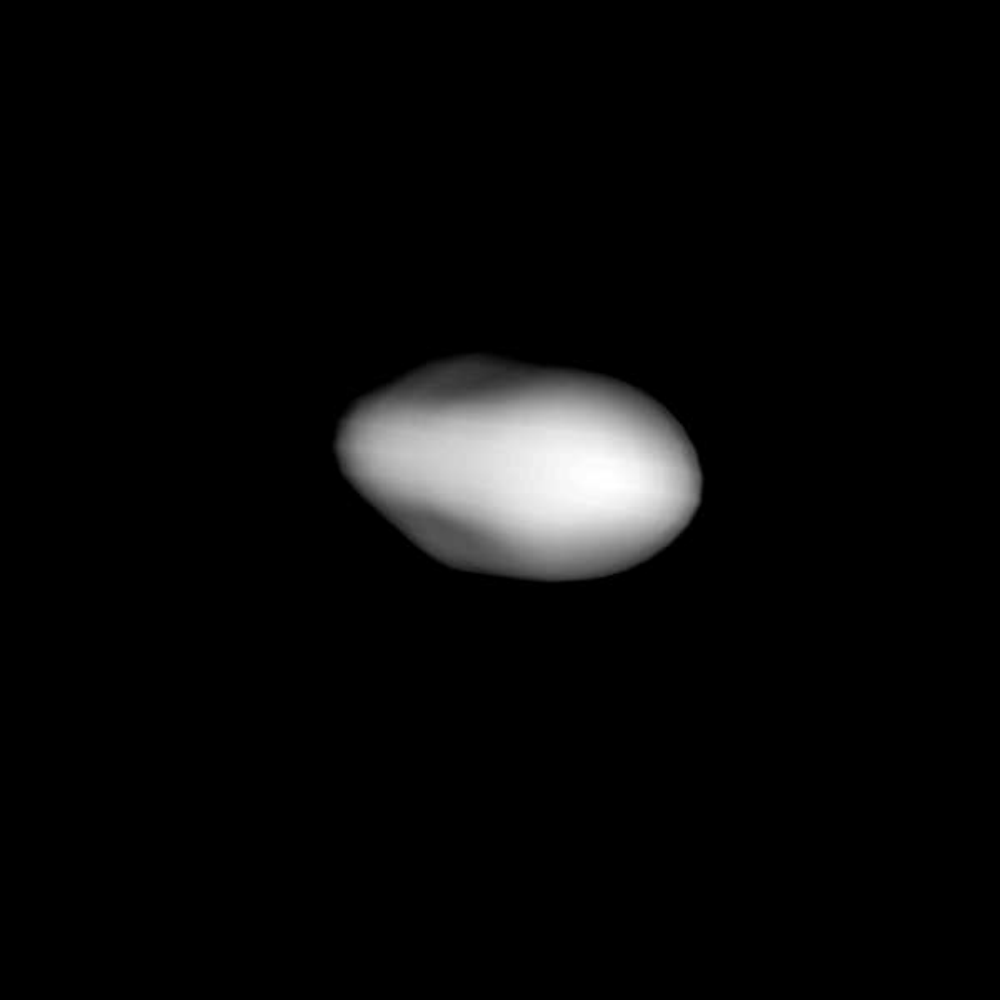}}\resizebox{0.24\hsize}{!}{\includegraphics{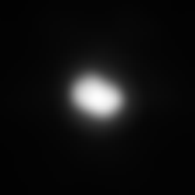}}\resizebox{0.24\hsize}{!}{\includegraphics{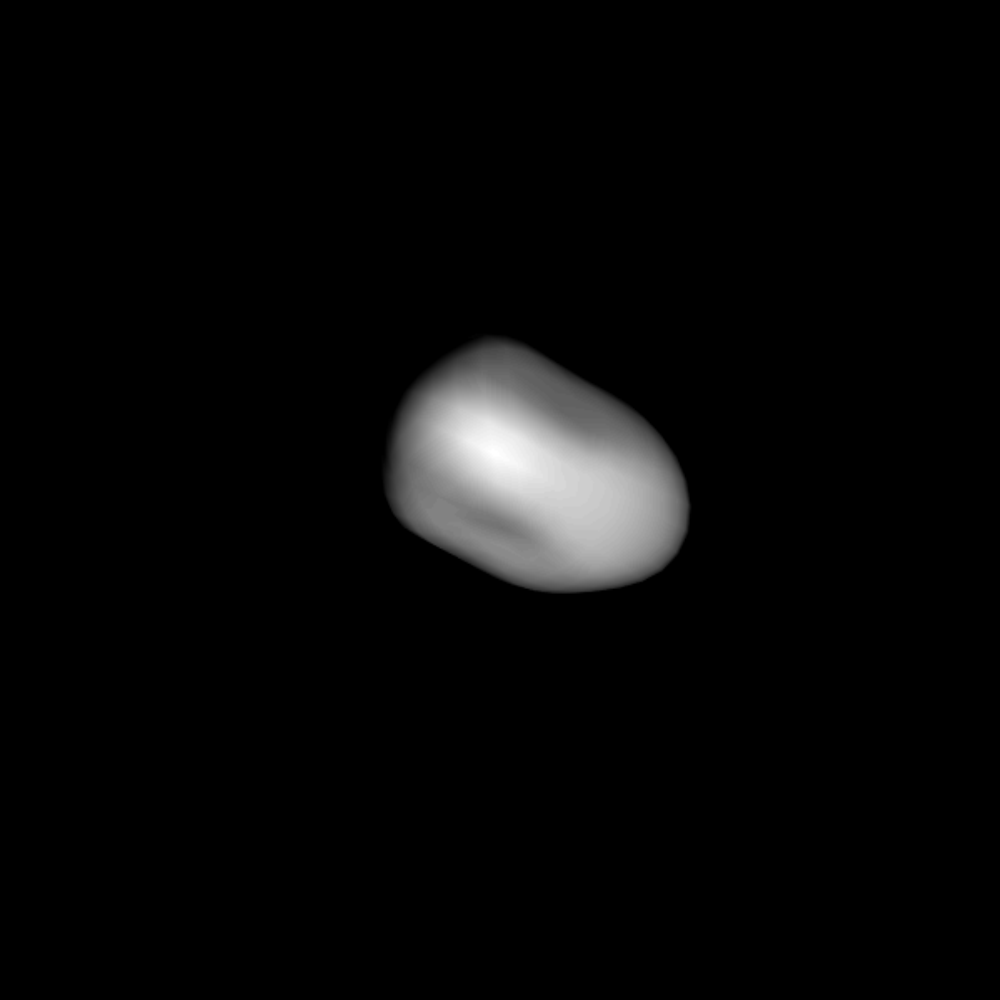}}\\
        \resizebox{0.24\hsize}{!}{\includegraphics{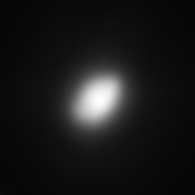}}\resizebox{0.24\hsize}{!}{\includegraphics{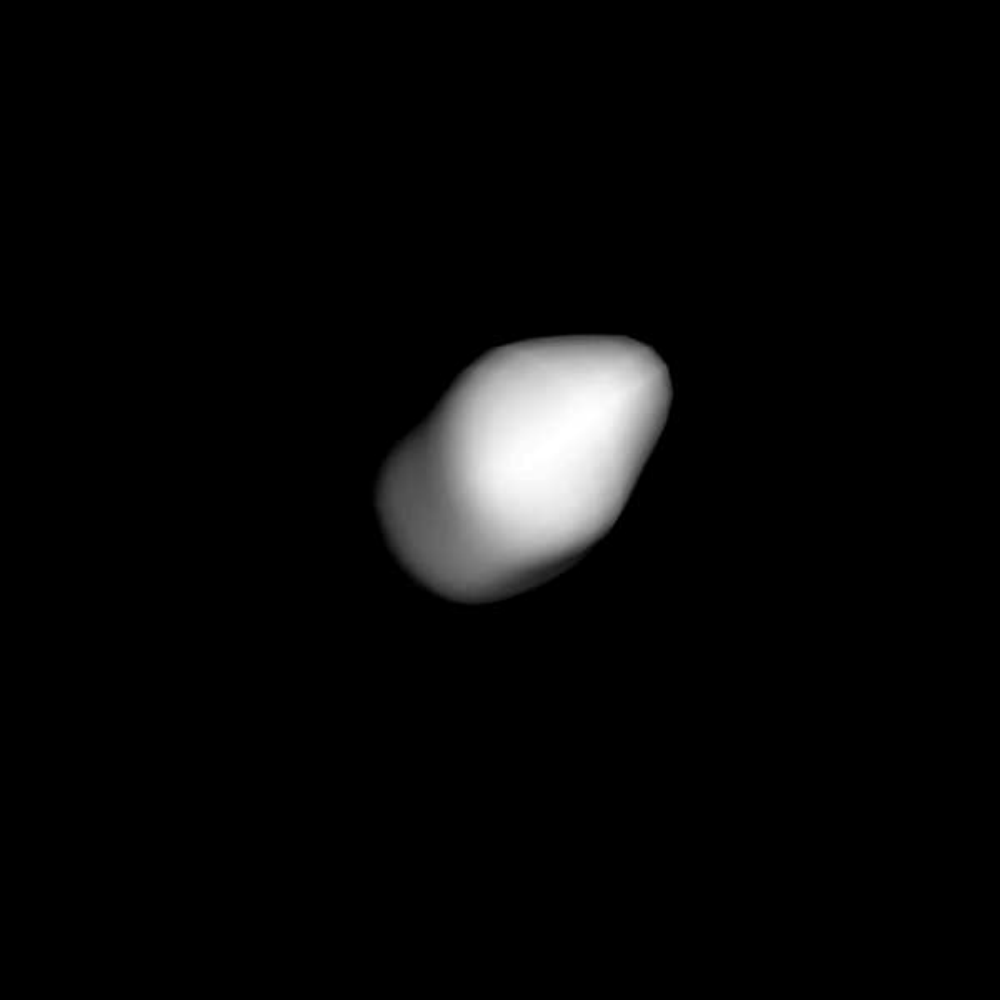}}\resizebox{0.24\hsize}{!}{\includegraphics{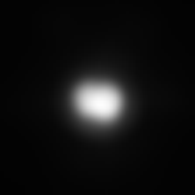}}\resizebox{0.24\hsize}{!}{\includegraphics{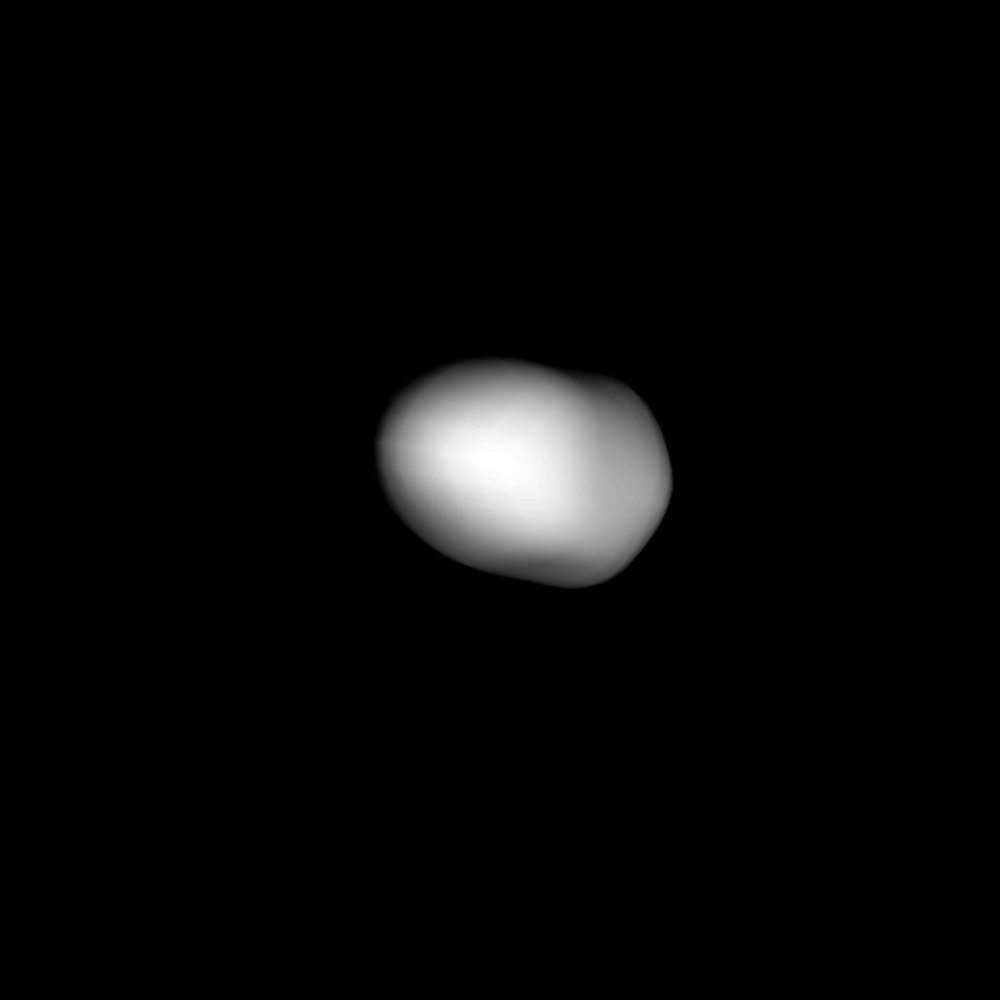}}\\
    \caption{\label{fig:87b}Comparison between model projections and corresponding AO images for asteroid (87) Sylvia (second part).}
\end{figure}

\begin{figure}[tbp]
    \centering
        \resizebox{0.24\hsize}{!}{\includegraphics{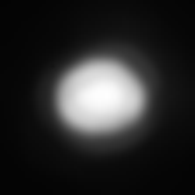}}\resizebox{0.24\hsize}{!}{\includegraphics{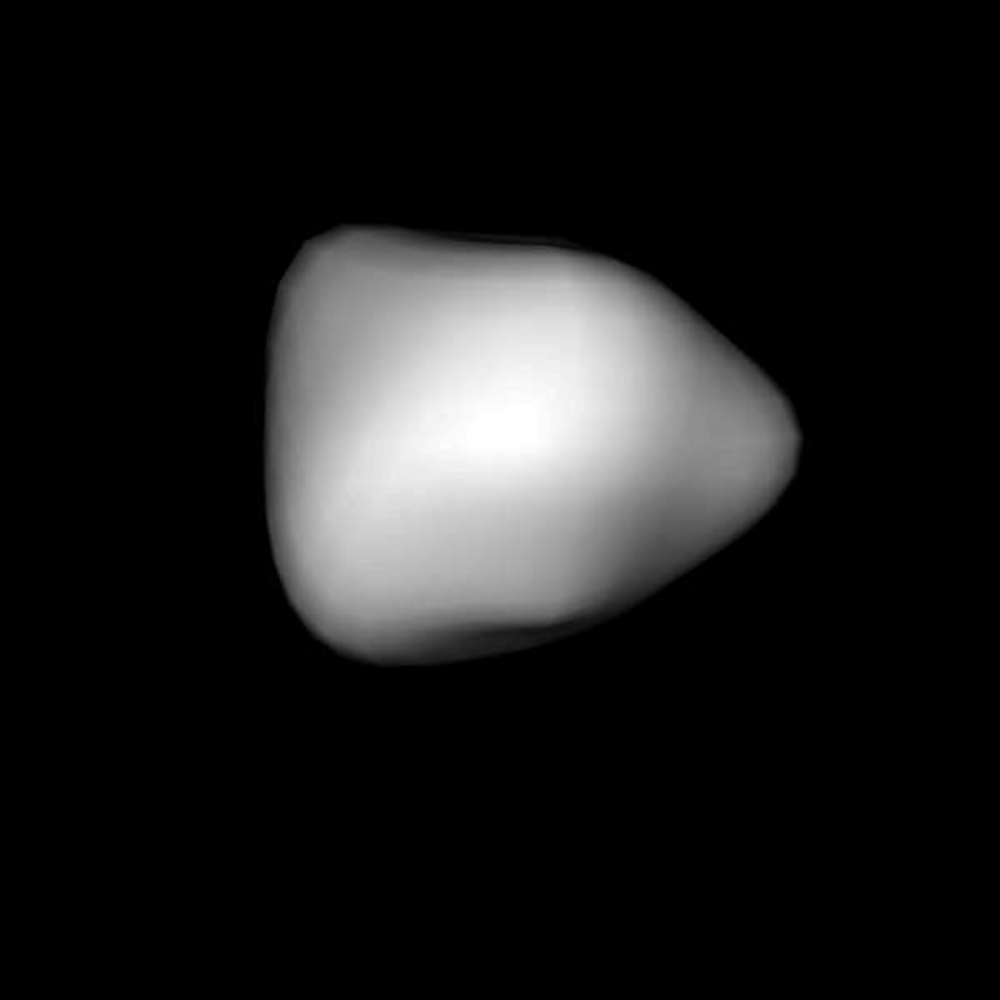}}\resizebox{0.24\hsize}{!}{\includegraphics{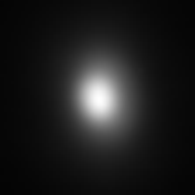}}\resizebox{0.24\hsize}{!}{\includegraphics{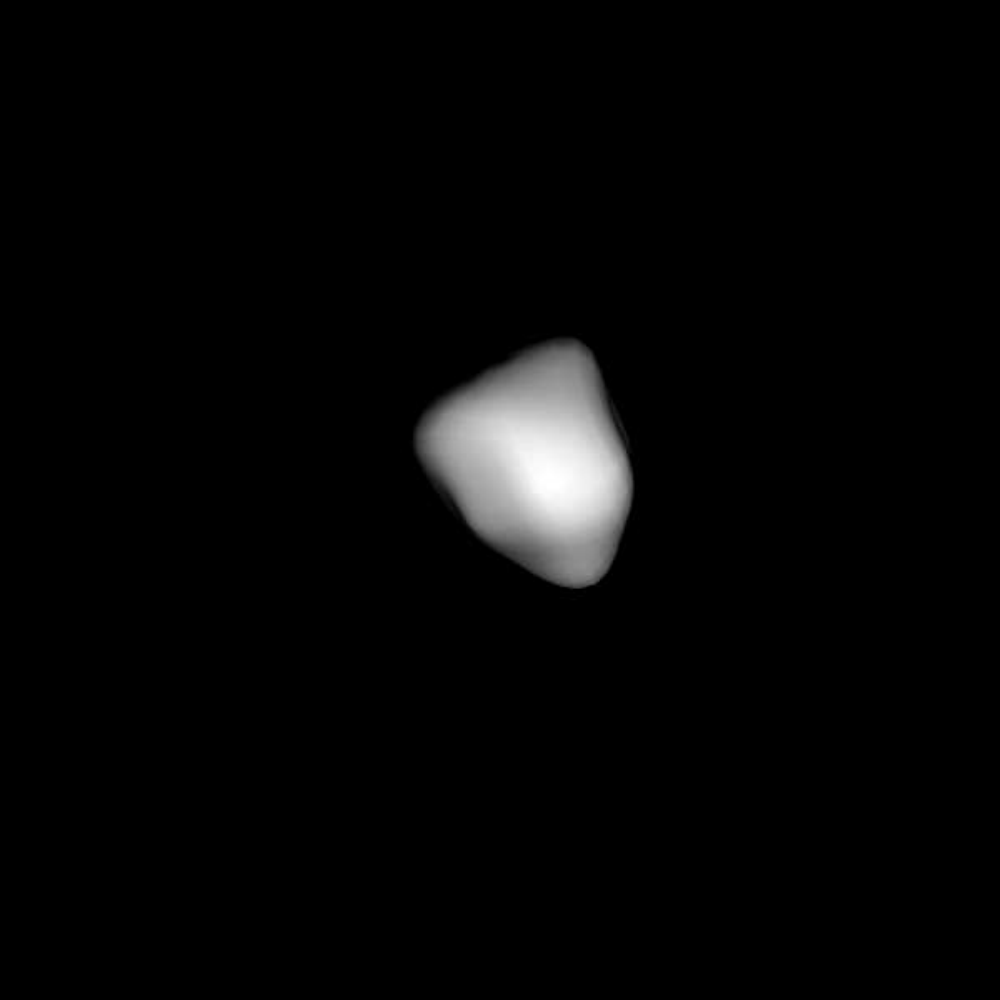}}\\
    \caption{\label{fig:88}Comparison between model projections and corresponding AO images for asteroid (88) Thisbe.}
\end{figure}

\begin{figure}[tbp]
    \centering
        \resizebox{0.24\hsize}{!}{\includegraphics{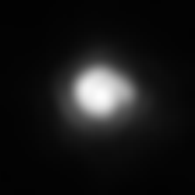}}\resizebox{0.24\hsize}{!}{\includegraphics{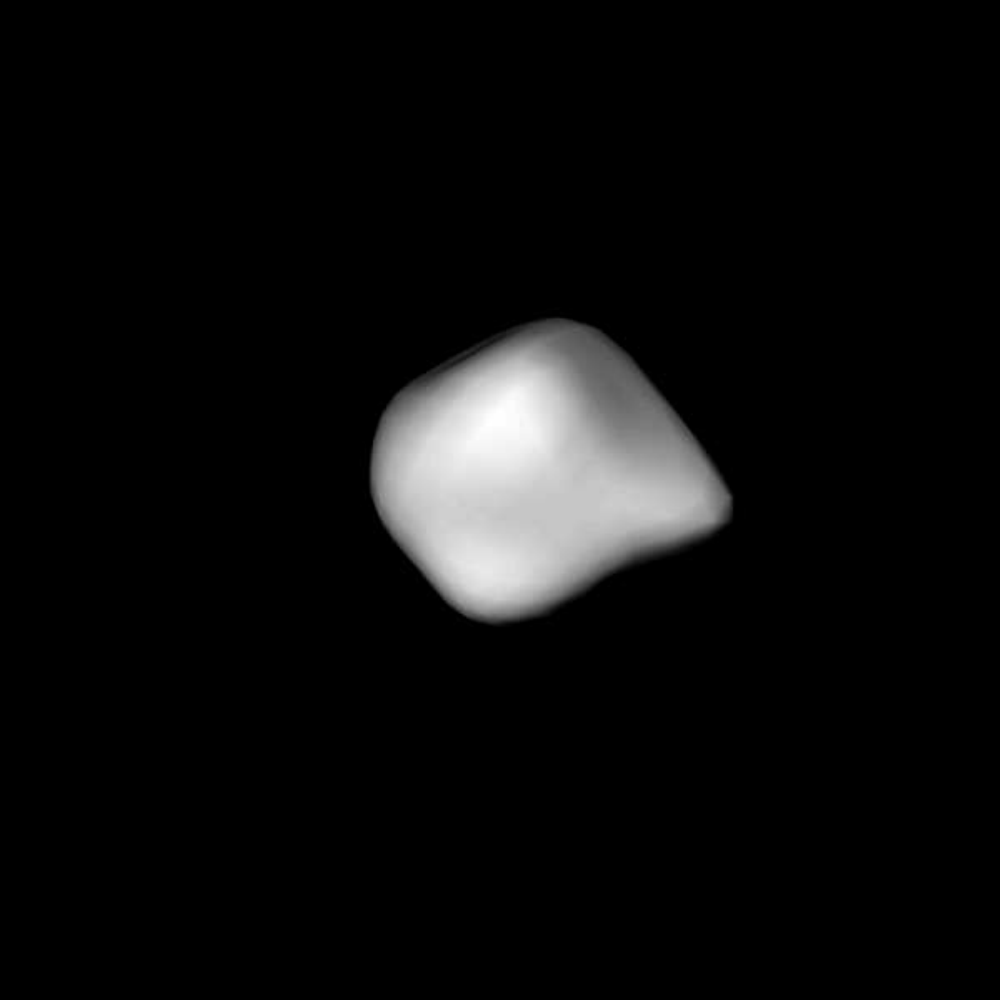}}\\
    \caption{\label{fig:89}Comparison between model projections and corresponding AO images for asteroid (89) Julia.}
\end{figure}

\begin{figure}[tbp]
    \centering
        \resizebox{0.24\hsize}{!}{\includegraphics{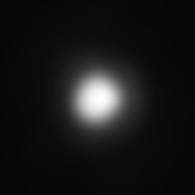}}\resizebox{0.24\hsize}{!}{\includegraphics{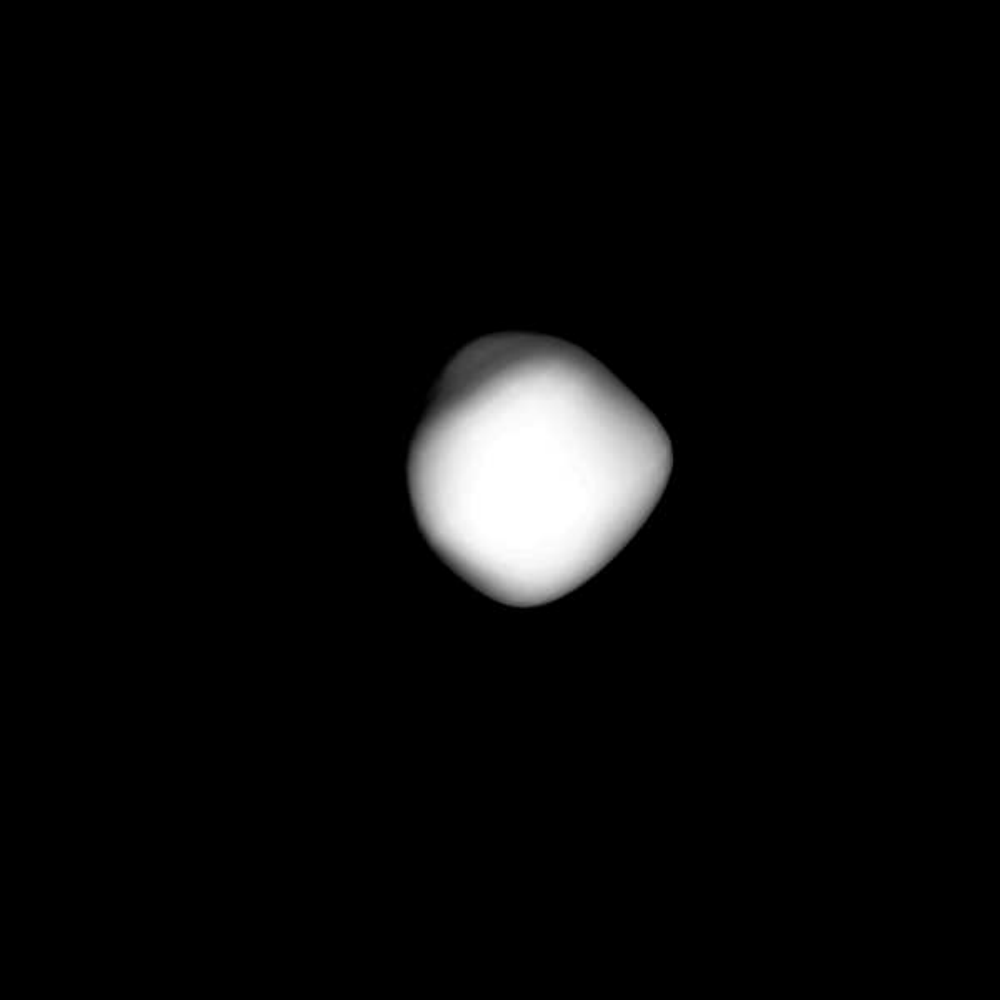}}\resizebox{0.24\hsize}{!}{\includegraphics{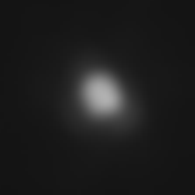}}\resizebox{0.24\hsize}{!}{\includegraphics{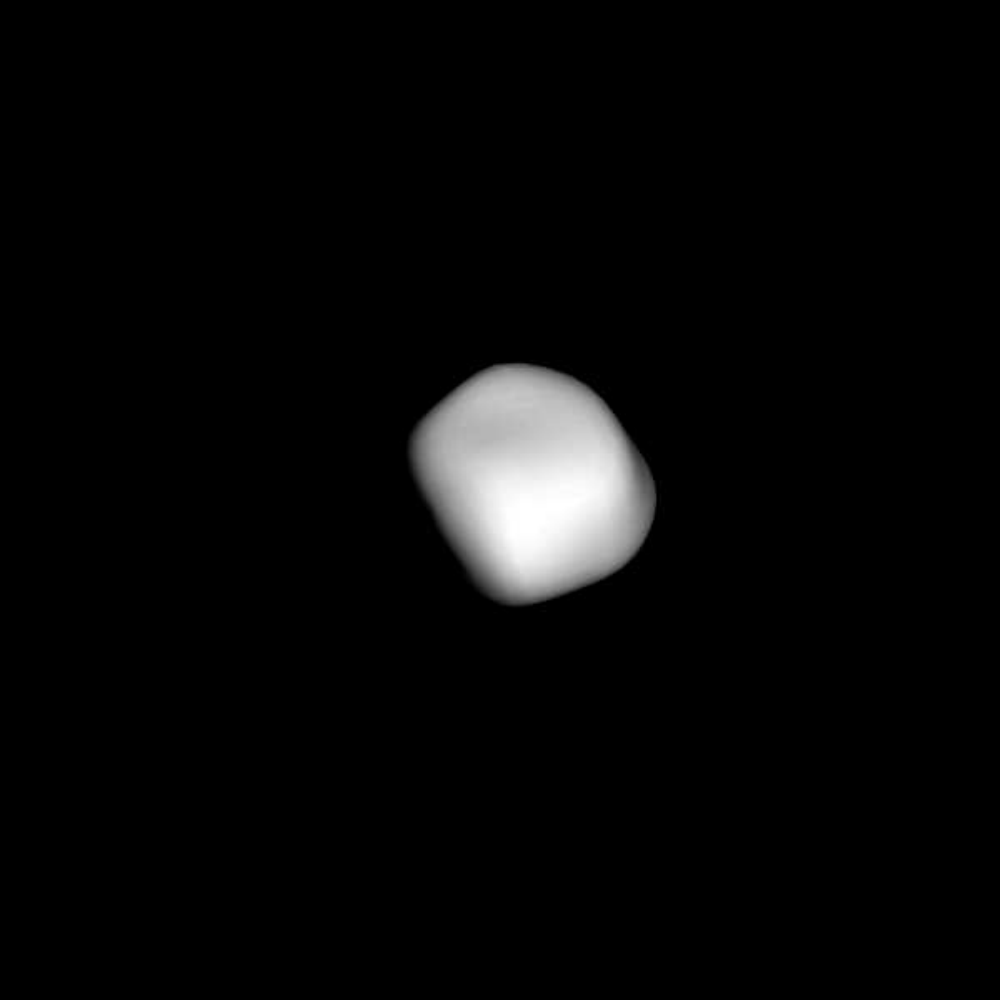}}\\
        \resizebox{0.24\hsize}{!}{\includegraphics{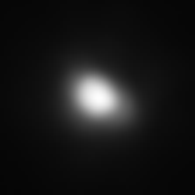}}\resizebox{0.24\hsize}{!}{\includegraphics{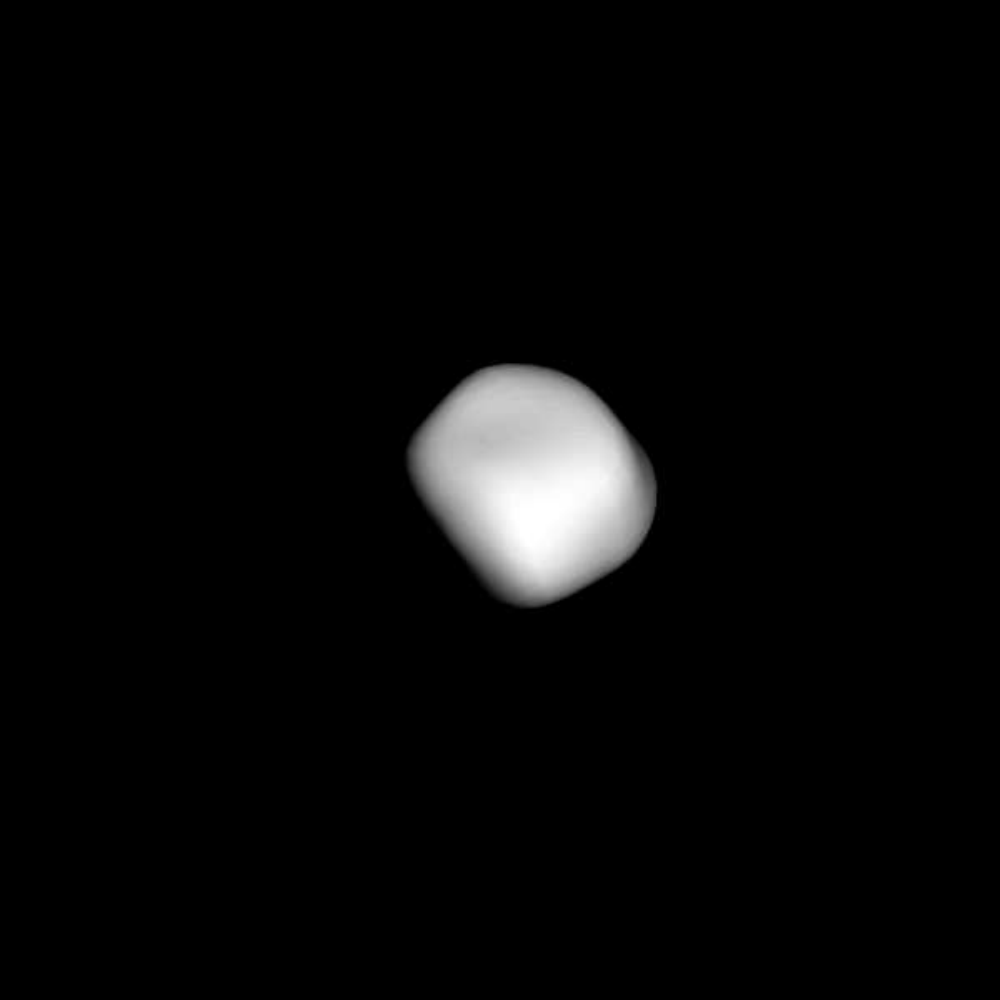}}\resizebox{0.24\hsize}{!}{\includegraphics{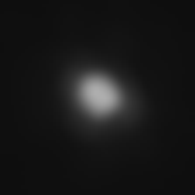}}\resizebox{0.24\hsize}{!}{\includegraphics{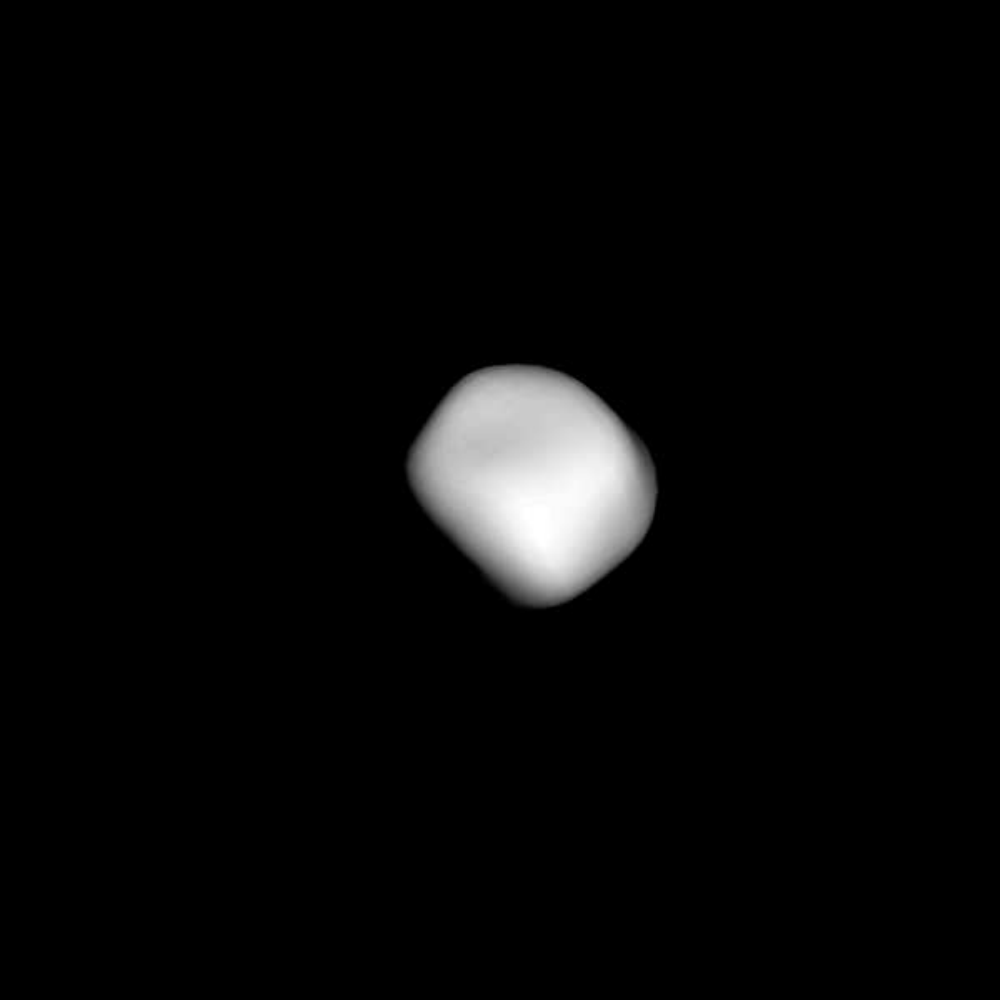}}\\
    \caption{\label{fig:93}Comparison between model projections and corresponding AO images for asteroid (93) Minerva.}
\end{figure}

\begin{figure}[tbp]
    \centering
        \resizebox{0.24\hsize}{!}{\includegraphics{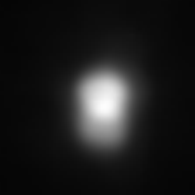}}\resizebox{0.24\hsize}{!}{\includegraphics{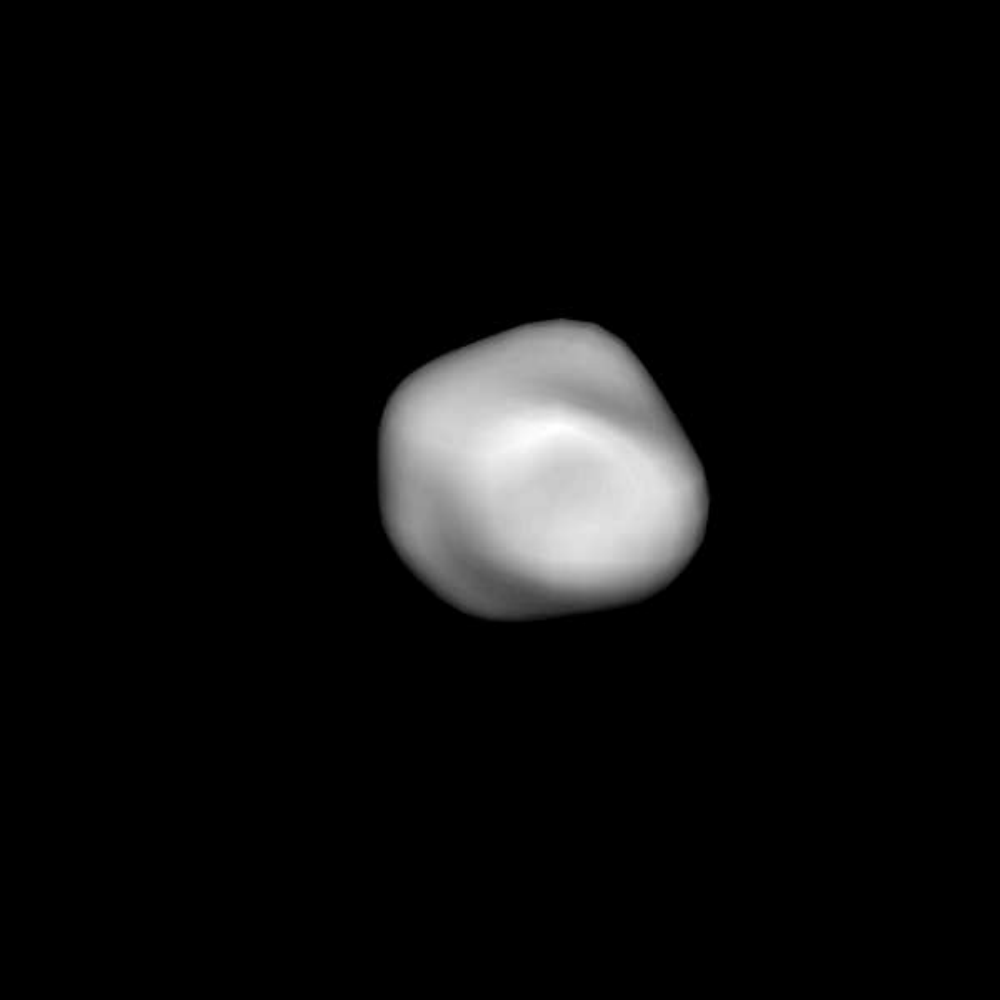}}\resizebox{0.24\hsize}{!}{\includegraphics{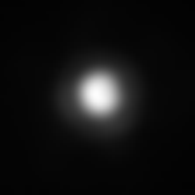}}\resizebox{0.24\hsize}{!}{\includegraphics{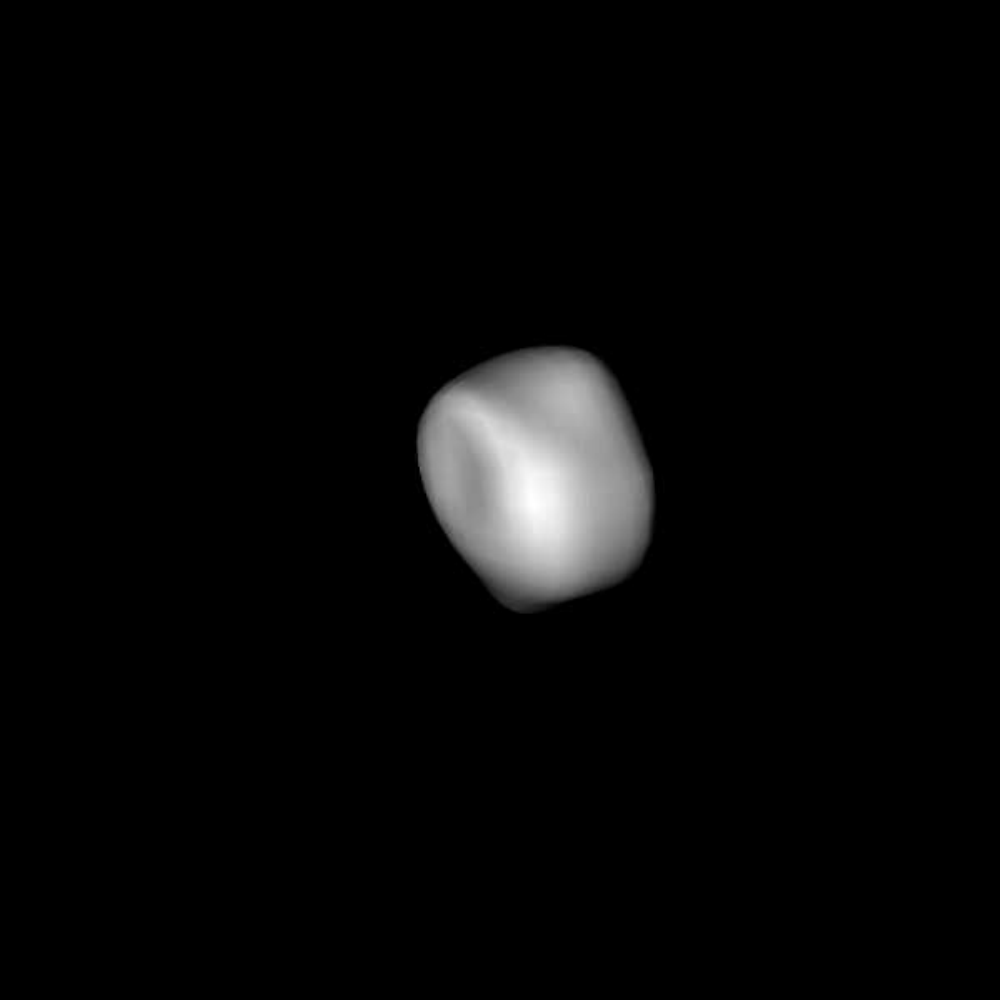}}\\
    \caption{\label{fig:94}Comparison between model projections and corresponding AO images for model 1 of asteroid (94) Aurora.}
\end{figure}

\clearpage

\begin{figure}[tbp]
    \centering
        \resizebox{0.24\hsize}{!}{\includegraphics{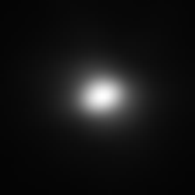}}\resizebox{0.24\hsize}{!}{\includegraphics{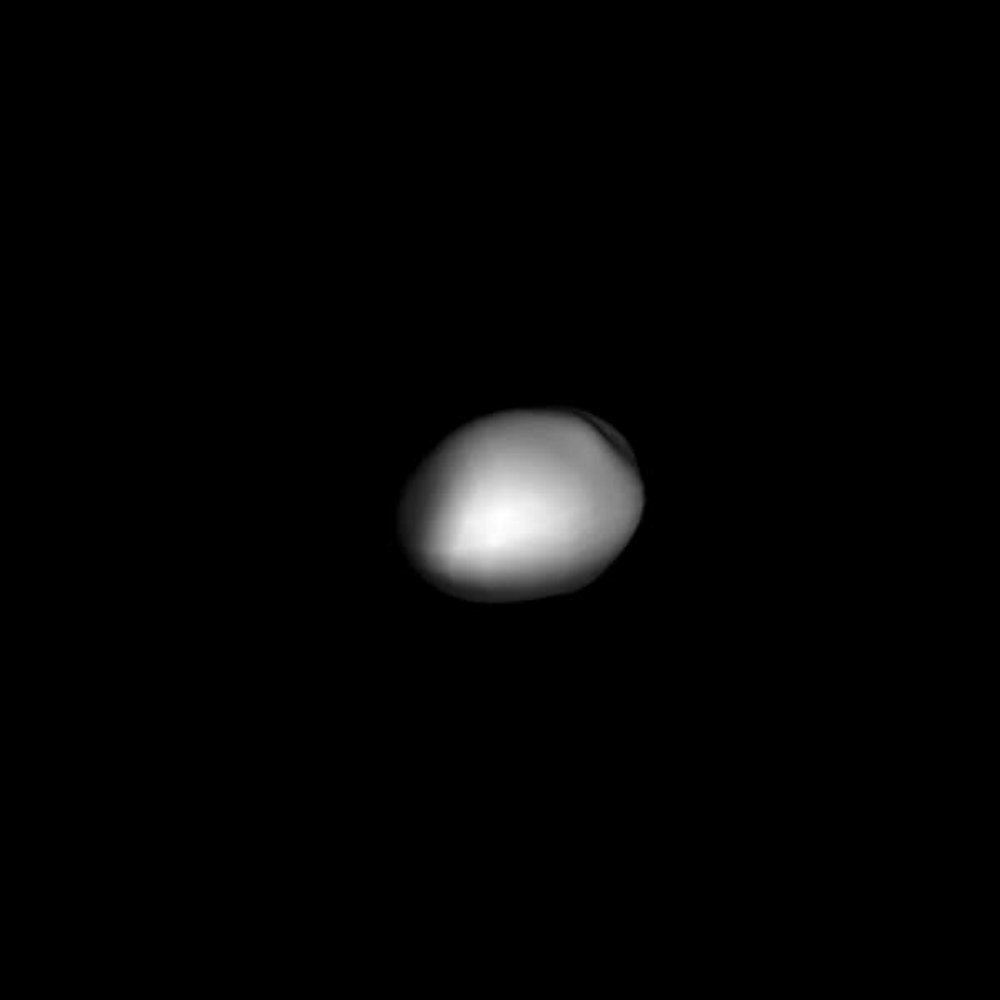}}\resizebox{0.24\hsize}{!}{\includegraphics{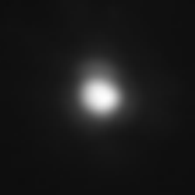}}\resizebox{0.24\hsize}{!}{\includegraphics{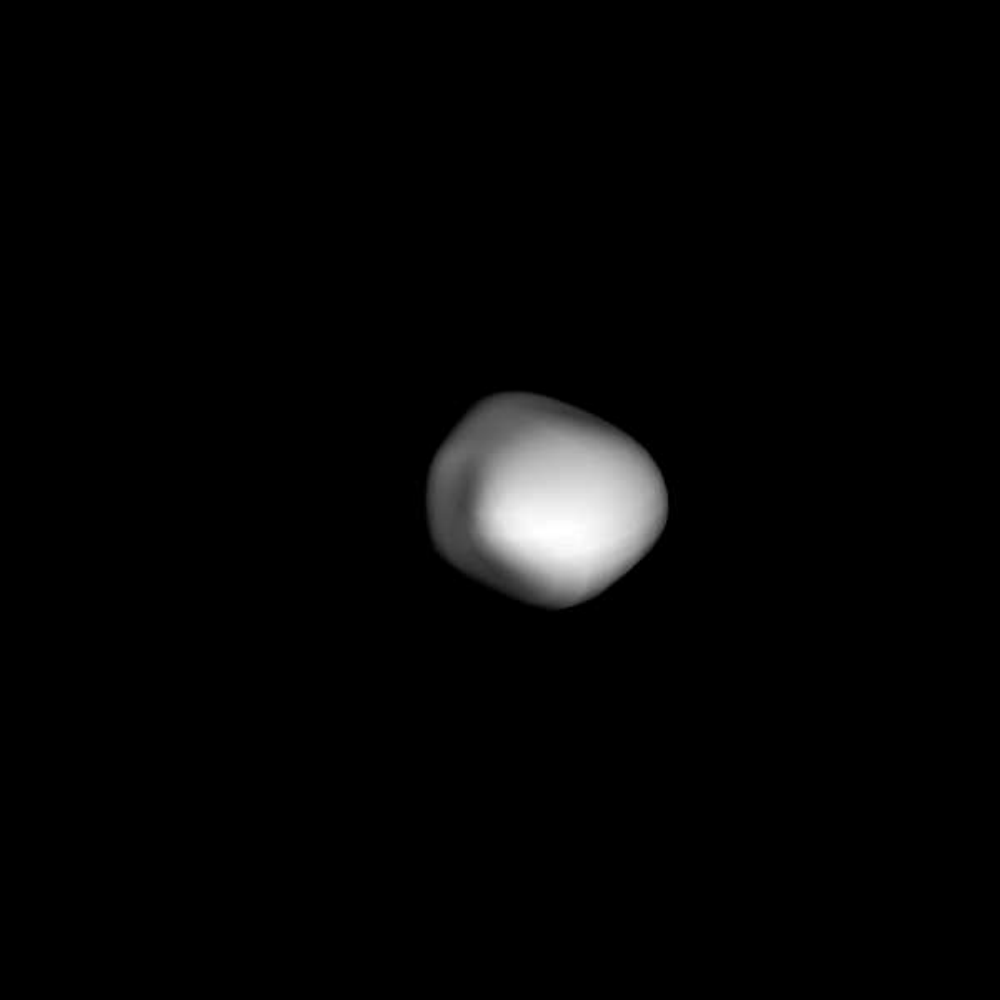}}\\
        \resizebox{0.24\hsize}{!}{\includegraphics{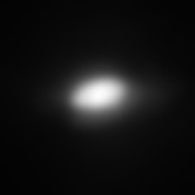}}\resizebox{0.24\hsize}{!}{\includegraphics{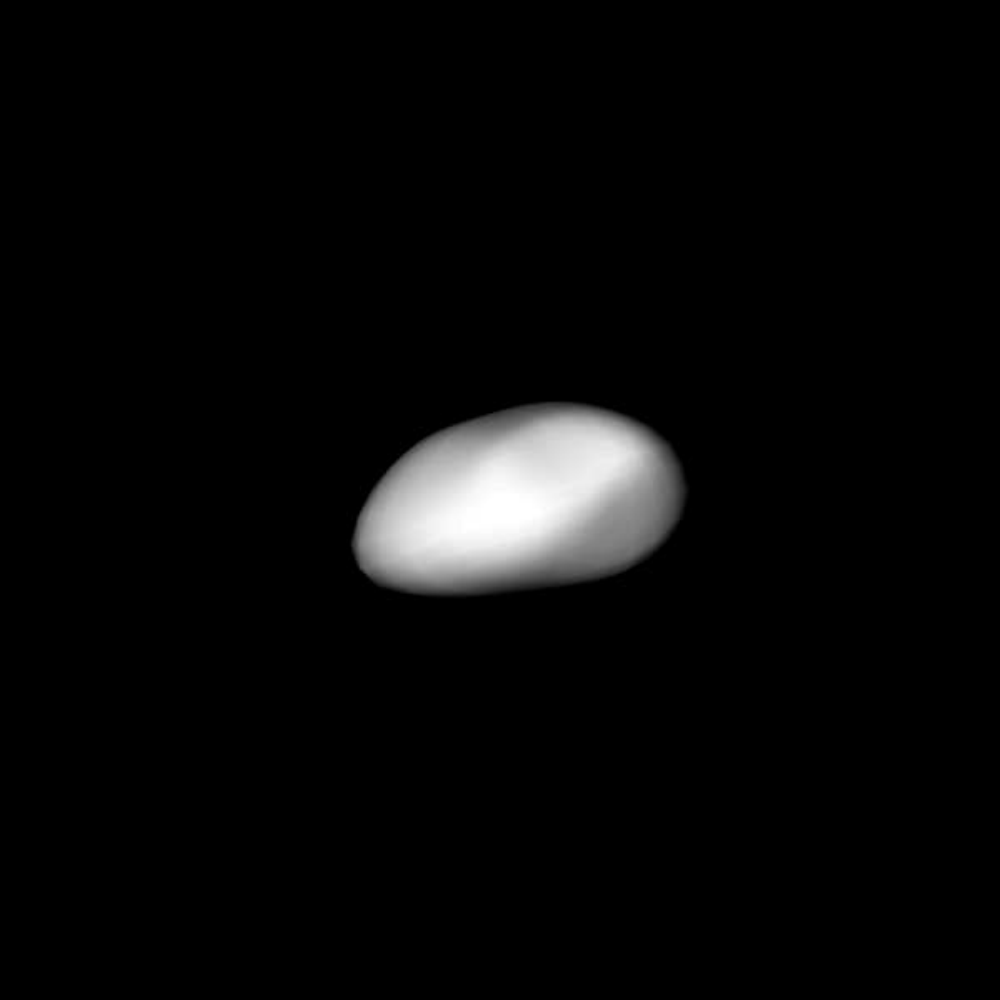}}\resizebox{0.24\hsize}{!}{\includegraphics{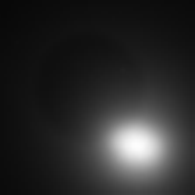}}\resizebox{0.24\hsize}{!}{\includegraphics{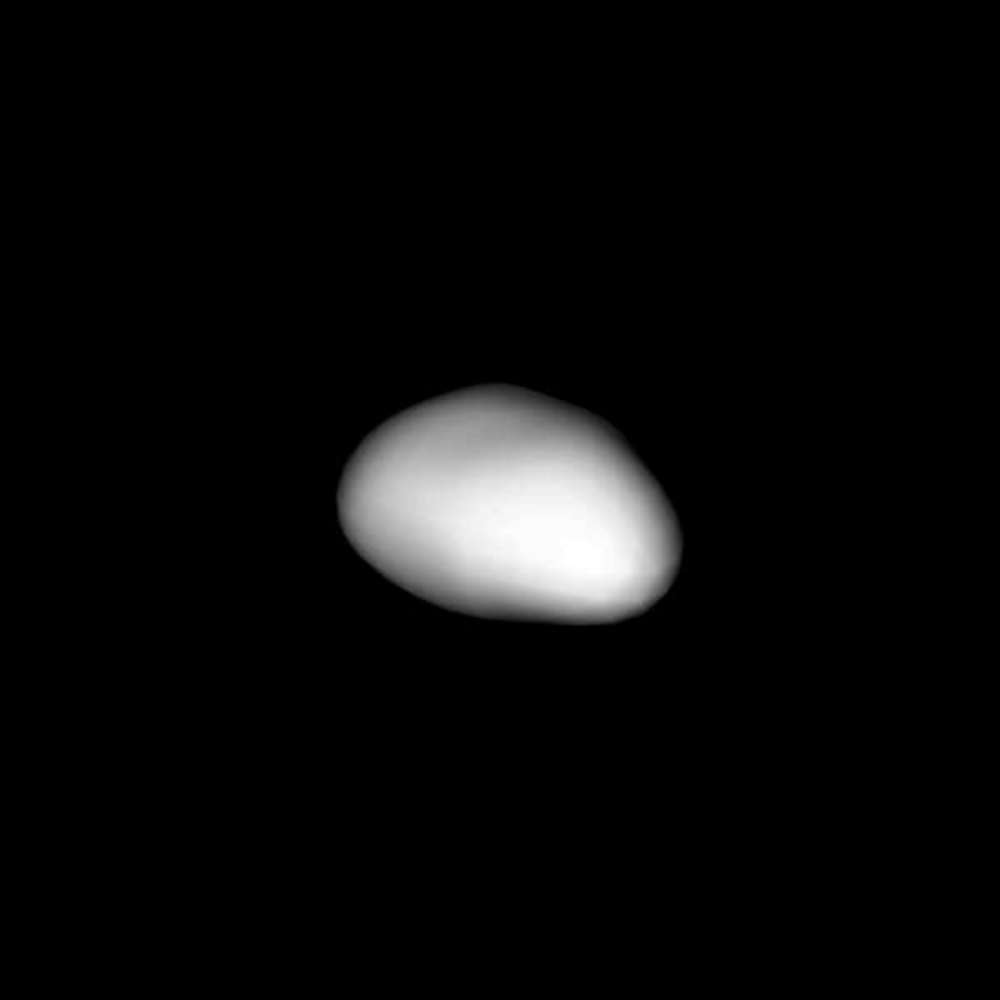}}\\
        \resizebox{0.24\hsize}{!}{\includegraphics{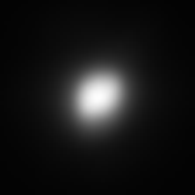}}\resizebox{0.24\hsize}{!}{\includegraphics{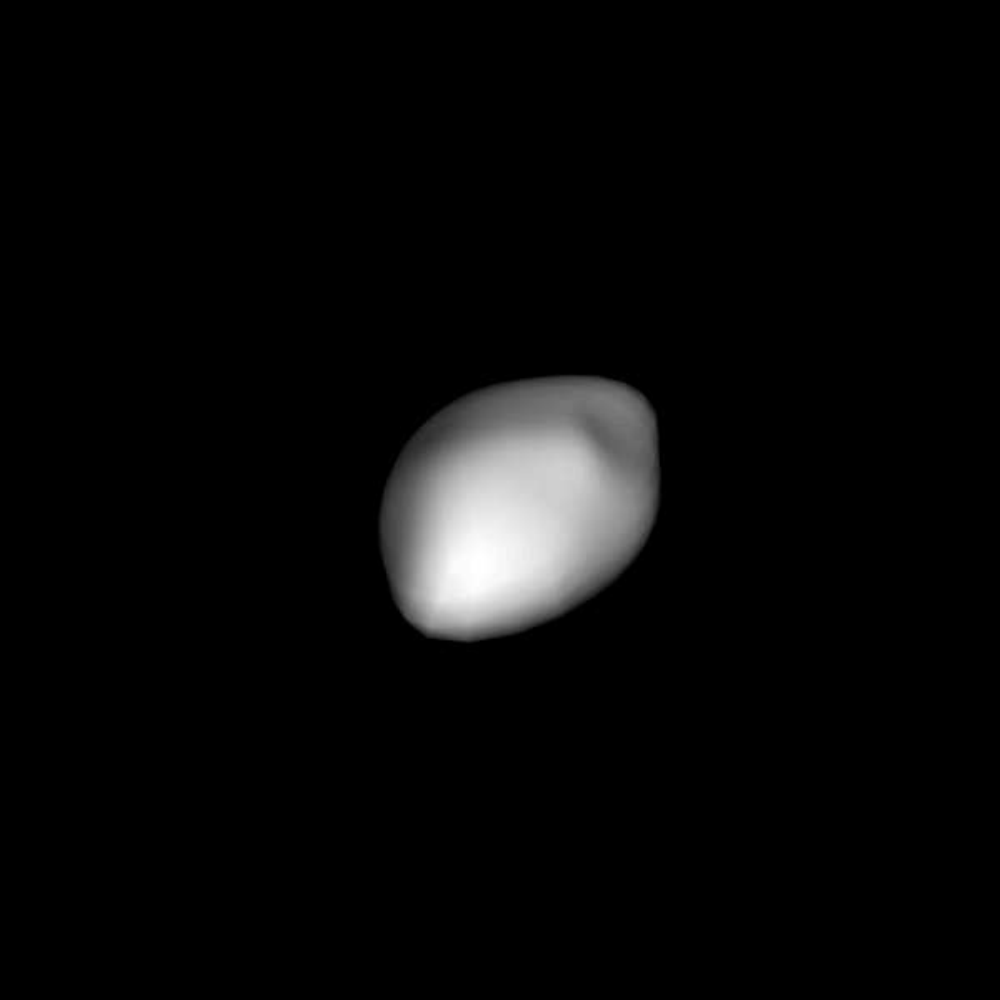}}\resizebox{0.24\hsize}{!}{\includegraphics{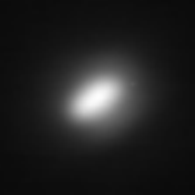}}\resizebox{0.24\hsize}{!}{\includegraphics{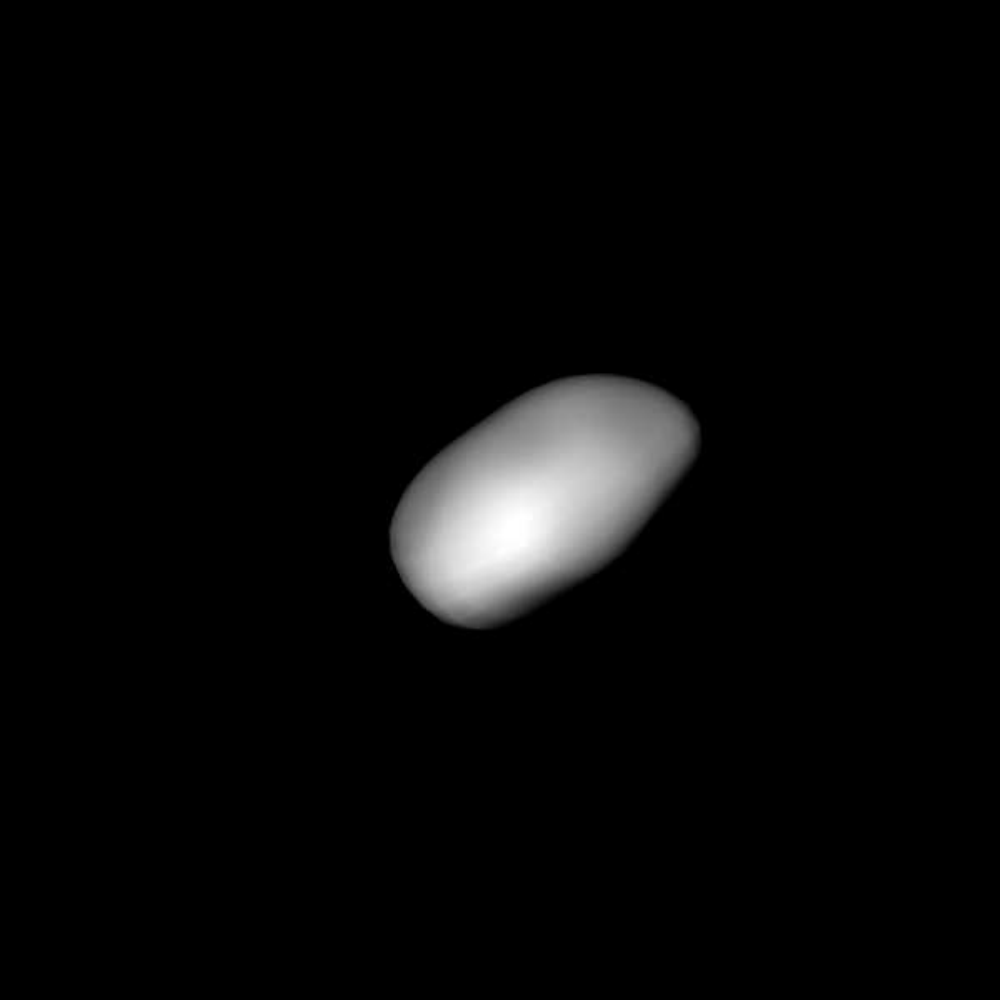}}\\
        \resizebox{0.24\hsize}{!}{\includegraphics{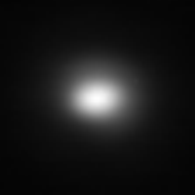}}\resizebox{0.24\hsize}{!}{\includegraphics{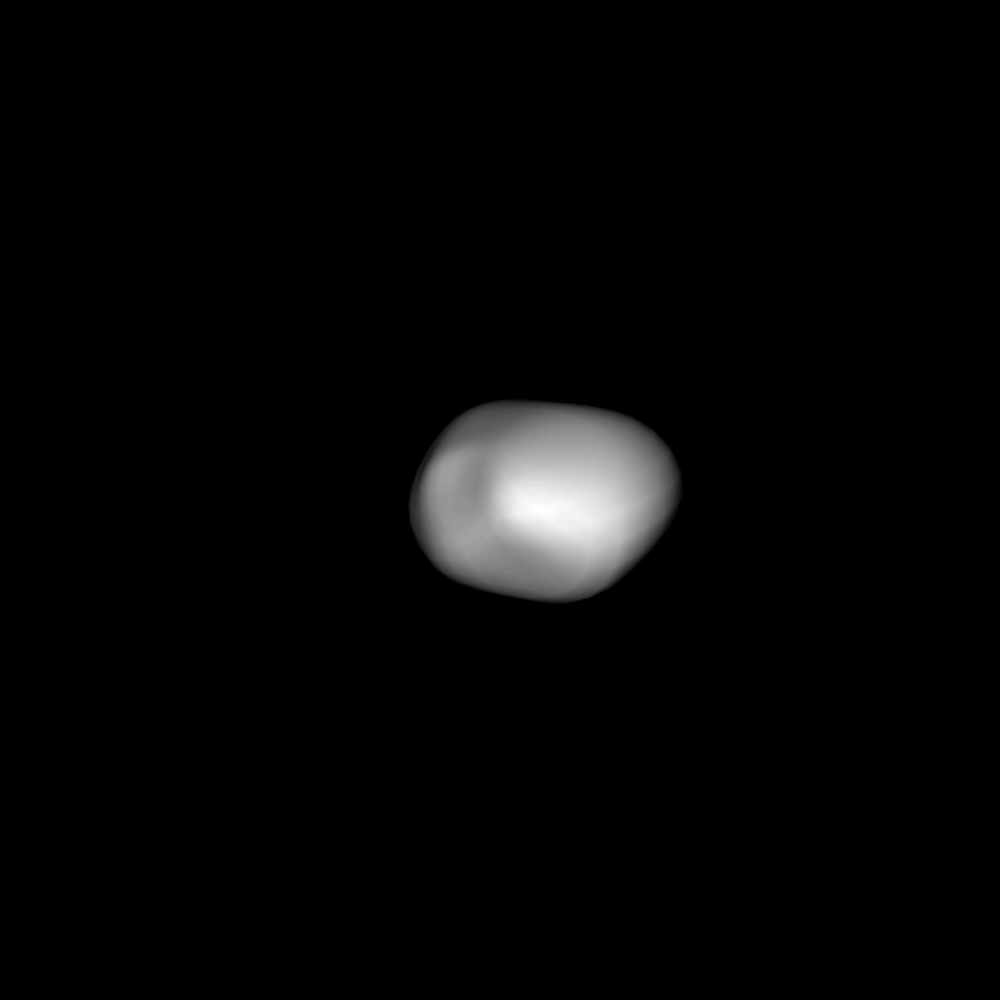}}\resizebox{0.24\hsize}{!}{\includegraphics{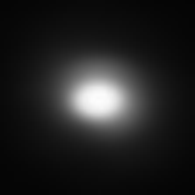}}\resizebox{0.24\hsize}{!}{\includegraphics{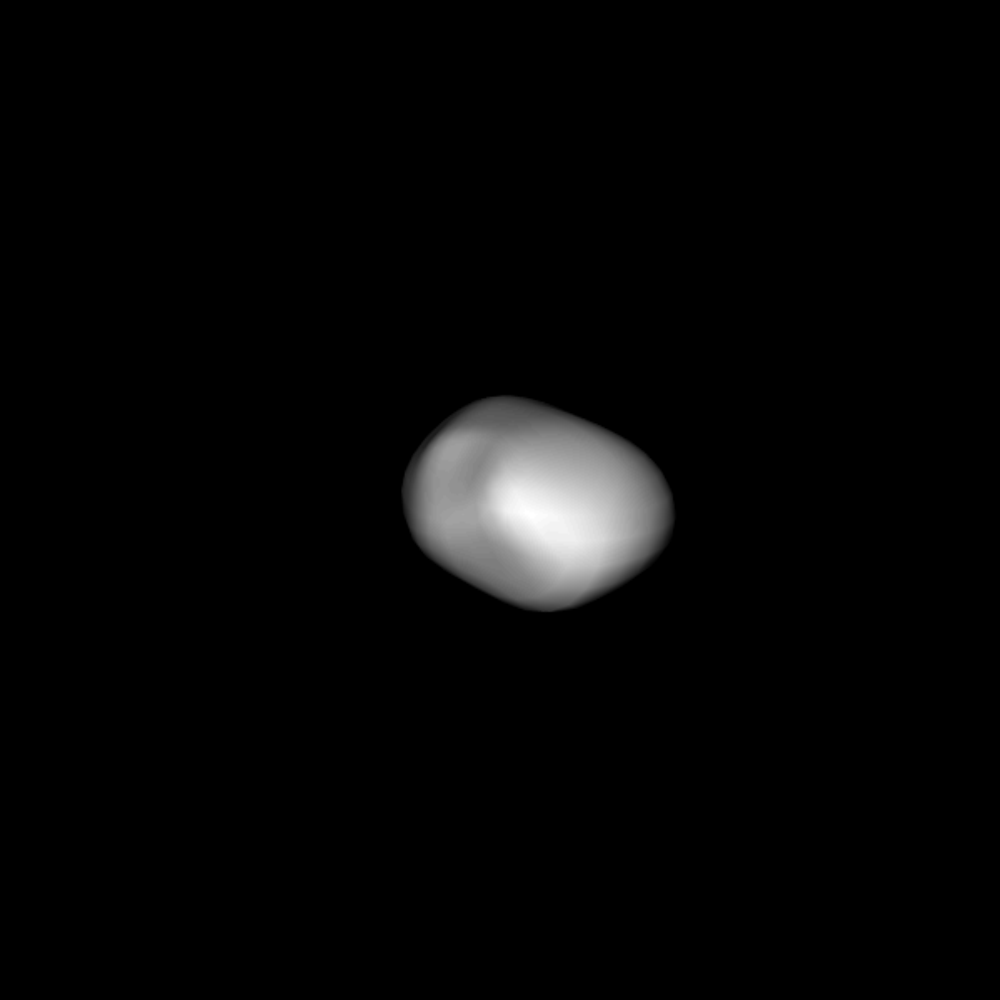}}\\
        \resizebox{0.24\hsize}{!}{\includegraphics{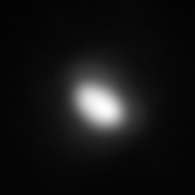}}\resizebox{0.24\hsize}{!}{\includegraphics{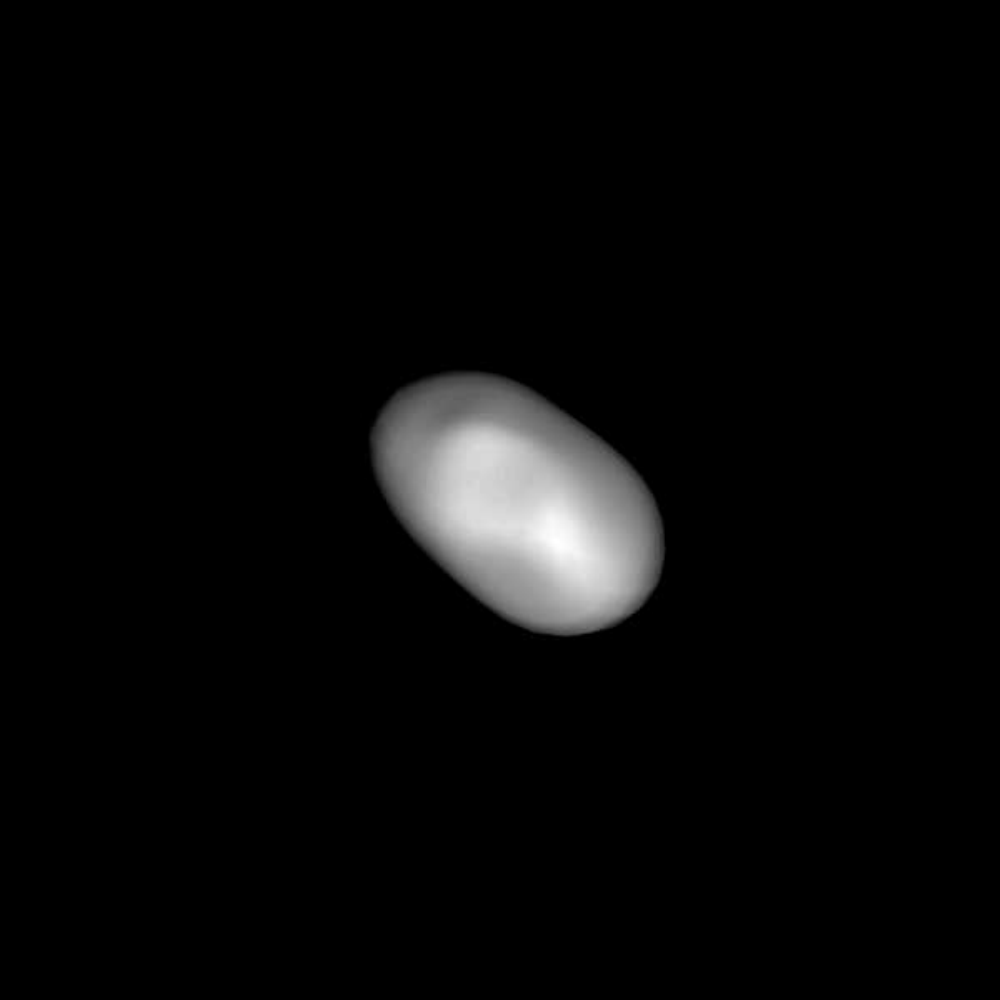}}\resizebox{0.24\hsize}{!}{\includegraphics{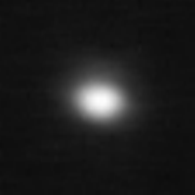}}\resizebox{0.24\hsize}{!}{\includegraphics{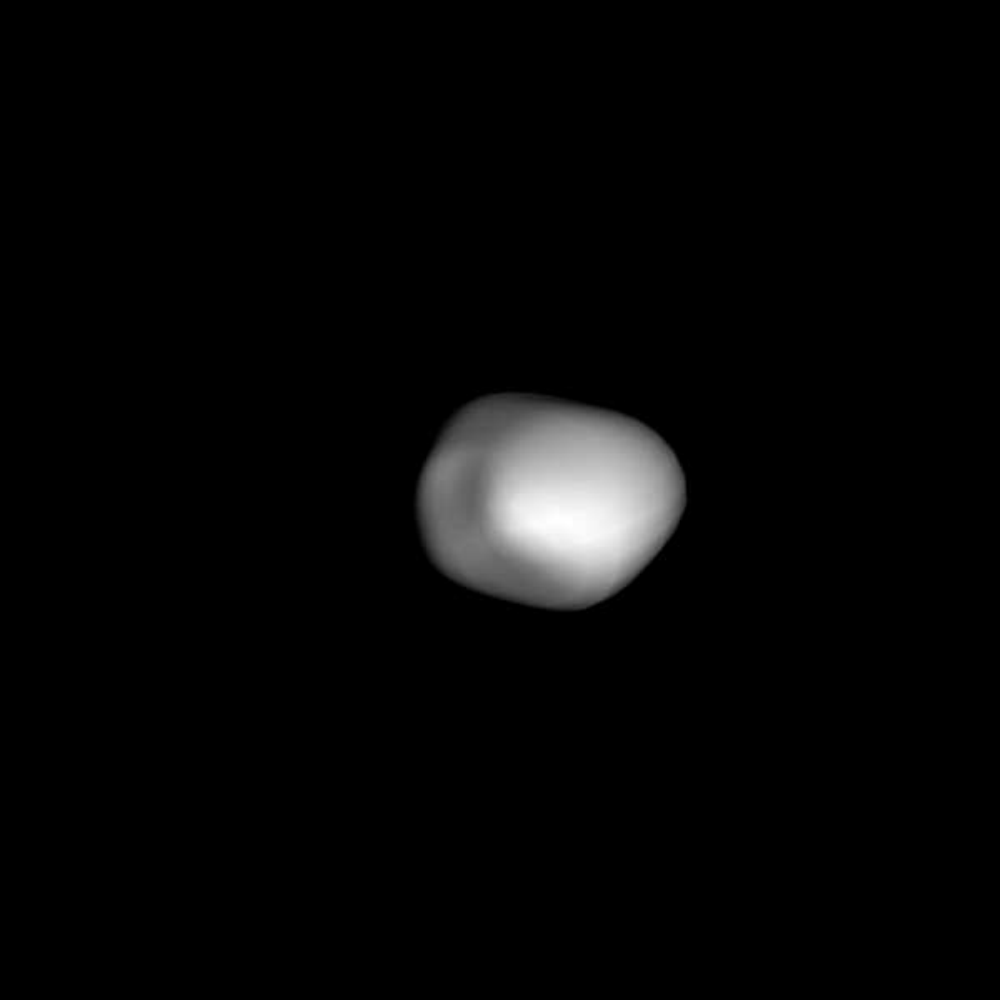}}\\
        \resizebox{0.24\hsize}{!}{\includegraphics{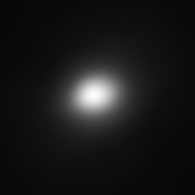}}\resizebox{0.24\hsize}{!}{\includegraphics{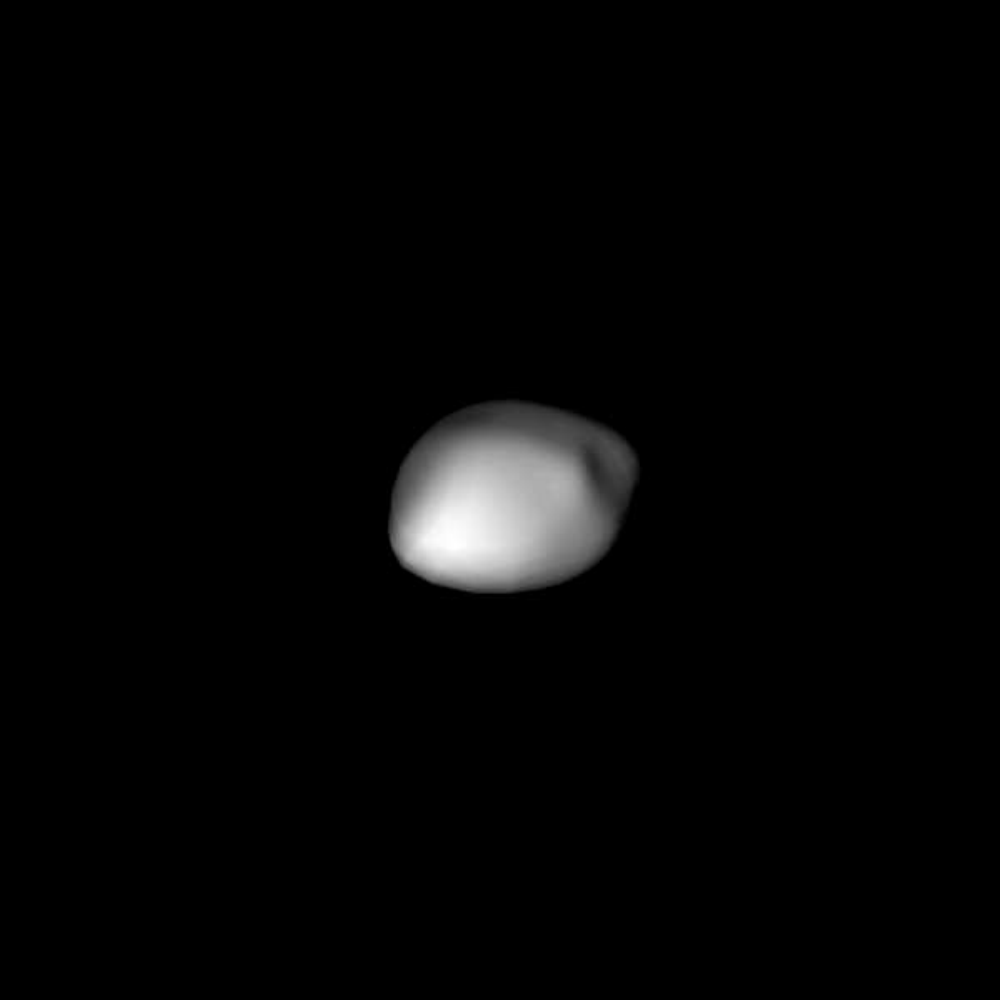}}\resizebox{0.24\hsize}{!}{\includegraphics{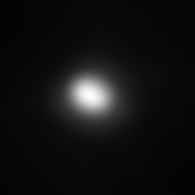}}\resizebox{0.24\hsize}{!}{\includegraphics{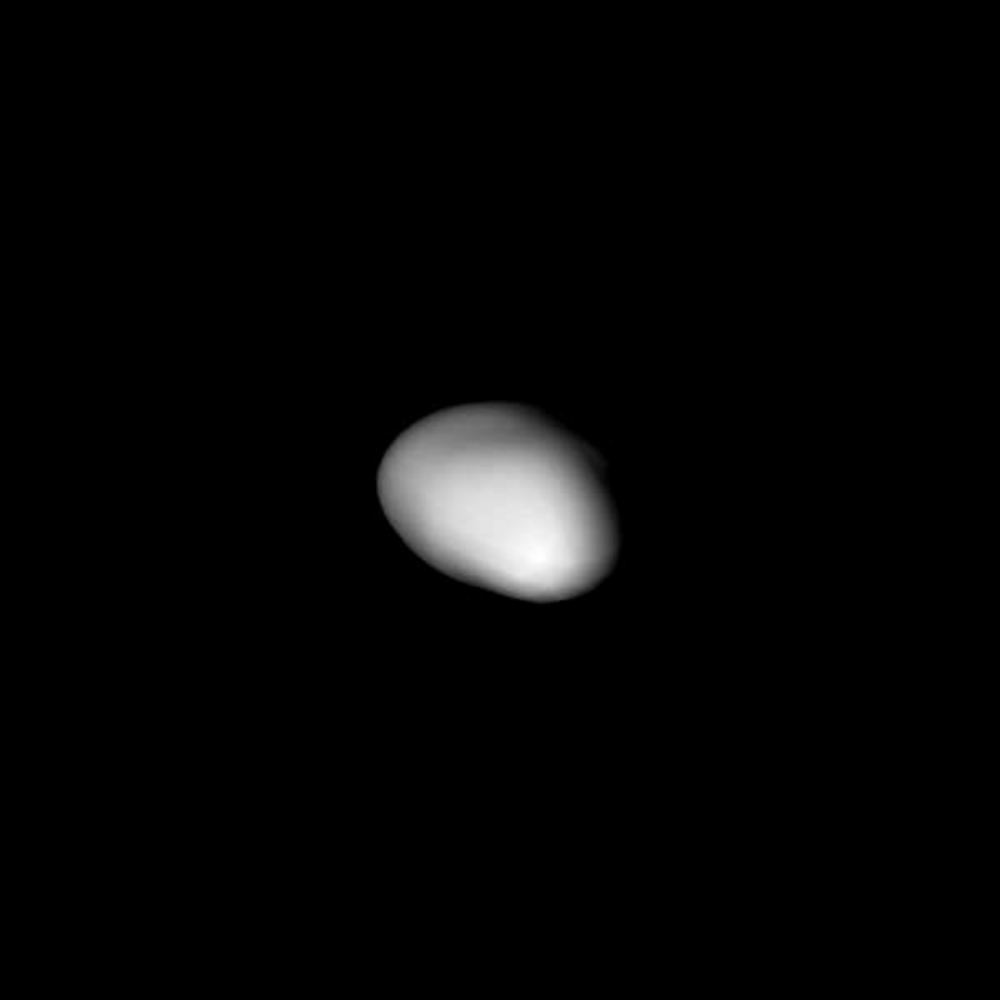}}\\
        \resizebox{0.24\hsize}{!}{\includegraphics{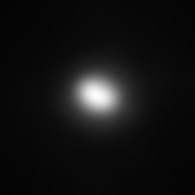}}\resizebox{0.24\hsize}{!}{\includegraphics{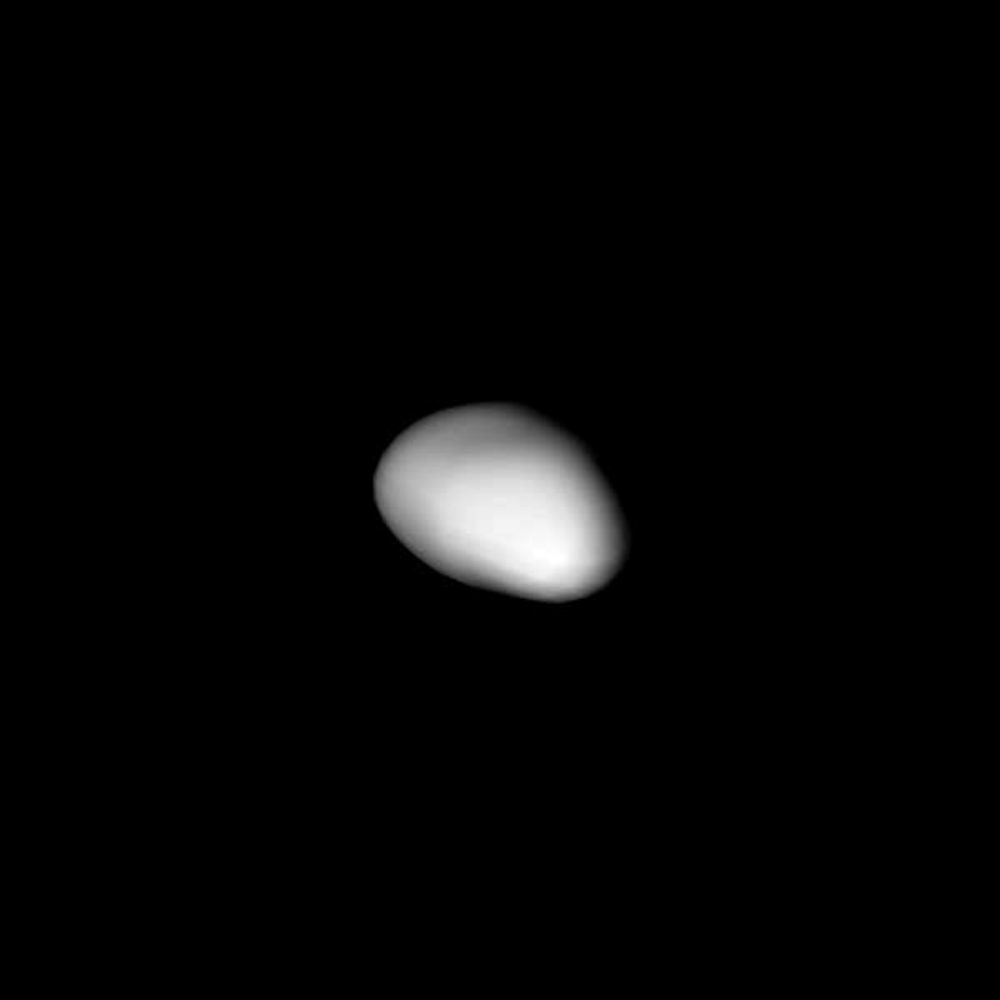}}\resizebox{0.24\hsize}{!}{\includegraphics{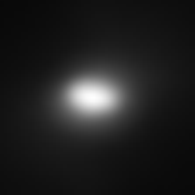}}\resizebox{0.24\hsize}{!}{\includegraphics{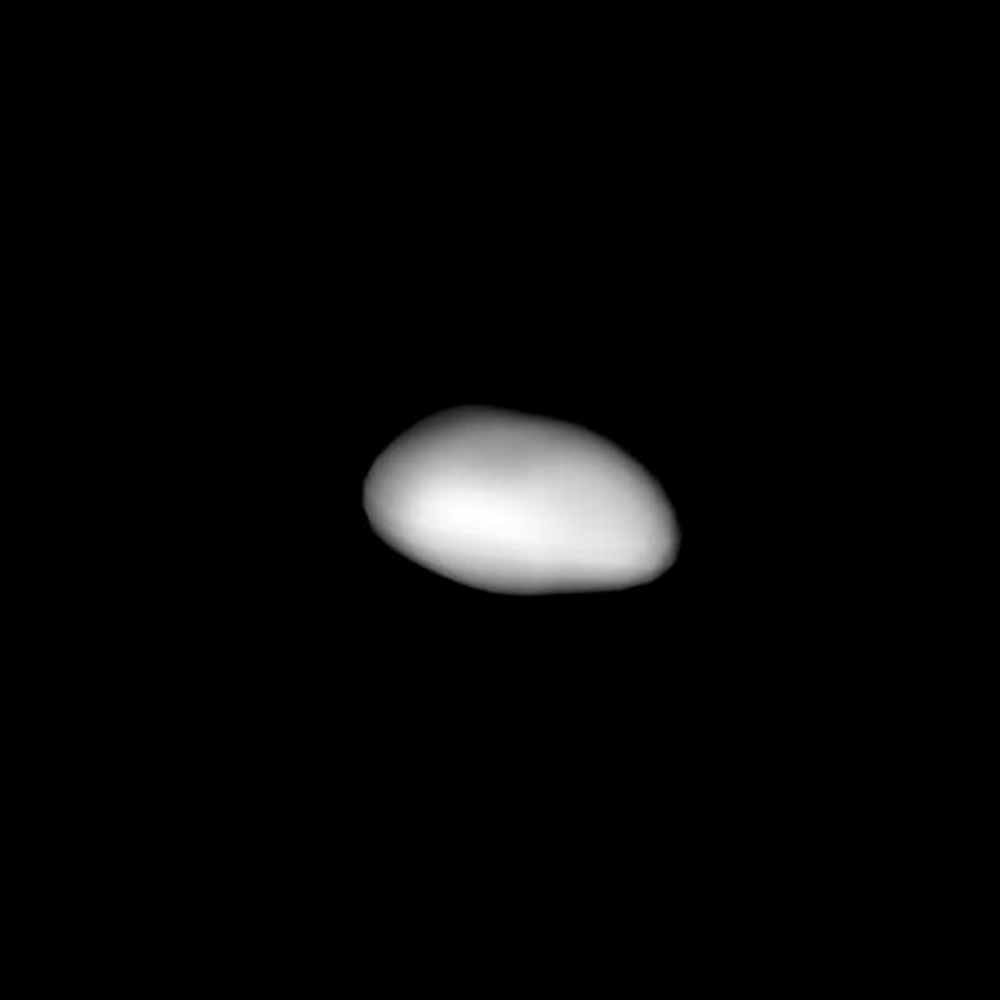}}\\
    \caption{\label{fig:107a}Comparison between model projections and corresponding AO images for asteroid (107) Camilla (first part).}
\end{figure}

\begin{figure}[tbp]
    \centering
        \resizebox{0.24\hsize}{!}{\includegraphics{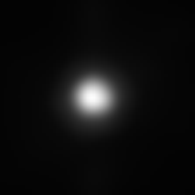}}\resizebox{0.24\hsize}{!}{\includegraphics{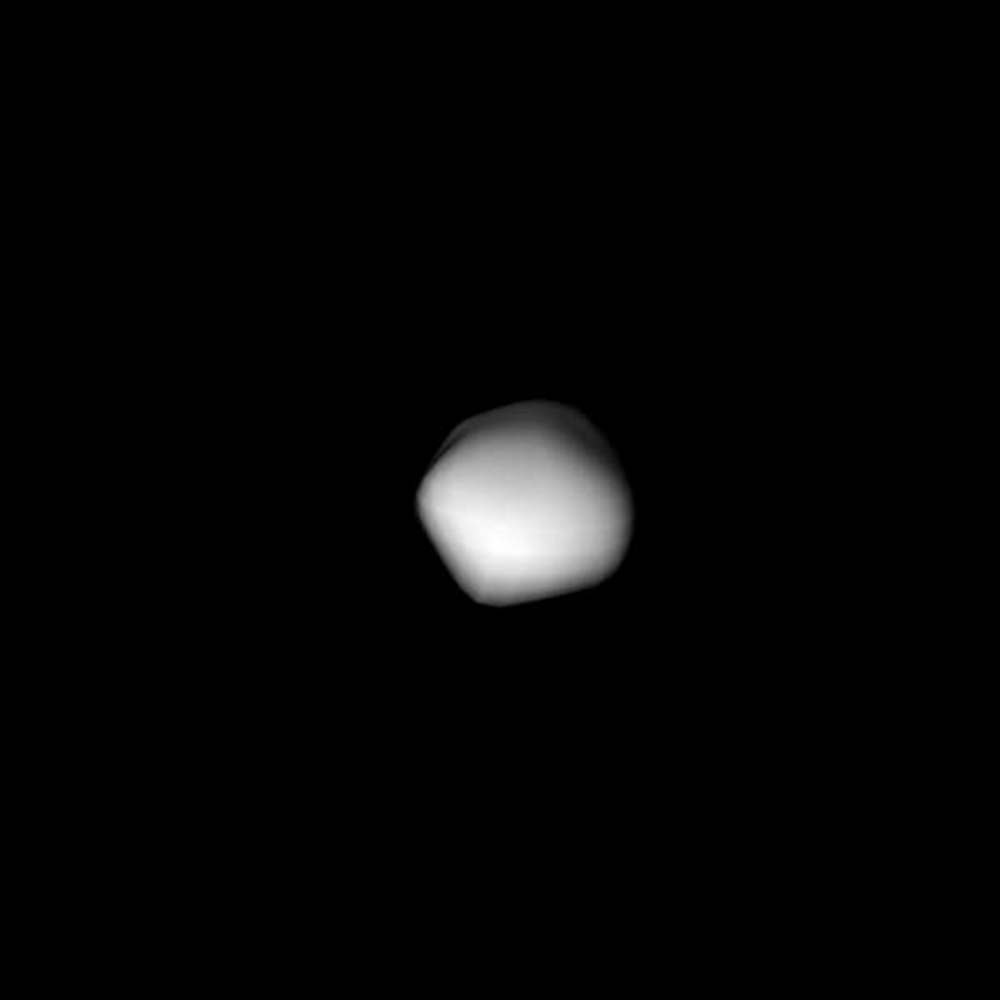}}\resizebox{0.24\hsize}{!}{\includegraphics{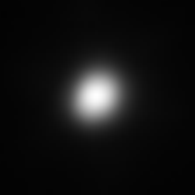}}\resizebox{0.24\hsize}{!}{\includegraphics{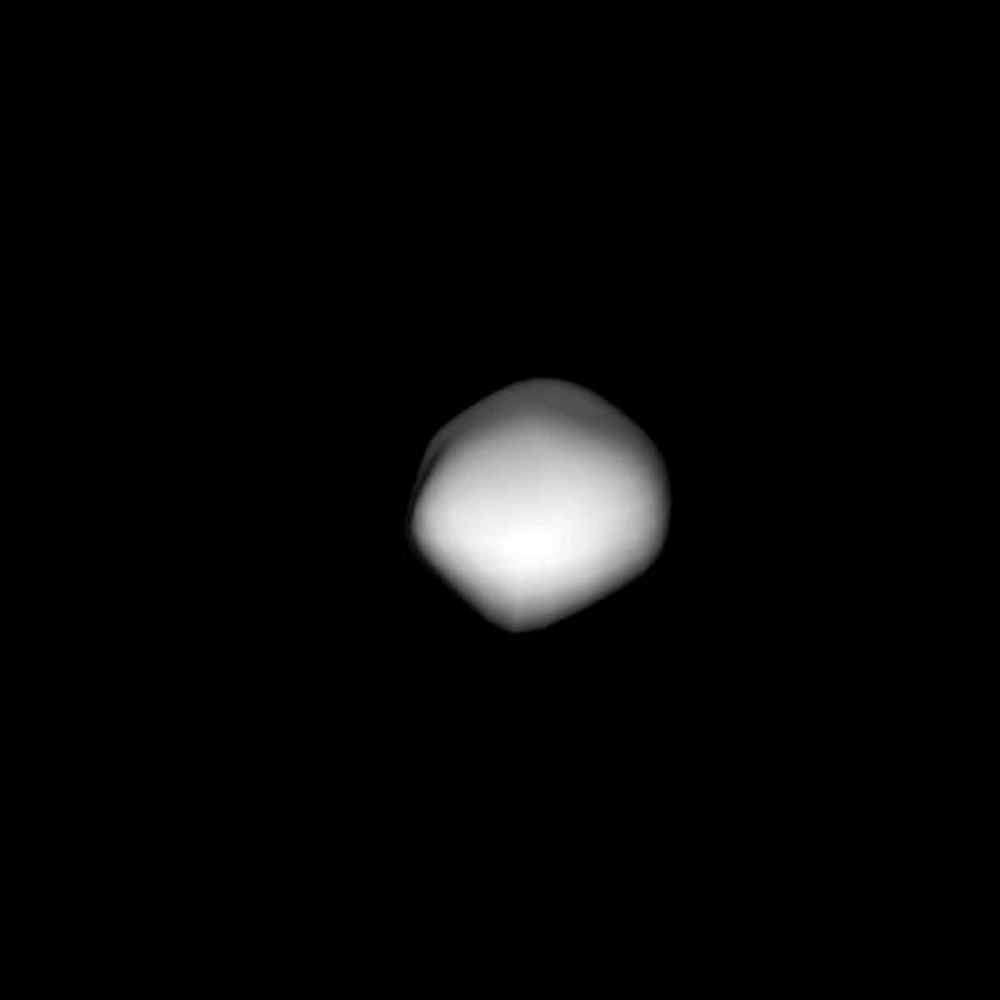}}\\
        \resizebox{0.24\hsize}{!}{\includegraphics{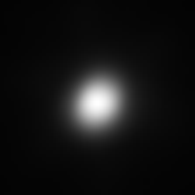}}\resizebox{0.24\hsize}{!}{\includegraphics{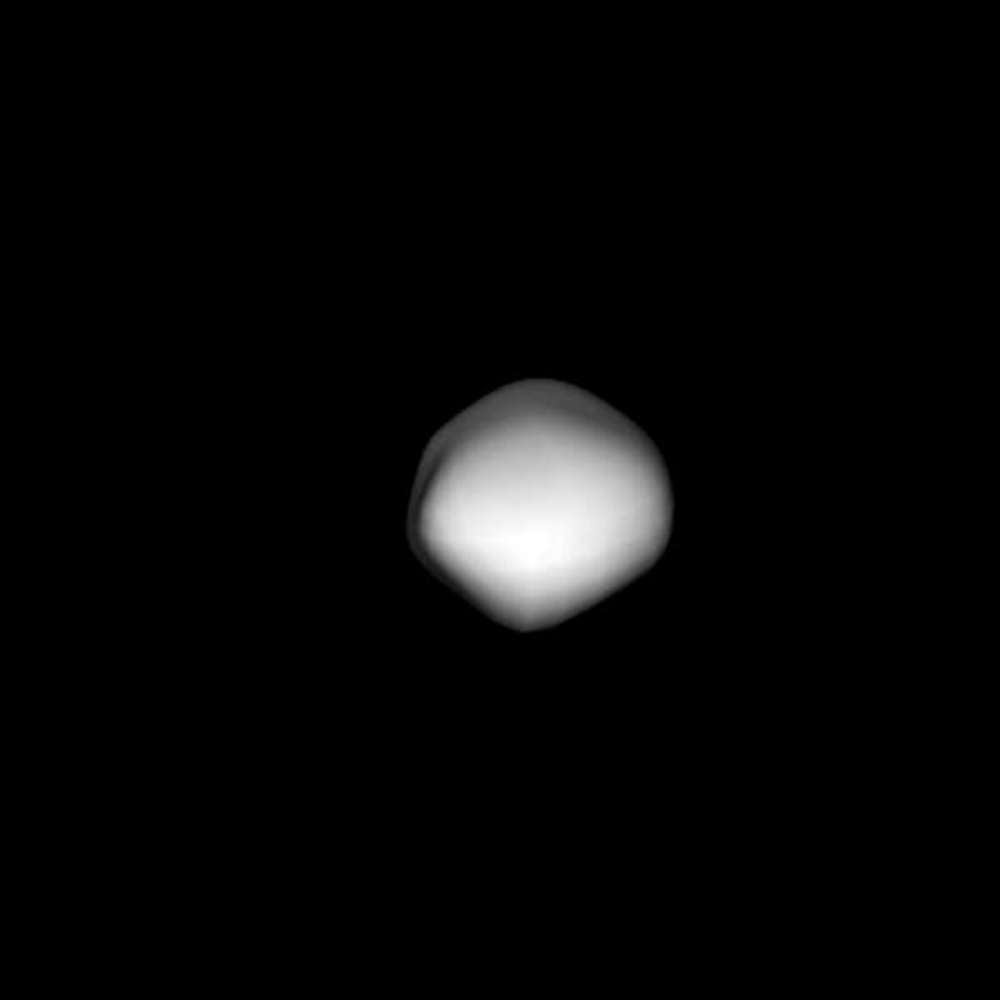}}\resizebox{0.24\hsize}{!}{\includegraphics{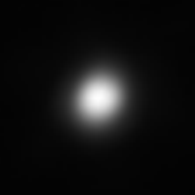}}\resizebox{0.24\hsize}{!}{\includegraphics{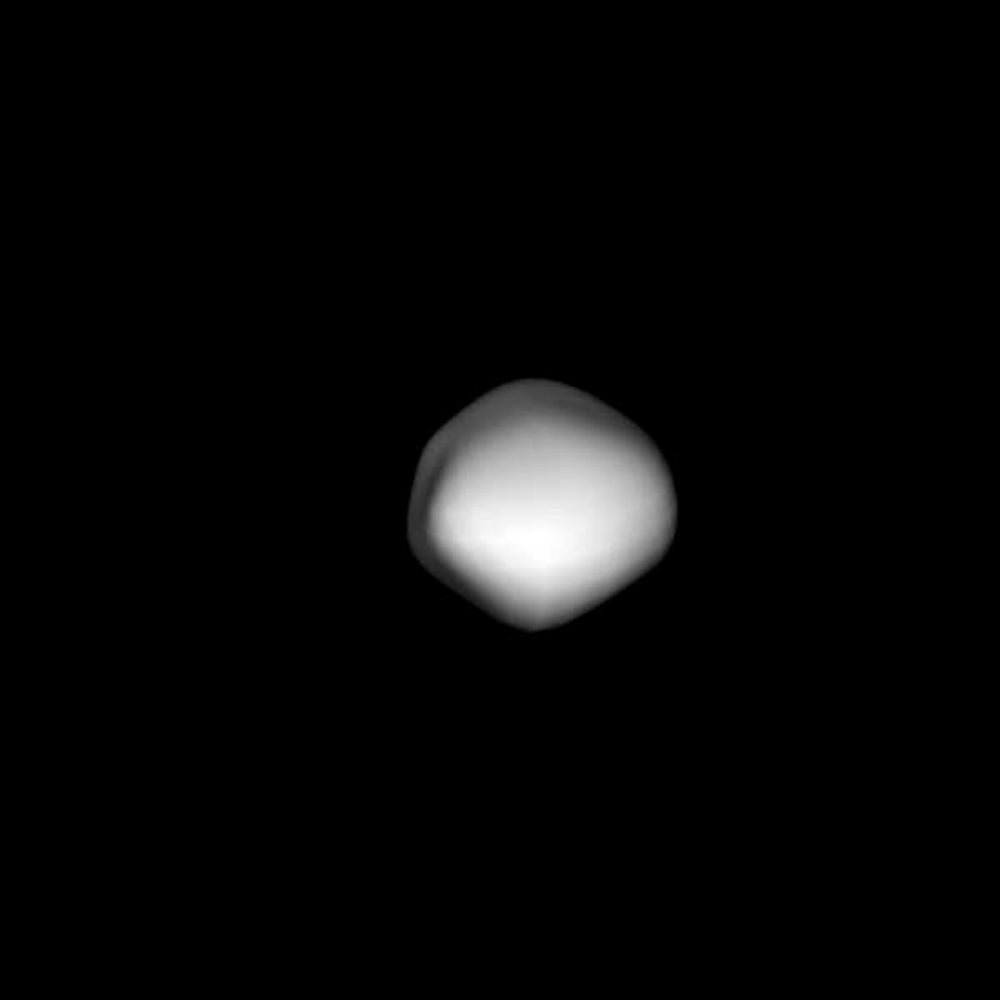}}\\
        \resizebox{0.24\hsize}{!}{\includegraphics{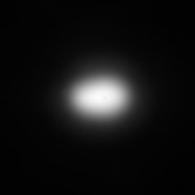}}\resizebox{0.24\hsize}{!}{\includegraphics{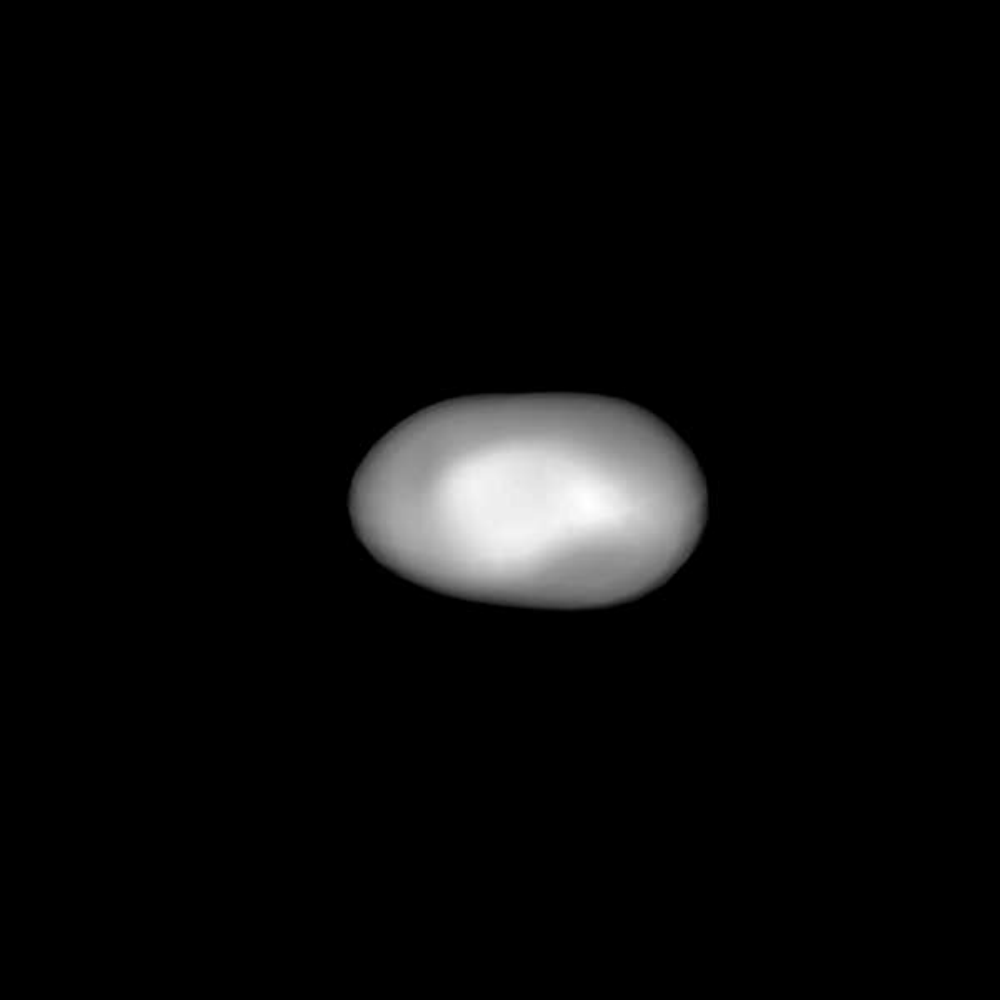}}\resizebox{0.24\hsize}{!}{\includegraphics{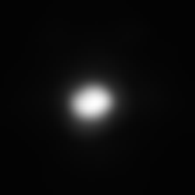}}\resizebox{0.24\hsize}{!}{\includegraphics{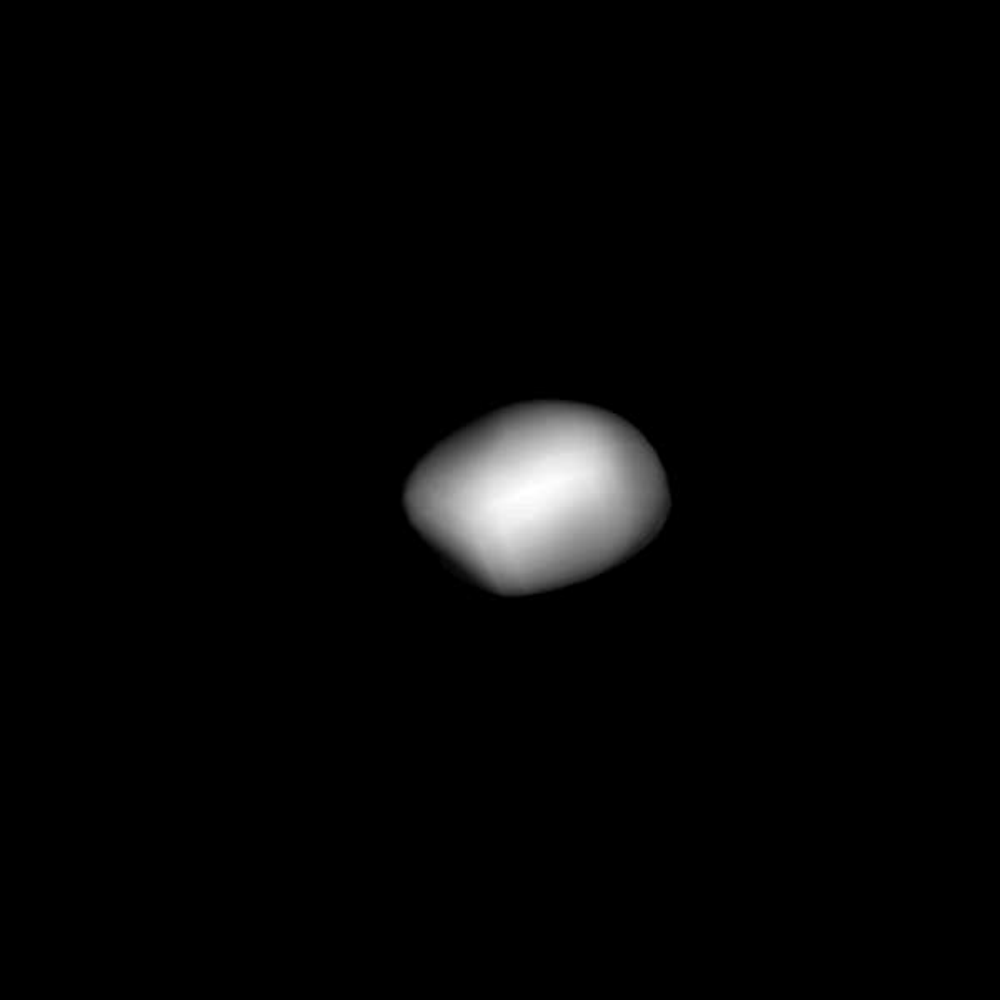}}\\
        \resizebox{0.24\hsize}{!}{\includegraphics{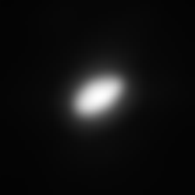}}\resizebox{0.24\hsize}{!}{\includegraphics{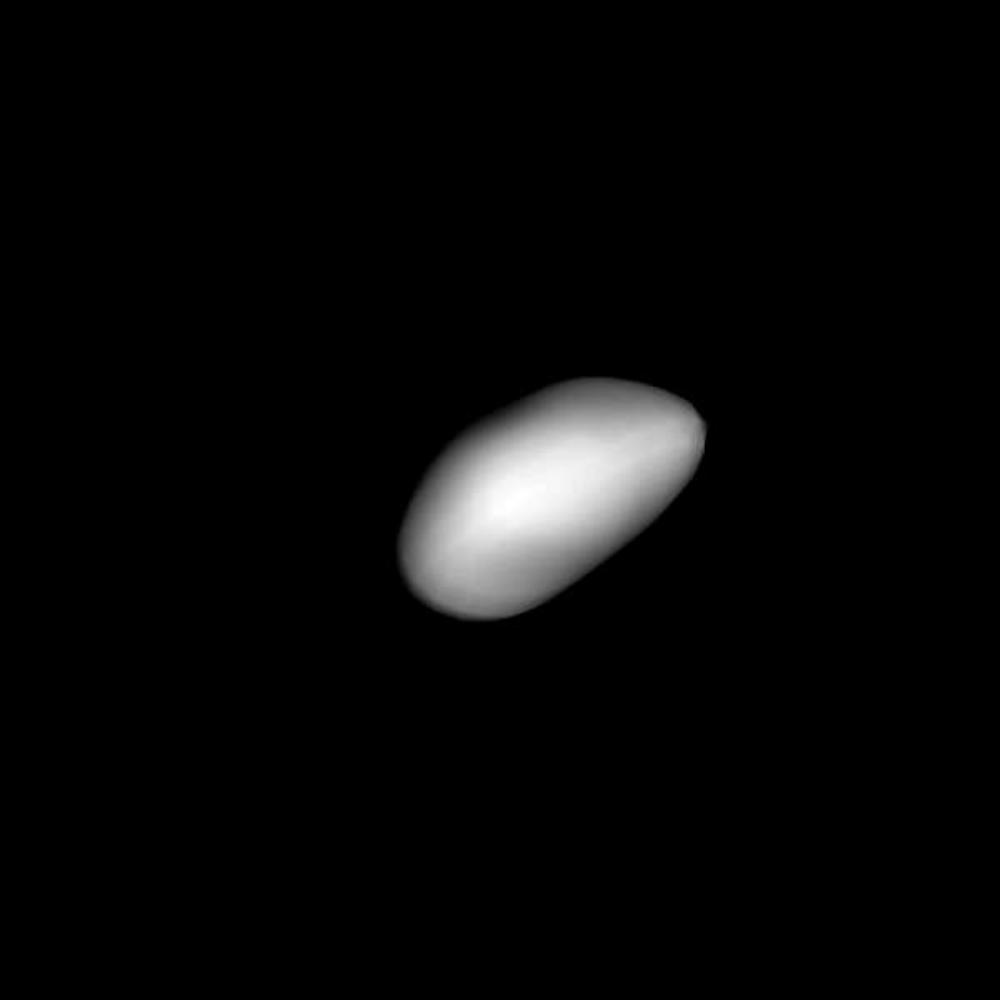}}\\
    \caption{\label{fig:107b}Comparison between model projections and corresponding AO images for asteroid (107) Camilla (second part).}
\end{figure}

\begin{figure}[tbp]
    \centering
        \resizebox{0.24\hsize}{!}{\includegraphics{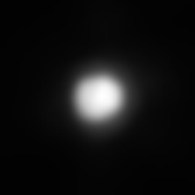}}\resizebox{0.24\hsize}{!}{\includegraphics{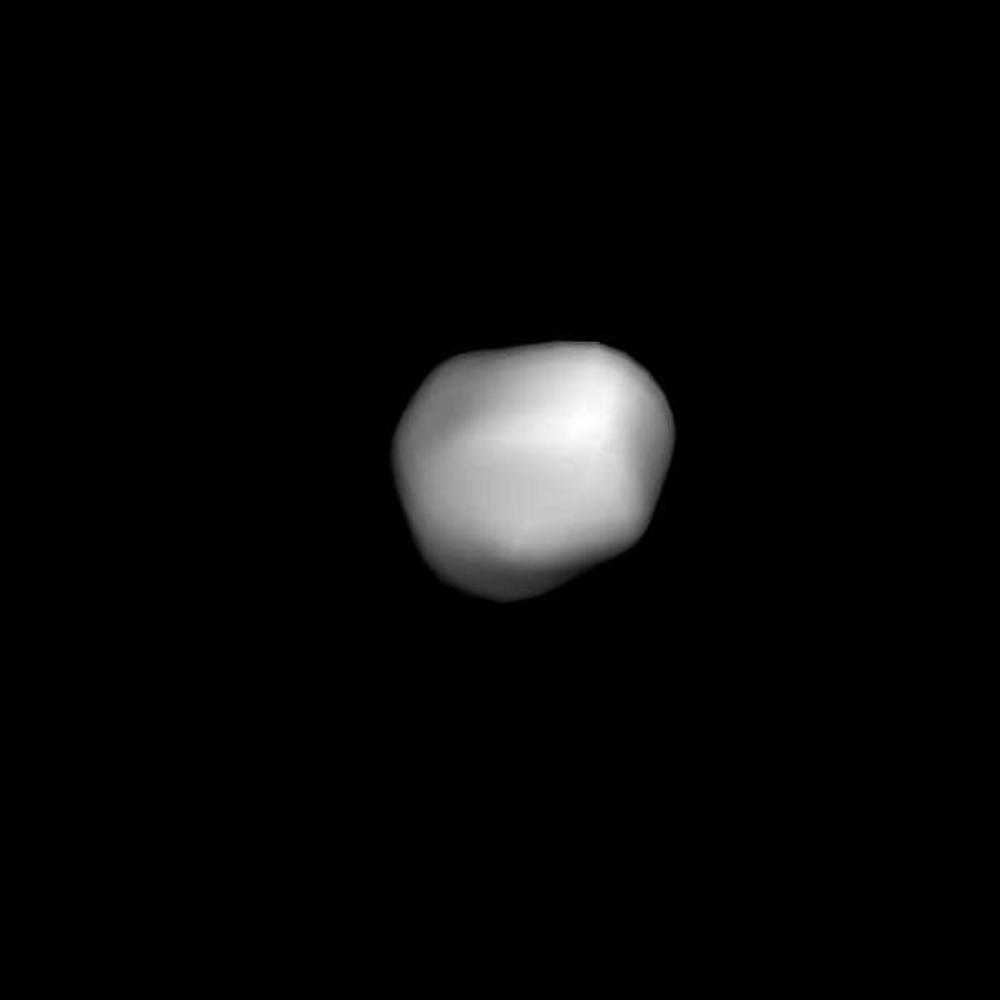}}\resizebox{0.24\hsize}{!}{\includegraphics{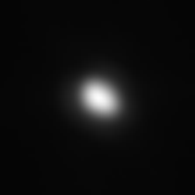}}\resizebox{0.24\hsize}{!}{\includegraphics{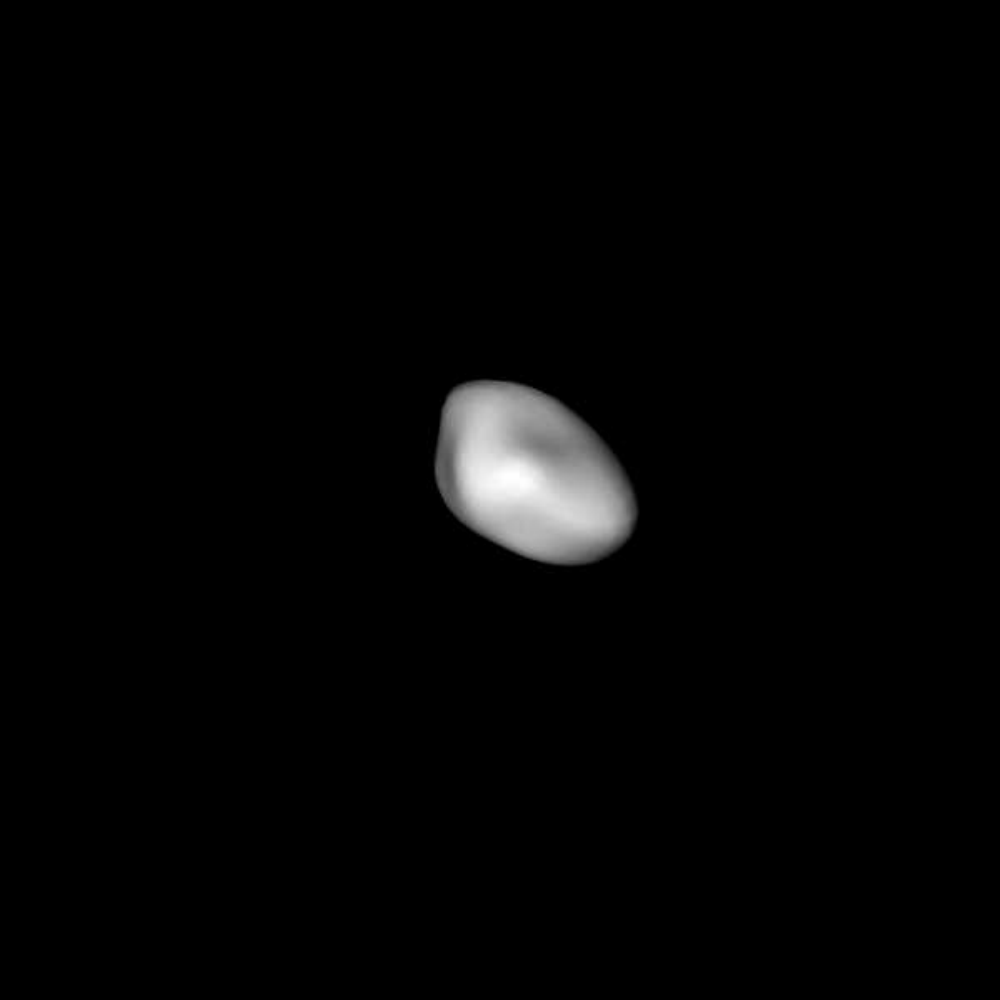}}\\
        \resizebox{0.24\hsize}{!}{\includegraphics{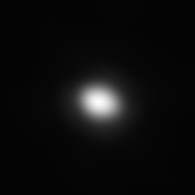}}\resizebox{0.24\hsize}{!}{\includegraphics{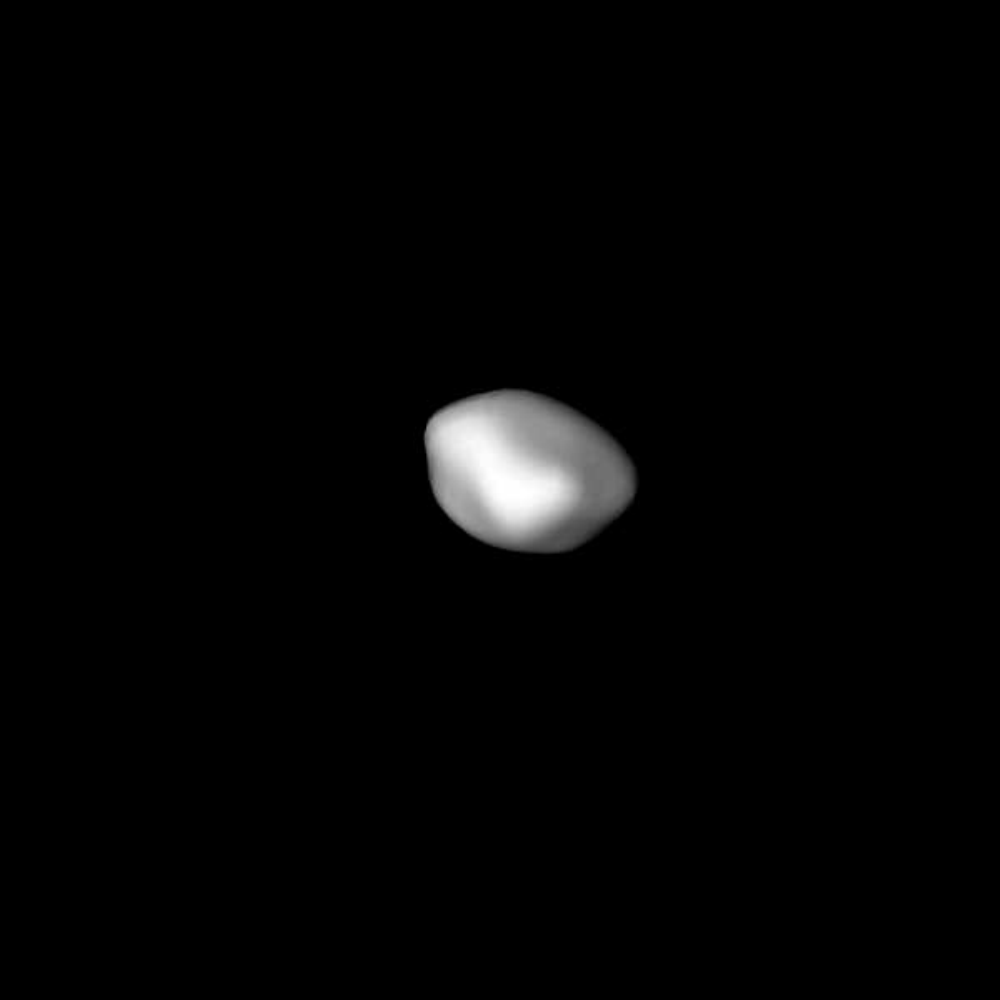}}\resizebox{0.24\hsize}{!}{\includegraphics{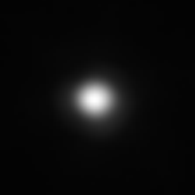}}\resizebox{0.24\hsize}{!}{\includegraphics{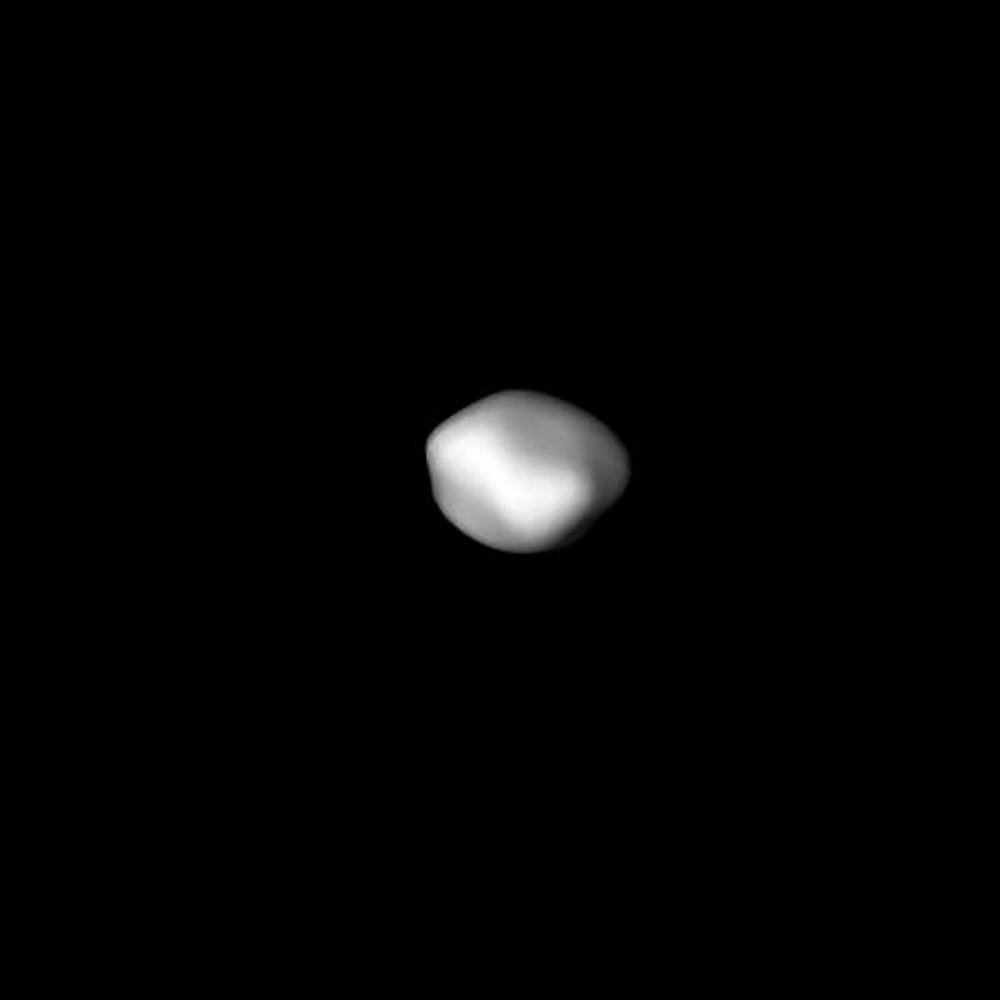}}\\
        \resizebox{0.24\hsize}{!}{\includegraphics{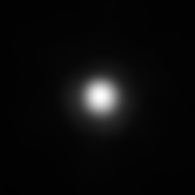}}\resizebox{0.24\hsize}{!}{\includegraphics{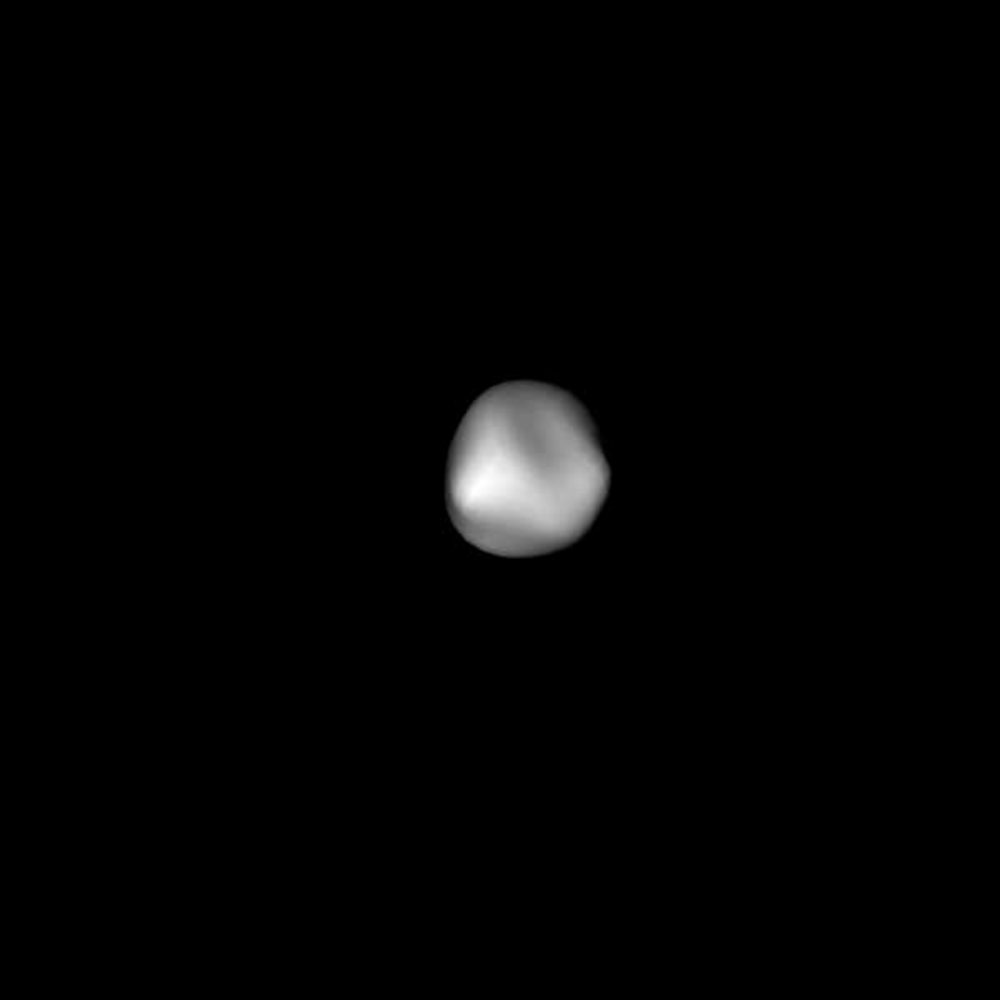}}\resizebox{0.24\hsize}{!}{\includegraphics{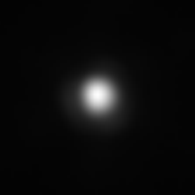}}\resizebox{0.24\hsize}{!}{\includegraphics{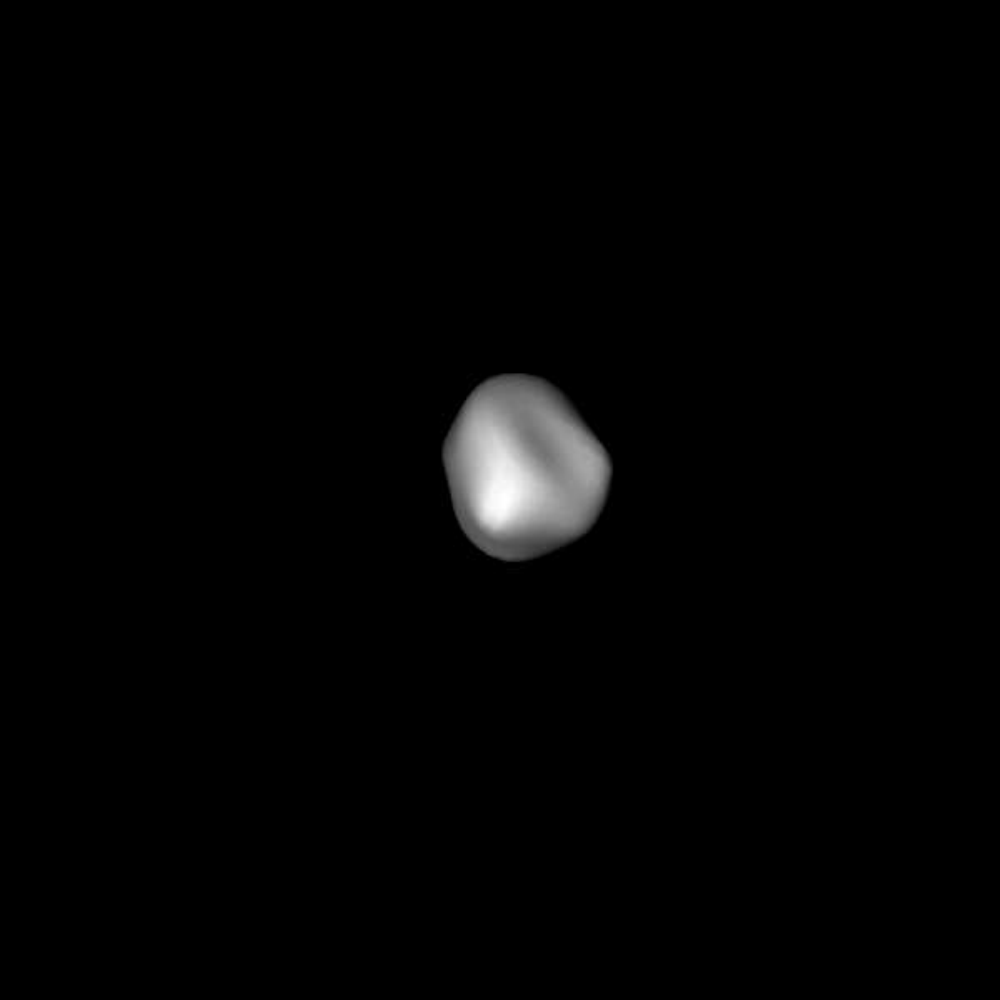}}\\
        \resizebox{0.24\hsize}{!}{\includegraphics{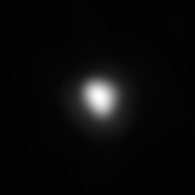}}\resizebox{0.24\hsize}{!}{\includegraphics{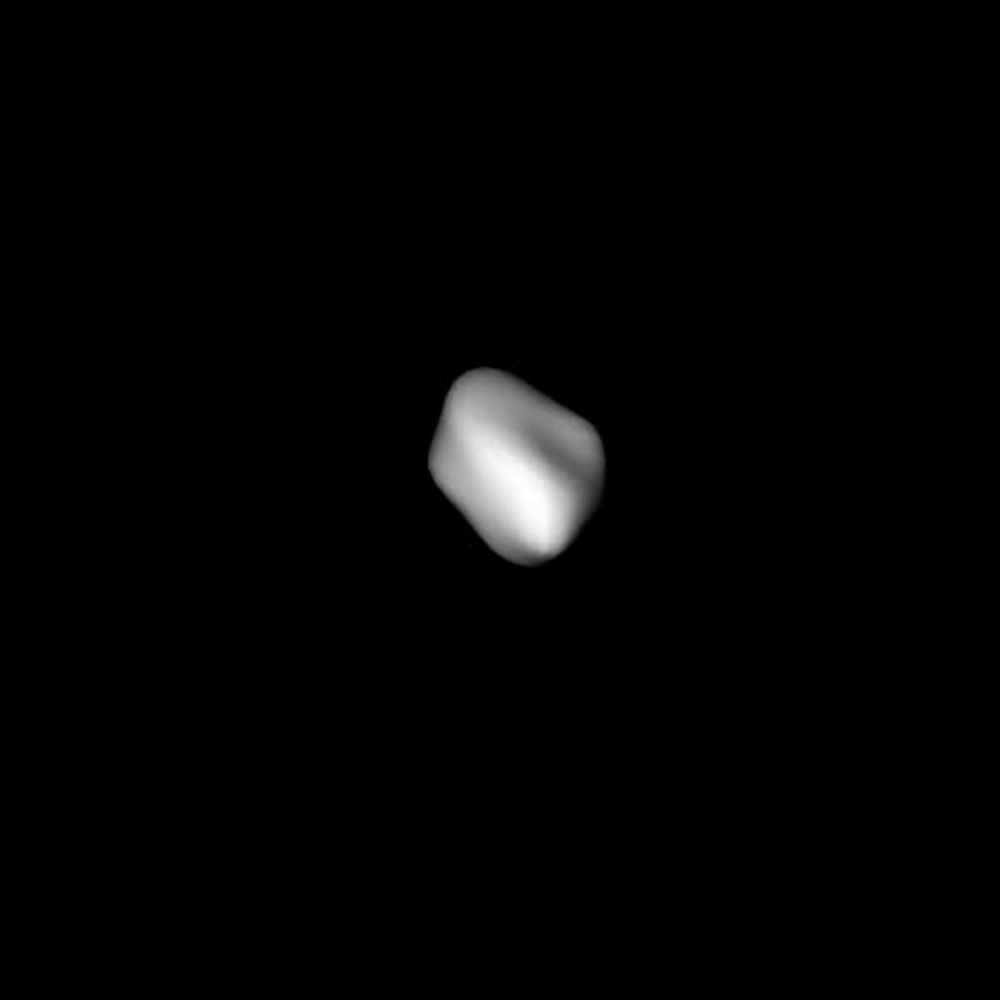}}\resizebox{0.24\hsize}{!}{\includegraphics{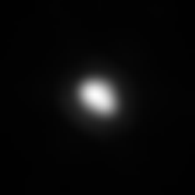}}\resizebox{0.24\hsize}{!}{\includegraphics{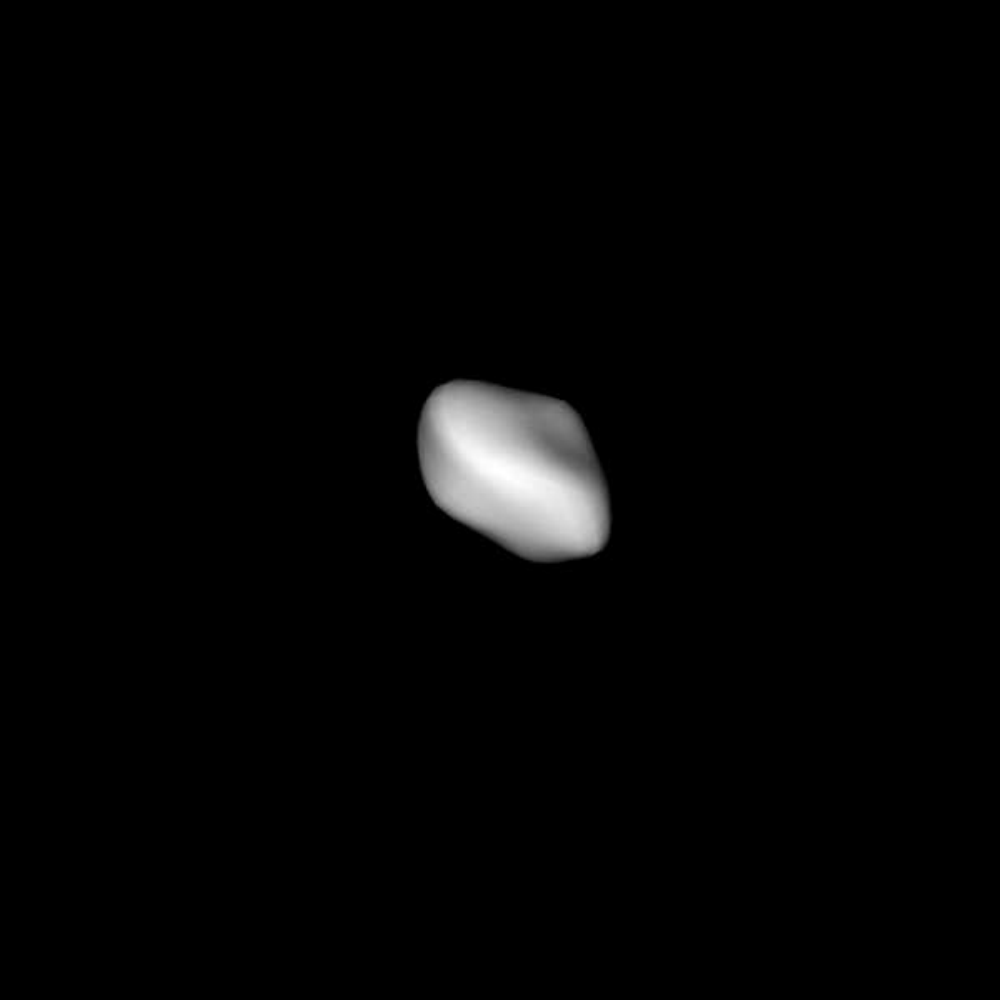}}\\
        \resizebox{0.24\hsize}{!}{\includegraphics{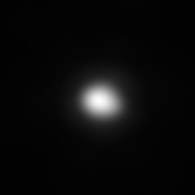}}\resizebox{0.24\hsize}{!}{\includegraphics{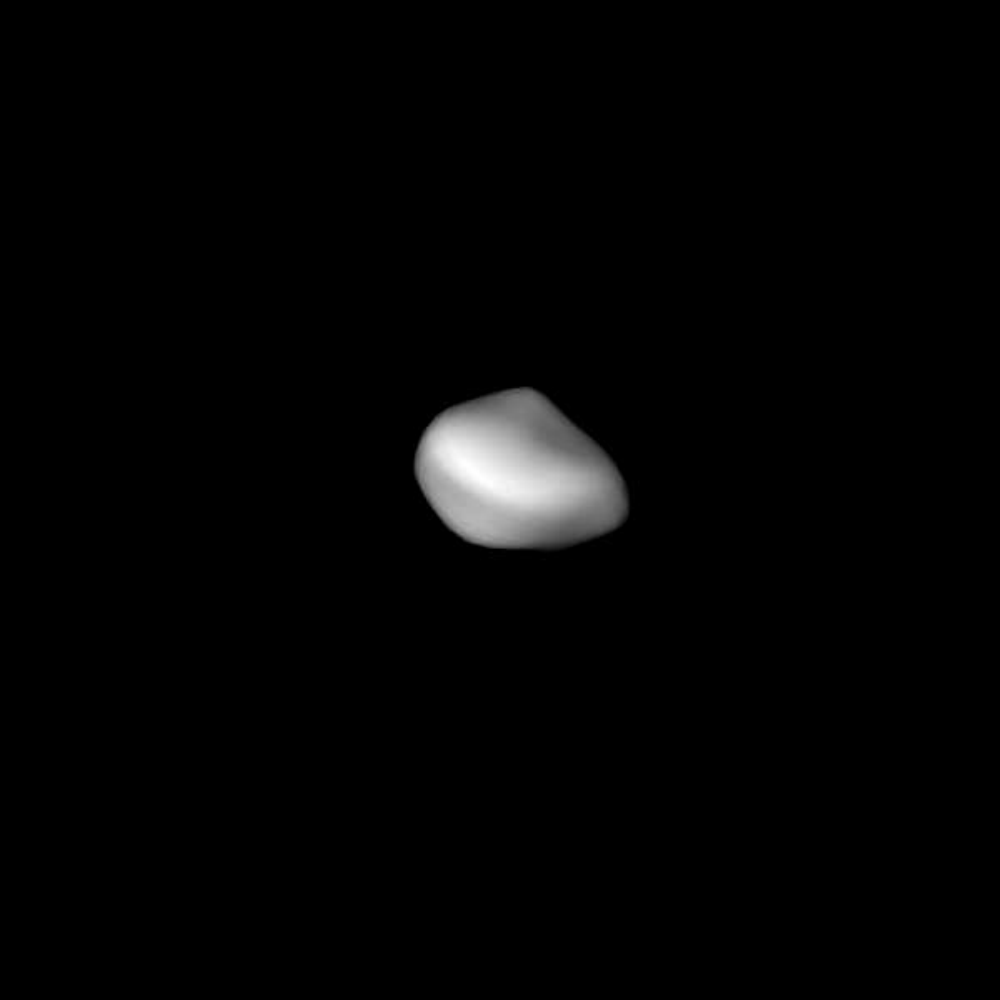}}\\
    \caption{\label{fig:129}Comparison between model projections and corresponding AO images for asteroid (129) Antigone.}
\end{figure}

\begin{figure}[tbp]
    \centering
        \resizebox{0.24\hsize}{!}{\includegraphics{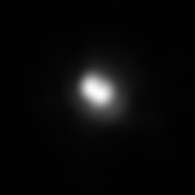}}\resizebox{0.24\hsize}{!}{\includegraphics{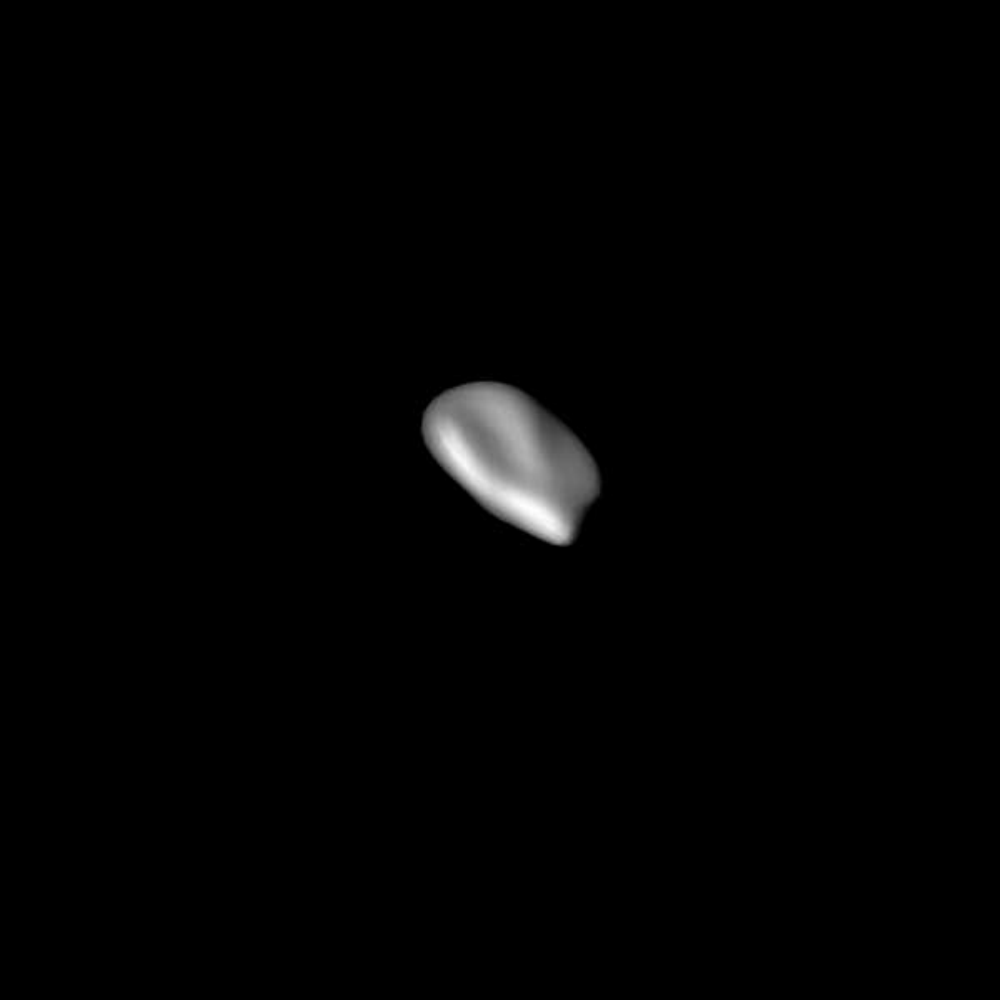}}\resizebox{0.24\hsize}{!}{\includegraphics{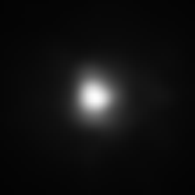}}\resizebox{0.24\hsize}{!}{\includegraphics{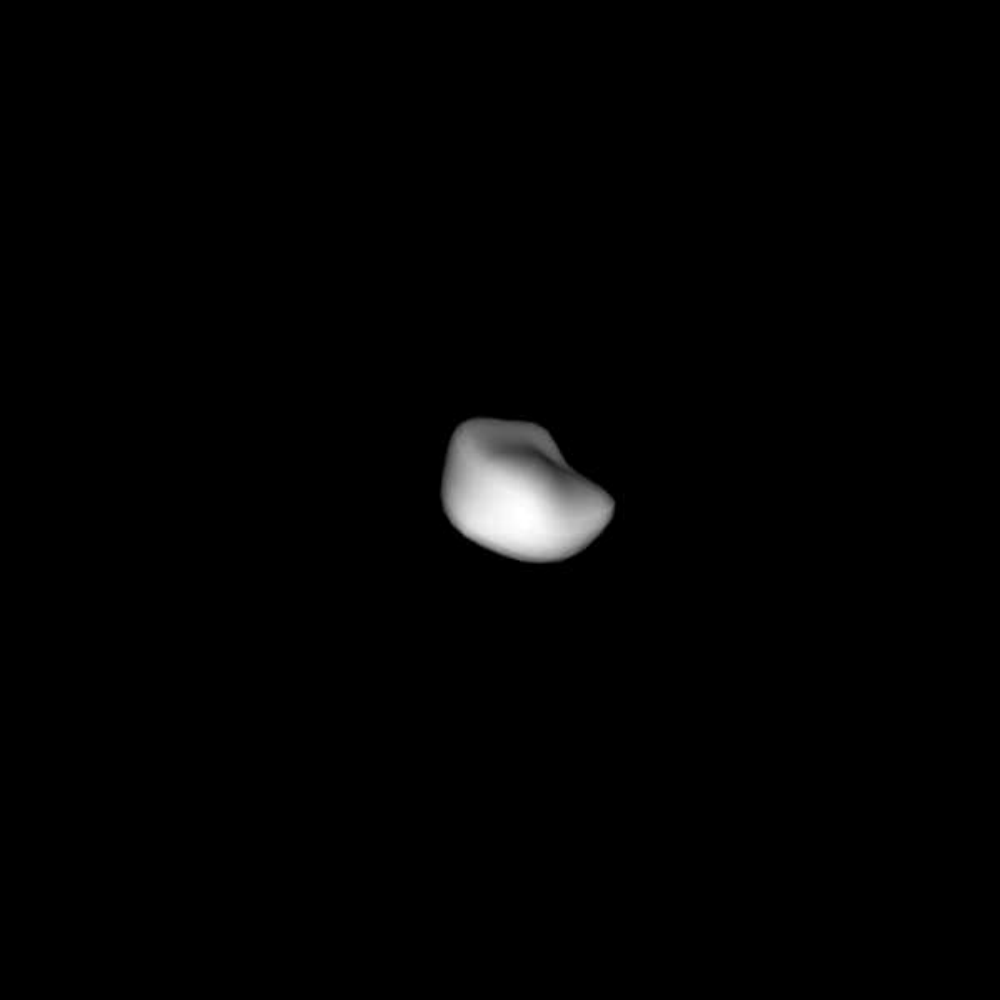}}\\
    \caption{\label{fig:135}Comparison between model projections and corresponding AO images for asteroid (135) Hertha.}
\end{figure}

\begin{figure}[tbp]
    \centering
        \resizebox{0.24\hsize}{!}{\includegraphics{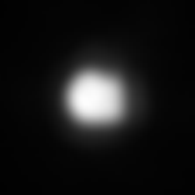}}\resizebox{0.24\hsize}{!}{\includegraphics{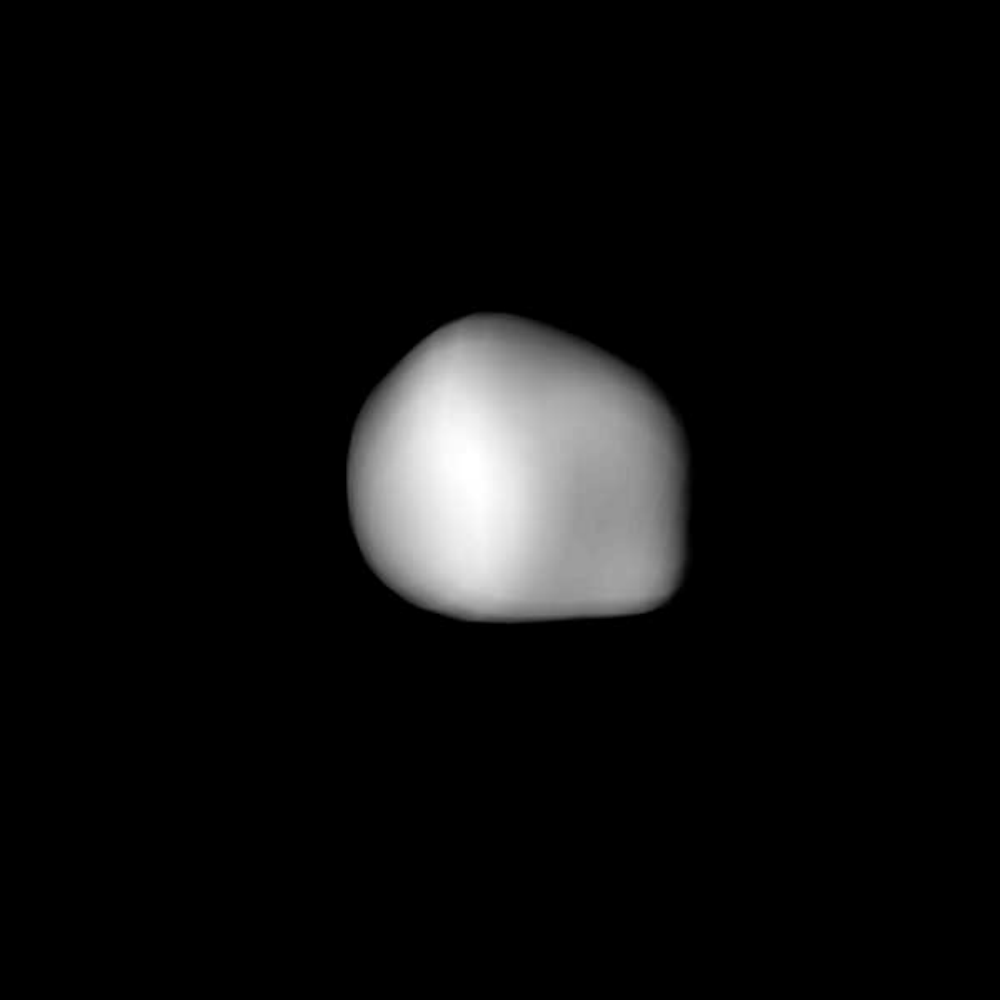}}\resizebox{0.24\hsize}{!}{\includegraphics{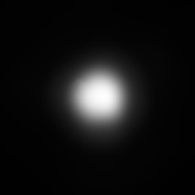}}\resizebox{0.24\hsize}{!}{\includegraphics{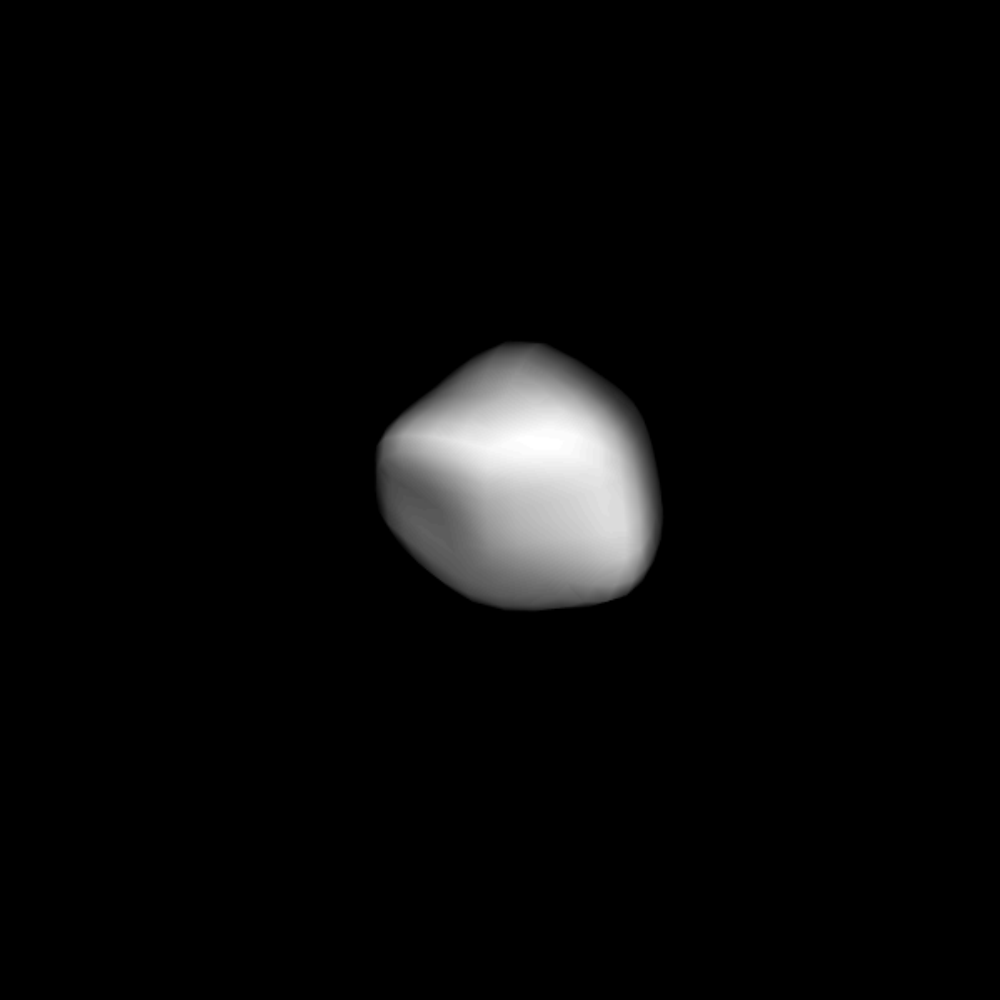}}\\
    \caption{\label{fig:144}Comparison between model projections and corresponding AO images for asteroid (144) Vibilia.}
\end{figure}

\begin{figure}[tbp]
    \centering
        \resizebox{0.24\hsize}{!}{\includegraphics{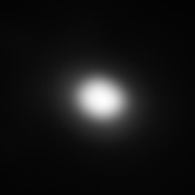}}\resizebox{0.24\hsize}{!}{\includegraphics{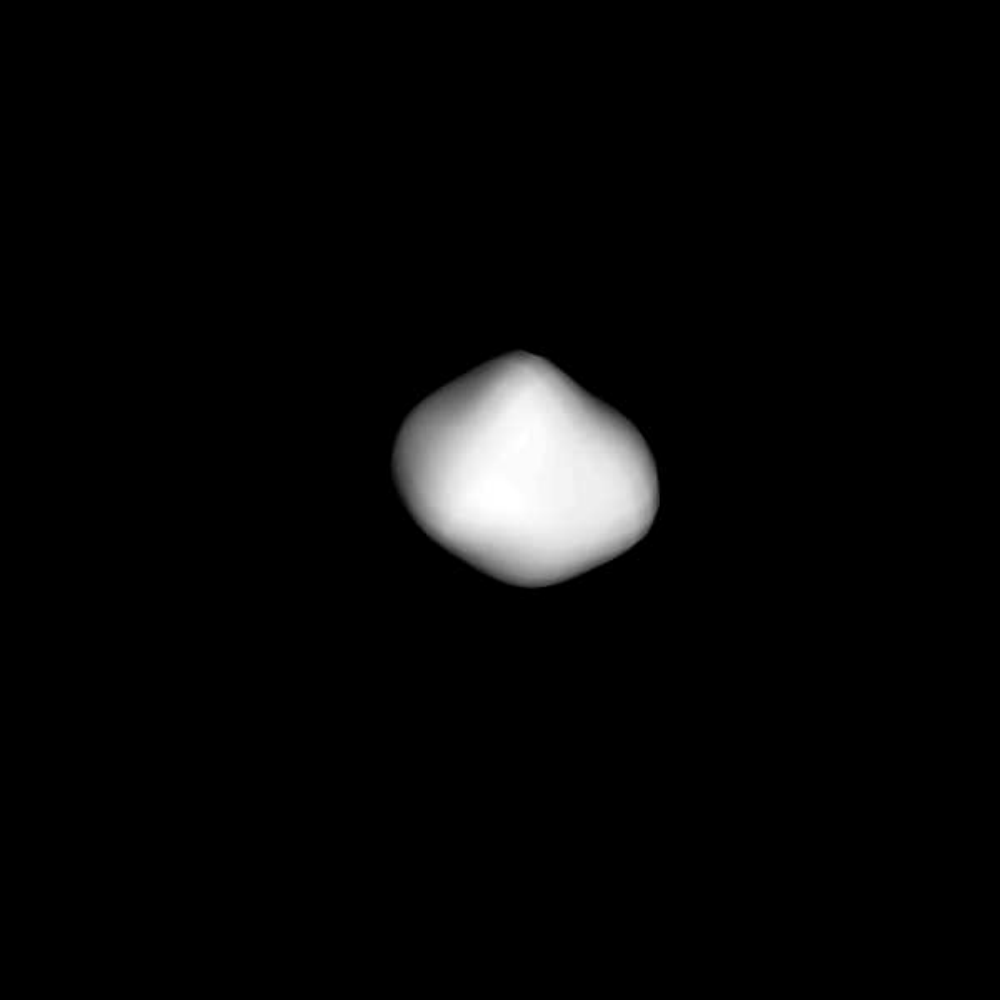}}\resizebox{0.24\hsize}{!}{\includegraphics{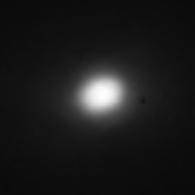}}\resizebox{0.24\hsize}{!}{\includegraphics{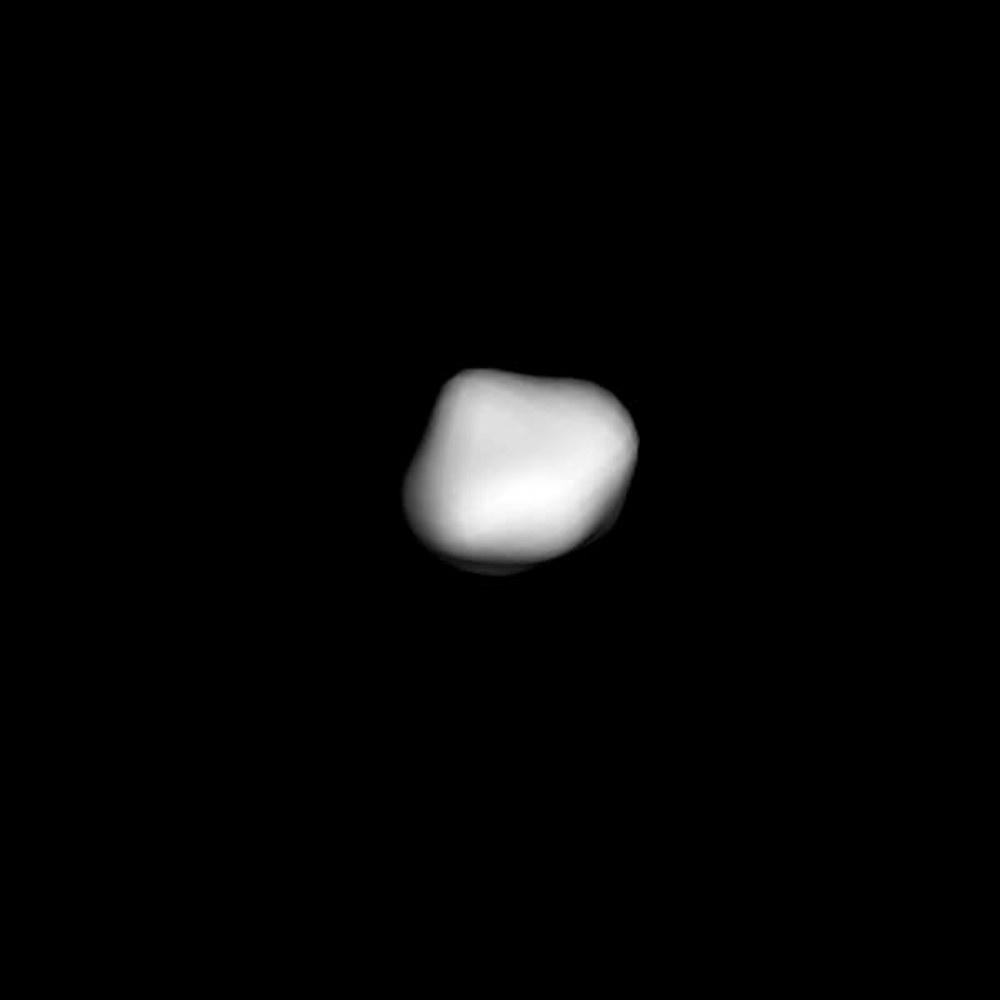}}\\
        \resizebox{0.24\hsize}{!}{\includegraphics{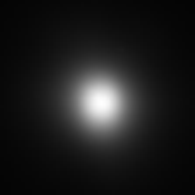}}\resizebox{0.24\hsize}{!}{\includegraphics{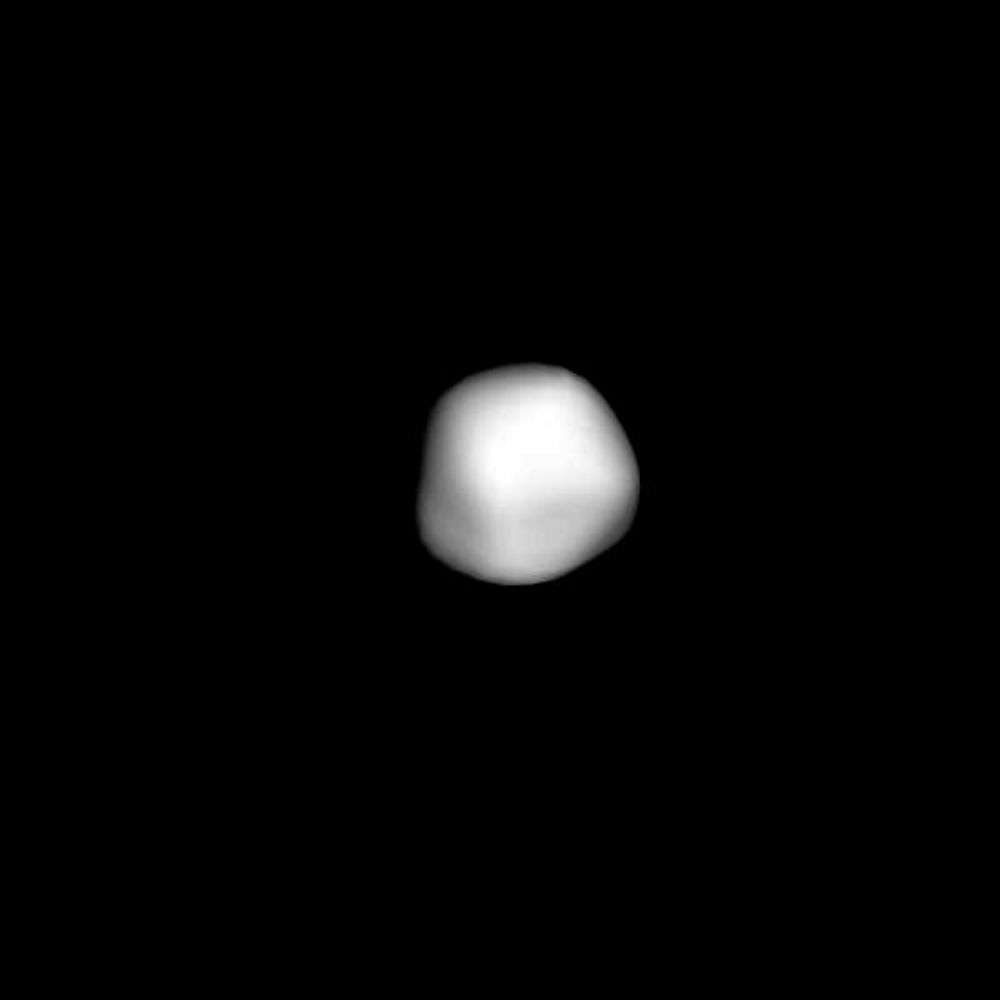}}\resizebox{0.24\hsize}{!}{\includegraphics{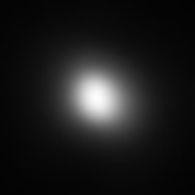}}\resizebox{0.24\hsize}{!}{\includegraphics{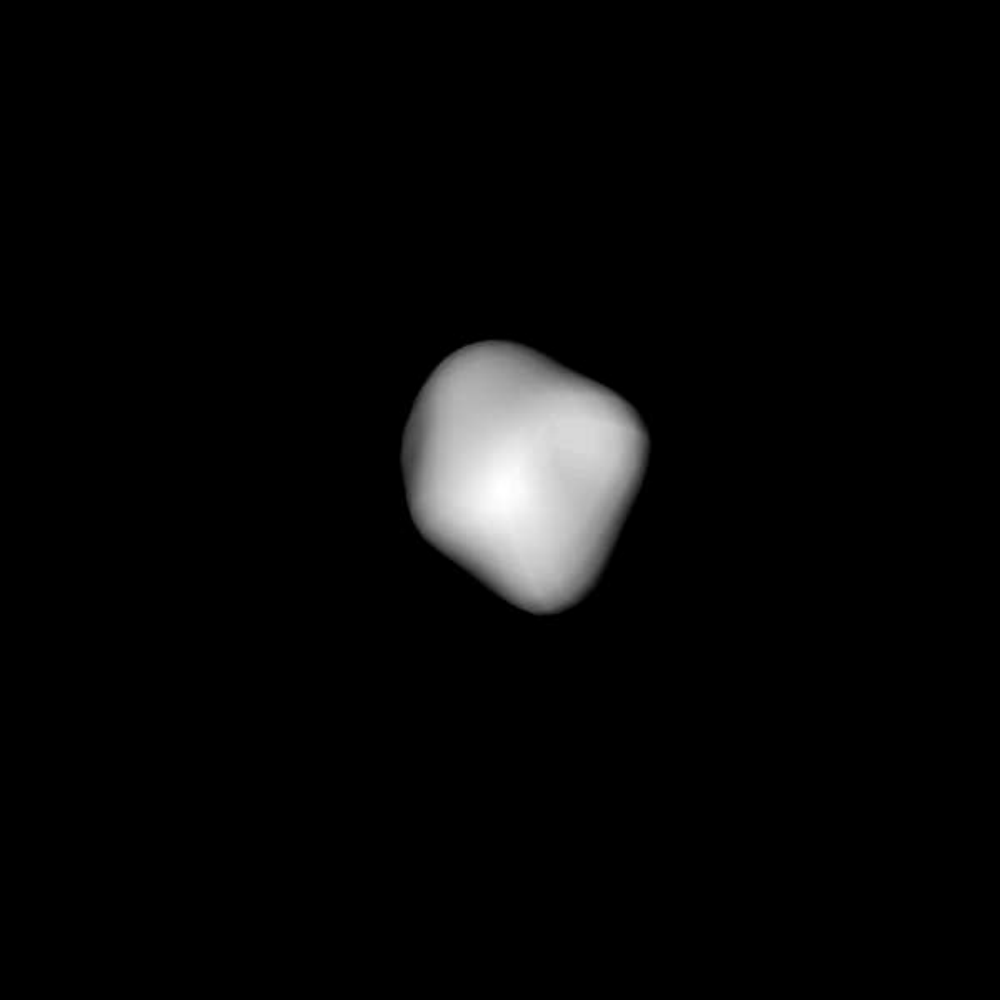}}\\
    \caption{\label{fig:165}Comparison between model projections and corresponding AO images for asteroid (165) Loreley.}
\end{figure}

\clearpage

\begin{figure}[tbp]
    \centering
        \resizebox{0.24\hsize}{!}{\includegraphics{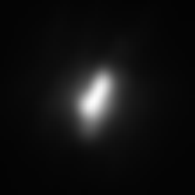}}\resizebox{0.24\hsize}{!}{\includegraphics{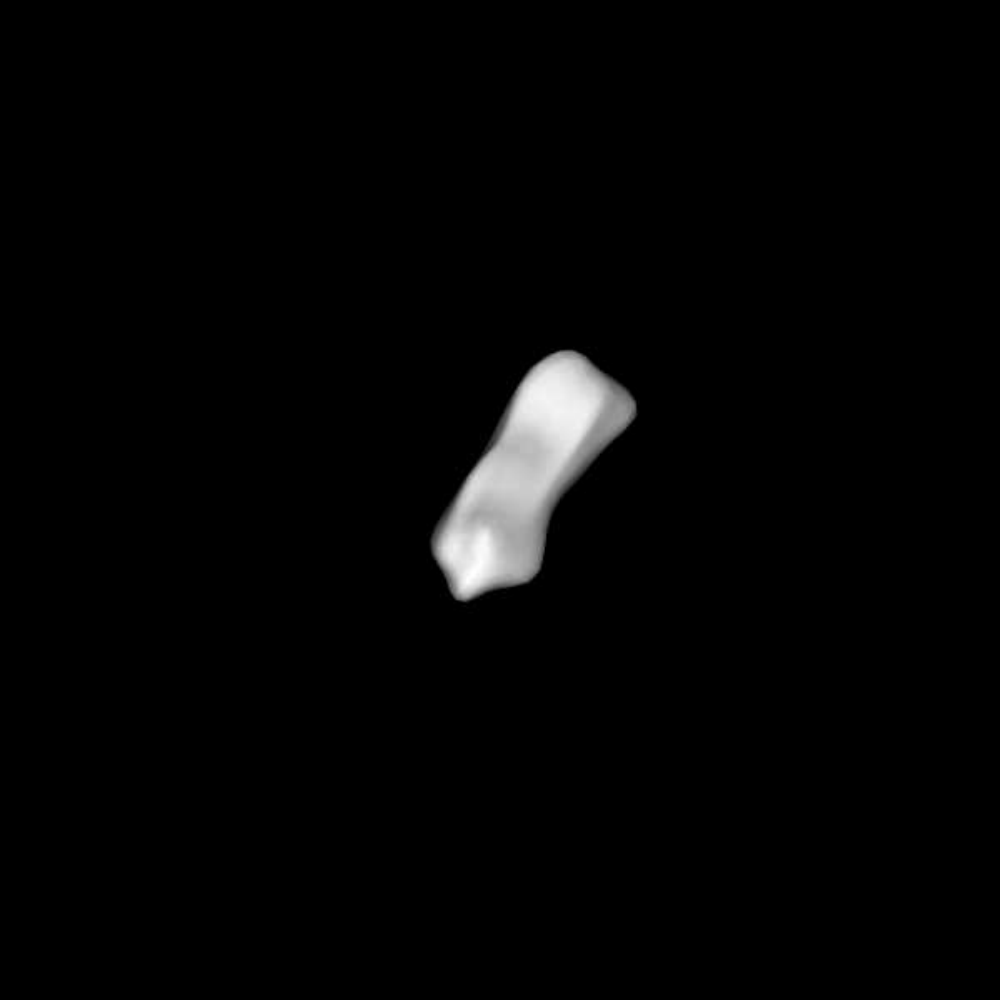}}\resizebox{0.24\hsize}{!}{\includegraphics{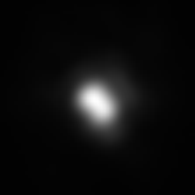}}\resizebox{0.24\hsize}{!}{\includegraphics{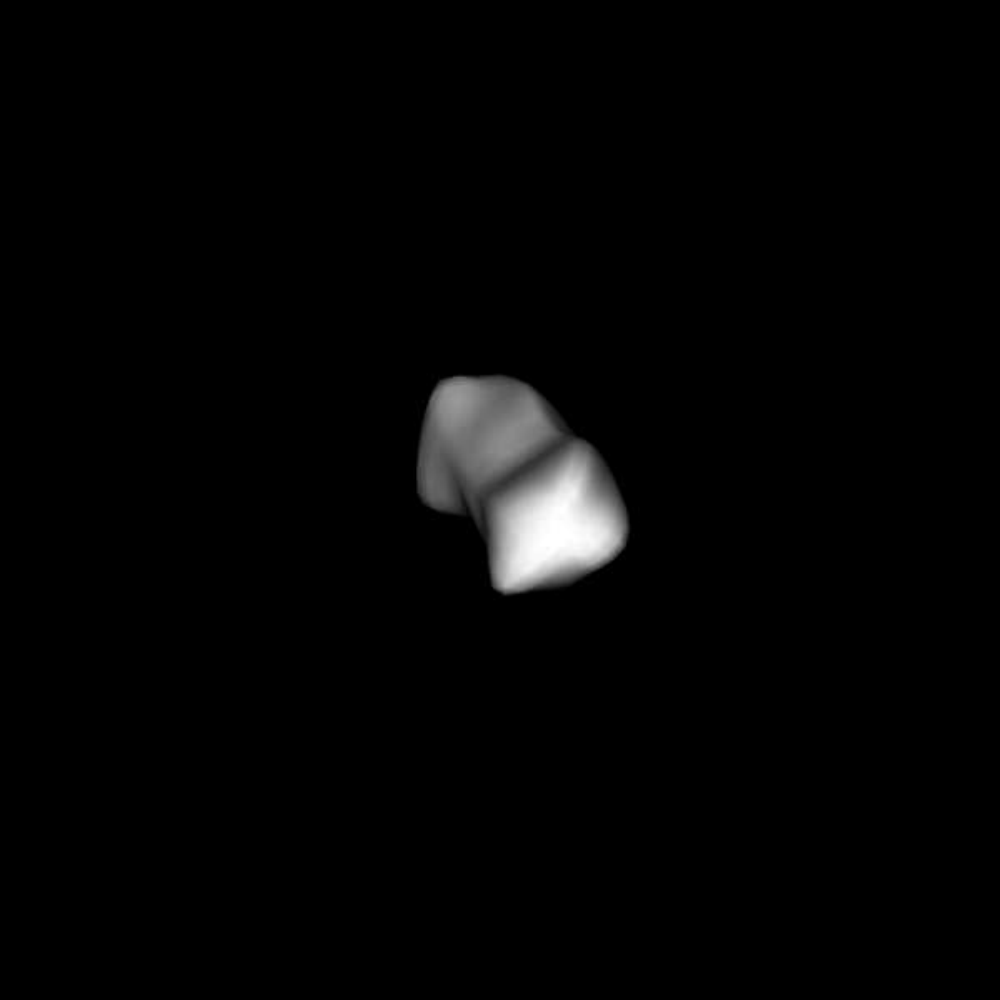}}\\
        \resizebox{0.24\hsize}{!}{\includegraphics{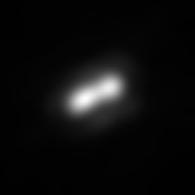}}\resizebox{0.24\hsize}{!}{\includegraphics{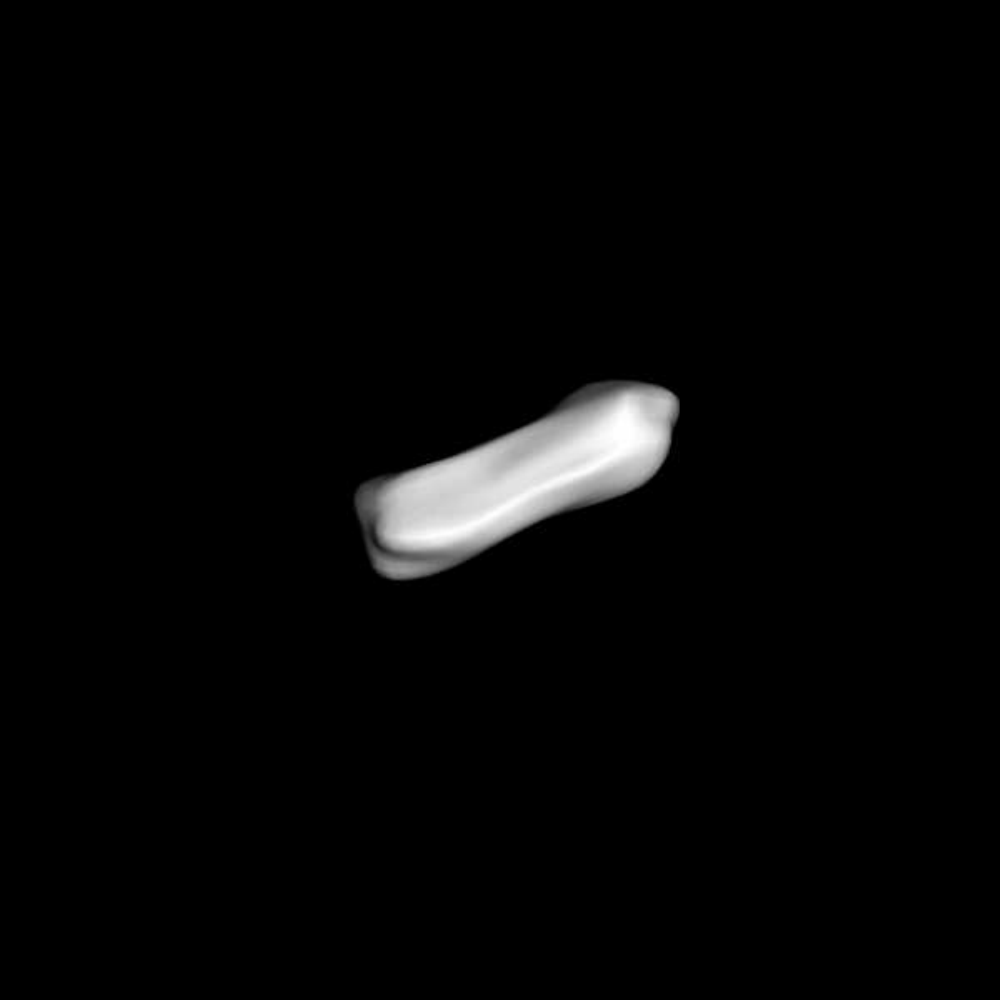}}\resizebox{0.24\hsize}{!}{\includegraphics{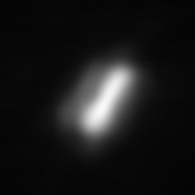}}\resizebox{0.24\hsize}{!}{\includegraphics{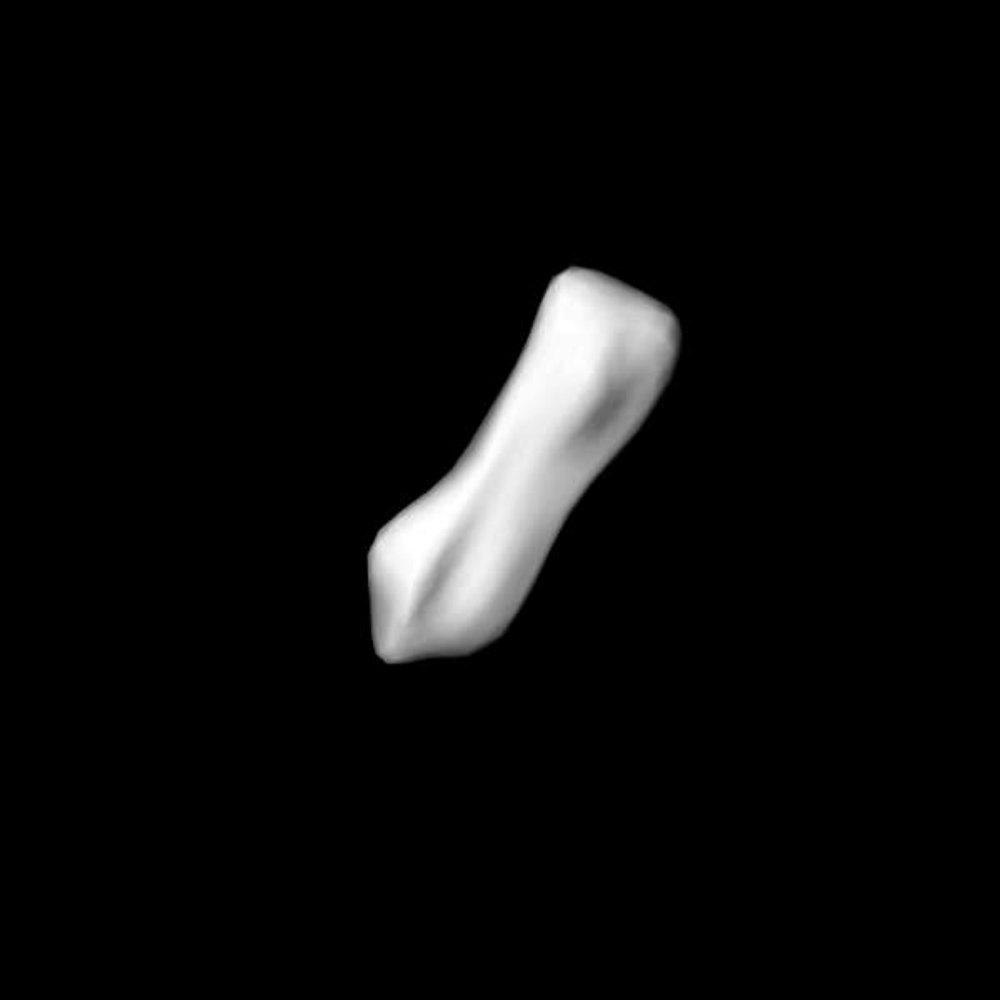}}\\
        \resizebox{0.24\hsize}{!}{\includegraphics{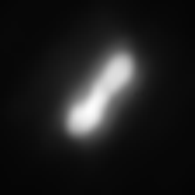}}\resizebox{0.24\hsize}{!}{\includegraphics{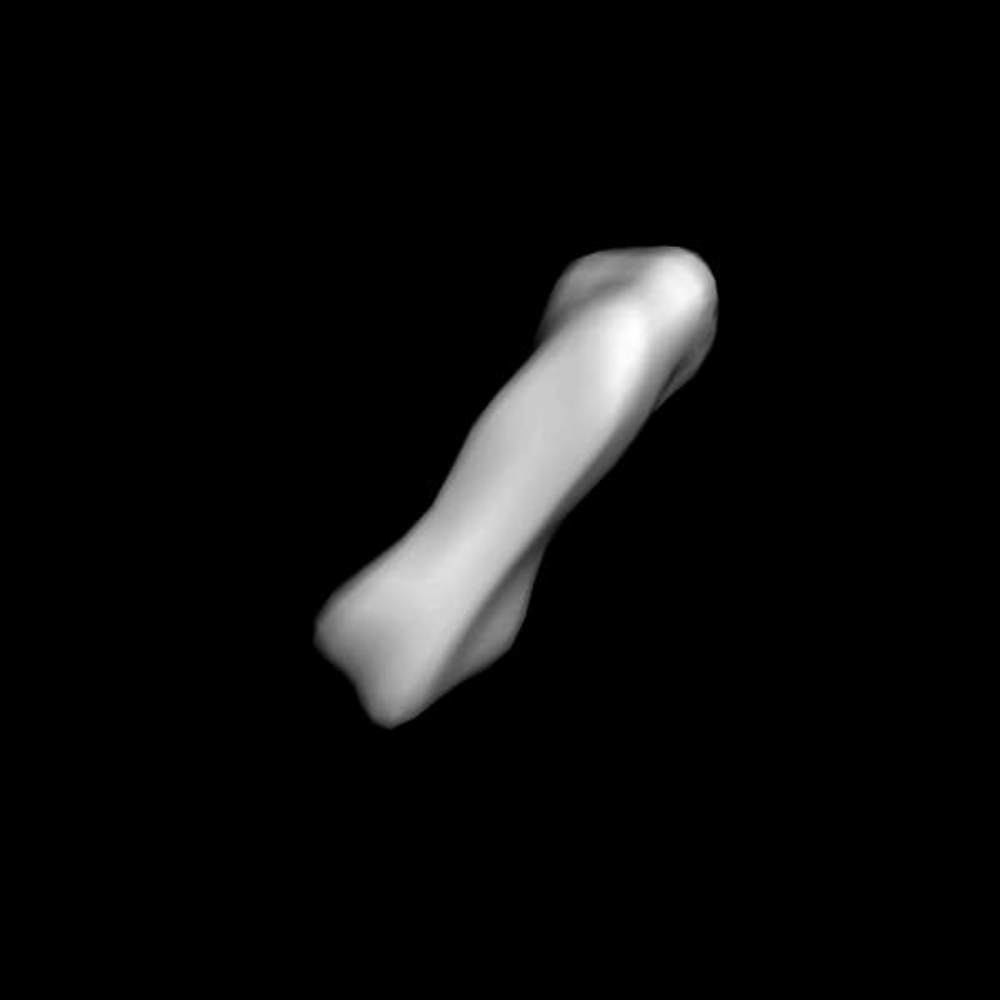}}\resizebox{0.24\hsize}{!}{\includegraphics{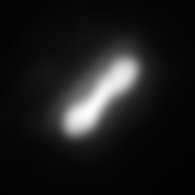}}\resizebox{0.24\hsize}{!}{\includegraphics{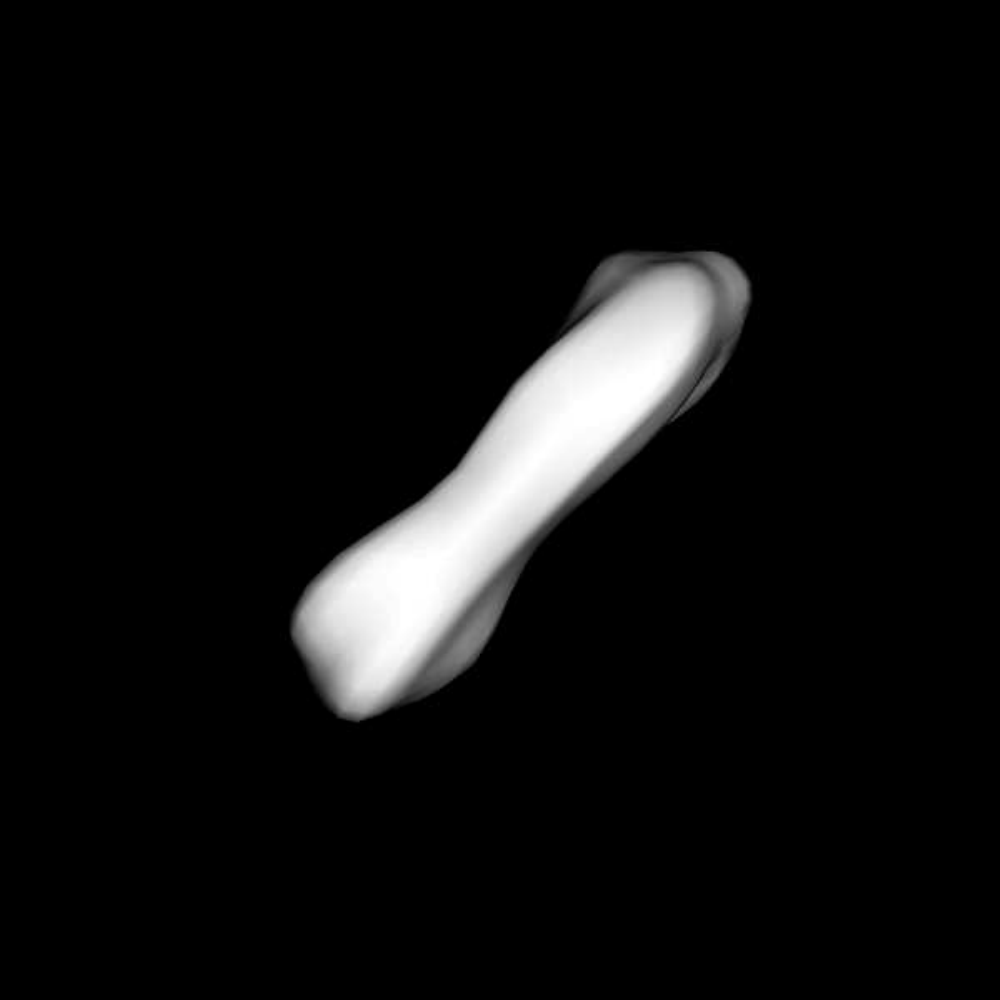}}\\
        \resizebox{0.24\hsize}{!}{\includegraphics{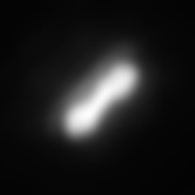}}\resizebox{0.24\hsize}{!}{\includegraphics{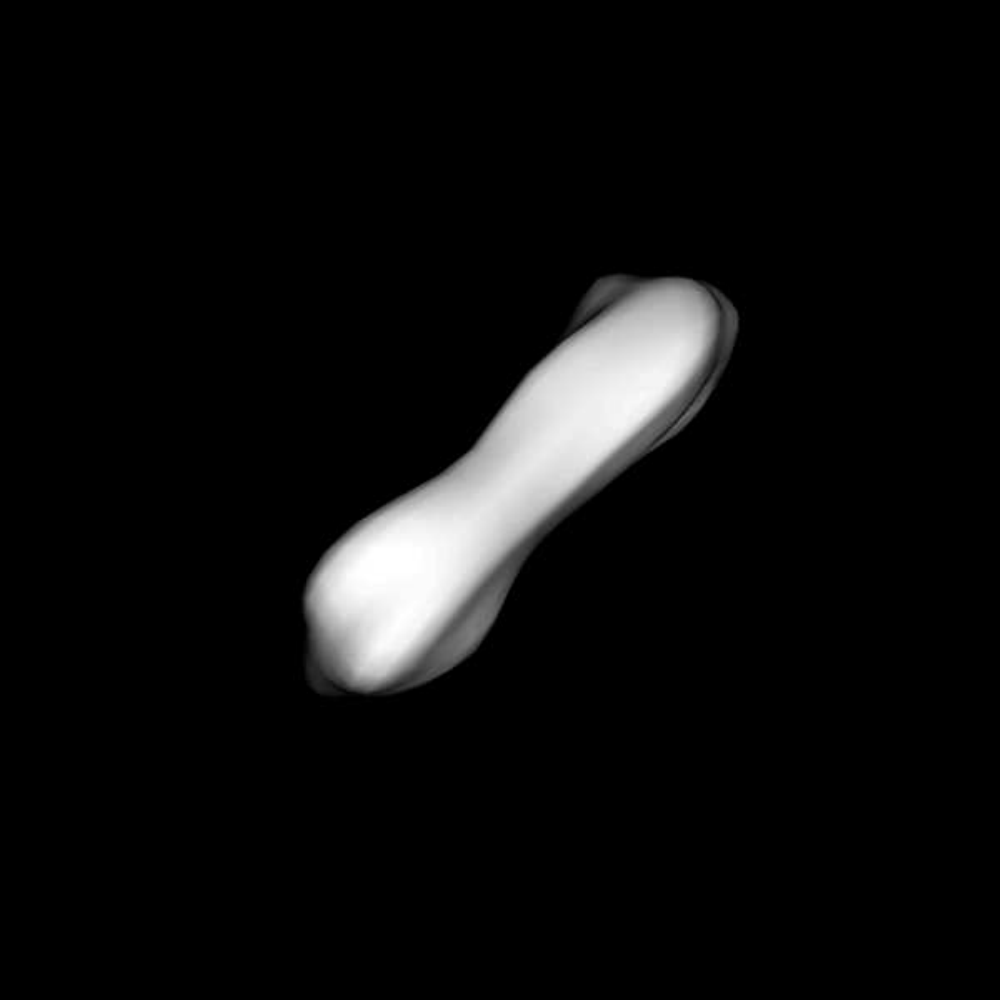}}\resizebox{0.24\hsize}{!}{\includegraphics{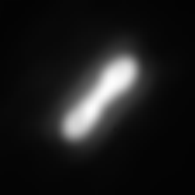}}\resizebox{0.24\hsize}{!}{\includegraphics{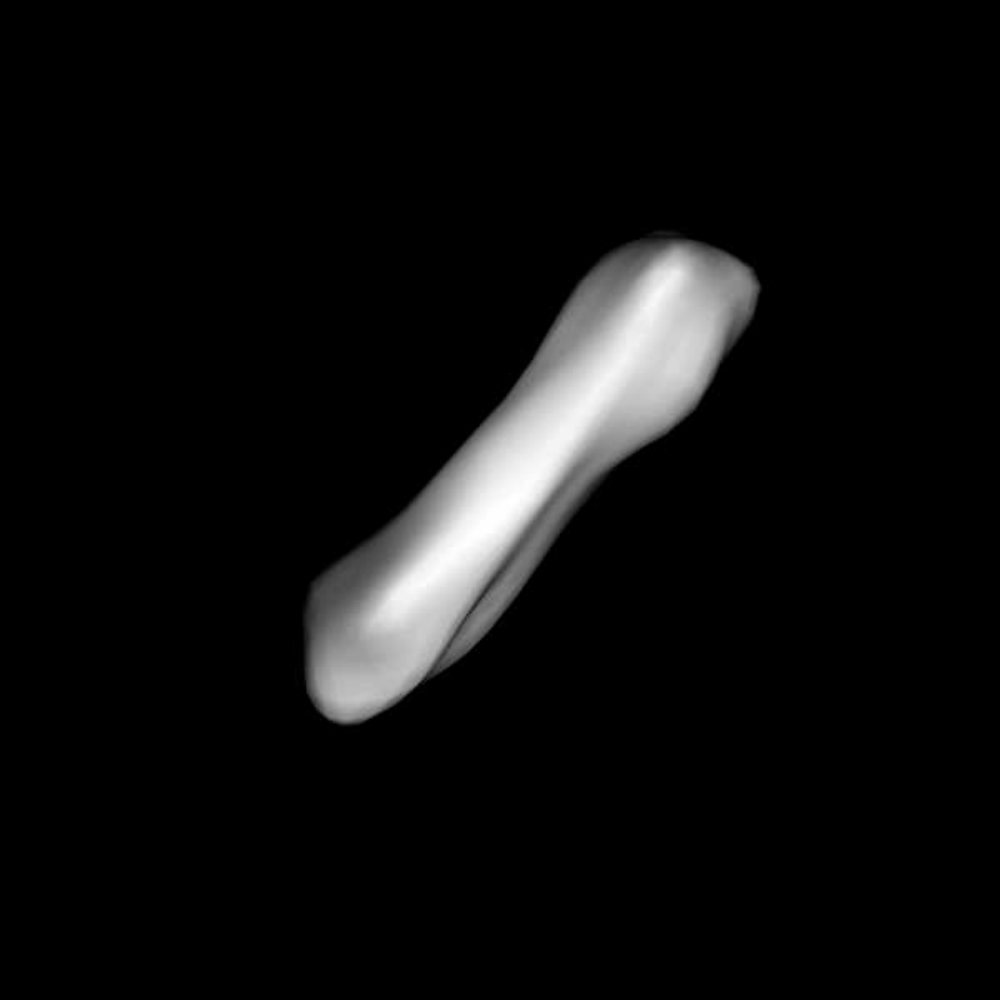}}\\
        \resizebox{0.24\hsize}{!}{\includegraphics{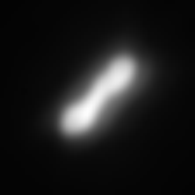}}\resizebox{0.24\hsize}{!}{\includegraphics{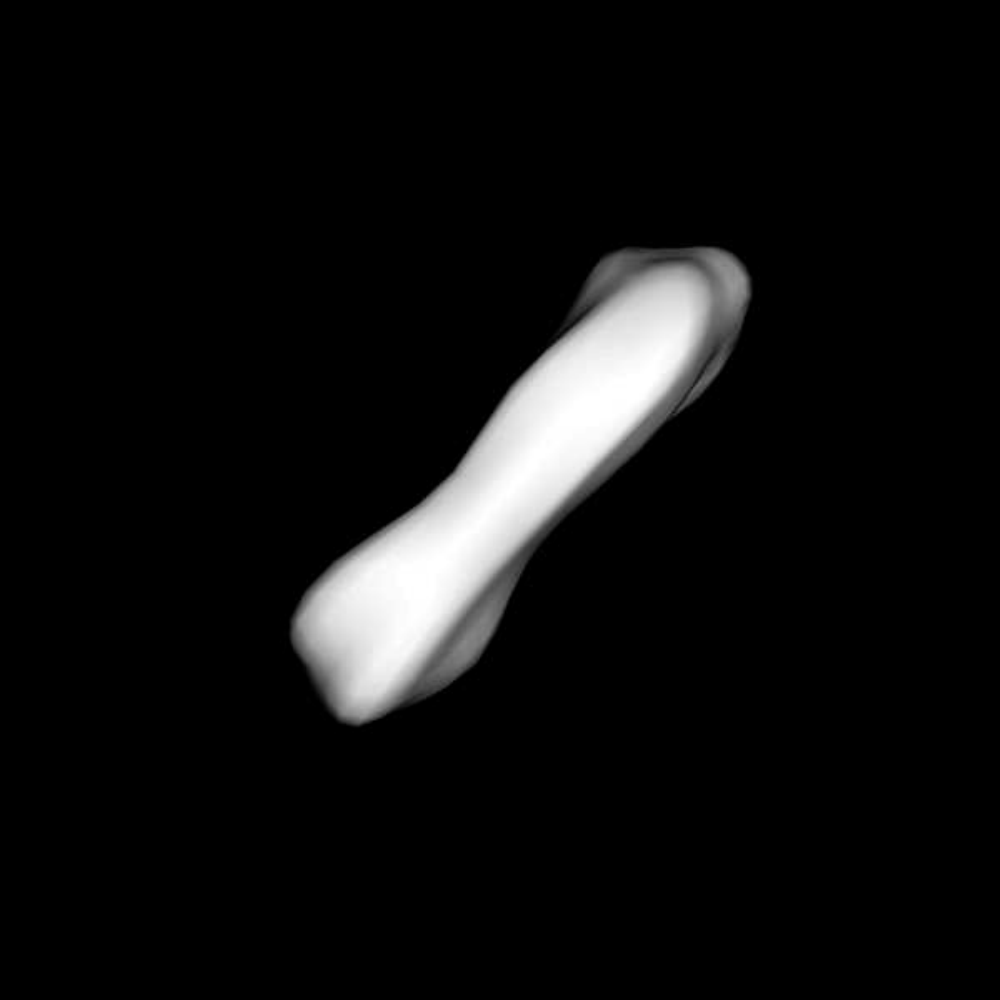}}\resizebox{0.24\hsize}{!}{\includegraphics{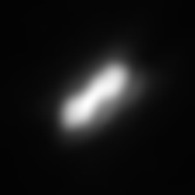}}\resizebox{0.24\hsize}{!}{\includegraphics{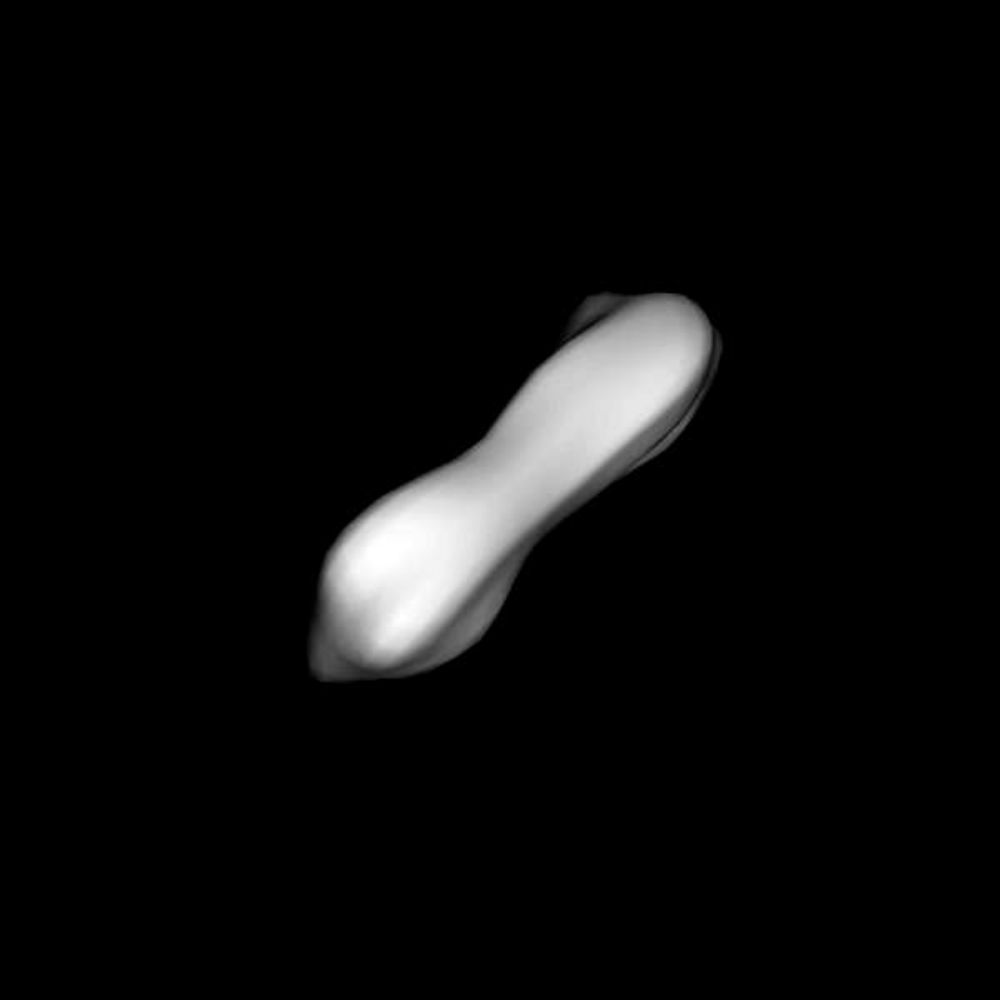}}\\
	\resizebox{0.24\hsize}{!}{\includegraphics{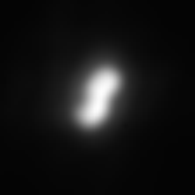}}\resizebox{0.24\hsize}{!}{\includegraphics{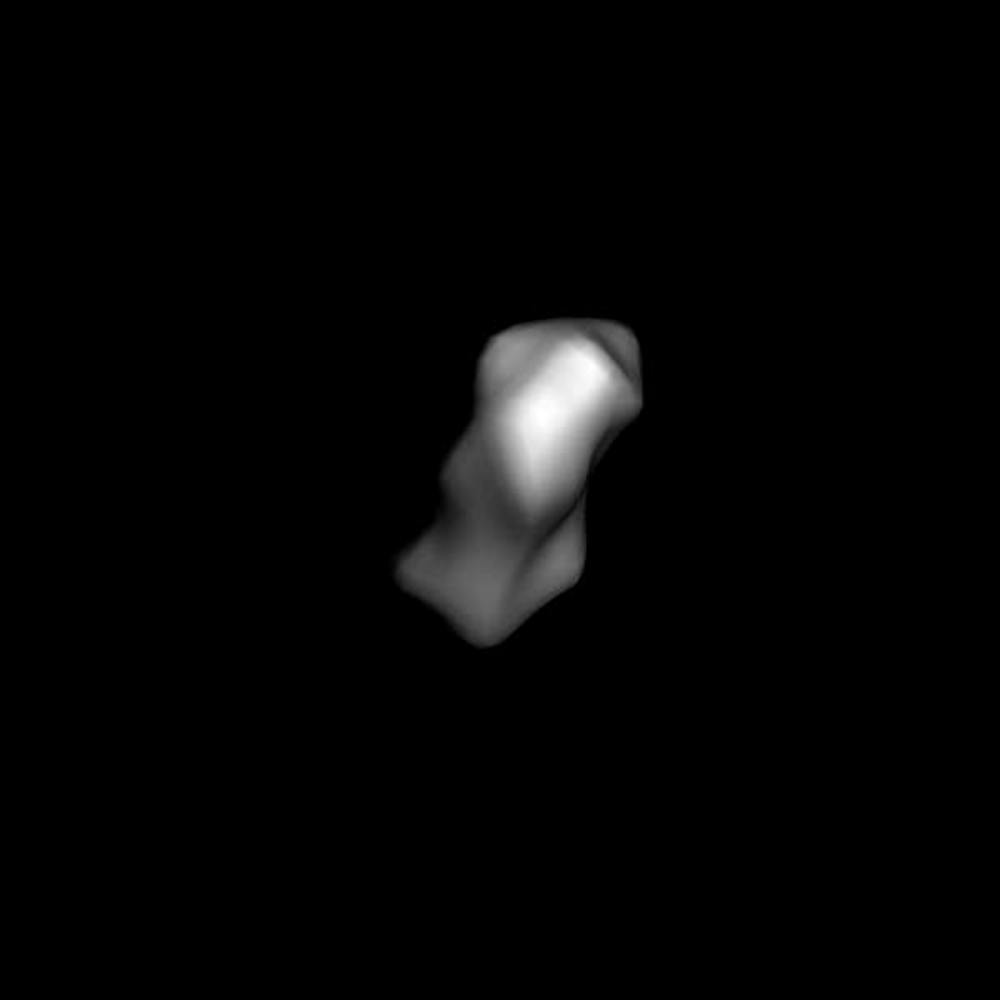}}\resizebox{0.24\hsize}{!}{\includegraphics{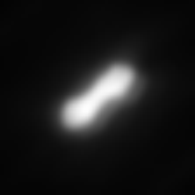}}\resizebox{0.24\hsize}{!}{\includegraphics{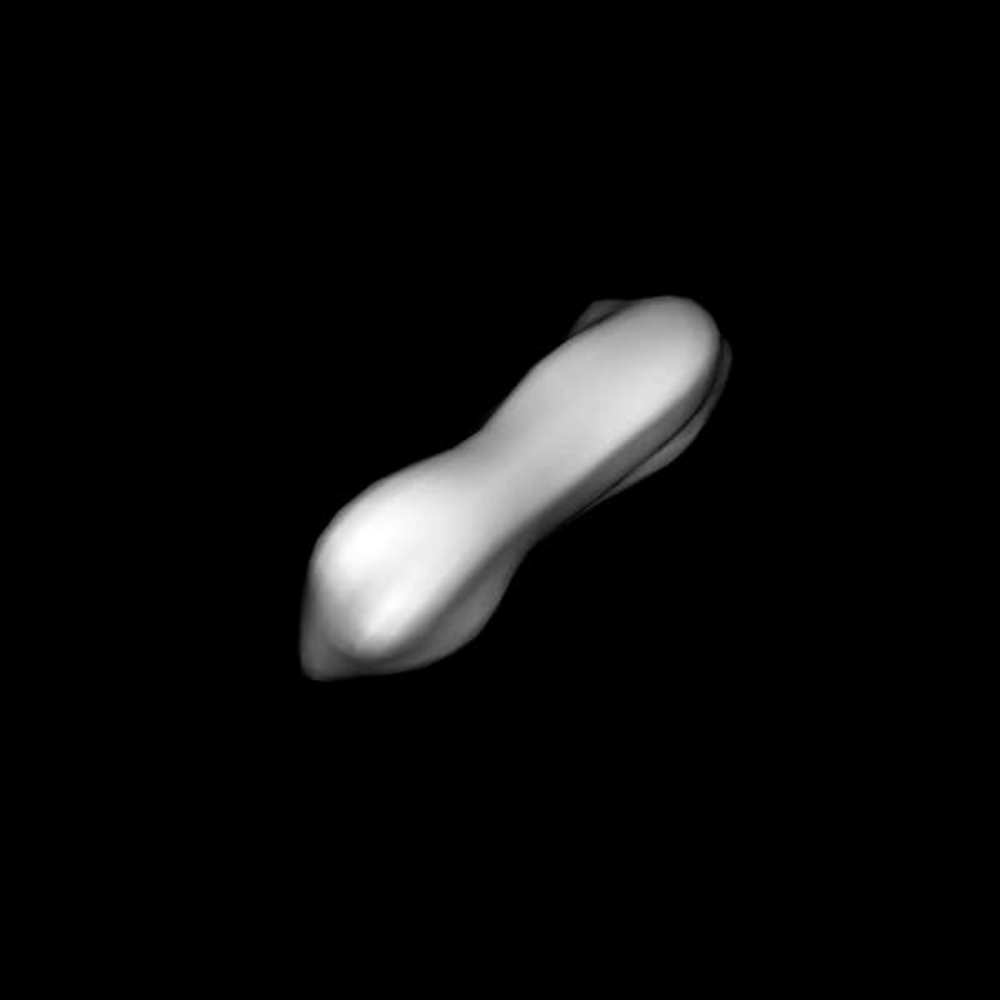}}\\
        \resizebox{0.24\hsize}{!}{\includegraphics{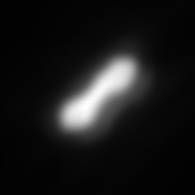}}\resizebox{0.24\hsize}{!}{\includegraphics{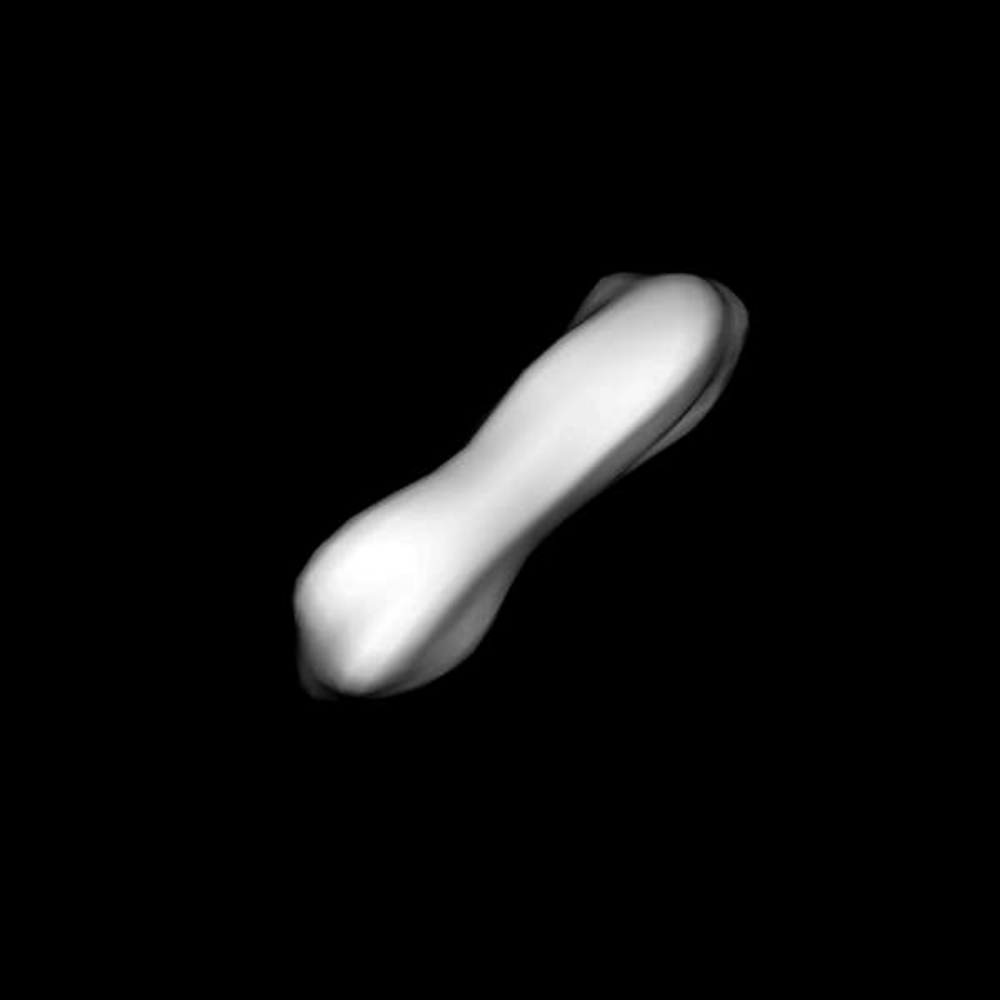}}\resizebox{0.24\hsize}{!}{\includegraphics{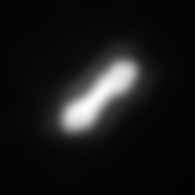}}\resizebox{0.24\hsize}{!}{\includegraphics{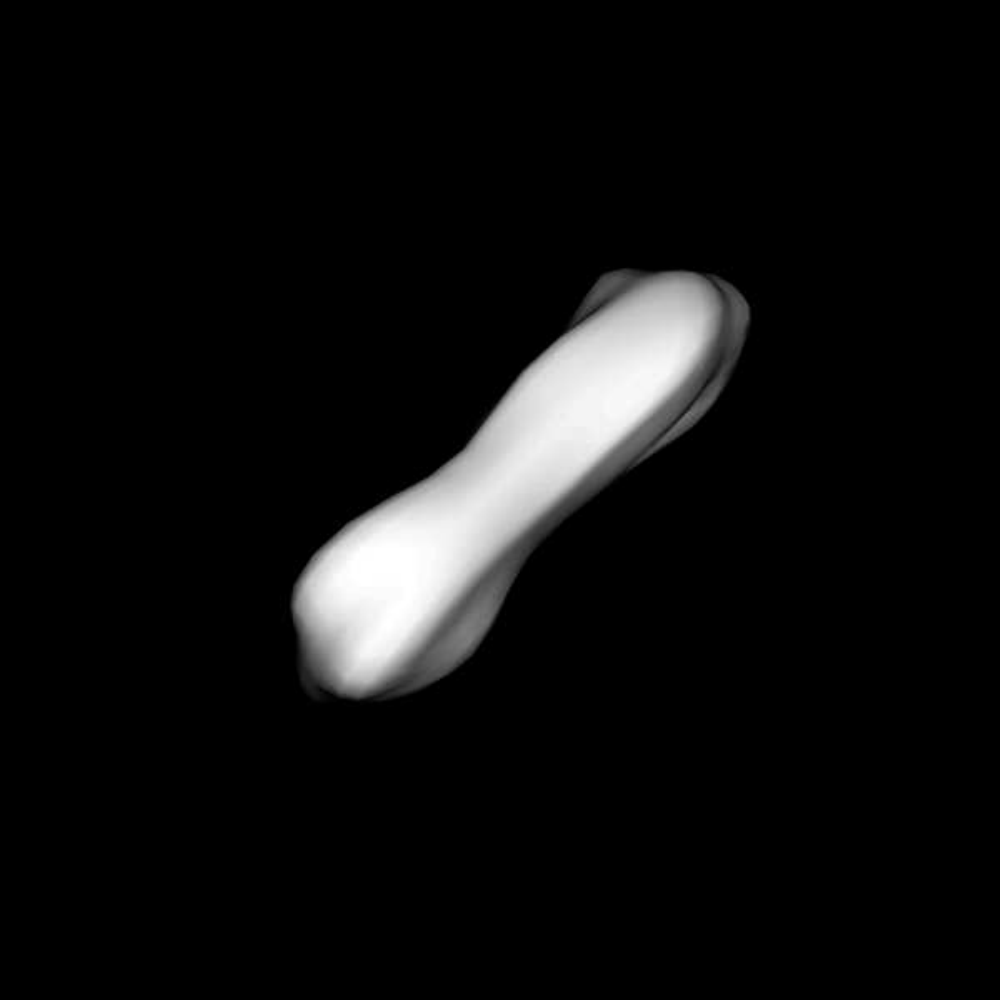}}\\
    \caption{\label{fig:216}Comparison between model projections and corresponding AO images for asteroid (216) Kleopatra.}
\end{figure}

\begin{figure}[tbp]
    \centering
        \resizebox{0.24\hsize}{!}{\includegraphics{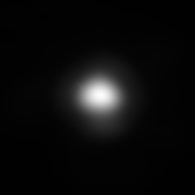}}\resizebox{0.24\hsize}{!}{\includegraphics{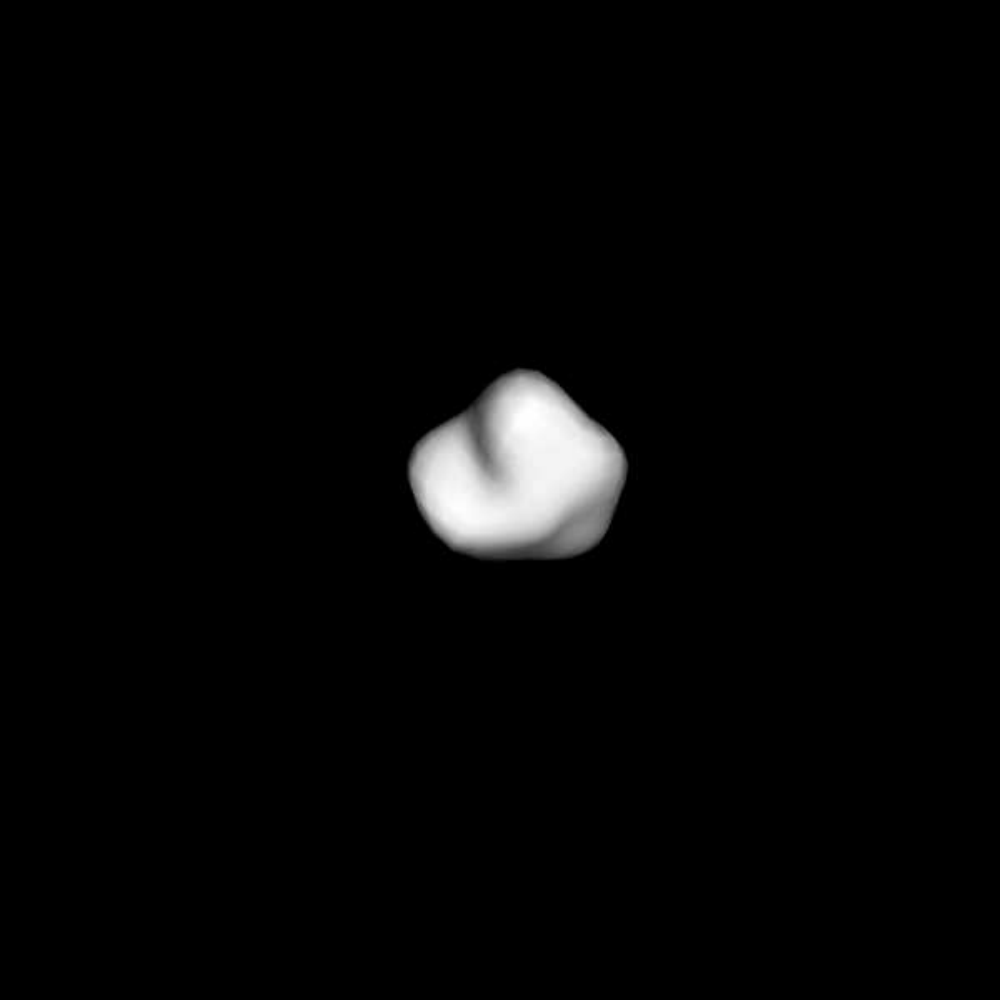}}\\
    \caption{\label{fig:233}Comparison between model projections and corresponding AO images for asteroid (233) Asterope.}
\end{figure}

\begin{figure}[tbp]
    \centering
        \resizebox{0.24\hsize}{!}{\includegraphics{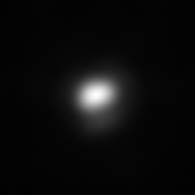}}\resizebox{0.24\hsize}{!}{\includegraphics{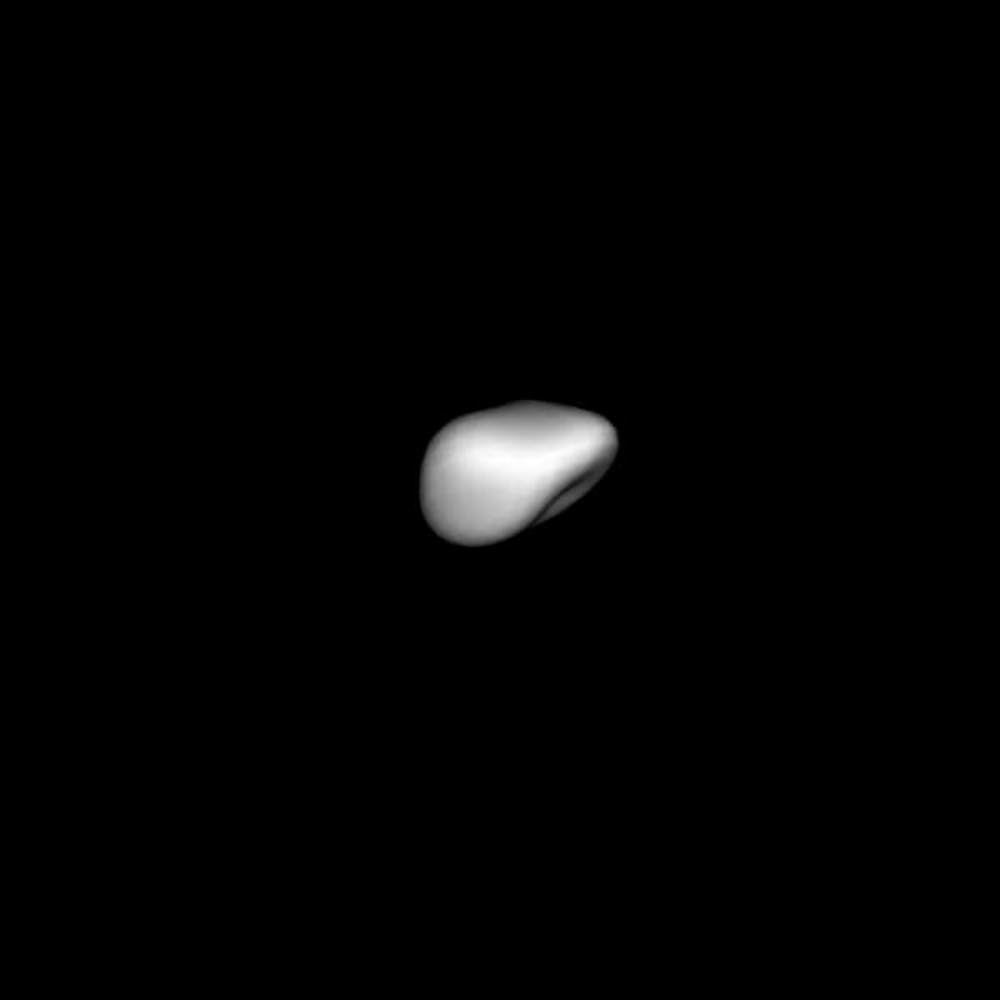}}\resizebox{0.24\hsize}{!}{\includegraphics{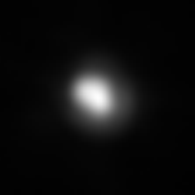}}\resizebox{0.24\hsize}{!}{\includegraphics{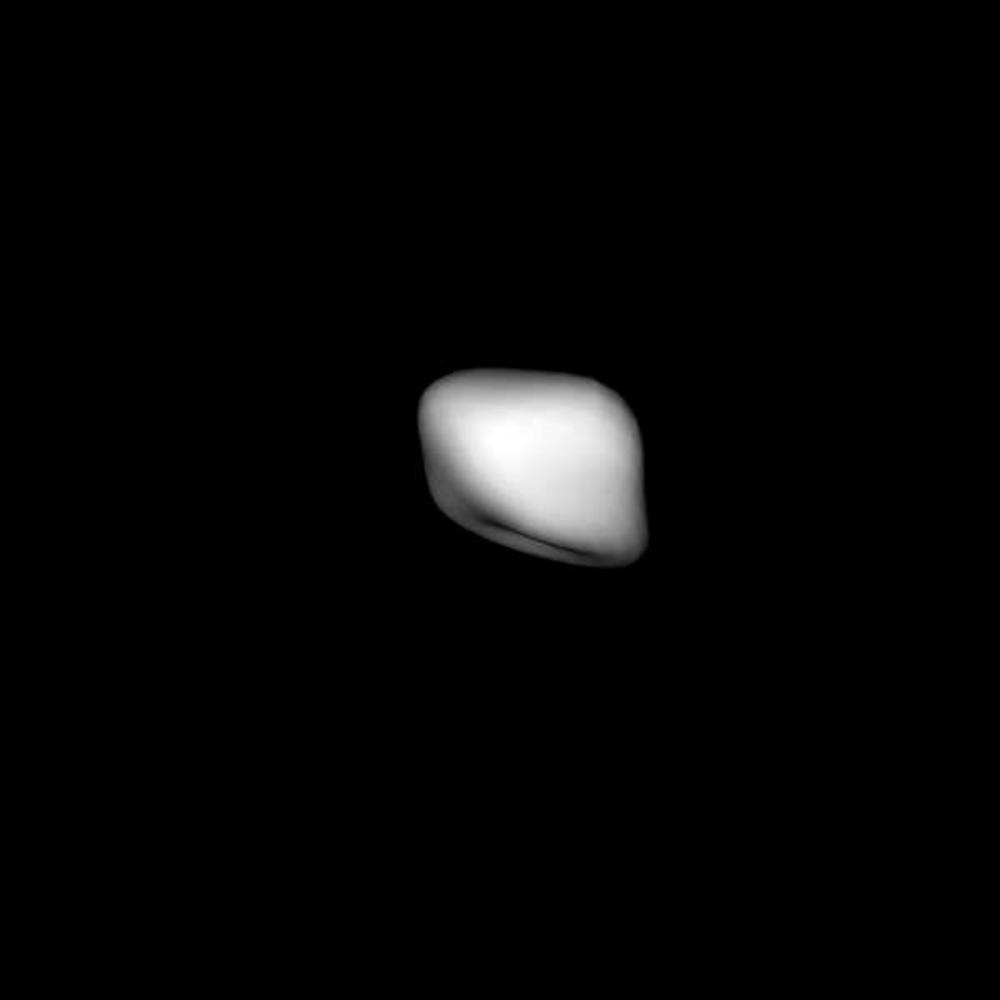}}\\
    \caption{\label{fig:360}Comparison between model projections and corresponding AO images for asteroid (360) Carlova.}
\end{figure}

\begin{figure}[tbp]
    \centering
        \resizebox{0.24\hsize}{!}{\includegraphics{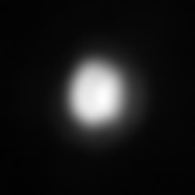}}\resizebox{0.24\hsize}{!}{\includegraphics{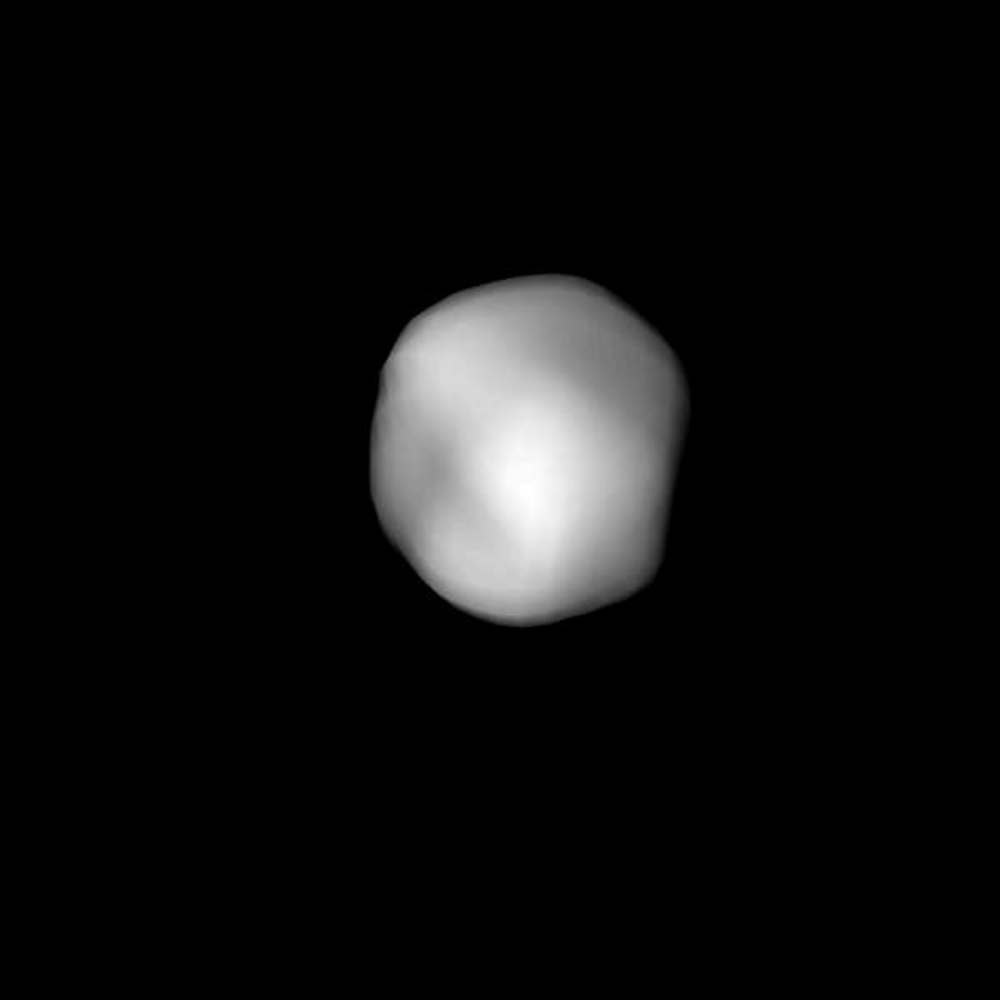}}\\
    \caption{\label{fig:386}Comparison between model projections and corresponding AO images for asteroid (386) Siegena.}
\end{figure}

\begin{figure}[tbp]
    \centering
        \resizebox{0.24\hsize}{!}{\includegraphics{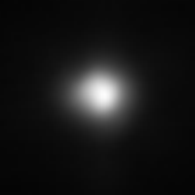}}\resizebox{0.24\hsize}{!}{\includegraphics{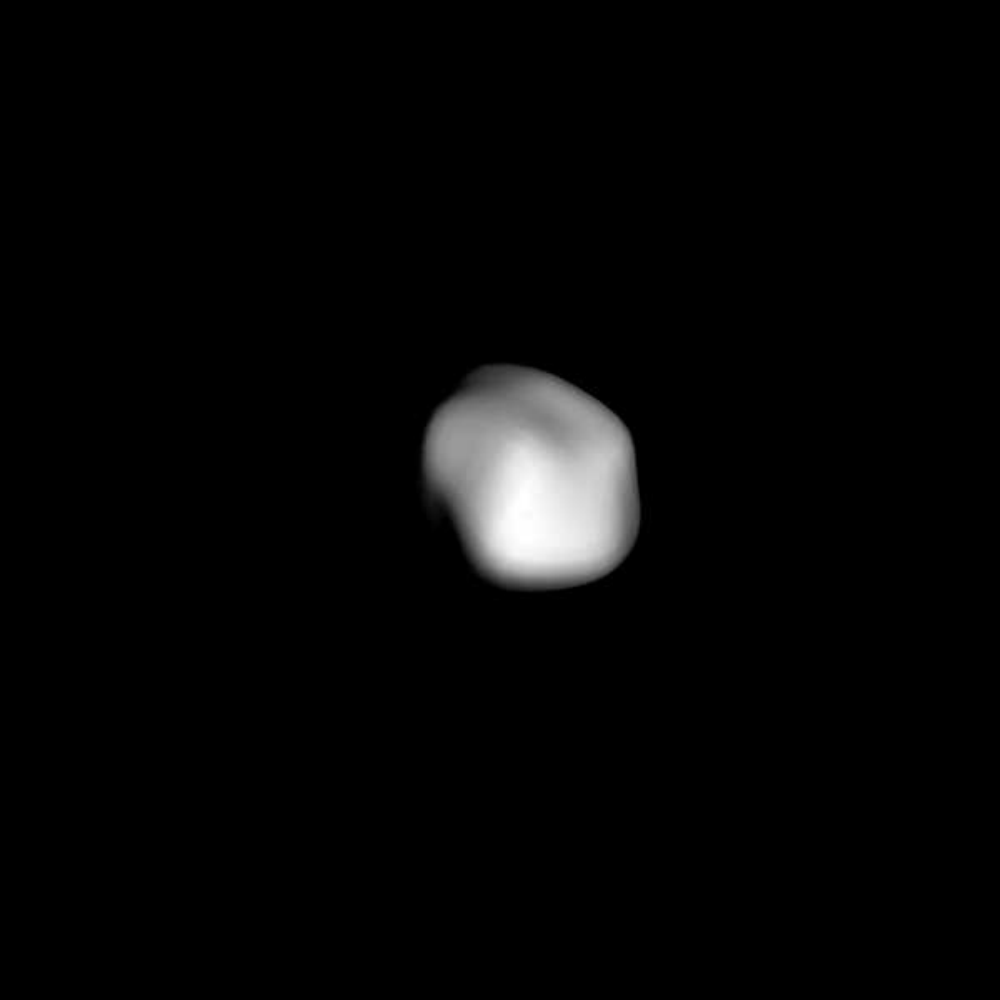}}\resizebox{0.24\hsize}{!}{\includegraphics{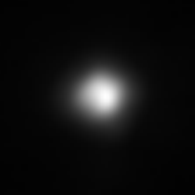}}\resizebox{0.24\hsize}{!}{\includegraphics{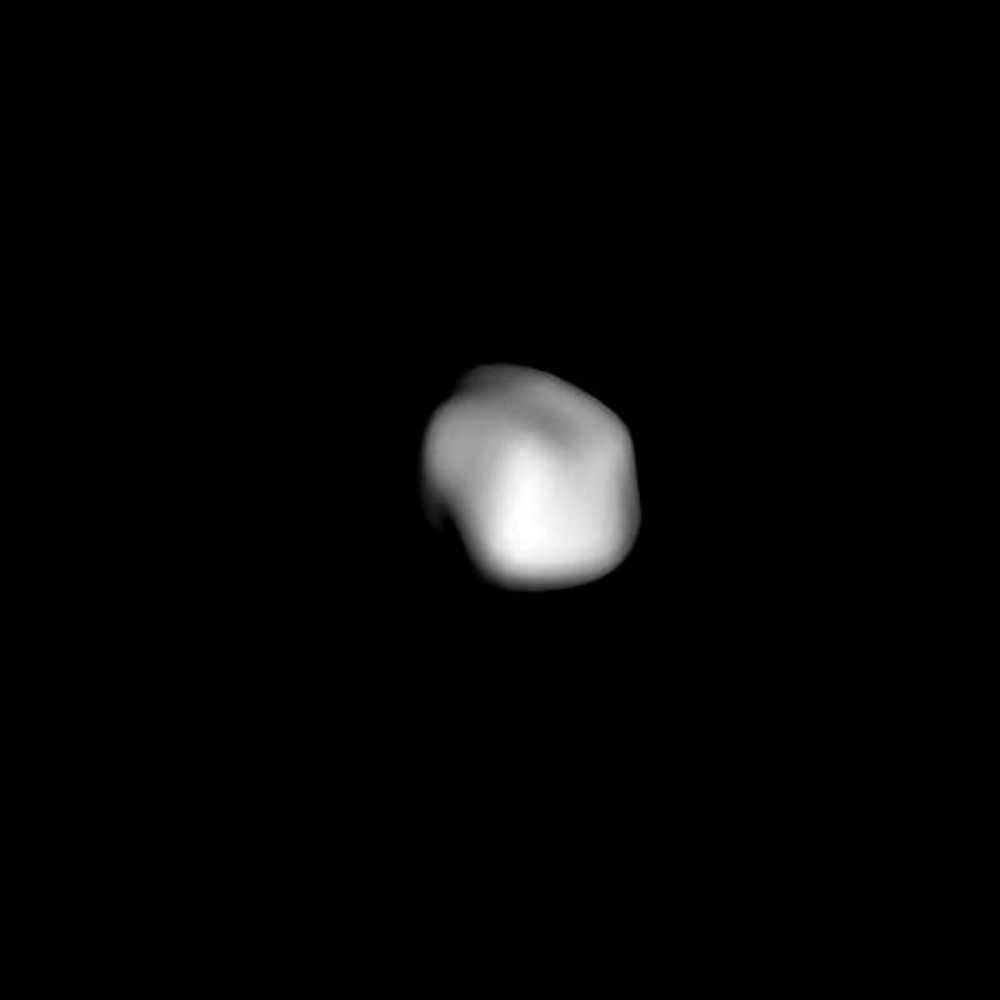}}\\
        \resizebox{0.24\hsize}{!}{\includegraphics{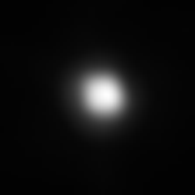}}\resizebox{0.24\hsize}{!}{\includegraphics{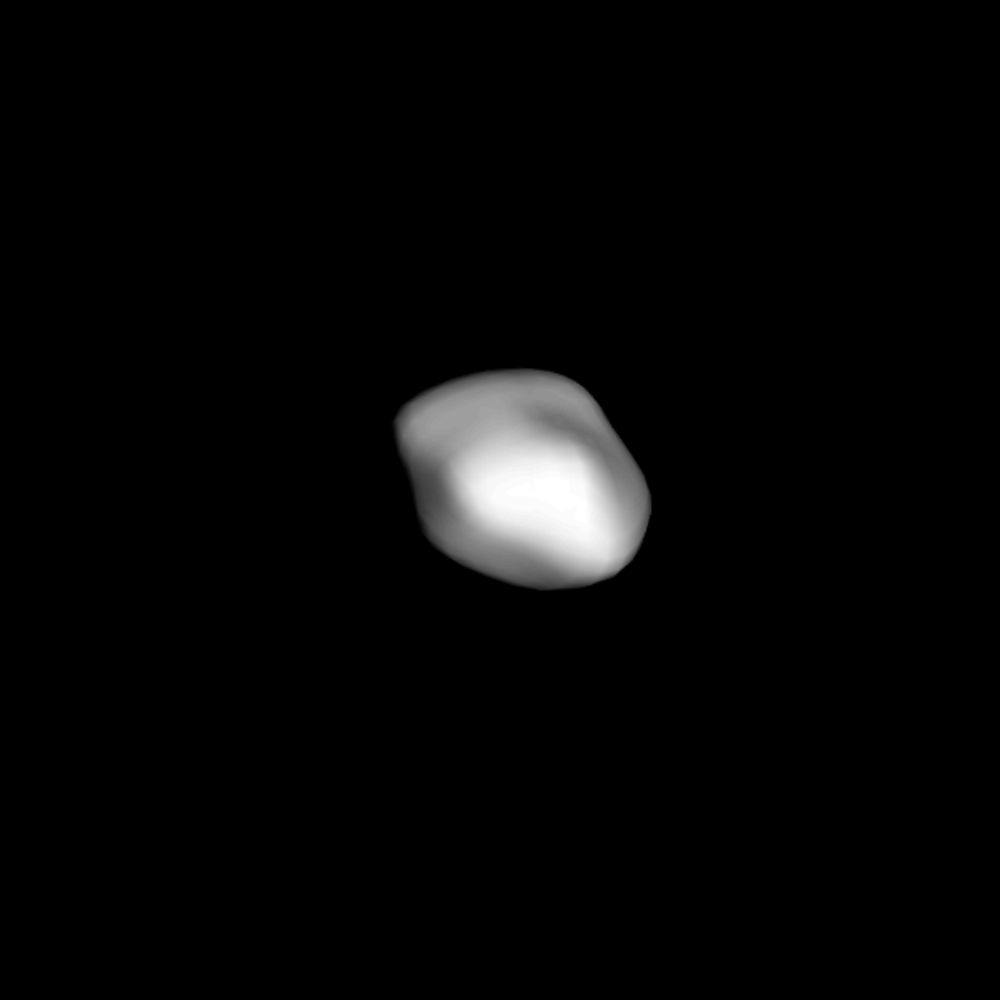}}\resizebox{0.24\hsize}{!}{\includegraphics{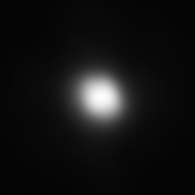}}\resizebox{0.24\hsize}{!}{\includegraphics{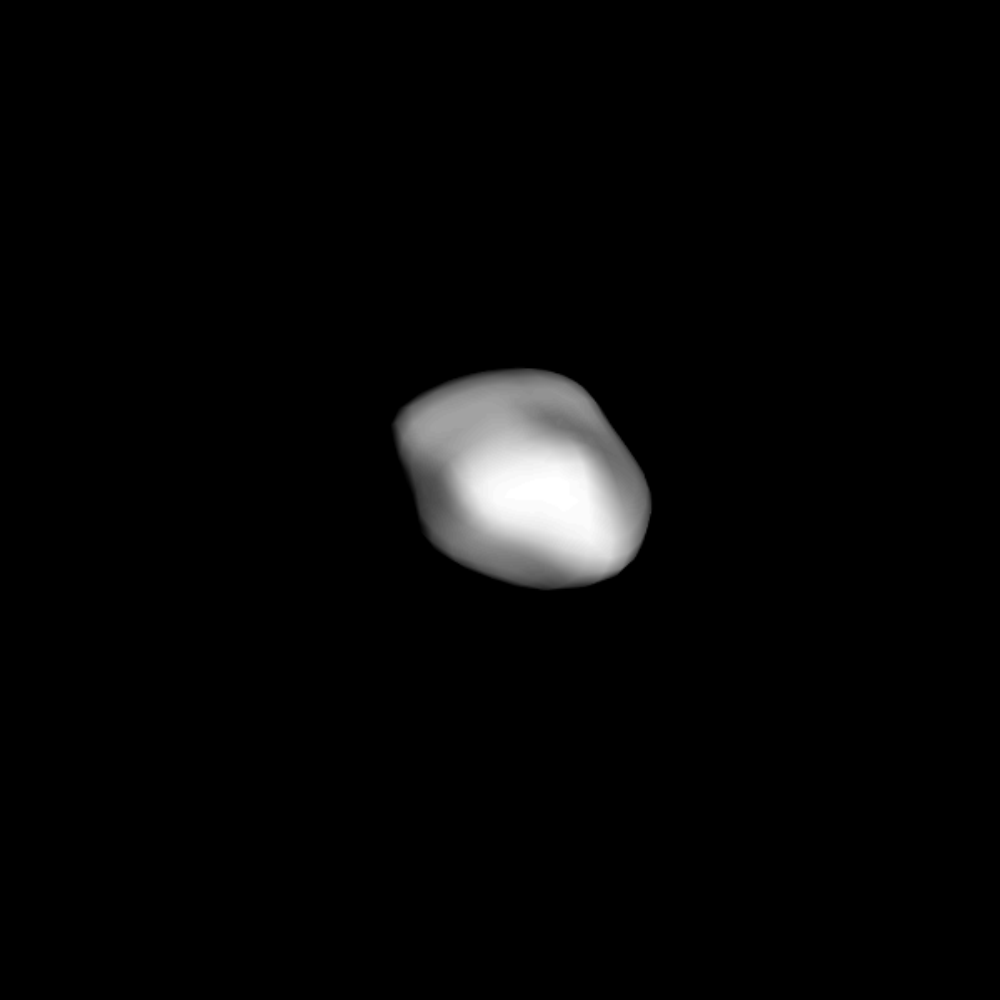}}\\
        \resizebox{0.24\hsize}{!}{\includegraphics{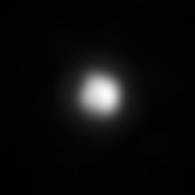}}\resizebox{0.24\hsize}{!}{\includegraphics{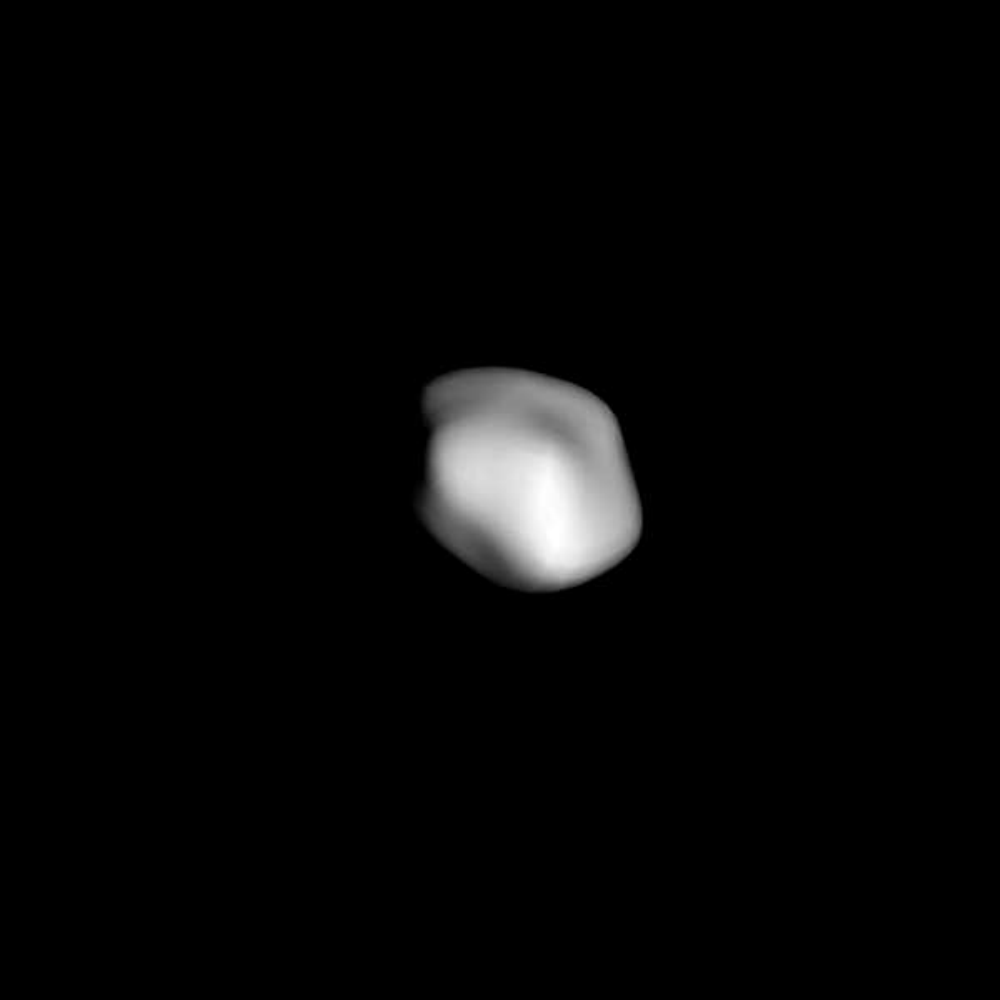}}\resizebox{0.24\hsize}{!}{\includegraphics{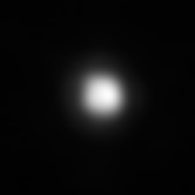}}\resizebox{0.24\hsize}{!}{\includegraphics{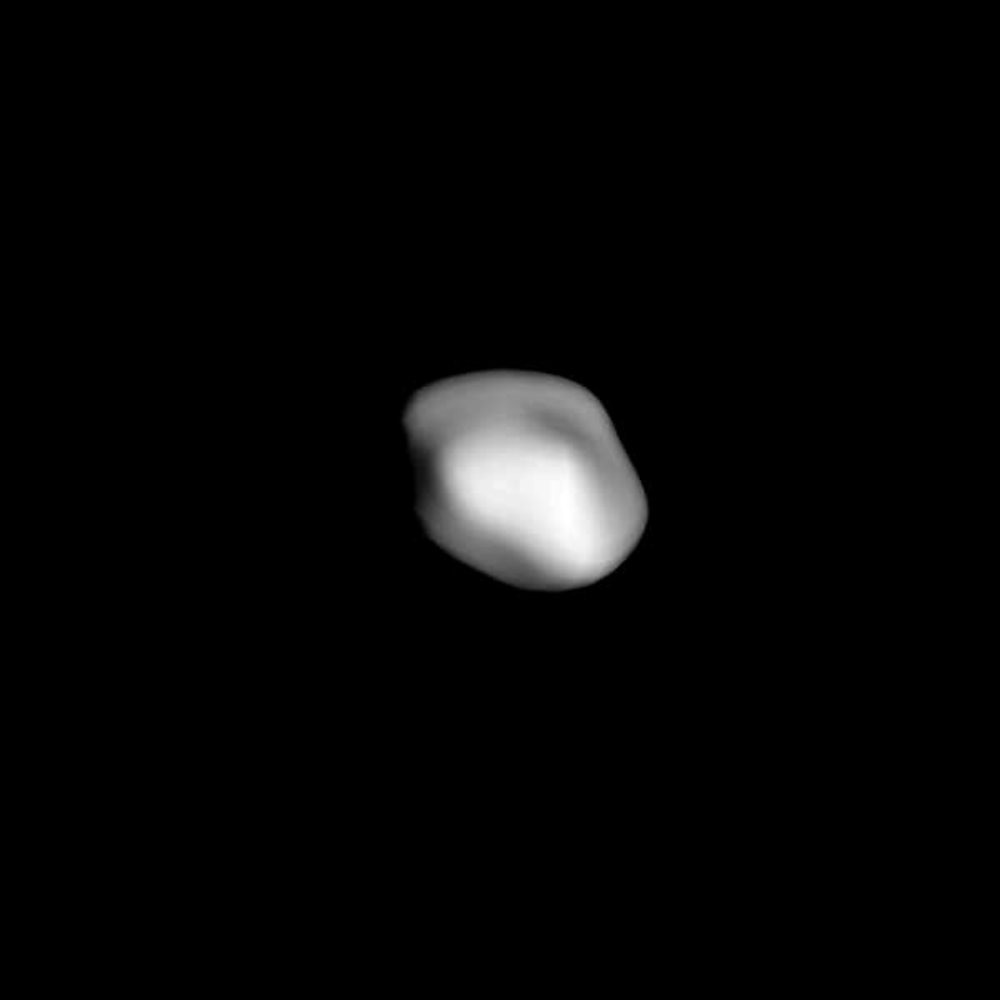}}\\
        \resizebox{0.24\hsize}{!}{\includegraphics{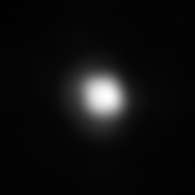}}\resizebox{0.24\hsize}{!}{\includegraphics{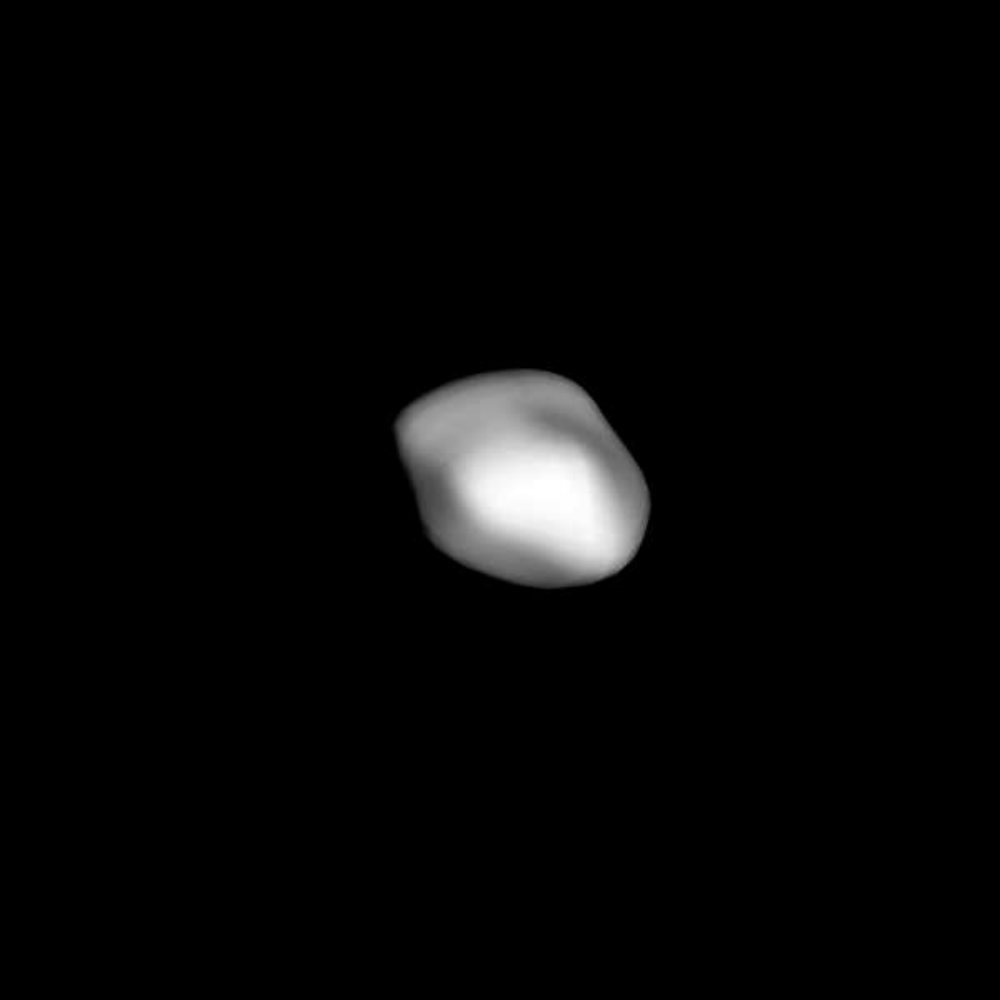}}\\
    \caption{\label{fig:387}Comparison between model projections and corresponding AO images for asteroid (387) Aquitania.}
\end{figure}

\begin{figure}[tbp]
    \centering
        \resizebox{0.24\hsize}{!}{\includegraphics{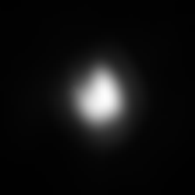}}\resizebox{0.24\hsize}{!}{\includegraphics{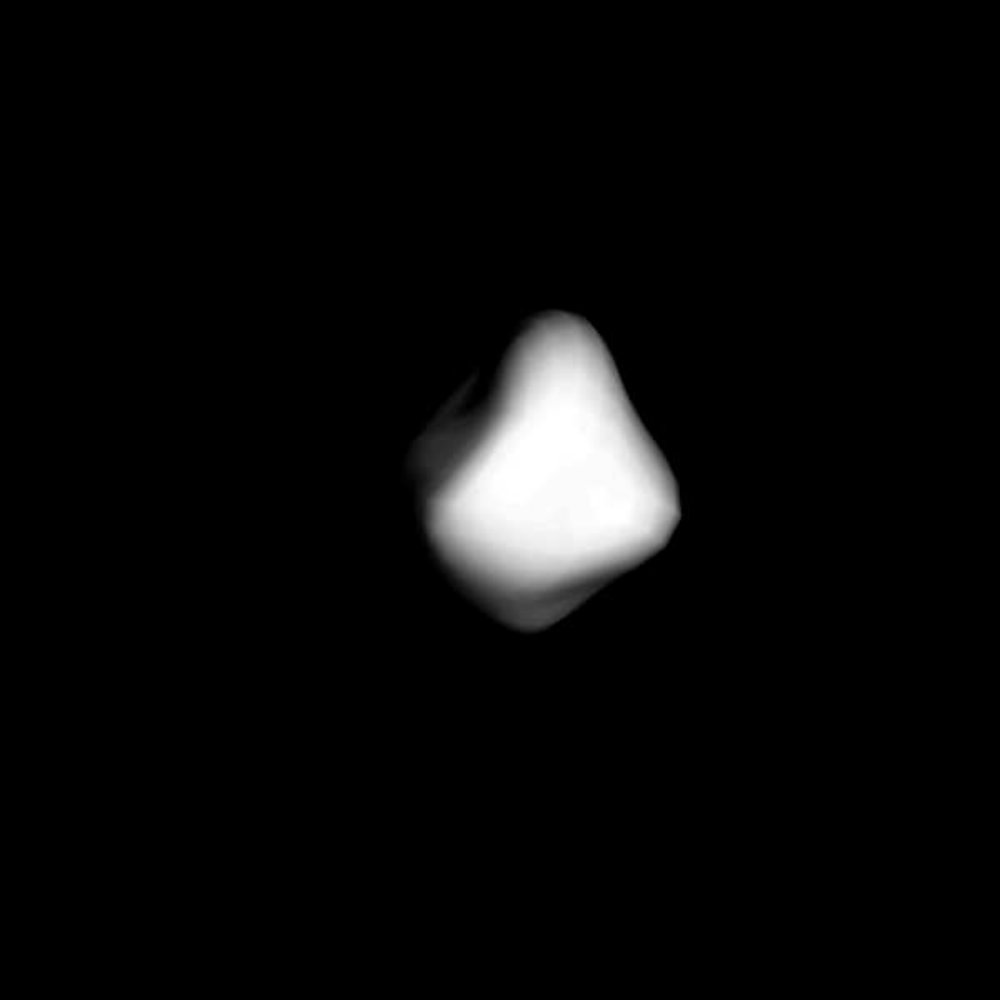}}\resizebox{0.24\hsize}{!}{\includegraphics{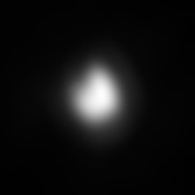}}\resizebox{0.24\hsize}{!}{\includegraphics{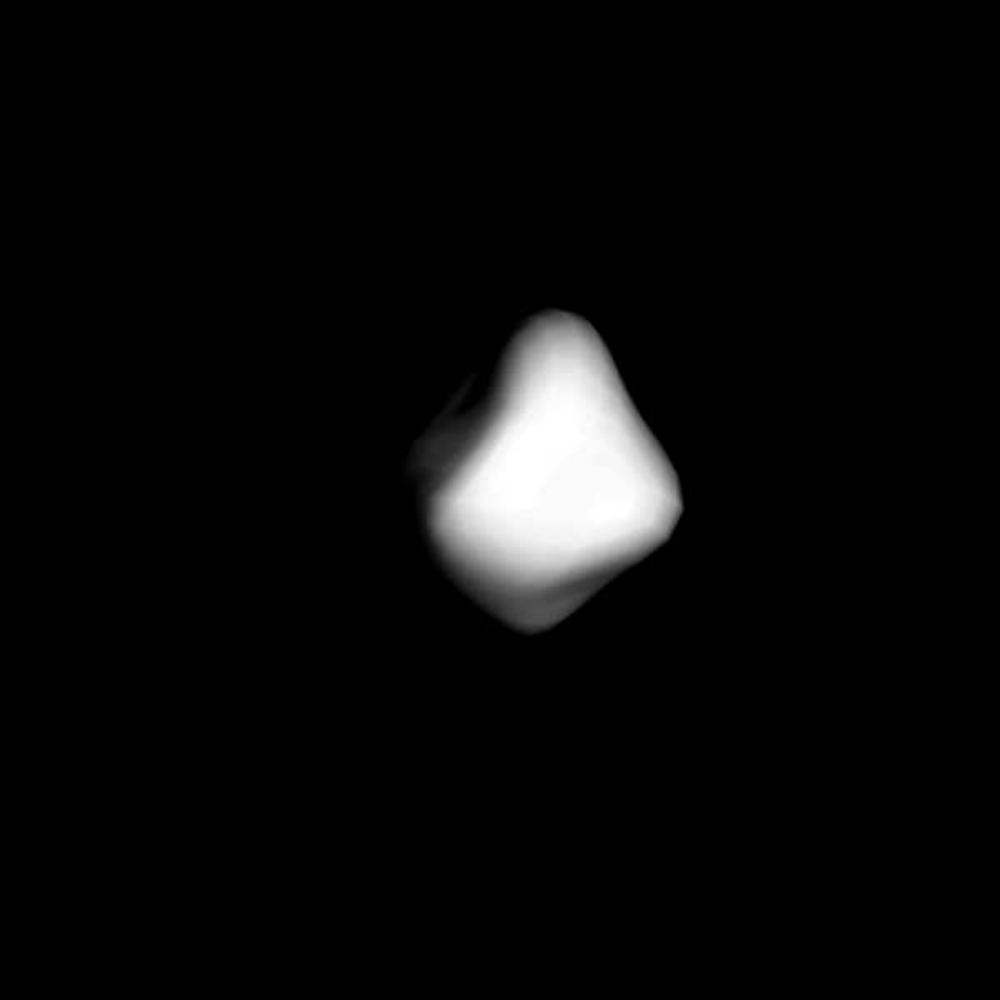}}\\
        \resizebox{0.24\hsize}{!}{\includegraphics{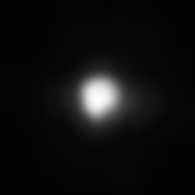}}\resizebox{0.24\hsize}{!}{\includegraphics{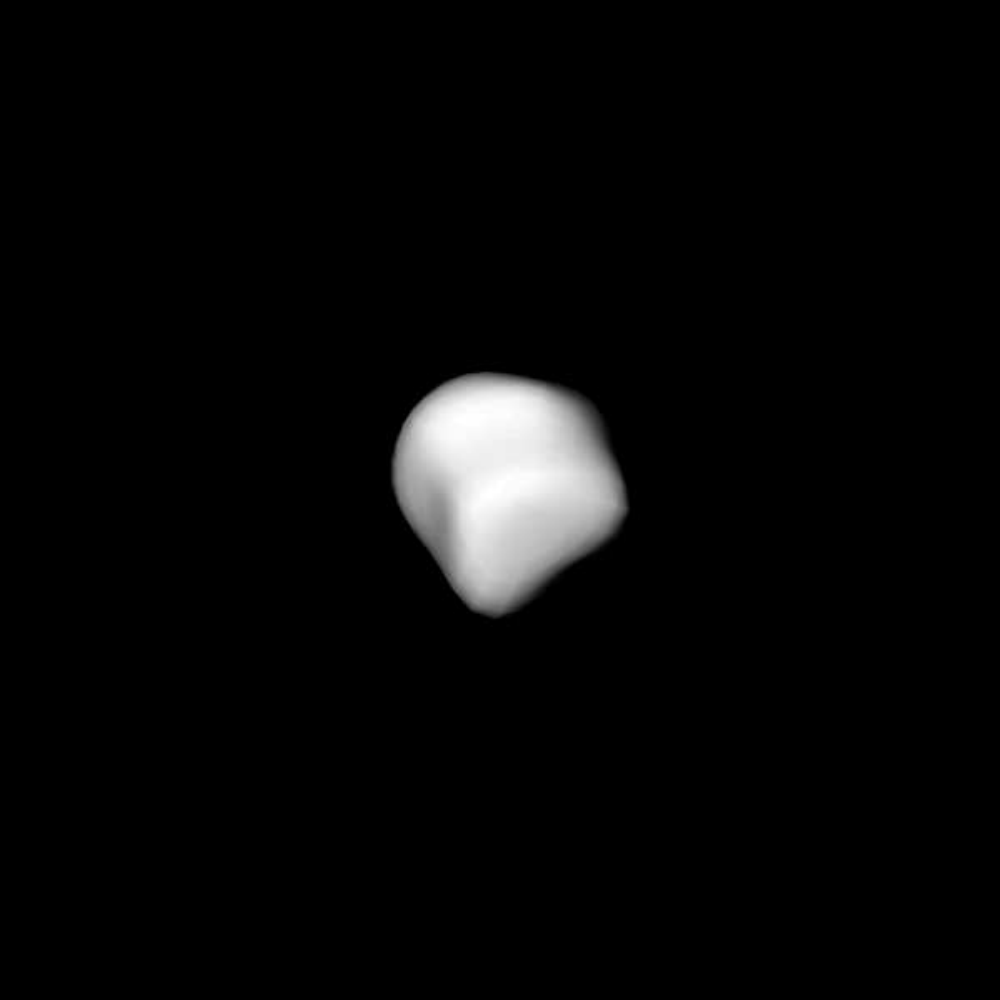}}\resizebox{0.24\hsize}{!}{\includegraphics{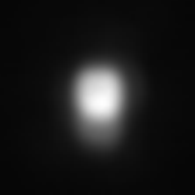}}\resizebox{0.24\hsize}{!}{\includegraphics{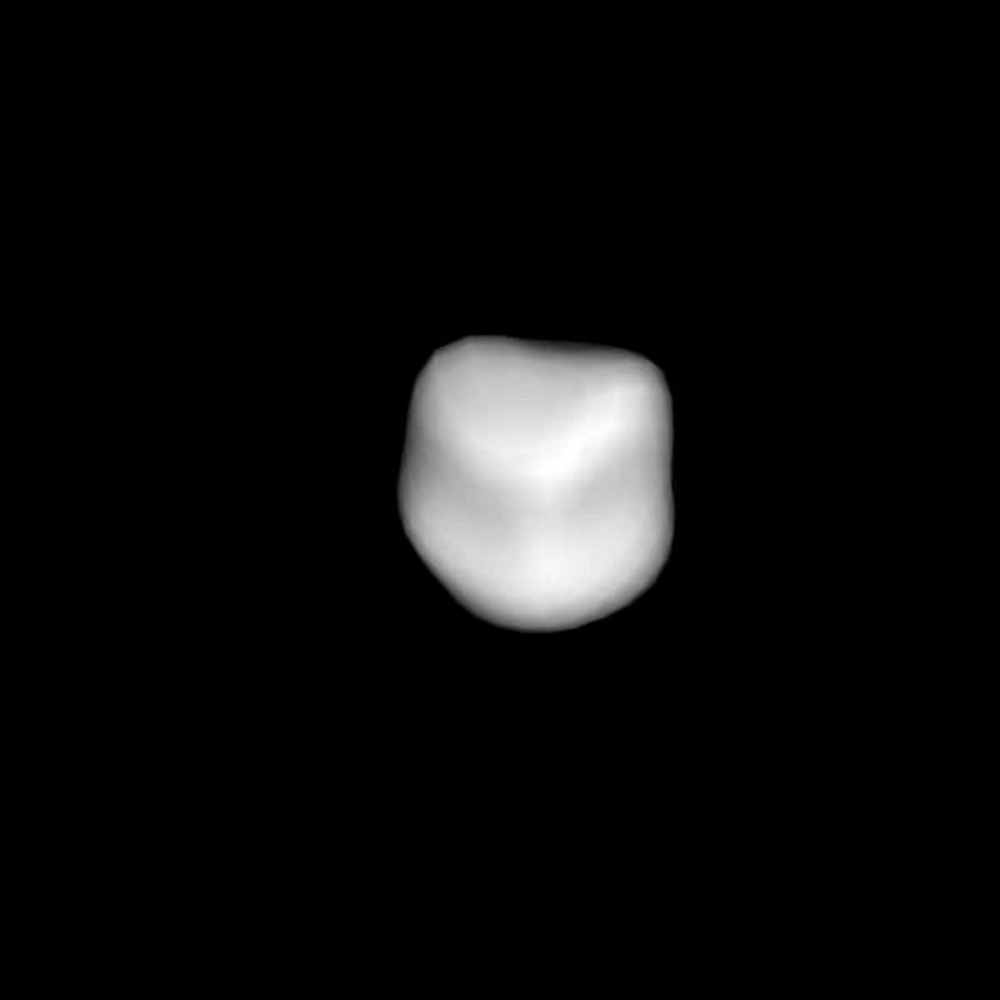}}\\
        \resizebox{0.24\hsize}{!}{\includegraphics{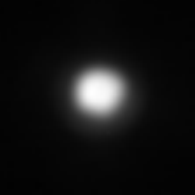}}\resizebox{0.24\hsize}{!}{\includegraphics{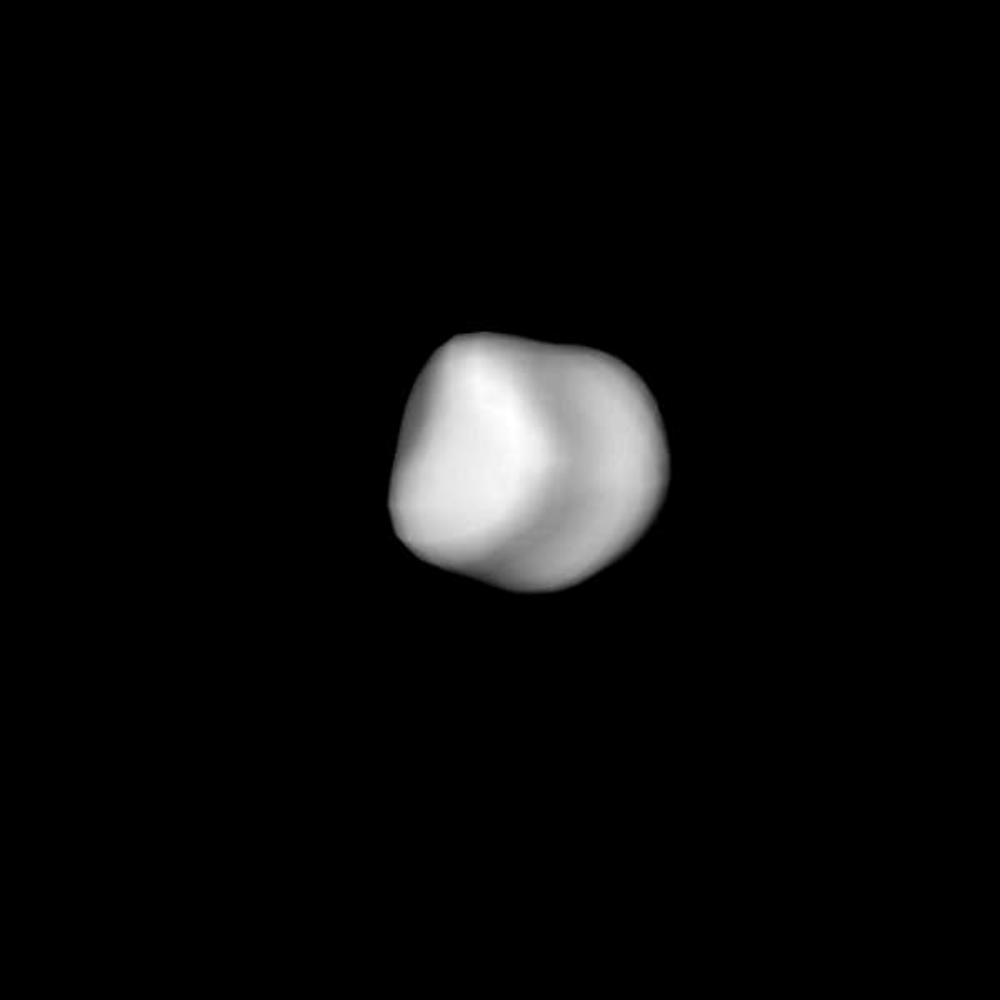}}\resizebox{0.24\hsize}{!}{\includegraphics{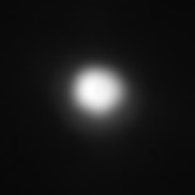}}\resizebox{0.24\hsize}{!}{\includegraphics{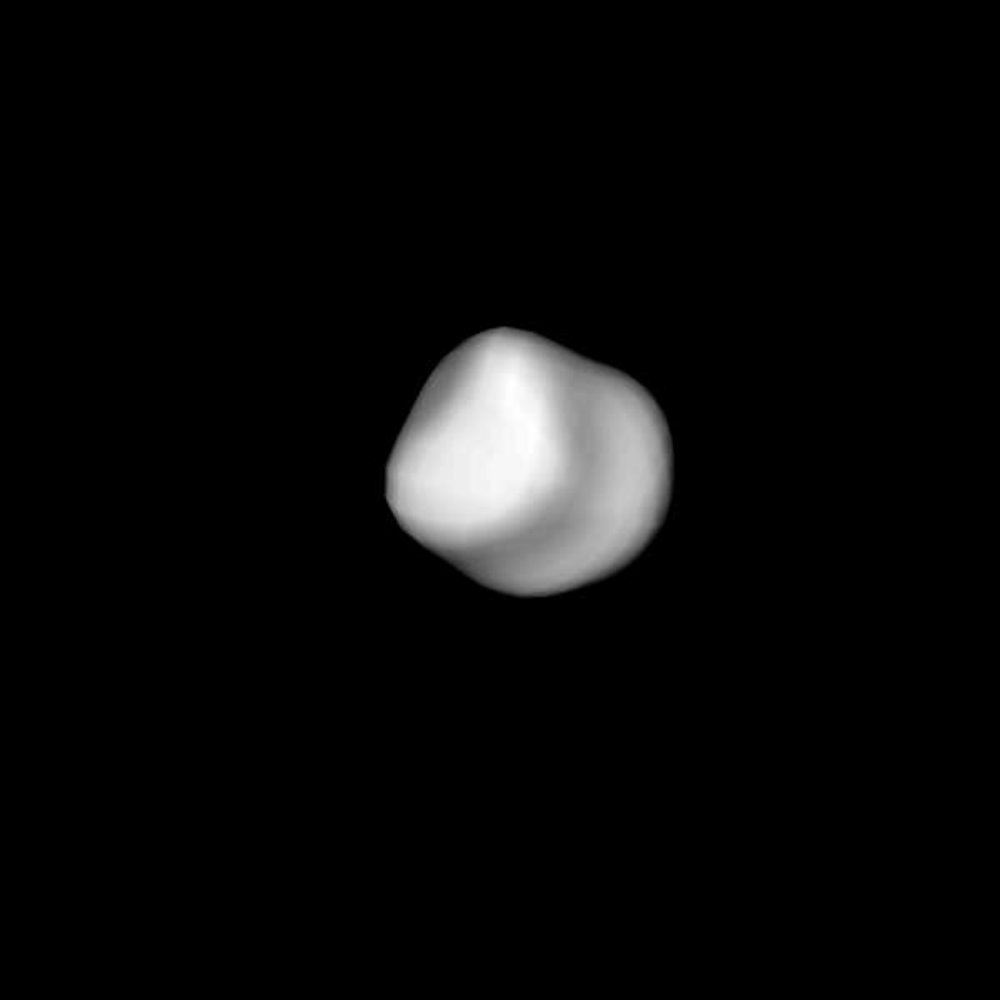}}\\
        \resizebox{0.24\hsize}{!}{\includegraphics{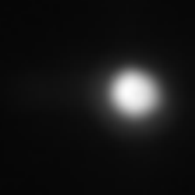}}\resizebox{0.24\hsize}{!}{\includegraphics{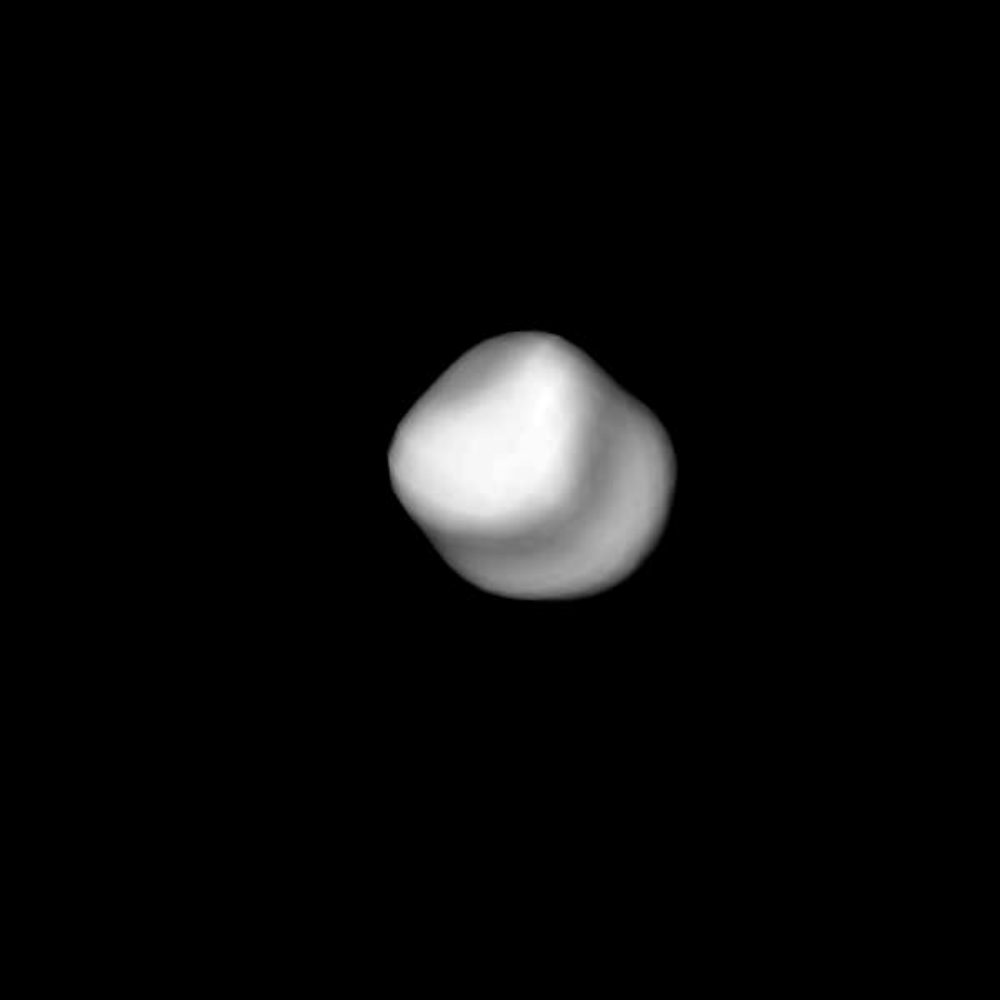}}\resizebox{0.24\hsize}{!}{\includegraphics{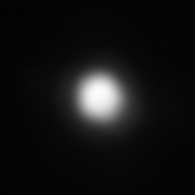}}\resizebox{0.24\hsize}{!}{\includegraphics{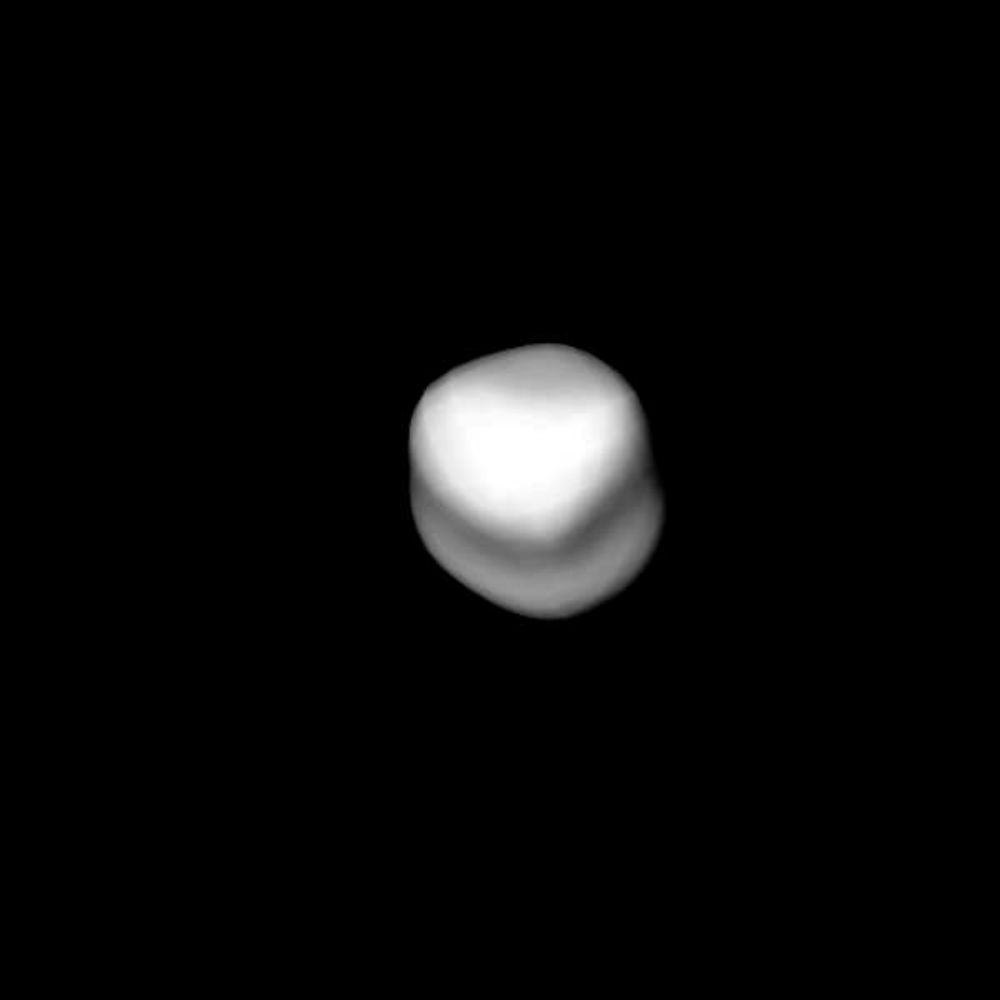}}\\
        \resizebox{0.24\hsize}{!}{\includegraphics{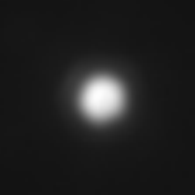}}\resizebox{0.24\hsize}{!}{\includegraphics{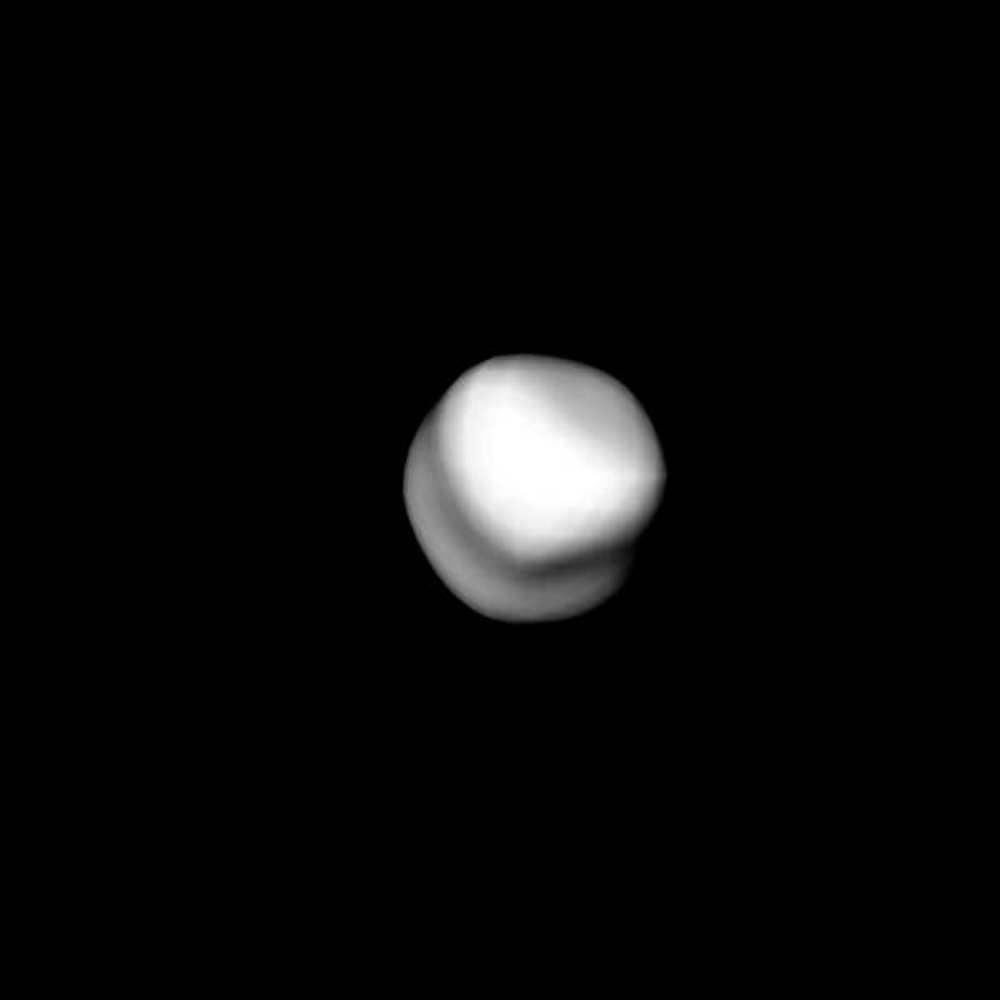}}\\
    \caption{\label{fig:409}Comparison between model projections and corresponding AO images for asteroid (409) Aspasia.}
\end{figure}

\begin{figure}[tbp]
    \centering
        \resizebox{0.24\hsize}{!}{\includegraphics{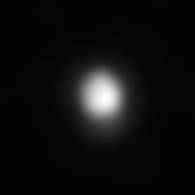}}\resizebox{0.24\hsize}{!}{\includegraphics{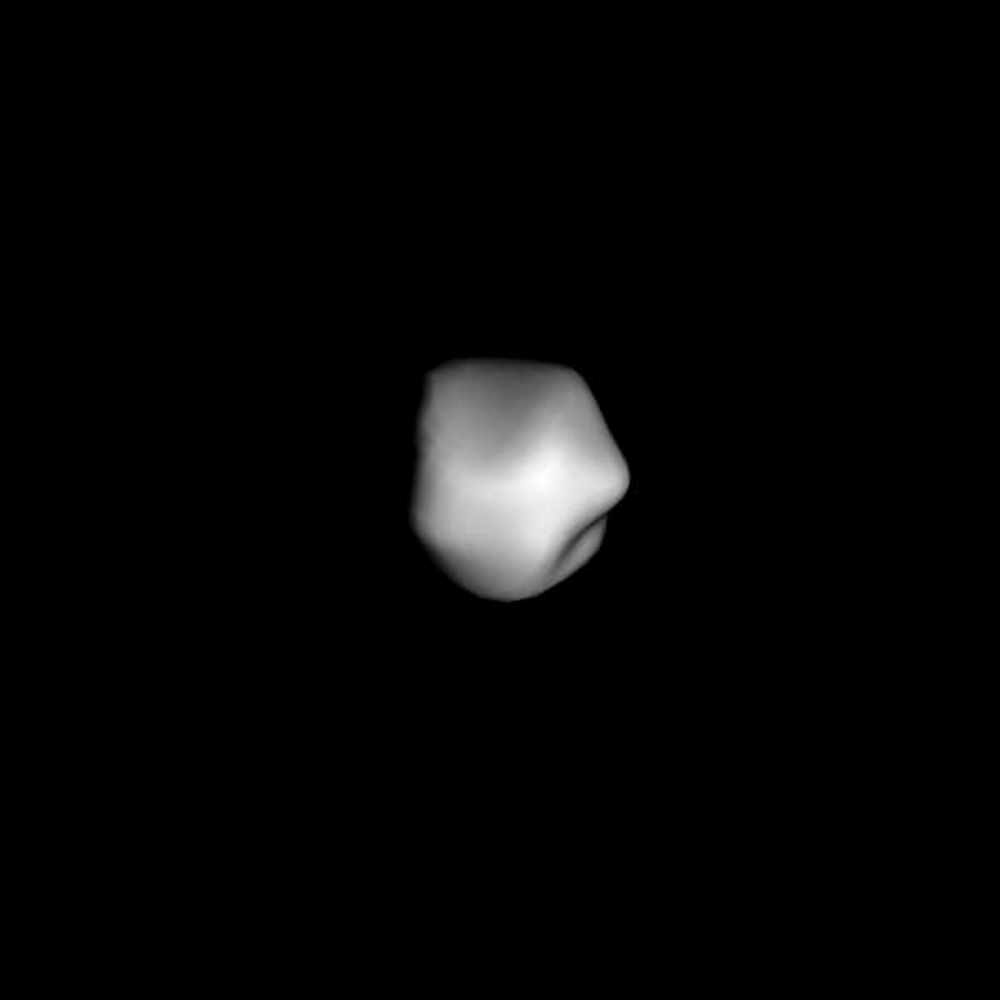}}\\
    \caption{\label{fig:419}Comparison between model projections and corresponding AO images for model 1 of asteroid (419) Aurelia.}
\end{figure}

\begin{figure}[tbp]
    \centering
        \resizebox{0.24\hsize}{!}{\includegraphics{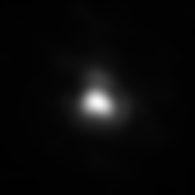}}\resizebox{0.24\hsize}{!}{\includegraphics{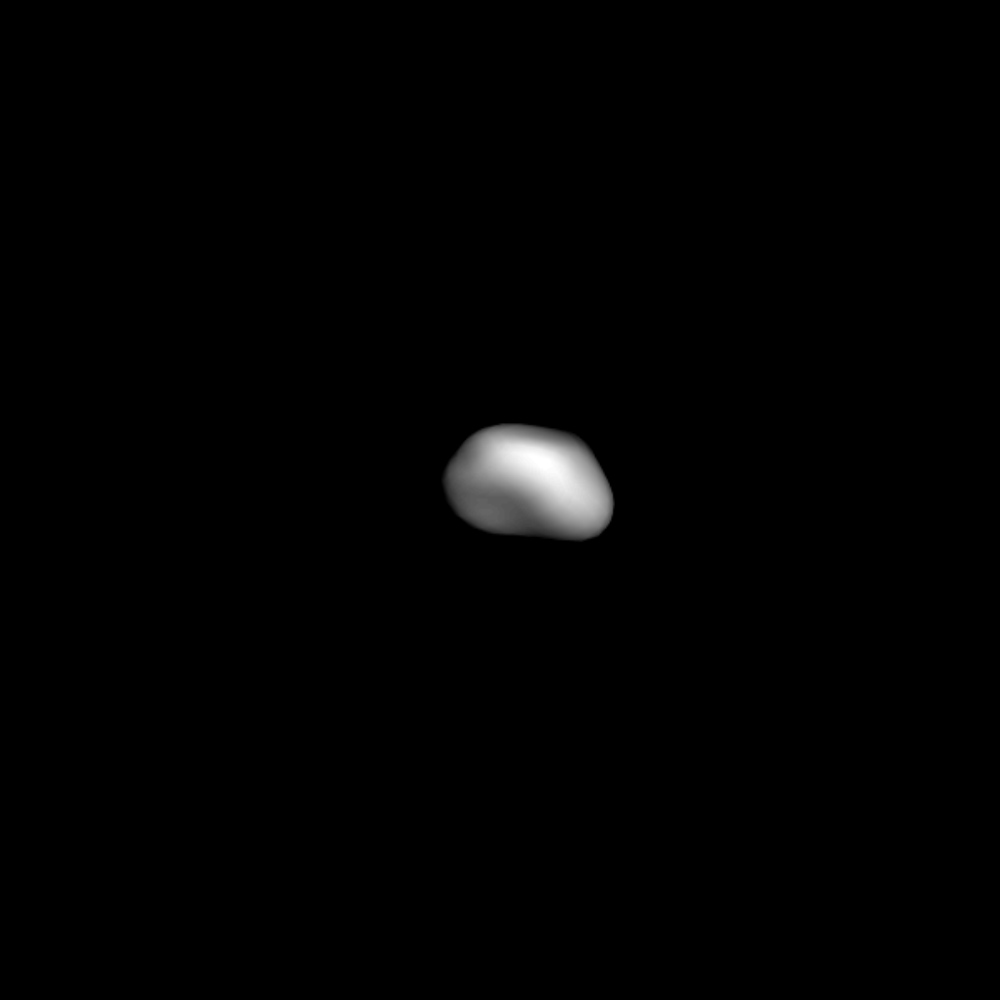}}\\
    \caption{\label{fig:471}Comparison between model projections and corresponding AO images for asteroid (471) Papagena.}
\end{figure}

\begin{figure}[tbp]
    \centering
        \resizebox{0.24\hsize}{!}{\includegraphics{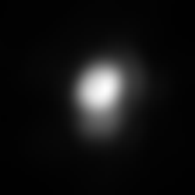}}\resizebox{0.24\hsize}{!}{\includegraphics{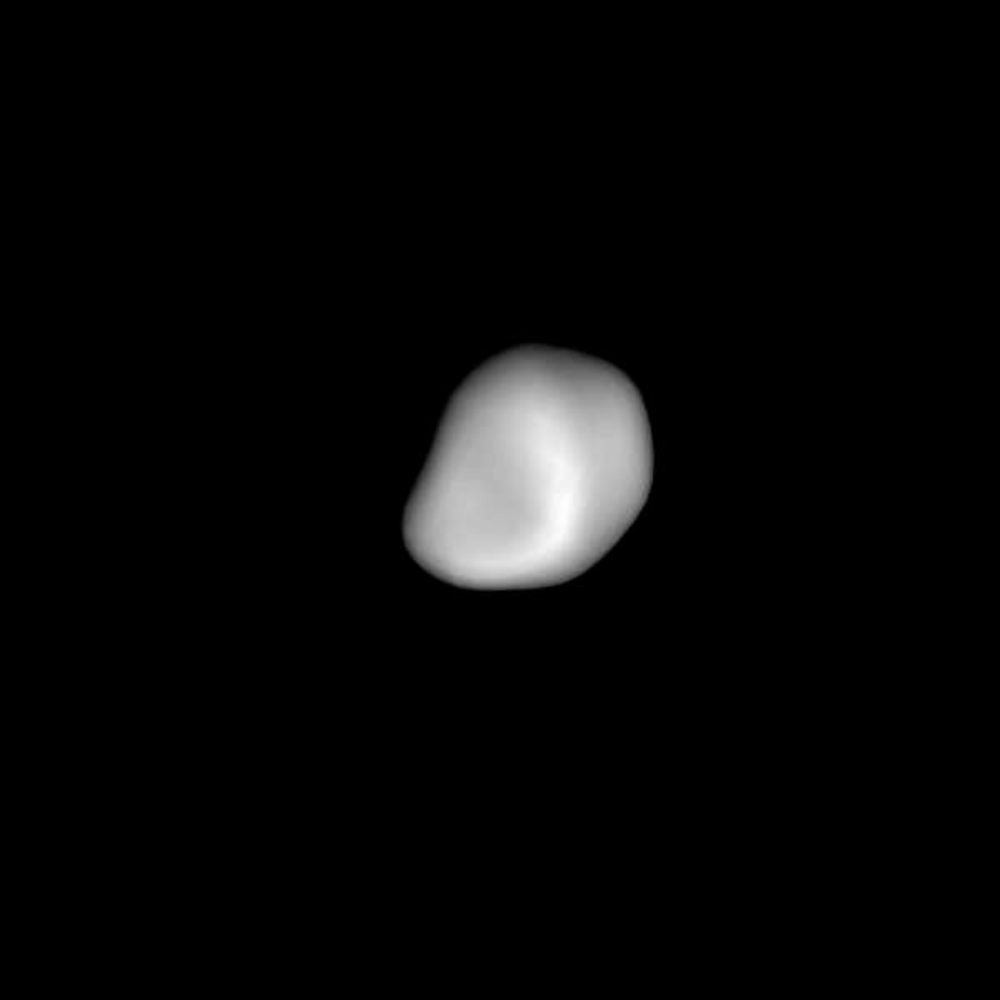}}\resizebox{0.24\hsize}{!}{\includegraphics{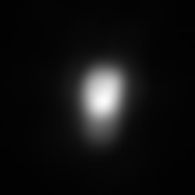}}\resizebox{0.24\hsize}{!}{\includegraphics{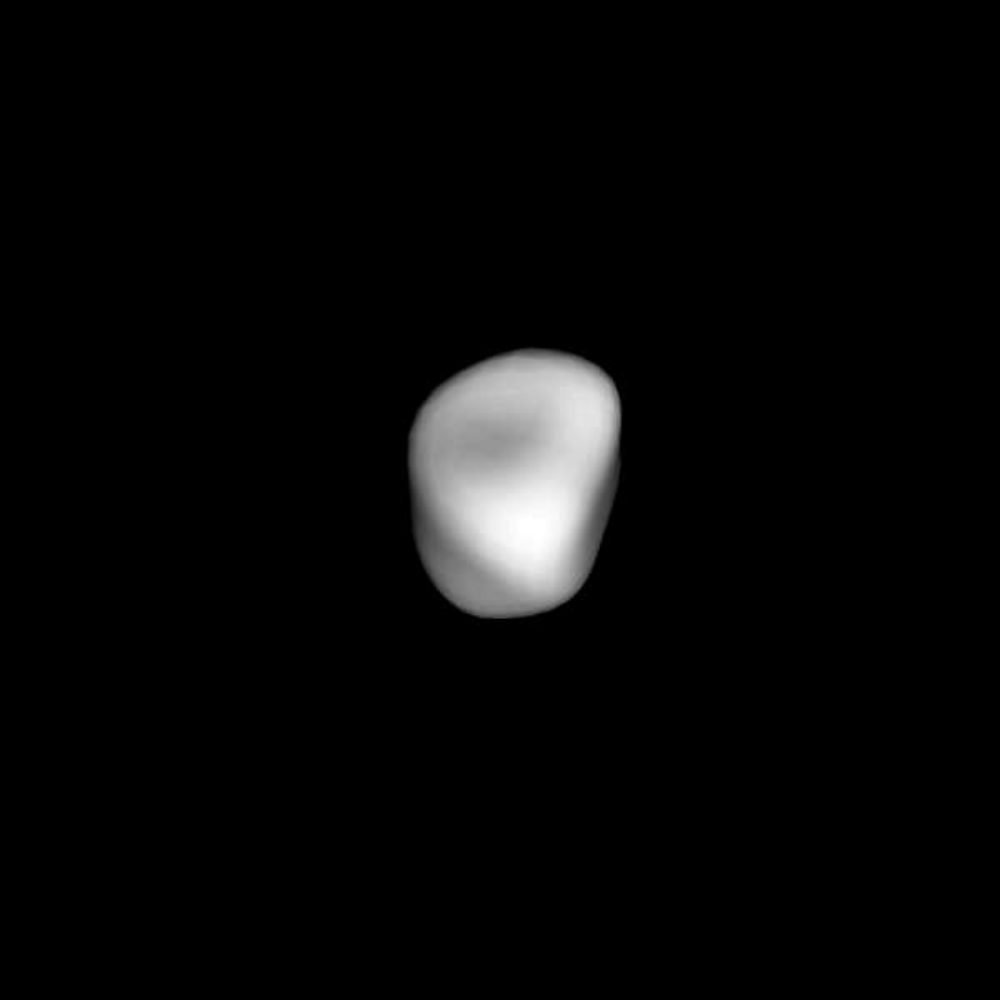}}\\
        \resizebox{0.24\hsize}{!}{\includegraphics{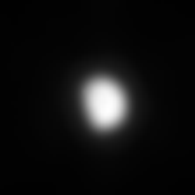}}\resizebox{0.24\hsize}{!}{\includegraphics{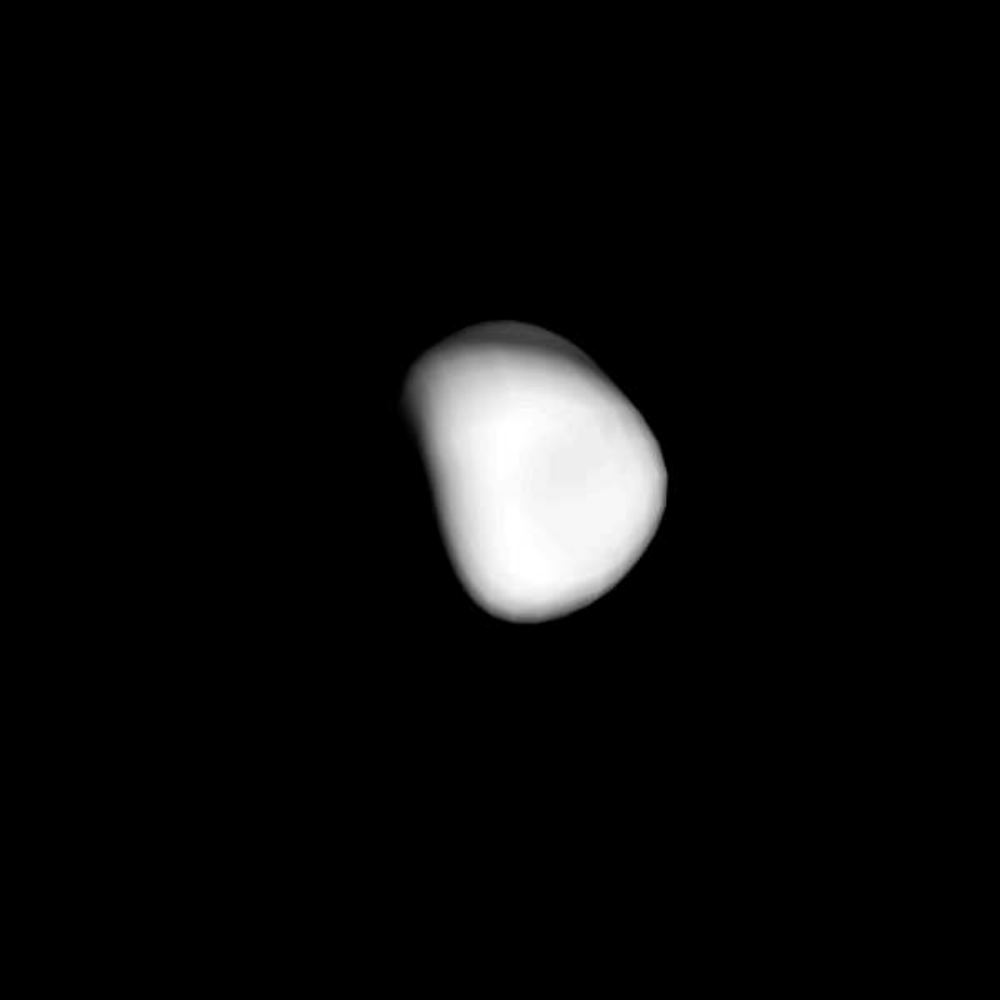}}\resizebox{0.24\hsize}{!}{\includegraphics{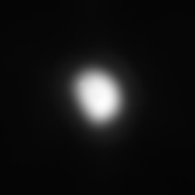}}\resizebox{0.24\hsize}{!}{\includegraphics{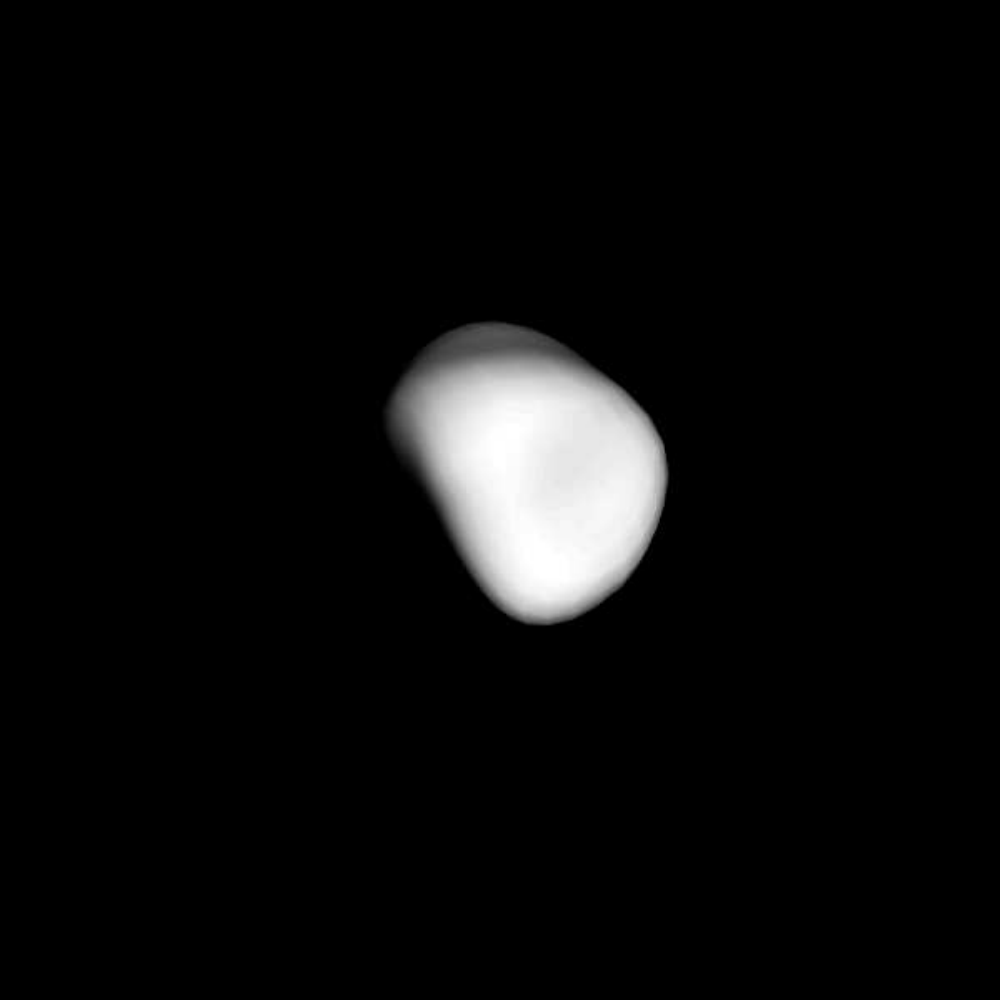}}\\
    \caption{\label{fig:532}Comparison between model projections and corresponding AO images for asteroid (532) Herculina.}
\end{figure}


\clearpage
\begin{figure}[tbp]
    \centering
 \resizebox{0.33\hsize}{!}{\includegraphics{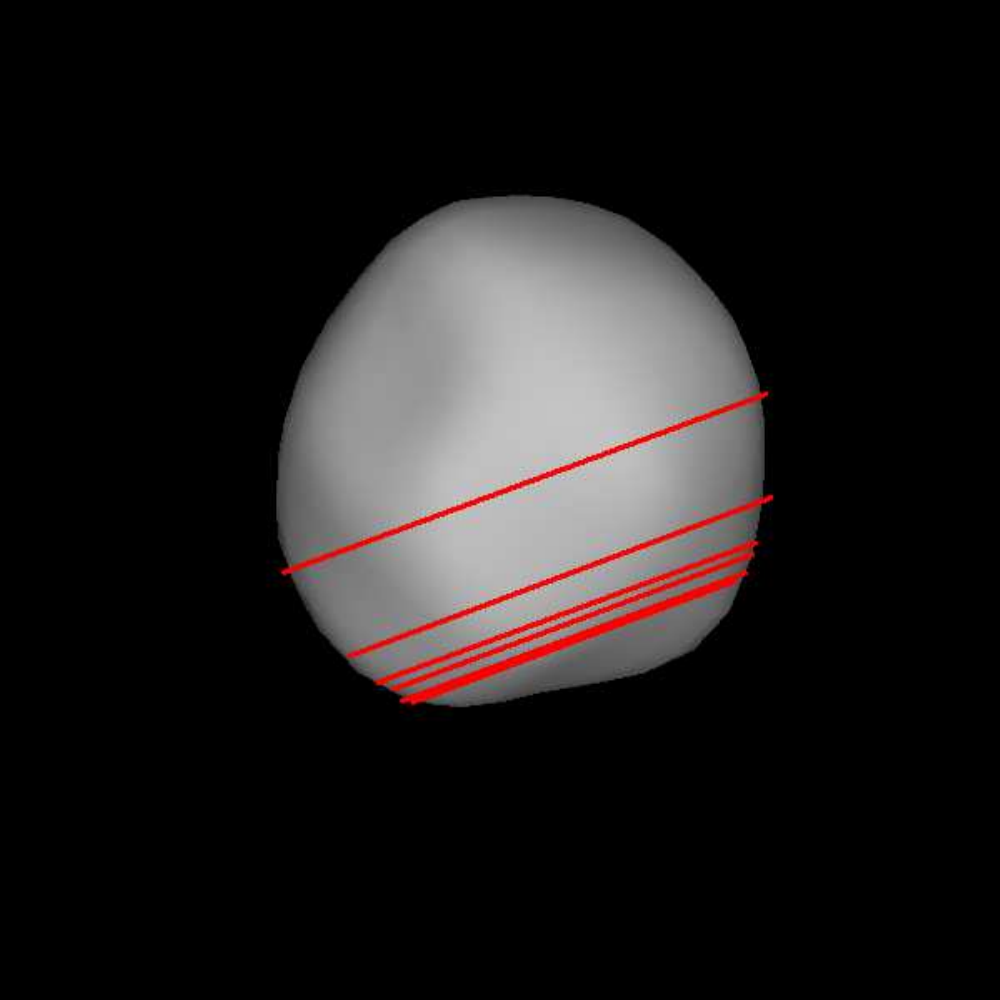}}\resizebox{0.33\hsize}{!}{\includegraphics{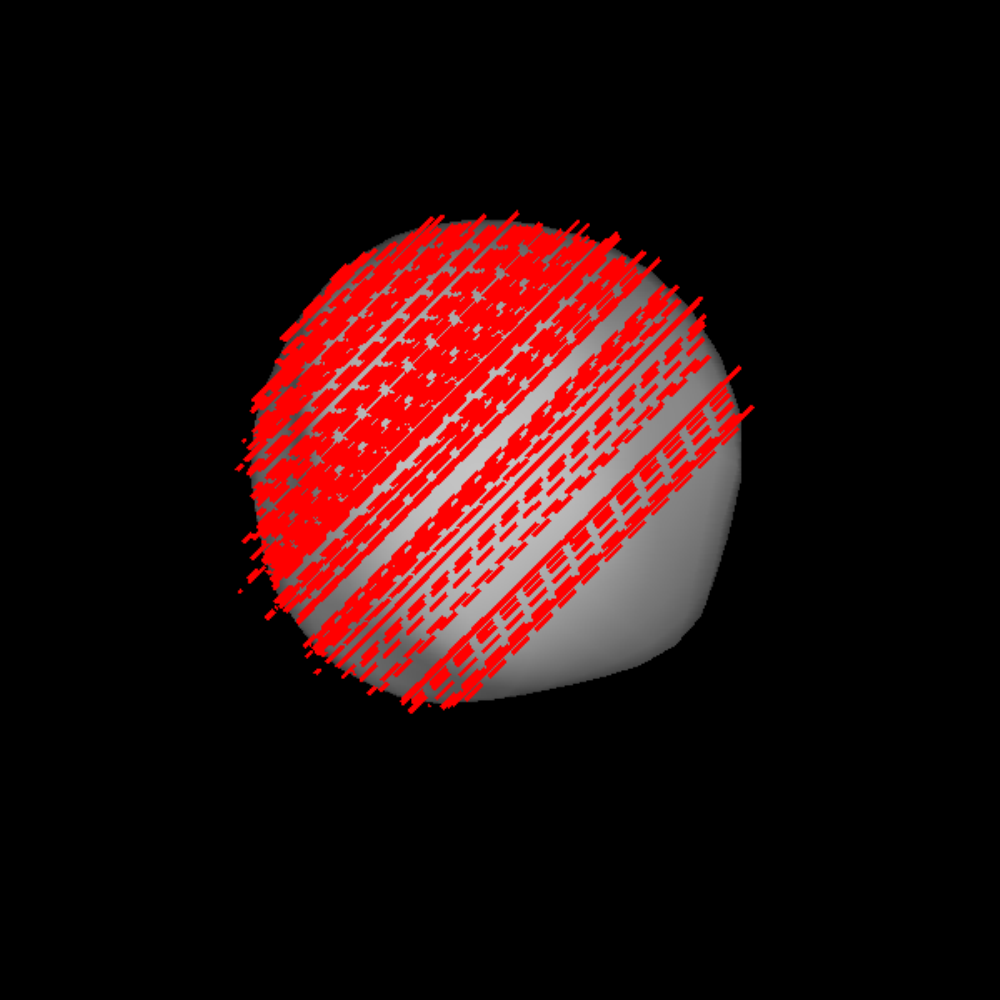}}\\
    \caption{\label{fig:2_occ}Comparison between model projections and corresponding stellar occultation(s) for asteroid (2) Pallas.}
\end{figure}

\begin{figure}[tbp]
    \centering
 \resizebox{0.33\hsize}{!}{\includegraphics{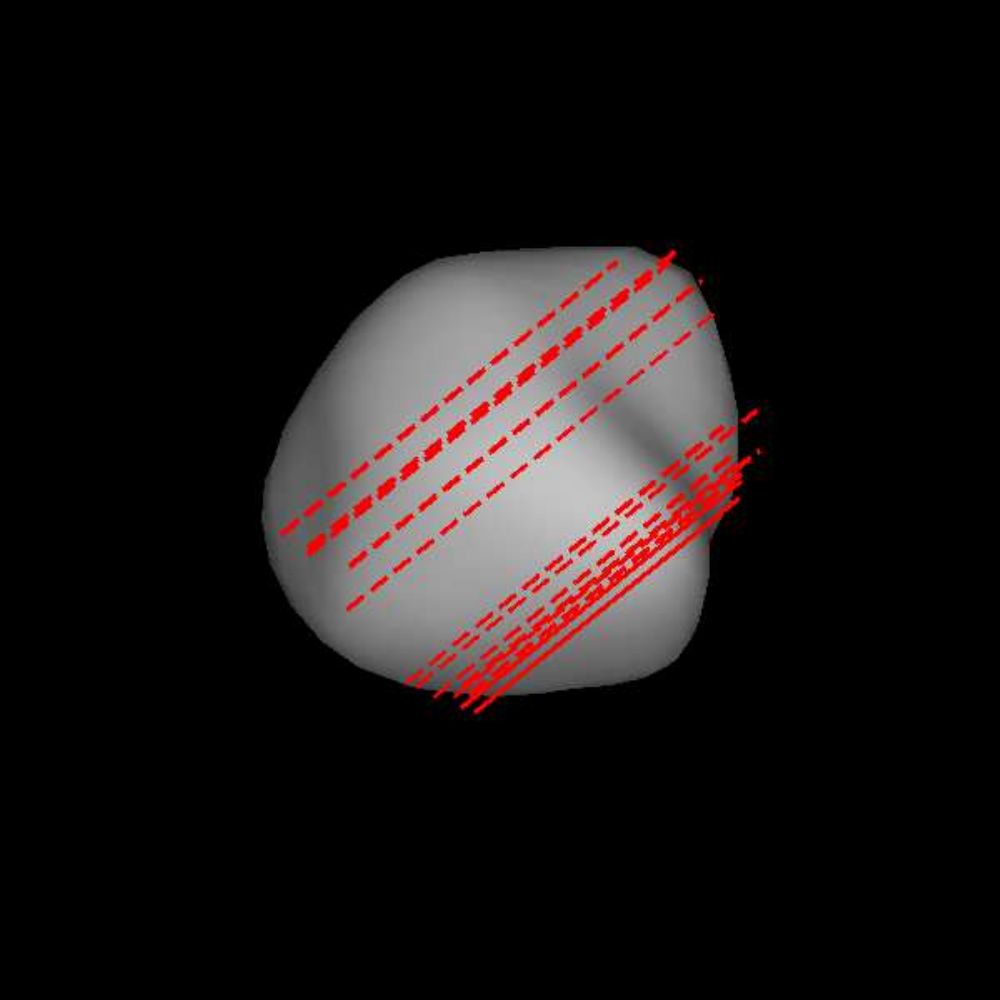}}\\
    \caption{\label{fig:5_occ}Comparison between model projections and corresponding stellar occultation(s) for asteroid (5) Astraea.}
\end{figure}

\begin{figure}[tbp]
    \centering
 \resizebox{0.33\hsize}{!}{\includegraphics{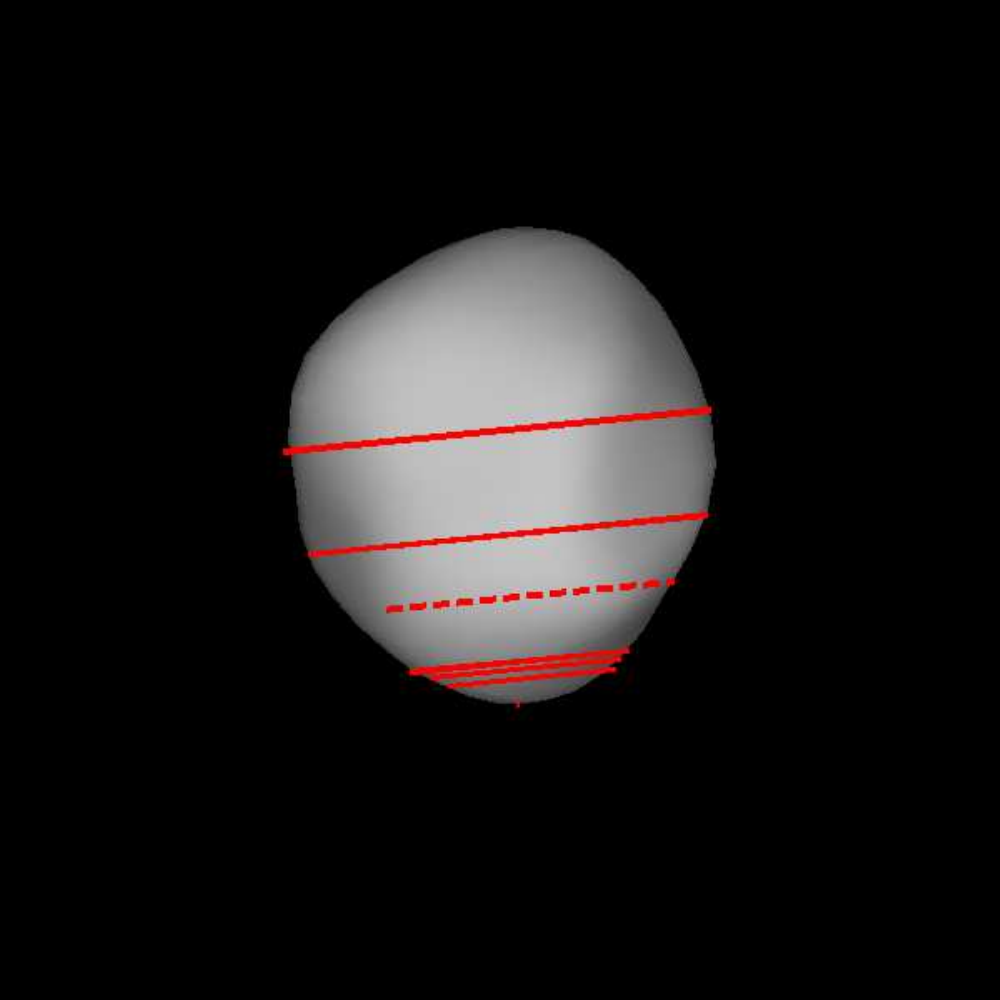}}\\
    \caption{\label{fig:8_occ}Comparison between model projections and corresponding stellar occultation(s) for asteroid (8) Flora.}
\end{figure}

\begin{figure}[tbp]
    \centering
 \resizebox{0.33\hsize}{!}{\includegraphics{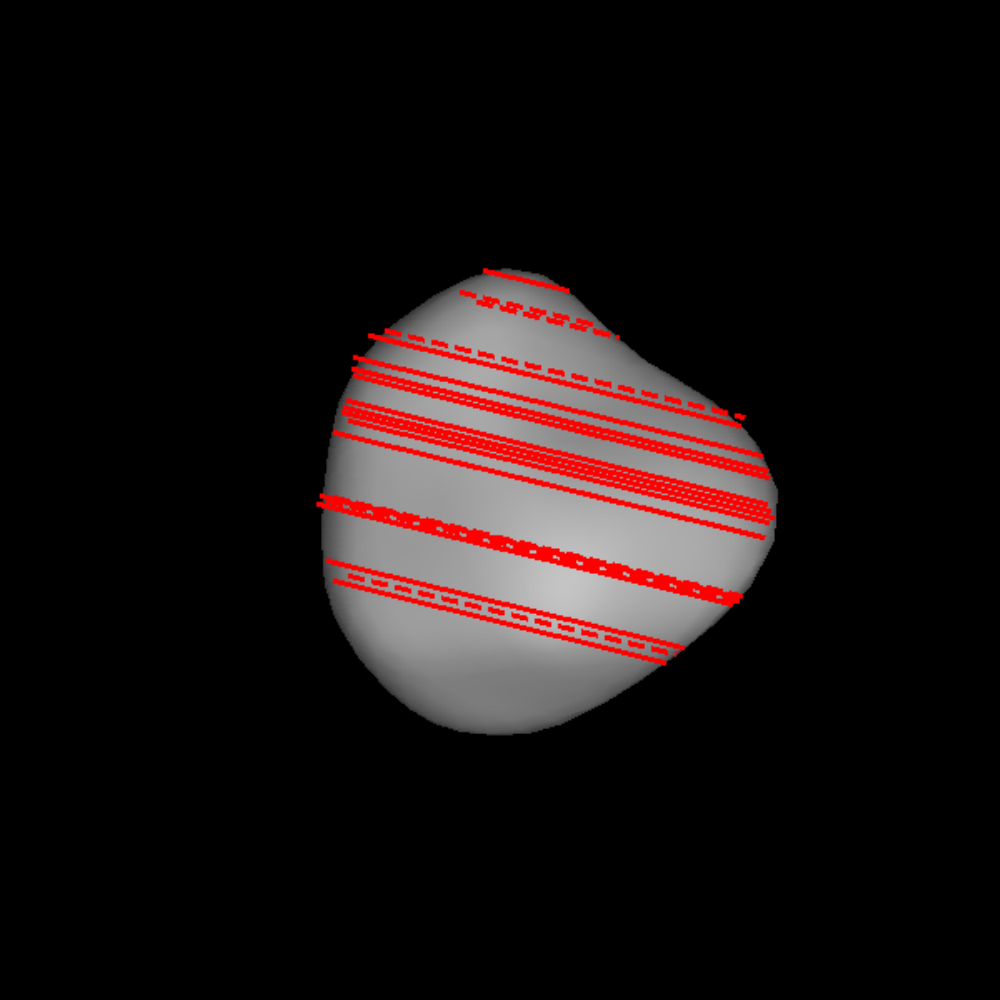}}\resizebox{0.33\hsize}{!}{\includegraphics{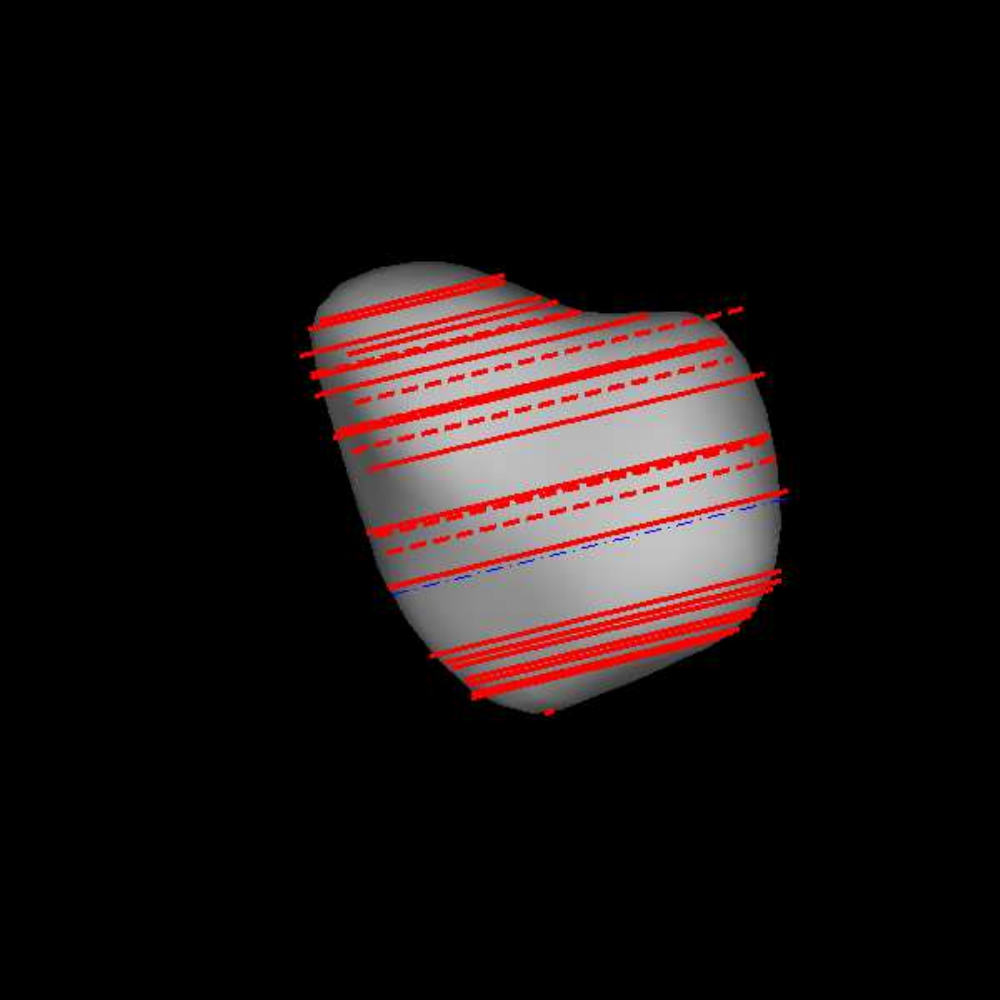}}\\
    \caption{\label{fig:9_occ}Comparison between model projections and corresponding stellar occultation(s) for asteroid (9) Metis.}
\end{figure}

\begin{figure}[tbp]
    \centering
 \resizebox{0.33\hsize}{!}{\includegraphics{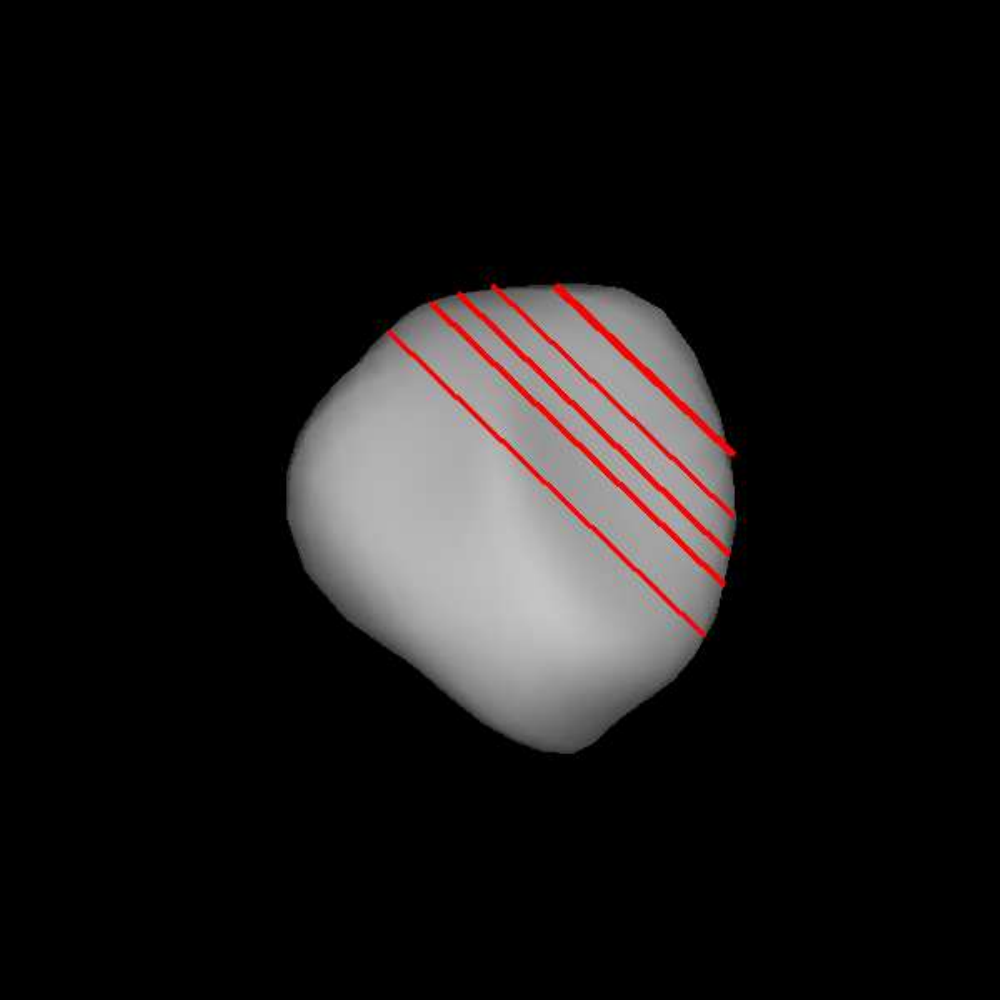}}\resizebox{0.33\hsize}{!}{\includegraphics{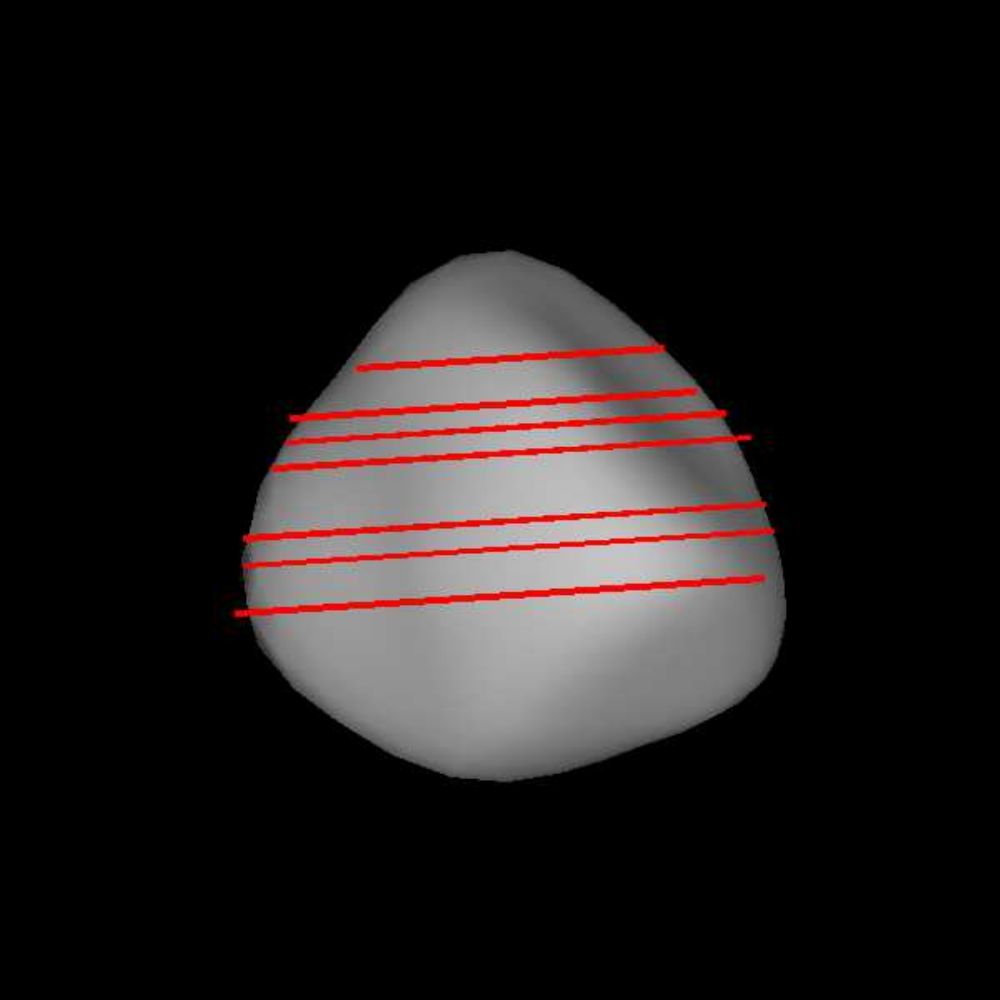}}\\
 \resizebox{0.33\hsize}{!}{\includegraphics{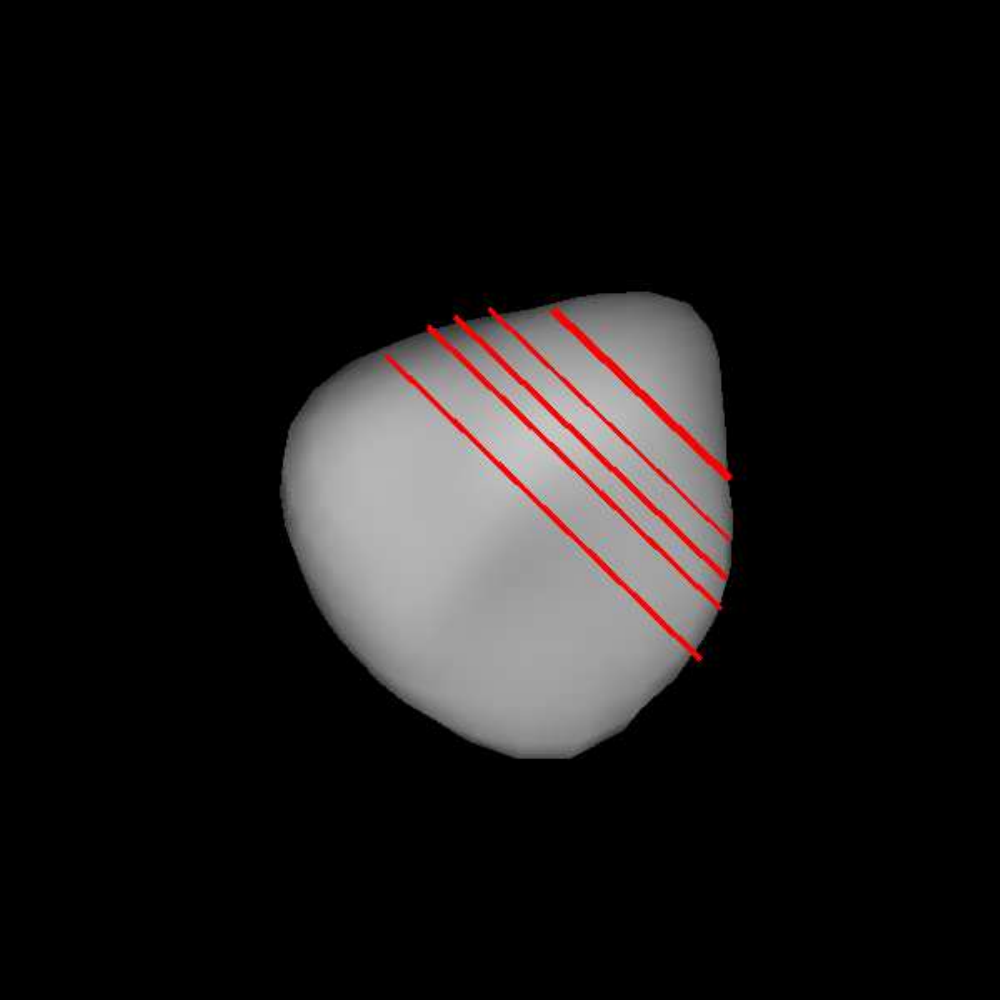}}\resizebox{0.33\hsize}{!}{\includegraphics{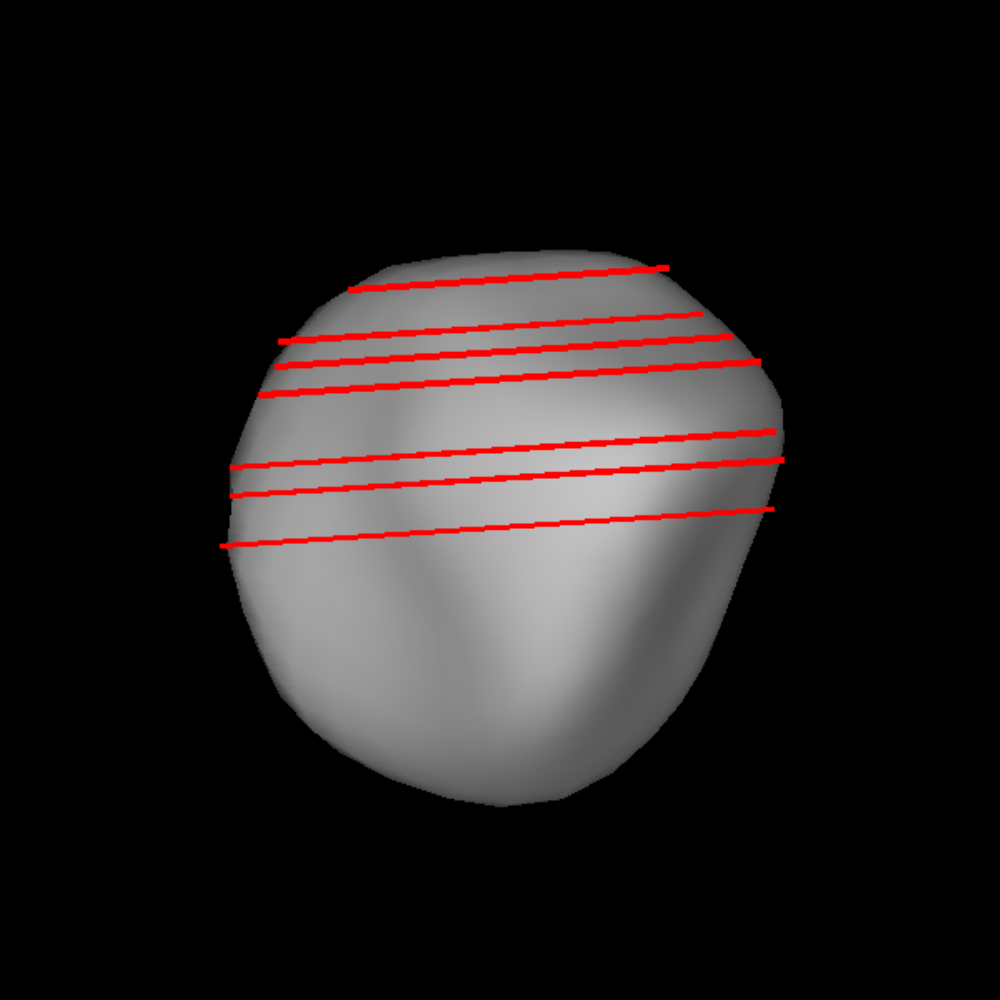}}\\
    \caption{\label{fig:10_occ}Comparison between model projections and corresponding stellar occultation(s) for asteroid (10) Hygiea. We show the fit for both pole solutions.}
\end{figure}

\begin{figure}[tbp]
    \centering
 \resizebox{0.33\hsize}{!}{\includegraphics{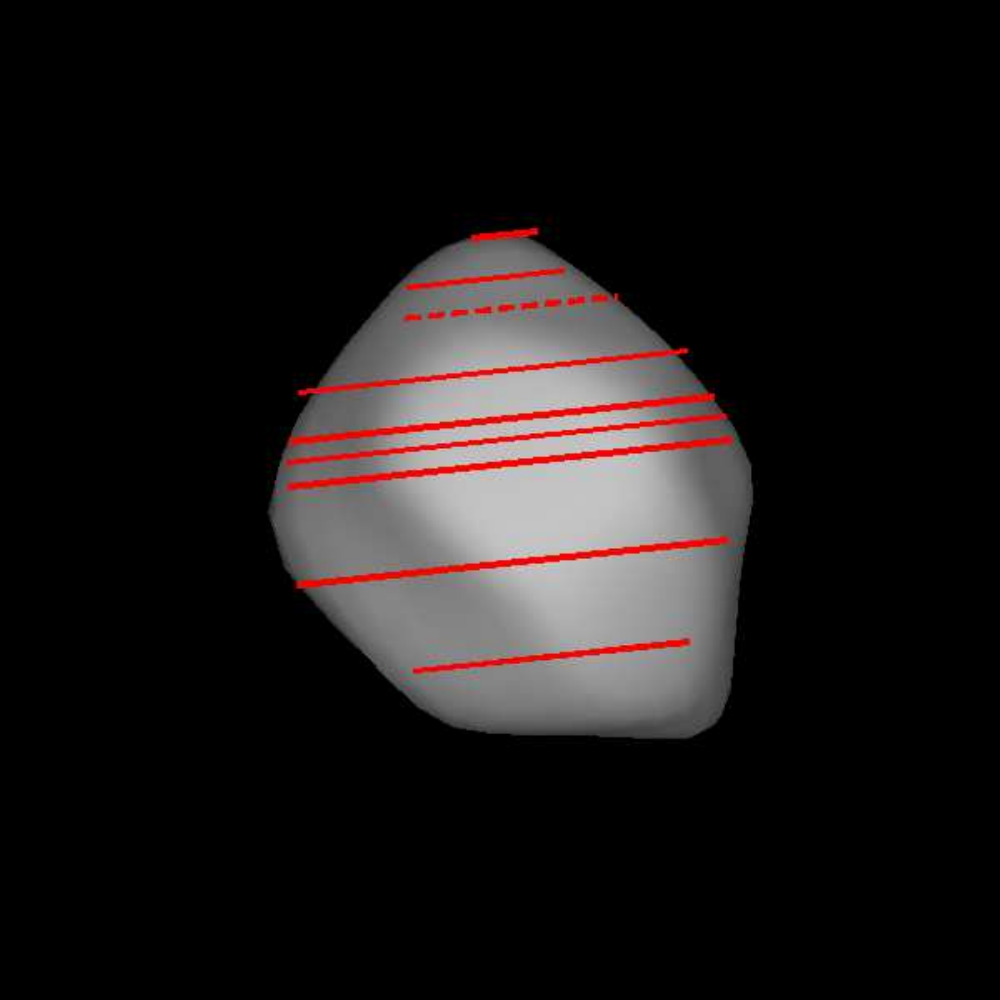}}\\
    \caption{\label{fig:11_occ}Comparison between model projections and corresponding stellar occultation(s) for asteroid (11) Parthenope.}
\end{figure}

\begin{figure}[tbp]
    \centering
 \resizebox{0.33\hsize}{!}{\includegraphics{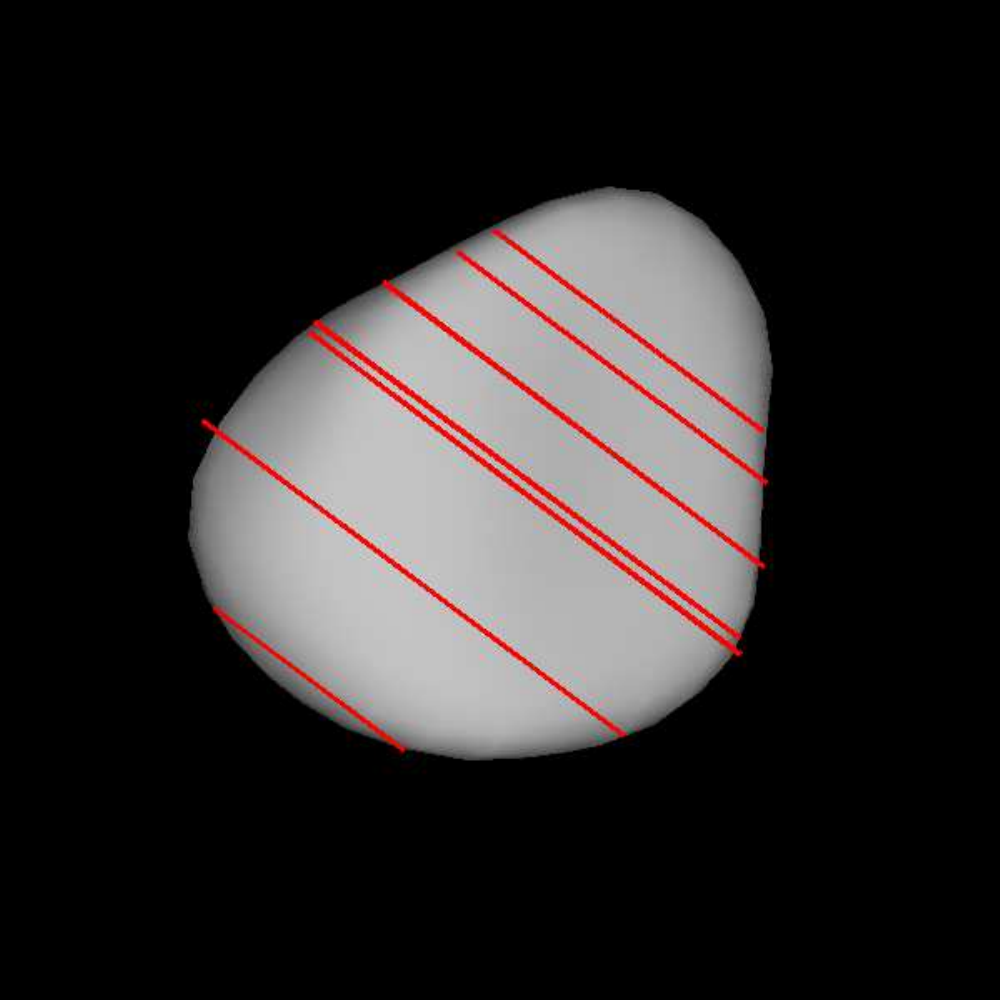}}\\
 \resizebox{0.33\hsize}{!}{\includegraphics{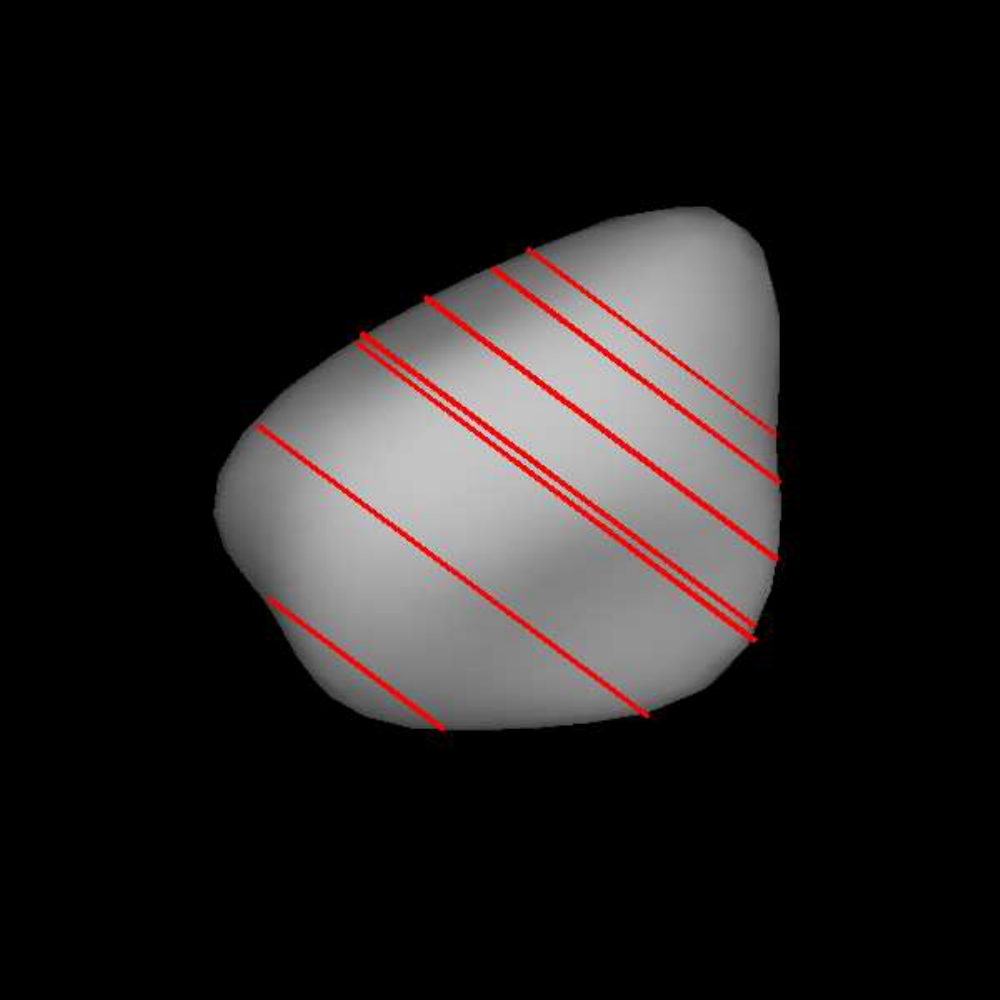}}\\
    \caption{\label{fig:13_occ}Comparison between model projections and corresponding stellar occultation(s) for asteroid (13) Egeria. We show the fit for both pole solutions.}
\end{figure}

\begin{figure}[tbp]
    \centering
 \resizebox{0.33\hsize}{!}{\includegraphics{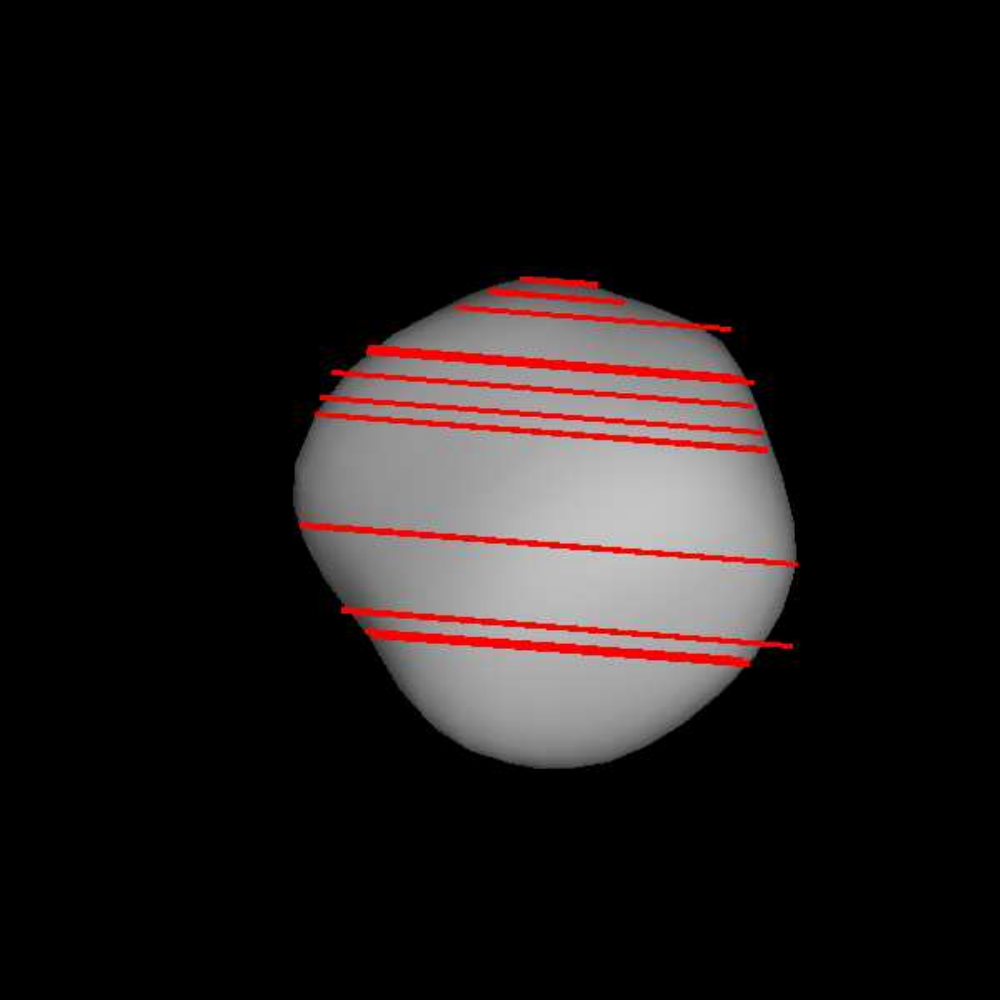}}\resizebox{0.33\hsize}{!}{\includegraphics{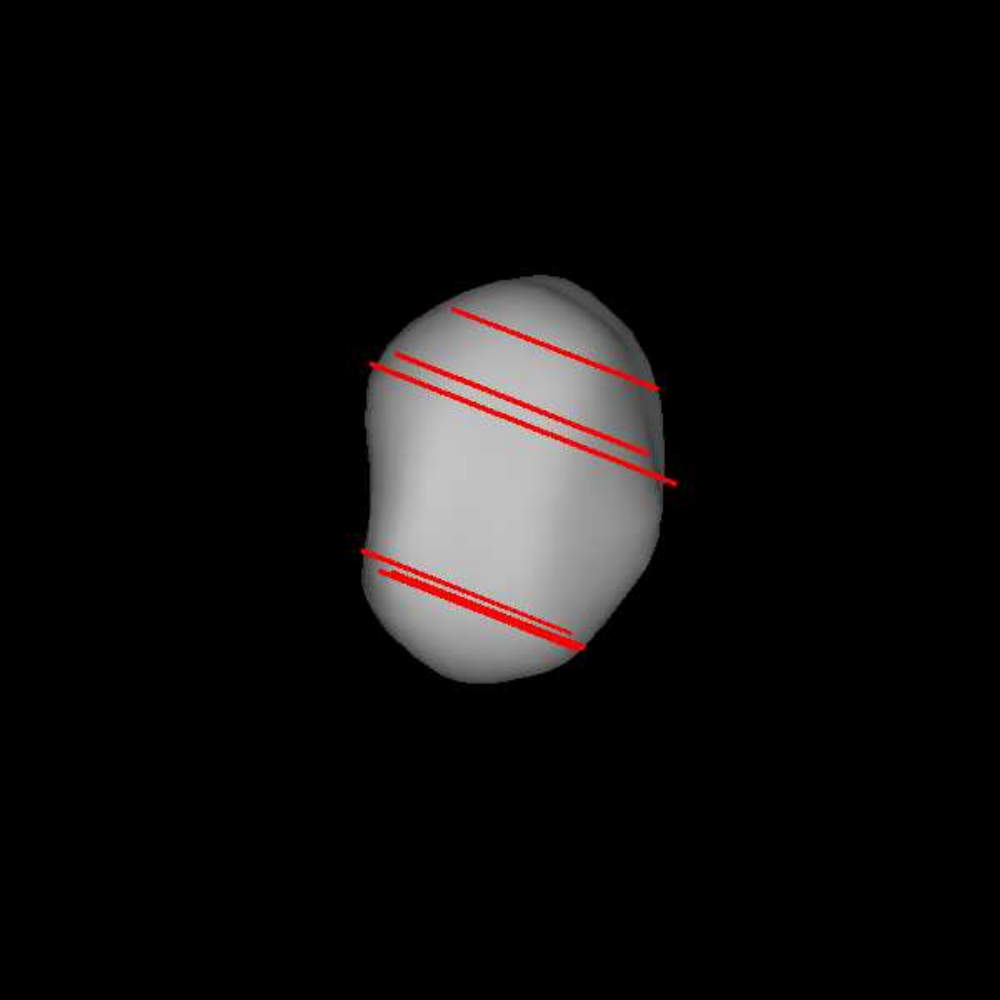}}\\
    \caption{\label{fig:16_occ}Comparison between model projections and corresponding stellar occultation(s) for asteroid (16) Psyche.}
\end{figure}

\begin{figure}[tbp]
    \centering
 \resizebox{0.33\hsize}{!}{\includegraphics{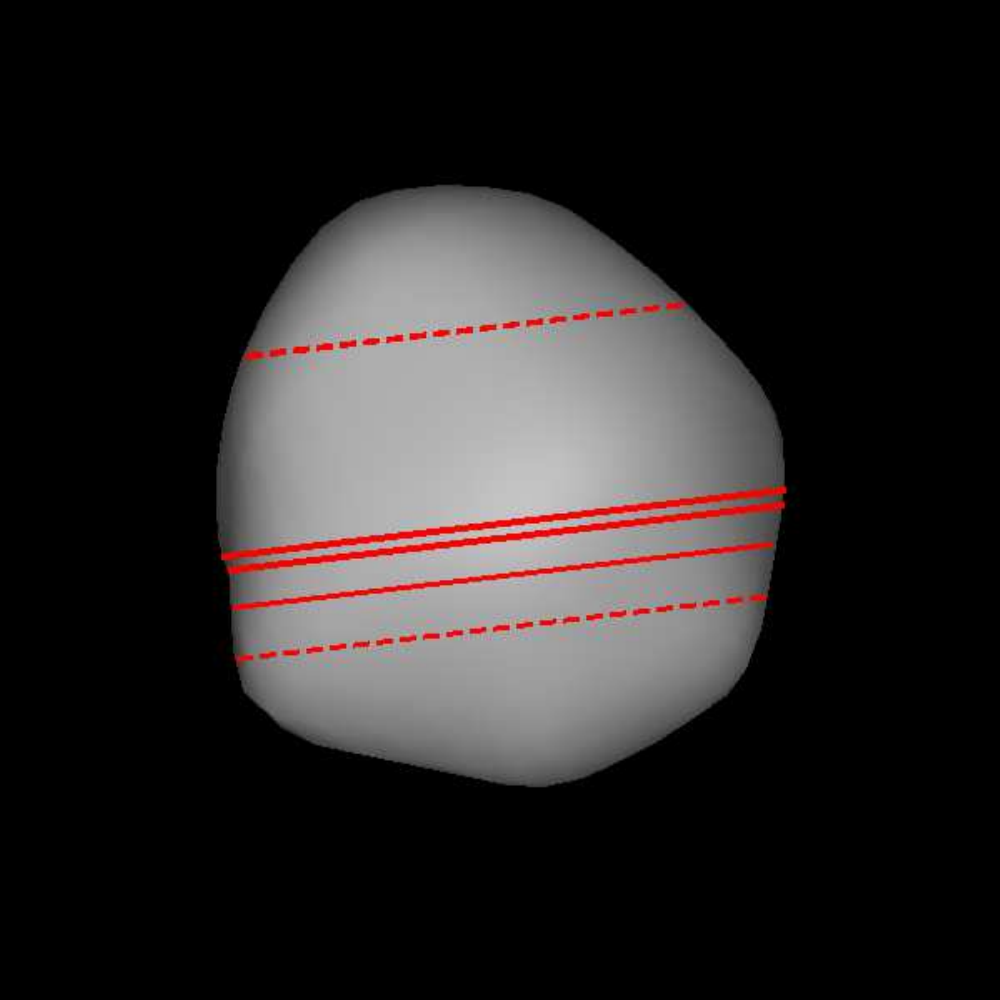}}\\
    \caption{\label{fig:18_occ}Comparison between model projections and corresponding stellar occultation(s) for asteroid (18) Melpomene.}
\end{figure}

\begin{figure}[tbp]
    \centering
 \resizebox{0.33\hsize}{!}{\includegraphics{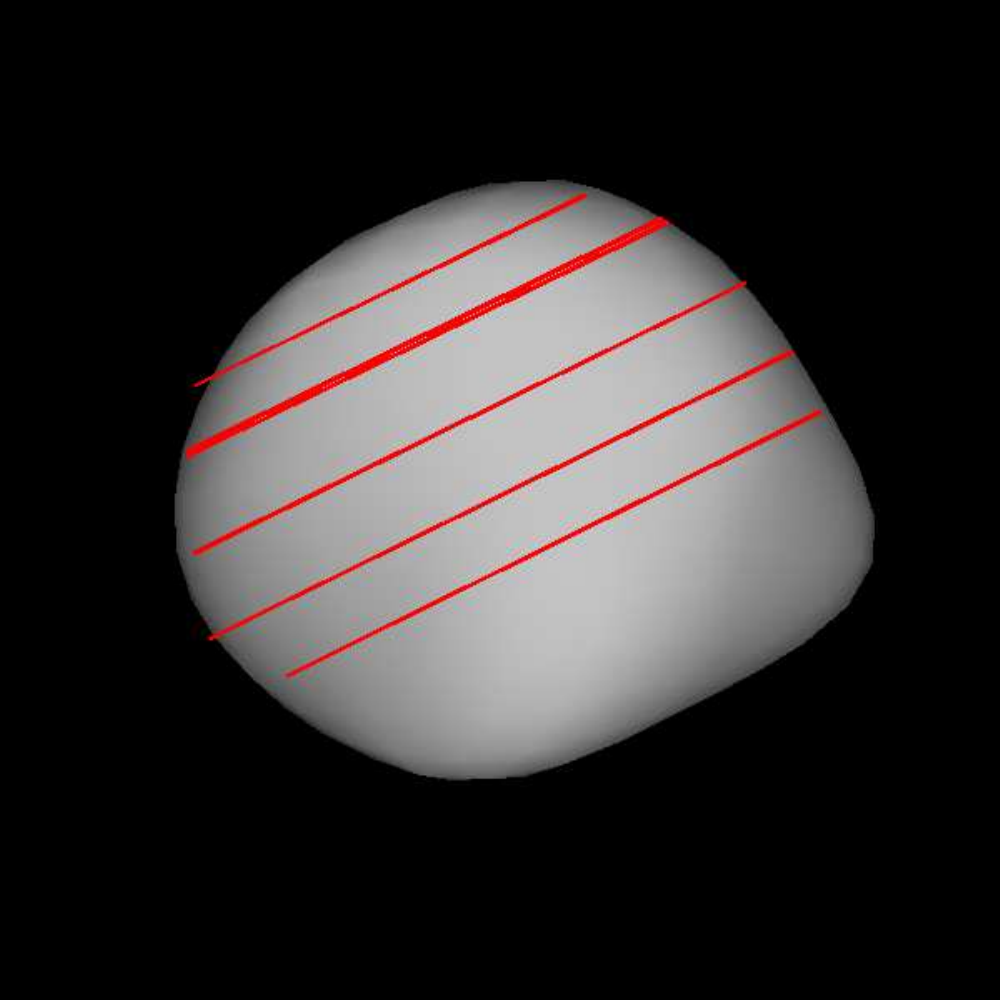}}\resizebox{0.33\hsize}{!}{\includegraphics{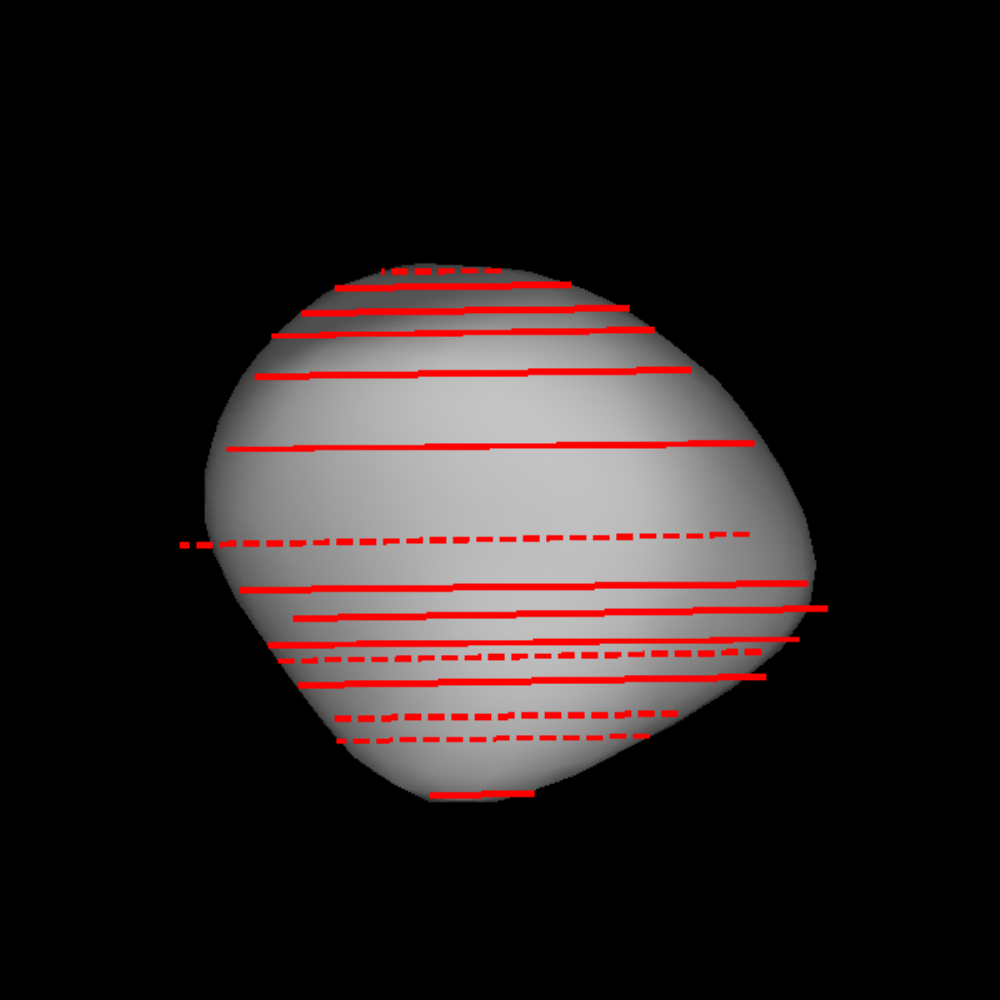}}\\
    \caption{\label{fig:19_occ}Comparison between model projections and corresponding stellar occultation(s) for asteroid (19) Fortuna.}
\end{figure}


\begin{figure}[tbp]
    \centering
 \resizebox{0.33\hsize}{!}{\includegraphics{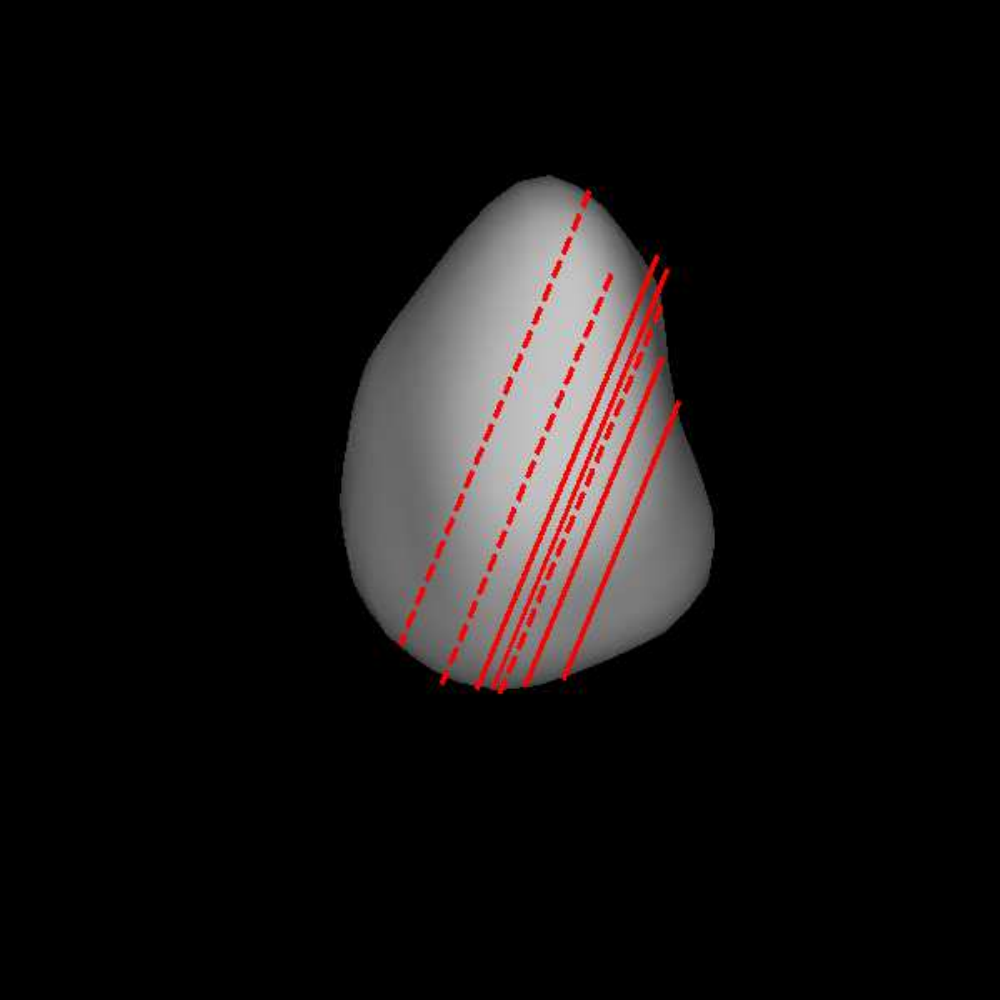}}\\
    \caption{\label{fig:22_occ}Comparison between model projections and corresponding stellar occultation(s) for asteroid (22) Kalliope.}
\end{figure}

\begin{figure}[tbp]
    \centering
 \resizebox{0.33\hsize}{!}{\includegraphics{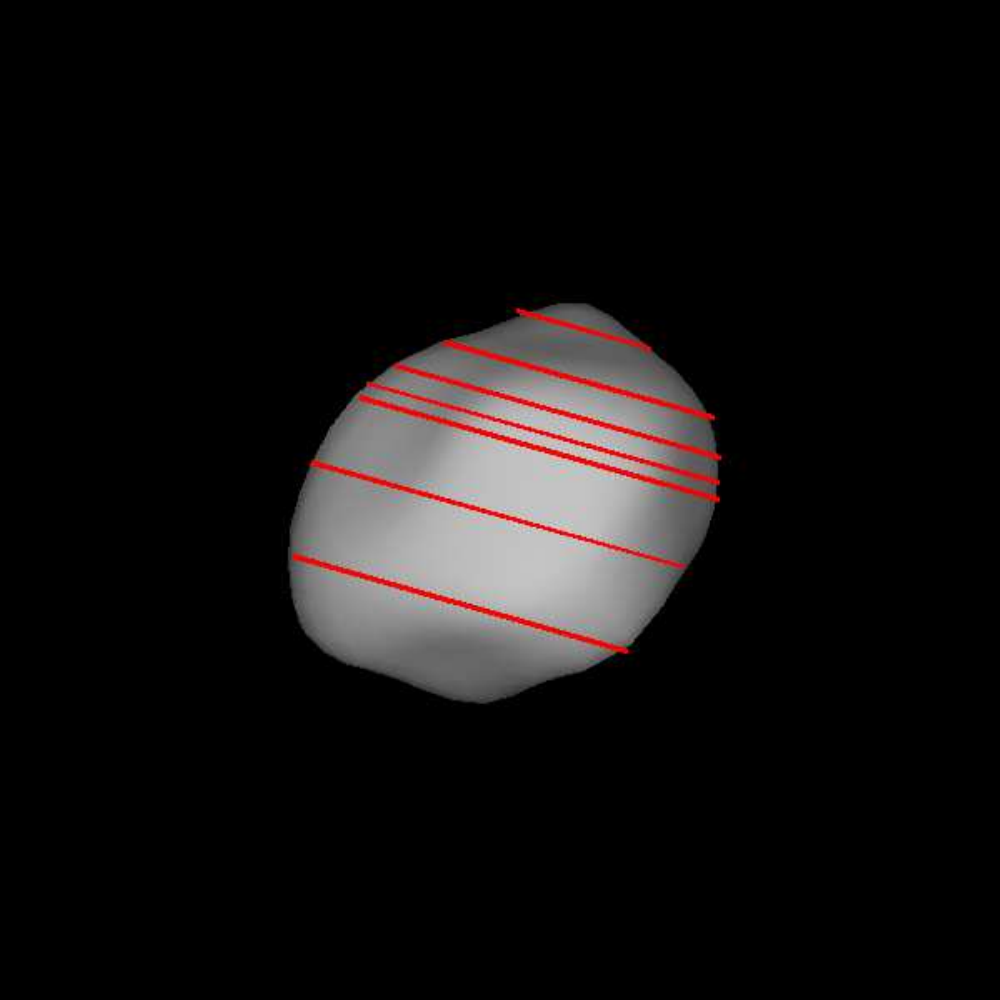}}\\
    \caption{\label{fig:29_occ}Comparison between model projections and corresponding stellar occultation(s) for asteroid (29) Amphitrite.}
\end{figure}

\clearpage

\begin{figure}[tbp]
    \centering
 \resizebox{0.33\hsize}{!}{\includegraphics{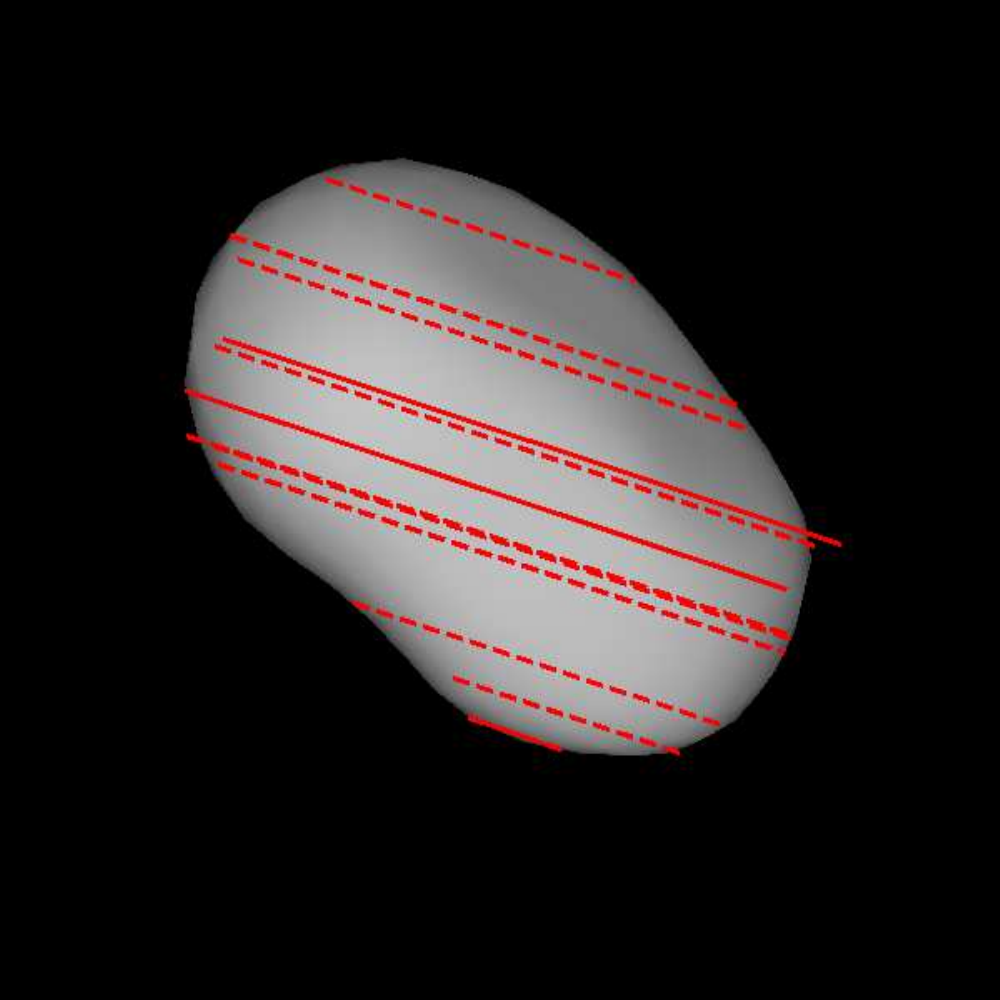}}\\
    \caption{\label{fig:39_occ}Comparison between model projections and corresponding stellar occultation(s) for asteroid (39) Laetitia.}
\end{figure}

\begin{figure}[tbp]
    \centering
 \resizebox{0.33\hsize}{!}{\includegraphics{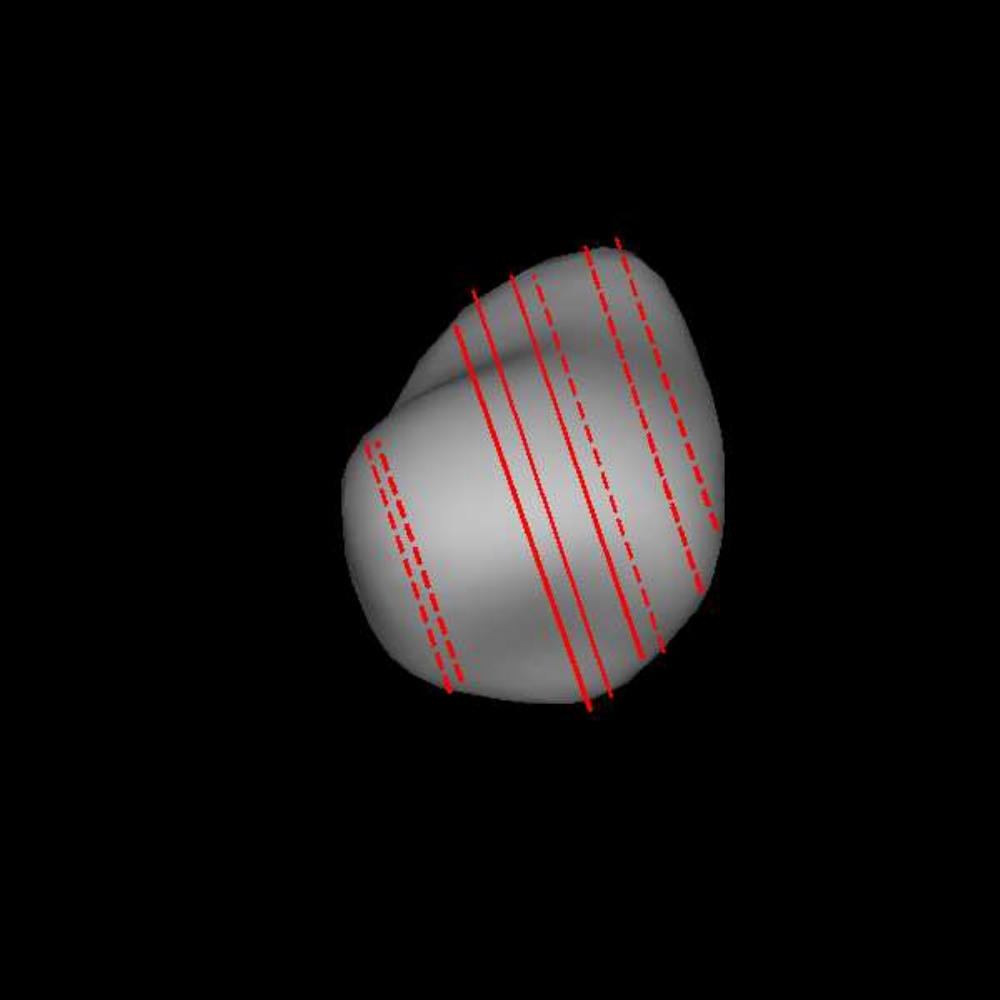}}\resizebox{0.33\hsize}{!}{\includegraphics{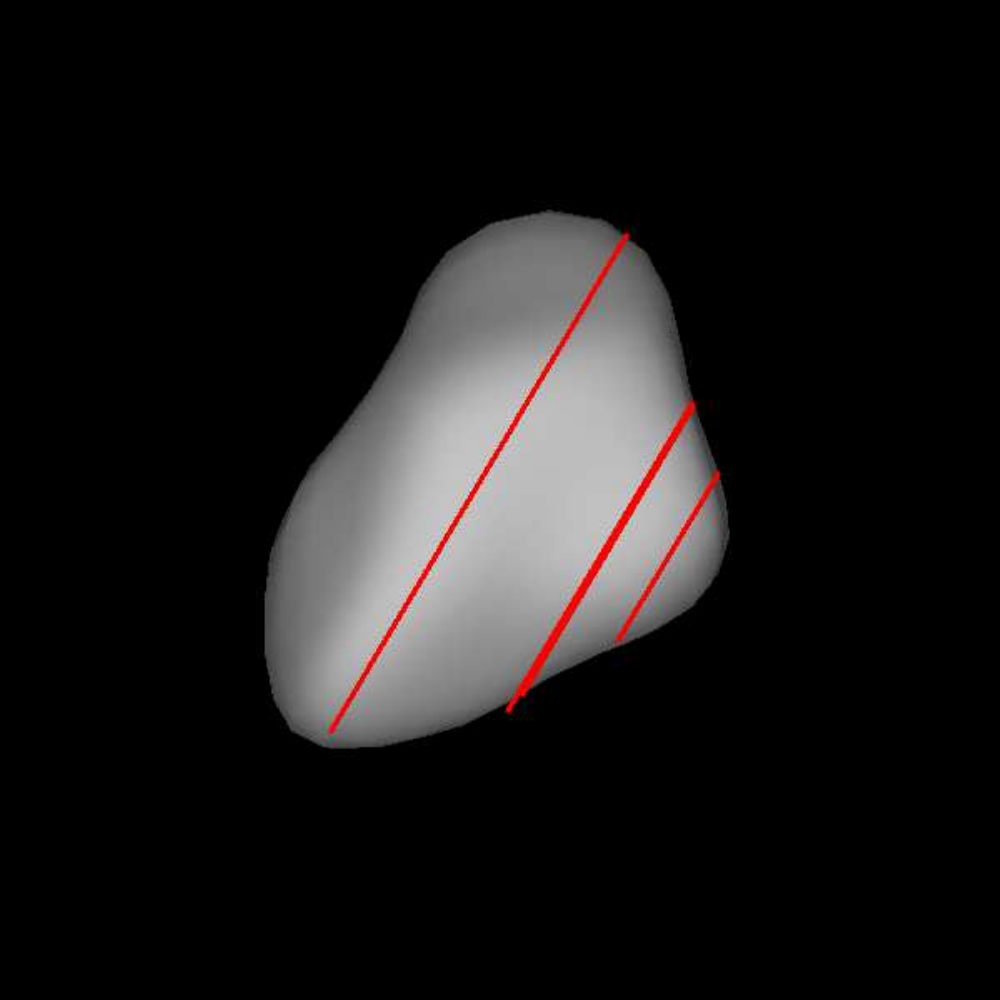}}\\
    \caption{\label{fig:41_occ}Comparison between model projections and corresponding stellar occultation(s) for asteroid (41) Daphne.}
\end{figure}

\begin{figure}[tbp]
    \centering
 \resizebox{0.33\hsize}{!}{\includegraphics{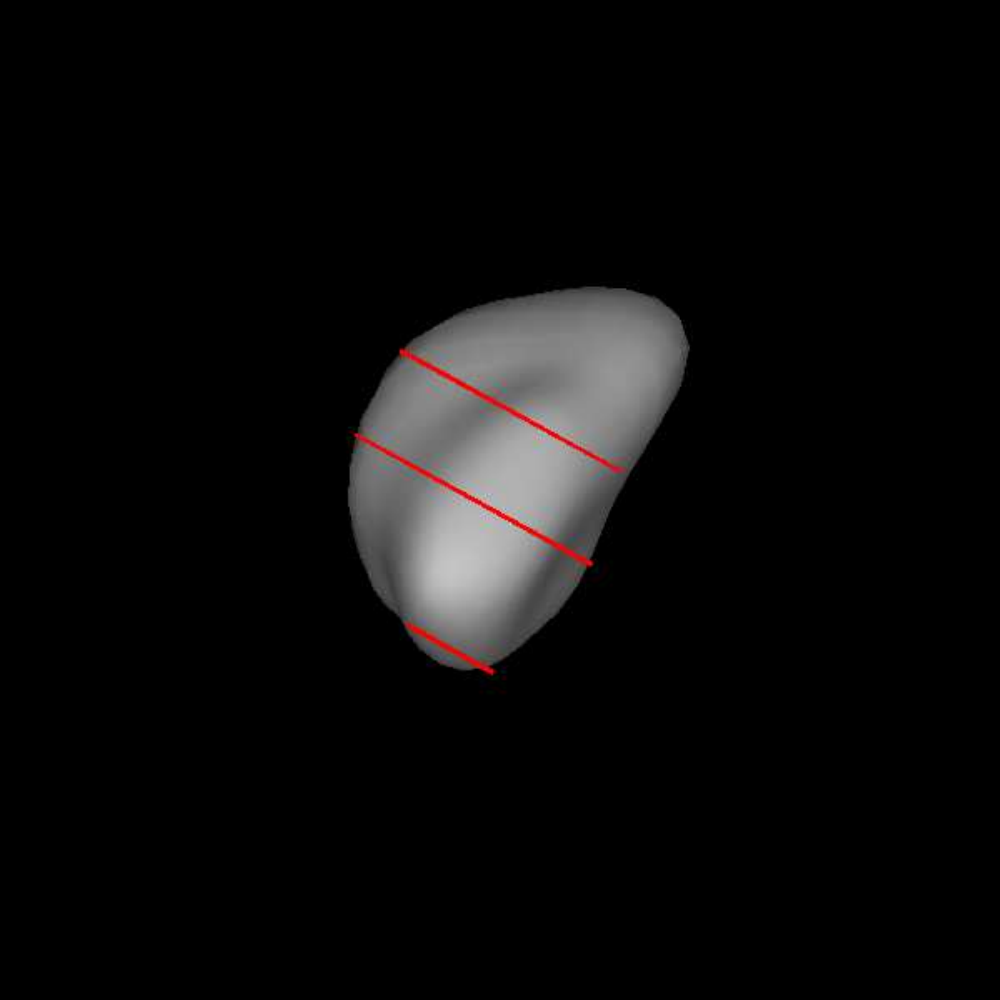}}\\
    \caption{\label{fig:43_occ}Comparison between model projections and corresponding stellar occultation(s) for asteroid (43) Ariadne.}
\end{figure}

\begin{figure}[tbp]
    \centering
 \resizebox{0.33\hsize}{!}{\includegraphics{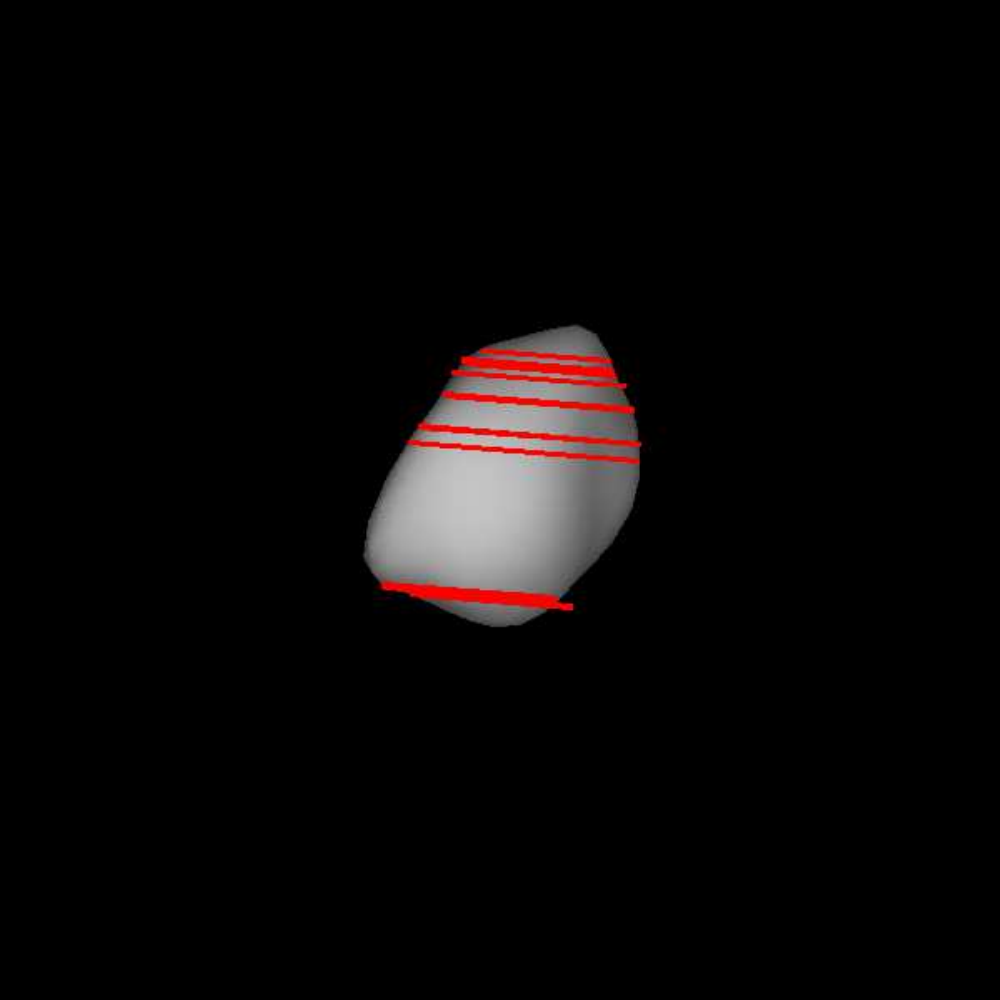}}\resizebox{0.33\hsize}{!}{\includegraphics{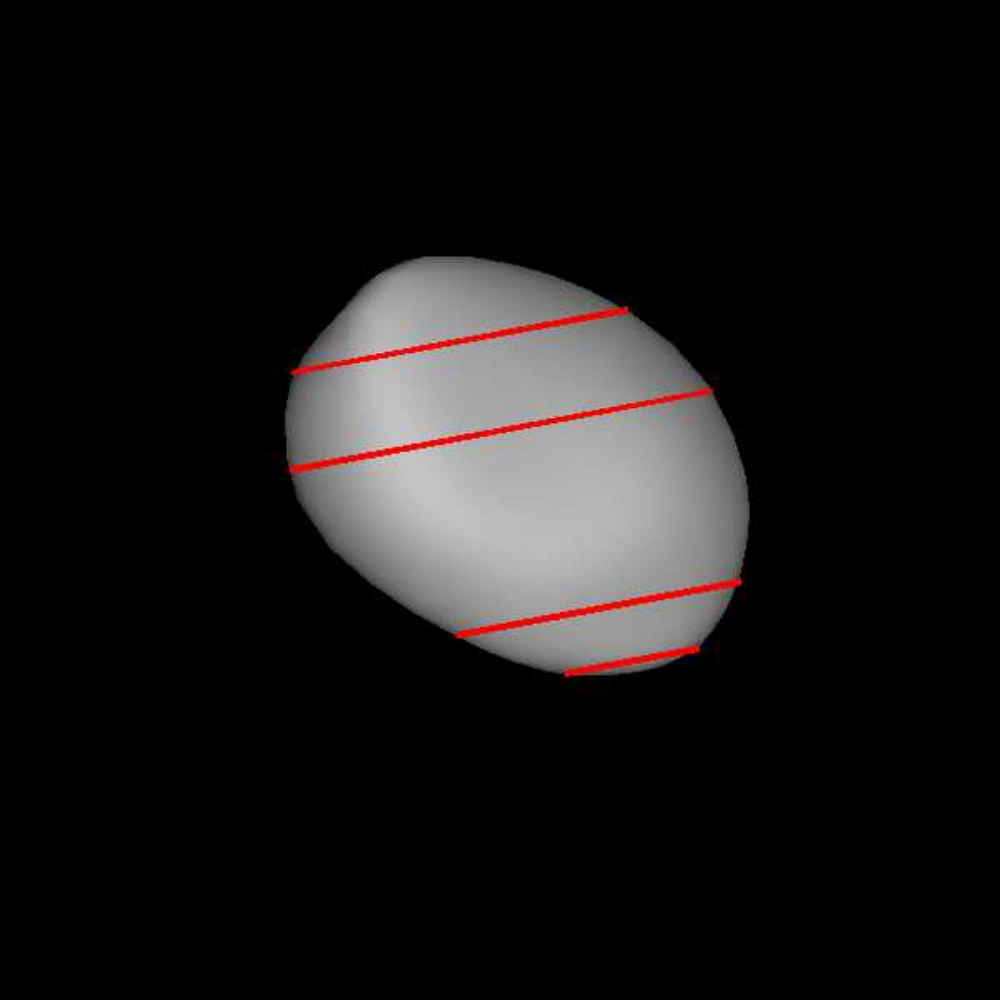}}\\
    \caption{\label{fig:45_occ}Comparison between model projections and corresponding stellar occultation(s) for asteroid (45) Eugenia.}
\end{figure}

\begin{figure}[tbp]
    \centering
 \resizebox{0.33\hsize}{!}{\includegraphics{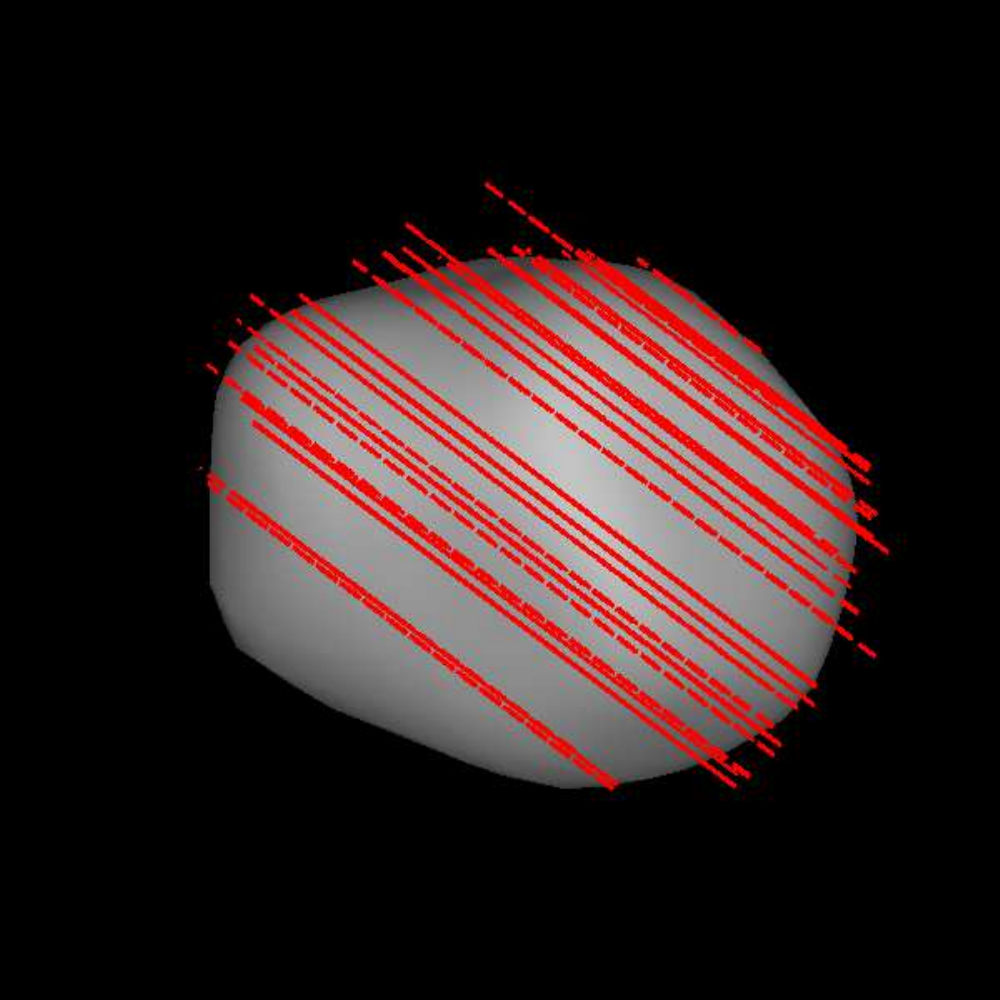}}\resizebox{0.33\hsize}{!}{\includegraphics{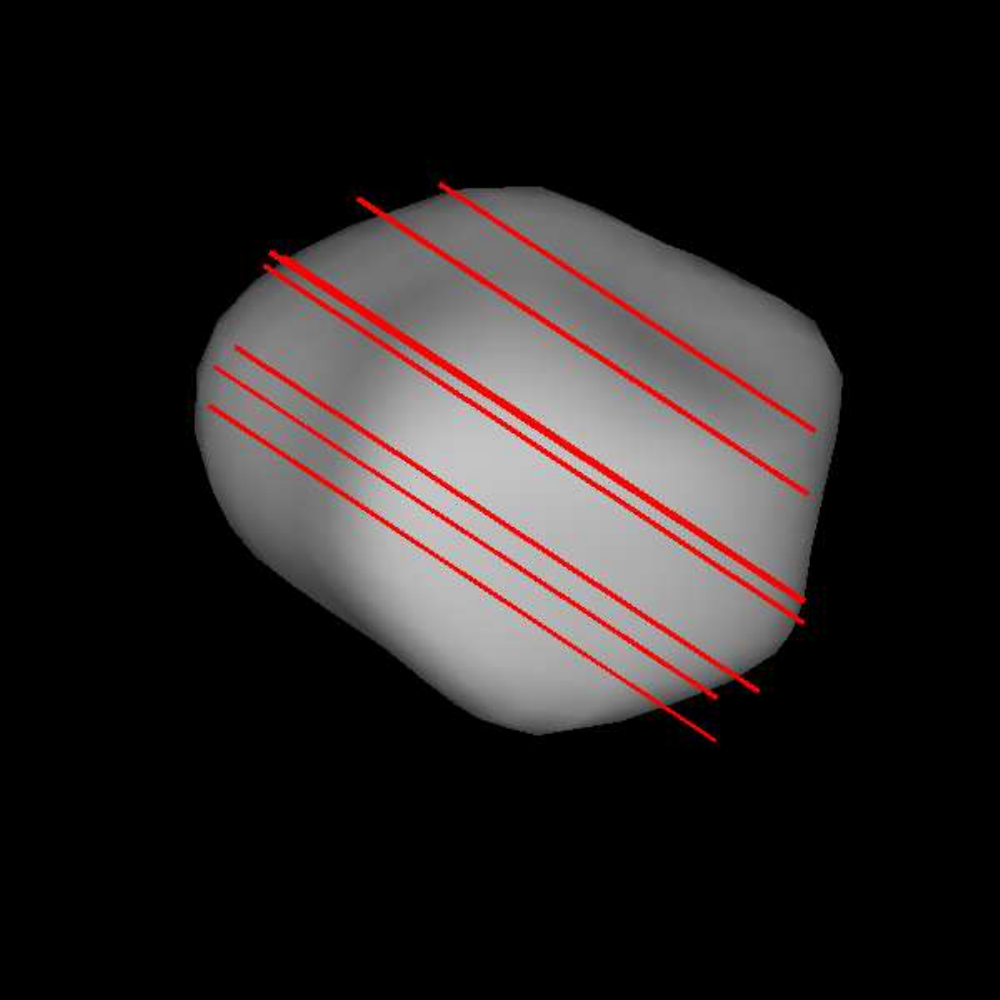}}\\
 \resizebox{0.33\hsize}{!}{\includegraphics{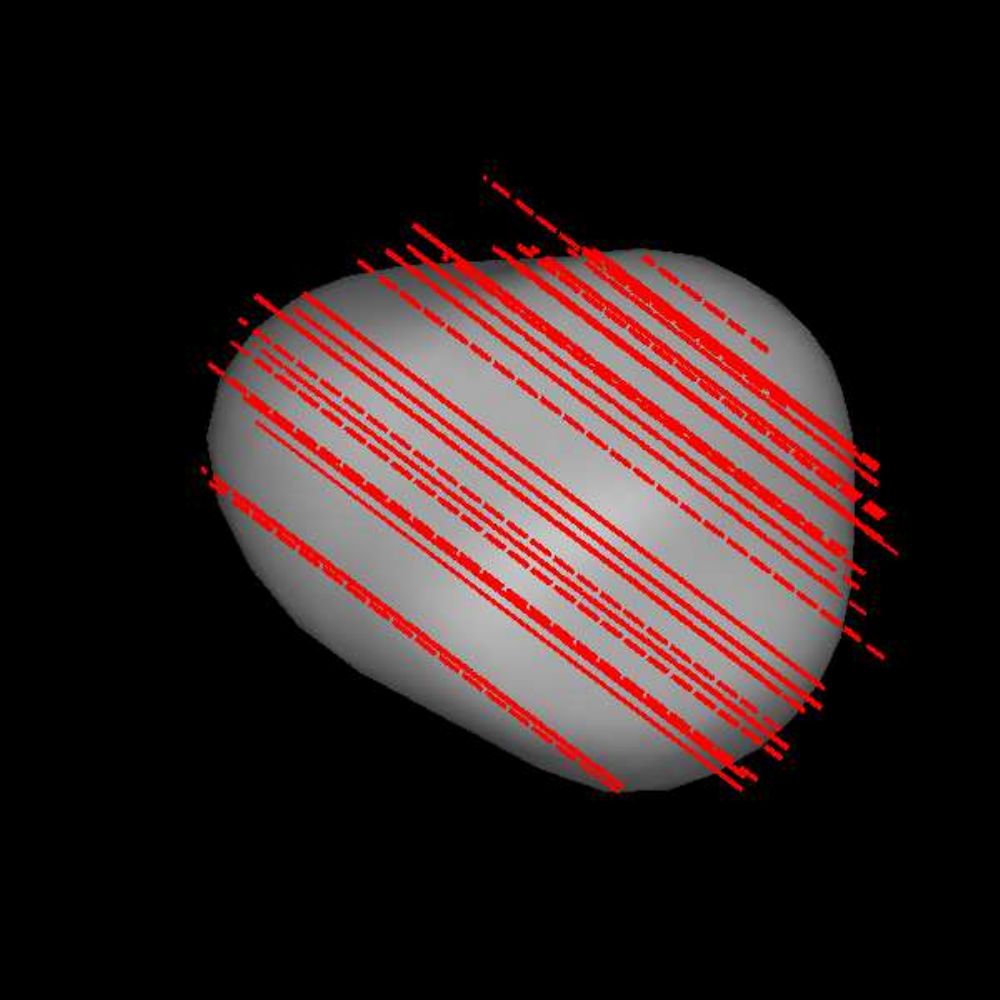}}\resizebox{0.33\hsize}{!}{\includegraphics{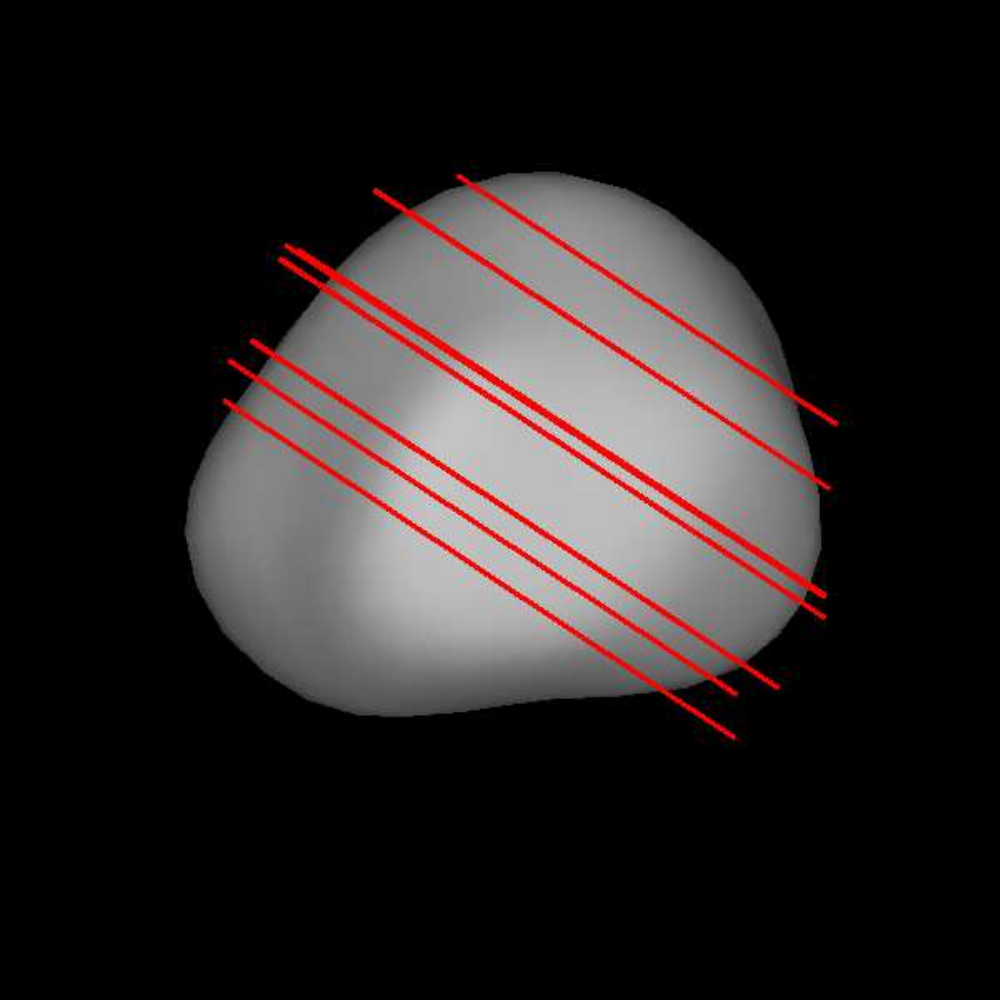}}\\
    \caption{\label{fig:51_occ}Comparison between model projections and corresponding stellar occultation(s) for asteroid (51) Nemausa. We also show the fit for the rejected pole solution (bottom panel).}
\end{figure}

\begin{figure}[tbp]
    \centering
 \resizebox{0.24\hsize}{!}{\includegraphics{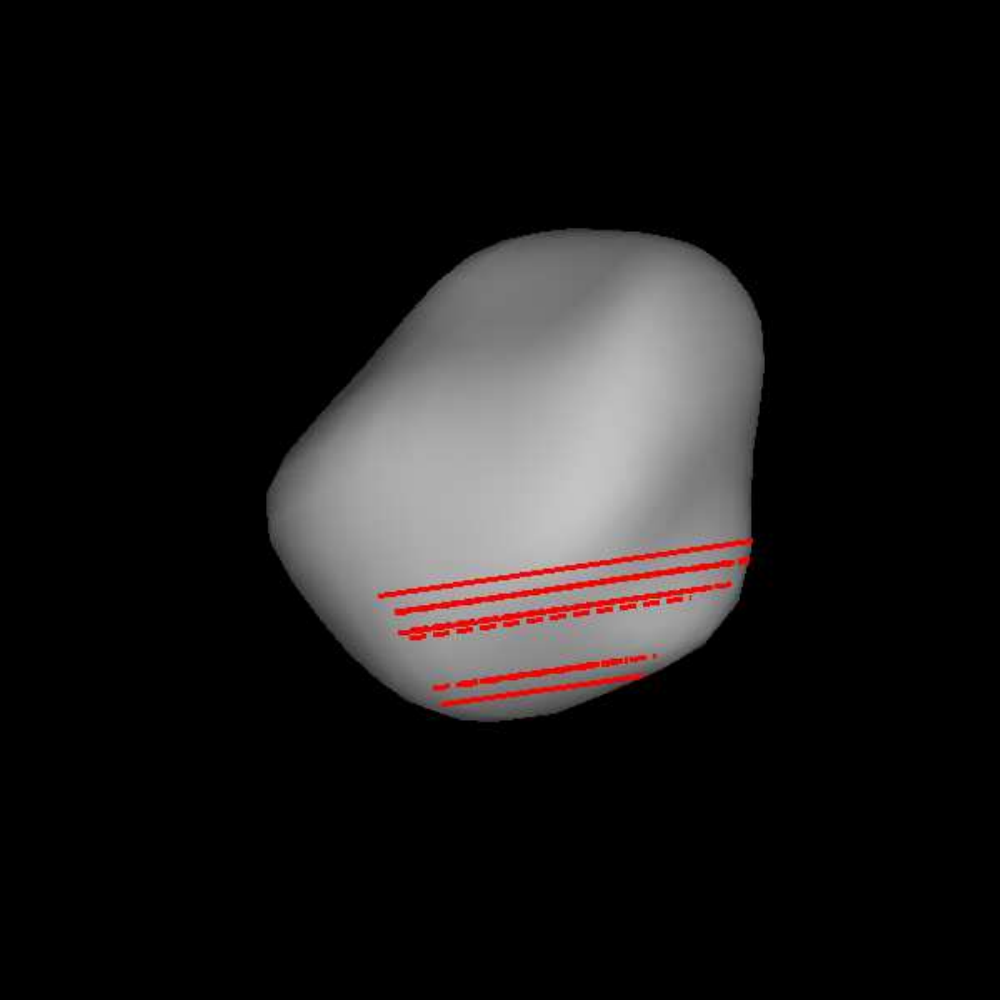}}\resizebox{0.24\hsize}{!}{\includegraphics{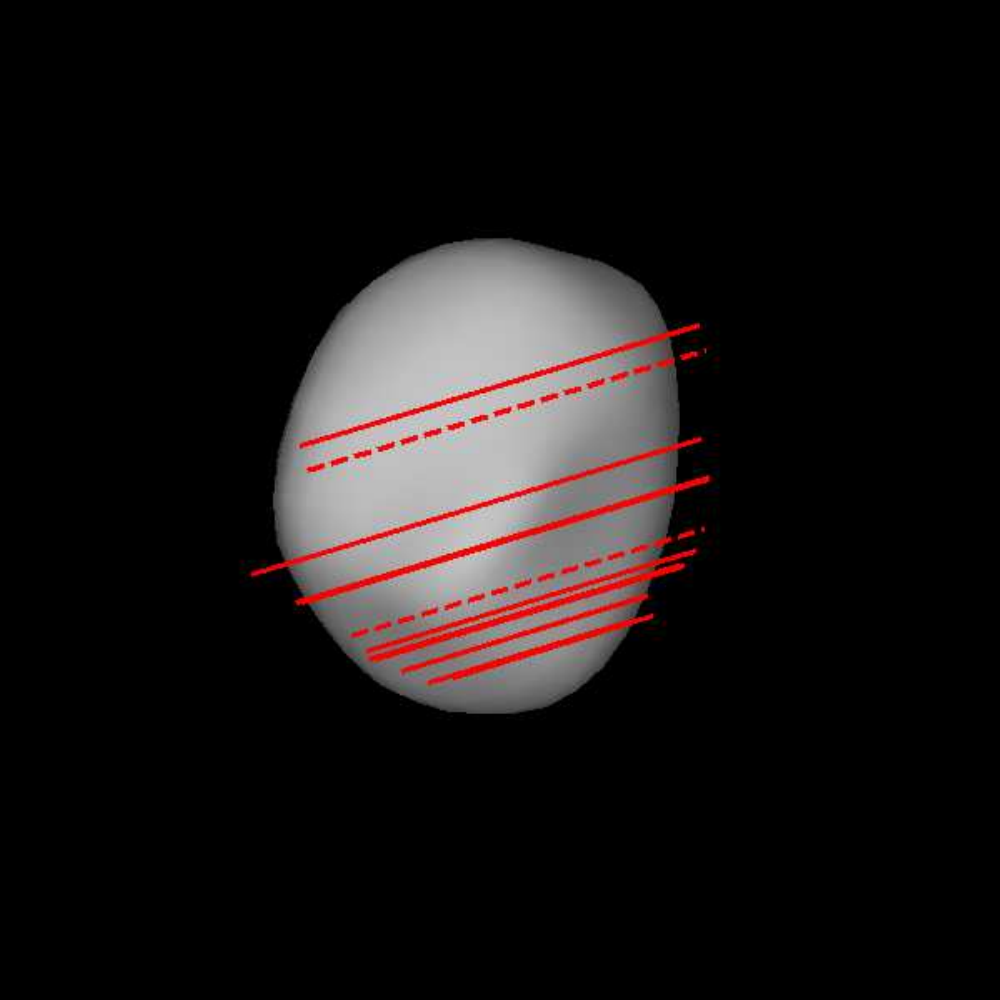}}\resizebox{0.24\hsize}{!}{\includegraphics{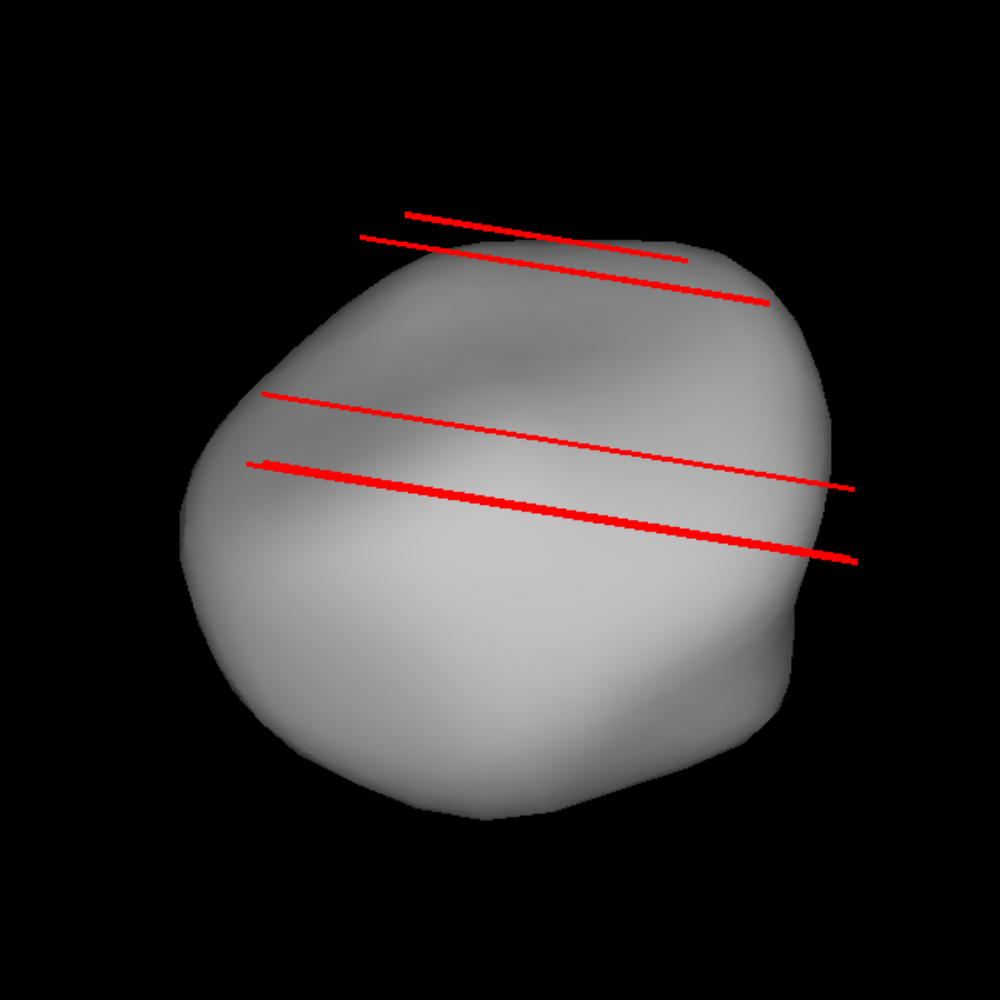}}\resizebox{0.24\hsize}{!}{\includegraphics{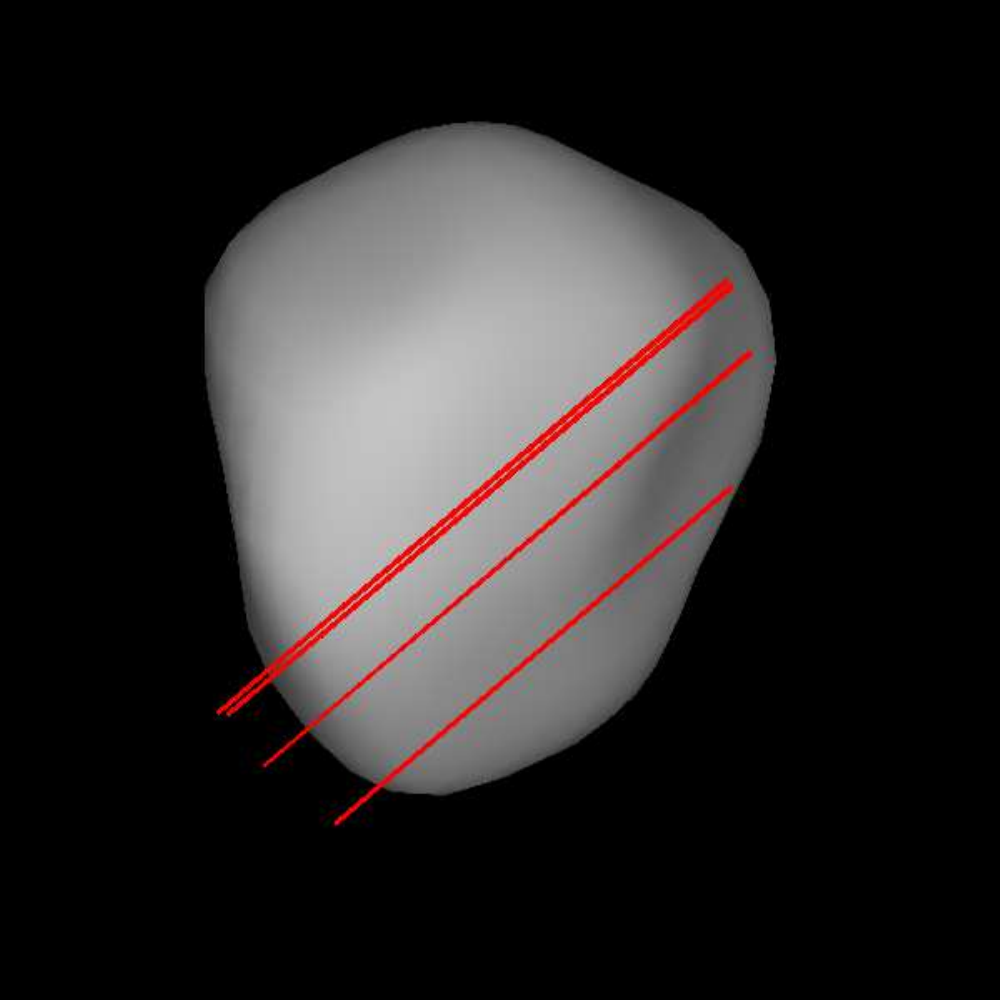}}\\
    \caption{\label{fig:52_occ}Comparison between model projections and corresponding stellar occultation(s) for asteroid (52) Europa.}
\end{figure}

\begin{figure}[tbp]
    \centering
 \resizebox{0.33\hsize}{!}{\includegraphics{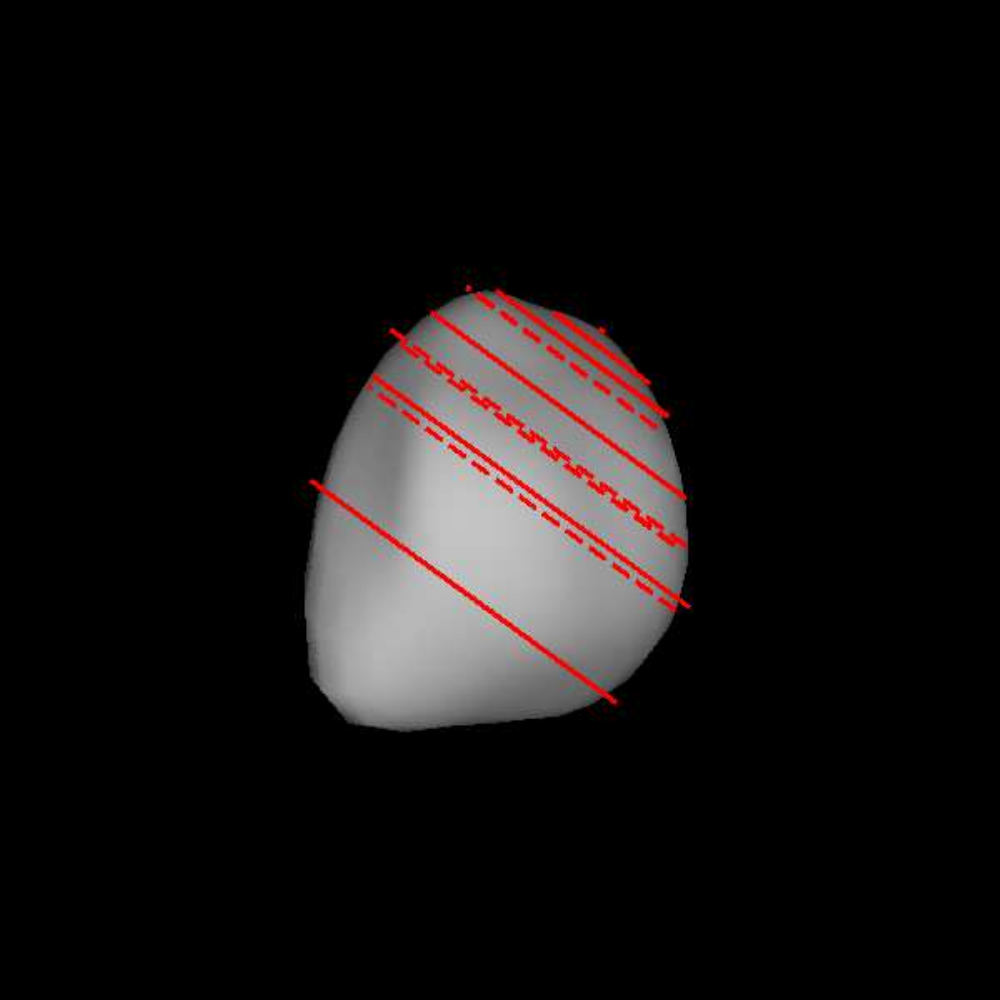}}\\
    \caption{\label{fig:54_occ}Comparison between model projections and corresponding stellar occultation(s) for asteroid (54) Alexandra.}
\end{figure}

\begin{figure}[tbp]
    \centering
 \resizebox{0.33\hsize}{!}{\includegraphics{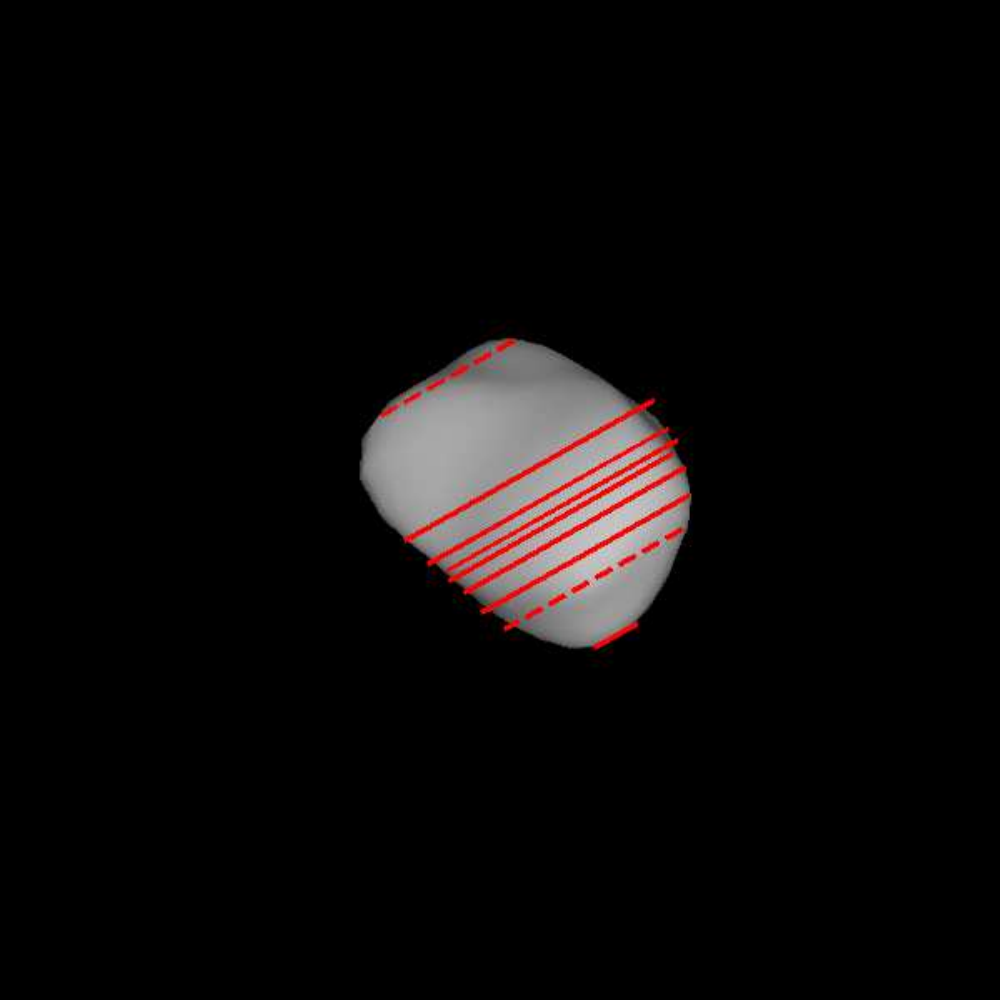}}\\
    \caption{\label{fig:80_occ}Comparison between model projections and corresponding stellar occultation(s) for asteroid (80) Sappho.}
\end{figure}

\begin{figure}[tbp]
    \centering
 \resizebox{0.33\hsize}{!}{\includegraphics{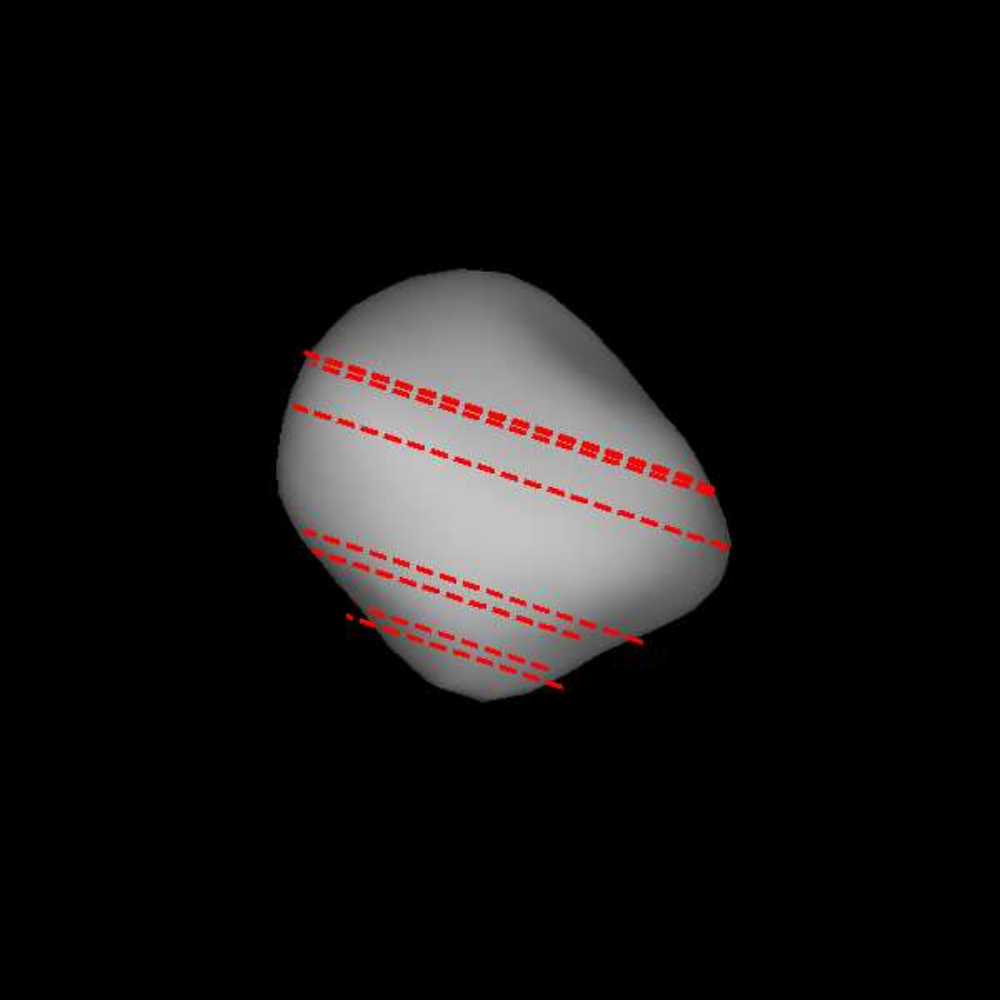}}\\
    \caption{\label{fig:85_occ}Comparison between model projections and corresponding stellar occultation(s) for asteroid (85) Io.}
\end{figure}

\begin{figure}[tbp]
    \centering
 \resizebox{0.33\hsize}{!}{\includegraphics{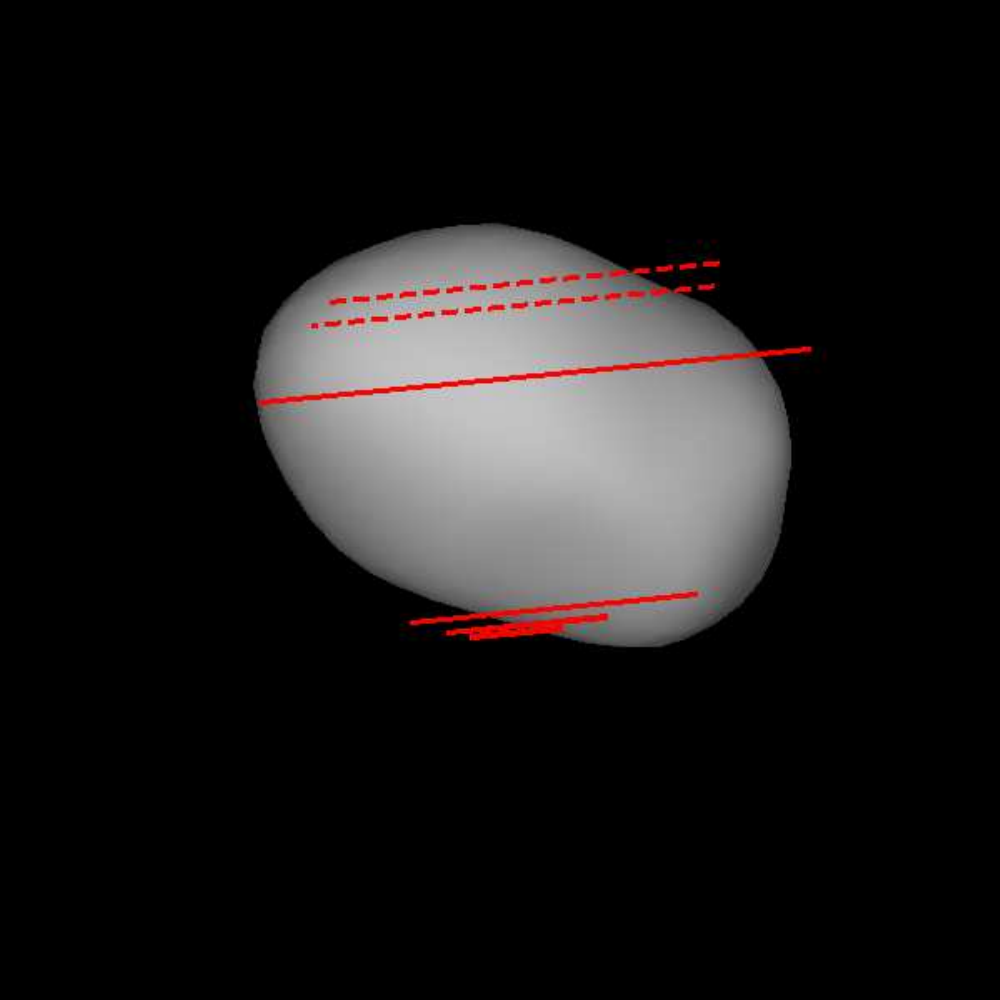}}\resizebox{0.33\hsize}{!}{\includegraphics{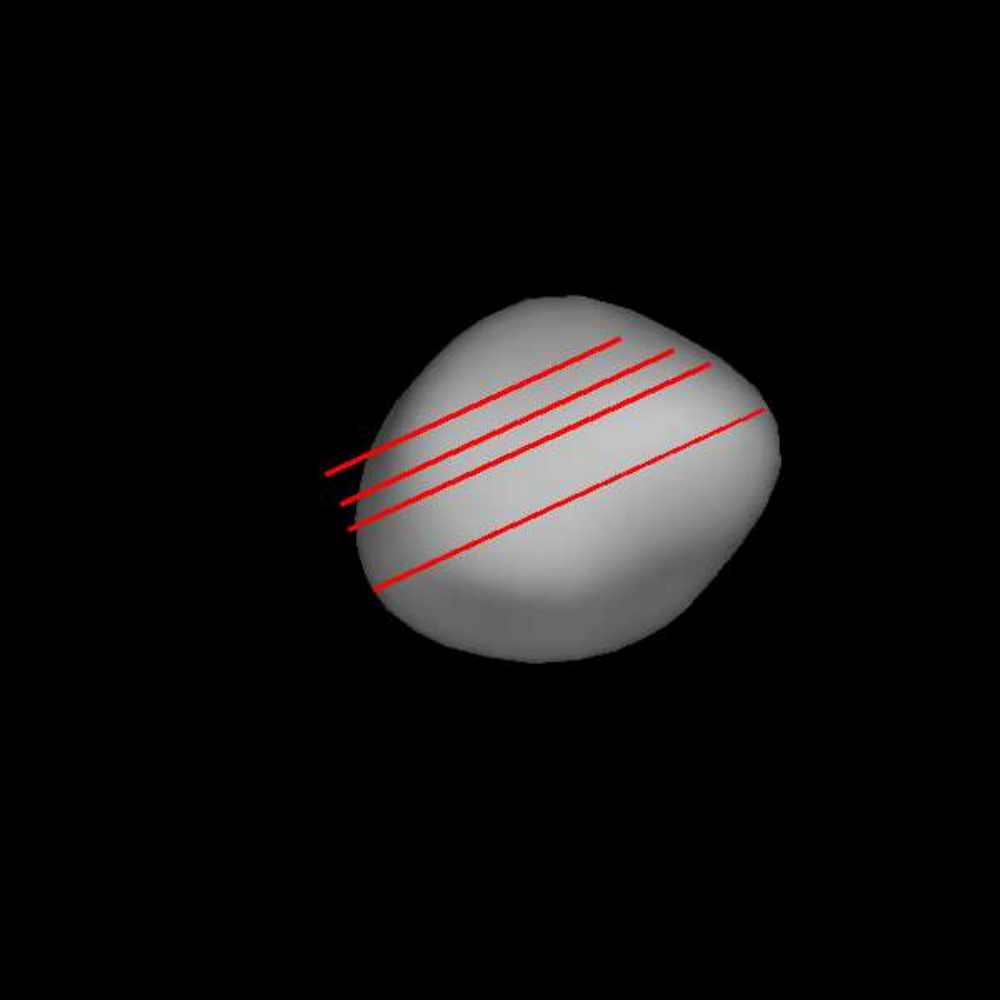}}\\
    \caption{\label{fig:87_occ}Comparison between model projections and corresponding stellar occultation(s) for asteroid (87) Sylvia.}
\end{figure}

\begin{figure}[tbp]
    \centering
 \resizebox{0.33\hsize}{!}{\includegraphics{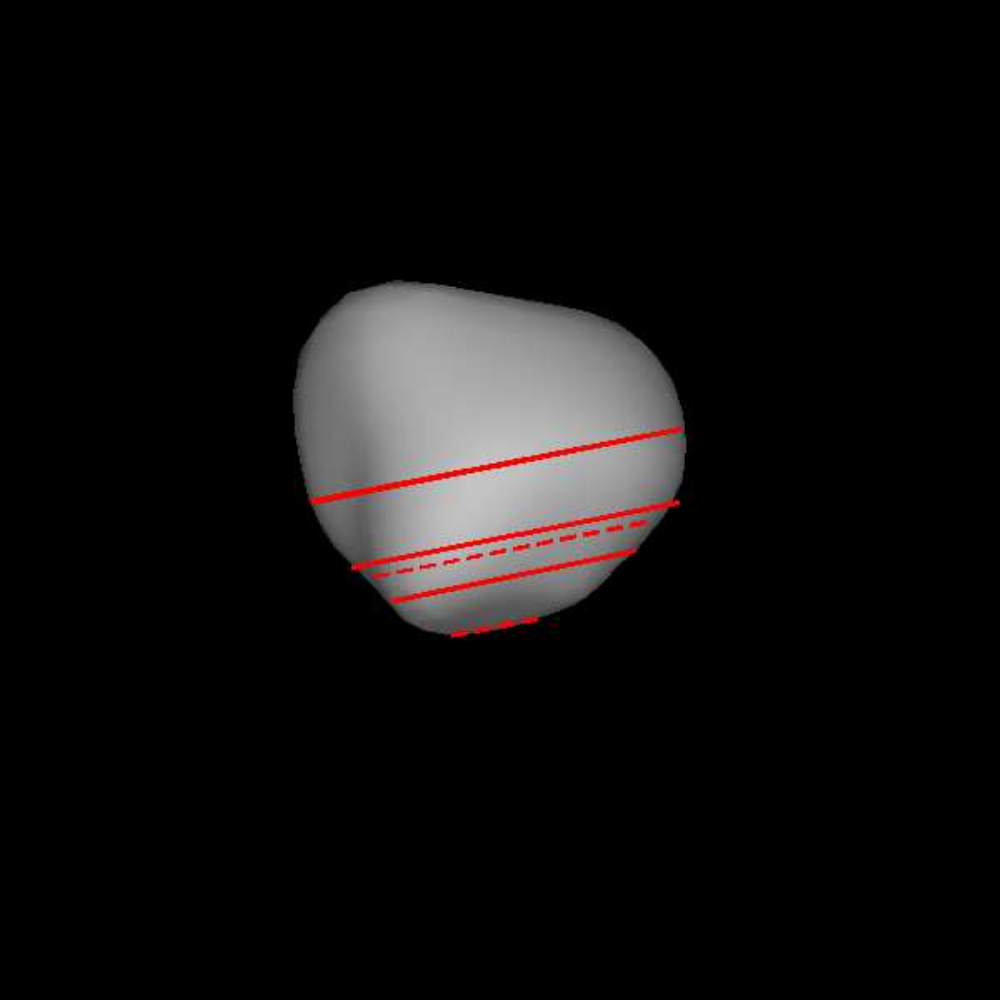}}\\
    \caption{\label{fig:88_occ}Comparison between model projections and corresponding stellar occultation(s) for asteroid (88) Thisbe.}
\end{figure}

\begin{figure}[tbp]
    \centering
 \resizebox{0.33\hsize}{!}{\includegraphics{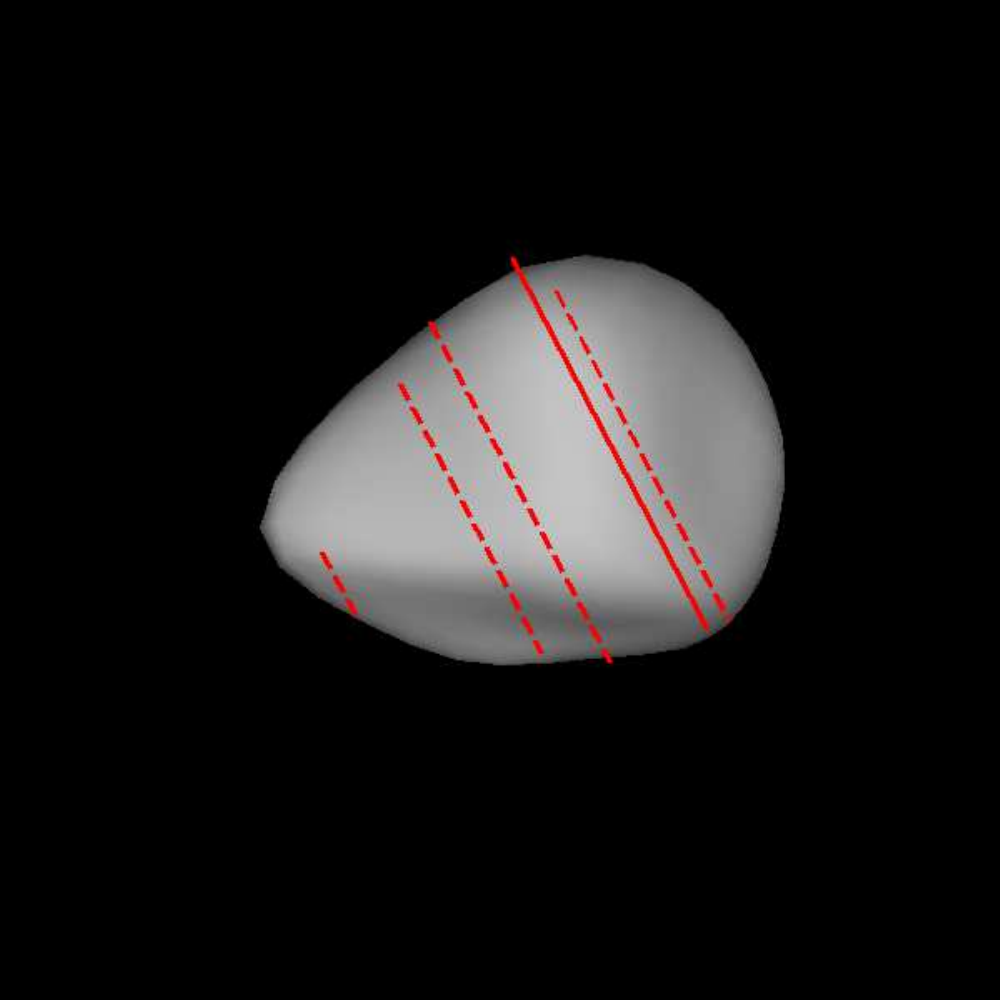}}\resizebox{0.33\hsize}{!}{\includegraphics{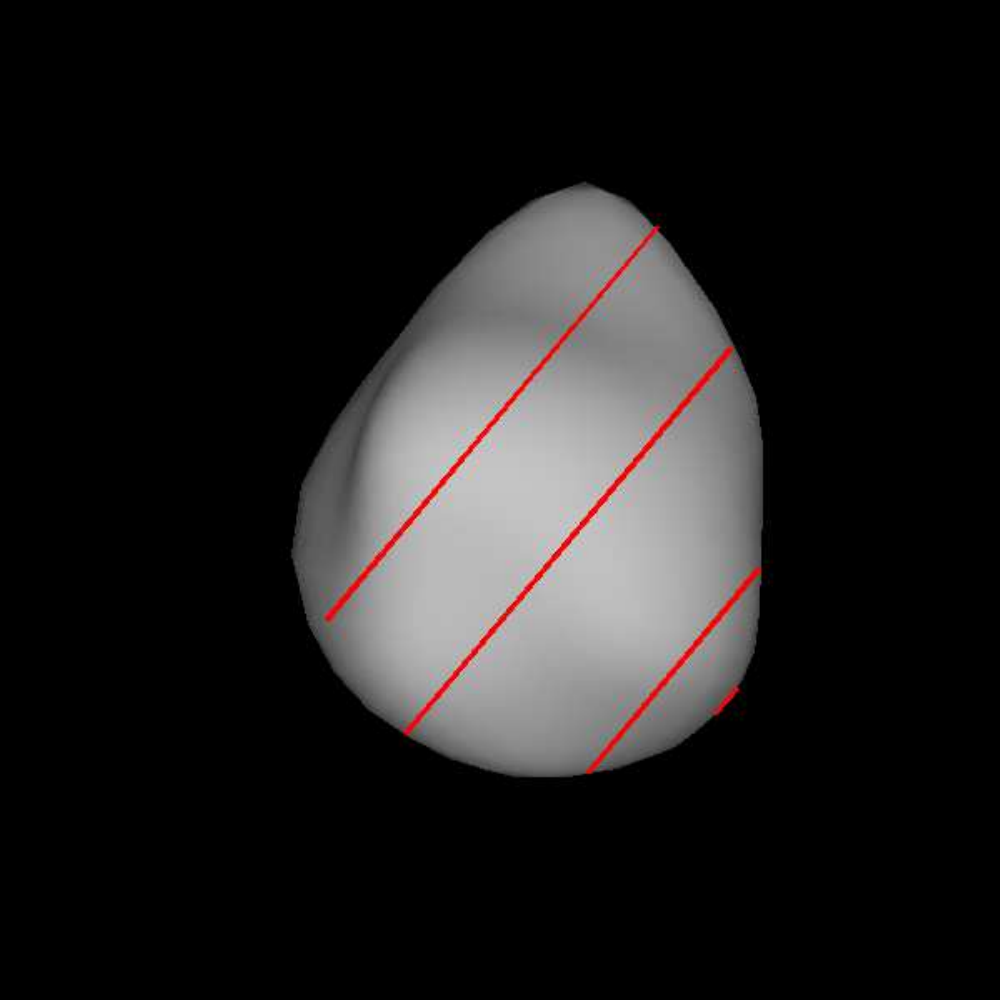}}\\
    \caption{\label{fig:89_occ}Comparison between model projections and corresponding stellar occultation(s) for asteroid (89) Julia.}
\end{figure}

\clearpage

\begin{figure}[tbp]
    \centering
 \resizebox{0.33\hsize}{!}{\includegraphics{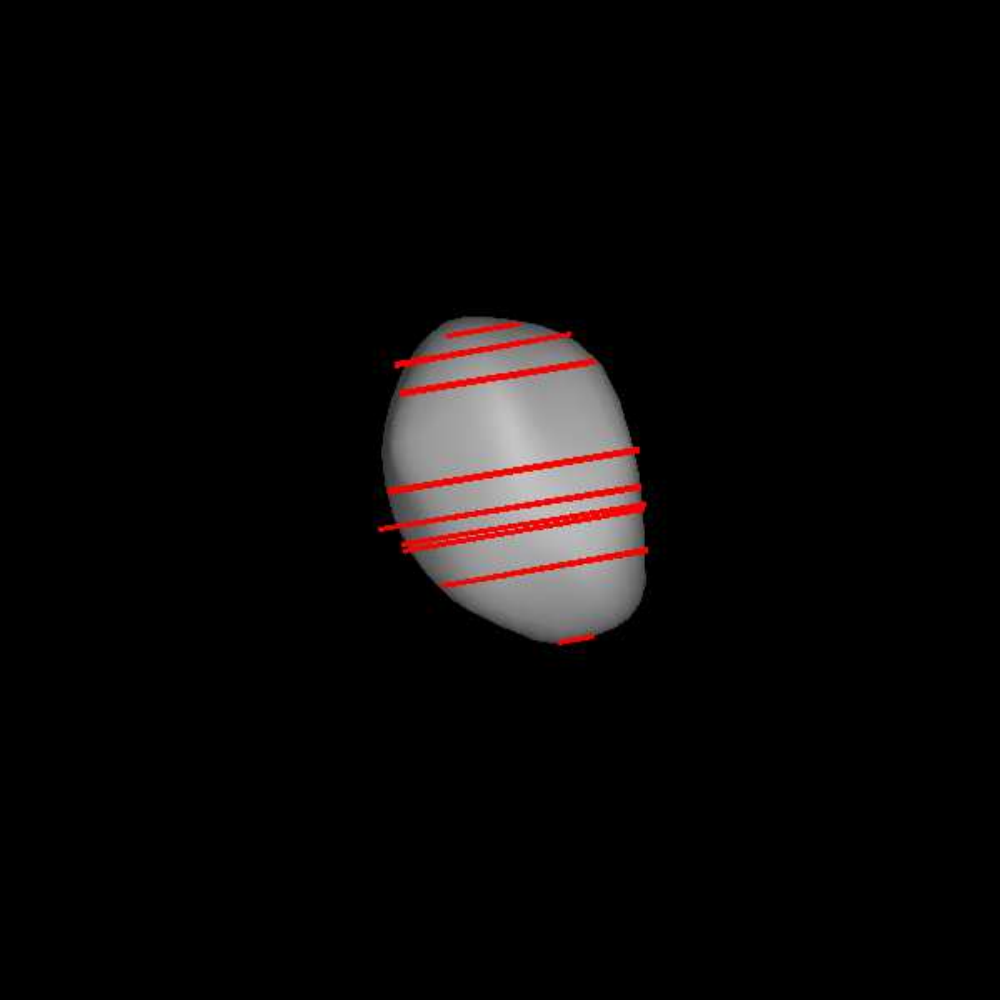}}\resizebox{0.33\hsize}{!}{\includegraphics{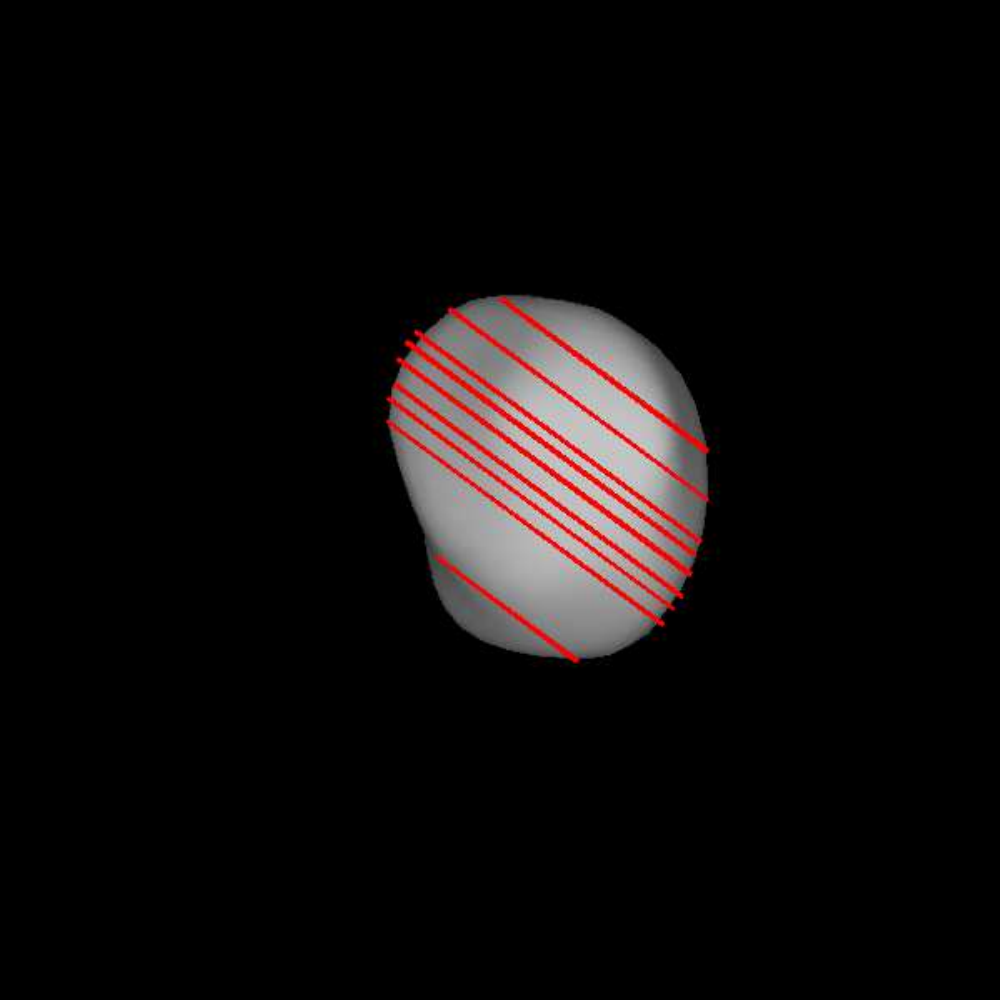}}\\
    \caption{\label{fig:93_occ}Comparison between model projections and corresponding stellar occultation(s) for asteroid (93) Minerva.}
\end{figure}

\begin{figure}[tbp]
    \centering
 \resizebox{0.33\hsize}{!}{\includegraphics{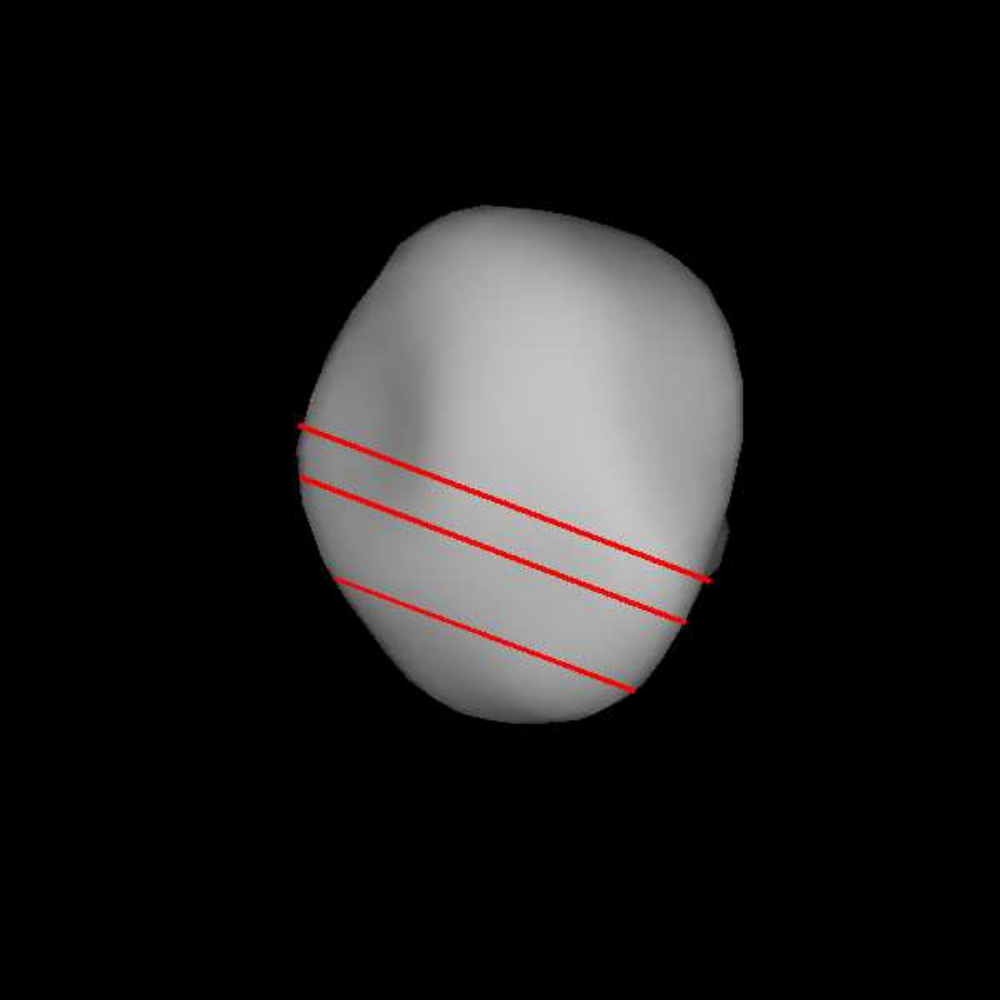}}\resizebox{0.33\hsize}{!}{\includegraphics{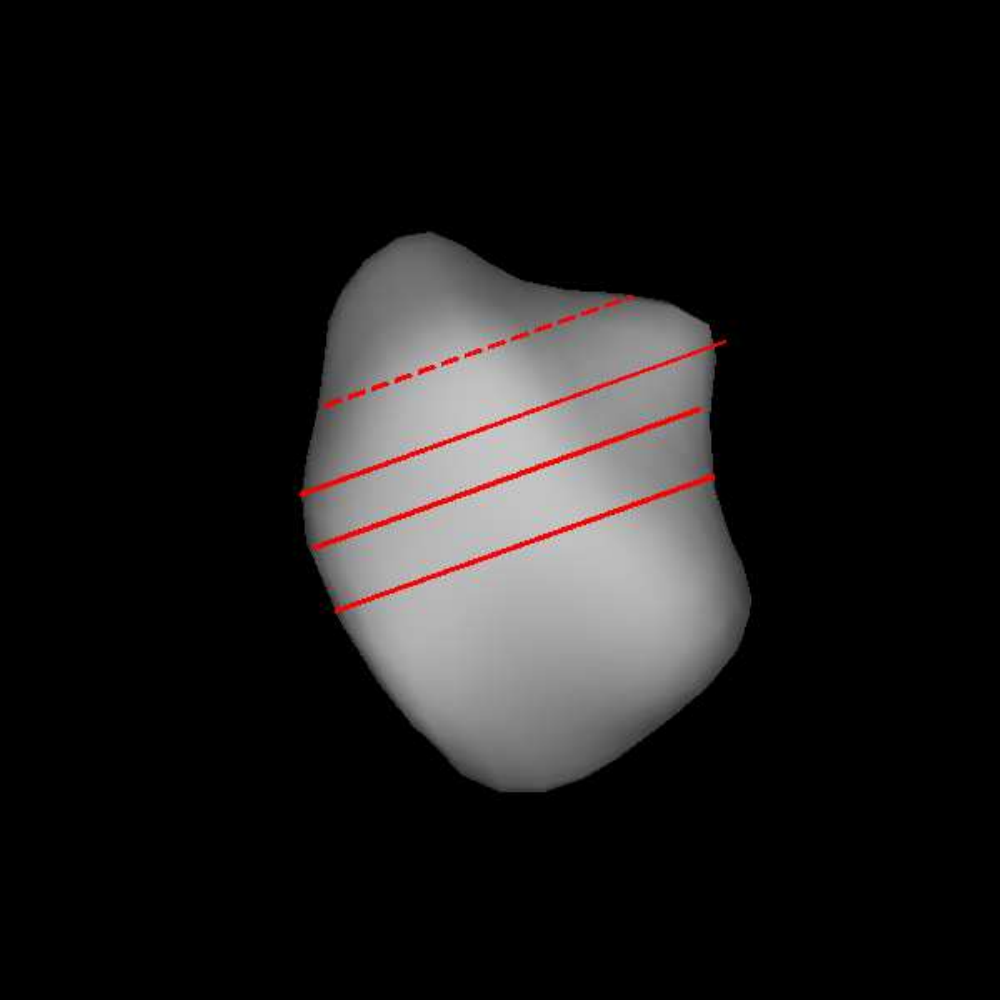}}\\
    \caption{\label{fig:94_occ}Comparison between model projections and corresponding stellar occultation(s) for asteroid (94) Aurora. We show the fit for both pole solutions.}
\end{figure}

\begin{figure}[tbp]
    \centering
 \resizebox{0.33\hsize}{!}{\includegraphics{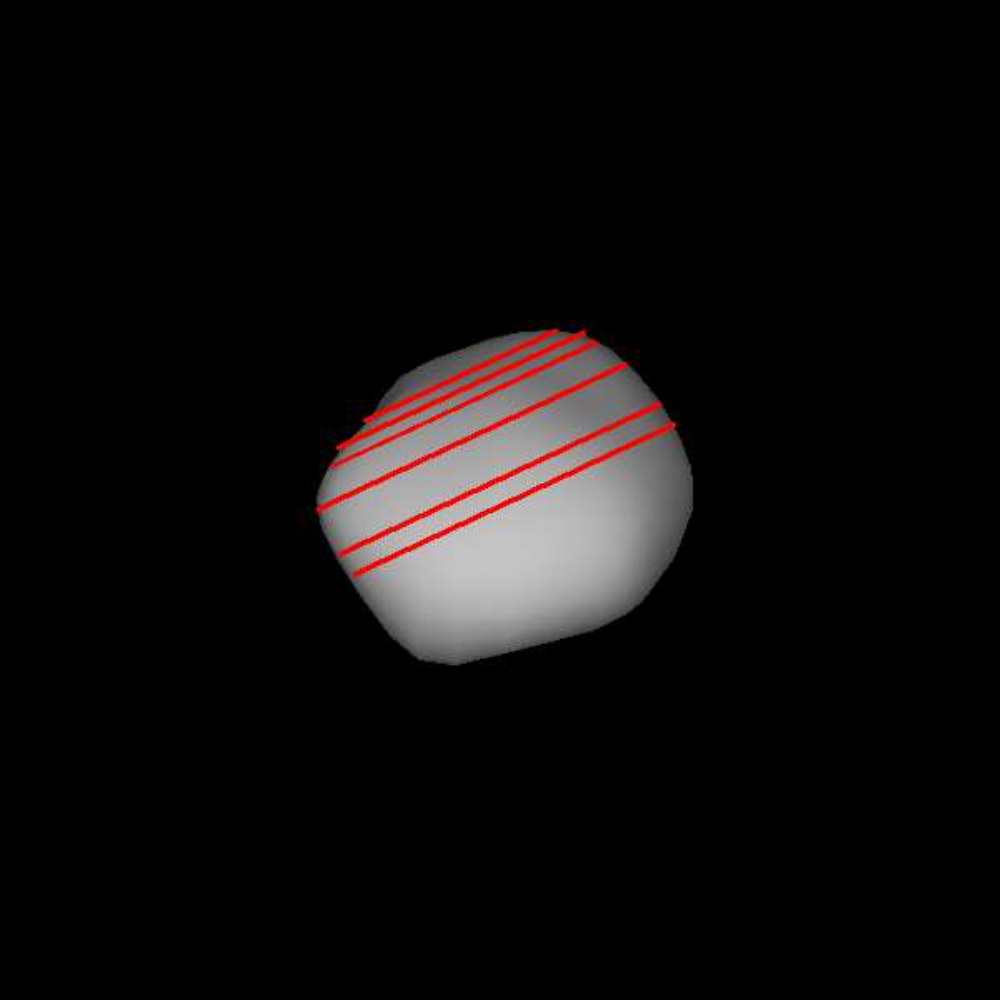}}\\
    \caption{\label{fig:107_occ}Comparison between model projections and corresponding stellar occultation(s) for asteroid (107) Camilla.}
\end{figure}

\begin{figure}[tbp]
    \centering
 \resizebox{0.33\hsize}{!}{\includegraphics{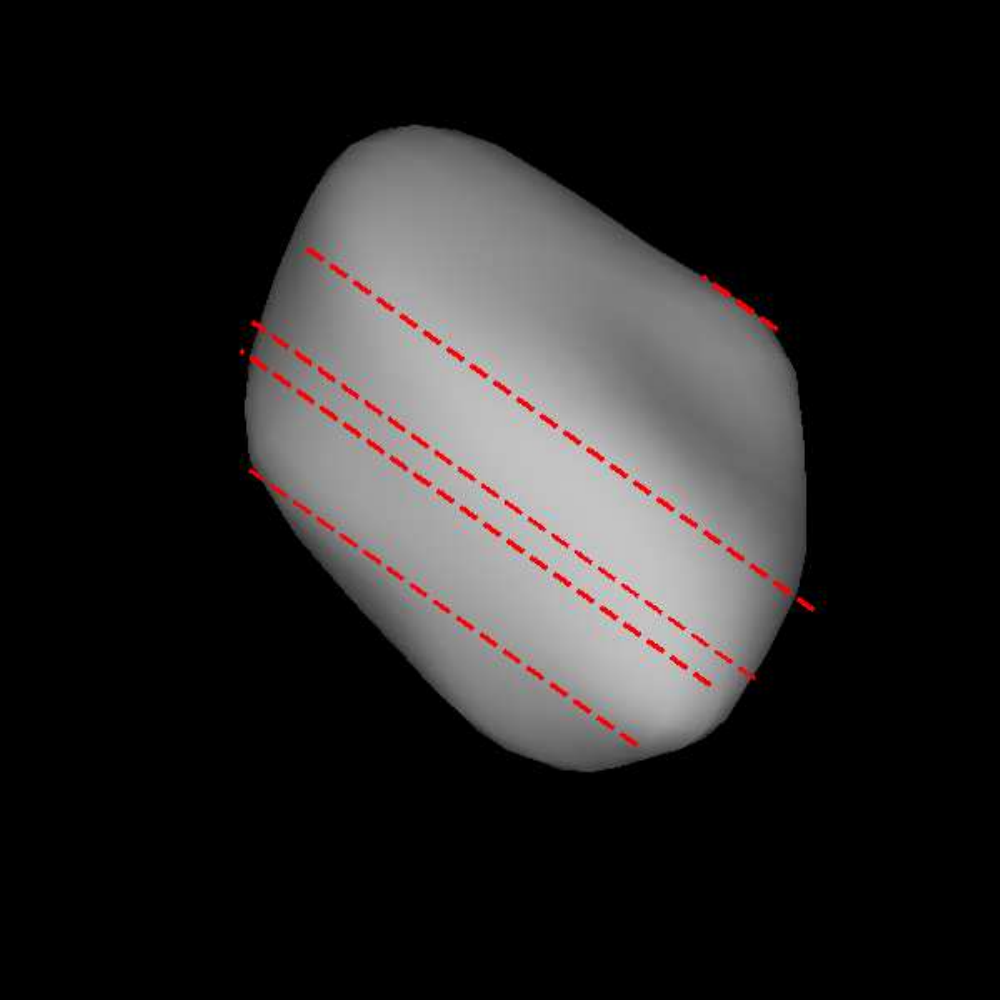}}\resizebox{0.33\hsize}{!}{\includegraphics{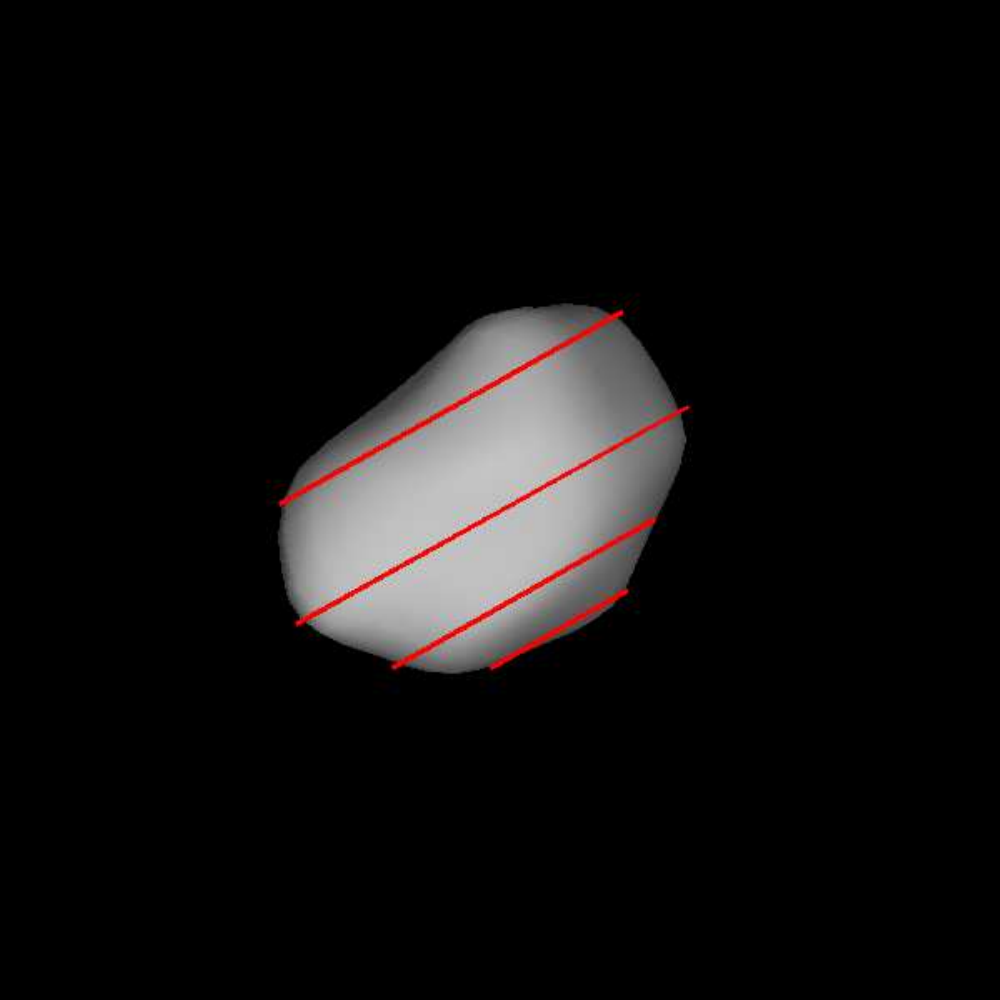}}\\
    \caption{\label{fig:129_occ}Comparison between model projections and corresponding stellar occultation(s) for asteroid (129) Antigone.}
\end{figure}

\begin{figure}[tbp]
    \centering
 \resizebox{0.33\hsize}{!}{\includegraphics{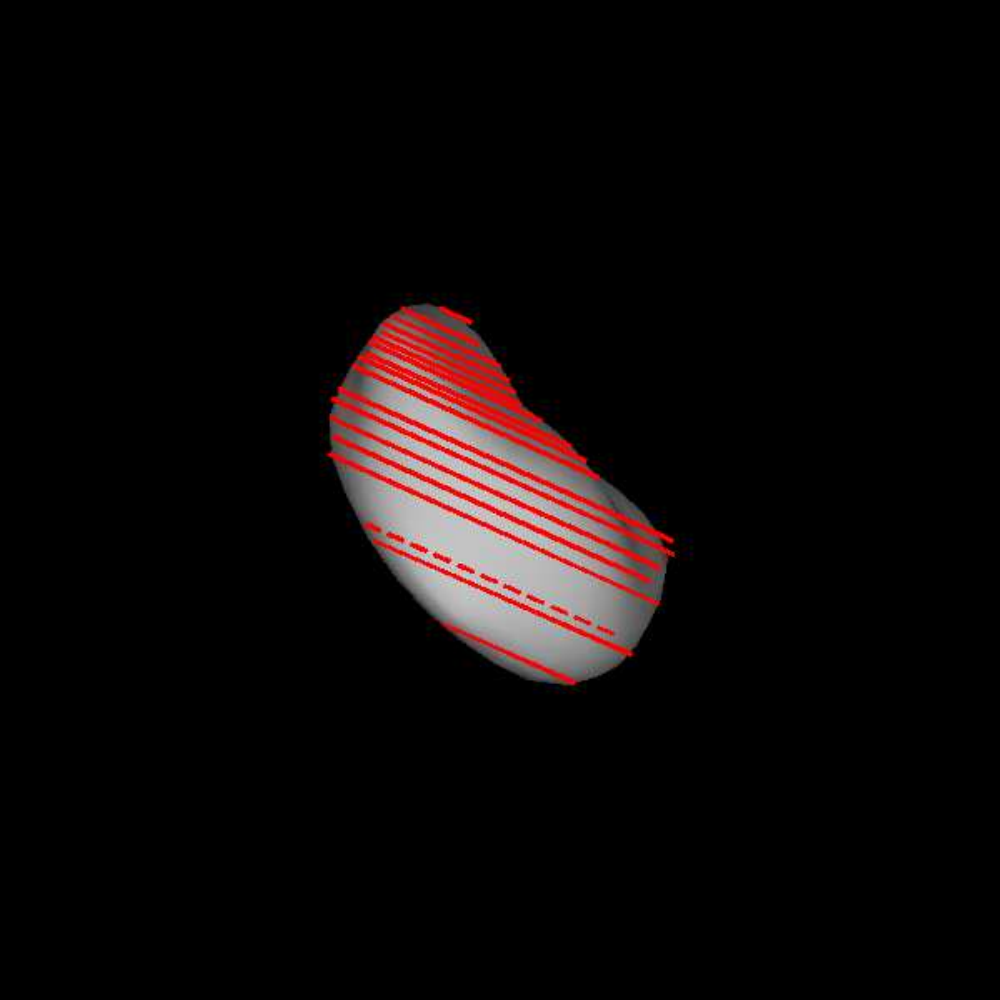}}\\
    \caption{\label{fig:135_occ}Comparison between model projections and corresponding stellar occultation(s) for asteroid (135) Hertha.}
\end{figure}

\begin{figure}[tbp]
    \centering
 \resizebox{0.33\hsize}{!}{\includegraphics{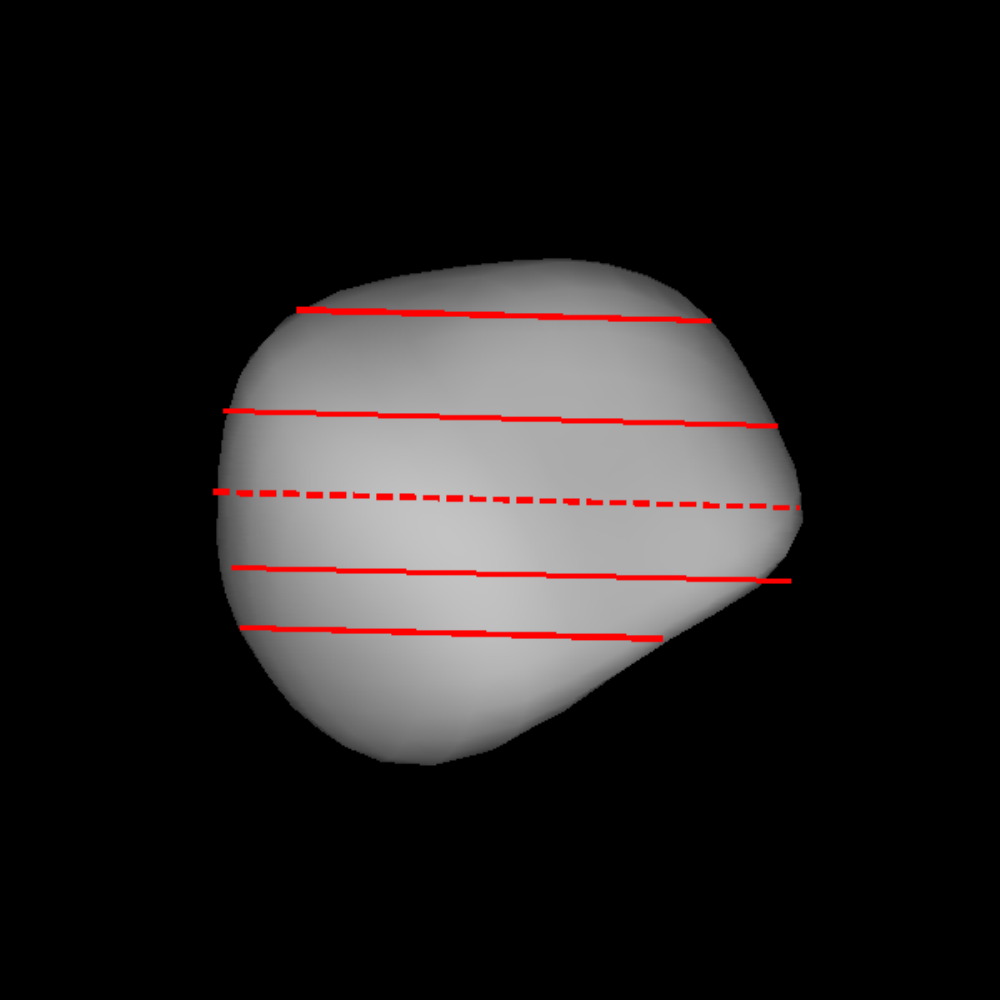}}\resizebox{0.33\hsize}{!}{\includegraphics{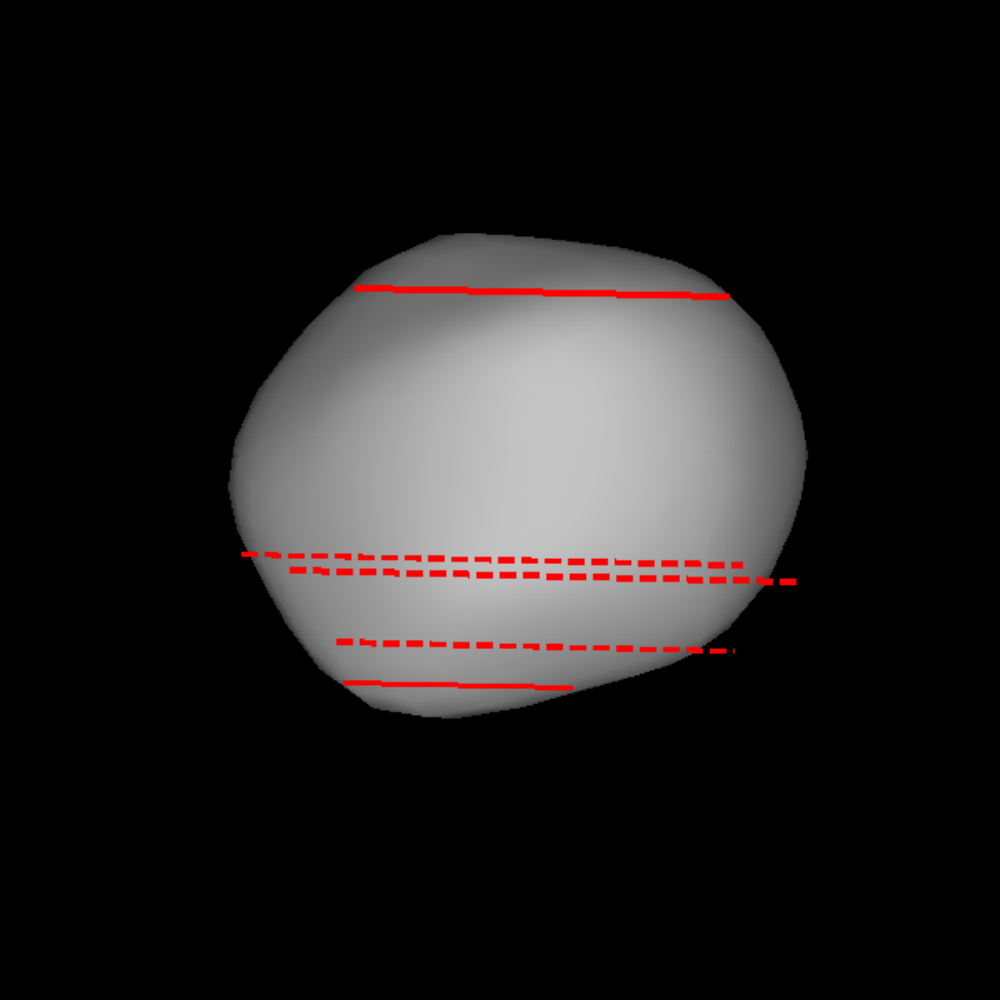}}\resizebox{0.33\hsize}{!}{\includegraphics{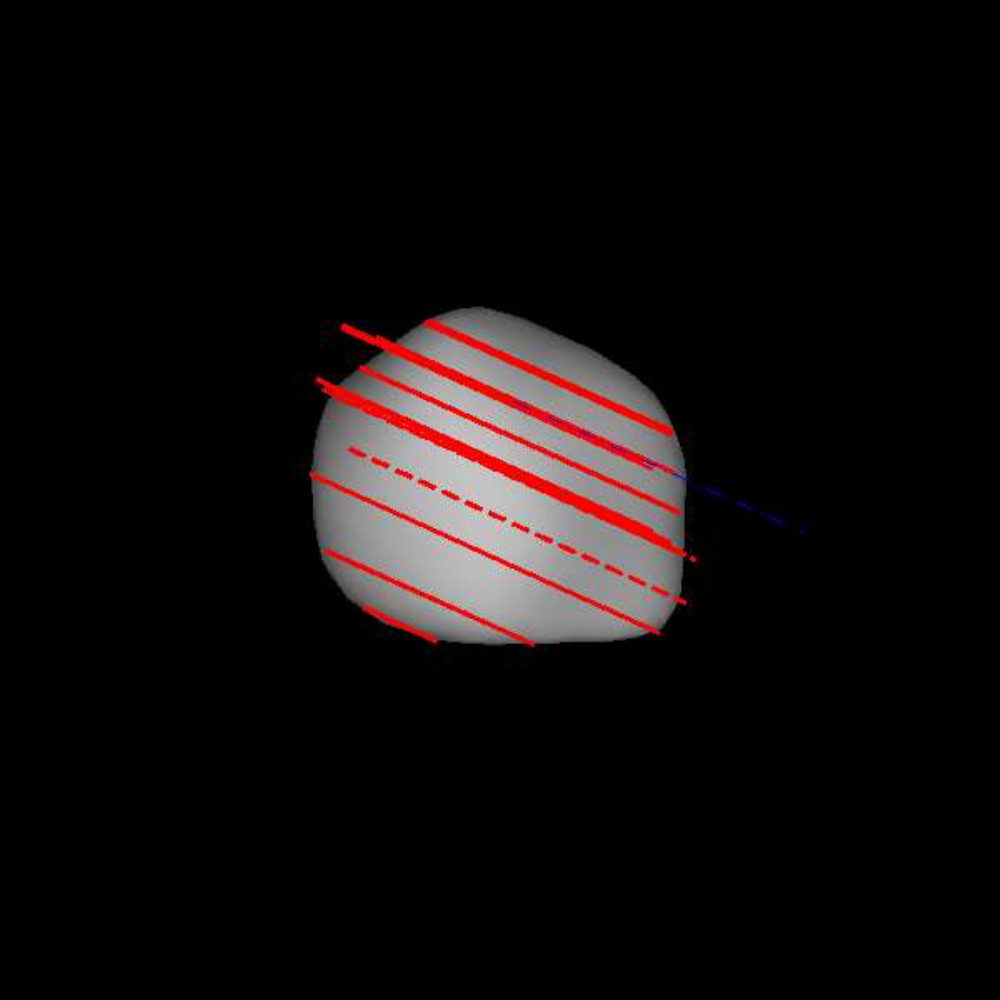}}\\
    \caption{\label{fig:144_occ}Comparison between model projections and corresponding stellar occultation(s) for asteroid (144) Vibilia.}
\end{figure}

\begin{figure}[tbp]
    \centering
 \resizebox{0.33\hsize}{!}{\includegraphics{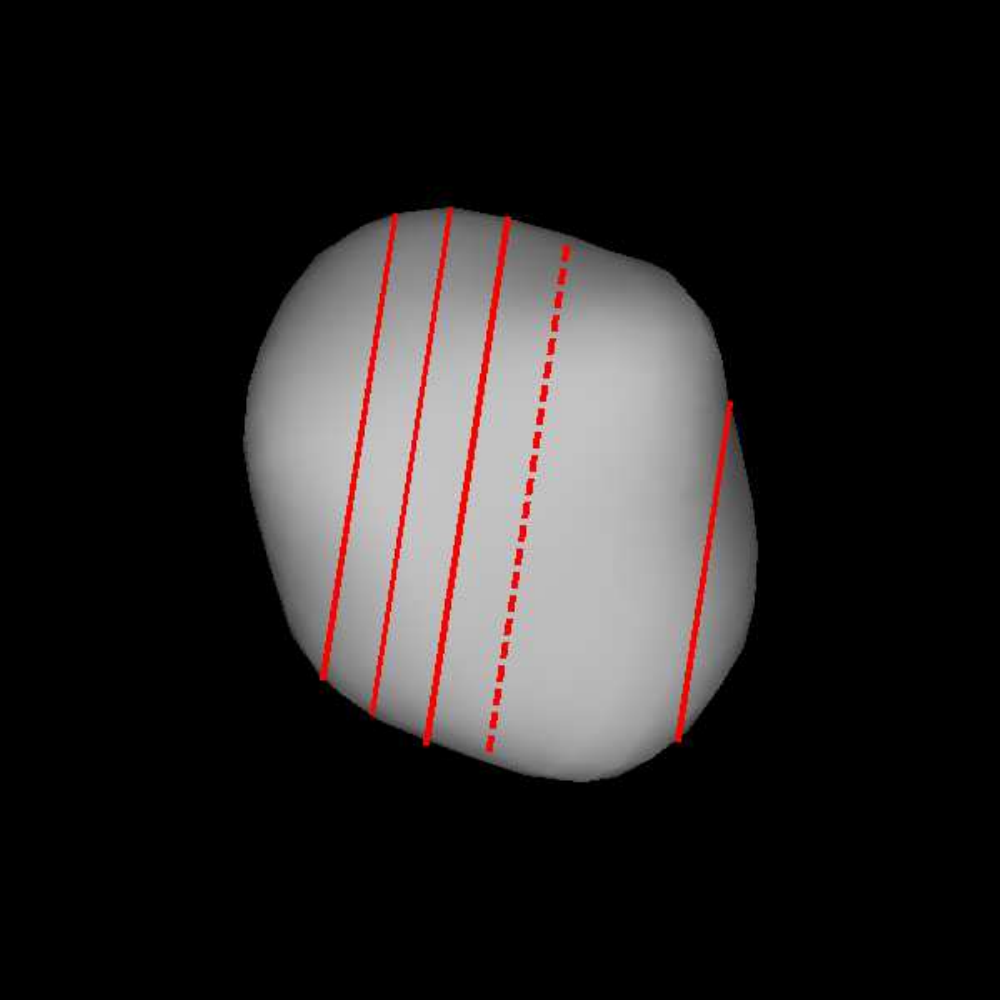}}\\
    \caption{\label{fig:165_occ}Comparison between model projections and corresponding stellar occultation(s) for asteroid (165) Loreley.}
\end{figure}

\begin{figure}[tbp]
    \centering
 \resizebox{0.33\hsize}{!}{\includegraphics{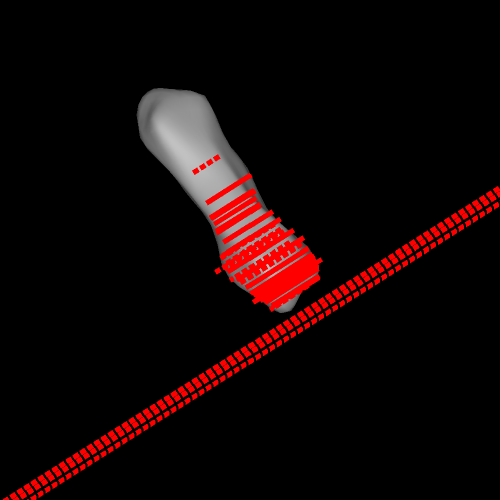}}\resizebox{0.33\hsize}{!}{\includegraphics{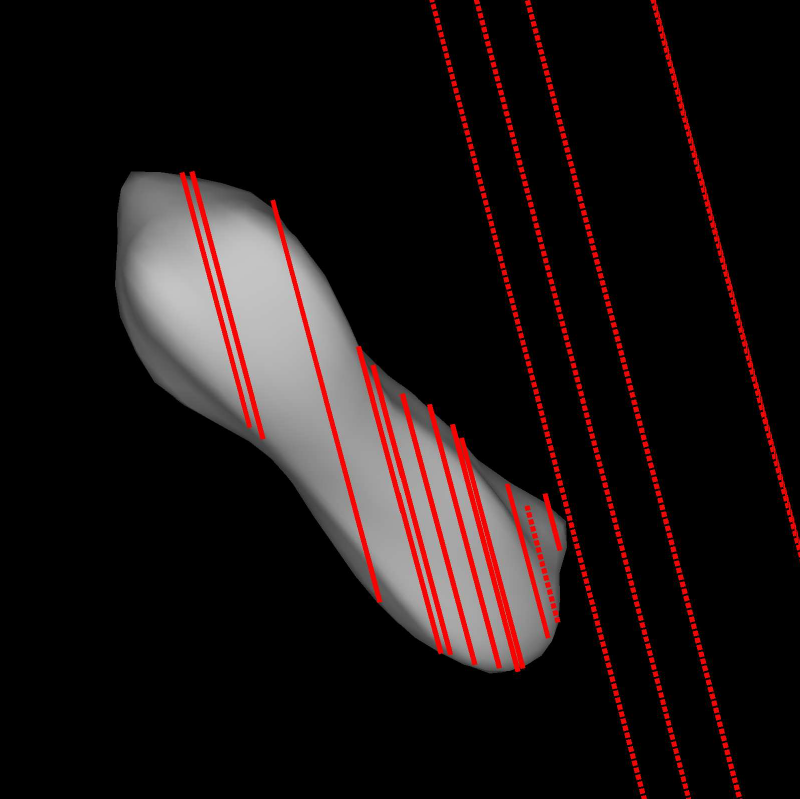}}\resizebox{0.33\hsize}{!}{\includegraphics{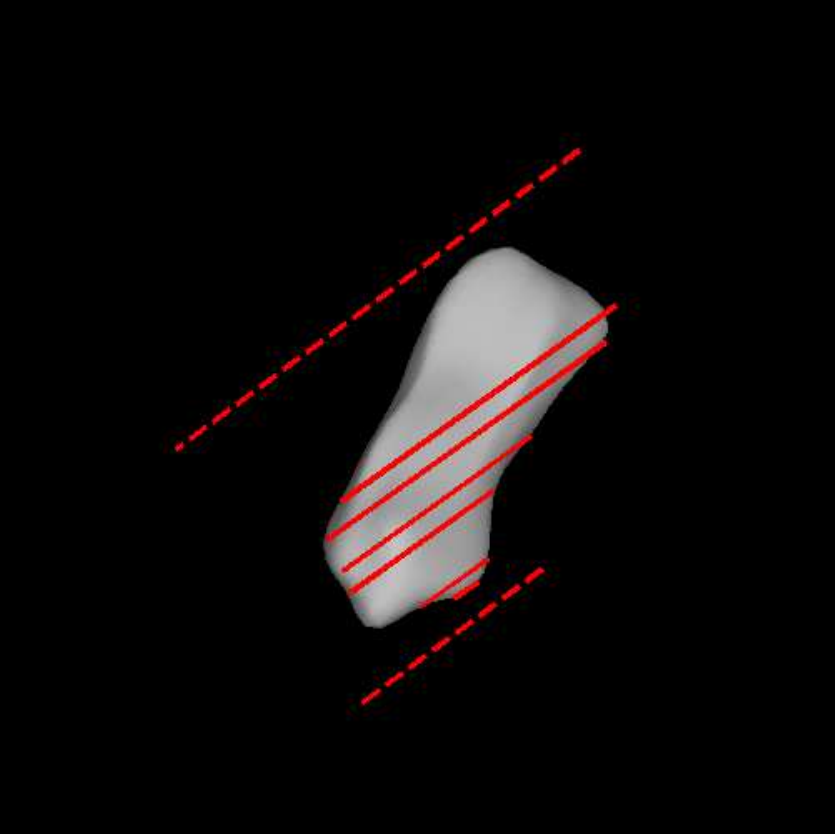}}\\
    \caption{\label{fig:216_occ}Comparison between model projections and corresponding stellar occultation(s) for asteroid (216) Kleopatra.}
\end{figure}

\begin{figure}[tbp]
    \centering
 \resizebox{0.33\hsize}{!}{\includegraphics{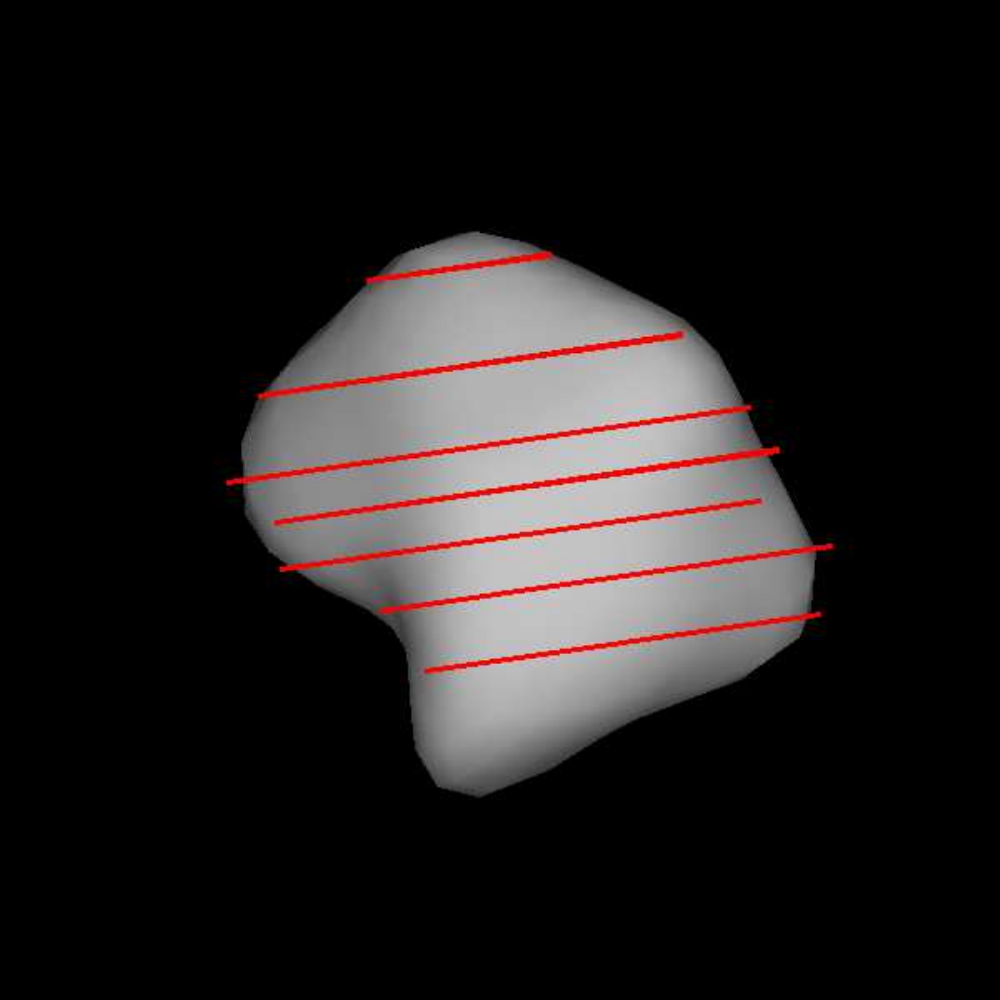}}\\
    \caption{\label{fig:233_occ}Comparison between model projections and corresponding stellar occultation(s) for asteroid (233) Asterope.}
\end{figure}

\begin{figure}[tbp]
    \centering
 \resizebox{0.33\hsize}{!}{\includegraphics{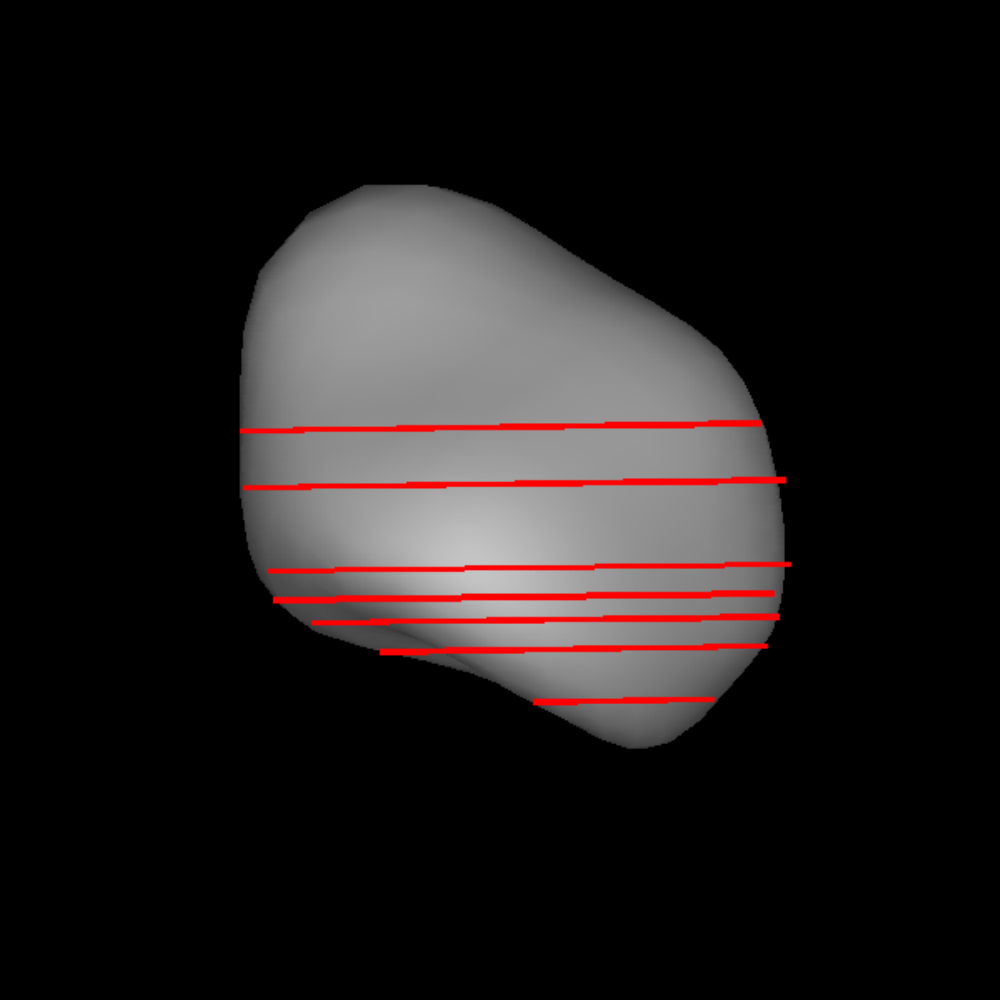}}\\
    \caption{\label{fig:360_occ}Comparison between model projections and corresponding stellar occultation(s) for asteroid (360) Carlova.}
\end{figure}

\begin{figure}[tbp]
    \centering
 \resizebox{0.33\hsize}{!}{\includegraphics{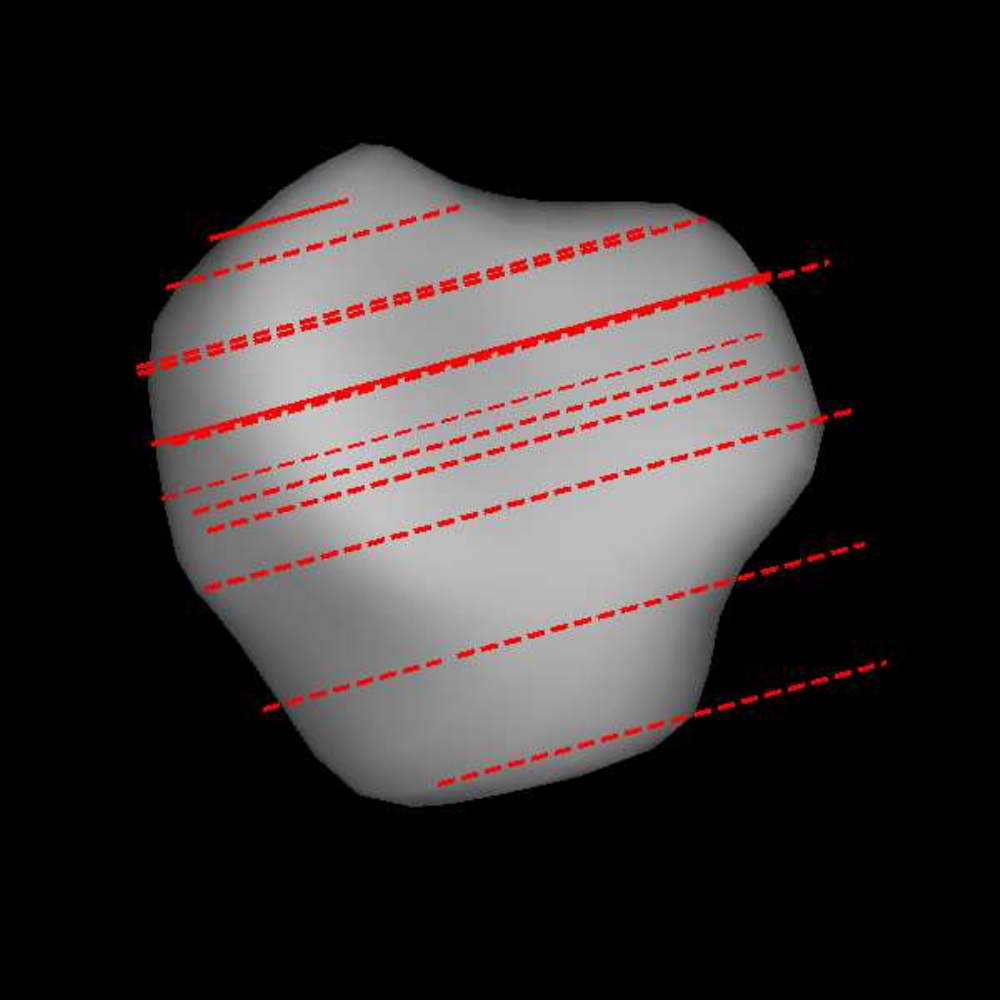}}\resizebox{0.33\hsize}{!}{\includegraphics{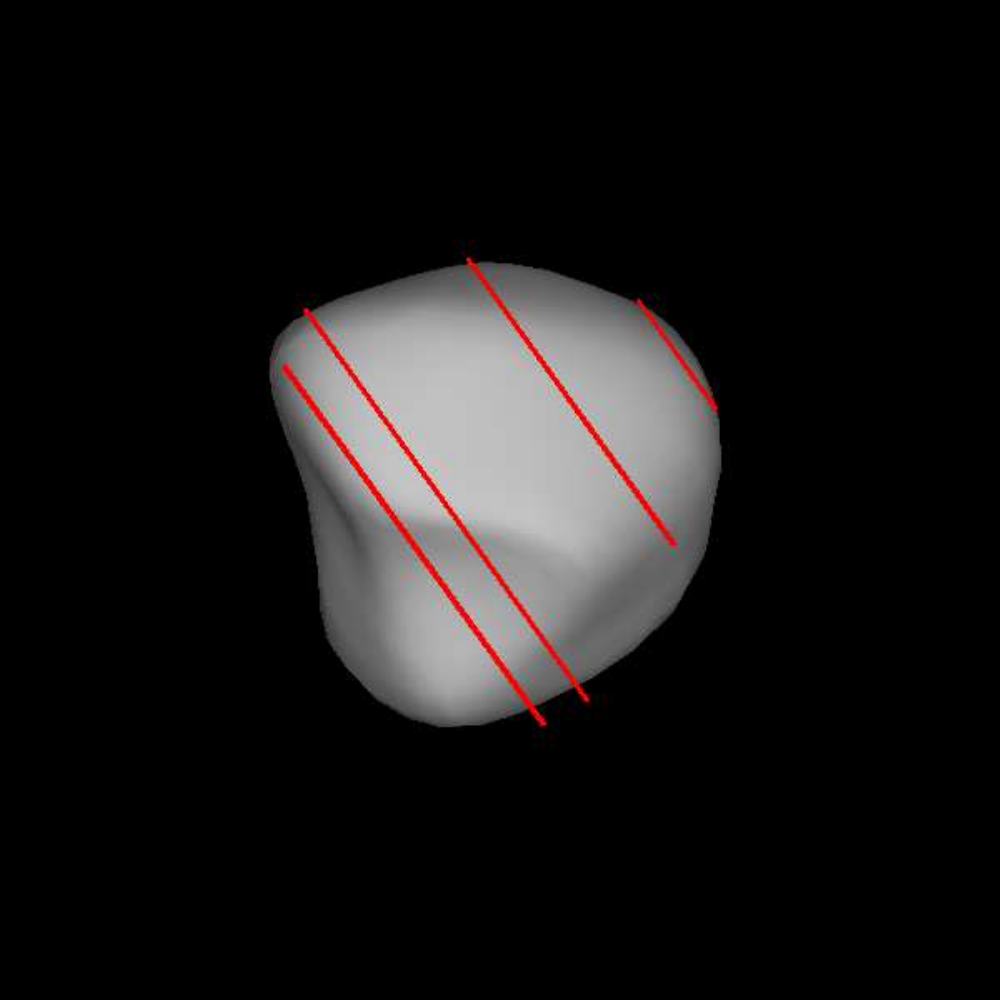}}\\
    \caption{\label{fig:386_occ}Comparison between model projections and corresponding stellar occultation(s) for asteroid (386) Siegena.}
\end{figure}

\begin{figure}[tbp]
    \centering
 \resizebox{0.33\hsize}{!}{\includegraphics{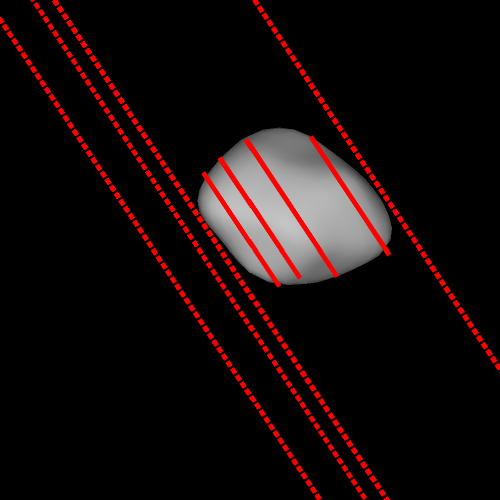}}\\
    \caption{\label{fig:387_occ}Comparison between model projections and corresponding stellar occultation(s) for asteroid (387) Aquitania.}
\end{figure}

\begin{figure}[tbp]
    \centering
 \resizebox{0.33\hsize}{!}{\includegraphics{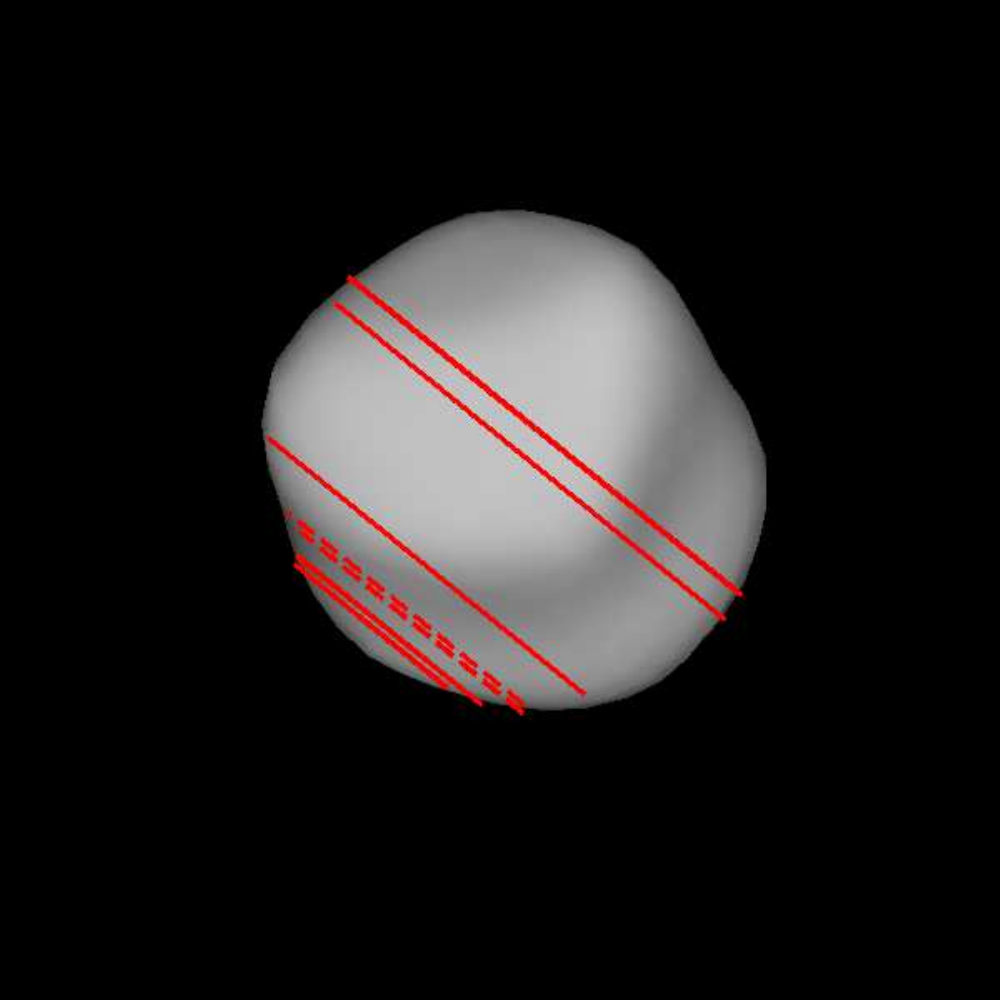}}\resizebox{0.33\hsize}{!}{\includegraphics{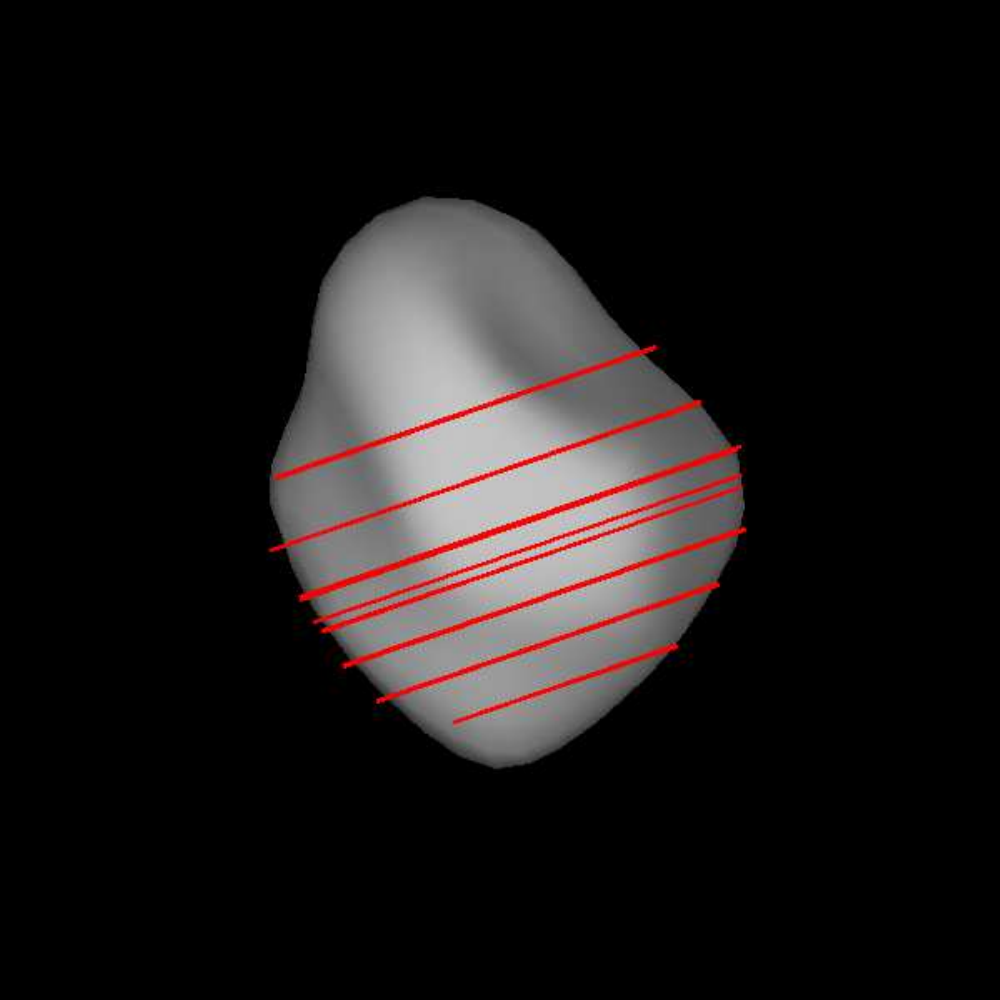}}\resizebox{0.33\hsize}{!}{\includegraphics{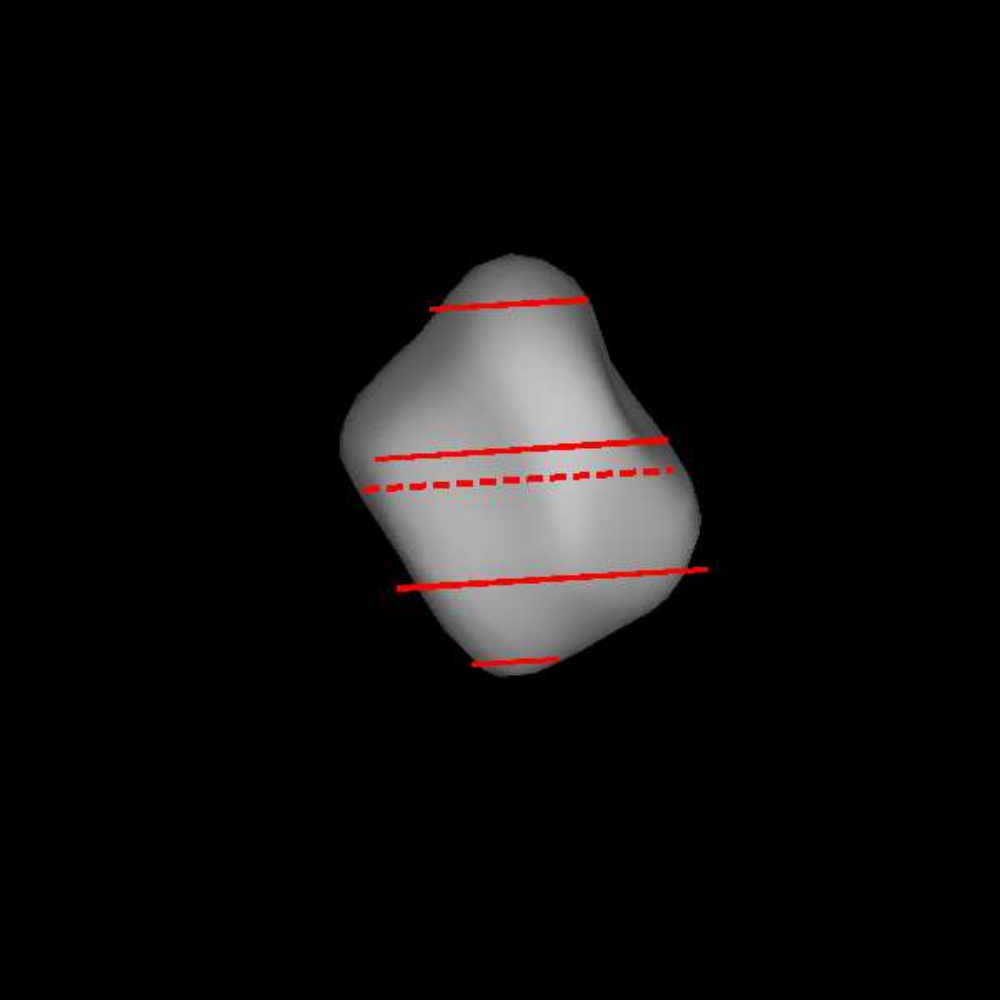}}\\
    \caption{\label{fig:409_occ}Comparison between model projections and corresponding stellar occultation(s) for asteroid (409) Aspasia.}
\end{figure}

\begin{figure}[tbp]
    \centering
 \resizebox{0.33\hsize}{!}{\includegraphics{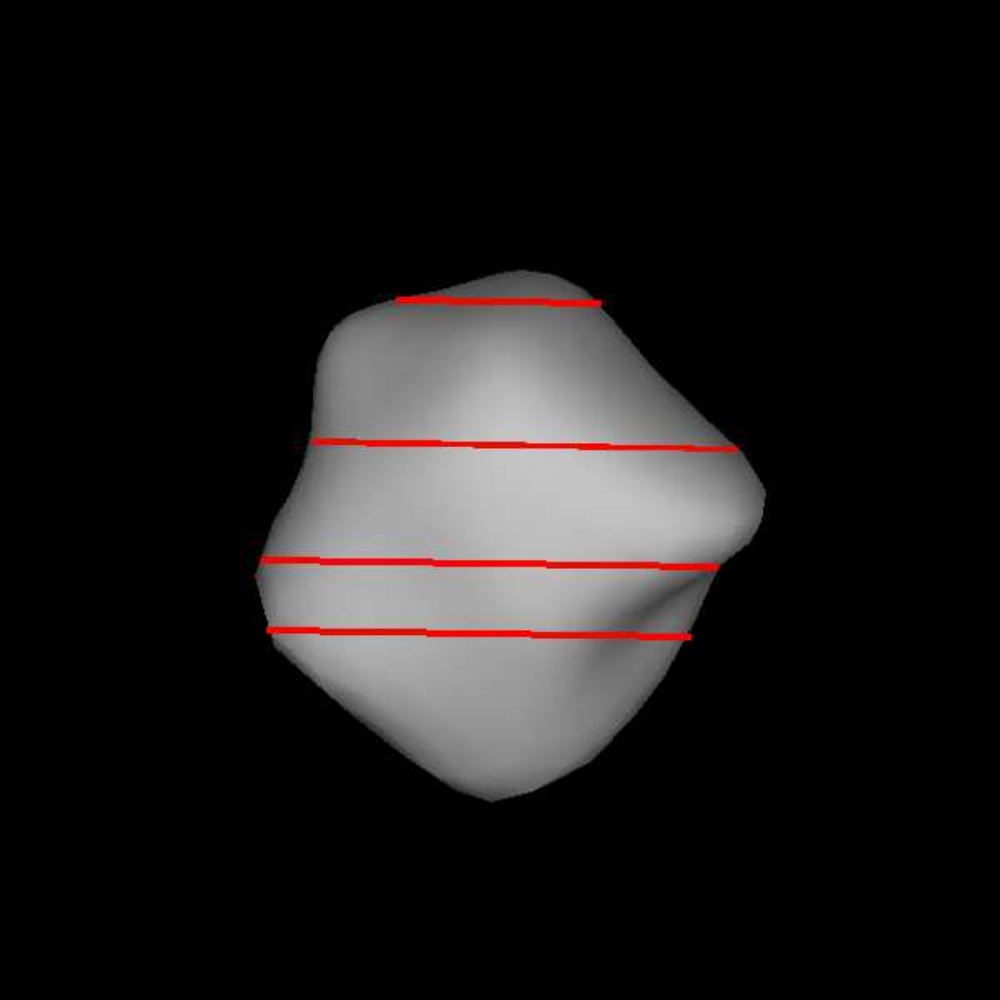}}\\
 \resizebox{0.33\hsize}{!}{\includegraphics{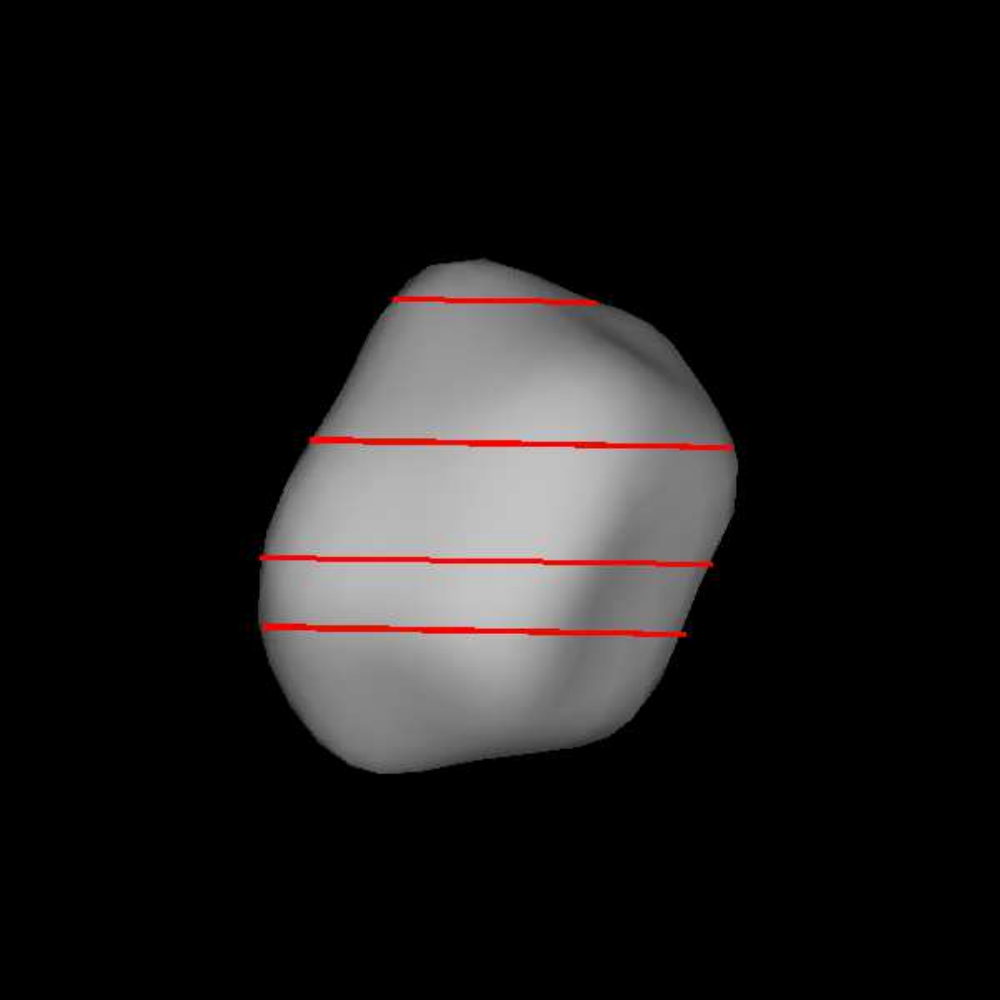}}\\
    \caption{\label{fig:419_occ}Comparison between model projections and corresponding stellar occultation(s) for asteroid (419) Aurelia. We show the fit for both pole solutions.}
\end{figure}

\begin{figure}[tbp]
    \centering
 \resizebox{0.33\hsize}{!}{\includegraphics{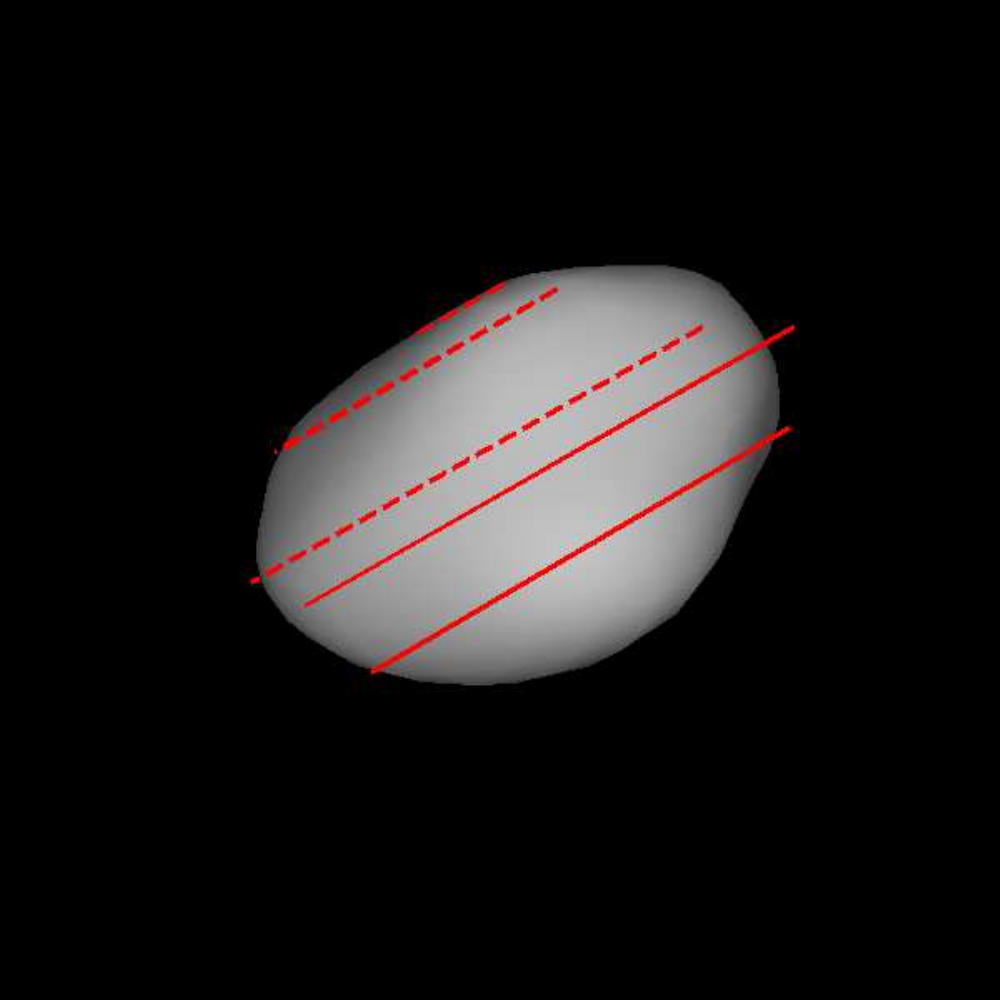}}\resizebox{0.33\hsize}{!}{\includegraphics{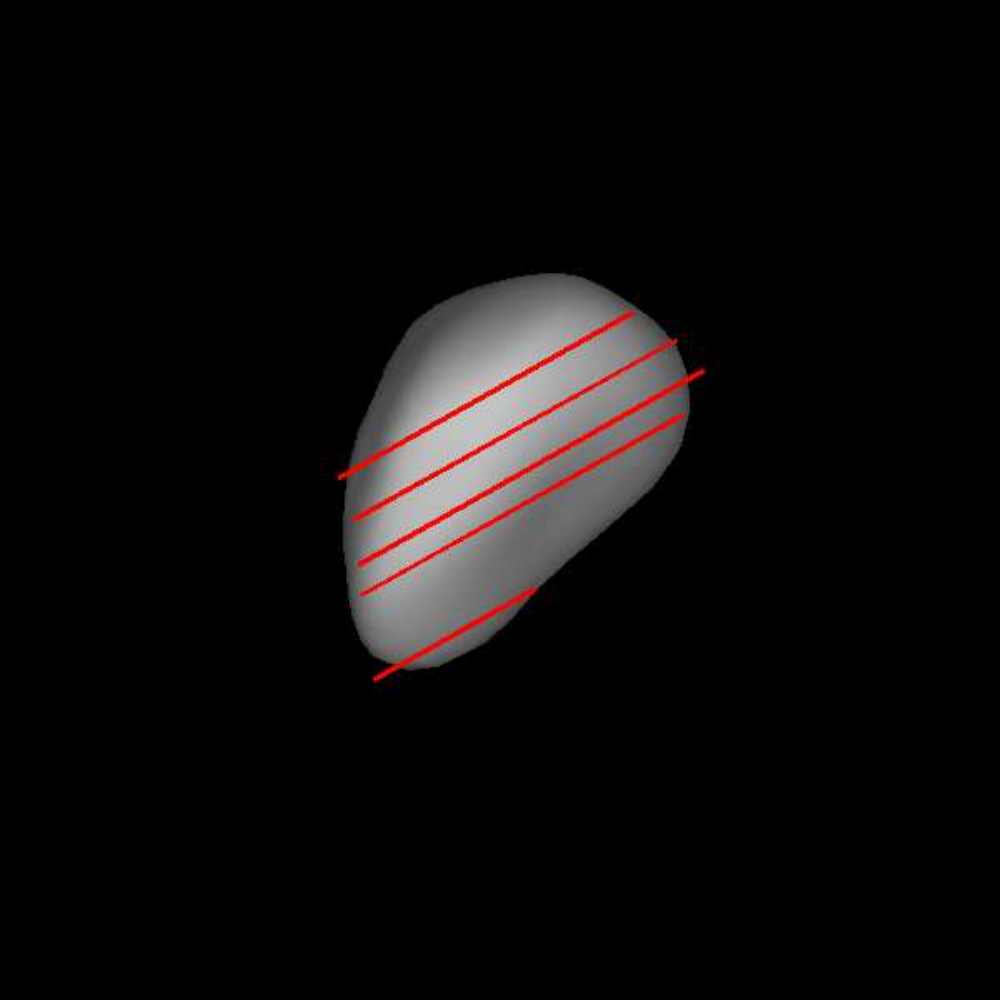}}\\
    \caption{\label{fig:471_occ}Comparison between model projections and corresponding stellar occultation(s) for asteroid (471) Papagena.}
\end{figure}

\begin{figure}[tbp]
    \centering
 \resizebox{0.33\hsize}{!}{\includegraphics{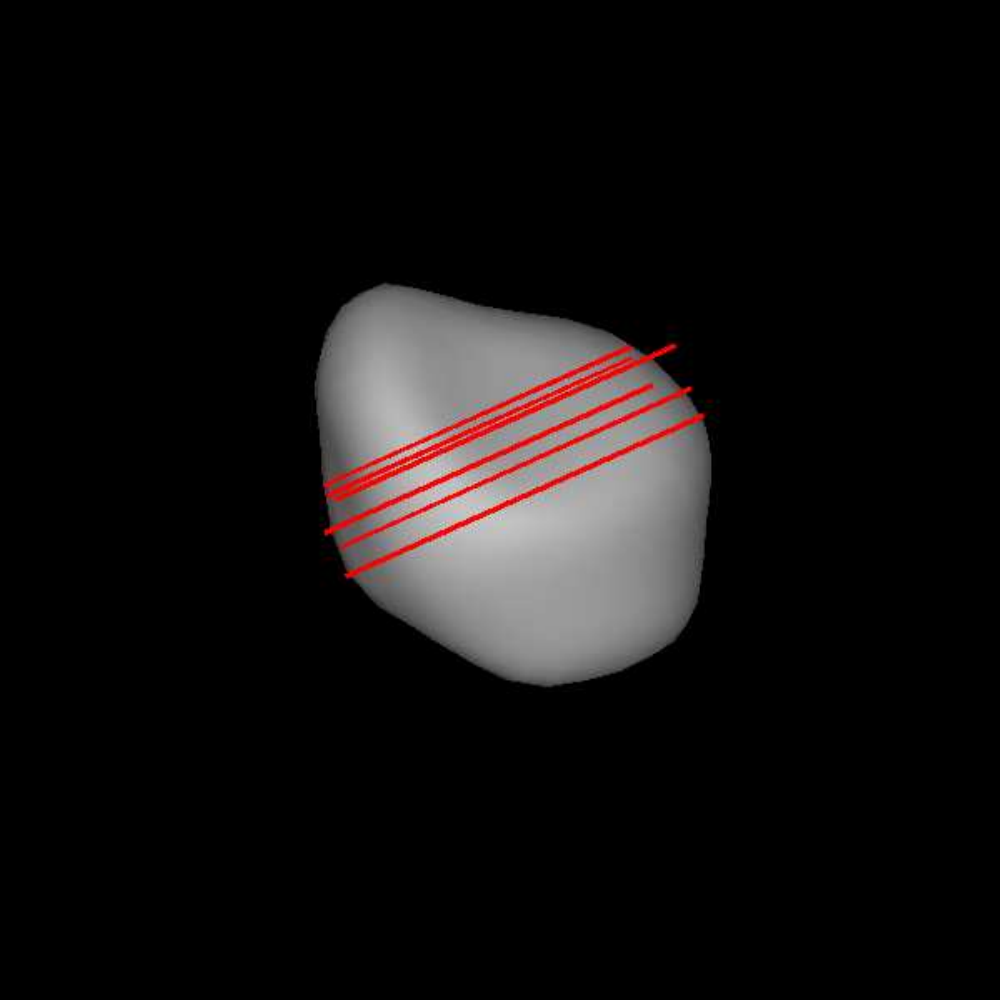}}\\
    \caption{\label{fig:532_occ}Comparison between model projections and corresponding stellar occultation(s) for asteroid (532) Herculina.}
\end{figure}

\begin{figure}[tbp]
    \centering
 \resizebox{0.33\hsize}{!}{\includegraphics{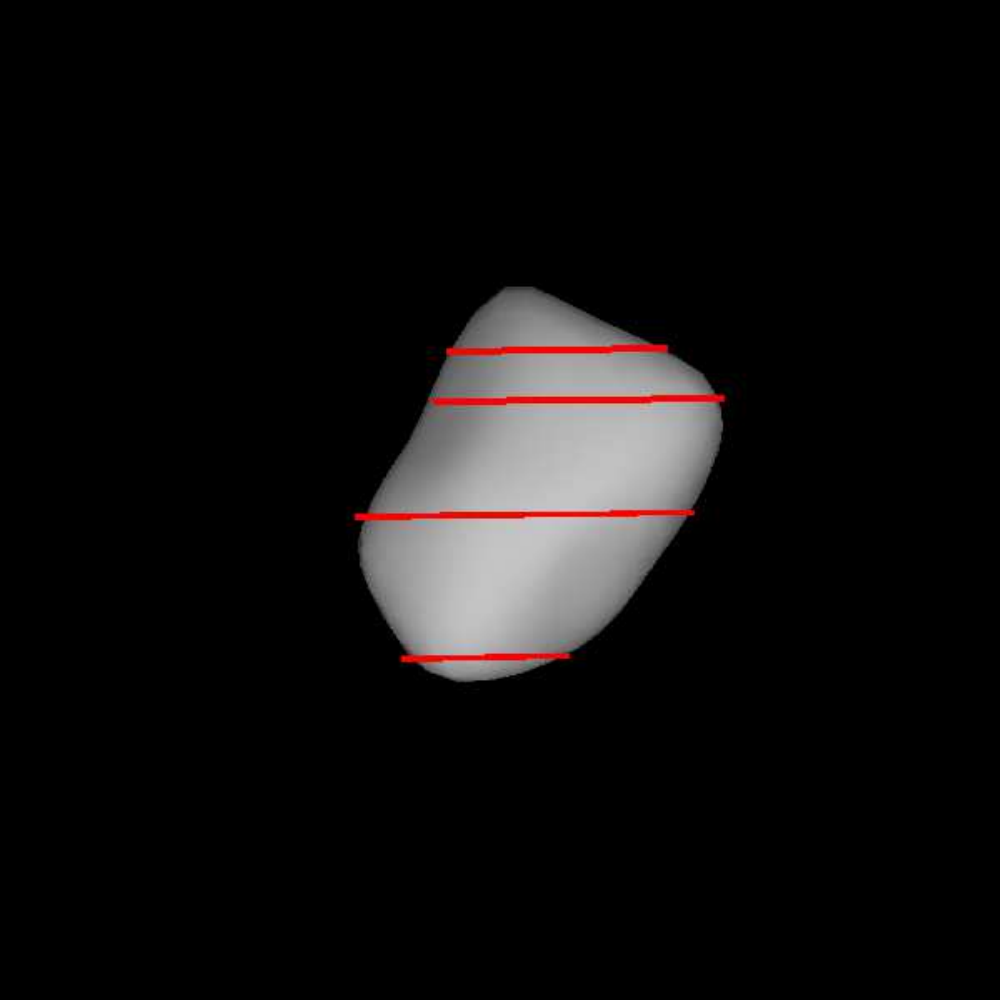}}\resizebox{0.33\hsize}{!}{\includegraphics{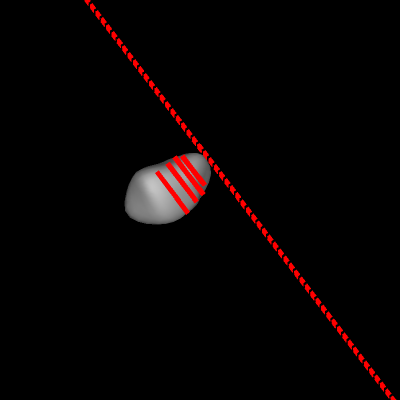}}\\
    \caption{\label{fig:849_occ}Comparison between model projections and corresponding stellar occultation(s) for asteroid (849) Ara.}
\end{figure}

\end{appendix}

\end{document}